\RequirePackage{fix-cm}
\documentclass[natbib,smallextended]{svjour3}       
\smartqed  
\usepackage{graphicx}

\usepackage{amsfonts}
\usepackage{amsmath}
\usepackage{amssymb}
\usepackage{bm}
\usepackage{subcaption}
\usepackage{booktabs}
\usepackage{mathrsfs}
\usepackage{ytableau}

\usepackage[colorlinks=true,citecolor=blue,urlcolor=blue,breaklinks]{hyperref}

\newcommand{\dd}{\mathrm{d}}
\newcommand{\de}{\mathrm{e}}
\newcommand{\dD}{\mathrm{D}}
\newcommand{\di}{\mathrm{i}}

\newcommand{\dI}{\mathrm{I}}
\newcommand{\dJ}{\mathrm{J}}
\newcommand{\dK}{\mathrm{K}}
\newcommand{\dL}{\mathrm{L}}
\newcommand{\dM}{\mathrm{M}}
\newcommand{\dE}{\mathrm{E}}
\newcommand{\dF}{\mathrm{F}}
\newcommand{\dMbar}{\overline{\mathrm{M}}}
\newcommand{\dS}{\mathrm{S}}
\newcommand{\dSbar}{\overline{\mathrm{S}}}
\newcommand{\dU}{\mathrm{U}}
\newcommand{\dV}{\mathrm{V}}
\newcommand{\dW}{\mathrm{W}}
\newcommand{\dX}{\mathrm{X}}
\newcommand{\dY}{\mathrm{Y}}
\newcommand{\dZ}{\mathrm{Z}}
\newcommand{\dP}{\mathrm{P}}
\newcommand{\dG}{\mathrm{G}}
\newcommand{\dQ}{\mathrm{Q}}
\newcommand{\dO}{\mathrm{O}}
\newcommand{\dC}{\mathrm{C}}

\newcommand\calO{\mathcal{O}}

\newcommand{\FP}{\mathop{\mathrm{FP}}_{B=0}}
\newcommand{\FPprop}{\mathop{\mathrm{FP}}_{B=0}\Box^{-1}_\text{ret}}

\newcommand{\Pf}{\mathop{\mathrm{Pf}}_{s_1, s_2}}
\newcommand{\pf}{\mathrm{Pf}}

\newcommand{\ua}{^{\alpha}}

\newcommand{\la}{_{\alpha}}

\newcommand{\ab}{^{\alpha\beta}}

\newcommand{\nn}{\nonumber}

\allowdisplaybreaks

\journalname{Living Reviews in Relativity}

\begin{document}

\title{Post-Newtonian theory for gravitational waves%
  \thanks{This article is a revised version of \url{https://doi.org/10.12942/lrr-2014-2}.\\
    \textbf{Change summary} Major revision, updated and expanded.\\
    \textbf{Change details} The main improvements, rewriting, and addition of new subsections have been done in Sects.~\ref{sec:intro}, \ref{sec:asympGW}, \ref{sec:EOM}, and \ref{sec:GW}. The number of total references has increased from 428 to 602.
    }
}


\author{Luc Blanchet}

\institute{L. Blanchet \at 
        Institut d'Astrophysique de Paris \\
         98bis Boulevard Arago \\
         75014 Paris, France \\
        \email{luc.blanchet@iap.fr}
}


\date{Received: date / Accepted: date}

\maketitle


\begin{abstract}
To be observed and analyzed by the network of current gravitational wave detectors (LIGO, Virgo, KAGRA), and in anticipation of future third generation ground based (Einstein Telescope, Cosmic Explorer) and space borne (LISA) detectors, inspiralling compact binaries -- binary star systems composed of neutron stars and/or black holes in their late stage of evolution prior the final coalescence -- require high-accuracy predictions from general relativity. The orbital dynamics and emitted gravitational waves of these very relativistic systems can be accurately modelled using state-of-the-art post-Newtonian theory. In this article we review the multipolar-post-Minkowskian approximation scheme, merged to the standard post-Newtonian expansion into a single formalism valid for general isolated matter system. This cocktail of approximation methods (called MPM-PN) has been successfully applied to compact binary systems, producing equations of motion up to the fourth-post-Newtonian (4PN) level, and gravitational waveform and flux to 4.5PN order beyond the Einstein quadrupole formula. We describe the dimensional regularization at work in such high post-Newtonian calculations, for curing both ultra-violet and infra-red divergences. Several landmark results are detailed: the definition of multipole moments, the gravitational radiation reaction, the conservative dynamics of circular orbits, the first law of compact binary mechanics, and the non-linear effects in the gravitational wave propagation (tails, iterated tails and non-linear memory). We also discuss the case of compact binaries moving on eccentric orbits, and the effects of spins (both spin-orbit and spin-spin) on the equations of motion and gravitational wave energy flux and waveform.
\end{abstract}

\setcounter{tocdepth}{3}
\tableofcontents


\section{Physical introduction}
\label{sec:intro}

The theory of gravitational radiation from isolated sources, is a fascinating science that can be explored by means of the analytical (i.e., mathematical) resolution of partial differential equations. Indeed, the \cite{E15a} field equations of general relativity (GR), when use is made of the harmonic-coordinate conditions, take the form of a quasi-linear hyperbolic differential system of equations, involving the \cite{Dalembert} wave operator. The resolution of that system of equations constitutes a \emph{probl\`eme bien pos\'e} in the sense of \cite{Hadamard, YCB62}, and which is amenable to an analytic solution using approximation methods.

Nowadays, the importance of the field lies in the exciting comparison of the theory with contemporary astrophysical observations, of binary pulsars like the historical \cite{HulseTaylor} pulsar PSR~1913+16, and of gravitational waves produced by massive and rapidly evolving systems of neutron stars and black holes. These are now routinely analyzed on Earth by the world-wide network of large-scale laser interferometers, comprising the interferometers LIGO, Virgo and KAGRA (with the smaller detector GEO in support). The first direct detection of the coalescence of two black holes has been achieved on September~14 by the advanced LIGO detectors \citep{GW150914}. Since 2017 joint observations were made by both LIGO and Virgo (with thus a good sky localization of the events), including the historical event of the merger of two neutron stars detected on August~17, 2017 \citep{GW170817}, which constituted the start of multi messenger astronomy with gravitational waves, with the almost simultaneous detection of a $\gamma$-ray burst by the satellites Fermi and Integral \citep{LIGOmultmess}. Further ahead, the space-based laser interferometer detector LISA -- an outstanding mission of the European Space Agency to be launch in 2034 -- should be able to detect the collision and merger of supermassive black-hole binaries at cosmological distances, and learning about the formation of large scale structure, stellar evolution and the early Universe.

To prepare these experiments, the required theoretical work consists of carrying out a sufficiently general solution of the Einstein field equations, valid for a large class of matter systems, and describing the physical processes of the emission and propagation of the gravitational waves from the source to the distant detector, as well as their back-reaction onto the source. The solution should then be applied to specific sources like inspiralling compact binaries -- i.e. binary sources loosing energy by gravitational waves prior to merger.

For general sources it is hopeless to solve the problem via a rigorous deduction within the exact theory of general relativity, and we have to resort to approximation methods. Of course the ultimate aim of approximation methods is to extract from the theory some firm predictions to be compared with the outcome of experiments. However, we have to keep in mind that such methods are often missing a clear theoretical framework and are sometimes not related in a very precise mathematical way to the first principles of the theory \citep{FS83, Rend92}. 

The flagship of approximation methods is the \emph{post-Newtonian} approximation, which has been developed from the early days of general relativity by \cite{Droste17, LD17}. This approximation is at the origin of many of the great successes of general relativity, and it gives wonderful answers to the problems of motion and gravitational radiation of systems of compact objects. Three crucial applications are:
\begin{enumerate}
\item The motion of point-like objects at the first post-Newtonian approximation level \citep{EIH}, is routinely taken into account to describe the solar system dynamics (motion of the centers of mass of planets);
\item The gravitational radiation-reaction force, which appears in the equations of motion at the second-and-a-half post-Newtonian (2.5PN) order \citep{DD81a, DD81b, D83, Dhouches}, has been experimentally verified by the observation of the secular acceleration of the orbital motion of the Hulse--Taylor binary pulsar \citep{TFMc79, TW82, T93} and for instance the double pulsar \citep{KW09};
\item The analysis of gravitational waves emitted by inspiralling compact binaries -- two neutron stars or black holes driven into coalescence by emission of gravitational radiation -- necessitates the prior knowledge of the equations of motion and radiation field up to very high post-Newtonian order \citep{3mn,CF94}. 
\end{enumerate}
This article reviews the current status of the post-Newtonian approach to the problems of the motion of inspiralling compact binaries and their emission of gravitational waves. When the two compact objects approach each other toward merger, the post-Newtonian expansion will lose accuracy and should be taken over by numerical-relativity computations \citep{Pret05, Camp06, Bak06}. We refer to review articles \citep{GN09, FR12, LP14, DZ18} for numerical-relativity methods. Despite very intensive developments of numerical relativity, the post-Newtonian approximation remains indispensable for describing the inspiral phase of compact binaries to high accuracy, and for providing a powerful benchmark against which the numerical computations are tested.

The Sect.~\ref{sec:PNsource} of the article deals with general post-Newtonian matter sources. The exterior field of the source is investigated by means of a combination of analytic multipolar and post-Minkowskian (MPM) approximations. The physical observables in the far-zone of the source are described by a specific set of radiative multipole moments. By matching the exterior solution to the metric of the post-Newtonian (PN) source in the near-zone the explicit expressions of the source multipole moments are obtained. The relationships between the radiative and source moments involve many non-linear multipole interactions, among them those associated with the tails (and tails-of-tails, etc.) of gravitational waves. This is the so-called MPM-PN formalism.

The Sect.~\ref{sec:compactbinary} is devoted to the application to compact binary systems, with particular emphasis on black hole binaries with spins. We present the equations of binary motion, and the associated Lagrangian and Hamiltonian, at the fourth post-Newtonian (4PN) order beyond the Newtonian acceleration. The gravitational-wave energy flux, taking consistently into account the relativistic corrections in the binary's moments as well as the various tail effects, is derived through 4PN order with respect to the quadrupole formalism. The binary's orbital phase, whose prior knowledge is crucial for searching and analyzing the signals from inspiralling compact binaries, is deduced from an energy balance argument at the 4.5PN order (in the simple case of circular orbits).


\subsection{Analytic approximations and wave generation formalism}
\label{sec:analapprox}

\subsubsection{Cocktail of approximation methods}
\label{sec:cocktail}

The basic problem we face is to relate the asymptotic gravitational-wave form $h_{ij}$ generated by some isolated source (in a suitable asymptotic coordinate system), at the location of a detector in the wave zone of the source, to the material content of the source, i.e., its stress-energy tensor $T^{\alpha\beta}$, using approximation methods in general relativity.\footnote{In this article Greek indices $\alpha\beta\cdots\mu\nu\cdots$ take space-time values 0, 1, 2, 3 and Latin indices $ab\cdots ij\cdots$ spatial values 1, 2, 3. Cartesian coordinates are assumed throughout and boldface notation is often used for ordinary Euclidean vectors. Our signature is +2; hence the Minkowski metric reads $\eta_{\alpha\beta} = \text{diag}(-1,+1,+1,+1)$. As usual $G$ and $c$ are Newton's constant and the speed of light, consistently included unless otherwise specified.} Therefore, a general wave-generation formalism must solve the field equations, and the non-linearity therein, by imposing some suitable approximation series in one or several small physical parameters. Some important approximations that we shall use in this article are the post-Newtonian method (or non-linear $1/c$-expansion), the post-Minkowskian method or non-linear iteration ($G$-expansion), the multipole decomposition in irreducible representations of the rotation group (or equivalently $a$-expansion in the source radius), the far-zone expansion ($1/R$-expansion in the distance to the source), and the perturbation in the small mass limit ($\nu$-expansion in the mass ratio of a binary system). In particular, the post-Newtonian expansion has provided us in the past with our best insights into the problems of motion and radiation. The most successful wave-generation formalisms make a \emph{gourmet} cocktail of these approximation methods. For reviews on analytic approximations and applications to the motion and the gravitational-wave generation see \cite{Thhouches, Dhouches, Dcargese, D300, Th300, W94, Bhouches, Borleans, Schaferorleans, Maggiore, BuonSathya15}. For reviews on black-hole pertubations and the self-force approach see \cite{PoissonLR, SasakiLR, Detweilerorleans, Barackorleans, BP18}.

The post-Newtonian approximation is valid under the assumptions of a weak gravitational field inside the source (we shall see later how to model neutron stars and black holes), and of slow internal motions.\footnote{Establishing the post-Newtonian expansion rigorously has been the subject of numerous mathematically oriented works, see e.g., \cite{Rend90, Rend92, Rend94}.} The main problem with this approximation, is its domain of validity, which is limited to the near zone of the source -- the region surrounding the source that is of small extent with respect to the wavelength of the gravitational waves. A serious consequence is the \emph{a priori} inability of the post-Newtonian expansion to incorporate the boundary conditions at infinity, which determine the radiation reaction force in the source's local equations of motion.

The post-Minkowskian expansion, by contrast, is uniformly valid, as soon as the source is weakly self-gravitating, over all space-time. In a sense, the post-Minkowskian method is more fundamental than the post-Newtonian one; it can be regarded as an ``upstream'' approximation with respect to the post-Newtonian expansion, because each coefficient of the post-Minkowskian series can in turn be re-expanded in a post-Newtonian fashion. Therefore, a way to take into account the boundary conditions at infinity in the post-Newtonian series is to control \emph{first} the post-Minkowskian expansion. Notice that the post-Minkowskian method is also upstream (in the previous sense) with respect to the multipole expansion, when considered outside the source, and with respect to the far-zone expansion, when considered far from the source.

The most ``downstream'' approximation that we shall use in this article is the post-Newtonian one; therefore this is the approximation that dictates the allowed physical properties of our matter source. We assume mainly that the source is at once \emph{slowly moving} and \emph{weakly stressed}, and we abbreviate this by saying that the source is \emph{post-Newtonian}. For post-Newtonian sources, the parameter defined from the components of the matter stress-energy tensor $T^{\alpha\beta}$ and the source's Newtonian potential $U$ by
\begin{equation}\label{epsPN}
  \epsilon \equiv \max \left\{\left|\frac{T^{0i}}{T^{00}}\right|,
  \left|\frac{T^{ij}}{T^{00}}\right|^{1/2}\!\!,
  \left|\frac{U}{c^2}\right|^{1/2}\right\}\,,
\end{equation}
is much less than one. This parameter represents essentially a slow motion estimate $\epsilon \sim v/c$, where $v$ denotes a typical internal velocity. By a slight abuse of notation, following, we shall henceforth write formally $\epsilon\equiv 1/c$, even though $\epsilon$ is dimensionless whereas $c$ has the dimension of a velocity. Thus, $1/c \ll 1$ in the case of post-Newtonian sources. The small post-Newtonian remainders will be denoted $\calO(1/c^n)$. Furthermore, following \cite{C65}, we shall refer to a small post-Newtonian term with formal order $\calO(1/c^n)$ relative to the Newtonian acceleration in the equations of motion, as being ``$\frac{n}{2}$PN''. 

We have $\vert U/c^2\vert^{1/2} \ll 1/c$ for sources with negligible self-gravity, and whose dynamics are therefore driven by non-gravitational forces. However, we shall generally assume that the source is self-gravitating; in that case we see that it is necessarily \emph{weakly} (but not negligibly) self-gravitating, i.e.,  $\vert U/c^2\vert^{1/2}=\calO(1/c)$.\footnote{Note that in the case of Newtonian binary systems, for high eccentricities (with say $e\to 1^{-}$) the Newtonian potential $U$ can be numerically much larger than the estimate $\calO(1/c^2)\sim v^2/c^2$ around the apastron of the orbit.} Note that the adjective ``slow-motion'' is a bit clumsy because we shall in fact consider \emph{very} relativistic sources such as inspiralling compact binaries, for which $v/c$ can be as large as 50\% in the last rotations, and whose description necessitates the control of high post-Newtonian approximations.

At the lowest-order in the Newtonian limit $1/c\to 0$, the gravitational waveform of a post-Newtonian matter source is generated by the time variations of the quadrupole moment of the source. We shall review in Sect.~\ref{sec:quadform} the utterly important ``Newtonian'' quadrupole moment formalism of \cite{E18, LL}. Taking into account higher post-Newtonian corrections in a wave generation formalism will mean including into the waveform the contributions of higher multipole moments, beyond the mass quadrupole moment. Post-Newtonian corrections of order $\calO(1/c^n)$ beyond the quadrupole formalism will still be denoted as $\frac{n}{2}$PN. But notice that the quadrupole formalism corresponds itself to a small radiation reaction effect of order 2.5PN $=\calO(1/c^5)$ relative to the Newtonian acceleration. The lesson here is that building a post-Newtonian wave generation formalism must be concomitant to understanding the multipole expansion in general relativity.

The multipole expansion is one of the most useful tools of physics, but its use in general relativity is difficult because of the non-linearity of the theory and the tensorial character of the gravitational interaction. In the stationary case, the multipole moments are determined by the expansion of the metric at spatial infinity \citep{G70, H74, SB83}, while, in the case of non-stationary fields, the moments, starting with the quadrupole, are defined at future null infinity. The multipole moments have been extensively studied in the linearized theory, which ignores the gravitational forces inside the source. Early studies have extended the Einstein quadrupole formula [given by Eq.~\eqref{fluxE} below] to include the current-quadrupole and mass-octupole moments \citep{Papa62, Papa71}, and obtained the corresponding formulas for linear momentum \citep{Papa62, Papa71, Bek73, Press77} and angular momentum \citep{Pe64, CB69}. The general structure of the infinite multipole series in the linearized theory was investigated by several works \citep{SB58, S61, Pi64, Th80}, from which it emerged that the expansion is characterized by two and only two sets of moments: Mass-type and current-type moments. Below we shall use a particular multipole decomposition of the linearized (vacuum) metric, parametrized by symmetric and trace-free (STF) mass and current moments, as given by \cite{Th80}. The expressions of the multipole moments, valid in the linearized theory but irrespective of a slow motion hypothesis, have been worked out by \cite{M62, CM71, CMM77} and culminated with \cite{DI91b} obtaining the closed-form expressions of the time-dependent STF mass and current multipole moments as integrals over the source in linearized gravity.

In the full non-linear theory, the (radiative) multipole moments can be read off the coefficient of $1/R$ in the expansion of the metric when $R\to +\infty$, with a null coordinate $U = \mathrm{const}$. The solutions of the field equations in the form of a far-field expansion (power series in $1/R$) have been constructed, and their properties elucidated, by \cite{BBM62} and \cite{Sachs62}. The precise way under which such radiative space-times fall off asymptotically has been formulated geometrically by \cite{P63, P65, GH78} in the concept of an asymptotically simple space-time. The interest of this structure of the asymptotic field arises also from the fact that it is preserved under an infinite set of residual symmetries, the Bondi-Metzner-Sachs (BMS) group, generated by supertranslations and arbitrary diffeomorphisms on the two-sphere at infinity (see \citealt{ACL18} for a review). The Bondi-Sachs-Penrose approach is very powerful, but it can answer \emph{a priori} only a part of the problem, because it gives information on the field only in the limit where $R\to +\infty$, which cannot be connected in a direct way to the actual matter content and dynamics of the source. In particular the multipole moments that one considers in this approach are those measured at infinity -- we call them the \emph{radiative} multipole moments. These moments are distinct, because of non-linearities, from some more natural \emph{source} multipole moments, which are operationally given by explicit integrals extending over the matter and gravitational field of the source.

The problem of non-linearities can be attacked by the post-Minkowskian expansion, valid at once inside and outside the source \citep{ThK75, CTh77}. This approximation has been developed in the pionneering work of \cite{BertottiP60}. \cite{LSB08} obtained a closed form expression for the Hamiltonian of $N$ particles in the first post-Minkowskian (1PM) approximation (see also \citealt{BFok18}). The equations of motion of systems of particles were derived up to 2PM order by \cite{WG79, WH80, BeDD81, Westpf85}. An important revival of the post-Minkowskian approximation took place recently using scattering amplitudes and the effective field theory (EFT): \cite{Bern19a, Bern19b} derived the 3PM corrections to the conservative two-body Hamiltonian for spinless compact objects; this was extended to 4PM order by \cite{Bern21, DKLP22} (but for the potential modes only). The problem of gravitational radiation has also progressed using the EFT post-Minkowskian method \citep{MRV21, Brandhuber23, Herderschee23, elkhidir2023, Georgoudis2023}. The amplitude-based derivation of the gravitational waveform generated by the scattering of two scalar black holes has been successfully compared to the result of the MPM-PN formalism up to 2.5PN order \citep{BDG23, BDDGH24, GHR24a, GHR24b}.

A different way of defining the multipole expansion within the complete non-linear theory was proposed by \cite{BD86, B87}, following works by \cite{Bo59, BoR61, BoR66, HR69} and \cite{Th80}. In this approach the basic multipole moments are the \emph{source} moments, rather than the radiative ones.\footnote{An alternative definition of the source multipole moments, in terms of canonical Noether charges associated with residual gauge transformations in harmonic coordinates, has been proposed by \cite{COS18}.} In a first stage, the moments are left unspecified, as being some arbitrary functions of time, supposed to describe an actual physical source. They are iterated by means of a post-Minkowskian expansion of the vacuum field equations (valid in the source's exterior). Technically, the post-Minkowskian approximation scheme is greatly simplified by the assumption of a multipolar expansion, as one can consider separately the iteration of the different multipole pieces composing the exterior field. In this multipolar-post-Minkowskian (MPM) formalism, which is physically valid over the entire weak-field region outside the source, and in particular in the wave zone (going up to future null infinity), the radiative multipole moments are obtained in the form of some non-linear functionals of the more basic source moments. \emph{A priori}, the method is not limited to post-Newtonian sources; however, in practice, the \emph{closed-form} expressions of the source multipole moments can be established only in the case where the source is post-Newtonian \citep{BD89, B95, B98mult}. The reason is that in this case the domain of validity of the post-Newtonian iteration (viz. the near zone) overlaps the exterior weak-field region, so that there exists an intermediate zone in which the post-Newtonian and multipolar expansions can be matched together. This is an application of the method of matched asymptotic expansions in general relativity \citep{BuTh70, Bu71, AKKM82, PB02, BFN05}. The previous MPM-PN construction, i.e., the MPM expansion and its asymptotic matching to the near zone of the PN source, constitutes a powerful gravitational-wave generation formalism for arbitrary isolated sources, able to take into account, in principle, any post-Newtonian correction in the waveform generated by the source. The relationships between the radiative moments and the source moments include many non-linear multipole interactions, as the source moments mix with each other as they ``propagate'' from the source to the detector. Such multipole interactions include the famous tail effect, corresponding to the coupling between the non-static moments with the mass of the source, and the non-linear memory effect \citep{BD88, BD92}. Furthermore \cite{B87} proved that this construction, when considered in a neighbourhood of future null infinity, recovers the findings of the \cite{BBM62, Sachs62} approach to gravitational radiation.

A different wave-generation formalism has been devised by \cite{WW96} (see also \citealt{W99, PW00, PW02}). This formalism, called ``Direct Integration of the Relaxed Equations -- DIRE'', has exactly the same scope as the MPM-PN method, i.e., it applies to any isolated post-Newtonian sources, in principle up to any PN order. However it differs in the definition of the source multipole moments, that are some well-defined (compact-support) versions of the \cite{EW75, Th80} moments rather than the \cite{BD89} moments, and in the calculation of tails and related non-linear effects. In both formalisms the source multipole moments, which involve a whole series of relativistic PN corrections, must be coupled together in a complicated way in the true non-linear solution; such non-linear couplings form an integral part of the radiative moments at infinity and thereby of the observed signal. The equivalence, at the most general level, between the MPM-PN framework and the DIRE formalism, has been proved in Sect.~4.3 of \cite{BlanchetLR}.


\subsubsection{The quadrupole moment formalism}
\label{sec:quadform}

The lowest-order wave generation formalism is the famous \cite{E18} quadrupole moment formalism. In his original paper Einstein derived it in the case where the dynamics of the source is driven by non gravitational forces. However a simple derivation of the quadrupole formalism due to \cite{LL} showed that it is actually also valid for self-gravitating sources, whose dynamics is due to the gravitational force -- for instance a Newtonian binary star system.\footnote{The fact that the Newtonian conservation laws forbid the emission of dipole gravitational waves was pointed out by Abraham in 1914 (see \citealt{Kennefick}).}

The quadrupole formalism applies to a general isolated matter source which is post-Newtonian in the sense of existence of the small post-Newtonian parameter $\epsilon$ defined by Eq.~\eqref{epsPN}. However, the quadrupole formalism is valid in the Newtonian limit $\epsilon\to 0$; it can rightly be qualified as ``Newtonian'' because the quadrupole moment of the matter source is Newtonian and its evolution obeys Newton's laws of gravity. In this formalism the gravitational field $h^\mathrm{TT}_{ij}$ (written here as a perturbation of the ``gothic'' spatial metric $\mathfrak{g}^{ij}\equiv\sqrt{-g}g^{ij}$) is expressed in a transverse and traceless (TT) coordinate system covering the far zone of the source at retarded times,\footnote{The TT coordinate system can be extended to the near zone of the source as well; see, e.g., \cite{KSGE}.} as
\begin{equation}\label{hijquad}
  h^\mathrm{TT}_{ij} = - \frac{2G}{c^4R} \!\perp_{ijab}
  \!(\bm{N})\biggl\{\frac{\dd^2\dQ_{ab}}{\dd T^2}(T-R/c)+
  \calO\left(\frac{1}{c}\right)\biggr\}+
  \calO\left(\frac{1}{R^2}\right)\,,
\end{equation}
where $R=|\bm{X}|$ is the distance to the source, $T-R/c$ is the retarded time, $\bm{N}=\bm{X}/R$ is the unit direction from the source to the far away observer, and 
\begin{equation}\label{operatorTT}
\perp_{ijab} \,\equiv\, \frac{1}{2}\Bigl[\perp_{ai}\perp_{jb} + \perp_{aj}\perp_{ib} - \perp_{ij}\perp_{ab}\Bigr]\,,
\end{equation}
is the TT projection operator, with $\perp_{ij} \,\equiv\, \delta_{ij}-N_iN_j$ being the projector onto the plane orthogonal to $\bm{N}$. The source's quadrupole moment takes the familiar Newtonian form (see the footnote \ref{fnote:notation} for further notation)
\begin{equation}\label{Qij}
  \dQ_{ij}(t) = \int_\mathrm{source} \!\!\!
  \dd^3\mathbf{x} \, \rho
  (\mathbf{x},t)\left(x_ix_j-\frac{1}{3}\delta_{ij}
  \mathbf{x}^2\right)\,,
\end{equation}
where $\rho$ is the Newtonian (essentially baryonic) mass density, obeying the usual continuity equation $\partial_t\rho+\partial_i(\rho v^i)=0$. The total gravitational power emitted by the source in all directions around the source is given by the Einstein quadrupole formula
\begin{equation}\label{fluxE}
  \mathcal{F} \equiv \left(\frac{\dd \dE}{\dd T}\right)^\text{GW} \!\!=
  \frac{G}{c^5}\left\{\frac{1}{5}\frac{\dd^3\dQ_{ab}} {\dd
    T^3}\frac{\dd^3\dQ_{ab}}{\dd T^3}+
  \calO\left(\frac{1}{c^2}\right)\right\}\,.
\end{equation}
Our notation $\mathcal{F}$ stands for the total gravitational energy flux or gravitational ``luminosity'' of the source. Similarly, the total angular momentum flux is \citep{Papa71, Th80}
\begin{equation}\label{fluxAM}
  \mathcal{G}_i \equiv \left(\frac{\dd \dJ_i}{\dd T}\right)^\text{GW}
  \!\!= \frac{G}{c^5}\left\{\frac{2}{5}\epsilon_{iab}
  \frac{\dd^2\dQ_{ac}} {\dd T^2}\frac{\dd^3\dQ_{bc}}{\dd
    T^3}+ \calO\left(\frac{1}{c^2}\right)\right\}\,,
\end{equation}
where $\epsilon_{abc}$ denotes the standard Levi-Civita symbol with $\epsilon_{123}=1$. We shall give in Eqs. \eqref{balanceP}--\eqref{balanceG} the related fluxes of linear momentum and position of the center of mass.

Associated with the latter energy and angular momentum fluxes, there is also a quadrupole formula for the radiation reaction force, which reacts on the source's dynamics in consequence of the emission of waves. This force will inflect the time evolution of the orbital phase of the binary pulsar and is responsible for the inspiral of compact binaries. At the position $(\mathbf{x},t)$ in a particular coordinate system covering the source, the reaction force density can be written as \citep{BuTh70, Bu71, MTW}
\begin{equation}\label{reac}
  F_i^{\text{reac}} = \frac{G}{c^5}\rho \left\{-\frac{2}{5} x^a
  \frac{\dd^5\dQ_{ia}} {\dd t^5} +
  \calO\left(\frac{1}{c^2}\right)\right\}\,.
\end{equation}
This is the gravitational analogue of the damping force of electromagnetism. However, notice that gravitational radiation reaction is inherently coordinate dependent, so the expression of the force depends on the coordinate system which is used. Consider now the energy and angular momentum of an isolated matter system made of some perfect fluid, say
\begin{subequations}\label{EJ}
\begin{align}
\dE &= \int\dd^3\mathbf{x}\,\rho\left[\frac{\mathbf{
      v}^2}{2}+\Pi-\frac{U}{2}\right] +
\calO\left(\frac{1}{c^2}\right)\,,
  \label{En}\\
\dJ_i &= \int\dd^3\mathbf{x}\,\rho\,\epsilon_{iab}\,x_a\,v_b +
\calO\left(\frac{1}{c^2}\right)\,.
  \label{J}
\end{align}\end{subequations}
The specific internal energy of the fluid is denoted $\Pi$, and obeys the usual thermodynamic relation $\dd\Pi=-P\dd(1/\rho)$ where $P$ is the pressure; the gravitational potential obeys the Poisson equation $\Delta U = -4\pi G \rho$. We compute the mechanical losses of energy and angular momentum from the time derivatives of $\dE$ and $\dJ_i$. We employ the Eulerian equation of motion $\rho\,\dd v^i/\dd t = -\partial_i P + \rho \partial_i U + F_i^{\text{reac}}$ and continuity equation $\partial_t\rho+\partial_i(\rho\,v^i)=0$. Note that we add the small dissipative radiation-reaction contribution $F_i^{\text{reac}}$ in the equation of motion but neglect all conservative post-Newtonian corrections. The result is
\begin{subequations}\label{balanceEJ0}
\begin{align}
\frac{\dd \dE}{\dd t} &= \int\dd^3\mathbf{x} \, v^i\,F_i^{\text{reac}} = -
\mathcal{F} + \frac{\dd f}{\dd t}\,,
  \label{balanceE0}\\
\frac{\dd \dJ_i}{\dd t} &= \int\dd^3\mathbf{ x} \,
\epsilon_{iab}\,x_a\,F_b^{\text{reac}} = - \mathcal{G}_i + \frac{\dd
  g_i}{\dd t}\,,
\label{balanceJ0}
\end{align}\end{subequations}
where one recognizes the fluxes at infinity given by Eqs. \eqref{fluxE}--\eqref{fluxAM}, and where the second terms denote some total time derivatives made of quadratic products of derivatives of the quadrupole moment. Looking only for secular effects, we apply an average over time on a typical period of variation of the system; the latter time derivatives will be in average numerically small in the case of quasi-periodic motion \citep{BR81}. Hence we obtain
\begin{subequations}\label{balanceEJ}
\begin{align}
\langle \frac{\dd \dE}{\dd t}\rangle &= - \langle \mathcal{F}\rangle \,,
  \label{balanceE}\\
\langle \frac{\dd \dJ_i}{\dd t}\rangle &= - \langle \mathcal{G}_i\rangle
\,,
  \label{balanceJ}
\end{align}\end{subequations}
where the brackets denote the time averaging over an orbit.\footnote{In the most general case, the time average $\langle f \rangle$ of a function $f(t)$ is defined as
\begin{align*}
	\langle f \rangle \equiv \lim_{T\to
		+\infty}\frac{1}{2T}\int_{-T}^{T}\dd t\,f(t) \,.
\end{align*}
For periodic functions this reduces to the usual average over the period $P$,
\begin{align*}
	\langle f \rangle = \frac{1}{P} \int_0^P\dd t\,f(t)\,.
\end{align*}
}
These balance equations encode the secular decreases of energy and angular momentum by gravitational radiation emission.

The cardinal virtues of the \cite{E18,LL} quadrupole formalism are: Its generality -- the only restrictions are that the source be Newtonian and bounded; its simplicity, as it necessitates only the computation of the time derivatives of the Newtonian quadrupole moment (using the Newtonian laws of motion); and, most importantly, its agreement with the observation of the dynamics of the binary pulsar PSR~1913+16 by \cite{TFMc79, TW82, T93}. Indeed let us apply the balance equations \eqref{balanceEJ} to a system of two point masses moving on an eccentric orbit modelling the binary pulsar -- the classic references are \cite{PM63, Pe64}; see also \cite{Dyson, EH75, Wag75}. We use the binary's Newtonian energy and angular momentum,
\begin{subequations}\label{EJexpr}
\begin{align}
\dE &= - \frac{G m_1 m_2}{2a}\,,\\ 
\dJ &= m_1 m_2 \sqrt{\frac{G a(1-e^2)}{m_1 + m_2}}\,,
\end{align}\end{subequations}
where $a$ and $e$ are the semi-major axis and eccentricity of the orbit and $m_1$ and $m_2$ are the two masses. From the energy balance equation \eqref{balanceE} (averaged over the orbital period $P$) we obtain first the secular evolution of $a$; next changing from $a$ to the orbital period $P$ using Kepler's third law,\footnote{Namely $(G m)^1=\Omega^2a^3$, where $m=m_1+m_2$ is the total mass and $\Omega=2\pi/P$ is the orbital frequency. This law is also appropriately called the 1-2-3 law \citep{MTW}.} we get the secular evolution of the orbital period $P$ as
\begin{equation}\label{Pdot}
\langle\frac{\dd P}{\dd t}\rangle = - \frac{192\pi}{5
  c^5}\left(\frac{2\pi G}{P}\right)^{5/3}\!\!\frac{m_1
  m_2}{(m_1+m_2)^{1/3}}\,
\frac{1+\frac{73}{24}e^2+\frac{37}{96}e^4}{(1-e^2)^{7/2}}\,.
\end{equation}
The last factor, depending on the eccentricity, comes out from the orbital average and is known as the \cite{PM63} ``enhancement'' factor,\footnote{See also Eqs. \eqref{PM}--\eqref{gneBessel} below for more details on its derivation.} so designated because in the case of the binary pulsar, which has a rather large eccentricity $e\simeq 0.617$, it enhances the effect by a factor $\sim 12$. Numerically, one finds $\langle\dd P/\dd t\rangle = - 2.4\times 10^{-12}$, a dimensionless number in excellent agreement with the observations of the binary pulsar \citep{TFMc79, TW82, T93}. More recently even more impressive tests were performed with the double pulsar \citep{KW09, Kramer21}. On the other hand the secular evolution of the eccentricity $e$ is deduced from the angular momentum balance equation \eqref{balanceJ} [together with the previous result \eqref{Pdot}], as
\begin{equation}\label{edot}
\langle\frac{\dd e}{\dd t}\rangle = - \frac{608\pi}{15
  c^5}\,\frac{e}{P}\left(\frac{2\pi G}{P}\right)^{5/3}\!\!\frac{m_1
  m_2}{(m_1+m_2)^{1/3}}
\,\frac{1+\frac{121}{304}e^2}{(1-e^2)^{5/2}}\,.
\end{equation}
Interestingly, the system of equations \eqref{Pdot}--\eqref{edot} can
be thoroughly integrated in closed analytic form. This yields the
law of evolution of the eccentricity $e(t)$ in terms of the period $P(t)$ at any instant \citep{Pe64}:
\begin{equation}\label{peters}
\frac{e^2}{(1-e^2)^{19/6}}\left(1+\frac{121}{304}e^2\right)^{145/121}
= c_0\,P^{19/9}\,,
\end{equation}
where $c_0$ denotes an integration constant to be determined by the initial conditions at the start of the binary evolution. When $e\ll 1$ the latter relation gives approximately $e^2\simeq c_0\,P^{19/9}$.

Inspiralling compact binaries, containing neutron stars and/or black holes, constitute the bread-and-butter sources of gravitational waves for the detectors LIGO, Virgo and KAGRA on ground. The two compact objects steadily lose their orbital binding energy by emission of gravitational radiation; as a result, the orbital separation between them decreases, and the orbital frequency increases. Thus, the frequency of the gravitational-wave signal, which equals twice the orbital frequency for the dominant harmonics, ``chirps'' in time (i.e., the signal becomes higher and higher pitched) until the two objects collide and merge.

The orbit of most inspiralling compact binaries can be considered to be circular, apart from the gradual inspiral, because the gravitational radiation reaction forces tend to rapidly circularize the motion. This effect is due to the emission of angular momentum by gravitational waves, resulting in a secular decrease of the eccentricity of the orbit, which has been computed within the quadrupole formalism in Eq.~\eqref{edot}. For instance, suppose that the inspiralling compact binary was long ago (a few hundred million years ago) a system similar to the binary pulsar system, with an orbital frequency $\Omega_0\equiv 2\pi/P_0\sim 10^{-4}\,\mathrm{rad}/\mathrm{s}$ and a rather large orbital eccentricity $e_0\sim 0.6$. When it becomes visible by the detectors on ground, i.e., when the gravitational-wave signal frequency reaches about $f\equiv\Omega/\pi\sim 10\mathrm{\ Hz}$, the eccentricity of the orbit should be $e \sim 10^{-6}$ according to the formula \eqref{peters}. This is a very small eccentricity, even when compared to high-order relativistic corrections. Only non-isolated binary systems could have a non negligible eccentricity. For instance, the \cite{vonZeipel, Lidov62, Kozai62} (ZLK) mechanism\footnote{See \cite{IO19} for an historical discussion, and justification of the appelation ZLK for designating this effect.} is one important scenario that produces eccentric binaries and involves the interaction between the binary and the environment of the dense cores of globular clusters \citep{MH02}. If the mutual inclination angle of the inner compact binary is strongly tilted with respect to the outer compact star (in a ``hierarchical'' triplet), then a resonance occurs and can increase the eccentricity of the inner binary to large values. This is one motivation for looking at the waves emitted by inspiralling binaries in non-circular, quasi-elliptical orbits (see Sect.~\ref{sec:eccentric}).

For a long while, it was thought that the quadrupole formula would be sufficient for sources of gravitational radiation to be observed directly on Earth -- as it had proved to be amply sufficient in the case of the binary pulsar. However, the work of \cite{3mn}\footnote{This work entitled: ``The last three minutes: Issues in gravitational-wave measurements of coalescing compact binaries'' is sometimes coined the ``3mn Caltech paper''.} showed that this is not true, as one has to include post-Newtonian corrections up to high order beyond the quadrupole formula in order to prepare for the data analysis of future detectors. This expectation has been confirmed by many measurement-analyses \citep{CFPS93, FCh93, CF94, TNaka94, P95, PW95, KKS95, DIS98, DIS00, DIS02, DIJS03, BCV03a, BCV03b, AIRS05, AISS05, BIO09} which have demonstrated that the post-Newtonian precision needed for the LIGO/Virgo detectors corresponds at least, in the case of neutron-star binaries, to the 3PN $\sim 1/c^6$ approximation beyond the quadrupole formula. 

In practice, the post-Newtonian prediction for the inspiral phase has to be matched to numerical-relativity results for the subsequent merger and ringdown phases. The match proceeds essentially through two routes: Either the Effective-One-Body (EOB) waveform \citep{BuonD99, BuonD00, DJSisco, DNorleans} that builds on post-Newtonian results and extend their domain of validity to facilitate the comparison with numerical relativity \citep{Buon09, Pan10}, or the Hybrid waveform obtained by direct matching between the PN expanded prediction and the numerical computations \citep{Ajith08, Ajith10}.

\begin{figure}[htbp]
\centering
	\includegraphics[width=0.9\textwidth]{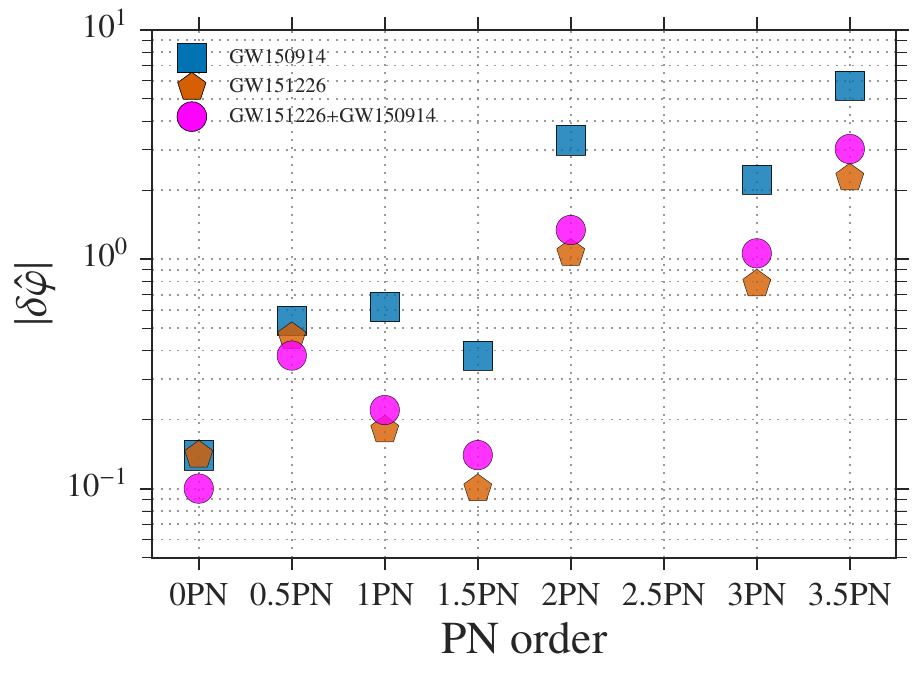}
	\includegraphics[width=0.9\textwidth]{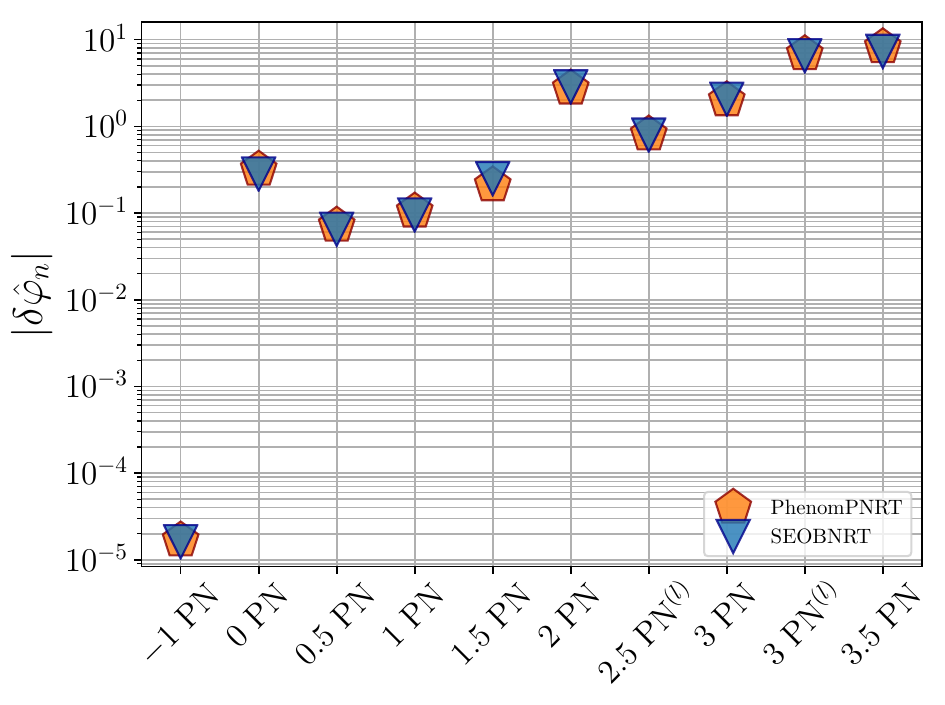}
\caption{Observational constraints on the post-Newtonian parameters, i.e. the coefficients in the phasing formula \eqref{phiSPA}, from measurements of the black hole events GW150914 and GW151226 (top panel) and from the neutron star event GW170817 (bottom panel) \citep{LIGOtestGR, LIGOtestGR2}. The limits are obtained by assuming the values predicted by general relativity for all the PN parameters but for one. This one is allowed to vary and its deviation with respect to GR is measured by the technique of matched filtering. The 1.5PN parameter agrees with the GR prediction within a fractional accuracy of the order of 10\%, which constitutes an interesting test of the tail effect. Images reproduced with permission from \cite{LIGOtestGR, LIGOtestGR2}.}
\label{fig:testGR}
\end{figure}

Strategies to detect and analyze the very weak signals from compact binary inspiral are based on the standard matched filtering technique. The raw output of the detector $o(t)$ consists of the superposition of the real gravitational-wave signal $h_{\mathrm{real}}(t)$ and of noise $n(t)$. The noise is assumed to be a stationary Gaussian random variable, with zero expectation value, and with (supposedly known) frequency-dependent power spectral density $S_n(\omega)$. The experimenters construct the correlation between $o(t)$ and a filter $q(t)$, i.e.,
\begin{equation}
		c(t) = \int^{+\infty}_{-\infty} \dd t' \, o (t') q(t+t')\,,
		\label{ct}
\end{equation}
and divide $c(t)$ by the square root of its variance, or correlation noise. The expectation value of this ratio defines the filtered signal-to-noise ratio (SNR). Looking for the useful signal $h_{\mathrm{real}}(t)$ in the detector's output $o(t)$, the data analysists adopt the \cite{Wiener} filter
\begin{equation}
		\tilde{q} (\omega) = \frac{\tilde{h} (\omega)}{S_n (\omega)}\,,
		\label{qtilde}
\end{equation}
where $\tilde{q} (\omega)$ and $\tilde{h} (\omega)$ are the Fourier transforms of $q(t)$ and of the \emph{theoretically computed} template $h(t)$. By the matched filtering theorem, \eqref{qtilde} maximizes the SNR if $h(t)=h_{\mathrm{real}}(t)$. The maximum SNR is then the best achievable with a linear filter. In practice, because of systematic errors in the theoretical modelling, the template $h(t)$ will not exactly match the real signal $h_{\mathrm{real}} (t)$; however if the template is to constitute a realistic prediction the errors will be small. This is of course the motivation for computing high order post-Newtonian templates, in order to reduce as much as possible the systematic errors. As we shall see, the signal at such high PN order contains the signature of several non-linear effects which are specific to general relativity. We thus have the possibility of probing, via matched filtering, some aspects of the non-linear structure of Einstein's theory \citep{BSat94, BSat95, AIQS06a, AIQS06b}. See Fig.~\ref{fig:testGR}.

\subsubsection{Influence of the internal structure of compact bodies}
\label{sec:intstructure}

The main point about modelling the inspiralling compact binary is that a model made of two \emph{structureless} point particles, characterized solely by mass parameters $m_\text{a}$ and possibly the spins $S_\text{a}$ (with $\text{a}$ labelling the particles), is sufficient in first approximation. Indeed, most of the non-gravitational effects usually plaguing the dynamics of binary star systems, such as the effects of a magnetic field, of an interstellar medium, of the internal structure of extended bodies, are dominated by gravitational effects. In particular the effects due to the finite size of the compact bodies are small, although not negligibly small. 

Consider the influence of the Newtonian quadrupole moments $q^{ij}_\text{a}$ through the tidal interaction between extended neutron stars without spins during the inspiral phase. It is known that for inspiralling compact binaries the neutron stars are not co-rotating because the tidal synchronization time is much larger than the time left till the coalescence. As shown by \cite{Kochanek} the best models for the fluid motion inside the two neutron stars are the so-called Roche-Riemann ellipsoids, which have tidally locked figures (the quadrupole moments facing each other at any instant during the inspiral), but for which the fluid motion has zero circulation in the inertial frame.

Here we perform a simple calculation to Newtonian order within such a model. The equations of motion of $N$ extended spinless bodies, to linear order in the quadrupole moments, are (with $\text{a},\text{b}=1,\cdots, N$)
\begin{equation}\label{EOM}
	m_\text{a} \frac{\dd v_\text{a}^i}{\dd t} = G \sum_{b \not= a} \left[m_\text{a} m_\text{b} \frac{\partial}{\partial y_\text{a}^i}\left(\frac{1}{r_\text{ab}}\right) + \frac{1}{2}\left( m_\text{a} \,q_\text{b}^{jk} + m_\text{b} \,q_\text{a}^{jk} \right)\frac{\partial^3}{\partial y_\text{a}^i\partial y_\text{a}^j\partial y_\text{a}^k}\left(\frac{1}{r_\text{ab}}\right)\right] \,,
\end{equation}
where $m_\text{a}$ are the masses, and we denote the position and velocity of the center of mass of the bodies by $\bm{y}_\text{a}(t)$ and $\bm{v}_\text{a}(t)=\dd \bm{y}_\text{a}/\dd t$, with the Euclidean separation between centers of mass being $r_\text{ab}=\vert \bm{y}_\text{a}-\bm{y}_\text{a}\vert$. The quadrupole moments of the bodies, supposed to be made of a perfect fluid, read
\begin{equation}\label{qaij}
	q_\text{a}^{ij} = \int_{\mathcal{V}_\text{a}} \dd^3\bm{z}_\text{a} \,\rho_\text{a}\Bigl( z_\text{a}^i z_\text{a}^j - \frac{1}{3} \delta^{ij} \bm{z}_\text{a}^2 \Bigr)\,,
\end{equation}
with $\mathcal{V}_\text{a}$ the volume of the body, $\bm{z}_\text{a}=\mathbf{x}-\bm{y}_\text{a}(t)$ the distance between a generic point $\mathbf{x}$ inside the body and the center of mass, $\rho_\text{a}=\rho(\bm{y}_\text{a}+\bm{z}_\text{a},t)$ the Newtonian mass density of the body, $\rho(\mathbf{x},t)$ being the usual Eulerian density. The mass-centred condition reads
\begin{equation}\label{CM}
	\int_{\mathcal{V}_\text{a}} \dd^3\bm{z}_\text{a} \,\rho_\text{a}\,z_\text{a}^i = 0\,.
\end{equation}
The conserved energy of the $N$-body system is the sum of the internal (Newtonian) energies $e_\text{a}$ and of the orbital contributions, including the effect of the quadrupoles:
\begin{equation}\label{E}
	\dE = \sum_\text{a} \biggl\{ e_\text{a} + \frac{1}{2}m_\text{a} \bm{v}_\text{a}^2 - \frac{G}{2} \sum_{b \not= a} \frac{m_\text{a} m_\text{b}}{r_\text{ab}} - \frac{1}{2} \,q_\text{a}^{ij} \mathcal{E}_\text{a}^{ij}\biggr\}\,,
\end{equation}
where we have introduced the tidal field acting on body $\text{a}$ and due to the other bodies $\text{b} \not= \text{a}$:
\begin{equation}\label{tidalfield}
	\mathcal{E}_\text{a}^{ij} \equiv  G \sum_{\text{b} \not= \text{a}} m_\text{b} \frac{\partial^2}{\partial y_\text{a}^i\partial y_\text{a}^j}\!\left(\frac{1}{r_\text{ab}}\right)\,.
\end{equation}
Posing $\bm{w}_\text{a}=\dd \bm{z}_\text{a}/\dd t$ for the internal velocity field of body $\text{a}$, $\Pi_\text{a}=\Pi(\bm{y}_\text{a}+\bm{z}_\text{a},t)$ for the specific internal energy satisfying the thermodynamical relation $\dd\Pi=- P\,\dd(1/\rho)$ (with $P$ the pressure), and $u_\text{a}$ for the internal self-gravity given by the Poisson integral over the volume of the body, we have
\begin{equation}\label{ea}
	e_\text{a} = \int_{\mathcal{V}_\text{a}} \dd^3\bm{z}_\text{a} \,\rho_\text{a}\left( \frac{1}{2}\bm{w}_\text{a}^2 + \Pi_\text{a} - \frac{u_\text{a}}{2} \right)\,.
\end{equation}
The coupling of the quadrupole moment $q_\text{a}^{ij}$ with the external tidal field $\mathcal{E}_\text{a}^{ij}$ of the other bodies implies a variation of the internal energy given by
\begin{equation}\label{dedt}
	\frac{\dd e_\text{a}}{\dd t} = \frac{1}{2} \frac{\dd q_\text{a}^{ij}}{\dd t} \mathcal{E}_\text{a}^{ij} \,.
\end{equation}
We consider the case where the quadrupole moment is induced by the tidal field of the other bodies. To linear order, we can introduce a coefficient $\lambda_\text{a}$ characterizing the response of the body under the influence of the external field, such that
\begin{equation}\label{lambda}
	q_\text{a}^{ij} = \lambda_\text{a} \,\mathcal{E}_\text{a}^{ij}\,.
\end{equation}
In the case of the induced quadrupole moments \eqref{lambda}, the total energy of the system becomes
\begin{equation}\label{Einduced}
	\dE = \sum_\text{a} \biggl\{ \frac{1}{2}m_\text{a} \bm{v}_\text{a}^2 - \frac{G}{2} \sum_{b \not= a} \frac{m_\text{a} m_\text{b}}{r_\text{ab}} - \frac{1}{4} \lambda_\text{a} \,\mathcal{E}_\text{a}^{ij} \mathcal{E}_\text{a}^{ij}\biggr\}\,.
\end{equation}

The response coefficient $\lambda_\text{a}$ is conventionally written as $\lambda_\text{a}=\frac{2}{3G} \,k_\text{a} r_\text{a}^5$, where $r_\text{a}$ is the radius of the body and $k_\text{a}$ is the body's dimensionless second \cite{Love11} number which depends on the internal structure of the body, and is, typically, of the order unity. Next we characterize the internal structure by the dimensionless ``deformability'' or ``polarizability'' parameter
\begin{equation}\label{Lambdaa}
	\Lambda_\text{a} = \frac{c^{10}}{G^4 m_\text{a}^5}\lambda_\text{a} = \frac{2}{3} k_\text{a} \left(\frac{c^2 r_\text{a}}{G m_\text{a}}\right)^5 = \frac{2}{3} \frac{k_\text{a}}{K_\text{a}^5}\,.
\end{equation}
where the dimensionless ratio $K_\text{a} \equiv \frac{G m_\text{a}}{r_\text{a} c^2}$ is the ``compactness'' parameter and typically equals $\sim 0.2$ for neutron stars (depending on their equation of state). 

Consider a compact binary system ($N=2$) moving on an exact circular orbit. From Eq.~\eqref{lambda} we see that the two quadrupole moments face each other, and remain constant along the circular orbit. The equation of the relative motion reduces to $\dd\bm{v}/\dd t=-\Omega^2 \bm{x}$, where $\bm{x}=\bm{y}_1-\bm{y}_2$ and $\bm{v}=\dd \bm{x}/\dd t$ (with $r \equiv r_{12}$). We find from \eqref{EOM} the orbital frequency
\begin{equation}\label{omegatidal}
	\Omega^2 = \frac{G m}{r^3} \biggl[ 1 + 9 \nu \bigl( X_1^3 \Lambda_1 + X_2^3 \Lambda_2\bigr) \gamma^5\biggr]\,.
\end{equation}
We pose $X_\text{a}=m_\text{a}/m$ with $m=m_1+m_2$ so that $\nu=X_1 X_2$ is the symmetric mass ratio \eqref{nu}, denote $\gamma=\frac{G m}{r c^2}$ and employ the notation \eqref{Lambdaa}. In turn the conserved energy \eqref{Einduced} for circular orbits reduces to
\begin{equation}\label{Ecirctidal}
	\dE = - \frac{G m^2 \nu}{2 r}\biggl[ 1 - 6 \nu \bigl( X_1^3 \Lambda_1 + X_2^3 \Lambda_2\bigr) \gamma^5\biggr] \,.
\end{equation}
As the effect of the deformation of the bodies computed here is purely ``Newtonian'', we see that the $c$'s we have introduced into our definitions naturally cancel out in Eqs. \eqref{omegatidal} and \eqref{Ecirctidal}.

The above dynamics is \textit{conservative}, as we have neglected the dissipative radiation reaction effect on the orbit. This effect is taken into account when we impose the flux-balance equation \eqref{balanceE}; there is no need to impose the angular momentum balance equation \eqref{balanceJ} for circular orbits. The total quadrupole moment of the system is the sum of the orbital one and of the intrinsic moments of the bodies, given by \eqref{qaij}, hence
\begin{equation}\label{Qijtot}
	Q^{ij} = m \nu \Bigl( x^i x^j - \frac{1}{3} \delta^{ij} r^2\Bigr) + q_1^{ij} + q_2^{ij} \,.
\end{equation}
Plugging this into the flux formula \eqref{fluxE}, computing the time derivatives using the equations of motion including the contributions from the quadrupole moments, see Eq.~\eqref{omegatidal}, and keeping only the terms linear in these quadrupoles, yields the flux (still for exact circular orbits) as
\begin{equation}\label{fluxr}
	\mathcal{F} = \frac{32 G}{5 c^5} r^4 \omega^6 m^2 \nu^2 \biggl[ 1 + 6 \bigl( X_1^4 \Lambda_1 + X_2^4 \Lambda_2 \bigr) \gamma^5 \biggr]\,.
\end{equation}
\begin{figure}[ht]
	\centering
		\includegraphics[width=10cm]{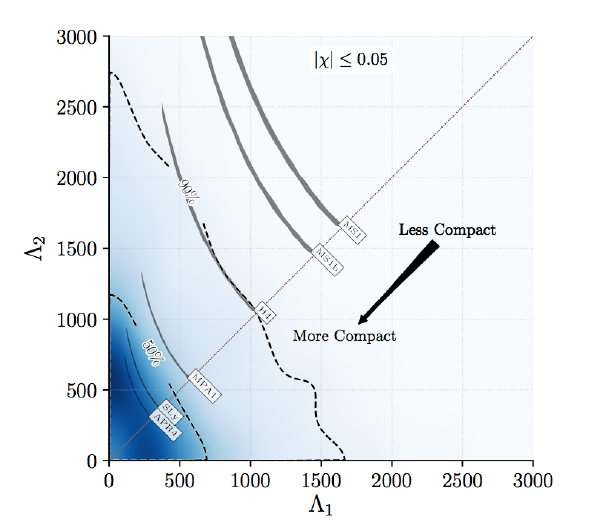}
	\caption{Observational constraints on the tidal deformability and the inner equation of state of neutron stars obtained with the GW170817 event \citep{GW170817}. The parameters $\Lambda_\text{a}$ are defined by \eqref{Lambdaa}. Contours enclosing 90\% and 50\% of the probability density are shown with dashed lines. The predictions for tidal deformability given by a set of representative equations of state are given with grey lines. For a stiff equation of state the pressure increases a lot for a given increase in density (for instance $P\propto\rho^\gamma$ with a large polytropic index $\gamma$), thus it gives more resistance to the gravitational force and the neutron star is less compact. The stiffest equations of state are excluded, while the softest (which predict more compact neutron stars) are still allowed; they appear in the dark blue region. The constraints are shown for a low-spin scenario, with dimensionless spin parameter $\vert\chi\vert\leqslant 0.05$, probably favored for neutron stars. Image reproduced with permission from \cite{GW170817}.}
        \label{fig:intstruct}
\end{figure}
Next we reexpress the invariants $\dE$ and $\mathcal{F}$ in terms of the orbital frequency $\omega$ instead of the coordinate-dependent separation distance $r$ using Eq.~\eqref{omegatidal}. Posing $x=(G m \Omega/c^3)^{2/3}$ we obtain
\begin{subequations}\label{EFcircxtidal}
	\begin{align}
		\dE &= - \frac{1}{2}m \nu c^2 x \biggl[ 1 - 9 \nu \bigl( X_1^3 \Lambda_1 + X_2^3 \Lambda_2\bigr) x^5\biggr]\,,\label{Ecircxtidal}\\
		\mathcal{F} &= \frac{32 c^5}{5 G} x^5 \nu^2 \biggl\{ 1 + 6 \Bigl[ \left(X_1+2\nu\right)X_1^3 \Lambda_1 + \left(X_2+2\nu\right)X_2^3 \Lambda_2\Bigr] x^5\biggr\}\,.\label{Fcircxtidal}
	\end{align}
\end{subequations}
At this stage we can draw a firm conclusion: The effect of the internal structure of non-spinning bodies is proportional to $x^5$, and is thus comparable to a relativistic effect occuring at the 5PN order. This is a higher post-Newtonian order than the one at which we shall obtain the phasing formula in Sect.~\ref{sec:orbevol}. Therefore, in that sense the internal structure of the two compact bodies is ``effaced'' and their dynamics and radiation depend only, in first approximation, on the masses and possibly the spins \citep{Dhouches}.

But the numerical value of the coefficient involving the $\Lambda_\text{a}$'s is to be taken into account. For instance $\frac{G m_\text{a}}{c^2 r_\text{a}}\sim 0.15$ for neutron stars (hence $x^5\sim 8\,10^{-5}$ at the merger), and the numerical estimates of the Love numbers for neutron stars are of the order of one or say, a tenth \citep{Hind08, DN09tidal, BinnP09}. Therefore the deformability parameters \eqref{Lambda} for compact bodies should be of the order of $\Lambda_\text{a} \sim 1000$, depending on the equation of state, see Fig.~\ref{fig:intstruct}.

The phase and frequency evolution follow from the balance equation \eqref{baleq} (without need to average), where both $\dE$ and $\mathcal{F}$ have been computed for the conservative dynamics in Eqs. \eqref{EFcircxtidal}. This approximation is justified as we are interested in the secular, adiabatic evolution of the orbit over a radiation reaction time scale. The secular variation of the energy $\dE$ is immediately deduced from \eqref{Ecircxtidal} as
\begin{equation}\label{dEdx}
	\frac{\dd \dE}{\dd t} = -\frac{1}{2} m \nu c^2 \biggl[ 1 - 54 \nu \Bigl( X_1^3 \Lambda_1 + X_2^3 \Lambda_2 \Bigr) x^5\biggr] \dot{x}\,.
\end{equation}
Combining this with \eqref{Fcircxtidal}, we get an ordinary differential equation for $x$. Solving for $x(t)$ and the phase evolution $\phi=\int\dd t\,\Omega(t)$, we thereby obtain the quadrupole finite size effect due to the internal structure as \citep{FHind08, DNV12, F14}\footnote{The leading-order 5PN tidal phase was extended to higher order by \cite{VHF11, BiniDF12, VF13, AGP18, BV18, Landry18, HFB20a, HFB20b} and is now complete up to the 7.5PN order, including mass quadrupole, current quadrupole and mass octupole tidal contributions, with tail terms arising at 6.5PN and 7.5PN orders.}
\begin{subequations}\label{xphisolintstruct}
	\begin{align}
		x &= \frac{1}{4}\tau^{-1/4}\left[ 1 + \frac{39}{8192} \tilde{\Lambda} \,\tau^{-5/4}\right]\,,\\
		\phi &= \phi_0 - \frac{x^{-5/2}}{32\nu} \left[ 1 + \frac{39}{8} \tilde{\Lambda} \,x^{5}\right]\,,
	\end{align}
\end{subequations}
where $\tau=\frac{\nu c^3}{5G m}(t_0-t)$ and $t_0$ is the instant of coalescence. The effect depends on the following combination of the two deformability parameters:
\begin{equation}\label{Lamdatilde}
	\tilde{\Lambda} = \frac{16}{13}\,\biggl[ \bigl( X_1 + 11 \nu\bigr) X_1^3 \Lambda_1 + \bigl( X_2 + 11 \nu\bigr) X_2^3 \Lambda_2 \biggr]\,,
\end{equation}
so normalized that in the case of two identical neutron stars (with the same mass, $X_1=X_2=\frac{1}{2}$, and the same equation of state) it reduces to $\tilde{\Lambda}=\Lambda_1=\Lambda_2$. The 5PN finite size effect \eqref{xphisolintstruct}--\eqref{Lamdatilde} is to be contrasted with the point-mass result, ignoring the internal structure, which will be obtained in Sect.~\ref{sec:orbevol}. Remarkably, it has been possible to put a bound on the tidal deformability of neutron stars, and a constraint on several possible equations of state, with the binary neutron star event GW170817 \citep{GW170817}, see Fig.~\ref{fig:intstruct}.


\subsubsection{The recoil by gravitational waves}
\label{sec:recoil}

A related topic is the loss of \emph{linear} momentum by gravitational radiation and the resulting gravitational recoil (or ``kick'') of black-hole binary systems. This phenomenon has potentially important astrophysical consequences \citep{Me04}. In models of formation of massive black holes involving successive mergers of smaller ``seed'' black holes, a recoil with sufficient velocity could eject the system from dwarf galaxies or globular clusters and terminate the process. Even in galaxies whose potential wells are deep enough to confine the recoiling system, displacement of the system from the center could have important dynamical consequences for the galactic core.

To compute the recoil we need the flux formula for the linear momentum. Integrating the leading 2.5PN radiation-reaction force \eqref{reac} over the source results in a total time derivative, so there is no net recoiling force at that order, and the effect is a subdominant 3.5PN one. We have \citep{BoR61, Peres62, Papa62, Papa71, Bek73, Th80}
\begin{equation}\label{balanceP}
	\mathcal{F}^i_{\dP} \equiv \left(\frac{\dd \dP_i}{\dd T}\right)^\text{GW} =  \frac{G}{c^7}\left[ \frac{2}{63}\dO_{iab}^{(4)}\dQ_{ab}^{(3)} + \frac{16}{45}
	\epsilon_{iab}  \dQ_{ac}^{(3)}\dC_{bc}^{(3)} \right] + \calO\left(\frac{1}{c^9}\right)\,,
\end{equation}
where the Newtonian source moments are $\dQ_{ij}=\int\dd^3\bm{x}\,\rho\,\hat{x}_{ij}$ (mass quadrupole), $\dO_{ijk}=\int\dd^3\bm{x}\,\rho\,\hat{x}_{ijk}$ (mass octupole) and $C_{ij}=\int\dd^3\bm{x}\,\rho\,\epsilon_{ab\langle i}\hat{x}_{j\rangle a}v_b$ (current quadrupole); see the footnote \ref{fnote:notation} for notation. Furthermore, besides the fluxes associated to energy $\dE$, angular momentum $\dJ_i$ and linear momentum $P_i$, there is also a flux associated to the center of mass $\dG_i = \dP_i\,t + \dZ_i$, where $\dZ_i$ denotes the initial ``position'' of the center of mass. The effect is also subdominant at 3.5PN order and reads \citep{KQ16, KNQ18, N18, BF19, COS20}\footnote{We provide the full multipole expansions of the various fluxes in terms of radiative-type moments in Eqs. \eqref{FluxFG} and \eqref{FluxPC} below.}
\begin{equation}\label{balanceG}
	\mathcal{F}^i_{\dG} \equiv \left(\frac{\dd \dG_i}{\dd T}\right)^\text{GW} = \frac{2G}{21 c^7} \dO_{iab}^{(3)}\dQ_{ab}^{(3)} + \calO\left(\frac{1}{c^9}\right) \,,
\end{equation}

The meaning of the latter flux formulas is that the secular evolution by gravitational wave emission of the linear momentum $\bm{\dP}$ and of the center-of-mass position $\bm{\dG}$ attributable to an isolated (post-Newtonian) system obey
\begin{subequations}\label{fluxeqs}
	\begin{align}
		\frac{\dd \bm{\dP}}{\dd t} &= - \bm{\mathcal{F}}_\dP\,,\label{fluxeqsa}\\ 
		\frac{\dd \bm{\dG}}{\dd t} &= \bm{\dP} - \bm{\mathcal{F}}_\dG\,,\label{fluxeqsb}
	\end{align}
\end{subequations}
where $\bm{\mathcal{F}}_\dP$ and $\bm{\mathcal{F}}_\dG$ are the fluxes given by \eqref{balanceP} and \eqref{balanceG}. Consider the case where the source is stationary before some instant $t_0$, then emits a pulse of gravitational waves with finite duration between times $t_0$ and $t_1$, and finally comes back to a stationary state at later times $t>t_1$. In this situation, it is straightforward to find the form of the solution to Eqs. \eqref{fluxeqs}. Initially, the linear momentum is constant, so that, by applying a Lorentz boost, we can put ourselves in the rest frame of the system, thus achieving $\bm{\dP}_0=0$ (for $t<t_0$). Furthermore, we can translate the origin of our coordinate system in such a way that it coincides with the center of mass of the system, hence $\bm{\dG}_0=0$ initially. Then, by integrating \eqref{fluxeqs}, we get (for $t_0<t<t_1$)
\begin{subequations}\label{GWem}
	\begin{align}
		\bm{\dP}(t) &= - \int_{t_0}^{t} \dd t' \,\bm{\mathcal{F}}_\dP(t')\,,\\ \bm{\dG}(t) &=
		- \int_{t_0}^{t} \dd t' \,(t-t')\,\bm{\mathcal{F}}_\dP(t') - \int_{t_0}^{t} \dd t'
		\,\bm{\mathcal{F}}_\dG(t')\,.
	\end{align}
\end{subequations}
After the period of emission (for $t>t_1$), the source is again stationary but has acquired a net constant linear momentum $\bm{\dP}_1$, and the motion of its center of mass is uniform, $\bm{\dG}_1=\bm{\dP}_1\,t+\bm{\dZ}_1$. We find
\begin{subequations}\label{afterGW}
	\begin{align}
		\bm{\dP}_1 &= - \int_{t_0}^{t_1} \dd t' \,\bm{\mathcal{F}}_\dP(t')\,,\\ \bm{\dZ}_1 &=
		\int_{t_0}^{t_1} \dd t' \Bigl[ t'\,\bm{\mathcal{F}}_\dP(t') -
		\bm{\mathcal{F}}_\dG(t')\Bigr]\,.
	\end{align}
\end{subequations}
The final value of the linear momentum $\bm{\dP}_1$ yields the total gravitational recoil velocity of the source as measured in the asymptotic Minkowskian frame. On the other hand, the cumulative effect of the center-of-mass flux $\bm{\mathcal{F}}_\dG$ results in a modification of the position $\bm{\dZ}_1$ of the center of mass after the GW emission. 

Consider the case of a Newtonian binary system with no spins and moving on a quasi-circular orbit. The right-side of the linear momentum flux equation \eqref{fluxeqsa} is straightforward to evaluate with result \citep{Fit83}
\begin{equation}\label{dPdtbin}
	\frac{\dd \bm{\dP}}{\dd t} =
	\frac{464}{105}\,\frac{G^4m^5\Omega}{c^7r^4}\,\Delta\,\nu^2\,
	\bm{\lambda} + \calO\left(\frac{1}{c^9}\right)\,.
\end{equation}
Here $m=m_1+m_2$ and $\nu=m_1m_2/m^2$, assuming $m_1\geqslant m_2$ (thus $\Delta=\frac{m_1-m_2}{m}=\sqrt{1-4\nu}$), the orbital frequency of the circular orbit is $\Omega=\sqrt{G m/r^3}$, and the unit vector $\bm{\lambda}$ is orthogonal to the orbital separation $\bm{n}=(\bm{y}_1-\bm{y}_2)/r$ in the orbital plane, such that $\dot{\bm{n}}=\Omega\bm{\lambda}$ and $\dot{\bm{\lambda}}=-\Omega\bm{n}$. The relation \eqref{dPdtbin} holds at any time along the orbit and can be integrated, yielding
\begin{equation}\label{Pbin}
	\bm{\dP} = \frac{464}{105}\,\frac{G^4m^5}{c^7r^4}\,\Delta\,\nu^2\,\bm{n} + \calO\left(\frac{1}{c^9}\right)\,.
\end{equation}
We assume from now on that a Lorentz boost and a shift of the origin of the coordinate system have been applied to set $\bm{\dP}$ and $\bm{\dG}$ to zero when averaged over an orbit (neglecting the radiation-reaction decay). Then, we use the center-of-mass balance equation \eqref{fluxeqsb}, which leads to
\begin{equation}\label{dGdtbin}
	\frac{\dd \bm{\dG}}{\dd t} = \bm{\dP} +
	\frac{544}{105}\,\frac{G^4m^5}{c^7r^4}\,\Delta\,\nu^2\,\bm{n} + \calO\left(\frac{1}{c^9}\right) \,.
\end{equation}
Combining the results \eqref{Pbin}--\eqref{dGdtbin} and integrating \citep{BF19},
\begin{equation}\label{Gbin}
	\bm{\dG} = -
	\frac{48}{5}\,\frac{G^4m^5}{c^7r^4\Omega}\,\Delta\,\nu^2\,\bm{\lambda} + \calO\left(\frac{1}{c^9}\right)	\,.
\end{equation}
The results give the instantaneous values of the linear momentum and center-of-mass position of a circular binary (neglecting the orbital decay); these are equal to minus those that can be attributed to the gravitational wave field.
\begin{figure}[ht!]
	\centering
		\includegraphics[width=10.0cm,angle=0]{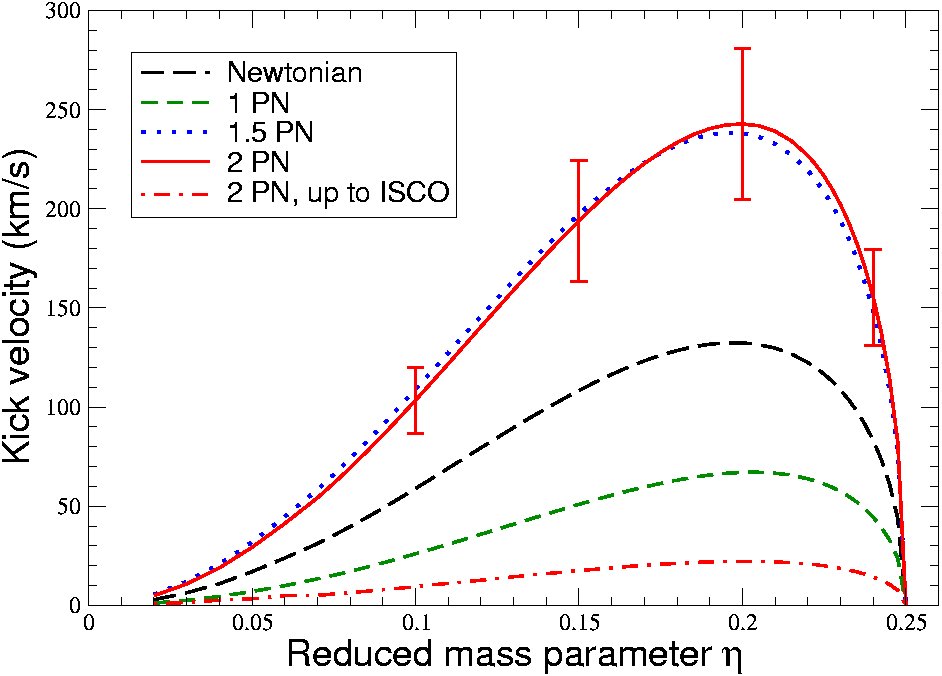}
		\caption{The gravitational recoil of non-spinning black-hole binaries generated by the inspiral + merger phases (up to the horizon), as a function of the symmetric mass ratio $\eta\equiv\nu$. The maximum recoil due to the inspiral phase up to the innermost stable circular orbit (ISCO) is of the order of $22 \, \mathrm{km} \, \mathrm{s}^{-1}$. The recoil accumulated during the plunge phase, from the ISCO up to the horizon, is obtained by integrating the 2PN momentum flux formula \eqref{dPdtbin2PN} along a plunge geodesic of the Schwarzschild metric within an effective one-body approach. The maximum recoil due to the inspiral + merger phases (ignoring the ringdown) amounts to about $250 \, \mathrm{km} \, \mathrm{s}^{-1}$ (red curve). Image reproduced with permission from \cite{BQW05}, copyright by AAS.}
                \label{fig:recoil1}
\end{figure}
\begin{figure}[ht!]
	\centering
		\includegraphics[angle=-90,width=\textwidth]{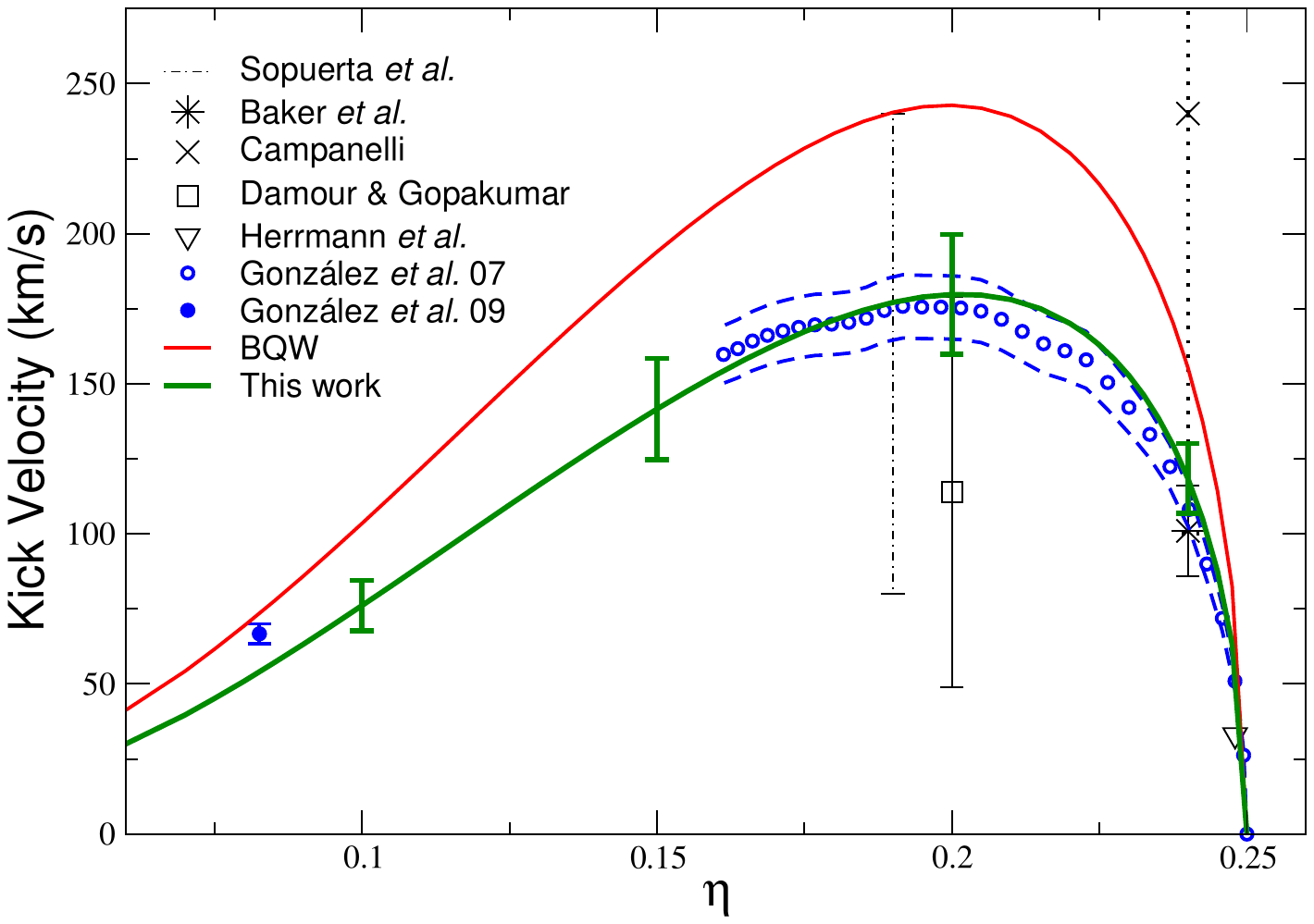}
		\caption{The total recoil of non-spinning black hole binaries generated by the inspiral + merger + ringdown phases (green curve). The recoil up to the merger is reproduced from Fig.~\ref{fig:recoil1} (red curve). The ringdown contribution is computed using a ``close-limit approximation'' for black hole binaries that uses 2PN-accurate initial data \citep{LB10}. The effect of the ringdown phase on the recoil velocity is to produce an ``anti-kick'', i.e. to reduce the value of the total recoil with respect to that computed up to the horizon. Thus the maximum recoil of non-spinning black-hole binaries is around $180 \, \mathrm{km} \, \mathrm{s}^{-1}$ at a mass ratio of $\nu_{\max}\approx 0.2$. Image reproduced with permission from \cite{LBW10}, copyright by IOP.}
                \label{fig:recoil2}
\end{figure}

The recoil of binary black hole systems has been estimated by \cite{Fit83, Wi92, K95, BQW05, RBK09, LBW10} using post-Newtonian methods,\footnote{Notably \cite{K95} gives the dominant PN contribution of the spins to the recoil.} and by \cite{OoNaka83,FitDet84,NakaOo87,FHH04} using perturbation methods. In parallel the problem of gravitational recoil of coalescing binaries has attracted considerable attention from the numerical relativity community. The numerical computations led to increasingly accurate estimates of the kick velocity from the merger along quasicircular orbits of binary black holes without spins \citep{Camp05, Bak06b} and with spins \citep{Camp07}. In particular they revealed the interesting result that very large kick velocities can be generated in the case of spinning black holes for particular spin configurations.

Unfortunately the post-Newtonian approach is not ideally suited for this problem because most of the recoil is generated in the strong field regime close to the merger. As a result the contribution of the ``plunge'' (difficult to model by post-Newtonian theory) dominates over that of the inspiral phase, as shown by the curve ``up to the ISCO'' in Fig.~\ref{fig:recoil1}. The post-Newtonian corrections in the recoil formula, up to 2PN order beyond the dominant 3.5PN effect given by Eq.~\eqref{dPdtbin}, have been computed using the gravitational-wave generation formalism of Sect.~\ref{sec:PNsource}, with result [where $x\equiv(\frac{G m \Omega}{c^3})^{2/3}$] \citep{BQW05}
\begin{align}\label{dPdtbin2PN}
	\frac{\dd \bm{\dP}}{\dd t} &= \frac{464}{105}\,\Delta\,\nu^2 x^{11/2} \left[1+\left(-\frac{452}{87} -\frac{1139}{522}\nu\right)x +
	\frac{309}{58}\,\pi\,x^{3/2} \right. \nonumber\\ &\qquad \left.
	+\left(-\frac{71345}{22968}+\frac{36761}{2088}\nu
	+\frac{147101}{68904}\nu^2\right)x^2\right] \bm{\lambda} + \calO\left(\frac{1}{c^{12}}\right)\,.
\end{align}
This result has been extended to 2.5PN order by \cite{MAI12}.

The recoil of non-spinning black-hole binaries generated by the inspiral and merger phases to 2PN relative order is shown in Fig.~\ref{fig:recoil1}. The final result, however, must include also the recoil generated during the ringdown phase, and is presented in Fig.~\ref{fig:recoil2}. The result is in good agreement with numerical computations for non-spinning black hole binaries \citep{Camp05, Bak06b, HHSL07, GSBHH07, GSB09}. See notably the sequence of dots and accompanying dashed lines in Fig.~\ref{fig:recoil2} (in blue), which are from an exhaustive series of numerical simulations by \cite{GSBHH07, GSB09}. Also shown in Fig.~\ref{fig:recoil2} are the comparisons with other analytic or semi-analytic methods: an application of the effective-one-body formalism \citep{DG06}; a close-limit calculation with Bowen--York type initial conditions \citep{SYL06}. An empirical fit to the final result from \cite{LBW10} in Fig.~\ref{fig:recoil2} is
\begin{equation}
	V_\text{recoil} \approx 9.5 \,\Delta\,\nu^2\,\bigl(1 + 0.3\,\nu\bigr) \times 10^3 \, {\rm km/s} \,,
\end{equation}
where the leading $\Delta\,\nu^2$ dependence on the mass ratio comes from the lowest-order ``Newtonian'' expression, see Eq.~\eqref{dPdtbin2PN}.


\subsection{Overview on motion and radiation}
\label{sec:Overview}

\subsubsection{Post-Newtonian equations of motion}
\label{sec:PNeom}

By equations of motion we mean the explicit expression of the accelerations of the bodies in terms of the positions and velocities. In Newtonian gravity, writing the equations of motion for a system of $N$ particles is trivial; in general relativity, even writing the equations in the case $N=2$ is difficult. The first relativistic terms, at the 1PN order, were derived by \cite{Droste17, LD17}. Subsequently, \cite{EIH} obtained the 1PN corrections by means of their famous ``surface-integral'' method, in which the equations of motion are deduced from the \emph{vacuum} field equations, and are therefore applicable to any compact objects (be they neutron stars, black holes, or, perhaps, naked singularities). The 1PN-accurate equations were also obtained, for the motion of the centers of mass of compact bodies, by \cite{Fock39} (see also \citealt{Petrova, Papa51}).

The 2PN approximation was tackled by \cite{OO73a, OO73b, OO74a, OO74b}, who considered the post-Newtonian iteration of the Hamiltonian of point-particles. We refer here to the Hamiltonian as a ``Fokker-type'' Hamiltonian, which is obtained from the matter-plus-field Arnowitt--Deser--Misner (ADM) Hamiltonian by eliminating the field degrees of freedom. The 2.5PN equations of motion were obtained in harmonic coordinates by \cite{DD81a, DD81b, D82, Dhouches}, building on a non-linear (post-Minkowskian) iteration of the metric of two particles initiated by \cite{BeDD81}. The corresponding result for the ADM-Hamiltonian of two particles at the 2PN order was given by \cite{DS85} (see also \citealt{S85, S86}). The 2.5PN equations of motion have also been derived in the case of two \emph{extended} compact objects by \cite{Kop85, GKop86}. The 2.5PN equations of two point masses as well as the near zone gravitational field in harmonic-coordinate were computed by \cite{BFP98}.

Up to the 2PN level the equations of motion are conservative. Only at the 2.5PN order does the first non-conservative effect appear, associated with the gravitational radiation emission. The equations of motion up to that level have been used for the study of the radiation damping of the binary pulsar -- its orbital ${\dot P}$ \citep{Dhouches, D83, DT91}. The result is in agreement with the prediction of the quadrupole formalism given by Eq.~\eqref{Pdot}. An important point is that the 2.5PN equations of motion have been proved to hold in the case of binary systems of strongly self-gravitating bodies \citep{Dhouches}. This is via the effacing principle for the internal structure of the compact bodies. As a result, the equations depend essentially only on the ``Schwarzschild'' masses $m_\text{a}$ of the neutron stars (and their spins). Notably the compactness parameters $K_\text{a}=\frac{G m_\text{a}}{r_\text{a} c^2}$ do not enter the equations of motion before the 5PN level (see the discussion in Sect.~\ref{sec:intstructure}). This fact has been explicitly verified up to the 2.5PN order by \cite{Kop85, GKop86}, who made a physical computation \emph{\`a la} Fock, taking into account the internal structure of two self-gravitating extended compact bodies. The 2.5PN equations of motion have also been obtained by \cite{IFA00, IFA01} in harmonic coordinates, using a variant (but, much simpler and more developed) of the surface-integral approach of \cite{EIH}, that is valid for compact bodies, independently of the strength of the internal gravity.

At the 3PN order the equations of motion have been worked out by several groups, using different methods, and with equivalent results:
\begin{enumerate}
	\item \cite{JaraS98, JaraS99, JaraS00, DJSpoinc, DJSequiv, DJSdim}, employ the ADM-Hamiltonian canonical formalism of general relativity; 
	\item \cite{BF00, BFeom, BFreg, BFregM, ABF01, BI03CM, BDE04} compute directly the equations of motion (instead of a Hamiltonian) in harmonic coordinates; 
	\item \cite{itoh1, itoh2, ItohLR} obtain the complete 3PN equations of motion in harmonic coordinates, without need of a self-field regularization; 
	\item \cite{FS3PN} derive the 3PN Lagrangian in harmonic coordinates within the effective field theory.
\end{enumerate}

The 3PN equations of motion contained, at some point, some unspecified numerical coefficient, called an ``ambiguity'', which is due to an incompleteness of the Hadamard self-field (ultra-violet, UV) regularization method, see Sect.~\ref{sec:had}. This coefficients has been fixed by means of dimensional regularization, both within the ADM-Hamiltonian formalism \citep{DJSdim}, and the harmonic-coordinates equations of motion \citep{BDE04}, while dimensional regularization is routinely employed by the effective field theory \citep{FS3PN}. All these works have demonstrated the power of dimensional regularization and its adequateness to the problem of interacting point masses in classical general relativity. By contrast, notice, interestingly, that the surface-integral method \citep{itoh1, itoh2} by-passes the need of a UV regularization. The regularizations -- both UV and infra-red (IR) -- are reviewed in Sect.~\ref{sec:reg}.

The effective field theory (EFT) approach -- sometimes coined the non-relativistic general relativity (NRGR) -- to the problems of motion and radiation of compact binaries, has been pioneered by \cite{GR06} and extensively developed since then (see \citealt{FSrevue, Portorevue, Levirevue} for reviews). It borrows techniques from quantum field theory and treats the gravitational interaction between point particles as a classical limit of a quantum field theory, i.e., in the ``tree level'' approximation. The theory is based on the effective action, defined from a Feynman path integral that integrates over the degrees of freedom that mediate the gravitational interaction. The phase factor in the path integral is built from the standard Einstein--Hilbert action for gravity, augmented by a harmonic gauge fixing term and by the action of particles. The Feynman diagrams naturally show up as a perturbative technique for solving iteratively the Green's functions.

By itself, computing the equations of motion and radiation field using Feynman diagrams in classical general relativity is not a new idea: \cite{BertottiP60} defined the diagrammatic tree-level perturbative expansion of the Green's functions in classical general relativity; \cite{HariDass}\footnote{This reference has an eloquent title: ``Feynman graph derivation of the Einstein quadrupole formula''.} showed how to derive the classical energy-loss formula at Newtonian approximation using tree-level propagating gravitons; Feynman diagrams have been used for the equations of motion up to 2PN order in general relativity \citep{OO73a, OO73b, OO74a, OO74b} and in scalar-tensor theories \citep{Dgef96}. Nevertheless, the systematic EFT treatment has proved to be powerful and innovative for the field, e.g., with the introduction of a decomposition of the metric into ``Kaluza-Klein type'' potentials \citep{KolS08}, the interesting link with the renormalization group equation \citep{GRoss10, GRR12}, and the systematization of the generation of diagrams in high PN approximations \citep{FS3PN}.

The 3.5PN terms in the equations of motion correspond to the 1PN relative corrections in the radiation reaction force. They were derived by \cite{IW93, IW95} for point-particle binaries in a general gauge and center-of-mass frame, relying on energy and angular momentum balance equations and the known expressions of the 1PN fluxes. These works have been extended to 2PN order \citep{GII97} and to include the leading spin-orbit effects \citep{ZW07}. The result has been established from ``first principles'' (i.e., not relying on balance equations) in various works at 1PN order \citep{JaraS97, PW02, KFS03, NB05, itoh3} and 2PN order \citep{LPY23}. The 1PN radiation reaction force has also been obtained for general extended fluid systems in the extended \cite{BuTh70, Bu71} gauge by \cite{B93, B97}. Known also is the contribution of gravitational-wave tails in the equations of motion, which arises at the 4PN order and contains both a conservative part and a 1.5PN relative modification of the radiation damping force \citep{BD88}. This 4PN tail-induced correction to the equations of motion has also been derived within the EFT approach \citep{FStail, GLPR16}.

The state-of-the-art on equations of motion is the 4PN approximation, with result in complete mutual agreement between the different derivations:
\begin{enumerate}
\item After partial results have been reported by \cite{JaraS12, JaraS13, JaraS15} using the ADM Hamiltonian formalism, and by \cite{FS4PN} using the EFT, the first derivation of the complete 4PN dynamics was obtained by \cite{DJS14}, combining the local contributions with the non-local (in time) term due to gravitational-wave tails \citep{BD88}. However this calculation contains one ``ambiguity'' parameter, due to the absence of control of the IR divergences, which has been fixed by matching the near-zone computation to auxiliary results imported from the gravitational self-force (GSF) theory (\citealt{BiniD13}, see also \citealt{BiniDG15, KOW15, HKO16}). The non-local 4PN Hamiltonian has been transformed into a local one using an order reduction and the dynamics has been transcribed into the effective one-body (EOB) formalism \citep{DJS15eob};
\item The 4PN equations of motion were obtained independently by \cite{BBBFMa, BBBFMb, BBBFMc, MBBF17, BBFM17} using the \cite{Fokker} Lagrangian approach in harmonic coordinates. This second derivation was actually the first one to derive the result from scratch without ambiguity parameter and without resorting to an auxiliary GSF calculation. This was achieved by using the dimensional regularization for both the UV and IR type divergences.\footnote{In fact a specific regularization procedure (called the ``$B\varepsilon$'' regularization) was used for the IR divergences, proven to be equivalent to dimensional regularization \citep{MBBF17}. We comment more on this procedure in Sects.~\ref{sec:DRrad} and \ref{sec:IRreg}.} The invariants of the motion (including periastron advance) and radiation-reaction terms were given by \cite{BBFM17};
\item The effective field theory (EFT) approach also obtained the complete, first-principle and ambiguity-free results at 4PN order \citep{FS4PN, FStail, GLPR16, FMSS17, PR17,FS19, FPRS19, BlumMMS20a}.
\end{enumerate}

Much efforts have been spent trying to extend the validity of the equations of motion beyond the 4PN order. The EFT approach in harmonic coordinates has been pushed to 5PN and even 6PN orders (including tails and hereditary effects) using \textit{ab initio} brute force calculations \citep{BlumMMS20b, BlumMMS21, BlumMMS22a}, and, in particular, by exploiting some property of factorisation of Feynman diagrams \citep{FS20, FS21, FST21, AFS23}. On the other hand a new methodology for deriving the equations of motion was introduced by \cite{BiniDG19}, combining several formalisms: post-Newtonian, post-Minkowskian, multipolar-post-Minkowskian, gravitational self-force, and effective one-body. This method (coined as ``Tutti-Frutti'') has been applied to the rederivation of the 3PM conservative Hamiltonian \citep{Bern19a} and to the computation of new coefficients in the equations of motion at 5PN and 6PN orders \citep{BiniDG20a, BiniDG20b, BiniDG20c, BiniDGLM21}. Unfortunately, despite these efforts the results are still incomplete: The 5PN Hamiltonian is determined except for two unknown numerical coefficients, in front of terms proportional to the square of the symmetric mass ratio $\nu$ and to the fifth and sixth power of the gravitational constant $G$. Such undetermined terms $\propto G^6\nu^2$ and $G^5\nu^2$ thus belong to a high post-Minkowskian approximation, and at the same time cannot be determined from the comparison with linear GSF; one would need at least the second-order GSF which is the current challenge in self-force calculations \citep{Poundetal20, WarburtonPound21, WPWMDL23}. Similarly, at the 6PN order the Hamiltonian contains four unknown numerical coefficients in factor of terms $\propto G^7\nu^3$, $G^7\nu^2$, $G^6\nu^2$ and $G^5\nu^2$. More work should be done to compute these coefficients.

Most of the works reviewed in this article concern general relativity. However, let us mention that the equations of motion of compact binaries in scalar-tensor theories have been developed up to 3PN order \citep{MW13, BernardST1, BernardST2}.

An important body of works in GR concerns the effects of spins on the equations of motion of compact bodies. The leading spin-orbit (SO) effect arises at the 1.5PN order while the leading spin-spin (SS) effect appears at 2PN order (see Sect.~\ref{sec:spins}). They have been computed using the traditional PN approach by \cite{KWW93, K95}. The next-to-leading SO effect, i.e., 1PN relative order corresponding to 2.5PN order, was obtained by \cite{TOO01, FBB06}. The results were also retrieved by two subsequent calculations, one using the ADM Hamiltonian \citep{DJSspin} and the other with the EFT method \citep{Le10so, Po10}. The ADM calculation was generalized by \cite{HaS11} to the $N$-body case and extended by \cite{SHS08a, SHS08b, SHS08c, HSS10, HaS11ss} to the next-to-leading spin-spin effects, including both the coupling between different spins and spin square terms, and the next-to-next-to-leading SS interactions between different spins at the 4PN order was obtained by \cite{HaS11ss}. Using the EFT method the next-to-leading 3PN SS and spin-squared contributions were derived by \cite{PoR06, PoR08a, PoR08b, Le10ss, LS15a}, and the next-to-next-to-leading 4PN SS interactions for different spins by \cite{Le12ss} and for spin-squared terms by \cite{LS16}. The next-to-next-to-leading order SO effects, corresponding to 3.5PN order equivalent to 2PN relative order, were obtained in the ADM-coordinates Hamiltonian by \cite{HaS11so, HaSS13} and in the harmonic-coordinates equations of motion by \cite{MBFB13, BMFB13}, with complete equivalence between the results. Comparisons between the EFT and ADM Hamiltonian schemes for high-order SO and SS couplings can be found in \cite{LS14, LS15a, LS15b}. The next-to-next-to-next-to-leading order (NNNLO) has been worked out by \cite{Antonelli2020a} (SO) and \cite{Antonelli2020b} (SS for different spins) with traditional method, and by \cite{Kim2023a, Mandal2023a} (SO) and \cite{Kim2023b, Mandal2023b} (SS) from the EFT.


\subsubsection{Post-Newtonian gravitational radiation}
\label{sec:PNrad}

The problem of the computation of the gravitational waveform (and the energy flux) is solved by application of a wave generation formalism valid for general isolated matter systems (see Sect.~\ref{sec:cocktail}). The earliest computation at the 1PN level beyond the quadrupole moment formalism was achieved by \cite{WagW76},\footnote{At a time when PN corrections to the emission of gravitational waves had only a purely academic interest.} based on the \cite{EW75} multipole moments, applied to compact binaries moving on eccentric orbits. This 1PN level calculation was redone and confirmed by \cite{BS89} using the \cite{BD86} multipole moments. 

At the 1.5PN order in the radiation field (beyond the quadrupole formula), appears the first ``hereditary'' contribution, which is \emph{a priori} sensitive to the entire past history of the source, i.e., depends on the source at all previous times up to $-\infty$ in the past up to current time \citep{BD88, BD92}. At 1.5PN order this hereditary term represents the dominant contribution of gravitational-wave tails in the wave zone -- a quadratic coupling between the quadrupole moment and the static mass of the source. It has been evaluated for compact binaries by \cite{P93a} in the small mass ratio limit, and in the general case by \cite{Wi93, BS93}. The 1.5PN tail term (and the radiation reaction contribution therein) is also known from \cite{FStail,GLPR16}.

Applying the general multipole moments at 2PN level \citep{B95}, the energy flux of compact binaries was completed to the 2PN order by \cite{BDI95, GI97}, and, independently, by \cite{WW96, W99}, using the DIRE formalism; see \cite{BDIWW95} for the joint report of these calculations. The waveform to 2PN order has been computed by \cite{BIWW96}. The energy flux and waveform have also been computed to 2PN order using the EFT approach, by \cite{GRoss10, LMRY20} with results agreeing with those of ``traditional'' methods.

Higher-order tail effects arise at the 2.5PN and have been incorporated in the 2.5PN wave generation formalism \citep{B96}. In the waveform at this order appears the leading contribution of a different hereditary effect called the ``non-linear memory'' \citep{B90mem, Chr91, WW91, Th92, BD92, B98quad}. The non-linearity is interpreted as due to the gravitational wave emission of gravitons themselves, following formulas for the emission by massless particles \citep{BragTh76, Th92}. The non-linear memory takes the form of a simple anti-derivative of an ``instantaneous'' term, and therefore becomes instantaneous in the energy flux, involving the time derivative of the waveform. In principle the memory contribution must be computed using some model for the evolution of the binary system in the past. Because of the cumulative effect of integration over the whole past, the memory term, though originating from 2.5PN order, finally contributes in the waveform at Newtonian order \citep{WW91, ABIQ04}. It represents a part of the waveform whose amplitude grows on a radiation reaction time scale, but is nearly constant over one orbital period; it is therefore essentially a \emph{zero-frequency} effect (or DC effect). The non-linear memory in the waveform of inspiralling compact binaries has been computed to high post-Newtonian order by \cite{F09, F11}. We give more details on this effect in Sect.~\ref{sec:memory}.

At the 3PN order appears the first cubic process, which is due to tails generated by tails themselves -- the so-called ``tails of tails'' \citep{B98tail, AFS21a}. The 3PN approximation also involves, besides the tails of tails, many non-tail contributions coming from the relativistic corrections in the (source type) multipole moments of the compact binary. The mass quadrupole moment at 3PN order has been obtained by \cite{BIJ02, BI04mult, BDEI05dr}. This calculation crucially requires dimensional regularization for treating UV divergences, see Sect.~\ref{sec:DRrad}. The wave generation from compact binaries is then complete to 3.5PN order \citep{BFIJ02, BDEI04}. The formalism is valid for general orbits, but often for applications to compact binaries the results are presented for quasi-circular orbits. In the case of non-circular orbits the 3PN energy and angular momentum fluxes, and the associated balance equations, are known \citep{ABIQ08tail, ABIQ08, ABIS09, LY17}, as well as the gravitational-wave modes, taking into account memory and tail contributions \citep{EBFMIJ19, BMCFGI19}. We discuss the problem of eccentric orbits in Sect.~\ref{sec:eccentric}.

In recent years the gravitational wave generation by compact binaries was developed up to still higher 4.5PN order. A central part of this program is the control of the mass quadrupole moment of compact binaries (without spins) with 4PN precision. In a preliminary calculation at 4PN order by \cite{MHLMFB20}, the UV divergences were properly treated by means of dimensional regularization, but the IR ones were regularized with the Hadamard partie finie regularization. This calculation was then completed by \cite{LHBF22, LBHF22} with: (i) the contributions from the dimensional regularization of the IR divergences; (ii) the non-local (in time) effect in the source quadrupole moment, due to the radiation modes associated with propagating tails at infinity. The latter effect is the analogue of the 4PN tail term in the conservative equations of motion and Lagrangian, see Eq.~\eqref{Ltail4PN}. The 3PN mass octupole moment and 3PN current quadrupole moment were also needed and derived by \cite{FBI15, HFB21}. By contrast, in scalar-tensor theory the waveform and flux have been so far developed to much lower order, see \cite{Lang14, Lang15, SMB16, BBT22}.

Interestingly, at the 4PN order in GR there is a new type of cubic effect, corresponding to the coupling between the static mass of the source and \textit{two} varying quadrupole moments. Physically it can be seen as a combination between the tails produced by the non-linear memory and the memory associated with the tail; thus we coin this effect ``tails-of-memory''. A thorough computation of the cubic tails of memory was performed by \cite{TB23} by means of the MPM formalism, together with a simpler cubic interaction between mass, spin and quadrupole moment, also arising at 4PN order. In particular it was shown that the tail of memory contains a DC contribution in perfect agreement with the general formalism for memory in Sect.~\ref{sec:memory}. An essential feature of this calculation is that it uses the MPM algorithm in radiative, non-harmonic coordinates as described in Sect.~\ref{sec:MPMrad}. A minor price to be paid is that the multipole moments in the radiative MPM construction are to be properly adjusted with respect to those in the corresponding harmonic MPM algorithm \citep{TLB22}. Finally at the 4.5PN order there is the first \textit{quartic} non-linear multipole interaction, between three static masses and the quadrupole; this effect, rightly called ``tail-of-tail-of-tail'' has been computed by \cite{MBF16, MNagar17}. 

These calculations lead to the complete 4.5PN energy flux for circular orbits, and the corresponding 4.5PN gravitational-wave frequency and phase evolution, reported in Sects.~\ref{sec:gravflux} and \ref{sec:orbevol} following \cite{BFHLT23a, BFHLT23b}. Concerning the gravitational-wave modes (reported in Sect.~\ref{sec:sphharm}), the dominant quadrupolar mode $(\ell,m)=(2,2)$ is derived at 4PN order while, all the other modes are computed to 3.5PN order in the waveform \citep{BFIS08, H23}.

The post-Newtonian results for the waveform and energy flux are in complete agreement (up to the 4.5PN order) with the results given by the very different technique of linear black-hole perturbations, valid when the mass of one of the bodies is small compared to the other. Linear black-hole perturbations in the radiation field, triggered by the geodesic motion of the small particle around the black hole, have been applied to this problem by \cite{P93a} at the 1.5PN order (following pioneering work by \citealt{Galtsov}), by \cite{TNaka94} using a numerical code up to the 4PN order, and by \cite{Sasa94, TSasa94, TTS96} (see also \citealt{MSSTT97}), analytically up to the 5.5PN order. They have been generalized to a rotating black hole by \cite{TSTS96}. More recently the method has been improved and extended up to extremely high 22PN order by \cite{Fuj14PN, Fuj22PN} using the \cite{MST96a, MST96b} method for analytical solution of the homogeneous Teukolsky equation -- but still for linear black-hole perturbations. The polarization modes and associated factorized resummed waveforms for circular orbits are known to 5.5PN order \citep{FI10}. For eccentric orbits (still in the small mass ratio limit) the energy and angular momentum fluxes are known to 9PN order \citep{FEH16, MEHF20}.

The spins (SO and SS) affect the gravitational waves through a modulation of their amplitude, phase and frequency. The orbital plane precesses in the case where the spins are not aligned or anti-aligned with the orbital angular momentum \citep{ACST94}. The leading SO and SS contributions in the waveform and flux of compact binaries are known from \cite{KWW93, K95, MVGer05}; the next-to-leading SO terms at order 2.5PN were obtained by \cite{BBF06} after a previous attempt by \cite{OTO98}; the 3PN SO contribution is due to tails and was computed by \cite{BBF11}, after intermediate results at the same order (but including SS terms) given by \cite{PRR10}. The next-to-next-to-leading SO contributions in the multipole moments and the energy flux, corresponding to 3.5PN order, and the next-to-leading SO tail corresponding to 4PN order, have been obtained by \cite{BMB13, MBBB13, CPY22}. The next-to-leading 3PN SS and spin-squared contributions in the radiation field were derived by \cite{BFMP15, CPP21}. The SO and SS effects in the spherical harmonic modes for non-circular orbits with non-precessing components have been computed by \cite{PM23} through 2PN order. In Sect.~\ref{sec:spins} we shall give the results for the SO contributions up to the 4PN order, and for the SS contributions up to the 3PN order, in the energy flux and gravitational wave phase evolution.

\section{Post-Newtonian isolated sources}
\label{sec:PNsource}


\subsection{Non-linear iteration of the field equations outside the source}
\label{sec:nonlinitere}


\subsubsection{Einstein's field equations}
\label{sec:EFE}

The field equations of general relativity are obtained by varying the action
\begin{equation}\label{EH}
	S = \frac{c^3}{16\pi G}\int \dd^4x\,\sqrt{-g}\,R +
	S_{m}\bigl[\Psi, g_{\alpha\beta}\bigr]\,,
\end{equation}
with respect to the space-time metric $g_{\alpha\beta}$, where the first term is the famous Einstein--Hilbert action for the gravitational field, and the second term is the matter action. The field equations form a system of ten second-order partial differential equations obeyed by the metric,
\begin{equation}\label{EinsteinG}
  G^{\alpha\beta}[g,\partial g,\partial^2g] = \frac{8\pi G}{c^4}
  T^{\alpha\beta}[\Psi, g]\,,
\end{equation}
where the Einstein curvature tensor $G^{\alpha\beta}\equiv R^{\alpha\beta}-\frac{1}{2}R \, g^{\alpha\beta}$ is generated, through the gravitational coupling constant $\kappa=8\pi G/c^4$, by the stress-energy tensor $T^{\alpha\beta}\equiv\frac{2}{\sqrt{-g}}\delta S_{m}/\delta g_{\alpha\beta}$ of the matter fields $\Psi$. Among these ten equations, four govern, via the contracted Bianchi identity, the evolution of the matter system, i.e. the covariant conservation of the matter stress-energy tensor,
\begin{equation}\label{Bianchi}
  \nabla_\mu G^{\alpha\mu} = 0 \quad \Longrightarrow \quad \nabla_\mu
  T^{\alpha\mu}=0\,.
\end{equation}
The matter field equations can be obtained by varying the matter action $S_{m}$ in \eqref{EH} with respect to each of the matter fields $\Psi$. Satisfying the matter field equations implies the conservation law \eqref{Bianchi} for the matter stress-energy tensor. The space-time geometry is constrained by the six remaining equations, which place six independent constraints on the ten components of the metric $g_{\alpha\beta}$, leaving four of them to be fixed by a choice of the coordinate system.

A widely adopted choice of coordinates are the so-called \emph{harmonic coordinates}, also known as \cite{deDonder} coordinates. We  define, as a basic variable, the gravitational-field amplitude (or ``gothic'' metric deviation from Minkowski space-time) as
\begin{equation}\label{hgothic}
  h^{\alpha\beta} \equiv \sqrt{-g}\, g^{\alpha\beta} -
  \eta^{\alpha\beta}\,,
\end{equation}
where $g^{\alpha\beta}$ denotes the contravariant metric (satisfying $g^{\alpha\mu}g_{\mu\beta}=\delta^\alpha_\beta$), where $g$ is the determinant of the covariant metric, $g \equiv \mathrm{det}(g_{\alpha\beta})$, and where $\eta^{\alpha\beta}$ represents an auxiliary Minkowskian metric $\eta^{\alpha\beta} \equiv \text{diag}(-1,1,1,1)$. The harmonic-coordinate condition, which accounts exactly for the four equations \eqref{Bianchi} corresponding to the conservation of the matter tensor, reads\footnote{Considering the coordinates $x^\alpha$ as a set of four \emph{scalars}, a simple calculation shows that
\begin{align*}
H^\alpha = \sqrt{-g}\,\Box_g x^\alpha\,,
\end{align*}
where $\Box_g\equiv g^{\mu\nu}\nabla_\mu\nabla_\nu$ denotes the curved d'Alembertian operator. Hence the harmonic-coordinate condition $H^\alpha = 0$ tells that the coordinates $x^\alpha$ themselves, considered as scalars, are harmonic, i.e., obey the vacuum (curved) d'Alembertian equation.} 
\begin{equation}\label{harmcond}
  H^\alpha \equiv \partial_\mu h^{\alpha\mu} = 0\,.
\end{equation}
Equation \eqref{harmcond} introduces into the definition of our coordinate system a preferred Minkowskian structure, with Minkowski metric $\eta_{\alpha\beta}$. Of course, this is not contrary to the spirit of general relativity, where there is only one physical metric $g_{\alpha\beta}$ without any flat prior geometry, because the coordinates are not governed by geometry (so to speak), but rather can be chosen at convenience, depending on physical phenomena under study. The coordinate condition \eqref{harmcond} is especially useful when studying gravitational waves as perturbations of space-time propagating on the fixed background metric $\eta_{\alpha\beta}$. This view is perfectly legitimate and represents a fruitful and rigorous way to think of the problem using approximation methods. Indeed, the metric $\eta_{\alpha\beta}$, originally introduced in the coordinate condition \eqref{harmcond}, does exist at any \emph{finite} order of approximation (neglecting higher-order terms), and plays the role of some physical ``prior'' flat geometry at any order of approximation. The harmonic gauge condition will be used for many investigations in this article;\footnote{It is conveniently imposed by adding a gauge-fixing term in the Einstein-Hilbert action, see e.g. \cite{Dgef96, FSrevue} and Sect.~\ref{sec:Fokker} below.} however for some applications it should be avoided, see in particular Sect.~\ref{sec:MPMrad}.

The Einstein field equations for the gothic metric deviation \eqref{hgothic} in a general coordinate system can be written in the following form,
\begin{equation}\label{EFE}
  \Box h^{\alpha\beta} - \partial H^{\alpha\beta} = \frac{16\pi G}{c^4} \tau^{\alpha\beta}\,,
\end{equation}
where $\Box\equiv\Box_\eta = \eta^{\mu\nu}\partial_\mu\partial_\nu$ is the flat d'Alembertian operator, and we have posed as useful shorthand (where $H^\alpha = \partial_\mu h^{\alpha\mu}$ is non-zero in general)
\begin{equation}\label{dHab}
	\partial H^{\alpha\beta} \equiv \partial^\alpha H^\beta + \partial^\beta H^\alpha - \eta^{\alpha\beta} \partial_\mu H^\mu\,.
\end{equation}
The source term $\tau^{\alpha\beta}$ can rightly be interpreted as the stress-energy \emph{pseudo}-tensor (actually, $\tau^{\alpha\beta}$ is a Lorentz-covariant tensor) of the matter fields, described by $T^{\alpha\beta}$, \emph{and} the gravitational field, given by the gravitational source term $\Lambda^{\alpha\beta}$, i.e.,
\begin{equation}\label{tauab}
  \tau^{\alpha\beta} = |g| T^{\alpha\beta}+ \frac{c^4}{16\pi
    G}\Lambda^{\alpha\beta}\,.
\end{equation}
The pseudo-tensor is conserved in an ordinary sense as a consequence of the Einstein field equations \eqref{EFE}, and this in turn is equivalent to the matter equations of motion \eqref{Bianchi}. Applying an ordinary divergence on \eqref{EFE}, and using the fact that $\partial_\mu\partial H^{\alpha\mu}=\Box H^\alpha$, we obtain
\begin{equation} \label{dtau0}
	\partial_\mu \tau^{\alpha\mu}=0
	\quad \Longleftrightarrow \quad
	\nabla_\mu T^{\alpha\mu}=0\,.
\end{equation}

The exact expression of $\Lambda^{\alpha\beta}$ reads
\begin{equation}
	\label{LambdaSource}
	\Lambda^{\alpha\beta} = h^{\mu\alpha}\partial_{\mu} H^{\beta} + h^{\mu\beta}\partial_{\mu} H^{\alpha} - \partial_\mu (h^{\alpha\beta} H^{\mu}) + \Lambda^{\alpha\beta}_\text{harm}\,,
\end{equation}
where $ \Lambda^{\alpha\beta}_\text{harm}$ is the source term  when assuming the harmonic gauge condition \eqref{harmcond}, and given, including all non-linearities, by\footnote{In $D=d+1$ space-time dimensions, only one coefficient in this expression is modified; see Eq.~\eqref{Lambdad} below.}
\begin{align}\label{Lambdadef}
  \Lambda^{\alpha\beta}_\text{harm} &= - h^{\mu\nu} \partial^2_{\mu\nu}
  h^{\alpha\beta}+\partial_\mu h^{\alpha\nu} \partial_\nu h^{\beta\mu}
  +\frac{1}{2}g^{\alpha\beta}g_{\mu\nu}\partial_\lambda h^{\mu\tau}
  \partial_\tau h^{\nu\lambda} \nn\\& -
  g^{\alpha\mu}g_{\nu\tau}\partial_\lambda h^{\beta\tau} \partial_\mu
  h^{\nu\lambda} -g^{\beta\mu}g_{\nu\tau}\partial_\lambda
  h^{\alpha\tau} \partial_\mu h^{\nu\lambda}
  +g_{\mu\nu}g^{\lambda\tau}\partial_\lambda h^{\alpha\mu}
  \partial_\tau h^{\beta\nu} \nn\\& +
  \frac{1}{8}\bigl(2g^{\alpha\mu}g^{\beta\nu}-g^{\alpha\beta}g^{\mu\nu}\bigr)
  \bigl(2g_{\lambda\tau}g_{\epsilon\pi}-g_{\tau\epsilon}g_{\lambda\pi}\bigr)
  \partial_\mu h^{\lambda\pi} \partial_\nu h^{\tau\epsilon}\,.
\end{align}

As is clear from its expression, $\Lambda^{\alpha\beta}$ is made of terms at least quadratic in the gravitational-field strength $h$ and its first and second space-time derivatives, while the left-hand side of \eqref{EFE} is linear in $h$. It looks hawkward, from a mathematical point of view, notably for setting the initial value problem, to include second order space-time (and in particular time) derivatives in the right-hand side of the differential equation \eqref{EFE}. But of course this is because these second time derivatives appear in terms quadratic in $h$ and can be treated perturbatively in the approximation scheme. In the following, for the highest post-Newtonian order that we shall consider, we will need the quadratic, cubic and quartic pieces of $\Lambda^{\alpha\beta}$; with obvious notation, we can write them as
\begin{equation}\label{LambdaNML}
  \Lambda^{\alpha\beta}_\text{harm} = N^{\alpha\beta} [h, h] + M^{\alpha\beta} [h,
    h, h] 
    + \calO(h^4)\,.
\end{equation}
These various terms can be straightforwardly computed from expanding
Eq.~\eqref{Lambdadef}; The leading quadratic and cubic pieces are explicitly given by\footnote{See Eqs. (3.8c) in \cite{BFeom} for the quartic terms. We denote e.g., $h^\alpha_\mu=\eta_{\mu\nu}h^{\alpha\nu}$, $h=\eta_{\mu\nu}h^{\mu\nu}$, and $\partial^\alpha=\eta^{\alpha\mu}\partial_\mu$. A parenthesis around a pair of indices denotes the usual symmetrization: $T^{(\alpha\beta)}=\frac{1}{2}(T^{\alpha\beta}+T^{\beta\alpha})$.}
\begin{subequations}\label{NMab}
	\begin{align}\label{Nab}
  N^{\alpha\beta} =& -h^{\mu\nu} \partial_{\mu\nu}^2
  h^{\alpha\beta}+\frac{1}{2} \partial^\alpha h_{\mu\nu}
  \partial^\beta h^{\mu\nu}-\frac{1}{4} \partial^\alpha h
  \partial^\beta h+ \partial_\nu h^{\alpha \mu} \bigl(\partial^\nu
  h^\beta_\mu+ \partial_\mu h^{\beta \nu} \bigr) \nn\\& -2
  \partial^{(\alpha} h_{\mu\nu} \partial^\mu h^{\beta) \nu} +
  \eta^{\alpha\beta} \Bigl[ -\frac{1}{4} \partial_\tau h_{\mu\nu}
    \partial^\tau h^{\mu\nu} +\frac{1}{8} \partial_\mu h \partial^\mu
    h+\frac{1}{2} \partial_\mu h_{\nu\tau} \partial^\nu h^{\mu \tau}
    \Bigr]\,.
    \\
    M^{\alpha\beta} =& - h^{\mu\nu}
    \Big(\partial^\alpha h_{\mu\tau} 
    \partial^\beta h^\tau_\nu+\partial_\tau h^\alpha_\mu \partial^\tau
    h^\beta_\nu -\partial_\mu h^\alpha_\tau \partial_\nu h^{\beta \tau}\Big)
    \nn\\& + h^{\alpha\beta} \bigg[
    -\frac{1}{4} \partial_\tau h_{\mu\nu} \partial^\tau
    h^{\mu\nu}+\frac{1}{8} \partial_\mu h \partial^\mu h+
    \frac{1}{2} \partial_\mu h_{\nu\tau} \partial^\nu h^{\mu\tau}
    \bigg]+ \frac{1}{2} h^{\mu\nu} \partial^{(\alpha} h_{\mu\nu}
    \partial^{\beta)} h \nn\\& + 2h^{\mu\nu} \partial_\tau
    h_\mu^{(\alpha} \partial^{\beta)} h_\nu^\tau+h^{\mu(\alpha} \bigg[
    \partial^{\beta)} h_{\nu\tau}\partial_\mu h^{\nu \tau}-2 \partial_\nu
    h_\tau^{\beta)} \partial_\mu h^{\nu\tau}-\frac{1}{2} \partial^{\beta)}
    h \partial_\mu h \bigg] \nn\\& + \eta^{\alpha\beta} \bigg[
    \frac{1}{8} h^{\mu\nu} \partial_\mu h \partial_\nu
    h-\frac{1}{4} h^{\mu\nu} \partial_\tau h_{\mu \nu}
    \partial^\tau h -\frac{1}{4} h^{\tau\lambda}
    \partial_\tau h_{\mu\nu} \partial_\lambda h^{\mu\nu}
    \nn
    \\ & - \frac{1}{2} h^{\tau\lambda} \partial_\mu h_{\tau\nu} 
    \partial^\nu h^\mu_\lambda+\frac{1}{2} h^{\tau\lambda}
    \partial_\mu h^\nu_\tau \partial^\mu h_{\lambda\nu} \bigg] \,,
\end{align}
\end{subequations}

In this article, we shall look for approximate solutions of the field equations \eqref{harmcond}--\eqref{EFE} under the following four hypotheses:
\begin{enumerate}
\item The matter stress-energy tensor $T^{\alpha\beta}$ is of spatially compact support, i.e., can be enclosed into some time-like world tube, say $r\leqslant a$, where $r=\vert\mathbf{x}\vert$ is the coordinate radial distance. Outside the domain of the source, when $r> a$, the gravitational source term, according to \eqref{dtau0}, is divergence-free,
  \begin{equation}\label{dLambda}
    \partial_\mu \Lambda^{\alpha\mu} = 0 \qquad (\text{when } r>a) \,;
  \end{equation} 
\item The matter distribution inside the source is smooth: $T^{\alpha\beta}\in C^\infty ({\mathbb{R}}^3)$.\footnote{$\mathbb{N}$, $\mathbb{Z}$, $\mathbb{R}$, and $\mathbb{C}$ are the usual sets of non-negative integers, integers, real numbers, and complex numbers; $C^p (\Omega)$ is the set of $p$-times continuously differentiable functions on the open domain $\Omega$ ($p\leqslant +\infty$).} We have in mind a smooth hydrodynamical fluid system, without any singularities nor shocks (\emph{a priori}), that is described by some Euler-type equations including high relativistic corrections. In particular, we \textit{a priori} exclude the presence of any black hole singularities; however, we shall return to this question in Sect.~\ref{sec:compactbinary} when we look for a model describing compact objects;
\item The source is \emph{post-Newtonian} in the sense of the existence of the small parameter defined by Eq.~\eqref{epsPN}. For such a source we assume the legitimacy of the method of matched asymptotic expansions for identifying the inner post-Newtonian field and the outer multipolar decomposition in the source's exterior near zone;
\item The gravitational field has been independent of time (stationary) in some remote past, i.e., before some finite instant $-\mathcal{T}$ in the past, namely
  \begin{equation}\label{statpast}
    \frac{\partial}{\partial t} \left[h^{\alpha\beta}(\mathbf{x},
      t)\right] = 0 \quad \text{when}\quad t\leqslant -\mathcal{T}\,.
  \end{equation}
\end{enumerate}
The latter condition is a means to impose, by ``brute force'', the \emph{no-incoming} radiation condition \citep{Dhouches}, ensuring that the matter source is isolated from the rest of the Universe and does not receive any radiation from infinity. Ideally, the no-incoming radiation condition should be imposed at past null infinity. As we shall see, this condition entirely fixes the radiation reaction forces inside the isolated source. We shall later argue (in Sect.~\ref{sec:gravtails}) that the condition of stationarity in the past \eqref{statpast}, although weaker than the ideal no-incoming radiation condition, does not entail essential physical restriction on the general validity of the formulas we derive. Furthermore, the condition \eqref{statpast} is actually better suited in the case of real astrophysical sources like inspiralling compact binaries, for which we do not know the details of the initial formation and remote past evolution. In practice the initial instant $-\mathcal{T}$ can be set when the two supernov{\ae} of the progenitor stars yielded two compact objects, perhaps right after the ``common enveloppe'' (CE) phase starting the close evolution of the compact binary system.


\subsubsection{Linearized vacuum equations}
\label{sec:linvac}

In what follows we shall solve the field equations \eqref{EFE}, in the \emph{vacuum} region outside the compact-support source, in the form of a formal non-linearity or \emph{post-Minkowskian} expansion, considering the field variable $h^{\alpha\beta}$ as a non-linear metric perturbation of Minkowski space-time, and further assuming formal multipole expansions. At the linearized level (or first-post-Minkowskian approximation), we write:
\begin{equation}\label{hMPM1}
  h^{\alpha\beta}_\text{MPM} = G \,h^{\alpha\beta}_{(1)} +
  \calO(G^2)\,,
\end{equation}
where the subscript MPM refers to the formal multipolar-post-Minkowskian expansion, which is physically meaningful only in the exterior of the source, and where we have introduced Newton's constant $G$ as a book-keeping parameter, enabling one to label conveniently the successive post-Minkowskian approximations. Since $h^{\alpha\beta}$ is a dimensionless variable, with our convention the linear coefficient $h^{\alpha\beta}_{(1)}$ in Eq.~\eqref{hMPM1} has the dimension of the inverse of $G$ (which should be a mass squared in a system of units where $\hbar=c=1$). In vacuum, further assuming the harmonic coordinate condition \eqref{harmcond}, the linearized metric coefficient $h^{\alpha\beta}_{(1)}$ satisfies
\begin{subequations}\label{eqh1}
	\begin{align}
		\Box h_{(1)}^{\alpha\beta} &= 0\,,
		\label{eqh1a}\\
		\partial_\mu h_{(1)}^{\alpha\mu} &= 0\,.
		\label{eqh1b}
\end{align}\end{subequations}
We want to solve those equations by means of an infinite multipolar series valid outside a time-like world tube containing the source. Indeed the multipole expansion is the appropriate method for describing the physics of the source as seen from its exterior ($r>a$). On the other hand, the post-Minkowskian series is physically valid in the weak-field region, which surely includes the exterior of any source, starting at a sufficiently large distance. For post-Newtonian sources the exterior weak-field region, where both multipole and post-Minkowskian expansions are valid, simply coincides with the exterior region $r>a$. It is therefore quite natural, and even, one would say inescapable when considering general sources, to combine the post-Minkowskian approximation with the multipole decomposition. This is the original idea of the ``double-expansion'' series of \cite{Bo59,BoR61,BoR66,HR69}, which combines the $G$-expansion (or $m$-expansion in their notation) with the $a$-expansion (equivalent to the multipole expansion, since the $\ell$-th order multipole moment scales with the source radius like $a^\ell$).

The multipolar-post-Minkowskian (MPM) method will be implemented systematically, using symmetric-trace-free (STF) harmonics to describe the multipole expansion \citep{Th80}, and looking for a definite ``\emph{algorithm}'' for implementing the approximation scheme \citep{BD86}. The solution of the system of equations \eqref{eqh1} takes the form of a series of retarded multipolar waves\footnote{Our notation is the following: $L=i_1i_2\cdots i_\ell$ denotes a multi-index, made of $\ell$ (spatial) indices. Similarly, we write for instance $P=j_1\cdots j_p$ (in practice, we generally do not need to write explicitly the ``carrier'' letter $i$ or $j$), or $aL-1=ai_1\cdots i_{\ell-1}$. Always understood in expressions such as Eq.~\eqref{h1solK} are $\ell$ summations over the indices $i_1, \cdots, i_\ell$ ranging from 1 to 3. The derivative operator $\partial_L$ is a short-hand for $\partial_{i_1}\cdots\partial_{i_\ell}$. The function $\mathrm{K}_L$ (for any space-time indices $\alpha\beta$) is \emph{symmetric and trace-free} (STF) with respect to the $\ell$ indices composing $L$. This means that for any pair of indices $i_p, i_q\in L$, we have $\mathrm{K}_{\cdots i_p\cdots i_q\cdots}=\mathrm{K}_{\cdots i_q\cdots i_p\cdots}$ and that $\delta_{i_pi_q}\mathrm{K}_{\cdots i_p\cdots i_q\cdots}=0$ (see \citealt{Th80} and Appendices A and B in \citealt{BD86} for reviews about the STF formalism). The STF projection is denoted with a hat, so $\mathrm{K}_L\equiv \hat{\mathrm{K}}_L$, or sometimes with carets around the indices, $\mathrm{K}_L\equiv \mathrm{K}_{\langle L \rangle}$. In particular, $\hat{n}_L=n_{\langle L \rangle}$ is the STF projection of the product of unit vectors $n_L=n_{i_1}\cdots n_{i_\ell}$, for instance $\hat{n}_{ij}=n_{\langle ij\rangle}=n_{ij}-\frac{1}{3}\delta_{ij}$ and $\hat{n}_{ijk} =n_{\langle ijk\rangle} = n_{ijk} - \frac{1}{5} (\delta_{ij} n_k + \delta_{ik} n_j + \delta_{jk} n_i)$; an expansion into STF tensors $\hat{n}_L=\hat{n}_L(\theta,\phi)$ is equivalent to the usual expansion in spherical harmonics $Y_{lm}=Y_{lm}(\theta,\phi)$, see Eqs. \eqref{NY} below. Similarly, we denote $x_L=x_{i_1}\cdots x_{i_\ell}=r^l n_{L}$ where $r=\vert\mathbf{x}\vert$, and $\hat{x}_L=x_{\langle L \rangle}=\text{STF}[x_L]$. The Levi-Civita antisymmetric symbol is denoted $\epsilon_{ijk}$ (with $\epsilon_{123}=1$). Parenthesis refer to symmetrization, $T_{(ij)}=\frac{1}{2}(T_{ij}+T_{ji})$. Superscripts $(q)$ indicate $q$ successive time derivations.\label{fnote:notation}}
\begin{equation}\label{h1solK}
  h_{(1)}^{\alpha\beta}=\sum_{\ell=0}^{+\infty}
  \partial_L\left(\frac{\mathrm{K}_L^{\alpha\beta}(t-r/c)}{r}\right)\,,
\end{equation}
where $r=\vert\mathbf{x}\vert$, and where the functions $\mathrm{K}_L^{\alpha\beta}\equiv \mathrm{K}_{i_1\cdots i_\ell}^{\alpha\beta}$ are smooth functions of the retarded time $u\equiv t-r/c$ [i.e., $\mathrm{K}_L(u)\in C^\infty (\mathbb{R})$], which become constant in the past, when $t\leqslant -\mathcal{T}$, see Eq.~\eqref{statpast}. Since a monopolar wave satisfies $\Box (\mathrm{K}_L (u) / r) = 0$ and the d'Alembertian commutes with the multi-derivative $\partial_L$, it is evident that Eq.~\eqref{h1solK} represents the most general solution of the wave equation \eqref{eqh1a}; but see Sect.~2 of \cite{BD86} for a rigorous proof based on the Euler-Poisson-Darboux equation. The gauge condition \eqref{eqh1b}, however, is not fulfilled in general, and to satisfy it we must algebraically decompose the set of functions $\mathrm{K}^{00}_L$, $\mathrm{K}^{0i}_L$, $\mathrm{K}^{ij}_L$ into ten tensors which are STF with respect to all their indices, including the spatial indices $i$, $ij$. Imposing the condition \eqref{eqh1b} reduces the number of independent tensors to six, and we find that the solution takes an especially simple ``canonical'' form, parametrized by only two moments, plus some arbitrary linearized gauge transformation.

\begin{theorem} \citep{SB58, Pi64, Th80, BD86} The most general solution of the linearized field equations \eqref{eqh1} outside some time-like world tube enclosing the source ($r>a$), and stationary in the past [see Eq.~\eqref{statpast}], reads
  \begin{equation}\label{hk1}
    h^{\alpha\beta}_{(1)} = h^{\alpha\beta}_{\text{can}\,(1)} +
    \partial\varphi^{\alpha\beta}_{(1)}\,.
  \end{equation}
The first term is referred to as a ``canonical'' solution and depends on two STF-tensorial multipole moments, $\dI_L(u)$ and $\dJ_L(u)$, which are arbitrary functions of time except for the laws of conservation of the monopole: $\dI = \mathrm{const}$, and dipoles: $\dI_i = \mathrm{const}$, $\dJ_i = \mathrm{const}$. It is given by
  \begin{subequations}\label{k1sol}
    \begin{align}
      h^{00}_{\text{can}\,(1)} &= -\frac{4}{c^2}\sum_{\ell\geqslant 0}
      \frac{(-)^\ell}{\ell !} \partial_L \left( \frac{1}{r}
      \dI_L (u)\right)\,, \\ h^{0i}_{\text{can}\,(1)} &=
      \frac{4}{c^3}\sum_{\ell\geqslant 1} \frac{(-)^\ell}{\ell !}
      \left\{ \partial_{L-1} \left( \frac{1}{r}
      \dI_{iL-1}^{(1)} (u)\right) + \frac{\ell}{\ell+1}
      \epsilon_{iab} \partial_{aL-1} \left( \frac{1}{r}
      \dJ_{bL-1} (u)\right)\right\}\,, \\ h^{ij}_{\text{can}\,(1)} &=
      -\frac{4}{c^4}\sum_{\ell\geqslant 2} \frac{(-)^\ell}{\ell !}
      \left\{ \partial_{L-2} \left( \frac{1}{r}
      \dI_{ijL-2}^{(2)} (u)\right) + \frac{2\ell}{\ell+1}
      \partial_{aL-2} \left( \frac{1}{r} \epsilon_{ab(i}
      \dJ_{j)bL-2}^{(1)} (u)\right)\right\}\,.
    \end{align}
  \end{subequations}
The second term, $\partial\varphi^{\alpha\beta}_{(1)} \equiv \partial^\alpha\varphi^\beta_{(1)} + \partial^\beta\varphi^\alpha_{(1)} - \eta^{\alpha\beta}\partial_\mu\varphi^\mu_{(1)}$, represents a linearized gauge transformation, with gauge vector $\varphi^\alpha_{(1)}$ parametrized by four other multipole moments, say $\dW_L(u)$, $\dX_L(u)$, $\dY_L(u)$ and $\dZ_L(u)$ [see Eqs. \eqref{phi1expr}].
  \label{th:linsol}
\end{theorem}
The conservation of the lowest-order moments gives the constancy of the total mass of the source, $\dM \equiv \dI = \mathrm{const}$, center-of-mass position, $\dX_i \equiv \dI_i / \dI = \mathrm{const}$, total linear momentum $\mathrm{P}_i \equiv \dI_i^{(1)} = 0$,\footnote{The constancy of the center of mass $\dX_i$ -- rather than a linear variation with time -- results from our assumption of stationarity before the date $-\mathcal{T}$, see Eq.~\eqref{statpast}. Hence, $\mathrm{P}_i = 0$.}  and total angular momentum, $\dJ_i = \mathrm{const}$.  It is always possible to achieve $\dX_i=0$ by translating the origin of our coordinates to the center of mass. The total mass $\dM$ is the ADM mass of the Hamiltonian formulation of general relativity. Note that the quantities $\dM$, $\dX_i$, $\mathrm{P}_i$ and $\dJ_i$ include the contributions due to the waves emitted by the source. They describe the initial state of the source, before the emission of gravitational waves.

The multipole functions $\dI_L(u)$ and $\dJ_L(u)$, which thoroughly encode the physical properties of the source at the linearized level (because the other moments $\dW_L, \cdots, \dZ_L$ parametrize a gauge transformation), will be referred to as the \emph{mass-type} and \emph{current-type} source multipole moments. Beware, however, that at this stage the moments are not specified in terms of the stress-energy tensor $T^{\alpha\beta}$ of the source: Theorem \ref{th:linsol} follows merely from the algebraic and differential properties of the vacuum field equations outside the source.

For completeness, we give the expressions of the components of the gauge-vector $\varphi^\alpha_{(1)}$ entering Eq.~\eqref{hk1}:
\begin{subequations}\label{phi1expr}
  \begin{align}
    \varphi^0_{(1)} &= {4\over c^3}\sum_{\ell\geqslant 0}
           {(-)^\ell\over \ell !} \partial_L \left( {1\over r}
           \dW_L (u)\right)\,, \\ \varphi^i_{(1)} &= -{4\over
             c^4}\sum_{\ell\geqslant 0} {(-)^\ell\over \ell !}
           \partial_{iL} \left( {1\over r} \dX_L (u)\right)
           \\ & -{4\over c^4}\sum_{\ell\geqslant 1} {(-)^\ell\over
             \ell !} \left\{ \partial_{L-1} \left( {1\over r}
           \dY_{iL-1} (u)\right) + {\ell\over \ell+1}
           \epsilon_{iab} \partial_{aL-1} \left( {1\over r}
           \dZ_{bL-1} (u)\right)\right\}\,.\nn
  \end{align}
\end{subequations}
Because the theory is covariant with respect to non-linear diffeomorphisms and not merely with respect to linear gauge transformations, the moments $\dW_L, \cdots, \dZ_L$ do play a physical role starting at the non-linear level, in the following sense. If one takes these moments equal to zero and continues the post-Minkowskian iteration [see Sect.~\ref{sec:MPMsolution}] one ends up with a metric depending on $\dI_L$ and $\dJ_L$ only, but that metric will not describe the same physical source as the one which would have been constructed starting from the six moments $\dI_L, \dJ_L, \cdots, \dZ_L$ altogether. In other words, the two non-linear metrics associated with the sets of multipole moments $\{\dI_L, \dJ_L, 0, \cdots, 0\}$ and $\{\dI_L, \dJ_L, \dW_L, \cdots, \dZ_L\}$ are not physically equivalent -- they are not isometric. We shall point out in Sect.~\ref{sec:sourcecan} below that the full set of moments $\{\dI_L, \dJ_L, \dW_L, \cdots, \dZ_L\}$ is in fact physically equivalent to some other reduced set of moments $\{\dM_L, \dS_L, 0, \cdots, 0\}$, but with some moments $\dM_L$, $\dS_L$ that differ from $\dI_L$, $\dJ_L$ by non-linear corrections, see e.g. Eq.~\eqref{relationMijIij}. The moments $\dM_L$, $\dS_L$ are called ``canonical'' moments; they play an important role when describing the radiation emitted by the isolated source. All the multipole moments $\dI_L$, $\dJ_L$, $\dW_L$, $\dX_L$, $\dY_L$, $\dZ_L$ will be computed in Sect.~\ref{sec:sourcemoments}.


\subsubsection{The multipolar post-Minkowskian solution in harmonic coordinates}
\label{sec:MPMsolution}

By Theorem \ref{th:linsol} we know the most general solution of the linearized equations in the exterior of the source. We then tackle the problem of the post-Minkowskian iteration of that solution. We consider the full post-Minkowskian series
\begin{equation}\label{hextMPM}
  h^{\alpha\beta}_\text{MPM}=\sum_{n=1}^{+\infty}G^n\,h^{\alpha\beta}_{(n)}\,,
\end{equation}
where the first term is composed of the result given by Eqs. \eqref{hk1}--\eqref{phi1expr}. In this article, we shall always understand the infinite sums such as the one in Eq.~\eqref{hextMPM} in the sense of \emph{formal} power series, i.e., as an ordered collection of coefficients, $(h^{\alpha\beta}_{(n)})_{n\in \mathbb{N}}$. We do not attempt to control the mathematical nature of the series and refer to the mathematical-physics literature for discussion in the present context \citep{CS79, DS90, Rend90, Rend92, Rend94}.

We substitute the post-Minkowski ansatz \eqref{hextMPM} into the vacuum harmonic coordinate Einstein field equations \eqref{harmcond}--\eqref{EFE}, i.e., with $\tau^{\alpha\beta}$ simply given by the gravitational source term $\Lambda^{\alpha\beta}=\Lambda^{\alpha\beta}_\text{harm}$, and we equate term by term the factors of the successive powers of our book-keeping parameter $G$. We get an infinite set of equations for each of the $h^{\alpha\beta}_{(n)}$'s: namely, $\forall n \geqslant 2$,
\begin{subequations}\label{eqhn}
\begin{align}
  \Box h_{(n)}^{\alpha\beta}&=\Lambda_{(n)}^{\alpha\beta}\bigl[h_{(1)},
    h_{(2)},\cdots, h_{(n-1)}\bigr]\,,
  \label{eqhna}
  \\ \partial_\mu h_{(n)}^{\alpha\mu}&=0\,.
  \label{eqhnb}
\end{align}
\end{subequations}
The right-hand side of the wave equation \eqref{eqhna} is obtained from inserting the previous iterations, known up to the order $n-1$, into the gravitational source term. In more details, the series of equations \eqref{eqhna} reads
\begin{subequations}\label{itere}
\begin{align}
  \Box h_{(2)}^{\alpha\beta} &= N^{\alpha\beta}\bigl[h_{(1)}, h_{(1)}\bigr]\,,
  \label{iterea}
  \\ \Box h_{(3)}^{\alpha\beta} &= M^{\alpha\beta}\bigl[h_{(1)}, h_{(1)},
    h_{(1)}\bigr]+ N^{\alpha\beta}\bigl[h_{(1)},
    h_{(2)}\bigr]+N^{\alpha\beta}\bigl[h_{(2)}, h_{(1)}\bigr]\,.
  \label{itereb}
  \\ &~~\vdots& \nn
\end{align}
\end{subequations}
The quadratic and cubic pieces of $\Lambda^{\alpha\beta}_\text{harm}$ are defined by Eqs. \eqref{NMab}.

Let us now proceed by induction. Some $n\in\mathbb{N}$ being given, we assume that we succeeded in constructing, starting from the linearized solution $h_{(1)}$, the sequence of post-Minkowskian solutions $h_{(2)}, h_{(3)}, \cdots, h_{(n-1)}$, and from this we want to infer the next solution $h_{(n)}$. The right-hand side of Eq.~\eqref{eqhna}, $\Lambda_{(n)}^{\alpha\beta}$, is known by induction hypothesis. Thus the problem is that of solving a flat wave equation whose source is given. The point is that this wave equation, instead of being valid everywhere in ${\mathbb{R}}^3$, is physically correct only outside the matter source ($r>a$), and it makes no sense to solve it by means of the usual retarded integral. Technically speaking, the right-hand side of Eq.~\eqref{eqhna} is composed of the product of many multipole expansions, which are singular at the origin of the spatial coordinates $r=0$, and which make the retarded integral divergent at that point. This does not mean that there are no solutions to the wave equation, but simply that the retarded integral does not constitute the appropriate solution in that context.

What we need is a solution which takes the same structure as the source term $\Lambda_{(n)}^{\alpha\beta}$, i.e., is expanded into multipole contributions, with a singularity at $r=0$, and satisfies the d'Alembertian equation as soon as $r>0$. Such a particular solution can be obtained by means of a mathematical trick, in which one first ``regularizes'' the source term $\Lambda_{(n)}^{\alpha\beta}$ by multiplying it by the factor $r^B$, where $r=\vert\mathbf{x}\vert$ is the spatial radial distance and $B$ is a complex number, $B\in \mathbb{C}$ \citep{BD86}. 

Let us assume, for definiteness, that $\Lambda^{\alpha\beta}_{(n)}$ is composed of multipolar pieces with maximal multipolarity $\ell_\mathrm{max}$. This means that we start the iteration from the linearized metric \eqref{hk1}--\eqref{phi1expr} in which the multipolar sums are actually finite.\footnote{This assumption is justified because we are ultimately interested in the radiation field at some given \emph{finite} post-Newtonian precision like 4PN, and because only a finite number of multipole moments can contribute at any finite order of approximation. With a finite number of multipoles in the linearized metric \eqref{hk1}--\eqref{phi1expr}, there is a maximal multipolarity $\ell_\mathrm{max}(n)$ at any post-Minkowskian order $n$, which grows linearly with $n$.} The divergences when $r\to 0$ of the source term are typically power-like, say $1/r^k$ (there are also powers of the logarithm of $r$), and with the previous assumption there will exist a maximal order of divergency, say $k_\mathrm{max}$. Thus, when the real part of $B$ is large enough, i.e., $\Re (B) > k_\mathrm{max}-3$, the ``regularized'' source term $r^B \Lambda^{\alpha\beta}_{(n)}$ is regular enough when $r\to 0$ so that one can perfectly apply the retarded integral operator. This defines the $B$-dependent retarded integral, when $\Re (B)$ is large enough,
\begin{equation}\label{IB}
  \mathcal{I}^{\alpha\beta}(B) \equiv \Box^{-1}_{\mathrm{ret}} \left[ {\widetilde
      r}^B \Lambda^{\alpha\beta}_{(n)} \right]\,,
\end{equation}
where the symbol $\Box^{-1}_{\mathrm{ret}}$ stands for the usual retarded d'Alembertian integral, extending over all space $\mathbf{x}'\in\mathbb{R}^3$ and given by
\begin{equation}\label{dalembertian}
	(\Box^{-1}_{\mathrm{ret}} f)(\mathbf{x}, t) \equiv -\frac{1}{4\pi}
	\int \!\!\!\! \int \!\!\!\! \int
	\frac{\dd^3\mathbf{x}'}{|\mathbf{x}-\mathbf{ x}'|} f
	\bigl(\mathbf{x}', t-|\mathbf{x}-\mathbf{x}'|/c\bigr)\,.
\end{equation}
Subject to the past-stationarity condition \eqref{statpast}, it is clear that the retarded integral will always be convergent at infinity.

It is convenient to introduce together with the regularizing factor $r^B$ some arbitrary constant length scale $r_0$ in order to make it dimensionless. Everywhere in this article we pose
\begin{equation}\label{regfactor}
  \widetilde{r}\equiv \frac{r}{r_0}\,.
\end{equation}
The fate of the constant $r_0$ in a detailed calculation will be interesting to follow, as we shall see. Now the point for our purpose is that the function $\mathcal{I}^{\alpha\beta}(B)$ on the complex plane, which was originally defined only when $\Re (B) > k_\mathrm{max}-3$, admits a unique \emph{analytic continuation} to all values of $B\in \mathbb{C}$ except at some integer values. Furthermore, the analytic continuation of $\mathcal{I}^{\alpha\beta}(B)$ can be expanded, when $B\to 0$ (namely the limit of interest to us) into a Laurent expansion involving in general some multiple poles. The key idea, as we shall prove, is that the \emph{finite part}, or the coefficient of the zeroth power of $B$ in that expansion, represents the particular solution we are looking for. We write the Laurent expansion of $\mathcal{I}^{\alpha\beta}(B)$, when $B\to 0$, in the form
\begin{equation}\label{Laurent}
  \mathcal{I}^{\alpha\beta}(B) = \sum_{p=p_0}^{+\infty} I^{\alpha\beta}_p
  B^p\,,
\end{equation}
where $p\in \mathbb{Z}$, and the various coefficients $I^{\alpha\beta}_p$ are functions of the field point $(\mathbf{x}, t)$. When $p_0\leqslant -1$ there are poles; and $-p_0$, which depends on the post-Minkowskian order $n$, refers to the maximal order of these poles (we have $p_0\to-\infty$ when $n\to+\infty$). By applying the d'Alembertian operator onto both sides of Eq.~\eqref{Laurent}, and equating the different powers of $B$, we arrive at
\begin{subequations}\label{resp}
  \begin{align}
    p_0\leqslant p\leqslant -1 \quad \Longrightarrow \quad \Box
    I^{\alpha\beta}_p &= 0\,, \\ p\geqslant 0 \quad
    \Longrightarrow \quad \Box I^{\alpha\beta}_p &= \frac{(\ln
      r)^p}{p!}\Lambda^{\alpha\beta}_{(n)}\,.
  \end{align}
\end{subequations}
As we see, the case $p=0$ shows that the finite-part coefficient in Eq.~\eqref{Laurent}, namely $I^{\alpha\beta}_0$, is a particular solution of the requested equation: $\Box I^{\alpha\beta}_0=\Lambda^{\alpha\beta}_{(n)}$. Furthermore, we can prove that this solution, by its very construction, owns the same structure made of a multipolar expansion singular at $r=0$ as the corresponding source.

Let us forget about the intermediate name $I^{\alpha\beta}_0$, and denote, from now on, the latter solution by $u^{\alpha\beta}_{(n)}\equiv I^{\alpha\beta}_0$, or, in more explicit terms,
\begin{equation}\label{ungen}
	u^{\alpha\beta}_{(n)} = \FPprop
	\left[ \widetilde{r}^B \Lambda^{\alpha\beta}_{(n)} \right]\,,
\end{equation}
where the finite-part symbol $\FP$ means the previously detailed operations of considering the analytic continuation, taking the Laurent expansion, and picking up the finite-part coefficient when $B\to 0$. The story is not complete, however, because $u^{\alpha\beta}_{(n)}$ does not fulfill the constraint of harmonic coordinates \eqref{eqhnb}; its divergence, say $w_{(n)}^\alpha=\partial_\mu u_{(n)}^{\alpha\mu}$, is different from zero in general. From the fact that the source term is divergence-free in vacuum, $\partial_\mu \Lambda_{(n)}^{\alpha\mu}=0$ [see Eq.~\eqref{dLambda}], we find instead
\begin{equation}\label{wngen}
  w^\alpha_{(n)} = \FPprop \left[
    B \, \widetilde{r}^B \frac{n_i}{r} \Lambda^{\alpha i}_{(n)}
    \right]\,.
\end{equation}
The factor $B$ comes from the differentiation of the regularization factor $\widetilde{r}^B$. So, $w^\alpha_{(n)}$ is zero only in the special case where the Laurent expansion of the retarded integral in Eq.~\eqref{wngen} does not develop any simple pole when $B\to 0$. Fortunately, when it does, the structure of the pole is quite easy to control. We find that it necessarily consists of an homogeneous solution of the \emph{source-free} d'Alembertian equation, and, what is more (as consequence of its stationarity in the past), that solution is a retarded one. Hence, taking into account the index structure of $w^\alpha_{(n)}$, there must exist four STF-tensorial functions of $u=t-r/c$, say $N_L$, $P_L$, $Q_L$ and $R_L$, such that
\begin{subequations}\label{wnexpl}
  \begin{align}
    w^0_{(n)} =& \sum_{\ell = 0}^{+\infty}\partial_L \left[r^{-1}
      N_L(u)\right]\,, \\ w^i_{(n)} =& \sum_{\ell =
      0}^{+\infty}\partial_{iL} \left[ r^{-1} P_L (u) \right] \nn\\ +&
  \sum_{\ell = 1}^{+\infty} \Bigl\{ \partial_{L-1} \left[ r^{-1}
      Q_{iL-1} (u) \right] + \epsilon_{iab} \partial_{aL-1}
    \left[r^{-1} R_{bL-1} (u) \right] \Bigr\}\,.
  \end{align}
\end{subequations}
From that expression we are able to find a new object, say $v_{(n)}^{\alpha\beta}$, which takes the same structure as $w^\alpha_{(n)}$ (a retarded solution of the source-free wave equation) and, furthermore, whose divergence is exactly the opposite of the divergence of $u_{(n)}^{\alpha\beta}$, i.e. $\partial_\mu v_{(n)}^{\alpha\mu}=-w_{(n)}^\alpha$. Such a $v_{(n)}^{\alpha\beta}$ is not unique, but we shall see that it is just necessary to make a choice for $v_{(n)}^{\alpha\beta}$ (the simplest one) in order to obtain the general solution. The formulas that we adopt are
\begin{subequations}\label{vnexpl}
  \begin{align}
    v^{00}_{(n)} =& - r^{-1} N^{(-1)} + \partial_a \left[ r^{-1}
      \left(- N^{(-1)}_a+ Q^{(-2)}_a -3P_a\right) \right]\,,
    \\ v^{0i}_{(n)} =& r^{-1} \left( - Q^{(-1)}_i +3 P^{(1)}_i\right)
    - \epsilon_{iab} \partial_a \left[ r^{-1} R^{(-1)}_b \right] -
    \sum_{\ell = 2}^{+\infty}\partial_{L-1} \left[ r^{-1} N_{iL-1}
      \right]\,, \\ v^{ij}_{(n)} =& - \delta_{ij} r^{-1} P + \sum_{\ell =
      2}^{+\infty} \biggl\{ 2 \delta_{ij}\partial_{L-1} \left[ r^{-1}
      P_{L-1}\right] - 6 \partial_{L-2(i} \left[ r^{-1}
      P_{j)L-2}\right] \\ & +
    \partial_{L-2} \left[ r^{-1} (N^{(1)}_{ijL-2} + 3 P^{(2)}_{ijL-2}
      - Q_{ijL-2}) \right] - 2 \partial_{aL-2}\left[ r^{-1}
      \epsilon_{ab(i} R_{j)bL-2} \right] \biggr\}\,.\nn
  \end{align}
\end{subequations}
Notice the presence of anti-derivatives, denoted e.g. $N^{(-1)}(u)=\int_{-\infty}^u \dd v \,N(v)$; there is no problem with the limit $v\to -\infty$ since all the corresponding functions are zero when $t\leqslant -\mathcal{T}$. The choice made in Eqs. \eqref{vnexpl} is dictated by the fact that the $00$ component involves only some monopolar and dipolar terms, and that the spatial trace $ii$ is monopolar: $v^{ii}_{(n)} = -3 r^{-1} P$. In the following we shall refer to the previous operation $w_{(n)}^{\alpha} \longrightarrow v_{(n)}^{\alpha\beta}$ as the ``harmonicity'' algorithm and denote it by $\mathcal{V}^{\alpha\beta}$ so that 
\begin{equation}\label{algoV}
	v_{(n)}^{\alpha\beta} \equiv \mathcal{V}^{\alpha\beta}\bigl[w_{(n)}\bigr]\,.
\end{equation}
Finally we see that we solve at once the d'Alembertian equation \eqref{eqhna} \emph{and} the coordinate condition \eqref{eqhnb} if we pose
\begin{equation}\label{hnexpl}
  h_{(n)}^{\alpha\beta}=u_{(n)}^{\alpha\beta}+v_{(n)}^{\alpha\beta}\,.
\end{equation}
That is, we have succeeded in finding a solution of the field equations at the $n$-th post-Minkowskian order. By induction the same method applies to \emph{any} order $n$, and, therefore, we have constructed a complete post-Minkowskian series \eqref{hextMPM} based on the linearized approximation $h^{\alpha\beta}_{(1)}$ given by Eqs. \eqref{hk1}--\eqref{phi1expr}. The previous procedure constitutes an \emph{algorithm}, which has been implemented on a computer programme by \cite{BFIS08, FMBI12}.


\subsubsection{Source multipole moments versus canonical moments}
\label{sec:sourcecan}

The six sets of multipole moments $\dI_L(u), \cdots, \dZ_L(u)$ contain the physical information about \emph{any} isolated source as seen in its exterior. Indeed we have the following result.

\begin{theorem} \citep{BD86} The most general solution of the harmonic-coordinates Einstein field equations in the vacuum region outside an isolated source, admitting some post-Minkowskian and multipolar expansions, is given by the previous construction as
\begin{equation}\label{hIJZ}
		h_{\text{\rm{MPM}}}^{\alpha\beta}=\sum_{n=1}^{+\infty}
		G^nh_{(n)}^{\alpha\beta}[\dI_L,\dJ_L,\cdots,\dZ_L]\,.
\end{equation}
It depends on two sets of arbitrary STF-tensorial functions of time $\dI_L(u)$ and $\dJ_L(u)$ (satisfying the conservation laws) defined by Eqs. \eqref{k1sol}, and on four supplementary functions $\dW_L(u), \cdots, \dZ_L(u)$ parametrizing the gauge vector \eqref{phi1expr}.\label{th:gensol}
\end{theorem}
The proof is quite easy. With Eq.~\eqref{hnexpl} we obtained a \emph{particular} solution of the system of equations \eqref{eqhn}. To it we should add the most general solution of the corresponding \emph{homogeneous} system of equations, which is obtained by setting $\Lambda^{\alpha\beta}_{(n)}=0$ into Eqs. \eqref{eqhn}. But this homogeneous system of equations is nothing but the \emph{linearized} vacuum field equations \eqref{eqh1}, to which we know the most general solution $h_{(1)}^{\alpha\beta}$ given by Eqs. \eqref{hk1}--\eqref{phi1expr}. Thus, we must add to our particular solution $h^{\alpha\beta}_{(n)}$ a general homogeneous solution that is necessarily of the type $h^{\alpha\beta}_{(1)}[\delta \dI_L,\cdots,\delta \dZ_L]$, where $\delta \dI_L, \cdots, \delta \dZ_L$ denote some corrections to the multipole moments at the $n$-th post-Minkowskian order (with the monopole $\delta \dI$ and dipoles $\delta \dI_i$, $\delta \dJ_i$ being constant). It is then clear, since precisely the linearized metric is a linear functional of all these moments, that the previous corrections to the moments can be absorbed into a re-definition of the original ones $\dI_L,\cdots,\dZ_L$ by posing
\begin{subequations}\label{redef}
	\begin{align}
		\dI_L^\mathrm{new}&=\dI_L+G^{n-1}\delta
		\dI_L\,, \\ &\vdots& \nn
		\\ \dZ_L^\mathrm{new}&=\dZ_L+G^{n-1}\delta
		\dZ_L\,.
	\end{align}
\end{subequations}
After re-arranging the metric in terms of these new moments, taking into account the fact that the precision of the metric is limited to the $n$-th post-Minkowskian order, and dropping the superscript ``new'', we find exactly the same solution as the one we had before (indeed, the moments are arbitrary functions of time) -- hence the proof.

However, it is always possible to find \emph{two}, and only two, sets of STF multipole moments, $\dM_L(u)$ and $\dS_L(u)$, for parametrizing the most general isolated source as well. The route for constructing such a general solution is to get rid of the moments $\dW_L,\dX_L, \dY_L,\dZ_L$ at the linearized level by performing the linearized gauge transformation $\delta x^\alpha = \varphi^\alpha_{(1)}$, where $\varphi^\alpha_{(1)}$ is the gauge vector given by Eqs. \eqref{phi1expr}. So, at the linearized level, we have only the two types of moments $\dM_L$ and $\dS_L$, parametrizing $h^{\alpha\beta}_{\text{can}\,(1)}$ by the same formulas as in Eqs. \eqref{k1sol}, hence the new ``canonical'' algorithm starts with
\begin{equation}\label{hcanMPM1}
	h^{\alpha\beta}_\text{can\,MPM} = G \,h^{\alpha\beta}_{\text{can}\,(1)}[\dM_L,
	\dS_L] +
	\calO(G^2)\,.
\end{equation}
We must be careful to denote these moments with names different from $\dI_L$ and $\dJ_L$ because $\dM_L$ and $\dS_L$ will ultimately differ from $\dI_L$ and $\dJ_L$ for a given physical source (see Theorem \ref{th:cansol}). Then we apply exactly the same post-Minkowskian algorithm, following the formulas \eqref{ungen}--\eqref{hnexpl} as we did above, but starting from the gauge-transformed linear metric $h^{\alpha\beta}_{\text{can}\,(1)}$ instead of $h^{\alpha\beta}_{(1)}$. That is, we successively define
\begin{align}\label{hncan}
	&\left.\begin{array}{l} \displaystyle u_{\text{can}\, (n)}^{\alpha\beta} 
		=
		\FPprop\left[\widetilde{r}^B\Lambda_{\text{can}\, (n)}^{\alpha\beta} \right] \\[0.4cm]\displaystyle w_{\text{can}\, (n)}^{\alpha} = \partial_\mu u_{\text{can}\, (n)}^{\alpha\mu}  \\[0.4cm]v_{\text{can}\, (n)}^{\alpha\beta} = \mathcal{V}^{\alpha\beta}\bigl[w_{\text{can}\, (n)}\bigr]
	\end{array}\right\} \Longrightarrow ~h_{\text{can}\, (n)}^{\alpha\beta} = u_{\text{can}\, (n)}^{\alpha\beta} + v_{\text{can}\, (n)}^{\alpha\beta}\,,
\end{align}
where the source term $\Lambda_{\text{can}\,(n)}^{\alpha\beta}$ depends on previous iterations $h_{\text{can}\,(m)}^{\alpha\beta}$ (with $m\leqslant n-1$), and where the harmonicity algorithm $\mathcal{V}^{\alpha\beta}$ is defined in Eq.~\eqref{algoV}.

\begin{theorem}
The most general solution of the Einstein field equations in the vacuum region outside an isolated source is given by
\begin{equation}\label{hMS}
		h_{\text{can\,\rm{MPM}}}^{\alpha\beta}=\sum_{n=1}^{+\infty}
		G^n h_{\text{can}\,(n)}^{\alpha\beta}[\dM_L,\dS_L]\,.
\end{equation}
The metric \eqref{hMS} is isometric to the original one \eqref{hIJZ} (i.e., describe the same physical source) if and only if the canonical moments $\dM_L$ and $\dS_L$ are related to the source moments $\dI_P,\dJ_P,\cdots,\dZ_P$ by non-linear (post-Minkowskian) relations
\begin{subequations}\label{relationMLSL}
	\begin{align}
			\dM_L &= \dI_L + \sum_{n=2}^{+\infty} G^{n-1}\mathcal{M}_{n\,L}[\dI_P, \dJ_P, \cdots, \dZ_P]\,,\\[-0.1cm]
			\dS_L &= \dJ_L + \sum_{n=2}^{+\infty} G^{n-1}\mathcal{S}_{n\,L}[\dI_P, \dJ_P, \cdots, \dZ_P]\,.
	\end{align}
\end{subequations}
\label{th:cansol}
\end{theorem}
Obviously the canonical post-Minkowskian algorithm \eqref{hMS} yields some simpler calculations than the general algorithm \eqref{hIJZ} as we have only two multipole moments to iterate. We shall give in Eqs. \eqref{relationMijIij}--\eqref{cansourceMS} the most up to date relations between the canonical moments and source moments. 

So why not consider from the start that the best description of the isolated source is provided by only the two types of multipole moments, $\dM_L$ and $\dS_L$, instead of the six types, $\dI_L, \dJ_L, \cdots, \dZ_L$? The reason is that we shall determine in Theorem \ref{th:sourcemoments} the explicit closed-form expressions of the six source moments $\dI_L, \dJ_L, \cdots, \dZ_L$, but that, by contrast, it seems to be impossible to obtain some similar closed-form expressions for the canonical moments $\dM_L$ and $\dS_L$. The only thing we can do is to write down the explicit non-linear algorithm that computes $\dM_L$, $\dS_L$ starting from $\dI_L, \dJ_L, \cdots, \dZ_L$, see Eqs. \eqref{relationMLSL}. In consequence, it is better to view the moments $\dI_L, \dJ_L, \cdots, \dZ_L$ as more ``fundamental'' than $\dM_L$ and $\dS_L$, in the sense that they appear to be more tightly related to the description of the source, since they admit closed-form expressions as some explicit integrals over the source.  Hence, we choose to refer collectively to the six moments $\dI_L, \dJ_L, \cdots, \dZ_L$ as \emph{the} multipole moments of the source. This being said, the moments $\dM_L$ and $\dS_L$ are generally very useful in practical computations, not only because they yield a simpler post-Minkowskian iteration but also because they are closely related to the two independent degrees of freedom of gravitational waves propagating at infinity from the source. Then, one can come back to the more fundamental source-rooted moments by using the relations \eqref{relationMLSL}, in particular the fact that $\dM_L$ and $\dS_L$ differ from the corresponding $\dI_L$ and $\dJ_L$ only by high-order post-Newtonian terms starting at 2.5PN order; see Eqs. \eqref{relationMijIij}--\eqref{cansourceMS} below. Indeed, this is to be expected because the physical difference between both types of moments stems only from non-linearities.

Finally, let us point out that in our presentation of the post-Minkowskian algorithms, either the general one \eqref{hIJZ} or the canonical one \eqref{hMS}, we have for the moment omitted a crucial recursive hypothesis, which is required in order to prove that at each post-Minkowskian order $n$, the inverse d'Alembertian operator can be applied in the way we did -- notably the fact that the $B$-dependent retarded integral can be analytically continued down to a neighbourhood of $B=0$. This hypothesis is that the ``near-zone'' expansion, i.e., when $r\to 0$, of each one of the post-Minkowskian coefficients $h_{(n)}^{\alpha\beta}$ [or $h_{\text{can}\,(n)}^{\alpha\beta}$] has a certain structure; 
this hypothesis is established as a theorem once the mathematical induction succeeds.

\begin{theorem} \citep{BD86}
The general structure of the expansion of the post-Minkowskian exterior metric in the near-zone (when $r\to 0$) is of the type: $\forall N \in \mathbb{N}$,\footnote{We employ the Landau symbol $o$ for remainders with its standard meaning. Thus, $f(r)=o[g(r)]$ when $r\to 0$ means that $f(r)/g(r)\to 0$ when $r\to 0$. Furthermore, we generally assume some differentiability properties such as $\dd^n f(r)/\dd r^n=o[g(r)/r^n]$.\label{fnote:landau}}
\begin{equation}\label{hnNZ}
	h_{(n)}^{\alpha\beta}(\mathbf{x}, t) = \sum \hat{n}_L r^m (\ln r)^p F_{L, m, p,
	n}^{\alpha\beta}(t)+o(r^{N})\,,
\end{equation}
where $m\in \mathbb{Z}$, with $m_0 \leqslant m \leqslant N$ (and $m_0$ becoming more and more negative as $n$ grows), $p \in \mathbb{N}$ with $p \leqslant n-1$. The functions $F_{L, m, p, n}$ are multilinear functionals of the source multipole moments $\dI_L, \cdots ,\dZ_L$ or the canonical moments $\dM_L$, $\dS_L$ (depending on the construction under consideration, $h_{(n)}^{\alpha\beta}$ or $h_{\text{can}\,(n)}^{\alpha\beta}$).
\label{th:NZstruct}
\end{theorem}
As we see, the near-zone expansion involves, besides the simple powers of $r$, some integer powers of the logarithm of $r$, with a maximal power of $n-1$. As a corollary of that theorem, we find, by restoring all the powers of $c$ in Eq.~\eqref{hnNZ} and using the fact that each $r$ goes into the combination $r/c$, that the general structure of the post-Newtonian expansion ($c\to +\infty$) is of the type
\begin{equation}\label{hngenPN}
	h_{(n)}(c) \simeq \!\! \sum_{p, q \in \mathbb{N}}\frac{(\ln
		c)^p}{c^q}\,,
\end{equation}
where $p\leqslant n-1$ (and $q\geqslant 2$). The post-Newtonian expansion proceeds not only with the normal powers of $1/c$ but also with powers of the logarithm of $c$. 
It is remarkable that the structure of the post-Newtonian expansion (at high orders) is not more complicated than \eqref{hngenPN}; for instance there are no terms behaving like $\sim \ln(\ln c)$.


\subsubsection{The multipolar post-Minkowskian solution in radiative coordinates}
\label{sec:MPMrad}
 
We now tackle the important problem of the far-zone structure of this MPM construction and the link with observables quantities at infinity. Paralleling the structure of the near-zone expansion given by \eqref{hnNZ}, we have a similar result concerning the \emph{far-zone} expansion at Minkowskian future null infinity, i.e., when $r\to +\infty$ with $u=t-r/c= \mathrm{const}$: $\forall N \in \mathbb{N}$,
\begin{equation}\label{hngenFZ}
  h_{(n)}^{\alpha\beta}(\mathbf{x}, t) = \sum \frac{\hat{n}_L (\ln r)^p}{r^k}
  G_{L, k, p, n}^{\alpha\beta}(u)+o\left(\frac{1}{r^N}\right)\,,
\end{equation}
where $k, p\in \mathbb{N}$, with $1\leqslant k\leqslant N$, and where, likewise in the near-zone expansion \eqref{hnNZ}, some integer powers of logarithms, such that $p \leqslant n-1$, appear. 

The appearance of logarithms in the far-zone expansion of the harmonic-coordinates metric has been known since the work of \cite{Fock}. One knows also that this is a coordinate effect, because the study of the ``asymptotic'' structure of space-time at future null infinity has been consistently performed in a coordinate system that avoids the appearance of far zone logarithms \citep{BBM62, Sachs62, NU63}. This asymptotic structure was further elucidated thanks to the tools of the \cite{NP62} formalism and conformal compactifications leading to the concept of asymptotically ``simple'' spacetimes in the sense of \cite{P63, P65}. More generally, a large class of so-called \emph{radiative} coordinate systems exist \citep{Papa69,MadoreI,MadoreII}, in which the far-zone expansion of the metric proceeds with simple powers of the inverse radial distance. Hence, the logarithms are simply an artifact of the use of harmonic coordinates \citep{IW68, B87}. An algorithm to go from harmonic coordinates to \cite{NU63} coordinates has been implemented by \cite{BCFOS21, BCFOS23}. The following theorem shows that general MPM construction of the metric in the exterior of the source, when developed at future null infinity, is consistent with the \cite{BBM62, Sachs62, P63, P65} approach to gravitational radiation.

\begin{theorem} \citep{B87} The most general multipolar-post-Minkowskian solution, stationary in the past [see Eq.~\eqref{statpast}], admits some radiative coordinates $(T,\mathbf{X})$, for which the expansion at future null infinity, $R\to+\infty$ with $U\equiv T-R/c= \mathrm{const}$, takes the form
\begin{equation}\label{structBSP}
    h_{\text{rad}\,(n)}^{\alpha\beta}(\mathbf{X},T) = \sum_{k\geqslant 1} \frac{\hat{N}_L}{R^k} K_{L, k,
      n}^{\alpha\beta}(U)+\calO\left(\frac{1}{R^N}\right)\,.
\end{equation} 
The functions $K_{L, k, n}^{\alpha\beta}$ are computable functionals of the source multipole moments. In radiative coordinates the retarded time $U$ is a null coordinate in the asymptotic far zone limit. Furthermore the metric
\begin{equation}\label{asymptsimple}
  	h_{\text{rad\,MPM}}^{\alpha\beta} = \sum_{n=1}^{+\infty} G^n h^{\alpha\beta}_{\text{rad}\,(n)}\,,
\end{equation} 
is asymptotically simple in the sense of \cite{P63, P65, GH78},\footnote{If the assumption of stationarity in the past is relaxed, there may be a loss of smoothness of the metric at future null indinity \citep{D86MG}.} perturbatively to any post-Minkowskian order. \label{th:FZstruct}
\end{theorem}
The proof consists of performing the multipolar-post-Minkowskian construction directly in radiative coordinates, thus solving the Einstein field equations in vacuum, in a non-harmonic coordinate system:
\begin{equation}\label{EFEvac}
	\Box h^{\alpha\beta} - \partial H^{\alpha\beta} = \Lambda^{\alpha\beta}\,.
\end{equation}
The source term is defined by Eqs. \eqref{LambdaSource}--\eqref{Lambdadef}. In radiative coordinates the retarded time coordinate $u = t-r/c$ is null, i.e. satisfies\footnote{Since we are iteratively constructing a coordinate system order by order, it is convenient to consider the coordinates as dummy variables and denote them by $(\mathbf{x},t)$. In the end, when we obtain the full radiative metric, we shall denote the radiative coordinates by $(\mathbf{X},T)$ as in Theorem \ref{th:FZstruct}.}
\begin{equation}\label{null}
	\bigl(\eta^{\alpha\beta} + h_{\text{rad\,MPM}}^{\alpha\beta}\bigr) \partial_\alpha u \,\partial_\beta u = 0\,,
\end{equation} 
or at least, becomes a null coordinate in the asymptotic limit $r\to+\infty$ with $u=$ const. Even at the linearized level, it is necessary to correct the harmonic coordinate metric in order to satisfy the requirement of asymptotically null retarded time. Consequently, the radiative MPM algorithm starts by performing a linear gauge transformation of the harmonic-coordinate metric defined by Eq.~\eqref{hcanMPM1}. A crucial point is that the multipole moments that parametrize the radiative algorithm will differ from their counterparts in the harmonic algorithm (for a given physical source). Hence we construct the radiative metric using multipole moments $\{\dMbar_L,\dSbar_L\}$ which differ from the canonical moments $\{\dM_L,\dS_L\}$ used in harmonic coordinates.\footnote{Although the two sets of moments $\{\dM_L,\dS_L\}$ and $\{\dMbar_L,\overline{\dS}_L\}$ differ in general, the conserved mass monopole (as well as the mass and current dipoles) are in fact identical in the harmonic and radiative constructions. Hence, $\dMbar=\dM$ for the ADM mass. See Eq. (5.4) of \cite{TLB22} for the relationship between the quadrupole moment $\dMbar_{ij}$ in the radiative algorithm and the moments $\{\dM_L,\dS_L\}$ in harmonic coordinates.} At linear order we pose
\begin{equation}\label{hrad1}
	h^{\alpha\beta}_{\text{rad}\, (1)} = h^{\alpha\beta}_{\text{can}\, (1)}\bigl[\dMbar_L,\overline{\dS}_L\bigr]+ \partial \xi_{(1)}^{\alpha\beta}
	\,,
\end{equation}
where $h^{\alpha\beta}_{\text{can}\, (1)}$ takes the same functional form as in \eqref{k1sol}, but is now computed with the moments $\{\dMbar_L,\dSbar_L\}$. The linear gauge transformation $\partial \xi_{(1)}^{\alpha\beta}$ is defined by the gauge vector
\begin{equation}\label{xi1}
	\xi^\alpha_{(1)} = \frac{2 \dM}{c^2} \, \eta^{0\alpha}\ln
	\left(\frac{r}{b_0}\right) \,,
\end{equation}
where $\dM$ is the total ADM mass, $b_0$ denotes an arbitrary length scale, and we have $\eta^{0\alpha}=(-1,\bm{0})$ with our signature. 

The effect of this linear gauge transformation is to correct for the well-known logarithmic deviation of the retarded time in harmonic coordinates, with respect to the true light cone. After the change of gauge, the coordinate $u=t-r/c$ coincides (asymptotically when $r\to\infty$) with a null coordinate at the linearized level. The linear radiative metric is no longer harmonic, since
\begin{equation}\label{eq:divrad1}
	H^{\alpha}_{\text{rad}\, (1)} \equiv \partial_\mu h^{\alpha\mu}_{\text{rad}\, (1)} = \Box \xi^\alpha_{(1)} =
	\frac{2 \dM}{c^2 r^2}\,\eta^{0\alpha}\,.
\end{equation}

Given any $n\geqslant 2$, let us recursively assume that: (i) we have obtained all the previous radiative post-Minkowskian coefficients $h^{\alpha\beta}_{\text{rad}\, (m)}$ for any $m\leqslant n-1$; (ii) all of them admit an expansion as $r\to+\infty$ with $u=$ const in simple powers of $1/r$ (no logarithms); and (iii) all the previous coefficients satisfy the fall-off conditions ($\forall \,m\leqslant n-1$)
\begin{subequations}\label{kkhradm0}
	\begin{align}
k_\alpha k_\beta \,h^{\alpha\beta}_{\text{rad}\, (m)} &=
	\calO\left(\frac{1}{r^2}\right)\,,\\
	H^{\alpha}_{\text{rad}\, (m)} &=
	\calO\left(\frac{1}{r^2}\right)\,,
\end{align}
\end{subequations}
where $k^\alpha = \eta^{\alpha\mu}k_\mu = (1, \mathbf{n})$ denotes the outgoing Minkowskian null vector. Denoting $z^{\alpha\beta}_{(m)}$ the leading $1/r$ coefficient of the metric when $r\to+\infty$, i.e. 
\begin{equation}\label{coeff1/r}
	h^{\alpha\beta}_{\text{rad}\, (m)} = \frac{1}{r} \,z^{\alpha\beta}_{(m)}\bigl(u,\mathbf{n}\bigr) +
\calO\left(\frac{1}{r^2}\right)\,,
\end{equation}
we see that the two conditions \eqref{kkhradm0} say that $k_\alpha k_\beta z^{\alpha\beta}_{(m)} = 0$ while $k_\mu z^{\alpha\mu}_{(m)}$ is constant. For a stationary metric (independent of time) we know that the $m$-th post-Minkowskian coefficient decreases as $1/r^m$ when $r\to+\infty$. By our assumption of stationarity in the past, Eq.~\eqref{statpast}, this means that $h^{\alpha\beta}_{\text{rad}\, (m)} = \calO(r^{-m})$ when $t\leqslant -\mathcal{T}$, hence the latter constant is zero when $m\geqslant 2$. Hence the recursive assumptions \eqref{kkhradm0} are equivalent to
\begin{subequations}\label{kkhradm}
	\begin{align}
		k_\mu z^{\alpha\mu}_{(1)} &= \frac{2\dM}{c^2}\,k^\alpha\,,\\
		k_\mu z^{\alpha\mu}_{(m)} &= 0 \qquad\text{(for $2\leqslant m\leqslant n-1$)}\,.
	\end{align}
\end{subequations}
One can easily check that the condition is satisfied by the linear metric \eqref{hrad1}.

The dominant piece when $r\to +\infty$ of the non-linear source term at the $n$-th order will be of order $1/r^2$ and will only be made of quadratic products of $h^{\alpha\beta}_{\text{rad}\, (m)}$, since each of the $h^{\alpha\beta}_{\text{rad}\, (m)}$'s behaves like $1/r$. Under our recursive assumptions, in particular \eqref{kkhradm}, and from the structure of the source term at quadratic order, see Eq.~\eqref{Nab}, one can prove that the $n$-th post-Minkowskian source term takes the form at leading order when $r\to+\infty$:
%
\begin{equation}\label{Lambdan}
	\Lambda_{\text{rad}\, (n)}^{\alpha\beta} = \frac{k^\alpha
		k^\beta}{r^2}\,\sigma_{(n)}\bigl(u,\mathbf{n}\bigr) +
	\calO\left(\frac{1}{r^3}\right)\,.
\end{equation}
This is the form of the stress-energy tensor of massless particles, i.e. gravitons in our case, with $\sigma_{(n)}$ being proportional to the $n$-th order contribution in the total power emitted by the massless radiation. The explicit expression in terms of the metric coefficient \eqref{coeff1/r} is (the dot denoting the time derivative)
\begin{equation} \label{sigman}
	\sigma_{(n)} = \sum_{m=1}^{n-1}
	\left(\frac{1}{2}\eta_{\alpha\gamma}\eta_{\beta\delta} -
	\frac{1}{4}\eta_{\alpha\beta}\eta_{\gamma\delta}\right) \dot{z}^{\alpha\beta}_{(m)} \dot{z}^{\gamma\delta}_{(n-m)} \,.
\end{equation}

We know that logarithms in the asymptotic expansion when $r\to+\infty$  only arise due to the retarded integral of source terms that behave like $1/r^2$ (see Appendix A of \citealt{B98quad}). Hence the dominant term written in Eq.~\eqref{Lambdan} is the only piece of $\Lambda_{\text{rad}\, (n)}^{\alpha\beta}$ that can yield logarithms at order $n$; see indeed the integration formula \eqref{expltail2}, which behaves like $\ln r/r$ at infinity. But now, thanks to the particular structure of this term in \eqref{Lambdan}, which follows from our recursive assumptions, we can gauge it away, thus constructing a coordinate system valid at the $n$-th post-Minkowskian order which avoids the appearance of logarithms. We find that an adequate gauge vector is
\begin{equation}\label{xina}
	\xi^\alpha_{(n)} = \Box^{-1}_{\mathrm{ret}}
	\left[\frac{c\,k^\alpha}{2r^2} \int_{-\infty}^{u} \!\! \dd v \,
	\sigma_{(n)} (v,\mathbf{n})\right]\,.
\end{equation}
The retarded integral is convergent at the bound $r\to 0$ and so there is no need to include a finite part operation FP. With this choice of gauge vector, the logarithms that will be generated from the gauge transformation will cancel the logarithms coming from the retarded integral of the source term \eqref{Lambdan}. 
Hence, similarly to the corresponding steps \eqref{hncan} in the harmonic algorithm, we construct the $n$-th post-Minkowskian metric by correcting for the new logarithms using the above gauge transformation: 
\begin{align}\label{hnrad}
	&\hspace{-0.5cm}\left.\begin{array}{l} \displaystyle u_{\text{rad}\, (n)}^{\alpha\beta} 
		=
		\FPprop\left[\widetilde{r}^B\Lambda_{\text{rad}\, (n)}^{\alpha\beta} \right] \\[0.4cm]\displaystyle w_{\text{rad}\, (n)}^{\alpha} = \partial_\mu u_{\text{rad}\, (n)}^{\alpha\mu}  \\[0.4cm]v_{\text{rad}\, (n)}^{\alpha\beta} = \mathcal{V}^{\alpha\beta}\bigl[w_{\text{rad}\, (n)}\bigr]
	\end{array}\!\right\} \!\Rightarrow h_{\text{rad}\, (n)}^{\alpha\beta} = u_{\text{rad}\, (n)}^{\alpha\beta} + v_{\text{rad}\, (n)}^{\alpha\beta} + \partial\xi_{(n)}^{\alpha\beta}\,.
\end{align}
Note that the combination $u_{\text{rad}\, (n)}^{\alpha\beta}+v_{\text{rad}\, (n)}^{\alpha\beta}$ is divergenceless, so the radiative metric obeys by construction the non-harmonic gauge condition
\begin{equation}\label{Hnrad}
	H_{\text{rad}\, (n)}^{\alpha} = \Box \xi^\alpha_{(n)} =
	\frac{c\,k^\alpha}{2 r^2} \int_{-\infty}^{t-r/c} \!\!\dd v \, \sigma_{(n)}
	(v,\mathbf{n})\,,
\end{equation}
and the Einstein field equations \eqref{EFEvac} are trivially satisfied to order $n$. The far-zone expansion of the full non-linear radiative metric constucted by virtue of this procedure is free of any logarithms, and the retarded time $u=t-r/c$ in these coordinates tends asymptotically toward a null coordinate at future null infinity [i.e., satisfying \eqref{null}]. The property of asymptotic simplicity in the form given by \cite{GH78}, is proved by introducing the conformal factor $\Omega=1/r$ in radiative coordinates \citep{B87}. Finally, as stated in Theorem \ref{th:FZstruct}, the metric, as a general functional of the moments $\{\dMbar_L,\dSbar_L\}$, represents physically the most general solution to the vacuum field equations outside the source, as legitimate as the general harmonic-coordinate metric of Theorem \ref{th:gensol}.

Finally, having constructed the metric step by step in a radiative (or Bondi-type) coordinate system, we \emph{formally} sum up the whole post-Minkowskian series from $n=1$ up to $+\infty$, thus defining 
\begin{equation}\label{resum}
	h_{\text{rad}}^{\alpha\beta} \equiv \sum_{n=1}^{+\infty} G^n\,h_{\text{rad}\,(n)}^{\alpha\beta}\,,
\end{equation}
Posing then $z\ab$ to be the leading $1/R$ coefficient of the radiative metric [we restore the names $(T,\bm{X})$ appropriate for radiative coordinates],
\begin{align}\label{zresumdef}
	h_{\text{rad}}\ab = \frac{1}{R} \,z\ab(U,\bm{N}) + \calO\left(\frac{1}{R^{2}}\right)\,,
\end{align} 
where $z\ab = \sum_{n\geqslant 1} G^n\,z_{(n)}^{\alpha\beta}$, we obtain after resummation the leading $1/R^2$ coefficient of the source term in radiative coordinates as
\begin{align}\label{Lambdares}
	\Lambda_{\text{rad}}\ab = \frac{K^\alpha K^\beta}{R^2} \,\sigma(U,\bm{N}) + \calO\left(\frac{1}{R^{3}}\right)\,,
\end{align} 
where $K^\alpha=(1,\bm{N})$ and the resummed energy distribution reads
\begin{align}\label{sigmares}
\sigma(U,\bm{N}) &= \frac{1}{2} \dot{z}^{\alpha\beta}\dot{z}_{\alpha\beta} - \frac{1}{4}\left(\dot{z}^{\alpha}_{\alpha}\right)^2 = \frac{1}{2} \!\perp_{ijkl}\dot{z}_{ij}\dot{z}_{kl} = \frac{1}{2} \,\dot{z}^\text{TT}_{ij}\dot{z}^\text{TT}_{ij}\,.
\end{align} 
Reminding from \eqref{kkhradm} that $K_\mu \dot{z}^{\alpha\mu} = 0$, an easy calculation gives the second and third equalities in \eqref{sigmares}, where $\perp_{ijkl}(\bm{N})$ is the TT projection operator already introduced in Eq.~\eqref{operatorTT}, and we pose $\dot{z}^\text{TT}_{ij}=\perp_{ijab}\dot{z}_{ab}$. Finally the energy distribution in the waves is directly given by $\sigma(U,\bm{N})$ as
\begin{align}\label{energiedistr}
	\left(\frac{\dd E}{\dd U\dd\Omega}\right)^\text{GW} = \frac{c^5}{16\pi G} \,\sigma(U,\bm{N}) =\lim_{R\to+\infty\atop U=\text{const}}\frac{c^3 R^2}{32\pi G} \,\dot{h}^\text{TT}_{ij}\dot{h}^\text{TT}_{ij} \,,
\end{align}
where $h^\text{TT}_{ij} = \perp_{ijab} h_{\text{rad}\,ab}$ denotes the TT waveform. The energy (and angular momentum) in the waves can be defined using the \cite{Isaacson} stress-energy tensor 
\begin{equation}\label{Isaacson}
	T_{\alpha\beta}^\text{GW} = \frac{c^4}{32\pi G}\langle\partial_\alpha h^\text{TT}_{ij}\partial_\beta h^\text{TT}_{ij}\rangle\,,
\end{equation}
where the brackets denote an average over the wavelength. This definition is general, and reduces asymptotically to
\begin{equation}\label{asympIsaacson} 
	T_{\alpha\beta}^\text{GW} = \frac{c^4}{16\pi G}\frac{K_\alpha K_\beta}{R^2}\sigma + \calO\left(\frac{1}{R^{3}}\right)\,.
\end{equation}
We systematically investigate the properties of the TT waveform in Sect.~\ref{sec:asympGW}.


\subsection{Matching to a post-Newtonian source}
\label{sec:matching}

By Theorem \ref{th:gensol} we control the most general class of solutions of the vacuum equations outside the source, in the form of non-linear functionals of the source multipole moments. For instance, these solutions can describe the Schwarzschild and Kerr solutions for black holes, as well as all their perturbations, and neutron stars with arbitrary internal structure. By Theorem \ref{th:FZstruct} we learned how to control the far zone expansion of the solutions; this will yield the notion of radiative moments at infinity, which constitute the observables of the radiation field far from the source (see Sect.~\ref{sec:asympGW}). We now want to understand how a specific choice of matter stress-energy tensor $T^{\alpha\beta}$, i.e., a specific choice of some physical model describing the material source, selects a particular physical exterior solution among our general class, and therefore a given set of multipole moments for the source.


\subsubsection{The matching equation}
\label{sec:matchingeq}

We shall provide the answer to that problem in the case of a post-Newtonian source for which the post-Newtonian parameter $\epsilon\sim 1/c$ defined by Eq.~\eqref{epsPN} is small. The fundamental fact that permits the connection of the exterior field to the inner field of the source is the existence of a ``matching'' region, in which both the multipole expansion and the post-Newtonian expansion are valid. This region is nothing but the exterior part of the near zone, such that $r>a$ (exterior) \emph{and} $r\ll\lambda$ (near zone); it always exists around post-Newtonian sources whose radius is much less than the emitted wavelength, $\frac{a}{\lambda}\sim\epsilon\ll 1$, hence $a \ll \lambda$. In the present formalism the multipole expansion is defined by the multipolar-post-Minkowskian (MPM) solution; see Sect.~\ref{sec:nonlinitere}. Matching together the post-Newtonian and MPM solutions in this overlapping region is an application of the method of matched asymptotic expansions, which has intensively been applied in the present context, both for radiation-reaction \citep{BuTh70, Bu71, AKKM82, BD88, B93} and wave-generation \citep{BD89, DI91a, B95, B98mult} problems.

Let us denote by $\mathcal{M}(h)$ the multipole expansion of the true metric $h$, valid everywhere inside and outside the source.\footnote{For simplicity in this section, we do not write the space-time indices $\alpha\beta$ on the metric.} By $\mathcal{M}(h)$ we really mean the MPM exterior metric that we have constructed in Sect.~\ref{sec:nonlinitere}. Specifically we adopt the general metric of Theorem \ref{th:gensol}, see Eq.~\eqref{hIJZ}, which is parametrized by six types of source multipole moments $\dI_L$, $\dJ_L$, $\dW_L$, $\dX_L$, $\dY_L$, $\dZ_L$:
\begin{equation}\label{Mhext}
  \mathcal{M}(h)  \equiv h_{\text{MPM}} = \sum_{n=1}^{+\infty} G^n
  h_{(n)}[\dI_L,\cdots,\dZ_L]\,.
\end{equation}

An important point about the multipole expansion (not always well appreciated) is that it is a formal solution of the vacuum equations which is defined for any radius $r>0$.
Of course, the true solution $h$ agrees with its own multipole expansion in the
exterior of the source, i.e.
\begin{equation}\label{Mhh}
  r > a \quad\Longrightarrow\quad \mathcal{M}(h) = h\,.
\end{equation}
By contrast, inside the source, $h$ and $\mathcal{M}(h)$ disagree with each other because $h$ is a fully-fledged solution of the field equations within the matter source, while $\mathcal{M}(h)$ is a vacuum solution becoming singular at $r=0$. Note also that the multipole expansion, physically valid when $r > a$, is to be distinguished from the far zone expansion of the field, valid when $r \gg \lambda$.

Now let us denote by $\overline h$ the post-Newtonian expansion of $h$. We have already anticipated the general structure of this expansion which is given in Eq.~\eqref{hngenPN}. In the matching region, where both the multipolar and post-Newtonian expansions are valid, we can write the numerical equality
\begin{equation}\label{Mhhbar}
  a < r \ll\lambda \quad\Longrightarrow\quad \mathcal{M}(h) =
  \overline h\,.
\end{equation}
This ``numerical'' equality is viewed here in a sense of formal expansions, as we do not control the convergence of the series. In fact, we should be aware that such an equality, though quite natural and even physically obvious, has not really been justified mathematically within the approximation scheme, and we simply take it here as part of our fundamental assumptions.

We now transform Eq.~\eqref{Mhhbar} into a \emph{matching equation}, by replacing in the left-hand side $\mathcal{M}(h)$ by its near-zone re-expansion $\overline{\mathcal{M}(h)}$, and in the right-hand side $\overline h$ by its multipole expansion $\mathcal{M}(\overline h)$. The structure of the near-zone expansion ($r\to 0$) of the exterior multipolar field has been found in Theorem \ref{th:NZstruct}, see Eq.~\eqref{hnNZ}. We denote the corresponding infinite series $\overline{\mathcal{M}(h)}$ with the same overbar as for the post-Newtonian expansion because it is really an expansion when $r/c\to 0$, equivalent to an expansion when $c\to\infty$. Concerning the multipole expansion of the post-Newtonian metric, $\mathcal{M}(\overline h)$, we simply postulate for the moment its existence, but we shall show later how to construct it explicitly. Therefore, the matching equation is the statement that
\begin{equation}\label{matchingeq}
  \overline{\mathcal{M}(h)} = \mathcal{M}(\overline h)\,,
\end{equation}
by which we really mean an infinite set of \emph{functional} identities, valid $\forall (\mathbf{x}, t) \in {\mathbb{R}}^3_* \times \mathbb{R}$, between the coefficients of the series in both sides of the equation. Note that such a meaning is different from that of a \emph{numerical} equality like Eq.~\eqref{Mhhbar}, which is valid only when $\mathbf{x}$ belongs to some limited spatial domain. The matching equation \eqref{matchingeq} tells us that the formal \emph{near-zone} expansion of the multipole decomposition is \emph{identical}, term by term, to the multipole expansion of the post-Newtonian solution. However, the latter expansion is nothing but the formal \emph{far-zone} expansion, when $r\to\infty$, of each of the post-Newtonian coefficients. Most importantly, it is possible to write down, within the present formalism, the general structure of these identical expansions as a consequence of Eq.~\eqref{hnNZ}: 
\begin{equation}\label{struct}
  \overline{\mathcal{M}(h)} = \!\sum \hat{n}_L r^m (\ln r)^p F_{L, m,
    p}(t) = \mathcal{M}(\overline h)\,,
\end{equation}
where the functions $F_{L, m, p} = \sum_{n\geqslant 1} G^n F_{L, m, p, n}$.  The latter expansion can be interpreted either as the singular re-expansion of the multipole decomposition when $r\to 0$ -- i.e., the first equality in Eq.~\eqref{struct} --, or the singular re-expansion of the post-Newtonian series when $r\to +\infty$ -- the second equality.

We recognize the beauty of singular perturbation theory, where two asymptotic expansions, taken formally outside their respective domains of validity, are matched together. Of course, the method works because there exists, physically, an overlapping region in which the two approximation series are expected to be numerically close to the exact solution. As we shall detail in Sects.~\ref{sec:exprmult} and \ref{sec:radreac}, the matching equation \eqref{matchingeq}, supplemented by the condition of no-incoming radiation [say in the form of Eq.~\eqref{statpast}], permits determining all the unknowns of the problem: On the one hand, the external multipolar decomposition $\mathcal{M}(h)$, i.e.,  the explicit expressions of the multipole moments therein (see Sects.~\ref{sec:exprmult} and \ref{sec:sourcemoments}); on the other hand, the terms in the inner post-Newtonian expansion $\overline{h}$ that are associated with radiation-reaction effects, i.e., those terms which depend on the boundary conditions of the radiative field at infinity, and which correspond in the present case to a post-Newtonian source which is isolated from other sources in the Universe; see Sect.~\ref{sec:radreac}.


\subsubsection{General expression of the multipole expansion}
\label{sec:exprmult}

\begin{theorem} \citep{B95, B98mult} Under the hypothesis of matching, Eq.~\eqref{matchingeq}, the multipole expansion of the solution of the Einstein field equation outside a post-Newtonian source reads
  \begin{equation}    
    \mathcal{M}(h^{\alpha\beta}) = \FPprop \Bigl[\widetilde r^B \mathcal{
        M}(\Lambda^{\alpha\beta})\Bigr] - \frac{4G}{c^4}
    \sum^{+\infty}_{\ell=0} \frac{(-)^\ell}{\ell !} \partial_L \left\{
    \frac{1}{r} \mathcal{H}^{\alpha\beta}_L (t-r/c) \right\}\,,
    \label{Mhab}
  \end{equation}
where the ``multipole moments'' are given by
  \begin{equation}
    \mathcal{H}^{\alpha\beta}_L (u) = \FP \int \dd^3
    \mathbf{x} \, \widetilde{r}^B x_L \, {\overline
      \tau}^{\alpha\beta}(\mathbf{x}, u)\,.
    \label{HLab}
  \end{equation}
Here, ${\overline \tau}^{\alpha\beta}$ denotes the post-Newtonian expansion of the stress-energy pseudo-tensor in harmonic coordinates as defined by Eq.~\eqref{tauab}. \label{th:multexp}
\end{theorem}
First notice where the physical restriction of considering a post-Newtonian source enters this theorem: The multipole moments \eqref{HLab} depend on the \emph{post-Newtonian} expansion ${\overline \tau}^{\alpha\beta}$ of the pseudo-tensor, rather than on $\tau^{\alpha\beta}$ itself. Consider $\Delta^{\alpha\beta}$, namely the difference between $h^{\alpha\beta}$, which is a solution of the field equations everywhere inside and outside the source, and the first term in Eq.~\eqref{Mhab}, namely the finite part of the retarded integral of the multipole expansion $\mathcal{M}(\Lambda^{\alpha\beta})$:
%
\begin{equation}
  \Delta^{\alpha\beta} \equiv h^{\alpha\beta} - \FPprop \Bigl[\widetilde r^B \mathcal{
  	M}(\Lambda^{\alpha\beta})\Bigr]\,.
  \label{Deltaab}
\end{equation}
Now the true solution $h^{\alpha\beta}$, subject to the past-stationarity condition \eqref{statpast}, is given by the retarded integral of the pseudo-tensor $\tau^{\alpha\beta}$,
\begin{equation}\label{intdiff}
  h^{\alpha\beta} = \frac{16\pi G}{c^4} \Box^{-1}_{\mathrm{ret}}
  \tau^{\alpha\beta}\,,
\end{equation}
where we recall that the retarded inverse d'Alembertian operator is defined by Eq.~\eqref{dalembertian}. So we have,
\begin{equation}
  \Delta^{\alpha\beta} = \frac{16\pi G}{c^4} \Box^{-1}_{\mathrm{ret}}
  \tau^{\alpha\beta} - \FPprop \Bigl[\widetilde r^B \mathcal{
  	M}(\Lambda^{\alpha\beta})\Bigr]\,.
  \label{Deltaab1}
\end{equation}
In the second term the finite part plays a crucial role because the multipole expansion $\mathcal{M}(\Lambda^{\alpha\beta})$ is singular at $r=0$. By contrast, the first term in Eq.~\eqref{Deltaab1}, as it stands, is well-defined because we are considering only some smooth matter distribution: $\tau^{\alpha\beta}\in C^\infty({\mathbb{R}}^4)$. There is no need to include the finite part regularization in the first term, but \emph{a contrario} there is no harm to add one in front of it, because for convergent integrals the finite part simply gives back the value of the integral. The advantage of adding artificially the finite part in the first term is that we can re-write Eq.~\eqref{Deltaab1} into the more interesting form
\begin{equation}
  \Delta^{\alpha\beta} = \frac{16\pi G}{c^4} \FPprop\Bigl[ \widetilde r^B \Bigl(\tau^{\alpha\beta} - \mathcal{
  M}(\tau^{\alpha\beta})\Bigr)\Bigr]\,,
  \label{Deltaab2}
\end{equation}
in which we have also used the fact that $\mathcal{M} (\Lambda^{\alpha\beta})=\frac{16\pi G}{c^4} \mathcal{M}(\tau^{\alpha\beta})$ because $T^{\alpha\beta}$ has a compact support. The interesting point about Eq.~\eqref{Deltaab2} is that $\Delta^{\alpha\beta}$ appears now to be the (finite part of a) retarded integral of a source with spatially \emph{compact} support. This follows from the fact that the pseudo-tensor agrees numerically with its own multipole expansion when $r>a$ [by the same equation as Eq.~\eqref{Mhhbar}]. Therefore, $\mathcal{M}(\Delta^{\alpha\beta})$ can be obtained from the known formula for the multipole expansion of the retarded solution of a wave equation with compact-support source. This formula, given in Appendix~B of \cite{BD89}, yields the second term in Eq.~\eqref{Mhab},
\begin{equation}
  \mathcal{M}(\Delta^{\alpha\beta}) = - \frac{4G}{c^4}
  \sum^{+\infty}_{\ell=0} \frac{(-)^\ell}{\ell!} \partial_L \left\{
  \frac{1}{r} \mathcal{H}^{\alpha\beta}_L (u) \right\}\,,
  \label{MDeltaab0}
\end{equation}
but in which the moments do not yet match the result \eqref{HLab}; instead,
\begin{equation}
  \mathcal{H}^{\alpha\beta}_L = \FP \int \dd^3 \mathbf{x} \, \widetilde r^B x_L
  \Bigl[\tau^{\alpha\beta} - \mathcal{M}(\tau^{\alpha\beta})\Bigr]\,.
  \label{HLab0}
\end{equation}
The reason is that we have not yet applied the assumption of a post-Newtonian source. Such sources are entirely covered by their own near zone (i.e., $a\ll\lambda$), and, in addition, for them the integral \eqref{HLab0} has a compact support limited to the domain of the source. In consequence, we can replace the integrand in Eq.~\eqref{HLab0} by its post-Newtonian expansion, valid over all the near zone:
\begin{equation}
  \mathcal{H}^{\alpha\beta}_L = \FP \int \dd^3 \mathbf{x} \, \widetilde r^B x_L
  \Bigl[{\overline\tau}^{\alpha\beta} - \overline{\mathcal{
        M}(\tau^{\alpha\beta})}\Bigr]\,.
  \label{HLab1}
\end{equation} 
Strangely enough, we do not get the expected result because of the presence of the second term in Eq.~\eqref{HLab1}. Actually, this term is a bit curious, because the object $\overline{\mathcal{M}(\tau^{\alpha\beta})}$ it contains is only known in the form of the formal series whose structure is given by the first equality in Eq.~\eqref{struct} (indeed $\tau\ab$ and $h\ab$ have the same type of structure). Happily -- because we would not know what to do with this term in practice -- we are going to prove that the second term in Eq.~\eqref{HLab1} is \emph{identically zero}. The proof is based on the properties of the analytic continuation as applied to the formal structure \eqref{struct} of $\overline{\mathcal{M}(\tau^{\alpha\beta})}$. Each term of this series yields a contribution to Eq.~\eqref{HLab1} that takes the form, after performing the angular integration, of the integral $\FP \int_0^{+\infty} \dd r \, r^{B+b} (\ln r)^p$, and multiplied by some function of time. We want to prove that the radial integral $\int_0^{+\infty} \dd r \, r^{B+b} (\ln r)^p$ is zero by analytic continuation ($\forall B\in\mathbb{C}$). First we can get rid of the logarithms by considering some repeated differentiations with respect to $B$; thus we need only to consider the simpler integral $\int_0^{+\infty} \dd r \, r^{B+b}$. We split the integral into a ``near-zone'' integral $\int_0^\mathcal{R} \dd r \, r^{B+b}$ and a ``far-zone'' one $\int_\mathcal{R}^{+\infty} \dd r \, r^{B+b}$, where $\mathcal{R}$ is some constant radius. When $\Re (B)$ is a large enough \emph{positive} number, the value of the near-zone integral is $\mathcal{R}^{B+b+1}/(B+b+1)$, while when $\Re (B)$ is a large \emph{negative} number, the far-zone integral reads the opposite, $-\mathcal{R}^{B+b+1}/(B+b+1)$. Both obtained values represent the unique analytic continuations of the near-zone and far-zone integrals for any $B\in\mathbb{C}$ except $-b-1$. The complete integral $\int_0^{+\infty} \dd r \, r^{B+b}$ is equal to the sum of these analytic continuations, and is therefore identically zero ($\forall B\in\mathbb{C}$, including the value $-b-1$). At last we have completed the proof of Theorem \ref{th:multexp}:
\begin{equation}
  \mathcal{H}^{\alpha\beta}_L = \FP \int \dd^3 \mathbf{x} \, \widetilde r^B x_L
  {\overline\tau}^{\alpha\beta}\,.
  \label{HLabfinal}
\end{equation} 

The latter proof makes it clear how crucial the analytic-continuation finite part regularization is, which we recall is the same as in our iteration of the exterior post-Minkowskian field [see Eq.~\eqref{ungen}]. Without a finite part, the multipole moment \eqref{HLabfinal} would be strongly divergent, because the pseudo-tensor ${\overline \tau}^{\alpha\beta}$ has a non-compact support owing to the contribution of the gravitational field, and the multipolar factor $x_L$ behaves like $r^\ell$ when $r\to +\infty$. The latter divergence has plagued the field of post-Newtonian expansions of gravitational radiation for many years. In applications such as in Sect.~\ref{sec:compactbinary} of this article, we must carefully follow the rules for handling the FP regularization.

The two terms in the right-hand side of Eq.~\eqref{Mhab} depend separately on the length scale $r_0$ that we have introduced into the definition of the finite part, through the analytic-continuation factor $\widetilde{r}^B=(r/r_0)^B$ introduced in Eq.~\eqref{regfactor}. However, the sum of these two terms, i.e., the exterior multipolar field $\mathcal{M}(h\ab)$ itself, is independent of $r_0$. To see this, the simplest way is to differentiate formally $\mathcal{M}(h\ab)$ with respect to $r_0$; the differentiations of the two terms of Eq.~\eqref{Mhab} cancel each other. The independence of the field upon $r_0$ is quite useful in applications, since in general many intermediate calculations do depend on $r_0$, and only in the final stage does the cancellation of $r_0$ occurs. 


\subsubsection{The source multipole moments}
\label{sec:sourcemoments}

In principle, the bridge between the exterior gravitational field generated by the post-Newtonian source and its inner field is provided by Theorem \ref{th:multexp}; however, we still have to make the connection with the explicit construction of the general multipolar-post-Minkowskian metric in Sect.~\ref{sec:nonlinitere}. Namely, we must find the expressions of the six STF source multipole moments $\dI_L$, $\dJ_L, \cdots, \dZ_L$ parametrizing the linearized metric \eqref{hk1}--\eqref{phi1expr} at the basis of that construction. Recall that in actual applications we need mostly the mass-type moment $\dI_L$ and current-type one $\dJ_L$, because the other moments parametrize a linearized gauge transformation and give contributions to high PN order.

To do this we first find the equivalent of the multipole expansion given in Theorem \ref{th:multexp}, which was parametrized by non-trace-free multipole functions $\mathcal{H}^{\alpha\beta}_L$, in terms of new multipole functions $\mathcal{F}^{\alpha\beta}_L$ that are STF in all their indices $L$. The result is (introducing the intermediate notation $u\ab$ for convenience) 
\begin{subequations}\label{resmultSTF}
\begin{align}   
	u\ab &\equiv \FPprop \Bigl[\widetilde r^B \mathcal{
		M}(\Lambda^{\alpha\beta})\Bigr] \,, \label{defuab}\\
	\mathcal{M}(h^{\alpha\beta}) &= u\ab - \frac{4G}{c^4}
	\sum^{+\infty}_{\ell=0} \frac{(-)^\ell}{\ell!} \partial_L \left\{
	\frac{1}{r} \mathcal{F}^{\alpha\beta}_L (t-r/c) \right\}\,,
	\label{multSTF}
\end{align}
\end{subequations}
where the STF multipole functions (witness the multipolar factor $\hat{x}_L\equiv \mathrm{STF}[x_L]$ therein) read
\begin{equation}\label{FL}
	\mathcal{F}^{\alpha\beta}_L (u) = \FP \int \dd^3 \mathbf{x}
	\, \widetilde r^B \hat{x}_L \,\int^1_{-1} \dd z \,\delta_\ell(z) \,{\overline
		\tau}^{\alpha\beta}(\mathbf{x}, u+z r/c)\,.
\end{equation}
Notice the presence of an extra integration variable $z$, ranging from
$-1$ to $1$. The $z$-integration involves the weighting function
\begin{equation}
	\delta_\ell (z) = \frac{(2\ell+1)!!}{2^{\ell+1} \ell!}
	(1-z^2)^\ell\,,
	\label{deltal}
\end{equation}
which approaches the Dirac delta-function (hence its name) in the limit of large multipolarities, i.e. $\lim_{\ell\to +\infty}\delta_\ell(z)=\delta (z)$, and is normalized such that
\begin{equation}
	\int^1_{-1} \dd z \, \delta_\ell (z) = 1\,.
	\label{intdeltal}
\end{equation}
The next step is to impose the harmonic-gauge conditions \eqref{harmcond} onto the multipole decomposition \eqref{multSTF}, hence $\mathcal{M}(\partial_\mu h^{\alpha\mu})=0$. This implies
\begin{align}   
		\partial_\mu u^{\alpha\mu} &= \frac{4G}{c^4}
		\sum^{+\infty}_{\ell=0} \frac{(-)^\ell}{\ell!} \partial_L \left\{
		\frac{1}{r} \mathcal{G}^{\alpha}_L (t-r/c) \right\}\,,
		\label{multSTFharm}
\end{align}
with STF multipole functions therein given by
\begin{equation}\label{GL}
	\mathcal{G}^{\alpha}_L (u) = \FP B \int \dd^3 \mathbf{x}
	\, \widetilde r^B \frac{x_i}{r^2}\,\hat{x}_L \,\int^1_{-1} \dd z \,\delta_\ell(z) \,{\overline
		\tau}^{\alpha i}(\mathbf{x}, u+z r/c)\,.
\end{equation}
Note the explicit factor $B$ in front of the integral, so that $\mathcal{G}^{\alpha}_L$ is actually zero unless the integral develops a pole when $B\to 0$.

The final step is to decompose the multipole functions $\mathcal{F}^{\alpha\beta}_L(u)$ and $\mathcal{G}^{\alpha}_L(u)$ into STF irreducible pieces with respect to both $L$ and their spatial indices contained into $\alpha\beta=00,0i,ij$ or $\alpha=0,i$. This technical part of the calculation is identical to the one of the STF irreducible multipole moments of linearized gravity \citep{DI91b}. Thus the formulas needed in this decomposition read
\begin{subequations}\label{FLirred}
	\begin{align}
		\mathcal{F}^{00}_L &= R_L\,, \\  
		\mathcal{F}^{0i}_L &=
		\,^{(+)}T_{iL} + \epsilon_{ai<i_\ell} {}^{(0)}T_{L-1>a}
		+\delta_{i<i_\ell} {}^{(-)}T_{L-1>}\,, \\  
		\mathcal{F}^{ij}_L &= \,^{(+2)}U_{ijL} +
		{\mathop{\text{STF}}_L}\ {\mathop{\text{STF}}_{ij}}\,
		\Bigl[\epsilon_{aii_\ell} {}^{(+1)}U_{ajL-1} +\delta_{ii_\ell}
		{}^{(0)}U_{jL-1} \nn\\[-0.1cm] &~~ +\delta_{ii_\ell}
		\epsilon_{aji_{\ell-1}} {}^{(-1)}U_{aL-2}
		+\delta_{ii_\ell} \delta_{ji_{\ell-1}} {}^{(-2)}U_{L-2} \Bigr] +
		\delta_{ij} V_L\,,
	\end{align}
\end{subequations}
where the ten tensors $R_L, {}^{(+)}T_{L+1}, {}^{(0)}T_{L}, \cdots, {}^{(-2)}U_{L-2}, V_L$ are STF, and can be uniquely given in terms of the $\mathcal{F}^{\alpha\beta}_L$'s by some inverse formulas.
Similarly, with four suplementary STF tensors $P_L, \cdots, {}^{(-1)}Q_{L-1}$, we have
\begin{subequations}
	\begin{align}
		\mathcal{G}^{0}_L &= P_L\,, \\  
		\mathcal{G}^{i}_L &=
		\,^{(+)}Q_{iL} + \epsilon_{ai<i_\ell} {}^{(0)}Q_{L-1>a}
		+\delta_{i<i_\ell} {}^{(-)}Q_{L-1>}\,.
	\end{align}
	\label{GLirred}
\end{subequations}
Finally, the latter decompositions yield the following.

\begin{theorem} \citep{B98mult} The STF multipole moments $\dI_L$ and $\dJ_L$ of a post-Newtonian source are given, formally up to any post-Newtonian order, by ($\ell\geqslant 2$)\footnote{The monopole and dipole cases $\ell=0,1$ are discussed in \cite{B98mult}.}
\begin{subequations}\label{sourcemoments}
  \begin{align}
      \dI_L(u)&= \FP \int \dd^3\mathbf{x}\,\widetilde r^B \int^1_{-1}
      \dd z\biggl\{ \delta_\ell(z)\hat{x}_L\overline{\Sigma} -
      \frac{4(2\ell+1)}{c^2(\ell+1)(2\ell+3)} \delta_{\ell+1}(z)
      \hat{x}_{iL} \overline{\Sigma}^{(1)}_i \nn\\ & \qquad
      + \frac{2(2\ell+1)}{c^4(\ell+1)(\ell+2)(2\ell+5)}
      \delta_{\ell+2}(z) \hat{x}_{ijL} \overline{\Sigma}^{(2)}_{ij} \biggr\}
      (\mathbf{x}, u+z r/c)\,, \\
\dJ_L(u)&= \FP \int \dd^3\mathbf{x}\,\widetilde r^B\int^1_{-1} \dd z
\, \epsilon_{ab \langle i_\ell} \biggl\{ \delta_\ell(z) \hat{x}_{L-1
  \rangle a} \overline{\Sigma}_b \nn\\ & \qquad - \frac{2\ell+1}{c^2(\ell+2)(2\ell+3)}
\delta_{\ell+1}(z) \hat{x}_{L-1 \rangle ac} \overline{\Sigma}^{(1)}_{bc} \biggr\}
(\mathbf{x}, u+z r/c)\,.
  \end{align}\end{subequations}
These moments are the ones that are to be inserted into the linearized metric $h^{\alpha\beta}_{(1)}$ which represents the lowest approximation to the post-Minkowskian field $h_\mathrm{MPM}^{\alpha\beta}=\sum_{n\geqslant 1}G^nh^{\alpha\beta}_{(n)}$ defined in Eq.~\eqref{hIJZ}. \label{th:sourcemoments}
\end{theorem}
In Eqs. \eqref{sourcemoments} some convenient source densities are defined from the post-Newtonian expansion (denoted by an overbar) of the pseudo-tensor $\tau^{\alpha\beta}$ by
\begin{align}\label{Sigma}
  \overline{\Sigma} = \frac{\overline\tau^{00}+\overline\tau^{ii}}{c^2}\,,
  \qquad \overline{\Sigma}_i = \frac{\overline\tau^{0i}}{c}\,, \qquad \overline{\Sigma}_{ij} =
  \overline{\tau}^{ij}\,,
\end{align}
(where $\overline{\tau}^{ii} \equiv\delta_{ij}\overline\tau^{ij}$). As indicated in Eqs. \eqref{sourcemoments} all these quantities are to be evaluated at the spatial point $\mathbf{x}$ and at time $u+z r/c$. Since the source densities \eqref{Sigma} have a non-compact spatial support (because of the gravitational field contribution in the pseudo-tensor) the FP regularization in \eqref{sourcemoments} plays a crucial role in curing IR type divergences.

The source moments $\dI_L(u)$ and $\dJ_L(u)$ are physically valid for post-Newtonian sources and make sense only in the form of a post-Newtonian expansion, so in practice we need to know how to expand the $z$-integrals as series when $c\to +\infty$. Here is the appropriate formula:
\begin{equation}
	\int^1_{-1} \dd z \,\delta_\ell(z) \,\overline{\Sigma}(\mathbf{x}, u+z r/c) =
	\sum_{k=0}^{+\infty}\frac{(2\ell+1)!!}{2^k k!(2\ell+2k+1)!!}
	\biggl(\frac{r}{c}\frac{\partial}{\partial u}\biggr)^{\!2k}
	\!\overline{\Sigma}(\mathbf{x}, u)\,.
	\label{intdeltaexp}
\end{equation}
Since the right-hand side involves only even powers of $1/c$, the same result holds equally well for the advanced variable $u+z r/c$ or the retarded one $u-z r/c$. Of course, in the Newtonian limit, the moments $\dI_L$ and $\dJ_L$ 
reduce to the standard Newtonian expressions. For instance, $\dI_{ij}(u) = \mathrm{Q}_{ij}(u) + \calO(c^{-2})$ recovers the Newtonian quadrupole moment \eqref{Qij}; similarly for the current quadrupole $\dJ_{ij} = \dC_{ij} + \calO(c^{-2})$ and mass octupole $\dI_{ijk} = \dO_{ijk} + \calO(c^{-2})$ appearing in Eq.~\eqref{balanceP}.

An interesting fact is that at the 1PN order the mass type source moments $\dI_L$ extend up only over the \emph{compact support} of the source, hence we do not need the FP regularization in this case. We have \citep{BD89}
\begin{align}\label{IL1PN}
	\dI_L(u) &= \int \dd^3 \mathbf{x} \left\{ \hat{x}_L \sigma+
	\frac{r^2 \,\hat{x}_L}{2c^2(2\ell+3)} \frac{\partial^2\sigma}{\partial t^2}- \frac{4(2\ell+1) \,\hat{x}_{iL}}{c^2(\ell+1) (2\ell+3)}
	\,\frac{\partial\sigma_i}{\partial t} \right\}(\mathbf{x},u) \nn\\&+
	\calO\left(\frac{1}{c^4}\right)\,,
\end{align}
where $\sigma$ and $\sigma_i$ (and later $\sigma_{ij}$) denote the compact-support parts of the source densities \eqref{Sigma}, i.e. defined from the matter stress-energy tensor $T^{\alpha\beta}$ alone:
\begin{align}\label{sigma}
	\sigma \equiv \frac{T^{00}+T^{ii}}{c^2}\,,
	\qquad \sigma_{i} \equiv
	\frac{T^{0i}}{c}\,, \qquad \sigma_{ij} \equiv T^{ij}\,,
\end{align}
(with $T^{ii} \equiv\delta_{ij}T^{ij}$). 
Only from the 2PN order will the mass moments acquire some non-compact supported contributions -- i.e., with some integrals extending up to infinity \citep{B95}. Concerning the current type moments $\dJ_L$, they are given at Newtonian order by 
\begin{align}\label{JLN}
		\dJ_L &= \int \dd^3 \mathbf{x}\,\epsilon_{ab<i_\ell}
		\,\hat{x}_{L-1>a} \,\sigma_b +
		\calO\left(\frac{1}{c^2}\right)\,,
\end{align}
and the non compactness appearing already at the next 1PN order \citep{DI91a,B95}.

For completeness, we give also the formulas for the four auxiliary source moments $\dW_L, \cdots, \dZ_L$, which parametrize the gauge vector $\varphi^\alpha_{(1)}$ as defined in Eqs. \eqref{phi1expr}:
\begin{subequations}\label{gaugemoments}
\begin{align}
\dW_L(u)&= \FP \int \dd^3\mathbf{x}\,\widetilde r^B \int^1_{-1} \dd
z\biggl\{ {2\ell+1\over (\ell+1)(2\ell+3)} \delta_{\ell+1} \hat{x}_{iL}
\overline{\Sigma}_i \nn\\& \qquad \qquad \qquad \qquad -
      {2\ell+1\over2c^2(\ell+1)(\ell+2)(2\ell+5)} \delta_{\ell+2}
      \hat{x}_{ijL} {\overline{\Sigma}}_{ij}^{(1)} \biggr\}\,,\\
\dX_L(u) &= \FP \int \dd^3\mathbf{x}\,\widetilde r^B \int^1_{-1} \dd
z\biggl\{ {2\ell+1\over 2(\ell+1)(\ell+2)(2\ell+5)} \delta_{\ell+2}
\hat{x}_{ijL} \overline{\Sigma}_{ij} \biggr\}\,,\\
\dY_L(u)&=\FP \int \dd^3\mathbf{x}\,\widetilde r^B \int^1_{-1} \dd
z\biggl\{ -\delta_\ell \hat{x}_L \overline{\Sigma}_{ii} + {3(2\ell+1)\over
  (\ell+1)(2\ell+3)} \delta_{\ell+1} \hat{x}_{iL} {\overline{\Sigma}}_i^{(1)}
\nn\\& \qquad \qquad \qquad \qquad - {2(2\ell+1)\over
  c^2(\ell+1)(\ell+2)(2\ell+5)} \delta_{\ell+2} \hat{x}_{ijL}
          {\overline{\Sigma}}_{ij}^{(2)} \biggr\}\,,\\
\dZ_L(u) &= \FP \int \dd^3\mathbf{x}\,\widetilde r^B \int^1_{-1} \dd z \,
\epsilon_{ab \langle i_\ell}\biggl\{- {2\ell+1\over (\ell+2)(2\ell+3)}
\delta_{\ell+1} \hat{x}_{L-1 \rangle bc} \overline{\Sigma}_{ac} \biggr\}\,.
\end{align}\end{subequations}
As discussed in Sect.~\ref{sec:cansource}, one can always find two intermediate ``packages'' of multipole moments, namely the canonical moments $\dM_L$ and $\dS_L$, which are some non-linear functionals of the source moments \eqref{sourcemoments} and \eqref{gaugemoments}, and such that the exterior field depends only on them, modulo a change of coordinates. However, the canonical moments $\dM_L$, $\dS_L$ do not admit general closed-form expressions like \eqref{sourcemoments}--\eqref{gaugemoments}.\footnote{The work \cite{BDI04zeta} provided some alternative expressions for all the multipole moments \eqref{sourcemoments} and \eqref{gaugemoments}, useful for some applications, in the form of \emph{surface integrals} extending on the outer part of the source's near zone.}

Needless to say, the formalism becomes prohibitively difficult to apply at very high post-Newtonian approximations. Some PN order being given, we must first compute the relevant relativistic corrections to the pseudo stress-energy-tensor $\overline{\tau}^{\alpha\beta}$; this necessitates solving the field equations inside the matter source, which we shall investigate in the next Sect.~\ref{sec:intPN}. Then $\overline{\tau}^{\alpha\beta}$ is to be inserted into the source moments \eqref{sourcemoments} and \eqref{gaugemoments}, where the formula \eqref{intdeltaexp} permits expressing all the terms up to that post-Newtonian order by means of more tractable integrals extending over ${\mathbb{R}}^3$. Given a specific model for the matter source we then have to find a way to compute all these spatial integrals; this is done in Sect.~\ref{sec:binarymoments} for the case of point-mass binaries. Next, we must substitute the source multipole moments into the linearized metric \eqref{hk1}--\eqref{phi1expr}, and iterate them until all the necessary multipole interactions taking place in the radiative moments $\dU_L$ and $\dV_L$ are under control. We shall work out these multipole interactions for general sources in Sect.~\ref{sec:radcanonical} up to the 4.5PN order for the flux of circular orbits, and 4PN order for the dominant gravitational-wave $(2,2)$ mode. Only at this point does one have the physical radiation field at infinity, from which we can build the templates for the detection and analysis of gravitational waves. We advocate here that the complexity of the formalism simply reflects the complexity of the Einstein field equations. It is probably impossible to devise a different formalism, valid for general sources \textit{a priori} devoid of symmetries, that would be substantially simpler.


\subsection{Interior field of a post-Newtonian source}
\label{sec:intPN}

Theorem \ref{th:sourcemoments} solves the question of the generation of gravitational waves by extended post-Newtonian matter sources. However, notice that this result has still to be completed by the precise procedure, i.e., an explicit ``\emph{algorithm}'', for the post-Newtonian iteration of the near-zone field, analogous to the multipolar-post-Minkowskian algorithm we defined in Sect.~\ref{sec:nonlinitere}. Such procedure will permit the systematic computation of the source multipole moments, which contain the full post-Newtonian expansion of the pseudo-tensor $\overline{\tau}^{\alpha\beta}$, and of the radiation reaction effects occurring within the matter source.

Before proceeding, let us recall that the ``standard'' post-Newtonian approximation, as it was used until, say, the early 1980s,\footnote{See notably \cite{AD75, Ehl80, K80a, K80b, PapaL81}, and also the earlier works by \cite{PB59, C65, CN69, CE70}.} was plagued with some apparently inherent difficulties, which croped up at some high post-Newtonian order. Historically, these difficulties, even appearing at higher approximations, have cast a doubt on the actual soundness, from a theoretical point of view, of the post-Newtonian expansion. Practically speaking, they posed the question of the reliability of the approximation, when comparing the theory's predictions with very precise experimental results. This was one of the reason for the ``radiation-reaction controversy'' raging at the time of the binary pulsar data \citep{Ehletal76, WalkW80}. 

One can distinguish two main issues:
\begin{enumerate}
\item The first issue is that in higher approximations some \emph{divergent} Poisson-type integrals appear. Indeed the post-Newtonian expansion replaces the resolution of a hyperbolic-like d'Alembertian equation by a perturbatively equivalent hierarchy of elliptic-like Poisson equations. Rapidly it is found during the post-Newtonian iteration that the right-hand side of the Poisson equations acquires a non-compact support (it is distributed all over space $\mathbb{R}^3$), and that as a result the standard Poisson integral diverges at the bound of the integral at spatial infinity, i.e., when $r\equiv |{\mathbf x}|\to +\infty$, with $t=\mathrm{const}$ -- an IR problem;
\item The second issue is related to the limitation of the post-Newtonian approximation to the near zone -- the region surrounding the source of small extent with respect to the wavelength of the emitted radiation: $r\ll \lambda$. As we have seen, the PN expansion assumes from the start that all retardations $r/c$ are small, so it can rightly be viewed as a formal \emph{near-zone} expansion, when $r\to 0$. Note that the fact which makes the Poisson integrals to become typically divergent, namely that the coefficients of the post-Newtonian series blow up at spatial infinity, when $r\to +\infty$, has nothing to do with the actual behaviour of the field at infinity. However, the serious consequence is that it is \emph{a priori} impossible to implement within the post-Newtonian scheme alone the physical information that the matter system is isolated from the rest of the Universe. Most importantly, the no-incoming radiation condition, imposed at past null infinity, cannot be taken directly into account, \emph{a priori}, into the post-Newtonian scheme.  In this sense the post-Newtonian approximation is not ``self-supporting'', because it necessitates some information taken from outside its own domain of validity.
\end{enumerate}
The divergencies are linked to the fact that the PN expansion is actually a singular perturbation, in the sense that the coefficients of the successive powers of $1/c$ are not uniformly valid in space, since they typically blow up at spatial infinity like some powers of $r$. For instance the post-Newtonian metric cannot be ``asymptotically flat'' starting at the 2PN or 3PN level, depending on the adopted coordinate system \citep{Rend92}. The consequence is that the standard Poisson integrals are in general badly-behaving at infinity. Trying to solve the post-Newtonian equations by means of the Poisson integral does not make sense. However, this does not mean that there are no solutions to the problem, but simply that the Poisson integral does not constitute the appropriate solution of the Poisson equation in the PN context.

Here we review a solution of both issues \citep{PB02, BFN05}, in the form of a general expression for the near-zone gravitational field, developed to any post-Newtonian order, which is determined as a solution of the matching equation \eqref{matchingeq}. This solution is free of the divergences of Poisson-type integrals (issue 1), and yields, in particular, some general expression, valid up to any order, for the terms associated with the gravitational radiation reaction force inside the PN source (issue 2).

Other methods are possible in order to resolve the problems of the PN approximation. Notably, an alternative solution to the problem of divergencies has been proposed by \cite{FS83, F83}, based on the initial-value formulation. In this method the problem of the appearance of divergencies is avoided because of the finiteness of the causal region of integration, between the initial Cauchy hypersurface and the considered field point. On the other hand, a different approach to the problem of radiation reaction, which does not use a matching procedure, is to work only within a post-Minkowskian iteration scheme without expanding the retardations, and thus to naturally incorporate the correct boundary conditions at infinity \citep{CKMR01}.


\subsubsection{Post-Newtonian iteration in the near zone}
\label{sec:PNiter}

We perform the post-Newtonian iteration of the field equations in harmonic coordinates in the near zone of an isolated matter distribution. We deal with a general hydrodynamical fluid, whose stress-energy tensor is smooth, i.e., $T^{\alpha\beta}\in C^\infty(\mathbb{R}^4)$. Thus the scheme \emph{a priori} excludes the presence of singularities and black holes; these will be dealt with in Sect.~\ref{sec:compactbinary} of this article.

We shall now prove that the post-Newtonian expansion can be \emph{indefinitely} iterated without divergences. Like in Eq.~\eqref{HLab} we denote by means of an overline the formal (infinite) post-Newtonian expansion of the field inside the source's near-zone. The general structure of the post-Newtonian expansion is denoted as
\begin{equation}\label{hPNgen}
 \mathop{\overline{h}}_{}\!{}\ab({\mathbf x},t,c) = \sum_{m=2}^{+\infty}\,\frac{1}{c^m}\,
 \mathop{\overline{h}}_{m}\!{}\ab({\mathbf x},t;\ln c)\,.
\end{equation}
The $m$-th post-Newtonian coefficient is naturally the factor of the $m$-th power of $1/c$. However, we know from restoring the factors $c$'s in Theorem \ref{th:NZstruct} [see Eq.~\eqref{hnNZ}], that the post-Newtonian expansion also involves powers of the logarithm of $c$; these are included for convenience here into the definition of the coefficients $\mathop{\overline{h}}_{m}\!\!\!\!{}\ab$.\footnote{For this argument we assume the validity of the matching equation \eqref{matchingeq} and that the post-\emph{Minkowskian} series over $n=1,\cdots,+\infty$ in Eq.~\eqref{hnNZ} has been formally summed up.} For the stress-energy pseudo-tensor appearing in Eq.~\eqref{HLab} we have the same type of expansion,
\begin{equation}\label{tauPNgen}
	\mathop{\overline{\tau}}_{}\!{}\ab({\mathbf x},t,c) = \sum_{m=-2}^{+\infty}\,
	\frac{1}{c^m}\,\mathop{\overline{\tau}}_{m}\!{}\ab({\mathbf x},t;\ln c)\,,
\end{equation}
with the expansion starting with a term of order $c^2$ corresponding to the rest mass-energy ($\overline{\tau}\ab$ has the dimension of an energy density). As usual we shall understand the infinite sums such as \eqref{hPNgen}--\eqref{tauPNgen} in the sense of \emph{formal} series, i.e., merely as an ordered collection of PN coefficients. Because of our consideration of regular extended matter distributions the post-Newtonian coefficients are smooth functions of space-time, i.e., $\mathop{\overline{h}}_{m}\!\!\!\!{}\ab({\mathbf x},t)\in C^\infty(\mathbb{R}^4)$.

Inserting the post-Newtonian ansatz \eqref{hPNgen} into the harmonic-coordinates Einstein field equation \eqref{harmcond}--\eqref{EFE} and equating together the powers of $1/c$, results is an infinite set of Poisson-type equations ($\forall m\geqslant 2$),
\begin{equation}\label{Poisson}
\Delta\mathop{\overline{h}}_{m}\!{}\ab = 16\pi G
\!\!\mathop{\overline{\tau}}_{m-4}\!\!\!{}\ab
+\,\partial_{t}^{2}\!\!\mathop{\overline{h}}_{m-2}\!\!\!{}\ab\,, 
\end{equation}
where the second term comes from the split of the d'Alembertian operator into a Laplacian and a second time derivative: $\Box=\Delta-\frac{1}{c^2}\partial_t^2$ (this term is zero when $m=2$ and $3$). We proceed by induction, i.e., we work at some given but arbitrary post-Newtonian order $m$, assume that we succeeded in constructing the sequence of previous coefficients $\mathop{\overline{h}}_{p}\!\!\!{}\ab$ ($\forall p\leqslant m-1$), and from that show how to infer the next-order coefficient $\mathop{\overline{h}}_{m}\!\!\!\!{}\ab$.

To cure the problem of divergencies we introduce a generalized solution of the Poisson equation with non-compact support source, in the form of an appropriate \emph{finite part} of the usual Poisson integral obtained by regularization of the bound at infinity by means of the specific FP regularization process. For any source term like $\mathop{\overline{\tau}}_{m}\!\!\!\!{}\ab$, we multiply it by the regularization factor $\widetilde{r}^B$ already extensively used in the construction of the exterior field, thus $B\in\mathbb{C}$ and $\widetilde{r}=r/r_0$ as given by \eqref{regfactor}. Only then do we apply the usual Poisson integral, which therefore defines a certain function of $B$. The well-definedness of that integral heavily relies on the behaviour of the integrand at the bound at infinity. There is no problem with the vicinity of the origin inside the source because of the smoothness of the pseudo-tensor. Then the latter function of $B$ generates a (unique) analytic continuation down to a neighbourhood of the value of interest $B=0$, except at $B=0$ itself, at which value it admits a Laurent expansion with multiple poles up to some finite order (but growing with the post-Newtonian order $m$). Then, we consider the Laurent expansion of that function when $B\rightarrow 0$ and pick up the finite part, or coefficient of the zero-th power of $B$, of that expansion. This \emph{defines} our generalized Poisson integral:
\begin{equation}\label{genpoiss}
\FP\Delta^{-1}\big[\widetilde{r}^B\!\mathop{\overline{\tau}}_{m}\!{}\ab\big](\mathbf{x},t)
\equiv -\frac{1}{4\pi}\,\FP\,\int
\frac{\dd^3\mathbf{x}'}{\vert\mathbf{x}-\mathbf{x}'\vert}
\,\widetilde{r}'^B\, \mathop{\overline{\tau}}_{m}\!{}\ab(\mathbf{x}',t) \,.
\end{equation} 
The integral extends over all three-dimensional space but are endowed with the latter finite-part regularization. The main properties of this generalized Poisson operator is that it solves the Poisson equation,
\begin{equation}\label{checkpoiss}
\Delta\left(\FP\Delta^{-1}\big[\widetilde{r}^B\!\mathop{\overline{\tau}}_{m}\!{}\ab\big]\right)
= \mathop{\overline{\tau}}_{m}\!{}\ab\,,
\end{equation} 
and that the solution $\Delta^{-1}\overline{\tau}_m\!\!\!\!{}\ab$ owns the same properties as its source $\overline{\tau}_m\!\!\!\!{}\ab$, i.e., the smoothness and the same type of behaviour at infinity, as given by Eq.~\eqref{struct}. Similarly, we define the generalized iterated Poisson integral as
\begin{equation}\label{genpoissiter}
\FP\Delta^{-k-1}\big[\widetilde{r}^B\!\mathop{\overline{\tau}}_{m}\!{}\ab\big](\mathbf{x},t)
\equiv -\frac{1}{4\pi}\,\FP \int
\dd^3\mathbf{x}'\,\frac{\vert\mathbf{x}-\mathbf{x}'\vert^{2k-1}}{(2k)!}
\,\widetilde{r}'^B\, \mathop{\overline{\tau}}_{m}\!{}\ab(\mathbf{x}',t) \,.
\end{equation} 

The most general solution of the Poisson equation will be obtained by application of the previous generalized Poisson operator to the right-hand side of Eq.~\eqref{Poisson}, and augmented by the most general \emph{homogeneous} solution of the Poisson equation. Thus, we can write
\begin{equation}\label{hngen}
\mathop{\overline{h}}_{m}\!{}\ab = 16\pi
G\,\FP\Delta^{-1}\big[\widetilde{r}^B\!\mathop{\overline{\tau}}_{m-4}\!\!\!{}\ab\big]
+\partial_{t}^{2}\!\FP\Delta^{-1}\big[\widetilde{r}^B\!\mathop{\overline{h}}_{m-2}\!\!\!{}\ab\big] +
\sum_{\ell=0}^{+\infty}\mathop{\mathcal{B}}_{m}{}_{\!L}\!\!\ab(t)\,\hat{x}_L\,.
\end{equation}
The last term represents the most general solution of the Laplace equation that is regular at the origin $r=0$. It can be written in STF guise as a multipolar series of terms of the type $\hat{x}_L$, and multiplied by arbitrary STF-tensorial functions of time ${}_m\mathcal{B}_L\!\!{}\ab(t)$. These functions will be associated with the radiation reaction of the field onto the source; they will depend on which boundary conditions are to be imposed on the field at infinity from the source.

It is now trivial to iterate the process. We substitute for $\overline{h}_{m-2}\!\!\!\!\!\!\!{}\ab$ in the right-hand side of Eq.~\eqref{hngen} the same expression but with $m$ replaced by $m-2$, and similarly come down until we stop at either one of the coefficients $\overline{h}_0\!\!\!{}\ab=0$ or $\overline{h}_1\!\!\!{}\ab=0$. At this point $\overline{h}_m\!\!\!\!{}\ab$ is expressed in terms of the previous $\overline{\tau}_p\!\!{}\ab$'s and ${}_{p}\mathcal{B}_L\!\!{}\ab$'s with $p\leqslant m-2$. To finalize the process we introduce what we call the operator of the ``\emph{instantaneous}'' potentials and denote it by $\Box^{-1}_{\text{inst}}$. Our notation is chosen to contrast with the standard operator of the retarded potentials $\Box^{-1}_{\mathrm{ret}}$ defined by Eq.~\eqref{dalembertian}. However, beware of the fact that unlike $\Box^{-1}_{\mathrm{ret}}$ the operator $\Box^{-1}_{\text{inst}}$ will be defined only when acting on a post-Newtonian series such as $\overline{\tau}\ab$. Indeed, we pose
\begin{equation}\label{instpotentials}
\FP\Box^{-1}_{\text{inst}}\big[\widetilde{r}^B\overline{\tau}\ab\big] \equiv
\sum_{k=0}^{+\infty}\left(\frac{\partial}{c\partial
  t}\right)^{\!\!2k}\FP\Delta^{-k-1} \big[\widetilde{r}^B\overline{\tau}\ab\big]\,,
\end{equation}
where the $k$-th iteration of the generalized Poisson operator is defined by Eq.~\eqref{genpoissiter}. This operator is ``instantaneous'' in the sense that it does not involve any integration over time. It is readily checked that in this way we have a solution of the d'Alembertian equation,
\begin{equation}\label{BoxBoxI}
\Box\left(\FP\Box^{-1}_\text{inst} \big[\widetilde{r}^B \overline{\tau}\ab\big]\right) =
\overline{\tau}\ab \,.
\end{equation} 
On the other hand, the homogeneous solution in Eq.~\eqref{hngen} will yield by iteration an homogeneous solution of the d'Alembertian equation that is necessarily regular at the origin. Hence it should be of the \emph{anti-symmetric} type, i.e., be made of the difference between a retarded multipolar wave and the corresponding advanced wave. We shall therefore introduce a new definition for some STF-tensorial functions $\mathcal{A}_L\ab(t)$ parametrizing those advanced-minus-retarded free waves. It is very easy to relate if necessary the post-Newtonian expansion of $\mathcal{A}_L\ab(t)$ to the functions ${}_m \mathcal{B}_L\ab(t)$ previously introduced in Eq.~\eqref{hngen}. Finally the most general post-Newtonian solution, iterated \emph{ad infinitum} and without any divergences, is obtained into the form
\begin{equation}\label{hgen1}
\!\!\!\mathop{\overline{h}}_{}\!{}\ab = \frac{16\pi G}{c^4}
\FP\Box^{-1}_\text{inst}\big[\widetilde{r}^B \overline{\tau}\ab\big] -
\frac{4G}{c^4}\sum^{+\infty}_{\ell=0}
\frac{(-)^\ell}{\ell!}\hat{\partial}_L \left\{ \frac{\mathcal{A}_L\ab
  (t-r/c)-\mathcal{A}_L\ab (t+r/c)}{2r} \right\}\,.
\end{equation} 
We shall refer to the $\mathcal{A}_L\ab(t)$'s as the \emph{radiation-reaction} functions. If we stay at the level of the post-Newtonian iteration which is confined into the near zone we cannot do more than Eq.~\eqref{hgen1}: There are no means to compute the radiation-reaction functions $\mathcal{A}_L\ab(t)$. We are here touching the second problem faced by the standard post-Newtonian approximation.


\subsubsection{Post-Newtonian metric and radiation reaction effects}
\label{sec:radreac}
 
As we have understood this problem is that of the limitation to the near zone. Such limitation can be circumvented to the lowest post-Newtonian orders by considering \emph{retarded} integrals that are formally expanded when $c\to +\infty$ as series of ``instantaneous'' Poisson-like integrals, see e.g., \cite{AD75}. This procedure works well up to the 2.5PN level and has been shown to correctly fix the dominant radiation reaction term at the 2.5PN order \citep{Ehl80, K80a, K80b, PapaL81}. Unfortunately the procedure assumes fundamentally that the gravitational field, after expansion of all retardations $r/c\to 0$, depends on the state of the source at a single time $t$, in keeping with the instantaneous character of the Newtonian interaction. However, we know that the post-Newtonian field (as well as the source's dynamics) will cease at some stage to be given by a functional of the source parameters at a single time, because of the imprint of gravitational-wave tails in the near zone field, in the form of an hereditary contribution (including a modification of the radiation reaction force) at the 4PN relative order \citep{BD88}. Thus the formal post-Newtonian expansion of retarded Green functions is no longer valid starting at the 4PN order.

The solution of the problem resides in the matching of the near-zone field to the exterior field. We have already seen in Theorems \ref{th:multexp} and \ref{th:sourcemoments} that the matching equation \eqref{matchingeq} yields the expression of the multipole expansion in the exterior domain. Now we prove that it also permits the full determination of the post-Newtonian metric in the near-zone, i.e., the radiation-reaction functions $\mathcal{A}_L\ab$ which have been left unspecified in Eq.~\eqref{hgen1}. As a result we find that the functions $\mathcal{A}_L\ab$ are composed of the multipole moment functions $\mathcal{F}_L\ab$ defined by Eq.~\eqref{FL}, which will characterize ``linear-order'' radiation reaction effects starting at 2.5PN order, and of an extra piece $\mathcal{R}_L\ab$, which will be due to non-linear effets in the radiation reaction (arising from radiative modes propagating at infinity) and turn out to arise at the 4PN order. Thus,
\begin{equation} \label{AFR}
\mathcal{A}_L\ab(t) = \mathcal{F}_L\ab(t) +\mathcal{R}_L\ab(t)\,.
\end{equation}
The piece $\mathcal{R}_L\ab$ is obtained from the multipole expansion of the pseudo-tensor $\mathcal{M}(\tau\ab)$.\footnote{We mean the fully-fledge $\mathcal{M}(\tau\ab)$; i.e., \emph{not} the formal object $\mathcal{M}(\overline{\tau}\ab)$.} Hence the radiation-reaction functions do depend on the behaviour of the field far away from the matter source (as physical intuition already told us). The explicit expression reads \citep{PB02}
\begin{equation}\label{RL}
\mathcal{R}_L\ab(t) =
\FP \int\,\dd^3\mathbf{x}\,\widetilde r^B \hat{x}_L\int_{1}^{+\infty}\dd
z\, \gamma_\ell(z)\,\mathcal{M}(\tau\ab)\,\left(\mathbf{x},t -z
r/c\right) \,.
\end{equation}
The fact that the multipolar expansion $\mathcal{M}(\tau\ab)$ is the source term for the function $\mathcal{R}_L\ab$ is the consequence of the matching equation \eqref{matchingeq}. The specific contributions due to $\mathcal{R}_L\ab$ in the post-Newtonian metric \eqref{hgen1} are associated with gravitational wave tails. Notice that the FP regularization deals with the bound of the integral at $r=0$ in Eq.~\eqref{RL}, while in Eq.~\eqref{FL} it deals with the bound at $r=+\infty$. The weighting function $\gamma_\ell(z)$ therein, where $z$ extends up to infinity in contrast to the analogous function $\delta_\ell(z)$ in Eq.~\eqref{FL}, is simply related to it by $\gamma_\ell(z)\equiv-2\delta_\ell(z)$; such definition is motivated by the fact that the integral of that function is normalized to one: $\int_1^{+\infty} \dd z\,\gamma_\ell(z) = 1$.\footnote{Though the latter integral is \emph{a priori} divergent, its value can be determined by invoking complex analytic continuation in $\ell\in\mathbb{C}$.} The post-Newtonian metric \eqref{hgen1} is now fully determined. However, let us now prove an interesting alternative formulation of the metric.

\begin{theorem} \citep{BFN05} The expression of the post-Newtonian field in the near zone of a post-Newtonian source, satisfying correct boundary conditions at infinity (no incoming radiation), reads
  \begin{equation}
\mathop{\overline{h}}_{}\!{}\ab=\frac{16\pi
  G}{c^4}\,\Box_{\mathrm{ret}}^{-1}\big[\,\overline{\tau}\ab\big]
- \frac{4G}{c^4}\sum^{+\infty}_{\ell=0}
\frac{(-)^\ell}{\ell!}\hat{\partial}_L \left\{
\frac{\mathcal{R}\ab_L (t-r/c)-\mathcal{R}\ab_L
  (t+r/c)}{2r} \right\}\,.
    \label{hgen2}
  \end{equation}
The first term represents a particular solution of the hierarchy of PN equations, while the second one is a homogeneous multipolar solution of the wave equation, of the ``anti-symmetric'' type that is regular at the origin $r=0$ located inside the source, and parametrized by the multipole-moment functions \eqref{RL}.
  \label{th:PNexp}
\end{theorem}
Let us be more precise about the meaning of the first term in Eq.~\eqref{hgen2}. Indeed such term is made of the formal expansion of the standard retarded integral \eqref{dalembertian} when $c\rightarrow\infty$, but acting on a \emph{post-Newtonian} source term $\overline{\tau}\ab$,
\begin{equation} \label{BoxRreg}
\Box_{\mathrm{ret}}^{-1}\big[\,\overline{\tau}^{\alpha\beta}\big](\mathbf{x},t)
\equiv -\frac{1}{4\pi}\sum_{m=0}^{+\infty}\frac{(-)^m}{m!}
\left(\frac{\partial}{c\,\partial t}\right)^{\!m}\!\FP \int
\dd^3\mathbf{x}'\,\widetilde{r}'^B \vert\mathbf{x}
-\mathbf{x}'\vert^{m-1}\,\overline{\tau}^{\alpha\beta}(\mathbf{x}',t)
\,.
\end{equation}
We emphasize that \eqref{BoxRreg} constitutes the \emph{definition} of a (formal) PN expansion, each term of which being built from the post-Newtonian expansion of the pseudo-tensor. Crucial in the present formalism, is that each of the terms is regularized by means of the FP operation in order to deal with the bound at infinity at which the post-Newtonian expansion is singular. Because of the presence of this regularization, the object \eqref{BoxRreg} should carefully be distinguished from the ``global'' solution $\Box_{\mathrm{ret}}^{-1}[\tau\ab]$ defined by Eq.~\eqref{dalembertian}, with global non-expanded pseudo-tensor $\tau\ab$.

The definition \eqref{BoxRreg} is of interest because it corresponds to what one would intuitively think as the natural way of performing the PN iteration, i.e., by formally Taylor expanding the retardations in Eq.~\eqref{dalembertian}, as advocated by \cite{AD75}. Moreover, each of the terms of the series \eqref{BoxRreg} is mathematically well-defined thanks to the finite part operation, and can therefore be implemented in practical computations. The point is that Eq.~\eqref{BoxRreg} solves the wave equation in a perturbative post-Newtonian sense,
\begin{equation}\label{BoxBoxR}
\Box\left(\Box_{\mathrm{ret}}^{-1}\big[
  \,\overline{\tau}^{\alpha\beta}\big]\right) =
\overline{\tau}^{\alpha\beta} \,,
\end{equation} 
so constitutes a good prescription for a particular solution of the wave equation -- as legitimate as the solution \eqref{instpotentials}. Therefore the two solutions should differ by an homogeneous solution of the wave equation which is necessarily of the anti-symmetric type (regular inside the source). Detailed investigations yield
\begin{equation}\label{BoxRI}
\Box_{\mathrm{ret}}^{-1}\big[\,\overline{\tau}^{\alpha\beta}\big] =
\Box_\text{inst}^{-1}\big[\,\overline{\tau}^{\alpha\beta}\big] -
\frac{1}{4\pi}\sum^{+\infty}_{\ell=0}
\frac{(-)^\ell}{\ell!}\hat{\partial}_L \left\{
\frac{\mathcal{F}^{\alpha\beta}_L (t-r/c)-\mathcal{F}^{\alpha\beta}_L
  (t+r/c)}{2r} \right\}\,,
\end{equation} 
where the homogeneous solution is parametrized by the multipole-moments $\mathcal{F}_L\ab(t)$. By combining Eqs. \eqref{hgen2} and \eqref{BoxRI}, we indeed become reconciled with the previous expression of the post-Newtonian field found in Eq.~\eqref{hgen1}.

For computations limited to the 3.5PN order (level of the 1PN correction to the radiation reaction force), the first term in Eq.~\eqref{hgen2} with the ``intuitive'' prescription \eqref{BoxRreg} is sufficient. But because of the second term in \eqref{hgen2} there is a fundamental breakdown of this scheme at the 4PN order where it becomes necessary to take into account non-linear radiation effects associated with tails. The second term in \eqref{hgen2} constitutes a generalization of the tail-transported radiation mode arising at 4PN order \citep{BD88}. In particular, the tail-transported radiation reaction is required by energy conservation and the propagation of tails in the wave zone. The usual radiation reaction terms, up to 3.5PN order, are contained in the first term of Eq.~\eqref{hgen2}, and are parametrized by the same multipole-moment functions $\mathcal{F}_L\ab$ as the exterior multipolar field, as Eq.~\eqref{BoxRI} explicitly shows. In Sect.~\ref{sec:4PNradreac} we shall give an explicit expression of the radiation reaction force showing the usual radiation reaction terms to 3.5PN order, issued from $\mathcal{F}_L\ab$, and exhibiting the above tail-induced 4PN effect, issued from $\mathcal{R}_L\ab$.

Finally note that the post-Newtonian solution, in either \eqref{hgen1} or \eqref{hgen2} form, has been obtained without imposing the condition of harmonic coordinates \eqref{harmcond} in an explicit way. We have simply matched together the PN and multipolar expansions, satisfying the ``relaxed'' Einstein field equations \eqref{EFE} in their respective domains, and found that the matching determines uniquely the solution. An important check in \citealt{PB02, BFN05}, is to verify that the harmonic coordinate condition \eqref{harmcond} is indeed satisfied as a consequence of the conservation of the pseudo-tensor \eqref{dtau0}, so that we really grasp a solution of the full Einstein field equations.


\subsubsection{The 3.5PN metric for general matter systems}
\label{sec:3PNmetric}

The detailed calculations that are called for in applications necessitate having at one's disposal some explicit expressions of the metric coefficients $g_{\alpha\beta}$, in harmonic coordinates, at the highest possible post-Newtonian order. The 3.5PN metric that we present below can be viewed as an application of the formalism of the previous section. It is expressed by means of some particular retarded-type potentials, $V$, $V_i$, $\hat{W}_{ij}$, $\cdots$, whose main advantages are to somewhat minimize the number of terms, so that even at the 3.5PN order the metric is still tractable, and to delineate the different problems associated with the computation of different categories of terms. Of course, these potentials have no direct physical significance by themselves, but they offer a convenient parametrization of the 3.5PN metric.

The idea of this post-Newtonian parametrization comes from the following simple expression of the metric at the 1PN order.

\begin{theorem} \citep{BD89} The post-Newtonian metric in harmonic coordinates at the 1PN order is described by a scalar potential $V$ and a vector potential $V_i$ as
\begin{subequations}\label{gmunu1PN}
\begin{align}
			g_{00} &= - 1 + \frac{2}{c^2} V - \frac{2}{c^4} V^2 + \calO
			\left( \frac{1}{c^{6}} \right)\,, \\ 
			g_{0i} &= - \frac{4}{c^3} V_i + \calO\left(\frac{1}{c^5}\right)\,,
			\\
			g_{ij} &= \delta_{ij} \left[ 1 + \frac{2}{c^2} V 
			\right] +
			\calO\left(\frac{1}{c^4} \right)\,.
\end{align}
\end{subequations}
The potentials represent some retarded versions of the gravito-electric (Newtonian) and gravito-magnetic potentials,
\begin{align}\label{pot1PN}
			V = \Box^{-1}_{\mathrm{ret}} \bigl[ -4 \pi G \sigma \bigr]\,,\qquad 
			V_i = \Box^{-1}_{\mathrm{ret}} \bigl[ -4 \pi G \sigma_i \bigr]\,,
\end{align}
and the components of the matter stress-energy tensor $T^{\alpha\beta}$ are encoded into the convenient definitions [recalling Eqs. \eqref{sigma}],
\begin{align}\label{sigmadef}
	\sigma \equiv \frac{T^{00}+T^{ii}}{c^2}\,,
		\qquad \sigma_{i} \equiv
		\frac{T^{0i}}{c}\,, \qquad \sigma_{ij} \equiv T^{ij}\,,
\end{align}
(where $T^{ii} \equiv\delta_{ij}T^{ij}$). In particular $\sigma$ agrees up to 1PN order with the \cite{Tolman} mass density valid for stationary systems. \label{th:1PN}
\end{theorem}
The generalization to 3PN order -- actually 3.5PN since the radiation-reaction odd parity terms are included into the definition of the potentials -- reads \citep{BFeom}
\begin{subequations}\label{gmunu3PN}
  \begin{align}
  g_{00} &= - 1 + \frac{2}{c^2} V - \frac{2}{c^4} V^2 + \frac{8}{c^6}
  \left( \hat{X} + V_i V_i + \frac{V^3}{6} \right) \nn\\ &+
  \frac{32}{c^8} \left( \hat{T} - \frac{1}{2} V \hat{X} + \hat{R}_i
  V_i - \frac{1}{2} V V_i V_i - \frac{1}{48} V^4 \right) + \calO
  \left( \frac{1}{c^{10}} \right)\,, \\ 
g_{0i} &= - \frac{4}{c^3} V_i - \frac{8}{c^5} \hat{R}_i -
\frac{16}{c^7} \left( \hat{Y}_i + \frac{1}{2} \hat{W}_{ij} V_j +
\frac{1}{2} V^2 V_i \right) + \calO\left(\frac{1}{c^9}\right)\,,
\\
g_{ij} &= \delta_{ij} \left[ 1 + \frac{2}{c^2} V + \frac{2}{c^4} V^2 +
  \frac{8}{c^6} \left(\hat{X} + V_k V_k + \frac{V^3}{6} \right)
  \right] + \frac{4}{c^4} \hat{W}_{ij} \nn\\ &+ \frac{16}{c^6}
\left( \hat{Z}_{ij} + \frac{1}{2} V \hat{W}_{ij} - V_i V_j \right) +
     \calO\left(\frac{1}{c^8} \right)\,.
  \end{align}
\end{subequations}
In addition to \eqref{pot1PN}, from the 2PN order we have the potentials
\begin{subequations}\label{pot2PN}
  \begin{align}
    \hat{X} &= \Box^{-1}_{\mathrm{ret}} \left[ - 4 \pi G
      V \sigma_{ii}+\hat{W}_{ij} \partial_{ij} V \right.\nn\\& \left.\qquad\qquad +2 V_i \partial_t
      \partial_i V+ V \partial_t^2 V + \frac{3}{2} (\partial_t V)^2 -
      2 \partial_i V_j \partial_j V_i \right]\,, \\  
\hat{R}_i &= \Box^{-1}_{\mathrm{ret}} \left[ -4 \pi G (V \sigma_i-V_i
  \sigma)-2 \partial_k V \partial_i V_k-\frac{3}{2} \partial_t V
  \partial_i V \right], \\ \hat{W}_{ij} &= \Box^{-1}_{\mathrm{ret}}
\left[ -4 \pi G (\sigma_{ij}-\delta_{ij} \sigma_{kk})- \partial_i V
  \partial_j V \right]\,.
  \end{align}
\end{subequations}
As we see, some parts of these potentials are directly generated by compact-support matter terms, while other parts are made of non-compact-support products of $V$-type potentials. There exists also an important cubically non-linear term generated by the coupling between $\hat{W}_{ij}$ and $V$, see the second term in the $\hat{X}$-potential. Note the important point that here and below the retarded integral operator $\Box^{-1}_{\mathrm{ret}}$ is really meant to be the one given by Eq.~\eqref{BoxRreg}; thus it involves in principle the finite part regularization FP to deal with (IR-type) divergences occurring at high PN orders for non-compact-support integrals. For instance, such finite part regularization is important to take into account in the computation of the near zone metric at the 3PN order and for the radiation field at the 4PN order \citep{BDLW10a, BFHLT23b}.

At the next level, 3PN, we have the even more complicated potentials
\begin{subequations}\label{pot3PN}
  \begin{align}
    \hat{T} &= \Box^{-1}_{\mathrm{ret}} \biggl[ -4 \pi G \left(
      \frac{1}{4} \sigma_{ij} \hat{W}_{ij}+\frac{1}{2} V^2 \sigma_{ii}
      + \sigma V_i V_i \right) + \hat{Z}_{ij} \partial_{ij} V +
      \hat{R}_i \partial_t \partial_i V \nn\\ & \qquad\qquad - 2 \partial_i V_j \partial_j
      \hat{R}_i - \partial_i V_j \partial_t \hat{W}_{ij} + V V_i \partial_t \partial_i V + 2 V_i \partial_j V_i
      \partial_j V \nn\\ & \qquad\qquad + \frac{3}{2} V_i \partial_t V \partial_i
      V+\frac{1}{2} V^2 \partial^2_t V + \frac{3}{2} V (\partial_t
      V)^2 - \frac{1}{2} (\partial_t V_i)^2 \biggr]\,, \\
\hat{Y}_i &= \Box^{-1}_{\mathrm{ret}} \biggl[ -4 \pi G \left( - \sigma
  \hat{R}_i - \sigma V V_i + \frac{1}{2} \sigma_k \hat{W}_{ik} +
  \frac{1}{2} \sigma_{ik} V_k + \frac{1}{2} \sigma_{kk} V_i \right)
  \nn\\ & \qquad \:\: +\, \hat{W}_{kl} \partial_{kl} V_i -
  \partial_t \hat{W}_{ik} \partial_k V + \partial_i \hat{W}_{kl}
  \partial_k V_l - \partial_k \hat{W}_{il} \partial_l V_k - 2
  \partial_k V \partial_i \hat{R}_k \nn\\ & \qquad - \frac{3}{2} V_k \partial_i V
  \partial_k V - \frac{3}{2} V \partial_t
  V \partial_i V - 2 V \partial_k V \partial_k V_i + V \partial^2_t
  V_i + 2 V_k \partial_k \partial_t V_i \biggr]\,, \\
\hat{Z}_{ij} &= \Box^{-1}_{\mathrm{ret}} \biggl[ -4 \pi G V \left(
  \sigma_{ij} - \delta_{ij} \sigma_{kk} \right) - 2 \partial_{(i} V
  \partial_t V_{j)} + \partial_i V_k \partial_j V_k + \partial_k V_i
  \partial_k V_j \nn\\ &
  \qquad - 2 \partial_{(i} V_k \partial_k V_{j)} - \frac{3}{4} \delta_{ij} (\partial_t V)^2 - \delta_{ij}
  \partial_k V_m (\partial_k V_m-\partial_m V_k) \biggr]\,.
  \end{align}
\end{subequations}
These involve many types of compact-support contributions, as well as quadratic-order and cubic-order parts; but, surprisingly, there are no quartically non-linear terms. Indeed it has been possible to ``integrate directly'' all the quartic contributions in the 3PN metric; see the terms composed of $V^4$ and $V\hat{X}$ in the first of Eqs. \eqref{gmunu3PN}.

Note that the 3PN metric \eqref{gmunu3PN} does represent the inner post-Newtonian field of an \emph{isolated} system, because it contains, to this order, the correct radiation-reaction terms corresponding to outgoing radiation. These terms come from the expansions of the retardations in the retarded potentials \eqref{pot1PN}--\eqref{pot3PN}; we elaborate more on radiation-reaction effects in Sect.~\ref{sec:4PNradreac}.

The above potentials are not independent: They are linked together by some differential identities issued from the harmonic gauge conditions, which are equivalent, via the Bianchi identities, to the equations of motion of the matter fields; see Eq.~\eqref{dtau0}. These identities read
\begin{subequations}
  \begin{align}
& \partial_t \left\{ V +\frac{1}{c^2} \left[ \frac{1}{2} \hat{W}_{kk}+
      2 V^2 \right] + \frac{4}{c^4} \left[ \hat{X} + \frac{1}{2}
      \hat{Z}_{kk}+\frac{1}{2} V \hat{W}_{kk} + \frac{2}{3} V^3
      \right] \right\} \nn\\ &\quad + \partial_i \left\{ V_i + \frac{2}{c^2}
    \left[\hat{R}_i + V V_i \right] + \frac{4}{c^4} \left[\hat{Y}_i -
      \frac{1}{2} \hat{W}_{ij} V_j + \frac{1}{2} \hat{W}_{kk} V_i + V
      \hat{R}_i + V^2 V_i\right] \right\} \nn\\ &\qquad\qquad\qquad =
    \calO\left(\frac{1}{c^6}\right)\,, \\
& \partial_t \left\{ V_i + \frac{2}{c^2} \left[ \hat{R}_i + V V_i
      \right] \right\} + \partial_j \left\{ \hat{W}_{ij} - \frac{1}{2}
    \hat{W}_{kk} \delta_{ij} + \frac{4}{c^2} \left[\hat{Z}_{ij} -
      \frac{1}{2} \hat{Z}_{kk} \delta_{ij} \right] \right\} \nn\\ &\qquad\qquad\qquad =
    \calO\left(\frac{1}{c^4}\right)\,.
  \end{align}
\end{subequations}
The above 3PN metric and differential identities have been generalized to 4PN order in the Appendix~A of \cite{MHLMFB20}.

For later applications to systems of compact objects, let us give the geodesic equations of a particle moving in the 3.5PN metric \eqref{gmunu3PN}. Of course the geodesic equations are appropriate for the motion of particles without spins nor internal structure; for spinning particles one has also to take into account the coupling of the spin to the space-time curvature, see the Mathisson-Papapetrou equations \eqref{MathPapa}. It is convenient to write the geodesic equations in the form (where $t=x^0/c$ is the coordinate time)
\begin{equation}\label{dPdtF}
\frac{\dd P_i}{\dd t} = F_i \,,
\end{equation}
where the ``linear momentum density'' $P_i$ and the ``force density'' $F_i$ of the particle are given by
\begin{subequations}\label{PiFidef}
\begin{align} 
P_i &= \frac{g_{i\mu}(x) \,v^\mu}{ \sqrt{- g_{\rho\sigma}(x) \,\frac{v^\rho
    v^\sigma}{c^2}}}\,,\\ F_i &= \frac{1}{2}\frac{ \partial_i g_{\mu\nu}(x) \,v^\mu
  v^\nu}{ \sqrt{- g_{\rho\sigma}(x) \,\frac{v^\rho v^\sigma}{c^2}}}\,,
\end{align}\end{subequations}
where $v^\mu=(c, v^i)$ with $v^i=\dd x^i/\dd t$ being the particle's ordinary coordinate velocity, and where the metric components and their gradients are taken at the location of the particle. Notice that we are viewing here the particle as moving in the fixed background metric \eqref{gmunu3PN}. In Sect.~\ref{sec:compactbinary}, the metric will be generated by the system of particles itself, and we shall see that the same equations still hold in this case, provided that we supplement the computation of the metric at the location of one of these particles by a suitable self-field regularization.

Further denoting $P_i=v_i+\hat{P}_i$ the equations give the ordinary acceleration $a^i=\dd v^i/\dd t$ as
\begin{equation}\label{dPdt}
	a^i = F_i - \frac{\dd \hat{P}_i}{\dd t}\,,
\end{equation}
where the expressions of $F_i$ and $\hat{P}_i$ in terms of the non-linear potentials follow from insertion of the 3.5PN metric coefficients \eqref{gmunu3PN}. We obtain 
\begin{subequations}\label{PiFi}\begin{align}
		\label{Pi} \hat{P}_{i} &= \frac{1}{c^2} \left(\frac{1}{2} v^2v^i +3 V v^i-4 V_i
		\right) \nn\\ &+ \frac{1}{c^4} \left(\frac{3}{8} v^4 v^i+\frac{7}{2} V
		v^2 v^i-4 V_j v^i v^j -2 V_i v^2 \right.\nn\\&\left.\qquad\qquad + \frac{9}{2} V^2 v^i-4 V V_i +4
		\hat{W}_{ij} v^j-8 \hat{R}_i \right) \nn\\& + \frac{1}{c^6}
		\left(\frac{5}{16} v^6 v^i+\frac{33}{8} V v^4 v^i-\frac{3}{2} V_i v^4
		-6 V_j v^i v^j v^2+ \frac{49}{4} V^2 v^2 v^i \right.\nn\\&\left.\qquad\qquad +2 \hat{W}_{ij} v^j
		v^2-4 \hat{R}_i v^2 +2 \hat{W}_{jk} v^i v^j v^k-10 V
		V_i v^2-20 V V_j v^i v^j \right.\nn\\&\left.\qquad\qquad -8 \hat{R}_j v^i v^j+
		\frac{9}{2} V^3 v^i +12 V_j V_j v^i +12 \hat{W}_{ij} V v^j +12 \hat{X} v^i +16 \hat{Z}_{ij} v^j \right.\nn\\&\left.\qquad\qquad -10 V^2
		V_i -8 \hat{W}_{ij} V_j -8 V \hat{R}_i-16 \hat{Y}_i\right) +
		\calO\left(\frac{1}{c^8}\right) \,,\\
		\label{Fi} F_{i} &= \partial_i V \nn\\& +\frac{1}{c^2} \left( -V \partial_iV
		+\frac{3}{2} \partial_i V v^2-4 \partial_i V_j v^j \right)\nn\\ &+
		\frac{1}{c^4} \left(\frac{7}{8} \partial_i V v^4-2 \partial_i V_j v^j
		v^2 + \frac{9}{2} V \partial_i V v^2 +2 \partial_i \hat{W}_{jk} v^j
		v^k \right.\nn\\& \left.\qquad\qquad -4 V_j \partial_i V v^j-4 V
		\partial_i V_j v^j -8 \partial_i \hat{R}_j v^j+\frac{1}{2} V^2
		\partial_i V +8 V_j \partial_i V_j+4 \partial_i \hat{X} \right)
		\nn\\&+ \frac{1}{c^6} \left(\frac{11}{16} v^6 \partial_i
		V-\frac{3}{2} \partial_i V_j v^j v^4+\frac{49}{8} V \partial_i V v^4+
		\partial_i \hat{W}_{jk} v^2 v^j v^k -10 V_j \partial_i V v^2 v^j\right.\nn\\& \left.\qquad\qquad-10 V
		\partial_i V_j v^2 v^j -4 \partial_i
		\hat{R}_j v^2 v^j +\frac{27}{4} V^2 \partial_i V v^2 + 12 V_j
		\partial_i V_j v^2 \right.\nn\\&\left.\qquad\qquad +6 \hat{W}_{jk} \partial_i V v^j v^k +6 V \partial_i
		\hat{W}_{jk} v^j v^k +6 \partial_i
		\hat{X} v^2 +8 \partial_i \hat{Z}_{jk} v^j v^k\right.\nn\\&\left.\qquad\qquad -20 V_j V \partial_i V
		v^j -10 V^2 \partial_i V_j v^j -8 V_k \partial_i \hat{W}_{jk} v^j-8
		\hat{W}_{jk} \partial_i V_k v^j \right.\nn\\&\left.\qquad\qquad -8
		\hat{R}_j \partial_i V v^j -8 V \partial_i \hat{R}_j v^j -16
		\partial_i \hat{Y}_j v^j -\frac{1}{6} V^3 \partial_i V -4 V_j V_j
		\partial_i V \right.\nn\\&\left.\qquad\qquad +16 \hat{R}_j \partial_i V_j+16 V_j \partial_i \hat{R}_j -8 V V_j \partial_i V_j -4 \hat{X}
		\partial_i V -4 V \partial_i \hat{X} +16 \partial_i \hat{T} \right) \nn\\&+
		\calO\left(\frac{1}{c^8}\right) \,.
\end{align}\end{subequations}

Note that all the accelerations appearing in the potentials and in the final expression of the equations of motion should be order-reduced by means of the equations of motion themselves. For instance, we see that when computing the time-derivative of $\hat{P}_i$ we shall meet some accelerations at 1PN order which are therefore to be replaced by the explicit 2.5PN equations of motion. The order-reduction is a crucial aspect of the post-Newtonian method. It is justified by the fact that the matter equations of motion, say $\nabla_\mu T^{\alpha\mu}=0$, represent four out of the ten Einstein field equations, see Sect.~\ref{sec:EFE} for discussion. In the harmonic-coordinate approach the equations of motion are equivalent to the harmonic gauge conditions $\partial_\mu h^{\alpha\mu}=0$. Post-Newtonian predictions based on such consistent PN order-reduction have been very successful. Note, however, that the operation of order-reduction is illicit at the level of the Lagrangian. The elimination of accelerations in a Lagrangian by substituting the equations of motion derived from that Lagrangian, results in a different Lagrangian whose equations of motion differ from those of the original Lagrangian by a gauge transformation \citep{S84}.


\subsubsection{Radiation reaction potentials}
\label{sec:4PNradreac}

We said that the metric \eqref{gmunu3PN} contains the correct radiation-reaction terms appropriate for an isolated system up to the 3.5PN level included. As we shall see the metric can even be generalized to include the piece of the radiation reaction which occurs at the 4PN order, and which is due to gravitational wave tails. In Sect.~\ref{sec:quadform} we already discussed a particular non-harmonic coordinate system which is very convenient to describe the radiation reaction terms. In this \cite{BuTh70, Bu71} coordinate system, the radiation reaction force at the lowest 2.5PN order takes the simple form of Eq.~\eqref{reac}, which involves only a scalar potential $V^{\text{reac}}$ so that
\begin{subequations}\label{reacBT}
\begin{align}
	F_i^{\text{reac}} &= \rho\,\partial_i V^{\text{reac}} + \calO\left(\frac{1}{c^7}\right)\,,\\V^{\text{reac}}(\mathbf{x},t) &= - \frac{G}{5 c^5}\,x^i
	x^j\,\mathrm{Q}_{ij}^{(5)}(t) + \calO\left(\frac{1}{c^7}\right) \,.
\end{align}
\end{subequations}
At such dominant 2.5PN level (``Newtonian'' radiation reaction) the source quadrupole moment is given by the usual Newtonian expression \eqref{Qij}.

Let us first emphasize that the Burke-Thorne expression \eqref{reacBT} of the radiation reaction force is not unique as it depends on the coordinate system \citep{Schafer81, Schafer82}. The first complete derivation of the radiation reaction force for general systems at the lowest 2.5PN order was done by \cite{CE70}\footnote{See also further works by \cite{PapaL81, K80a, K80b}.} who obtained a different expression valid in a different coordinate system. In harmonic coordinates the radiation reaction force reads
\begin{equation}\label{reacharm}
	F_i^{\text{reac}}\big|_\text{harm} = \frac{G}{c^5}\rho \left\{\frac{3}{5} x^j
	\dQ_{ij}^{(5)} + 2\frac{\dd}{\dd t}\Bigl( v^j
	\dQ_{ij}^{(3)}\Bigr) - \dQ_{jk}^{(3)}\partial_{ijk}\chi \right\} + 
	\calO\left(\frac{1}{c^7}\right) \,,
\end{equation}
where we have introduced the ``super-potential'' $\chi$ of the Newtonian potential $U$, such that $\Delta\chi = 2U$ and $\Delta U = -4\pi G \rho$.\footnote{We have $\chi(\mathbf{x},t)=G\int\dd^3\mathbf{x}'\rho(\mathbf{x}',t)\vert\mathbf{x}-\mathbf{x}'\vert$ and the super-potential is linked to the Newtonian tensor potential 
\begin{align*}
U_{ij}(\mathbf{x},t)=\int\dd^3\mathbf{x}'\rho(\mathbf{x}',t)\frac{(x^i-x'^i)(x^j-x'^j)}{\vert\mathbf{x}-\mathbf{x}'\vert^3}\,,
\end{align*}
(such that $U_{ii}=U$) by $\partial_{ij}\chi=\delta_{ij}U - U_{ij}$.} Repeating the analysis \eqref{balanceEJ0}--\eqref{balanceEJ} with the different expression of the radiation reaction force \eqref{reacharm} we recover the total energy flux given by the quadrupole formula \eqref{fluxE}; the only difference is that the total time derivatives in \eqref{balanceEJ0} will be different, with $f$ and $g_i$ corresponding to harmonic coordinates. The different forms of the radiation reaction force were a source of confusion in the 1970's and fed the so-called radiation-reaction controversy at that time \citep{Ehletal76, WalkW80}; see \cite{Kennefick} for an historical account.

The novel feature when one extends the Newtonian radiation reaction \eqref{reacBT} to include the 1PN corrections is that it is no longer composed of a single scalar depending on the mass-type multipole moments, but involves also a vectorial component depending in particular on the \emph{current}-type quadrupole moment. This was noticed in the physically restricted case where the dominant quadrupolar radiation from the source is suppressed \citep{BD84}. The vectorial component of the reaction force could be important in some astrophysical situations like rotating neutron stars undergoing gravitational instabilities. Here we report the results of the extension to 1PN order (and even 1.5PN) of the lowest-order Burke-Thorne (BT) scalar radiation reaction potential \eqref{reacBT}, in some appropriate coordinate system, which can be called extended BT coordinate system \citep{B93, B97}.

At that level (corresponding to 3.5PN order in the metric), and in the extended BT coordinate system, it suffices to incorporate some radiation-reaction contributions into the scalar and vectorial potentials which parametrize the 1PN metric in Eq.~\eqref{gmunu1PN}. We thus pose
\begin{subequations}\label{radreacpot}
  \begin{align}
\mathcal{V} &= V^\text{cons} + V^{\text{reac}}\,, \\ \mathcal{V}_i &=
V_i^\text{cons} + V_i^{\text{reac}}\,.
  \end{align}
\end{subequations}
Then the metric in extended BT coordinates, accurate to 3.5PN order regarding the radiation-reaction contributions and to 1PN order for the conservative effects -- we indicate this by using the symbol $\calO^{\text{reac}}$ for the remainders -- takes the same form as the harmonic-coordinates 1PN metric \eqref{gmunu1PN},
\begin{subequations}\label{gradreac}
	\begin{align}
		g_{00}^\text{BT} &= - 1 + \frac{2}{c^2} \mathcal{V} - \frac{2}{c^4}
		\mathcal{V}^2 + \calO^{\text{reac}}\left( \frac{1}{c^{11}}
		\right)\,, \\
		g_{0i}^\text{BT} &= - \frac{4}{c^3} \mathcal{V}_i +
		\calO^{\text{reac}}\left(\frac{1}{c^{10}}\right)\,, \\
		g_{ij}^\text{BT} &= \delta_{ij} \left[ 1 + \frac{2}{c^2} \mathcal{V} \right] +
		\calO^{\text{reac}}\left(\frac{1}{c^{9}} \right)\,.
	\end{align}
\end{subequations}
The other contributions, which are conservative (i.e., non radiative), are given up to 3PN order by the metric \eqref{gmunu3PN} in which all the potentials take the same form as in Eqs. \eqref{pot1PN}--\eqref{pot3PN}, \emph{but} where one neglects all the retardations, which means that the retarded integral operator $\Box^{-1}_{\mathrm{ret}}$ should be replaced by the operator of the ``\emph{instantaneous}'' potentials $\Box^{-1}_{\text{inst}}$ defined by Eq.~\eqref{instpotentials}. As such the conservative parts of the potentials \eqref{radreacpot} are defined by
\begin{align}\label{potcons}
	V^\text{cons} = \Box^{-1}_{\mathrm{inst}} \bigl[ -4 \pi G \sigma \bigr]\,,\qquad 
	V^\text{cons}_i = \Box^{-1}_{\mathrm{inst}} \bigl[ -4 \pi G \sigma_i \bigr]\,.
\end{align}
Up to 3.5PN order, the effect of all these retardations gets replaced by the effect of the radiation-reaction potentials $V^{\text{reac}}$ and $V_i^{\text{reac}}$. These are explicitly given by \citep{B93,B97}\footnote{Recall the footnote \ref{fnote:notation} for our notation. In particular $\hat{x}^{ijk}$ in the vector potential denotes the STF combination $\hat{x}^{ijk} = x^{ijk}-\frac{r^2}{5}(x^i\delta^{jk}+x^j\delta^{ki}+x^k\delta^{ij})$ with $x^{ijk}=x^{i}x^{j}x^{k}$.}
\begin{subequations}\label{VVireac}
\begin{align}
V^{\text{reac}}(\mathbf{x},t) =& - \frac{G}{5c^5} x^{ij}
\dI^{(5)}_{ij}(t) \nn\\ &+ \frac{G}{c^7} \left[\frac{1}{189}x^{ijk}
  \,\dI^{(7)}_{ijk}(t) - \frac{1}{70} r^2 x^{ij}
  \,\dI^{(7)}_{ij}(t) \right] +
\calO\left(\frac{1}{c^8}\right) \,, \label{Vreac}\\
V^{\text{reac}}_i(\mathbf{x},t) &= \frac{G}{c^5}\left[\frac{1}{21}
  \hat{x}^{ijk} \,\dI^{(6)}_{jk}(t) - \frac{4}{45}
  \,\epsilon_{ijk} \,x^{jl} \,\dJ^{(5)}_{kl}(t) \right] +
\calO\left(\frac{1}{c^7}\right)\,,\label{Vireac}
\end{align}
\end{subequations}
where the multipole moments $\dI_L$ and $\dJ_L$ denote the source multipole moments defined in a general way by Eqs. \eqref{sourcemoments}. The scalar potential $V^{\text{reac}}$ will obviously reproduce Eq.~\eqref{reacBT} at the dominant order. However, note that it is crucial to include in Eq.~\eqref{Vreac} the 1PN correction in the source quadrupole moment. Following \eqref{IL1PN} the mass-type quadrupole moment $\dI_{ij}$ to 1PN order involves only \emph{compact support} contributions and reads
\begin{align}\label{Iij1PN}
 \dI_{ij} = \int \dd^3 \mathbf{x} \left\{ \hat{x}_{ij} \sigma +
        \frac{1}{14c^2} \,r^2 \hat{x}_{ij}\,\partial^2_t
        \sigma - \frac{20}{21c^2}\,\hat{x}_{ijk}
        \,\partial_t \sigma_k \right\} +
        \calO\left(\frac{1}{c^4}\right)\,,
\end{align}
where the compact-support matter source densities $\sigma$ and $\sigma_i$ are given in Eqs. \eqref{sigmadef}. Obviously to this order the other moments are Newtonian, $\dI_L = \int \dd^3 \mathbf{x} \,\hat{x}_L \sigma + \calO(c^{-2})$ and $\dJ_L = \int \dd^3 \mathbf{x}\,\epsilon_{ab<i_\ell} \,\hat{x}_{L-1>a} \,\sigma_b + \calO(c^{-2})$. 

The 3.5PN radiation reaction force in the equations of motion of compact binary systems has been derived by \cite{IW93, IW95} in an arbitrary gauge, in the frame of the center of mass, assuming the validity of the flux-balance equations for energy and angular momentum at the relative 1PN order. Their result is presented in terms of a set of arbitrary parameters reflecting the arbitrariness in the choice of the coordinate system. \cite{IW95} have shown that the extended BT coordinate system indeed corresponds to a unique set of these parameters. \cite{PW02, NB05, itoh3} obtained the 1PN radiation in harmonic coordinates (again, corresponding to a unique set of Iyer-Will parameters) and \cite{KFS03} obtained it in ADM coordinate (corresponding to another unique set of parameters). The method of balance equations has been extended to the 4.5PN order corresponding to 2PN relatively to lowest order by \cite{GII97}. To date the only calculation of radiation reaction from first principle to relative 2PN order was done using the EFT by \cite{LPY23}, however restricted to the center-of-mass frame [they match the \cite{GII97} parameters]. 

The previous formalism can be extended to 4PN order (1.5PN beyond the leading radiation reaction). Indeed at this order there is a modification of the scalar radiation-reaction potential due to reaction to gravitational-wave tails propagating at infinity. The tail term is a hereditary integral characterized by a logarithmic kernel (we shall discuss thoroughly the tails in Sect.~\ref{sec:gravtails}) and we have \citep{B97}\footnote{Actually one must consider the time-asymetric (dissipative) piece obtained by changing 
\begin{align*}
	\dI^{(7)}_{ij}(t-\tau)\longrightarrow\frac{1}{2}\Bigl[\dI^{(7)}_{ij}(t-\tau)+\dI^{(7)}_{ij}(t+\tau)\Bigr]\,.
\end{align*}}
\begin{align}\label{Vreactail}
 		V^{\text{reac}}\Big|_\text{tail} =& -
 		\frac{4G^2\dM}{5c^8} \,x^{ij} \int^{+\infty}_0 \dd\tau
 		\,\dI^{(7)}_{ij}(t-\tau)\left[\ln \left(c\tau\over 2 r_0
 		  \right) + \frac{11}{12}\right] 
 		+
 		\calO\left(\frac{1}{c^9}\right) \,, 
\end{align}
which has to be added to Eq.~\eqref{Vreac} in order to complete the radiation reaction to 4PN order (in the extended BT gauge).

With the radiation-reaction potentials in hands, one can \emph{prove} the flux-balance equation for energy up to 1.5PN order, namely
\begin{align}\label{balance15PN}
\frac{\dd E}{\dd t} &= - \frac{G}{5c^5}
\left(\dI^{(3)}_{ij} + \frac{2G\dM}{c^3}
\int^{+\infty}_0 \dd\tau \,\dI^{(5)}_{ij}(t-\tau)\left[\ln
  \left(c\tau\over 2 r_0 \right) + \frac{11}{12}\right]\right)^2
\nn\\&\quad - \frac{G}{c^7} \left[\frac{1}{189}
  \dI^{(4)}_{ijk} \dI^{(4)}_{ijk} +
  \frac{16}{45}\dJ^{(3)}_{ij}\dJ^{(3)}_{ij}
  \right]+\calO\left(\frac{1}{c^9}\right) \,,
\end{align}
where $E$ denotes a functional of the matter in the source made of conservative terms and also, consistently with the approximation, of 2.5PN, 3.5PN and 4PN dissipative terms. The constant $11/12$ depends on the gauge and corresponds to a total time-derivative in the right-hand side of \eqref{balance15PN}, which can thus be transferred to the left-hand side. In the right-hand side one recognizes (minus) the positive-definite expression for the energy flux at 1.5PN order. Indeed, as we shall prove in \ref{sec:gravtails} [see Eqs. \eqref{Uij15PN} and \eqref{relationMijIij}], the effective quadrupole moment which appears in the parenthesis of \eqref{balance15PN} agrees with the tail-modified \emph{radiative} quadrupole moment $\dU_{ij}$ parametrizing the field in the far zone.


\subsection{Asymptotic gravitational waveform}
\label{sec:asympGW}


\subsubsection{The radiative multipole moments}
\label{sec:radmult}

In Theorem \ref{th:FZstruct} we constructed the metric in a radiative (or Bondi-type) coordinate system $(T,\mathbf{X})$, such that the retarded time $U$ becomes asymptotically a null coordinate at future null infinity \citep{Papa69,MadoreI,MadoreII}. The leading-order term $1/R$ (when $R\to+\infty$) yields the operational definition of two sets of STF \emph{radiative} multipole moments, mass-type $\dU_L(U)$ and current-type $\dV_L(U)$, where $U\equiv T-R/c$. The radiative moments are defined from the spatial components $ij$ of the metric in a transverse-traceless (TT) radiative coordinate system. \emph{By definition}, we have \citep{Th80}
\begin{align}\label{hijTT}
	h^\mathrm{TT}_{ij} (U,\mathbf{X}) &= - \frac{4G}{c^2R}
	\!\perp_{ijab}\!(\bm{N}) \sum^{+\infty}_{\ell=2}\frac{1}{c^\ell
		\ell !}  \biggl\{ N_{L-2} \dU_{abL-2}(U) \nn\\&\qquad\quad\quad -
	\frac{2\ell}{c(\ell+1)} N_{cL-2} \epsilon_{cd(a}
	\dV_{b)dL-2}(U) \biggr\} + \calO\left(\frac{1}{R^2}\right)\,.
\end{align}
The metric is written here as a perturbation of the ``gothic'' metric $\sqrt{-g}g^{ij}$; a change of sign is required when considering a perturbation of the ordinary metric $g_{ij}$. As before we denote for instance $N_{L-2} = N_{i_1}\cdots N_{i_{\ell-2}}$ and so on, where $N_i=(\bm{N})_i$ and $\bm{N}=\bm{X}/R$, and the TT algebraic projection operator $\perp_{ijab}$ is given by Eq.~\eqref{operatorTT}. Obviously the multipole decomposition \eqref{hijTT} represents the generalization of the quadrupole moment formalism. Notice that the meaning of Eq.~\eqref{hijTT} is for the moment rather empty, because we do not yet know how to relate the radiative moments to the actual source parameters. Only at the Newtonian level do we know this relation, which is $\dU_{ij}(U) = \mathrm{Q}_{ij}^{(2)}(U) + \calO(c^{-2})$, where $\mathrm{Q}_{ij}$ is the Newtonian quadrupole moment \eqref{Qij}. We shall relate below the radiative moments to the source moments $\dI_L$ and $\dJ_L$ obtained in Theorem \ref{th:sourcemoments}.

Associated to the asymptotic waveform \eqref{hijTT} we can compute by standard methods the total energy flux $\mathcal{F}$ and angular momentum flux $\mathcal{G}_i$ in gravitational waves (see Table \ref{tab:coeffs}),
\begin{subequations}\label{FluxFG}
	\begin{align}
		&\!\!\mathcal{F} \equiv \left(\frac{\dd E}{\dd U}\right)^\text{GW} \!\!= \sum^{+\infty}_{\ell=2} \frac{G}{c^{2\ell+1}}
		\biggl\{a_\ell
		\,\dU^{(1)}_L \dU^{(1)}_L + \frac{b_\ell}{c^2}  \,\dV^{(1)}_L
		\dV^{(1)}_L \biggr\}\,,\label{FluxF}\\&\!\!\mathcal{G}_i \equiv
		\left(\frac{\dd \dJ_i}{\dd U}\right)^\text{GW} \!\!=\epsilon_{iab} \sum^{+\infty}_{\ell=2} \frac{G}{c^{2\ell+1}}
		\biggl\{a_\ell'\,\dU_{aL-1} \dU^{(1)}_{bL-1} + \frac{b_\ell'}{c^2}  \,\dV_{aL-1}
		\dV^{(1)}_{bL-1} \biggr\}\,.
	\end{align}
\end{subequations}
For instance the energy flux is readily obtained by plugging \eqref{hijTT} into \eqref{energiedistr} and integrating over angles. The above fluxes satisfy the time-averaged flux-balance equations [already used in Eqs. \eqref{balanceEJ}]
\begin{align}
		\langle \frac{\dd \dE}{\dd t}\rangle = - \langle \mathcal{F}\rangle \,,
		\qquad\quad
		\langle \frac{\dd \dJ_i}{\dd t}\rangle &= - \langle \mathcal{G}_i\rangle
		\,.
\end{align}
Similarly one can compute the linear momentum flux $\mathcal{P}_i$ which is responsible for the GW recoil of the source (see Sec. \ref{sec:recoil}), as well as the flux associated to the center of mass position $\dG_i$. For a conservative system (for instance, stationary), there are ten Noetherian conserved invariants associated with the ten symmetries of the Poincar\'e group. In addition to $\dE$, $\dJ_i$ and $\dP_i$, there is also the initial ``position'' of the center of mass, say $\dZ_i = \dG_i - \dP_i\,t$, where $\dG_i$ denotes the position of the center of mass multiplied by the total (constant) mass. The conservation of $\dZ_i$ is due to the invariance of the dynamics under Lorentz boosts. For self-gravitating systems, $\dG_i$ coincides with the mass-type dipole moment $\dI_i$. We have the fluxes
\begin{subequations}\label{FluxPC}
	\begin{align}
		&\!\!\mathcal{P}_i \equiv \left(\frac{\dd \dP_i}{\dd U}\right)^\text{GW} \!\!= \sum^{+\infty}_{\ell=2} \frac{G}{c^{2\ell+3}}
		\biggl\{a_\ell''
		\,\dU^{(1)}_{iL} \dU^{(1)}_L + \frac{b_\ell''}{c^2}  \,\dV^{(1)}_{iL}
		\dV^{(1)}_L + c_\ell'' \,\epsilon_{iab} \,\dU^{(1)}_{aL-1} \dV^{(1)}_{bL-1}\biggr\}\,,\\
		&\!\!\mathcal{C}_i \equiv \left(\frac{\dd \dG_i}{\dd U}\right)^\text{GW} \!\!= \sum^{+\infty}_{\ell=2} \frac{G}{c^{2\ell+3}}
		\biggl\{a_\ell'''
		\,\dU_{iL} \dU^{(1)}_L + \frac{b_\ell'''}{c^2}  \,\dV_{iL}
		\dV^{(1)}_L \biggr\}\,,
	\end{align}
\end{subequations}
which enter the following time-averaged flux-balance equations
\begin{align}
	 \langle \frac{\dd \dP_i}{\dd U}\rangle = - \langle \mathcal{P}_i\rangle\,,
	\qquad\quad
	\langle \frac{\dd \dG_i}{\dd U}\rangle = \dP_i - \langle \mathcal{C}_i\rangle
	\,.
\end{align}
While the general multipole expansions of the fluxes for energy, angular momentum and linear momentum in terms of radiative moments are well known (see notably \citealt{Th80}), the flux associated to the position of the center of mass has been computed more recently \citep{KQ16, KNQ18, N18, BF19, COS20}.\footnote{The result obtained by \cite{KNQ18, COS20} is
\begin{align*}
	\tilde{\mathcal{C}_i} = \frac{1}{2}\sum^{+\infty}_{\ell=2} \frac{G}{c^{2\ell+3}}
		\biggl\{a_\ell'''
		\Bigl(\dU_{iL} \dU^{(1)}_L - \dU_{iL}^{(1)} \dU_L\Bigr) + \frac{b_\ell'''}{c^2} \Bigl(\dV_{iL} \dV^{(1)}_L - \dV_{iL}^{(1)} \dV_L\Bigr) \biggr\}\,,
\end{align*} 
which differs from $\mathcal{C}_i$ obtained by \cite{BF19} by a total time derivative, and is therefore physically equivalent after time averaging. An advantage of the formulation $\tilde{\mathcal{C}}_i$, as shown by \cite{COS20}, is that it can be written by means of a covariant formula over the ``celestial'' sphere in Bondi gauge.\label{fnote:CM}
}
The coefficients in the flux formulas \eqref{FluxFG} and \eqref{FluxPC} are given in Table \ref{tab:coeffs}.
\begin{table}[ht]
\hspace{-1.0cm}	
\begin{tabular}{|l||l|l|l|}
		\hline
		energy flux $\mathcal{F}$ & $a_\ell = \frac{(\ell+1)(\ell+2)}{(\ell-1)\ell \ell!(2\ell+1)!!}$ & $b_\ell = \frac{4\ell
			(\ell+2)}{(\ell-1)(\ell+1)!(2\ell+1)!!}$ & \\[0.15cm]
		\hline
		angular momentum flux $\mathcal{G}_i$ & $a_\ell' = \frac{(\ell+1)(\ell+2)}{(\ell-1) \ell!(2\ell+1)!!}$ & $b_\ell' = \frac{4\ell^2
			(\ell+2)}{(\ell-1)(\ell+1)!(2\ell+1)!!}$ & \\[0.15cm]
		\hline
		linear momentum flux $\mathcal{P}_i$ & $a_\ell'' = \frac{2(\ell+2)(\ell+3)}{\ell (\ell+1)!(2\ell+3)!!}$ & $b_\ell'' = \frac{8(\ell+3)}{(\ell+1)!(2\ell+3)!!}$ & $c_\ell'' = \frac{8(\ell+2)}{(\ell-1)(\ell+1)!(2\ell+1)!!}$ \\[0.15cm]
		\hline
		center of mass flux $\mathcal{C}_i$ & $a_\ell''' = \frac{2(\ell+2)(\ell+3)}{\ell \ell!(2\ell+3)!!}$ & $b_\ell''' = \frac{8(\ell+3)}{\ell!(2\ell+3)!!}$ & \\[0.15cm]
		\hline
\end{tabular}
	\caption{The $\ell$-dependent coefficients in the flux formulas.}
	\label{tab:coeffs}
\end{table}

Next we introduce two unit polarization vectors $\bm{P}$ and $\bm{Q}$, orthogonal and transverse to the direction of propagation $\bm{N}$, such that $(\bm{N}, \bm{P}, \bm{Q})$ forms an oriented orthonormal triad (hence $N_iN_j+P_iP_j+Q_iQ_j=\delta_{ij}$). Our convention for the choice of $\bm{P}$ and $\bm{Q}$ is clarified in the footnote \ref{fnote:polar}. Then the two ``plus'' and ``cross'' polarization states of the asymptotic waveform are defined by
\begin{subequations}\label{hpc}\begin{align}
		h_+ &\equiv \frac{1}{2}\left(P_iP_j-Q_iQ_j\right)
		h_{ij}^\text{TT}\,,\\ h_\times &\equiv
		\frac{1}{2}\left(P_iQ_j+P_jQ_i\right) h_{ij}^\text{TT}\,.
\end{align}\end{subequations}

Although the multipole decomposition \eqref{hijTT} is completely general, it will also be important, having in view the comparison between the post-Newtonian and numerical results, to consider separately the various modes $(\ell,m)$ of the asymptotic waveform as defined with respect to a basis of spin-weighted spherical harmonics of weight $-2$. Those harmonics are function of the spherical angles $(\theta,\phi)$ defining the direction of propagation $\bm{N}$, and given by
\begin{subequations}\label{harm}\begin{align}
		Y^{\ell m}_{-2} &= \sqrt{\frac{2\ell+1}{4\pi}}\,d^{\,\ell
			m}(\theta)\,\de^{\di \,m \,\phi}\,,\\d^{\,\ell m} &=
		\sum_{k=k_1}^{k_2}\frac{(-)^k}{k!}\de^{\,\ell m}_{k}
		\left(\cos\frac{\theta}{2}\right)^{2\ell+m-2k-2}
		\!\!\!\left(\sin\frac{\theta}{2}\right)^{2k-m+2}\,,\\\de^{\,\ell m}_{k}
		&= \frac{\sqrt{(\ell+m)!(\ell-m)!(\ell+2)!(\ell-2)!}}
		{(k-m+2)!(\ell+m-k)!(\ell-k-2)!}\,,
\end{align}\end{subequations}
where $k_1=\mathrm{max}(0,m-2)$ and $k_2=\mathrm{min}(\ell+m,\ell-2)$. We thus decompose $h_+$ and $h_\times$ onto the basis of such spin-weighted spherical harmonics, which means \citep{BuonCP07, K07}
\begin{equation}\label{spinw}
	h_+ - \di h_\times = \sum^{+\infty}_{\ell=2}\sum^{\ell}_{m=-\ell}
	h^{\ell m}\,Y^{\ell m}_{-2}(\theta,\phi)\,.
\end{equation}
Using the orthonormality properties of these harmonics we can invert the latter decomposition and obtain the separate modes $h^{\ell m}$ from a surface integral,
\begin{equation}\label{decomp}
	h^{\ell m} = \int \dd\Omega \,\Bigl[h_+ - \di h_\times\Bigr]
	\,\overline{Y}^{\,\ell m}_{-2} (\theta,\phi)\,,
\end{equation}
where the overline refers to the complex conjugation. On the other hand, we can also relate $h^{\ell m}$ directly to the radiative multipole moments $\dU_L$ and $\dV_L$. The result is
\begin{equation}\label{inv}
	h^{\ell m} = -\frac{G}{\sqrt{2} R c^{\ell+2}}\left[\dU^{\ell
		m}-\frac{\di}{c}\dV^{\ell m}\right]\,,
\end{equation}
where $\dU^{\ell m}$ and $\dV^{\ell m}$ denote the radiative mass and current moments in standard (non-STF) guise. These are related to the STF moments by
\begin{subequations}\label{UV}\begin{align}
		\dU^{\ell m} &=
		\frac{4}{\ell!}\,\sqrt{\frac{(\ell+1)(\ell+2)}{2\ell(\ell-1)}}
		\,\alpha_L^{\ell m}\,\dU_L\,,\\ \dV^{\ell m} &=
		-\frac{8}{\ell!}\,\sqrt{\frac{\ell(\ell+2)}{2(\ell+1)(\ell-1)}}
		\,\alpha_L^{\ell m}\,\dV_L\,.
\end{align}\end{subequations}

Here $\alpha_L^{\ell m}$ denotes the STF tensor connecting the usual basis of spherical harmonics $Y^{\ell m}(\theta,\phi)$ to the set of STF tensors $\hat{N}_L=N_{\langle i_1}\cdots N_{i_\ell\rangle}$ with $N_i=(\sin\theta\cos\phi,\sin\theta\sin\phi,\cos\theta)$, where the brackets indicate the STF projection. Indeed both $Y^{\ell m}$ and $\hat{N}_L$ are basis of an irreducible representation of weight $\ell$ of the rotation group; the two basis are related by\footnote{With the usual conventions $Y^{\ell m}(\theta,\phi)=\sqrt{\frac{2\ell+1}{4\pi}\frac{(\ell-m)!}{(\ell+m)!}} P_\ell^m(\cos\theta)\,\de^{\di m \phi}$ where $P_\ell^m(z)$ is the associated Legendre function.}
\begin{subequations}\label{NY}\begin{align}
		\hat{N}_L(\theta,\phi) &= \sum_{m=-\ell}^{\ell} \alpha_L^{\ell
			m}\,Y^{\ell m}(\theta,\phi)\,,\\Y^{\ell m}(\theta,\phi) &=
		\frac{(2\ell+1)!!}{4\pi \ell!}\,\overline{\alpha}_L^{\ell
			m}\,\hat{N}_L(\theta,\phi)\,.
\end{align}\end{subequations}
The STF tensorial coefficient $\alpha_L^{\ell m}$ can be computed as $\alpha_L^{\ell m} = \int \dd\Omega\,\hat{N}_L\,\overline{Y}^{\,\ell m}$. To present it in the best way we introduce the orthonormal triad $(\bm{i},\bm{j},\bm{k})=(\delta^1_i,\delta^2_i,\delta^3_i)$ such that $\bm{N}=\bm{i}\sin\theta\cos\phi+\bm{j}\sin\theta\sin\phi+\bm{k}\cos\theta$. Posing $\bm{\mathfrak{m}}=(\bm{i}+ \di \,\bm{j})/\sqrt{2}$ we have $\bm{\mathfrak{m}}\cdot\bm{\mathfrak{m}}=0$, $\bm{\mathfrak{m}}\cdot\overline{\bm{\mathfrak{m}}}=1$ and $\bm{\mathfrak{m}}\cdot\bm{k}=0$. The indices $L=i_1\cdots i_\ell$ can be split into $M=i_1\cdots i_m$ and the complementary $L-M=i_{m+1}\cdots i_\ell$, and the result reads (with explicit STF projection)\footnote{See Eq.~(4.7) of \cite{HFB21} in which a factor $(-)^m$ should be removed. Our definition is related to that given in Eq.~(2.12) of \cite{Th80} (see also \citealt{K07}) by $\mathcal{Y}_L^{\ell m}=\frac{(2\ell+1)!!}{4\pi \ell!}\,(-)^m\,\overline{\alpha}_L^{\ell m}$.} 
\begin{equation}\label{alphaLlm}
	\alpha_L^{\ell m} = \ell! \sqrt{\frac{4\pi}{2\ell+1}\frac{2^m}{(\ell+m)!(\ell-m)!}} \,\,\overline{\mathfrak{m}}_{\langle M} \,k_{L-M\rangle}\,,
\end{equation}
for $m\geqslant 0$, and $\alpha_L^{\ell m}=(-)^m\,\overline{\alpha}_L^{\ell \vert m\vert}$ for $m<0$. Note that we have the normalization condition
\begin{equation}\label{normalphaLlm}
\alpha_L^{\ell m}\overline{\alpha}_L^{\ell m'}=\frac{4\pi\ell!}{(2\ell+1)!!}\delta_{m m'}\,.
\end{equation}
In Sect.~\ref{sec:sphharm} we shall present the dominant mode $(\ell,m)=(2,2)$ of gravitational waves from non spinning inspiralling compact binaries up to 4PN order, and all the other sub-dominant modes to 3.5PN order.


\subsubsection{Gravitational-wave tails and tails of tails}
\label{sec:gravtails}

We learned from Theorem \ref{th:FZstruct} the general method which permits the computation of the radiative multipole moments $\dU_L$, $\dV_L$ in terms of the source parameters. In the general case those are described by the six types of source moments $\dI_L, \dJ_L$ and $\dW_L, \dX_L, \dY_L, \dZ_L$ given by \eqref{sourcemoments} and \eqref{gaugemoments}. However we also pointed out that an equally general description of the source is by two sets of canonical moments $\dM_L, \dS_L$. We shall now show that the relation between $\dU_L$, $\dV_L$ and $\dM_L$, $\dS_L$ includes some hereditary tail effects starting at the relative 1.5PN order. The different sets of moments $\dMbar_L$, $\dSbar_L$ used in the radiative construction of the metric are connected to the moments $\dM_L$, $\dS_L$ in the harmonic metric, see Eq. (5.4) of \cite{TLB22}. All the results below are given for the moments $\dM_L$, $\dS_L$.

Tails are due to the back-scattering of multipolar waves off the Schwarzschild curvature generated by the total mass monopole $\dM$ of the source. They correspond to the non-linear interaction between $\dM$ and the multipole moments $\dM_L$ and $\dS_L$ for $\ell\geqslant 2$, and are given by some ``hereditary'' or non-local-in-time integrals, extending over the past history of the source. The effect arises at the 1.5PN order and we have \citep{BD92}
\begin{subequations}\label{radtail}
	\begin{align}
		\dU_L(U) &= \dM^{(\ell)}_L(U) +
		\frac{2G\dM}{c^3} \int^{+\infty}_0 \!\dd\tau \,
		\dM^{(\ell+2)}_L (U-\tau) \left[ \ln \left(
		\frac{c\tau}{2b_0} \right) + \kappa_\ell \right] +
		\calO\left(G^2\right)\,,\label{radtailU}
		\\ \dV_L(U) &= \dS^{(\ell)}_L(U) +
		\frac{2G\dM}{c^3} \int^{+\infty}_0 \!\dd\tau \,
		\dS^{(\ell+2)}_L (U-\tau) \left[ \ln \left(
		\frac{c\tau}{2b_0} \right) + \pi_\ell \right] +
		\calO\left(G^2\right)\,,\label{radtailV}
	\end{align}
\end{subequations}
where the constant length scale $b_0$ parametrizes the gauge transformation between radiative and harmonic coordinates, see Eq.~\eqref{Uu}, and the constants $\kappa_\ell$ and $\pi_\ell$ are given by \citep{B95, AFS21a}
\begin{align}\label{pikappa}
		\kappa_\ell = \frac{2\ell^2+5\ell+4}{\ell(\ell+1)(\ell+2)}
		+H_{\ell-2}\,,\qquad \pi_\ell =
		\frac{\ell-1}{\ell(\ell+1)}
		+H_{\ell-1}\,,
\end{align}
where $H_p=\sum_{k=1}^{p}\frac{1}{k}$ denotes the harmonic number.

From the expression of the linearized gauge vector $\xi^\alpha_{(1)}$ in Eq.~\eqref{xi1} the retarded time $U = T-R/c$ in radiative coordinates is related to the retarded time $u = t-r/c$ in harmonic coordinates by
\begin{equation}\label{Uu}
	U = u -\frac{2G\dM}{c^3}\ln\left(\frac{r}{b_0}\right) +
	\calO\left(G^2\right)\,.
\end{equation}
Changing $b_0$ simply means shifting the origin of time of the radiative coordinates with respect to that of the harmonic coordinates, which has clearly no physical implication. Inserting $U$ into Eqs. \eqref{radtail} we obtain the radiative moments expressed in terms of ``source-rooted'' harmonic coordinates, i.e.
\begin{equation}\label{intUu}
	\dU_L = \dM^{(\ell)}_L(u) +
	\frac{2G\dM}{c^3} \int^{+\infty}_0 \!\dd\tau \,
	\dM^{(\ell+2)}_L (u-\tau) \biggl[ \ln \left( \frac{c\tau}{2r}
	\right) + \kappa_\ell \biggr] +
	\calO\left(G^2\right) \,,
\end{equation}
and similarly for $\dV_L$. This expression no longer depends on the arbitrary constant $b_0$, i.e., we find that $b_0$ gets replaced by $r$, so it behaves logarithmically when $r\to+\infty$. If we now replace the harmonic coordinates $(t, r)$ to some new ones, like for instance ``Schwarzschild'' coordinates $(t', r')$ such that $t'=t$ and $r'=r+G\dM/c^2$ (and $u'=u-G\dM/c^3$), we get
\begin{equation}\label{intUu'}
	\dU_L = \dM^{(\ell)}_L(u') +
	\frac{2G\dM}{c^3} \int^{+\infty}_0 \! \dd\tau \,
	\dM^{(l+2)}_L (u'-\tau) \biggl[ \ln \left( \frac{c\tau}{2r'}
	\right) + \kappa'_\ell \biggr] +
	\calO\left(G^2\right) \,,
\end{equation}
where $\kappa'_\ell=\kappa_\ell+1/2$. This shows that the constant $\kappa_\ell$ (and $\pi_\ell$ as well) depends on the choice of source-rooted coordinates $(t, r)$: For instance, we have $\kappa_2=11/12$ in harmonic coordinates from Eq.~\eqref{pikappa}, but $\kappa'_2=17/12$ in Schwarzschild coordinates \citep{P93a}.

The tail integrals in Eqs. \eqref{radtail} involve all the instants from $-\infty$ in the past up to the current retarded time $U$. However, strictly speaking, they do not extend up to infinite past, since we have assumed in Eq.~\eqref{statpast} that the metric is stationary before the date $-\mathcal{T}$. The range of integration of the tails is therefore limited \emph{a priori} to the time interval $[-\mathcal{T}, U]$. But now, once we have derived the tail integrals, thanks to the latter technical assumption of stationarity in the past, we can argue that the results are in fact valid in more general situations for which the field has \emph{never} been stationary. We have in mind the case of two bodies moving initially on some unbound (hyperbolic-like) orbit, and which, because of the loss of energy by gravitational radiation, will capture each other to form a gravitationally bound system around time $-\mathcal{T}$, entering then into the inspiral regime till the final merger.

In this situation let us check, using a simple Newtonian model for the behaviour of the multipole moment $\dM_L(U-\tau)$ in the past, when $\tau\to +\infty$, that the tail integrals, when assumed to extend over the whole time interval $[-\infty, U]$, remain perfectly well-defined (i.e., convergent) at the integration bound $\tau=+\infty$. Indeed it was shown by \cite{Eder} that the motion of initially free particles interacting gravitationally is given by $x^i(U-\tau)=V^i\tau+W^i\ln\tau+X^i+o(1)$, where $V^i$, $W^i$ and $X^i$ denote constant vectors, with $W^i= G m V^i/V^3$, and $o(1)\to 0$ when $\tau\to+\infty$.\footnote{To simplify we consider the relative motion of two particles with total mass $m$.} From that physical assumption we find that the multipole moments behave when $\tau\to+\infty$ like
\begin{equation}\label{Ipast}
	\dM_{L}(U-\tau) = A_{L}\tau^\ell+B_{L}\tau^{\ell-1}\ln\tau+
	C_{L} \tau^{\ell-1} + o(\tau^{\ell-1})\,,
\end{equation}
where $A_{L}$, $B_{L}$ and $C_{L}$ are constant tensors. We used the fact that the moment $\dM_{L}$ will agree at the Newtonian level with the standard expression for the $\ell$-th mass multipole moment. The appropriate time derivatives of the moment appearing in Eq.~\eqref{radtailU} are therefore dominantly like
\begin{equation}\label{Ipast'}
	\dM^{(\ell+2)}_{L}(U-\tau) = \frac{D_{L}}{\tau^{3}}+
	o(\tau^{-3})\,,
\end{equation}
which ensures that the tail integral is convergent. This fact can be regarded as an \emph{a posteriori} justification of our \emph{a priori} too restrictive assumption of stationarity in the past. Thus, this assumption does not seem to yield any physical restriction on the applicability of the final formulas. However, once again, we emphasize that the past-stationarity is appropriate for real astrophysical sources of gravitational waves which have been formed and started to emit at a finite instant in the past.

The formulas \eqref{radtail} can be generalized to include \emph{two} mass monopoles $\dM$ interacting with the varying moments $\dM_L$ or $\dS_L$. We coin such cubic interactions as ``tails-of-tails''. They arise at the 3PN order in the waveform. For the mass moments, including the 1.5PN tail and 3PN tail-of-tail, we have \citep{B98tail, FBI15}
	\begin{align}\label{radtailtail}
		\dU_L(U) &= \dM^{(\ell)}_L(U) +
		\frac{2G\dM}{c^3} \int^{+\infty}_0 \!\dd\tau \,
		\dM^{(\ell+2)}_L (U-\tau) \left[ \ln \left(
		\frac{c\tau}{2b_0} \right) + \kappa_\ell \right] \nn\\& +
		\frac{2G^2\dM^2}{c^6} \int^{+\infty}_0 \!\dd\tau \,
		\dM^{(\ell+3)}_L (U-\tau) \left[ \ln^2 \left(
		\frac{c\tau}{2b_0} \right) + 2\kappa_\ell \ln \left(
		\frac{c\tau}{2b_0} \right) \right.\nn\\&\left.\qquad\qquad\qquad\qquad\qquad + \frac{1}{2}\beta^{(m)}_\ell \ln \left(
		\frac{c\tau}{2r_0} \right) + \xi_\ell \right] +
		\calO\left(G^3\right)\,,
	\end{align}
The novel feature at this order is that there is a dependence not only on the pure gauge constant $b_0$, but also on the regularization scale $r_0$ which was introduced in the regulator \eqref{regfactor} of the MPM algorithm. The constant $r_0$ cannot be removed by a coordinate transformation, but it must disappear from the radiative moments (and hence from the physical waveform) once the moments $\dM_L$ are explicitly related to the parameters of the matter source, e.g. the masses, trajectories and velocities of the particles in a binary system.\footnote{In Eq.~\eqref{radtailtail} $\kappa_\ell$ is given in \eqref{pikappa} and the numerical constant $\xi_\ell$ is irrelevant for the discussion; but we shall obtain $\xi_2=\frac{124627}{22050}$ in Eq.~\eqref{Uij3PN} below.}

Exactly as we have done for the tail term in Eq.~\eqref{intUu}, we can remove all the $b_0$'s by inserting \eqref{Uu} back into \eqref{radtailtail} and expanding the result keeping the necessary terms consistently.  In the end we find that there remains a $r_0$-dependent term at the 3PN order, namely
\begin{equation}\label{ULr0}
	\dU_L = \dM^{(\ell)}_{L}(u) + \beta^{(m)}_\ell
	\ln\left(\frac{r}{r_0}\right) \left( \frac{G\dM}{c^3} \right)^2
	\dM^{(\ell+2)}_{L}(u)\,,
\end{equation}
plus terms that are independent of $r_0$. It is known from \cite{GRoss10,GRR12,AFS21a} that the scale $r_0$ is a running scale in the sense of renormalization group theory. The coefficient in front of the logarithm $\ln(r/r_0)$ in \eqref{ULr0} is the beta-function coefficient $\beta^{(m)}_\ell$, where $(m)$ stands for ``mass-type'', which is associated to the renormalization of the mass-type multipole moment. The appearance of a logarithm and its associated constant $r_0$ at the 3PN order was pointed out by \cite{AKKM82}. The beta coefficient for general $\ell$ has been obtained from a long computation of the multipole interactions $\dM \times \dM \times \dM_L$ by \cite{BD88} [see Eq.~(A6) there], with result
\begin{equation}\label{betamell}
	\beta^{(m)}_\ell = -2\frac{15\ell^4+30\ell^3 +
		28\ell^2+13\ell+24}{\ell(\ell+1)(2\ell+3)(2\ell+1)(2\ell-1)}\,.
\end{equation}
In particular, $\beta^{(m)}_2 = -\frac{214}{105}$ and $\beta^{(m)}_3 = -\frac{26}{21}$, see Eqs. \eqref{Uij3PN} and \eqref{Uijk3PN}. This result has also been obtained using the EFT approach by \cite{GRoss10,GRR12} for $\ell=2$ and by \cite{AFS21a} for general $\ell$. The apparent dependence of the radiative moment \eqref{ULr0} on $r_0$ should \emph{in fine} disappear. The reason is that when we explicitly compute the mass moment $\dM_L$ for a given matter source, we will find an extra contribution depending on $r_0$ occurring at the 3PN order which will cancel out the one in Eq.~\eqref{ULr0}. This contribution is contained into the source moment $\dI_L$ defined in Theorem \ref{th:sourcemoments}, where the FP regularization will produce a $\ln r_0$ at the 3PN order. This has been explicitly verified for the quadrupole moment of compact binaries, and we observe on the result \eqref{Iijcirc}--\eqref{Iijrenorm_AB} below the requested terms depending on $\ln r_0$. Now the canonical moments $\dM_L$ differ from the source moments $\dI_L$ by small 2.5PN and 3.5PN terms, see e.g. Eq.~\eqref{relationMijIij}, hence the statement that
\begin{equation}\label{MLr0}
	\dM_L = \mu \,\hat{x}_L - \beta^{(m)}_\ell \ln\left(\frac{r_{12}}{r_0}\right) \left( \frac{G m}{c^3} \right)^2 \bigl(\mu \,\hat{x}_L\bigr)^{(2)}\,,  
\end{equation}
plus terms independent of $r_0$, where $\mu\,\hat{x}_{L}$ is the binary's Newtonian multipole moment, with $r_{12}$ the orbital separation and $m$ the total mass differing from the ADM mass $\dM$ by small post-Newtonian corrections; so indeed the $r_0$ cancels between \eqref{ULr0} and \eqref{MLr0}.

Concerning current type moments $\dV_L$ their expression in terms of $\dS_L$ is the same as \eqref{radtailtail} with $\pi_\ell$ in place of $\kappa_\ell$, see Eq.~\eqref{pikappa}, and with the general beta-function coefficient given by \citep{AFS21a}
\begin{equation}\label{betacell}
	\beta^{(c)}_\ell = -2\frac{60\ell^6+180\ell^5+135\ell^4-574\ell^3 - 368\ell^2+221\ell-360}{\ell(\ell+1)(2\ell+5)(2\ell+3)(2\ell+1)(2\ell-1)(2\ell-3)}\,,
\end{equation}
with $(c)$ meaning ``current-type''. We have $\beta^{(c)}_2 = -\frac{214}{105}$, see Eq.~\eqref{Vij3PN}. The previous reasonings leading to Eqs. \eqref{ULr0} and \eqref{MLr0} apply as well for the current type moments. The cancellation of the renormalization scale $r_0$ has been checked in the 3PN current quadrupole moment for compact binaries \citep{HFB21}. 

The quartic tails-of-tails-of-tails terms arising at the order 4.5PN in the radiative mass quadrupole moment will be given in Eq.~\eqref{tailtailtail} below.

\subsubsection{Formulas to integrate non-linearities}
\label{sec:MPMtails}

To obtain the explicit results like \eqref{radtail} and \eqref{radtailtail}, we must implement the MPM algorithm presented in Sect.~\ref{sec:MPMsolution}, together with some practical formulas for integrating the wave equation when the source term is a complicated (in general non-local) functional of the multipole moments. Without loss of generality we can consider the wave equation whose source term has a given multipolarity $\ell$ in STF guise, thus takes the form
\begin{equation}\label{generalWaveEquation}
	\Box \Psi_L = \hat{n}_L S(r,t-r/c)\,,
\end{equation}
where $S(r,u)$ is a function of $r=\vert\mathbf{x}\vert$ and $u=t-r/c$ which verifies straightforward smoothness properties, is zero when $u\leqslant -\mathcal{T}$ [cf. the hypothesis of stationarity in the past, Eq.~\eqref{statpast}] and diverges along the set of scale functions $(\ln r)^q/r^p$ ($p, q\in\mathbb{N}$) when $r\to 0$. To deal with the divergence when $r\to 0$ we multiply the right side of \eqref{generalWaveEquation} by the regulator $\widetilde{r}^B\equiv (r/r_0)^B$, apply the retarded d'Alembertian operator $\Box^{-1}_\text{ret}$ and select the finite part (FP) in the Laurent expansion when $B\to 0$; remind Eq.~\eqref{ungen}. Hence
\begin{equation}\label{solMPM}
	\Psi_L = \FPprop\Bigl[\widetilde{r}^B \hat{n}_L S(r,t-r/c)\Bigr]\,,
\end{equation}
which is the unique solution of \eqref{generalWaveEquation}, valid in the sense of distributions and stationary in the past, thus satisfying the no-incoming radiation condition when $r\to+\infty$ with $t+r/c=$ const.

\begin{theorem} \citep{BD86}
The solution \eqref{solMPM}, endowed with definite multipolarity $\ell$, can be explicitly written as
\begin{equation}\label{PsiLR} \Psi_L = \FP \int_{-\infty}^{t-r/c}c\dd s\, \hat{\partial}_L \Biggl[ \frac{R_B\Bigl(\frac{c(t-s)-r}{2}, s\Bigr)-R_B\Bigl(\frac{c(t-s)+r}{2}, s\Bigr)}{r}\Biggr]\,,
\end{equation}
in which we define the auxiliary $B$-dependent function
\begin{equation}\label{RB}
		R_B(\rho, s) \equiv \rho^\ell \int_\alpha^\rho \dd\sigma\,\frac{(\rho-\sigma)^\ell}{\ell!}\left(\frac{2}{\sigma}\right)^{\ell-1} \widetilde{\sigma}^B S(\sigma,s) \,,
\end{equation}
where the regulator is $\widetilde{\sigma}^B = (\sigma/r_0)^B$ and the analytic continuation in $B\in\mathbb{C}$ is understood throughout.
\end{theorem}
The function defined in \eqref{RB} depends also on the lower bound of the integral denoted $\alpha$; however $\alpha$ can be chosen to be an arbitrary function of $t$ and $s$ because the full solution $\Psi_L$ in \eqref{PsiLR} is actually independent of $\alpha$. This can be seen from the following alternative form of the solution, which is obtained by plugging Eq.~\eqref{RB} into \eqref{PsiLR}:\footnote{One uses the useful identity, Eq. (A36) of \cite{BD86}:
\begin{align*}
	\hat{\partial}_L
	\left\{ \frac{(\lambda-r)^{p} - (\lambda+r)^{p}}{r}\right\} = 0 \quad\text{when}\quad p=0, 2, \cdots, 2\ell\,.
\end{align*}
\label{fnote:A36}}
\begin{align}\label{PsiLS}
	\Psi_L &= - \frac{1}{\ell!} \FP \int_{-\infty}^{t-r/c} c\dd s \int_{\frac{c(t-s)-r}{2}}^{\frac{c(t-s)+r}{2}} \dd \sigma \, \left(\frac{2}{\sigma}\right)^{\ell-1} \widetilde{\sigma}^B S(\sigma, s) \nn\\&\qquad\qquad\qquad\quad\times\hat{\partial}_L\!\left[\frac{1}{r}\left(\frac{c(t-s)-r}{2}\right)^\ell \,\left(\frac{c(t-s)-r}{2}-\sigma\right)^\ell \right]\,.
\end{align}
However, the form of the solution given by \eqref{PsiLR}--\eqref{RB} is generally the best in practical applications.

At the quadratic non-linear level we have to consider the simple case where the source term is of the type $S(r,u)=\dF(u)/r^k$. Here $\dF(u)$ will be a product of two time derivatives of mutipole moments $\dM_L(u)$ and/or $\dS_L(u)$. The first case of interest is when the source term has $k=2$; this corresponds to the leading behaviour when $r\to\infty$, $u=$ const. In this case it is easy to check that the retarded integral is convergent -- see the standard form \eqref{dalembertian} of the integral -- therefore the limit $B\to 0$ must be finite. Working out this limit one gets (using footnote \ref{fnote:A36})
\begin{align}\label{exprnonlocal}
	\Box^{-1}_\text{ret}\biggl[ \frac{\hat n_L}{r^2} \,\dF(u) \biggr] &= \frac{(-)^\ell}{2\ell!}
	\int^{+\infty}_r \dd \lambda\, \dF(t-\lambda/c) 
	\nn\\ &\qquad\times\hat{\partial}_L
	\left\{ \frac{(\lambda-r)^{\ell} \ln(\lambda-r)
		-(\lambda+r)^{\ell} \ln (\lambda+r)}{r}\right\}\,, 
\end{align}
where we have removed the reference to the finite part at $B=0$, and we note that the length scale $r_0$ drops from the result. The expression \eqref{exprnonlocal} is clearly non-local in time, or ``hereditary'', and typically describes the tails. There is an interesting connection to the Legendre function of the second kind $Q_{\ell}(x)$,\footnote{The Legendre function $Q_\ell$ is given in terms of the Legendre polynomial $P_\ell$ by
\begin{align*}
	Q_{\ell} (x) = \frac{1}{2} \int^1_{-1} \frac{\dd z \, P_\ell
		(z)}{x- z} = \frac{1}{2} P_\ell (x) \, \mathrm{ln}
		\left(\frac{x+1}{x-1} \right)- \sum^{\ell}_{ j=1} \frac{1}{j}
	P_{\ell-j}(x) P_{j-1}(x)\,.
\end{align*}
In the complex plane there is a branch cut from $-\infty$ to 1. The first equality is the Neumann formula for the Legendre function. When $x\to 1^+$ we have $Q_\ell (x)= -\frac{1}{2} \ln \left( {x-1\over 2}\right) - H_\ell + o(1)$ where $H_\ell$ is the harmonic number. We also note that
\begin{align}\label{formuleQ}  
	\hat n_L Q_\ell (x) = \int \frac{\dd\Omega'}{4\pi}
\frac{\hat n'_L}{x - \bm{n}\cdot\bm{n}'} \,.  
\end{align}
\label{fnote:legendre}}
since we can rewrite the previous expression as
\begin{equation}\label{tailQl}
	\Box^{-1}_\text{ret}\biggl[ \frac{\hat n_L}{r^2} \,\dF(u) \biggr] = - \frac{\hat{n}_L}{r} \int^{+\infty}_r \dd \lambda \, Q_{\ell}\Bigl(\frac{\lambda}{r}\Bigr)
	\,\mathrm{F} (t-\lambda/c)\,.
\end{equation}
We have also the following alternative expression, valid for any function $\dF(\bm{n}, u)$ (not only with pure multipolarity $\ell$), and which can be retrieved directly from the retarded integral in its usual form \eqref{dalembertian},
\begin{align}\label{expltail2}
 	\Box^{-1}_\text{ret}\biggl[ \frac{\dF (\bm{n}, u)}{r^2} \biggr] =
 	- \int^{+\infty}_r \dd\lambda \int \frac{\dd\Omega'}{4\pi}
 	\,\frac{\dF (\bm{n}',t-\lambda/c)}{\lambda-r\,\bm{n}\cdot\bm{n}'} \,.
\end{align}

The expansion at retarded infinity $r\to+\infty$ with $u=$ const exhibits the usual logarithmic behaviour of harmonic coordinates (see Sect.~\ref{sec:MPMrad}), namely
\begin{align}\label{expltailinf}
	\Box^{-1}_\text{ret}\biggl[ \frac{\hat n_L}{r^2} \,\dF(u) \biggr] = \frac{\hat n_L}{2r} \int^{+\infty}_0 \dd\tau
	\,\dF (u-\tau)
	\biggl[\ln \left( \frac{c\tau}{2r} \right)
	+ 2H_\ell \biggr] + o\left(\frac{1}{r}\right) \,,     
\end{align}
where the remainder is indicated as $o(1/r)$ rather than $\calO(1/r^2)$ because it contains powers of the logarithm of $r$; it could be more accurately written as $o(r^{\epsilon-2})$ for some $\epsilon\ll 1$. The result \eqref{expltailinf} gives directly the tail integral in the form met in Eq.~\eqref{intUu}.

Consider next the source term $\dF(u)/r^k$ with $k\geqslant 3$. Two cases must be distinguished. When $3\leqslant k\leqslant \ell +2$ (thus necessarily $\ell\geqslant 1$ in this case), the limit $B\to 0$ turns out to be finite and one obtains a local, ``instantaneous'' expression, again independent of the scale $r_0$:
\begin{align}\label{solk}
	\Box^{-1}_\text{ret}\biggl[ \frac{\hat n_L}{r^k} \,\dF(u) \biggr] =& - \frac{2^{k-3} (k-3)!(\ell+2-k)!}{(\ell+k-2)!}
	\, \hat n_L   \sum^{k-3}_{j=0}  \frac{(\ell+j)!}{2^j j!(\ell-j)!}
	\frac{\dF^{(k-3-j)} (u)}{c^{k-3-j}r^{j+1}} \,.  
\end{align}
This expression can be checked by applying the d'Alembertian operator on both sides of \eqref{solk}. Then the case $k\geqslant \ell +3$ is where the FP procedure shows its power and flexibility. In this case one obtains a non-local expression, which admits a logarithmic-free expansion at retarded infinity, but which depends on the arbitrary scale $r_0$. We have \citep{B98quad}
\begin{align}\label{solkhered}
	&\FPprop \biggl[ \widetilde{r}^B \frac{\hat n_L}{r^k} \dF(u) \biggr] = \sum^{k-3}_{i=0} \alpha_{k\ell}^i \,\hat n_L
	\frac{F^{(i)} (u)}{c^i r^{k-i-2}} \\
	&\qquad\quad
	+ {(-)^{k+\ell+1} 2^{k-2} (k-3)! \over (k+\ell-2)!(k-\ell-3)!}  \frac{\hat{n}_L}{c^{k-2}r} \int^{+\infty}_r \dd \lambda \, \overline{Q}_{\ell}\Bigl(\frac{\lambda}{r},r\Bigr)
	\,\dF^{(k-2)} (t-\lambda/c)\,.\nn
\end{align}
In the first term the coefficients are given by\footnote{Together with the combination of harmonic numbers:
	\begin{align*}
		H_{k\ell}^i \equiv H_{k-\ell-3}-H_{\ell+i-k+3}-H_{k-3}+H_{k-3-i}+H_{k+\ell-2}-H_{k+\ell-i-3}\,,
	\end{align*}
}
\begin{subequations}\label{coeffsalphabeta}
\begin{align}
	\alpha_{k\ell}^i &= \frac{(-)^{k+\ell}2^i(k-3)!(k+\ell-i-3)!}{(k+\ell-2)!(k-\ell-3)!
		(k-i-3)!}\,\beta_{k\ell}^i\,,\\[0.2cm]
	\beta_{k\ell}^i &= \left\{\begin{array}{l} \displaystyle (-)^{k+\ell+i}(k-\ell-i-4)! \qquad\text{when $0\leqslant i\leqslant k-\ell-4$}\,, \\[0.2cm]\displaystyle \frac{H_{k\ell}^i}{(\ell-k+i+3)!} \qquad\qquad\text{when $k-\ell-3\leqslant i\leqslant k-3$}\,,
	\end{array}\right. 
\end{align}
\end{subequations}
while in the second term we have introduced the modification of the Legendre function defined by $\overline{Q}_m(x,r) \equiv Q_m(x)-\frac{1}{2}P_m(x)\ln(\frac{r}{r_0})$, and which explicitly depends on the scale $r_0$. The interest of this particular combination is that it does not produce any far zone logarithms when $r\to+\infty$ with $u=$ const \citep{TLB22}; hence the solution \eqref{solkhered} admits a far zone expansion in simple powers of $1/r$, although it is still non-local in time and depends on the scale $r_0$. This feature of the solution is very important in the construction of the metric in radiative coordinates, see Sect.~\ref{sec:MPMrad}.

The above formulas permit the control of any quadratic non-linear interactions (e.g., tails). To cubic order (e.g., tails-of-tails) one needs some more involved formulas. In particular we have to integrate the wave equation when the source term $S(r,u)$ is itself a non-local integral of the type \eqref{tailQl}. Let us show the most difficult of the integration formulas needed for the calculation of tails-of-tails:\footnote{The formula \eqref{cubictail} has been obtained using a not so well known mathematical relation between the Legendre functions and polynomials:
\begin{align*}
	\frac{1}{2} \int^1_{-1} \frac{\dd z \, P_\ell (z)}{\sqrt{(xy - z)^2 -
		(x^2 - 1) (y^2 - 1)}} = Q_\ell (x) P_\ell (y)\,,
\end{align*}
where $1 \leqslant y < x$ is assumed; see Appendix~A of \cite{B98tail} for the proof. This relation constitutes a generalization of the Neumann formula (see the footnote \ref{fnote:legendre}).}
\begin{align}\label{cubictail}
	& \FPprop \Biggl[ \widetilde{r}^B \frac{\hat{n}_L}{r^2} \int^{+\infty}_r \dd \lambda \, Q_{m}\Bigl(\frac{\lambda}{r}\Bigr)
	\,\mathrm{F} (t-\lambda/c)\Biggr] =
	\frac{\hat{n}_L}{r} \int^{+\infty}_r \dd \lambda \, \mathrm{F}^{(-1)} (t - \lambda/c)
	\nn\\& \qquad \times \left\{ Q_{\ell}\Bigl(\frac{\lambda}{r}\Bigr) \int^{\lambda/r}_1 \dd
	x \, Q_m (x) \frac{\dd P_\ell}{\dd x} (x) + P_{\ell}\Bigl(\frac{\lambda}{r}\Bigr)
	\int^{+\infty}_{\lambda/r} \!\!\dd x \, Q_m (x) \frac{\dd Q_\ell}{\dd x} (x)
	\right\}\,,
\end{align}
More formulas, even more involved, are requested for the computation of the quartic tails-of-tails-of-tails and cubic tails-of-memory interactions; these can be found in \cite{MBF16, TB23}.

\subsubsection{The non-linear memory effect}
\label{sec:memory}

The non-linear memory effect has its physical origin in the re-emission of gravitational waves by linearly generated waves \citep{B90mem, Chr91, WW91, Th92, BD92}.\footnote{The effect was initially computed using post-Newtonian theory by \cite{B90mem}, pointing out that it is dominantly of order 2.5PN, in contrast to tails at 1.5PN order. This was published later, together with physical interpretation, by \cite{BD92}.} It implies a permanent change in the wave amplitude from before to after the passage of a gravitational wave strain. The effect can be interpreted as due to a finite accumulation of the (mass type) radiative moments during the emission of gravitational waves, producing a transition between the initial and final asymptotic \cite{BBM62, Sachs62} (BMS) frames, which turn out to be related by a supertranslation \citep{NP66, Strom16}.

During the construction of the radiative metric in Sect.~\ref{sec:MPMrad}, we found that the non-linear source term for the metric takes to leading order $1/R^2$ in the distance the form of the stress-energy tensor for the massless gravitons, cf. \eqref{Lambdares}. Therefore the radiative metric will contain a piece which is the retarded integral of the stress-energy tensor of gravitational waves, see Eqs. \eqref{Isaacson}--\eqref{asympIsaacson}. This piece is at the origin of the memory-effect, hence we pose
\begin{align}\label{defmem}
	h\ab_\text{mem}(T,\bm{X}) = \Box^{-1}_{\mathrm{ret}} \left[ \frac{K^\alpha K^\beta}{R^2} \,\sigma(U,\bm{N})  \right]\,,
\end{align}
where $U=T-R/c$, $K^\alpha=(1,\bm{N})$ and we recall that the energy density in the waves is given by Eq.~\eqref{sigmares} and is related to the energy loss per steradian, see \eqref{energiedistr}. Following the integration formula \eqref{expltail2} we can write the memory term into the more explicit form
\begin{align}\label{defmem2}
	h\ab_\text{mem} = - \int_{-\infty}^U \dd v \int \frac{\dd\Omega(\bm{n})}{4\pi}
	\frac{k^\alpha k^\beta}{T-v-R\,\bm{N}\cdot\bm{n}} \,\sigma(\bm{n}, v) \,,
\end{align}
which involves the spherical average with respect to the ``dummy'' angles $\bm{n}$ and we denote $k^\alpha=(1,\bm{n})$, while $\bm{N}$ refers to the unit vector of the field point $\bm{X}=R\bm{N}$. As we know from from Sect.~\ref{sec:MPMtails}, the asymptotic expansion when $R\to +\infty$, $U={\rm const}$ of the retarded integral of a source with radial dependence $R^{-2}$ involves some far-zone logarithms of $R$. But we proved in Sect.~\ref{sec:MPMrad} that the logarithms can be gauged way, which permits to compute all observables in the asymptotic waveform (radiative moments). Here we shall treat the term \eqref{defmem2} directly, following the method of \cite{Th92, WW91}, which consists of applying first the TT projection operator onto \eqref{defmem2}. Because the TT projection kills any (linear) gauge term in the $1/R$ part of the metric, this method shortcuts the need of a transformation to radiative coordinates. However, one must be cautious in taking the limit $R\to +\infty$, $U=$const using \eqref{defmem2}, as it is not allowed to work out a leading $1/R$ term from \eqref{defmem2} because this term would involve a divergent integral, in accordance with the fact that the leading term is actually $\ln R/R$. But, the divergent parts of the integral cancel out after application of the TT projection, and in the end one recovers the correct result. Thus, we compute\footnote{Note that the TT projection as defined here is purely algebraic. Strictly speaking it agrees with the true TT projection only when acting on the leading $1/R$ term.} 
\begin{align}
	(h^{ij}_\text{mem})^\text{TT} = - \perp_{ijab} (\bm{N}) \int_{-\infty}^U \dd v \int \frac{\dd\Omega}{4\pi}
	\frac{n^a n^b}{T-v-R\,\bm{N}\cdot\bm{n}} \,\sigma(\bm{n}, v) \,.
\end{align}
Next we insert the STF multipole decomposition $\sigma(\bm{n}, v) = \sum_{\ell\geqslant 0} n_L \sigma_L (v)$ [with STF coefficients $\sigma_L(v)$] and obtain
\begin{align}
	(h^{ij}_\text{mem})^\text{TT} = - \perp_{ijab} (\bm{N}) \sum_{\ell=0}^{+\infty}
	\,\int^{U}_{-\infty} \dd v \, \sigma_L (v) \int \frac{\dd\Omega}{4\pi} \frac{n_{abL}}{T-v-R\,\bm{N}\cdot\bm{n}} \,.
\end{align}
We decompose the product of unit vectors $n_{abL}=n_a n_b n_L$ on the basis of STF tensors, we drop the terms having zero TT-projection, and we express the remaining terms using the Legendre function of the second kind $Q_\ell$ using \eqref{formuleQ}. Restoring the traces on the STF tensors, and droping further terms having zero TT-projection, we obtain
\begin{align}
(h^{ij}_\text{mem})^\text{TT} = - \frac{1}{R} \!\perp_{ijab} \sum_{\ell=2}^{+\infty}
	 N_{L-2}\int^{U}_{-\infty} \dd v\, \sigma_{abL-2} (v) \,Z_\ell\left( \frac{T-v}{R}\right) \,,
\end{align}
where $Z_\ell$ denotes the combination of Legendre functions given by
\begin{align}
	\frac{Z_\ell}{\ell (\ell-1)} = \frac{Q_{\ell+2}}{(2\ell+3)(2\ell+1)} - \frac{2Q_\ell}{(2\ell+3)(2\ell-1)} + \frac{Q_{\ell-2}}{(2\ell+1)(2\ell-1)} \,.
\end{align}
The limit at future null infinity can now be applied. Indeed it suffices to insert the expansion of $Q_\ell(x)$ when $x\to 1^+$ as given in the footnote \ref{fnote:legendre}. As expected the limit is finite, because the terms $\ln(x-1)/2$ in the expansions of the $Q_\ell$'s cancel out. We obtain $Z_\ell=\frac{2}{(\ell+1)(\ell+2)}$ in the limit and this yields \citep{BD92, B98quad}
\begin{align}\label{hijTTmemres}
	(h^{ij}_\text{mem})^\text{TT} = - \frac{2}{R} \!\perp_{ijab} \sum_{\ell=2}^{+\infty}
\frac{N_{L-2}}{(\ell+1)(\ell+2)} \int^{U}_{-\infty} \dd v \,
\sigma_{abL-2} (v) + \calO\left( \frac{1}{R^2} \right) \,.
\end{align}
Comparing with the TT waveform \eqref{hijTT} we see that the memory effect enters as a modification of the mass-type radiative multipole moments
\begin{align}\label{ULmemres}
	\dU_L^\text{mem}(U) = \frac{\ell!}{2(\ell+1)(\ell+2)} \frac{c^{\ell+3}}{G}\int^{U}_{-\infty} \dd v \,
	\sigma_{L} (v) \,,
\end{align}
but does not contribute to the current-type moments, i.e. $\dV_L^\text{mem}(U)=0$. It is to be remarked that the result \eqref{ULmemres} is exact; it gives the full memory contribution in non-linear GR without approximation \citep{F09,F11}.

The convergence of the integral \eqref{ULmemres} is ensured by the fact that the time derivatives of the moments are zero in the remote past, before the instant $-\mathcal{T}$. The memory property is evident from \eqref{ULmemres}. Even after the system has ceased to emit radiation, in the sense that the $(\ell+1)$-th time derivative of the source moments tend to zero at late time, the emission of gravitational waves will produce a constant (DC) contribution to the gravitational wave amplitude. More explicitly, one reads off for the total GW emission \citep{BD92}
\begin{align}\label{memoryUL}
	\dU_L(+\infty) - \dU_L(-\infty) &= \dM^{(\ell)}_L(+\infty) - \dM^{(\ell)}_L(-\infty) \nn\\ &+ \frac{\ell!}{2(\ell+1)(\ell+2)} \frac{c^{\ell+3}}{G}\int^{+\infty}_{-\infty} \dd v \,
	\sigma_{L} (v)\,,
\end{align}
where the first term represents the possible contribution of the ``linear'' memory which has been discussed by \cite{Zeldovich74,Braginsky87}. Generically the linear memory is present if the source contains freely moving masses in its initial or final state (scattering situation).

Let us now derive an interesting alternative expression for the non-linear memory effect \eqref{hijTTmemres}. We replace the $\ell$-th order STF coefficient $\sigma_L$ by its expression in terms of the energy distribution using Eq.~\eqref{energiedistr},
\begin{align}\label{sigmaLexpr}
	\sigma_L(U) = \frac{4(2\ell+1)!!}{\ell!}\int\dd\Omega \,\hat{n}_L\,\frac{\dd E}{\dd U\dd\Omega}(U,\bm{n})\,.
\end{align}
Hence we rewrite \eqref{hijTTmemres} in the following form, where the time integration has been explicitly performed,
\begin{align}\label{hijTTmemres2}
	(h^{ij}_\text{mem})^\text{TT} = - \frac{8}{R} \!\perp_{ijab} \sum_{\ell=2}^{+\infty}
	\frac{(2\ell+1)!!}{(\ell+2)!} N_{L-2} \int \dd\Omega \,\hat{n}_{abL-2} \,\frac{\dd E}{\dd\Omega}(U,\bm{n}) \,.
\end{align}

We shall now show that the infinite multipole series in the formula \eqref{hijTTmemres2} can be summed up in closed analytic form. Indeed consider the following TT projection with respect to the unit vector $\bm{N}=(N_i)$:
\begin{align}\label{proof1}
	A^\text{TT}_{ij} \equiv \,\perp_{ijab}\!(\bm{N}) \,\frac{n_{a}n_{b}}{1-\bm{N}\cdot\bm{n}}\,,
\end{align}
where the TT projector is given by \eqref{operatorTT}. We expand the denominator in \eqref{proof1} as a power series in $\bm{N}\cdot\bm{n}$ (which is convergent as soon as $\bm{N}\cdot\bm{n}<1$):
\begin{align}\label{proof2}
	A^\text{TT}_{ij} = \perp_{ijab}\!(\bm{N}) \sum_{\ell=2}^{+\infty} N_{L-2} \,n_{abL-2} \,.
\end{align}
The multi-index $L-2$ contains $\ell-2$ indices, namely $i_1\cdots i_{\ell-2}$. Using Eq. (A.21a) of \cite{BD86}, we transform the ordinary product of unit vectors $n_{abL-2}$ into a sum of STF products
\begin{subequations}\label{proof3}
	\begin{align}
		A^\text{TT}_{ij} &= \perp_{ijab}\!(\bm{N}) \sum_{\ell=2}^{+\infty} N_{L-2} \sum_{k=0}^{[\frac{\ell}{2}]} \alpha_k^\ell \,\delta_{\{2K}\,\hat{n}_{L-2K\}}\,,\label{proof3a}\\
		\text{with}\quad \alpha_k^\ell &\equiv \frac{(2\ell-4k+1)!!}{(2\ell-2k+1)!!}\,.\label{proof3b}
	\end{align}
\end{subequations}
The operation over indices $\{\}$ is defined as the un-normalized sum over the smallest set of permutations of $i_1\cdots i_\ell$ which makes the object symmetrical in $L=i_1\cdots i_\ell$. The object $\delta_{\{2K}\,\hat{n}'_{L-2K\}}$ contains $\ell$ indices $L=abL-2$ (with, say, $a = i_\ell$ and $b = i_{\ell-1}$) of which $2k$ are displayed onto the product of $k$ Kronecker symbols denoted $\delta_{2K}$. As an example we have $\delta_{\{ij}n_{k\}} \equiv \delta_{ij}n_k + \delta_{jk}n_i + \delta_{ki}n_j$. 

The point is that, among all the terms composing $\delta_{\{2K}\,\hat{n}_{L-2K\}}$, we can discard all those which contain either $\delta_{ab}$, $\delta_{a i_p}$ or $\delta_{b i_q}$, since such terms will be cancelled by the TT projection. This is obvious for $\delta_{ab}$; in the two other cases, this results from the fact that, after multiplication by $N_{L-2}$ in Eq.~\eqref{proof3a}, $\delta_{a i_p}$ or $\delta_{b i_q}$ will yield some $N_i$ or $N_j$. So, we can rewrite the expression \eqref{proof3a} by excluding the indices $ab$ from the operation $\{\}$, which we indicate by underlining the two indices $ab$: 
\begin{align}\label{proof4}
	A^\text{TT}_{ij} = \perp_{ijab}\!(\bm{N}) \sum_{\ell=2}^{+\infty} N_{L-2} \sum_{k=0}^{[\frac{\ell-2}{2}]} \alpha_k^\ell \,\delta_{\{2K}\,\hat{n}_{\underline{ab}L-2-2K\}}\,.
\end{align}
The next step is to notice from eq.~(A.19) of \cite{BD86} that the number of terms composing the object $\delta_{\{2K}\,\hat{n}_{\underline{ab}L-2-2K\}}$ is $\frac{(\ell-2)!}{2^k k!(\ell-2-2k)!}$. When contracted with $N_{L-2}$, all these terms will merge into a single one for each values of $\ell$ and $k$. Thus, we have 
\begin{align}\label{proof5a}
	A^\text{TT}_{ij} = \perp_{ijab}\!(\bm{N}) \sum_{\ell=2}^{+\infty} \sum_{k=0}^{[\frac{\ell-2}{2}]} \alpha_k^\ell \,\frac{(\ell-2)!}{2^k k!(\ell-2-2k)!}\, N_{L-2-2K}\,\hat{n}_{abL-2-2K}\,.
\end{align}
We change $\ell$ into $\ell+2k$ and rewrite the previous expression as
\begin{align}\label{proof5b}
	A^\text{TT}_{ij} &= \perp_{ijab}\!(\bm{N}) \sum_{\ell=2}^{+\infty} \sum_{k=0}^{+\infty} \alpha_k^{\ell+2k} \,\frac{(\ell+2k-2)!}{2^k k!(\ell-2)!}\, N_{L-2}\,\hat{n}_{abL-2} \nn\\ &= \perp_{ijab}\!(\bm{N}) \sum_{\ell=2}^{+\infty} \,S_\ell \,N_{L-2}\,\hat{n}_{abL-2}\,,
\end{align}
introducing the coefficient $S_\ell$ which is given as an infinite series over all integer values of $k$. However, we find, using the expression of the coefficients $\alpha_k^{\ell+2k}$ deduced from \eqref{proof3b}, that this series can actually be re-summed in closed analytic form:
\begin{align}\label{proof5c}
	S_\ell = \frac{(2\ell+1)!!}{(\ell-2)!} \sum_{k=0}^{+\infty} \,\frac{(\ell+2k-2)!}{2^k k!(2\ell+2k+1)!!} = 2^{\ell+2}\frac{\Gamma\bigl(\ell+\frac{3}{2}\bigr)}{\sqrt{\pi}\,\Gamma(\ell+3)} = 2\frac{(2\ell+1)!!}{(\ell+2)!}\,.
\end{align}
Hence we obtain the simple result
\begin{align}\label{proof5d}
	A^\text{TT}_{ij} = 2 \!\perp_{ijab}\!(\bm{N}) \sum_{\ell=2}^{+\infty} \frac{(2\ell+1)!!}{(\ell+2)!}\, N_{L-2}\,\hat{n}_{abL-2}\,,
\end{align}
where we recognize the infinite multipolar sum in Eq.~\eqref{hijTTmemres2}. 
Inserting thus the resummed form \eqref{proof1} into \eqref{hijTTmemres2} we obtain
\begin{align}\label{hijTTmemfinal}
	(h^{ij}_\text{mem})^\text{TT} = - \frac{4}{R} \!\perp_{ijab}\!(\bm{N}) \int \dd\Omega\,\frac{n_{a}n_{b}}{1-\bm{N}\cdot\bm{n}}\, \frac{\dd E}{\dd\Omega}(U,\bm{n}) \,,
\end{align}
where the factor $1/(1-\bm{N}\cdot\bm{n})$ is reminiscent of the Li\'enard-Wiechert potentials in the case of massless gravitons. The result \eqref{hijTTmemfinal}, which exactly reproduces the result of \cite{Th92}, constitutes the best interpretation of the non-linear memory effect as due to the re-radiation of gravitational waves by massless gravitons. Note that Eq.~\eqref{hijTTmemfinal} describes the gravitational memory produced not only by a burst of gravitons, but also by a burst of any other type of massless particles, e.g. massless neutrinos \citep{Epstein78}.

\subsubsection{Radiative versus canonical moments}
\label{sec:radcanonical}

We first give the result for the radiative quadrupole moment $\dU_{ij}$ expressed as a functional of the intermediate canonical moments $\dM_L$, $\dS_L$ up to 4PN order, plus a crucial hereditary contribution to 4.5PN order. Recalling that, at leading order, the radiative moments reduce to the $\ell$-th time derivatives of the canonical moments, we straightforwardly write  
\begin{equation}\label{Uijdef}
	\dU_{ij} = \dM_{ij}^{(2)}(U) + \sum_{n=3}^{9}\dU_{ij}^\text{$\frac{n}{2}$PN}(U) + \calO\left(\frac{1}{c^{10}}\right)\,,
\end{equation}
with small PN corrections up to 4.5PN, as indicated. Here $U\equiv T-R/c$ is the retarded time in harmonic coordinates. The leading correction at 1.5PN order is the quadratic quadrupolar tail $\dM\times \dM_{ij}$ (see Sect.~\ref{sec:gravtails}) given by
\begin{align}\label{Uij15PN}
	\dU_{ij}^\text{1.5PN} = \frac{2 G \dM}{c^3} \int_0^{+\infty} \dd\tau\, \dM_{ij}^{(4)}(U-\tau)\bigg[\ln \left(\frac{c \tau}{2 b_0}\right)+ \frac{11}{12}\biggl]\,.
\end{align}
The length scale $b_0$ is arbitrary and defined by Eq.~\eqref{Uu}. At the next 2.5PN order arises the leading non-local memory contribution (see Sect.~\ref{sec:memory}) due to the interaction between two quadrupole moments $\dM_{ij}\times \dM_{kl}$, together with associated instantaneous terms \citep{B98quad}:
\begin{align}\label{Uij25PN}
	\dU_{ij}^\text{2.5PN} = \frac{G}{c^5} \biggl\{ &- \frac{2}{7} \int_0^{+\infty} \dd\tau \left[\dM^{(3)}_{a\langle i}\dM^{(3)}_{j\rangle a}\right](U-\tau) \\ & + \frac{1}{7}\,\dM^{(5)}_{a\langle i}\dM^{}_{j\rangle a} - \frac{5}{7} \,\dM^{(4)}_{a\langle i}\dM^{(1)}_{j\rangle a} -\frac{2}{7}\,\dM^{(3)}_{a\langle i}\dM^{(2)}_{j\rangle a}  + \frac{1}{3}\epsilon^{}_{ab\langle i}\dM^{(4)}_{j\rangle a}\dS^{}_{b} \biggr\}\,.\nn
\end{align}
At the 3PN order appears the first cubic correction, called the tail-of-tail and consisting of the interaction between two masses $\dM$ and the quadrupole moment $\dM_{ij}$:
\begin{align}\label{Uij3PN}
	\dU_{ij}^\text{3PN} &= \frac{2 G^2 \dM^2}{c^6} \int_{0}^{+\infty} \dd\tau\,\dM_{ij}^{(5)}(U-\tau) \biggl[\ln^2\left(\frac{c \tau}{2b_0}\right) \nn\\& \qquad\qquad\qquad\qquad + \frac{11}{6} \ln\left(\frac{c \tau}{2b_0}\right)- \frac{107}{105} \ln\left(\frac{c \tau}{2r_0}\right) + \frac{124627}{44100}\biggr]\,.
\end{align}
We have discussed in Sect.~\ref{sec:gravtails} the occurence of the length scale $r_0$ (to be distinguished from the gauge constant $b_0$), which plays the role of renormalization scale in the EFT approach \citep{GRoss10} 
and originates from the Hadamard regularization scheme 
in the traditional post-Newtonian calculation \citep{B98tail}.

The next-order 3.5PN term  has a structure similar to the 2.5PN one, \emph{i.e.} with some memory type integrals and instantaneous terms. The interactions between moments, still of quadratic nature, are however more complicated: 
\begin{align}\label{Uij35PN}
	\dU_{ij}^\text{3.5PN} &= \frac{G}{c^7} \Biggl\{
	\int_0^{+\infty} \! \dd\tau\!  \left[ - \frac{5}{756} \dM_{ab}^{(4)} \dM_{ijab}^{(4)} - \frac{32}{63} \dS_{a \langle i}^{(3)} \dS_{j\rangle a}^{(3)}\right](U-\tau) \nn\\
	&\qquad
	- \frac{1}{432}	\dM_{ab} \dM_{ijab}^{(7)} + \frac{1}{432} \dM_{ab}^{(1)} \dM_{ijab}^{(6)} - \frac{5}{756} \dM_{ab}^{(2)} \dM_{ijab}^{(5)} \nn\\
	&\qquad + \frac{19}{648}\dM_{ab}^{(3)} \dM_{ijab}^{(4)} + \frac{1957}{3024} \dM_{ab}^{(4)} \dM_{ijab}^{(3)}  
	+ \frac{1685}{1008}\dM_{ab}^{(5)} \dM_{ijab}^{(2)} \nn\\
	&\qquad + \frac{41}{28} \dM_{ab}^{(6)}\dM_{ijab}^{(1)} + \frac{91}{216} \dM_{ab}^{(7)} \dM_{ijab} - \frac{5}{252} \dM_{ab \langle i} \dM_{j \rangle ab}^{(7)} \nn\\
	&\qquad + \frac{5}{189} \dM_{ab \langle i}^{(1)} \dM_{j \rangle ab}^{(6)} 
	+\frac{5}{126} \dM_{ab \langle i}^{(2)} \dM_{j \rangle ab}^{(5)} + \frac{5}{2268} \dM_{ab \langle i}^{(3)} \dM_{j \rangle ab}^{(4)}
	\nn\\
	&\qquad + \frac{5}{42} \dS_a \dS_{ija}^{(5)} + \frac{80}{63} \dS_{a \langle i} \dS_{j \rangle a}^{(5)} + \frac{16}{63} \dS_{a \langle i}^{(1)} \dS_{j \rangle a}^{(4)} - \frac{64}{63} \dS_{a\langle i}^{(2)} \dS_{j \rangle a}^{(3)} \nn \\ 
	& 
	\quad +\epsilon_{ac \langle i} \bigg( \int_0^{+\infty}\!\dd\tau \left[ \frac{5}{42} \dS_{j\rangle cb}^{(4)} \dM_{ab}^{(3)} - \frac{20}{189} \dM_{j \rangle cb}^{(4)} \dS_{ab}^{(3)} \right](U-\tau) \nn\\
	&\qquad
	 + \frac{1}{168} \dS_{j \rangle	bc}^{(6)} \dM_{ab} + \frac{1}{24} \dS_{j\rangle bc}^{(5)} \dM_{ab}^{(1)}	+ \frac{1}{28} \dS_{j \rangle bc}^{(4)} \dM_{ab}^{(2)}  - \frac{1}{6} \dS_{j \rangle bc}^{(3)} \dM_{ab}^{(3)} \nn\\
	 &\qquad + \frac{3}{56} \dS_{j \rangle bc}^{(2)} \dM_{ab}^{(4)} 
	  + \frac{187}{168} \dS_{j \rangle bc}^{(1)} \dM_{ab}^{(5)} + \frac{65}{84} \dS_{j \rangle bc} \dM_{ab}^{(6)} + \frac{1}{189} \dM_{j	\rangle bc}^{(6)} \dS_{ab} \nn\\
	  &\qquad - \frac{1}{189} \dM_{j \rangle bc}^{(5)}	\dS_{ab}^{(1)}
	 + \frac{10}{189} \dM_{j\rangle bc}^{(4)} \dS_{ab}^{(2)} + \frac{32}{189} \dM_{j \rangle bc}^{(3)} \dS_{ab}^{(3)} + \frac{65}{189} \dM_{j \rangle bc}^{(2)}\dS_{ab}^{(4)} \nn\\
	 &\qquad - \frac{5}{189} \dM_{j \rangle bc}^{(1)} \dS_{ab}^{(5)} - \frac{10}{63} \dM_{j \rangle bc}	\dS_{ab}^{(6)} \bigg)\Biggl\} \,.
\end{align}

\begin{figure}[ht]
  \centering
\begin{subfigure}[b]{0.25\textwidth}
		\includegraphics[width=\linewidth]{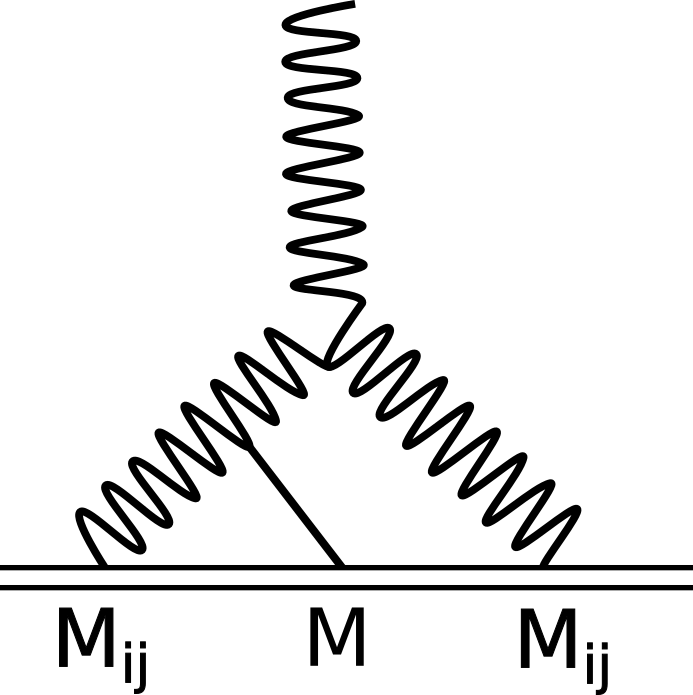}\caption{}\label{subfig:feynman1}
\end{subfigure}
	\qquad\qquad
\begin{subfigure}[b]{0.25\textwidth}
		\includegraphics[width=\linewidth]{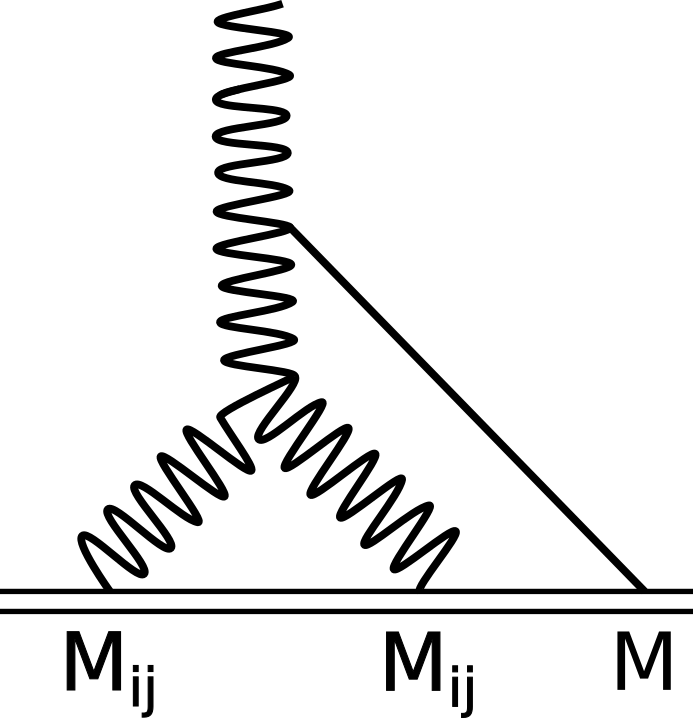}\caption{}\label{subfig:feynman2}
\end{subfigure}
	\qquad\qquad
\begin{subfigure}[b]{0.25\textwidth}
		\includegraphics[width=\linewidth]{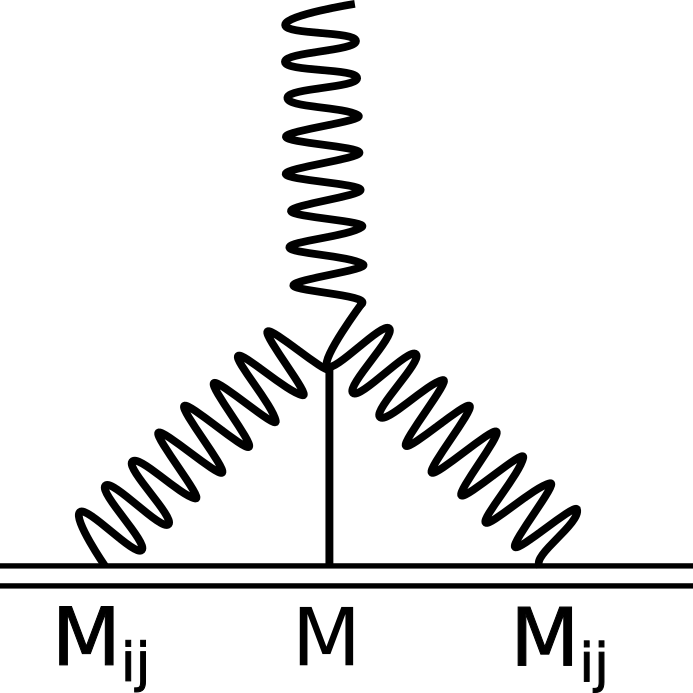}\caption{}\label{subfig:feynman3}
\end{subfigure}
\caption{From an effective field theory perspective, the tails-of-memory correspond to the three Feynman diagrams shown, which we consider here for illustrative purposes; see \cite{FSrevue, Portorevue, Levirevue} for their computational meaning within the EFT.}
\label{fig:feynman}
\end{figure}
At the 4PN order appear the tail-of-memory, made of a cubic interaction involving two time-varying quadrupole moments and the mass: $\dM \times \dM_{ij} \times \dM_{kl}$, see Fig.~\ref{fig:feynman}. In principle it comes with a double time integration over the two varying quadrupole moments, but involves also many tail-looking interactions with only one integration over the quadrupole moments. In addition there is also a cubic interaction involving the constant angular momentum, i.e. $\dM \times \dS_i \times \dM_{jk}$, see the last line of \eqref{Uij4PN}. We have \citep{TB23}
\begin{align}\label{Uij4PN}
	\dU_{ij}^\text{4PN} =\ & \frac{G^2 \dM}{c^8} \Bigg\{\!\frac{8}{7}\int_0^{+\infty} \!\!\!\!\!\!\!\dd\rho\,  \dM_{a \langle i}^{(4)}(U-\rho) 
	\!\int_0^{+\infty} \!\!\!\!\!\!\!\dd \tau \dM_{j \rangle a}^{(4)}(U-\rho-\tau)  \left[ \ln\left(\frac{c\tau}{2 r_0}\right) - \frac{1613}{270} \right]\nn\\
	&\quad  - \frac{20}{7} \int_0^{+\infty} \!\dd\tau \,  \left[\dM^{(3)}_{a \langle i} \dM^{(4)}_{j \rangle a}\right](U-\tau)\left[\ln\left(\frac{c\tau}{2 r_0}\right)+\frac{3}{2} \ln\left(\frac{c\tau}{2b_0}\right) \right]\nn\\
	&\quad  - \frac{24}{7}  \int_0^{+\infty} \!\dd\tau\, \left[\dM^{(2)}_{a \langle i}  \dM^{(5)}_{j \rangle a}\right](U-\tau) \left[\ln\left(\frac{c\tau}{2r_0}\right)  +\frac{11}{12} \ln\left(\frac{c\tau}{2b_0}\right)  \right]\nn\\
	&\quad  -\frac{20}{7} \int_0^{+\infty} \!\dd\tau \, \left[\dM^{(1)}_{a \langle i}\dM^{(6)}_{j \rangle a}\right](U-\tau) \left[\ln\left(\frac{c\tau}{2r_0}\right)  + \frac{3}{10} \ln\left(\frac{c\tau}{2 b_0}\right) \right]\nn\\
	&\quad    - \frac{8}{7}\int_0^{+\infty} \!\dd\tau\,\left[\dM^{}_{a \langle i}\dM^{(7)}_{j \rangle a}\right](U-\tau) \left[\ln\left(\frac{c\tau}{2r_0}\right) - \frac{1}{4}\ln\left(\frac{c\tau}{2 b_0}\right)\right]\nn\\
	&\quad   - \frac{16}{7}  \dM^{(2)}_{a \langle i} \int_0^{+\infty} \!\dd\tau\,  \dM^{(5)}_{j \rangle a}(U-\tau) \left[ \ln\left(\frac{c\tau}{2r_0}\right)+ \frac{27521}{5040} \right]\nn\\
	&\quad  - \frac{20}{7}\,  \dM^{(1)}_{a \langle i} \int_0^{+\infty} \!\dd\tau \, \dM^{(6)}_{j \rangle a}(U-\tau)  \left[\ln\left(\frac{c\tau}{2r_0}\right)+\frac{15511}{3150} \right]\nn\\
	&\quad  + \frac{4}{7} \, \dM_{a \langle i} \int_0^{+\infty} \!\dd\tau \,\dM^{(7)}_{j \rangle a}(U-\tau)  \left[ \ln\left(\frac{c\tau}{2r_0}\right)  - \frac{6113}{756}\right] \nn\\
	&\quad - \frac{2}{3}\,\dS_{a} \epsilon_{a b \langle i} \int_{0}^{+\infty} \!\!\!\!\!\!\dd\tau \dM_{j\rangle b}^{(6)}(U-\tau) \left[ \ln\left(\frac{c\tau}{2b_0}\right) +2 \ln\left(\frac{c\tau}{2r_0}\right)  + \frac{1223}{1890}  \right]\Bigg\}\,.
\end{align}
It is important to verify that the first line of \eqref{Uij4PN}, which corresponds to the ``genuine'' tail-of-memory, with two integrations over the quadrupole moments, can be retrieved from the general formalism of Sect.~\ref{sec:memory}. Indeed from Eq.~\eqref{ULmemres} together with the general result $\sigma = \frac{1}{2} \,\dot{z}^\text{TT}_{ij}\dot{z}^\text{TT}_{ij}$ where $z^\text{TT}_{ij}$ is nothing but the $1/R$ coefficient in the asymptotic waveform, see Eqs. \eqref{sigmares} and \eqref{energiedistr}, we have
\begin{align}\label{Uijmemres}
	\dU_{ij}^\text{mem}(U) = - \frac{2G}{7c^5} \int^{U}_{-\infty} \dd v \,\dU_{a\langle i}^{(1)}(v)\,\dU_{j\rangle a}^{(1)}(v)\,.
\end{align}
To lowest order $\dU_{ij}$ agrees with $\dM_{ij}^{(2)}$ hence \eqref{Uijmemres} recovers the leading memory term displayed in \eqref{Uij25PN}. To the next order $\dU_{ij}$ will be modified by the 1.5PN tail contribution, hence we have $\dU_{ij} = \dM_{ij}^{(2)} + \dU_{ij}^\text{1.5PN}$ where $\dU_{ij}^\text{1.5PN}$ is given by \eqref{Uij15PN}. Inserting this into \eqref{Uijmemres} and keeping only the leading 4PN correction
we obtain
\begin{align}\label{UijToM}
	\dU_{ij}^\text{mem} =& - \frac{2G}{7c^5} \int^{U}_{-\infty} \dd v \,\dM_{a\langle i}^{(3)}(v)\,\dM_{j\rangle a}^{(3)}(v) \nn\\ & - \frac{8 G^2 \dM}{7 c^8} \int_{-\infty}^U \dd v\,\dM_{a \langle i}^{(3)}(v) \int_0^{+\infty} \dd \tau\,  \dM_{j \rangle a}^{(5)}(v-\tau)  \ln\left(\frac{c\tau}{2}\right)\,,
\end{align}
where we have discarded the constant $b_0$ and the $11/12$ in Eq.~\eqref{Uij15PN}, which are irrelevant here. It is straightforward to check that (after an integration by parts) the 4PN term in \eqref{UijToM} agrees with the genuine tail-of-memory term given in the first line of \eqref{Uij4PN}.

At the next 4.5PN order only the hereditary, non-instantaneous quartic interaction $\dM^3\times \dM_{ij}$, naturally dubbed ``tail-of-tail-of-tail'', is known, but it is sufficient to control the full 4.5PN term in the energy flux for compact binaries on quasi-circular orbits. The main reason is that the contributions of instantaneous interactions entering the flux at half-integer PN order (e.g. 4.5PN order) vanish for quasi-circular orbits. The tail-of-tail-of-tail term in the radiative quadrupole moment is given by \citep{MBF16} 
\begin{align}\label{tailtailtail}
	\dU_{ij}^\text{4.5PN}{\bigg|}_{(\text{tail})^3} &=
	\frac{G^3 \dM^3}{c^9} \int_0^{+\infty} \dd\tau \,\dM_{ij}^{(6)}(U-\tau)\Bigg[
	\frac{4}{3}\ln^3 \left(\frac{c\tau}{2 b_0}\right)+ \frac{11}{3}\ln^2\left(\frac{c\tau}{2 b_0}\right) \nn\\&\qquad\qquad - \frac{428}{105}\ln\left(\frac{c\tau}{2 b_0}\right)\ln\left(\frac{c\tau}{2 r_0}\right)
	+ \frac{124627}{11025}\ln\left(\frac{c\tau}{2 b_0}\right) \nn\\&\qquad\qquad - \frac{1177}{315} \ln\left(\frac{c\tau}{2 r_0}\right) + \frac{129268}{33075}+ \frac{428}{315}\pi^2 \Bigg]\,.
\end{align}
We shall use this expression to extend the flux at 4.5PN order for circular orbits in Sect.\ref{sec:orbevol}. However note that the full $\dU_{ij}^\text{4.5PN}$ also contains quadratic memory interactions, like those entering at 2.5PN and 3.5PN, see \eqref{Uij25PN} and \eqref{Uij35PN}. If those are not needed to compute the flux for quasi-circular orbits, they will enter the expression of the gravitational wave $(2,2)$ mode, therefore restricting the accuracy we can reach when deriving the $(2,2)$ mode; this is why it will be presented in Sect.~\ref{sec:sphharm} at 4PN order and not 4.5PN order. 

Next we present the best known results concerning moments of higher multipolarity. Notably for the flux at 4PN order we need the mass octupole $\dU_{ijk}$ and current quadrupole $\dV_{ij}$ moments with a 3PN precision:
\begin{subequations}
	\begin{align}	
		\dU_{ijk} &= \dM_{ijk}^{(3)} + \dU_{ijk}^\text{1.5PN}  + \dU_{ijk}^\text{2.5PN} + \dU_{ijk}^\text{3PN} + \calO\left(\frac{1}{c^7}\right)\,,\\
		\dV_{ij} &= \dS_{ij}^{(2)} + \dV_{ij}^\text{1.5PN}  + \dV_{ij}^\text{2.5PN} + \dV_{ij}^\text{3PN} + \calO\left(\frac{1}{c^7}\right)\,.
	\end{align}
\end{subequations}
At the 1.5PN order there is the standard tail, at 2.5PN the memory, and at 3PN order the tail-of-tail contribution. We have \citep{FBI15}
\begin{subequations}
	\begin{align}
		\dU_{ijk}^\text{1.5PN} = 
		&
		\frac{2 G \dM}{c^3} \int_0^{+\infty} \dd\tau\, \dM_{ijk}^{(5)}(U-\tau)\bigg[\ln \left(\frac{c \tau}{2 b_0}\right)+ \frac{97}{60}\biggl]\,,\\
		\dU_{ijk}^\text{2.5PN} =
		&
		\frac{G}{c^5} \Biggl\{
		\int_0^{+\infty} \! \dd\tau\left[-\frac{1}{3}\dM^{(3)}_{a\langle i}\dM^{(4)}_{jk\rangle a}- \frac{4}{5}\epsilon_{ab\langle i}\dM^{(3)}_{j\underline{a}}\dS^{(3)}_{k\rangle b}\right]\!(U-\tau)\nn\\
		&\quad
		+ \frac{1}{4}\dM^{}_{a\langle i}\dM^{(6)}_{jk\rangle a}
		+ \frac{1}{4}\dM^{(1)}_{a\langle i}\dM^{(5)}_{jk\rangle a}
		+ \frac{1}{4}\dM^{(2)}_{a\langle i}\dM^{(4)}_{jk\rangle a}
		- \frac{4}{3}\dM^{(3)}_{a\langle i}\dM^{(3)}_{jk\rangle a}\nn\\
		&\quad
		- \frac{9}{4}\dM^{(4)}_{a\langle i}\dM^{(2)}_{jk\rangle a}
		- \frac{3}{4}\dM^{(5)}_{a\langle i}\dM^{(1)}_{jk\rangle a}
		+ \frac{1}{12}\dM^{(6)}_{a\langle i}\dM^{}_{jk\rangle a}
		+ \frac{12}{5}\dS^{}_{\langle i}\dS^{(4)}_{jk\rangle}\nn\\
		&\quad
		+ \epsilon_{ab\langle i}\bigg[
		\frac{9}{5}\dM^{}_{j\underline{a}}\dS^{(5)}_{k\rangle b}
		+ \frac{27}{5}\dM^{(1)}_{j\underline{a}}\dS^{(4)}_{k\rangle b}
		+ \frac{8}{5}\dM^{(2)}_{j\underline{a}}\dS^{(3)}_{k\rangle b}
		+ \frac{12}{5}\dM^{(3)}_{j\underline{a}}\dS^{(2)}_{k\rangle b}\nn\\
		&\quad\qquad\qquad
		+ \frac{3}{5}\dM^{(4)}_{j\underline{a}}\dS^{(1)}_{k\rangle b}
		+ \frac{1}{5}\dM^{(5)}_{j\underline{a}}\dS^{}_{k\rangle b}
		+ \frac{9}{20}\dM^{(5)}_{jk\rangle a}\dS^{}_b
		\bigg]
		\Biggr\}\,,\\
		\dU_{ijk}^\text{3PN} =
		&
		\frac{2 G^2 \dM^2}{c^6} \!\!\int_{0}^{+\infty} \!\!\!\dd\tau\,\dM_{ijk}^{(6)}(U-\tau) \! \biggl[\ln^2\left(\frac{c \tau}{2b_0}\right) + \frac{97}{30} \ln\left(\frac{c \tau}{2b_0}\right) \nn\\&\qquad\qquad\qquad - \frac{13}{21} \ln\left(\frac{c \tau}{2r_0}\right) + \frac{13283}{8820}\biggr]\,,\label{Uijk3PN}
\end{align}
where the underlined indices within angled brackets are excluded from the STF projection, and
	\begin{align}
		\dV_{ij}^\text{1.5PN} =
		&
		\frac{2 G \dM}{c^3} \int_0^{+\infty} \dd\tau\, \dS_{ij}^{(4)}(U-\tau)\bigg[\ln \left(\frac{c \tau}{2 b_0}\right)+ \frac{7}{6}\biggl]\,,\\
		\dV_{ij}^\text{2.5PN} =
		&
		\frac{G}{c^5} \Biggl\{
		- \frac{3}{7}\dM^{}_{a\langle i}\dS^{(5)}_{j\rangle a}
		- \frac{3}{7}\dM^{(1)}_{a\langle i}\dS^{(4)}_{j\rangle a}
		+ \frac{8}{7}\dM^{(2)}_{a\langle i}\dS^{(3)}_{j\rangle a}
		+ \frac{4}{7}\dM^{(3)}_{a\langle i}\dS^{(2)}_{j\rangle a}
		\nn\\&\qquad + \frac{17}{7}\dM^{(4)}_{a\langle i}\dS^{(1)}_{j\rangle a}
		+ \frac{9}{7}\dM^{(5)}_{a\langle i}\dS^{}_{j\rangle a}
		- \frac{1}{28}\dM^{(5)}_{ija}\dS^{}_a\nn\\
		&\quad
		+ \epsilon_{ab\langle i}\bigg[
		- \frac{15}{56}\dM^{}_{j\rangle ac}\dM^{(6)}_{bc}
		- \frac{113}{112}\dM^{(1)}_{j\rangle ac}\dM^{(5)}_{bc}
		- \frac{353}{336}\dM^{(2)}_{j\rangle ac}\dM^{(4)}_{bc}
		\nn\\
		&\qquad - \frac{3}{14}\dM^{(3)}_{j\rangle ac}\dM^{(3)}_{bc}
		+ \frac{5}{168}\dM^{(4)}_{j\rangle ac}\dM^{(2)}_{bc}
		+ \frac{3}{112}\dM^{(5)}_{j\rangle ac}\dM^{(1)}_{bc}
		\nn\\
		&\qquad - \frac{3}{112}\dM^{(6)}_{j\rangle ac}\dM^{}_{bc}
		+ \dS^{(4)}_{j\rangle a}\dS^{}_{b}\bigg]
		\Biggr\}\,,\\
		\dV_{ij}^\text{3PN} =
		&
		\frac{2 G^2 \dM^2}{c^6} \!\!\int_{0}^{+\infty} \!\!\!\dd\tau\,\dS_{ij}^{(5)}(U-\tau) \! \biggl[\ln^2\left(\frac{c \tau}{2b_0}\right) + \frac{7}{3} \ln\left(\frac{c \tau}{2b_0}\right)\nn\\&\qquad\qquad\qquad - \frac{107}{105} \ln\left(\frac{c \tau}{2r_0}\right) - \frac{13127}{11025}\biggr]\,.\label{Vij3PN}
	\end{align}
\end{subequations}
Finally we list all known results concerning higher order multipole moments, taken from \cite{FBI15}. For mass type moments:
\begin{subequations}
\begin{align}
\dU_{ijkl}^\text{1.5PN} =&
	\frac{G}{c^3}\Biggl\{2\dM\,\int_0^{+\infty}\!\! \dd\tau\, \dM_{ijkl}^{(6)}(U-\tau)\bigg[\ln \left(\frac{c \tau}{2 b_0}\right)+ \frac{59}{30}\biggl]
	\nn\\& + \frac{2}{5} \int_0^{+\infty} \!\! \dd\tau\,\left[\dM^{(3)}_{\langle ij}\dM^{(3)}_{kl\rangle}\right]\!(U-\tau)\nn\\
	&
	- \frac{21}{5}\dM^{}_{\langle ij}\dM^{(5)}_{kl\rangle}
	- \frac{63}{5}\dM^{(1)}_{\langle ij}\dM^{(4)}_{kl\rangle}
	- \frac{102}{5}\dM^{(2)}_{\langle ij}\dM^{(3)}_{kl\rangle}\Biggr\}\,,\\
\dU_{ijkl}^\text{2.5PN} =& \frac{G}{c^5} \Bigg\{ 
	\int_0^{+\infty} \dd \tau \bigg[ \frac{12}{55} \dM_{a \langle i}^{(4)}
	\dM_{jkl \rangle a}^{(4)} - \frac{14}{99} \dM_{a \langle
		ij}^{(4)} \dM_{kl \rangle a}^{(4)} + \frac{32}{45}
	\dS_{\langle ij}^{(3)} \dS_{kl \rangle}^{(3)}  \nn
	\\ &\qquad + \epsilon_{ab \langle i} \bigg(- \frac{4}{5} \dM_{j
		\underline{a}}^{(3)} \dS_{kl \rangle b}^{(4)} +
	\frac{32}{45} \dS_{j \underline{a}}^{(3)} \dM_{kl \rangle
		b}^{(4)} \bigg) \bigg] (U-\tau) \nn\\
	& + \frac{7}{55} \dM_{a \langle i} \dM_{jkl \rangle
		a}^{(7)} + \frac{7}{55} \dM_{a \langle i}^{(1)} \dM_{jkl \rangle
		a}^{(6)} + \frac{1}{25} \dM_{a \langle i}^{(2)} \dM_{jkl \rangle
		a}^{(5)} \nn \\ 
	&  - \frac{28}{11} \dM_{a \langle i}^{(3)} \dM_{jkl \rangle
		a}^{(4)} - \frac{273}{55} \dM_{a \langle
		i}^{(4)} \dM_{jkl \rangle a}^{(3)} - \frac{203}{55} \dM_{a \langle
		i}^{(5)} \dM_{jkl \rangle a}^{(2)} \nn \\ 
	&  - \frac{49}{55} \dM_{a \langle
		i}^{(6)} \dM_{jkl \rangle a}^{(1)} + \frac{14}{275} \dM_{a \langle
		i}^{(7)} \dM_{jkl \rangle a} +
	\frac{14}{33} \dM_{a \langle ij} \dM_{kl \rangle a}^{(7)} \nn \\ 
	&  +
	\frac{37}{33} \dM_{a \langle ij}^{(1)} \dM_{kl \rangle a}^{(6)} +
	\frac{9}{11} \dM_{a \langle ij}^{(2)} \dM_{kl \rangle a}^{(5)} +
	\frac{8}{33} \dM_{a \langle ij}^{(3)} \dM_{kl \rangle a}^{(4)} +
	\frac{9}{5} \dS_{\langle i} \dS_{jkl \rangle}^{(5)} \nn\\&
	 + \frac{16}{5} \dS_{\langle ij} \dS_{kl \rangle}^{(5)} +
	\frac{48}{5} \dS_{\langle ij}^{(1)} \dS_{kl \rangle}^{(4)} +
	\frac{32}{5} \dS_{\langle ij}^{(2)} \dS_{kl \rangle}^{(3)} \nn
	\\ & + \epsilon_{ab \langle i} \Big(- \frac{3}{5}
	\dM_{j\underline{a}} \dS_{kl \rangle b}^{(6)} - \frac{63}{25}
	\dM_{j\underline{a}}^{(1)} \dS_{kl \rangle b}^{(5)} + \frac{3}{5}
	\dM_{j\underline{a}}^{(2)} \dS_{kl \rangle b}^{(4)} + \frac{18}{5}
	\dM_{j\underline{a}}^{(3)} \dS_{kl \rangle b}^{(3)} \nn\\&
	 + \frac{9}{5} \dM_{j\underline{a}}^{(4)} \dS_{kl
		\rangle b}^{(2)} + \frac{3}{5} \dM_{j\underline{a}}^{(5)} \dS_{kl
		\rangle b}^{(1)} + \frac{3}{25} \dM_{j\underline{a}}^{(6)} \dS_{kl
		\rangle b} - \frac{8}{15} \dS_{j\underline{a}} \dM_{kl \rangle
		b}^{(6)} \nn\\&  - \frac{24}{25} \dS_{j\underline{a}}^{(1)} \dM_{kl \rangle
		b}^{(5)} - \frac{8}{5}
	\dS_{j\underline{a}}^{(2)} \dM_{kl \rangle b}^{(4)} + \frac{16}{3}
	\dS_{j\underline{a}}^{(3)} \dM_{kl \rangle b}^{(3)} + \frac{72}{5}
	\dS_{j\underline{a}}^{(4)} \dM_{kl \rangle b}^{(2)} \nn\\&  + \frac{56}{5}
	\dS_{j\underline{a}}^{(5)} \dM_{kl \rangle b}^{(1)} + \frac{232}{75} \dS_{j\underline{a}}^{(6)} \dM_{kl
		\rangle b} + \frac{29}{75} \dM_{jkl \rangle a}^{(6)} \dS_b \Big)
	\Bigg\} \,,\\
\dU_{ijklm}^\text{1.5PN} =\ 
	&
	\frac{G}{c^3}\Bigg\{ 2\dM\,\int_0^{+\infty} \dd\tau\, \dM_{ijklm}^{(7)}(U-\tau)\bigg[\ln \left(\frac{c \tau}{2 b_0}\right)+ \frac{232}{105}\biggl]
	\nn\\
	&\qquad + \frac{20}{21} \int_0^{+\infty} \! \dd\tau\,\left[\dM^{(3)}_{\langle ij}\dM^{(4)}_{klm\rangle}\right]\!(U-\tau)\nn\\
	&
	- \frac{15}{7}\dM^{}_{\langle ij}\dM^{(6)}_{klm\rangle}
	- \frac{41}{7}\dM^{(1)}_{\langle ij}\dM^{(5)}_{klm\rangle}
	- \frac{120}{7}\dM^{(2)}_{\langle ij}\dM^{(4)}_{klm\rangle}
	- \frac{710}{21}\dM^{(3)}_{\langle ij}\dM^{(3)}_{klm\rangle}\nn\\
	&
	- \frac{265}{7}\dM^{(4)}_{\langle ij}\dM^{(2)}_{klm\rangle}
	- \frac{155}{7}\dM^{(5)}_{\langle ij}\dM^{(1)}_{klm\rangle}
	- \frac{34}{7}\dM^{(6)}_{\langle ij}\dM^{}_{klm\rangle}
	\Bigg\}\,,\\
\dU_{ijklmn}^\text{1.5PN} &= \frac{G}{c^3} \Bigg\{ 2\dM\,\int_0^{+\infty} \dd\tau\, \dM_{ijklmn}^{(8)}(U-\tau)\bigg[\ln \left(\frac{c \tau}{2 b_0}\right)+ \frac{403}{168}\biggl] \nn\\
	&\qquad 
	+ \int_0^{+\infty} \dd \tau\bigg[
	\frac{5}{7} \dM_{\langle ijk}^{(4)} \dM_{lmn \rangle}^{(4)}
	 - \frac{15}{14} \dM_{\langle ij}^{(4)}  \dM_{klmn
		\rangle}^{(4)}  \bigg](U-\tau) \nn\\
	& 
	- \frac{45}{28} \dM_{\langle ij} \dM_{klmn
		\rangle}^{(7)} - \frac{111}{28} \dM_{\langle ij}^{(1)} \dM_{klmn
		\rangle}^{(6)} - \frac{561}{28} \dM_{\langle ij}^{(2)} \dM_{klmn
		\rangle}^{(5)} \nn\\&  - \frac{1595}{28}
	\dM_{\langle ij}^{(3)} \dM_{klmn \rangle}^{(4)} - \frac{2505}{28}
	\dM_{\langle ij}^{(4)} \dM_{klmn \rangle}^{(3)} - \frac{2115}{28}
	\dM_{\langle ij}^{(5)} \dM_{klmn \rangle}^{(2)} \nn\\&
	- \frac{909}{28} \dM_{\langle ij}^{(6)} \dM_{klmn
		\rangle}^{(1)} - \frac{159}{28} \dM_{\langle ij}^{(7)} \dM_{klmn
		\rangle} - \frac{15}{7} \dM_{\langle ijk} \dM_{lmn \rangle}^{(7)} 
	\nn
	\\ & -
	\frac{75}{7} \dM_{\langle ijk}^{(1)} \dM_{lmn \rangle}^{(6)} 
	 - \frac{135}{7} \dM_{\langle ijk}^{(2)} \dM_{lmn
		\rangle}^{(5)} - \frac{505}{21} \dM_{\langle ijk}^{(3)} \dM_{lmn
		\rangle}^{(4)} \Bigg\} \,.
\end{align}
For current type moments:
\begin{align}
\dV_{ijk}^\text{1.5PN} = 
	&
	\frac{G}{c^3}\Biggl\{2\dM\,\int_0^{+\infty}\!\! \dd\tau\, \dS_{ijk}^{(5)}(U-\tau)\bigg[\ln \left(\frac{c \tau}{2 b_0}\right)+ \frac{5}{3}\biggl]\nn\\
	&\qquad
	-2 \dM_{\langle ij}^{(4)}\dS_{k\rangle}
	- \frac{1}{10}\epsilon_{ab\langle i}\dM^{}_{j\underline{a}}\dM^{(5)}_{k\rangle b}
	+ \frac{1}{2}\epsilon_{ab\langle i}\dM^{(1)}_{j\underline{a}}\dM^{(4)}_{k\rangle b}
	\Biggr\}\,,\\
\dV_{ijk}^\text{2.5PN} &= \frac{G}{c^5} \Bigg\{ \frac{1}{12} \dM_{a \langle i} \dS_{jk \rangle
		a}^{(6)} + \frac{1}{12} \dM_{a \langle i}^{(1)} \dS_{jk \rangle
		a}^{(5)} + \frac{5}{12} \dM_{a \langle i}^{(2)} \dS_{jk \rangle
		a}^{(4)} + \frac{35}{12} \dM_{a \langle i}^{(4)} \dS_{jk \rangle
		a}^{(2)} \nn\\& \qquad + \frac{49}{12} \dM_{a \langle i}^{(5)} \dS_{jk \rangle
		a}^{(1)} + \frac{19}{12} \dM_{a \langle
		i}^{(6)} \dS_{jk \rangle a} + \frac{2}{27} \dS_{a \langle i} \dM_{jk
		\rangle a}^{(6)} + \frac{10}{27} \dS_{a \langle i}^{(1)} \dM_{jk
		\rangle a}^{(5)} \nn\\& \qquad + \frac{2}{27} \dS_{a \langle i}^{(2)} \dM_{jk
		\rangle a}^{(4)} + \frac{8}{9} \dS_{a \langle i}^{(3)} \dM_{jk \rangle
		a}^{(3)} - \frac{10}{27} \dS_{a \langle
		i}^{(4)} \dM_{jk \rangle a}^{(2)} \nn\\& \qquad - \frac{38}{27} \dS_{a \langle
		i}^{(5)} \dM_{jk \rangle a}^{(1)} - \frac{2}{3} \dS_{a \langle
		i}^{(6)} \dM_{jk \rangle a} - \frac{1}{60} \dS_a \dM_{ijka}^{(6)}
	\nn\\&  + \epsilon_{ab \langle i} \bigg(-
	\frac{1}{180} \dM_{jk \rangle ac}^{(7)} \dM_{bc} + \frac{11}{900} \dM_{jk
		\rangle ac}^{(6)} \dM_{bc}^{(1)} + \frac{7}{300} \dM_{jk \rangle
		ac}^{(5)} \dM_{bc}^{(2)} \nn\\& \qquad -
	\frac{37}{270} \dM_{jk \rangle ac}^{(4)} \dM_{bc}^{(3)} -
	\frac{191}{180} \dM_{jk \rangle ac}^{(3)} \dM_{bc}^{(4)} - \frac{65}{36}
	\dM_{jk \rangle ac}^{(2)} \dM_{bc}^{(5)} \nn\\& \qquad - \frac{367}{300} \dM_{jk \rangle
		ac}^{(1)} \dM_{bc}^{(6)} -
	\frac{133}{450} \dM_{jk \rangle ac} \dM_{bc}^{(7)} + \frac{1}{27} \dM_{j
		\underline{ac}} \dM_{k \rangle bc}^{(7)} \nn\\& \qquad + \frac{5}{162}
	\dM_{j\underline{ac}}^{(1)} \dM_{k \rangle bc}^{(6)} - \frac{5}{162}
	\dM_{j\underline{ac}}^{(2)} \dM_{k \rangle bc}^{(5)} - \frac{1}{81} \dM_{j\underline{ac}}^{(3)}
	\dM_{k \rangle bc}^{(4)} \nn\\& \qquad - \frac{11}{20} \dS_{jk \rangle b}^{(5)} \dS_a -
	\frac{88}{45} \dS_{j \underline{a}} \dS_{k \rangle b}^{(5)} -
	\frac{40}{9} \dS_{j\underline{a}}^{(1)} \dS_{k \rangle b}^{(4)}
	- \frac{32}{9}
	\dS_{j\underline{a}}^{(2)} \dS_{k \rangle b}^{(3)} \bigg)\Bigg\} \,,\\
\dV_{ijkl}^\text{1.5PN} = 
	&
	\frac{G}{c^3}\Biggl\{ 2\dM\,\int_0^{+\infty}\!\! \dd\tau\, \dS_{ijkl}^{(6)}(U-\tau)\bigg[\ln \left(\frac{c \tau}{2 b_0}\right)+ \frac{119}{60}\biggl]\nn\\
	&\qquad
	- \frac{11}{6}\dM^{}_{\langle ij}\dS^{(5)}_{kl\rangle}
	- \frac{25}{6}\dM^{(1)}_{\langle ij}\dS^{(4)}_{kl\rangle}
	- \frac{25}{3}\dM^{(2)}_{\langle ij}\dS^{(3)}_{kl\rangle}
	- \frac{35}{3}\dM^{(3)}_{\langle ij}\dS^{(2)}_{kl\rangle}
	\nn\\& \qquad - \frac{65}{6}\dM^{(4)}_{\langle ij}\dS^{(1)}_{kl\rangle}
	- \frac{19}{6}\dM^{(5)}_{\langle ij}\dS^{}_{kl\rangle}
	- \frac{11}{12}\dM^{(5)}_{\langle ijk}\dS^{}_{l\rangle}\nn\\
	&
	+ \epsilon_{ab\langle i}\bigg[
	\frac{1}{12}\dM^{}_{j \underline{a}}\dM^{(6)}_{kl\rangle b}
	+ \frac{37}{60}\dM^{(1)}_{j \underline{a}}\dM^{(5)}_{kl\rangle b}
	- \frac{5}{12}\dM^{(2)}_{j \underline{a}}\dM^{(4)}_{kl\rangle b}
	- \frac{5}{6}\dM^{(3)}_{j \underline{a}}\dM^{(3)}_{kl\rangle b}\nn\\
	& \qquad
	- \frac{11}{12}\dM^{(4)}_{j \underline{a}}\dM^{(2)}_{kl\rangle b}
	- \frac{1}{12}\dM^{(5)}_{j \underline{a}}\dM^{(1)}_{kl\rangle b}
	+ \frac{3}{60}\dM^{(6)}_{j \underline{a}}\dM^{}_{kl\rangle b}\bigg]
	\Biggr\}\,,\\
\dV_{ijklm}^\text{1.5PN} &= \frac{G}{c^3}
	\Bigg\{ 2\dM\,\int_0^{+\infty}\!\! \dd\tau\, \dS_{ijklm}^{(7)}(U-\tau)\bigg[\ln \left(\frac{c \tau}{2 b_0}\right)+ \frac{133}{60}\biggl]\nn\\
	&\qquad - \frac{3}{2} \dM_{\langle ij} \dS_{klm \rangle}^{(6)} -
	\frac{33}{10} \dM_{\langle ij}^{(1)} \dS_{klm \rangle}^{(5)} - 12
	\dM_{\langle ij}^{(2)} \dS_{klm \rangle}^{(4)} - 27 \dM_{\langle ij}^{(3)}
	\dS_{klm \rangle}^{(3)} \nn\\&\qquad - \frac{69}{2}
	\dM_{\langle ij}^{(4)} \dS_{klm \rangle}^{(2)} - \frac{39}{2} \dM_{\langle
		ij}^{(5)} \dS_{klm \rangle}^{(1)} - \frac{21}{5} \dM_{\langle
		ij}^{(6)} \dS_{klm \rangle} - \frac{4}{3} \dS_{\langle ij} \dM_{klm
		\rangle}^{(6)} \nn\\&\qquad  - \frac{76}{15}
	\dS_{\langle ij}^{(1)} \dM_{klm \rangle}^{(5)} - \frac{16}{3} \dS_{\langle
		ij}^{(2)} \dM_{klm \rangle}^{(4)} - 8 \dS_{\langle ij}^{(3)} \dM_{klm
		\rangle}^{(3)} - \frac{28}{3} \dS_{\langle ij}^{(4)} \dM_{klm
		\rangle}^{(2)} \nn\\& \qquad - \frac{20}{3}
	\dS_{\langle ij}^{(5)} \dM_{klm \rangle}^{(1)} - \frac{8}{5} \dS_{\langle
		ij}^{(6)} \dM_{klm \rangle} - \frac{3}{5} \dS_{\langle i} \dM_{jklm
		\rangle}^{(6)} \nn\\& + \epsilon_{ab
		\langle i} \Big( \frac{1}{14} \dM_{j\underline{a}} \dM_{klm \rangle
		b}^{(7)} + \frac{1}{2} \dM_{j\underline{a}}^{(1)} \dM_{klm \rangle
		b}^{(6)} - \frac{3}{5} \dM_{j\underline{a}}^{(2)} \dM_{klm \rangle
		b}^{(5)} \nn\\&\qquad  - \frac{4}{3} \dM_{j\underline{a}}^{(3)} \dM_{klm \rangle
		b}^{(4)} - \frac{3}{2}
	\dM_{j\underline{a}}^{(4)} \dM_{klm \rangle b}^{(3)} - \frac{1}{2}
	\dM_{j\underline{a}}^{(5)} \dM_{klm \rangle b}^{(2)} + \frac{1}{35}
	\dM_{j\underline{a}}^{(7)} \dM_{klm \rangle b} \nn\\&\qquad  + \frac{1}{7}
	\dM_{jk\underline{a}} \dM_{lm \rangle b}^{(7)} + \frac{2}{3} \dM_{jk\underline{a}}^{(1)} \dM_{lm
		\rangle b}^{(6)} + \frac{4}{3} \dM_{jk\underline{a}}^{(2)} \dM_{lm
		\rangle b}^{(5)} + \frac{1}{3} \dM_{jk \underline{a}}^{(3)} \dM_{lm
		\rangle b}^{(4)} \Big) \Bigg\} \,.
\end{align}
\end{subequations}
For all the other multipole moments, we can only state that there is agreement between the radiative and canonical moments up to 1.5PN order, namely
\begin{subequations}\label{ULVLnewtonian}
	\begin{align}
		\dU_L (U) &= \dM^{(\ell)}_L(U) +
		\calO\left(\frac{1}{c^3}\right)\,,\\ \dV_L (U) &=
		\dS^{(\ell)}_L(U) + \calO\left(\frac{1}{c^3}\right)\,.
\end{align}\end{subequations}

\subsubsection{Canonical versus source moments}
\label{sec:cansource}

In a second stage of the formalism, we must express the canonical moments $\{\dM_L, \dS_L\}$ in terms of the six types of source moments $\{\dI_L, \dJ_L, \dW_L, \dX_L, \dY_L, \dZ_L\}$. For the control of the flux and $(2,2)$ mode in the waveform up to 4PN order, we need to relate the canonical quadrupole moment $\dM_{ij}$ to the corresponding source quadrupole moment $\dI_{ij}$ up to that accuracy. The relation was known at 3.5PN order \citep{FMBI12} and has been extended at 4PN order \citep{BFL22, BFHLT23b} with result\footnote{This result consists of purely quadratic interactions. However, in a first calculation of this relation, performed in ordinary 3 dimensions using the Hadamard regularization scheme (for the IR divergences), some cubic interactions were found at the 4PN order. See Eq.~(1.1) in \cite{BFL22} where these interactions are of the types $\dM\times\dM\times\dW_{ij}$, $\dM\times\dM\times\dY_{ij}$ and $\dM\times\dW\times\dI_{ij}$. However further work \citep{BFHLT23b} showed that within the dimensional regularization scheme such interactions are cancelled, so that the relation \eqref{relationMijIij}, derived in the context of dimensional regularization, holds with exactly the same definition for the ``renormalized'' source quadrupole moment $\dI_{ij}$ as obtained by \cite{LHBF22, LBHF22}.}
\begin{align}\label{relationMijIij}
	\dM_{ij} &= 
	\dI_{ij}
	+ \frac{4G}{c^5}\bigg[\dW^{(2)}\dI^{}_{ij}-\dW^{(1)}\dI^{(1)}_{ij}\bigg]\nn\\
	&
	+ \frac{4G}{c^7}\Bigg\{
	\frac{4}{7}\dW^{(1)}_{a\langle i}\dI^{(3)}_{j\rangle a}
	+ \frac{6}{7}\dW^{}_{a\langle i}\dI^{(4)}_{j\rangle a}
	- \frac{1}{7}\dY^{(3)}_{a\langle i}\dI^{}_{j\rangle a}
	- \dY^{}_{a\langle i}\dI^{(3)}_{j\rangle a}
	- 2\dX\,\dI^{(3)}_{ij}\nn\\
	& \quad\quad
	- \frac{5}{21}\dW^{(4)}_{a}\dI^{}_{ija}
	+ \frac{1}{63}\dW^{(3)}_{a}\dI^{(1)}_{ija}
	- \frac{25}{21}\dY^{(3)}_{a}\dI^{}_{ija}
	- \frac{22}{63}\dY^{(2)}_{a}\dI^{(1)}_{ija}
	+ \frac{5}{63}\dY^{(1)}_{a}\dI^{(2)}_{ija}\nn\\
	& \quad\quad
	+ 2\dW^{(3)}\dW^{}_{ij}
	+ 2\dW^{(2)}\dW^{(1)}_{ij}
	- \frac{4}{3}\dW_{\langle i}\dW^{(3)}_{j\rangle}
	+ 2\dW^{(2)}\dY^{}_{ij}
	- 4\dW_{\langle i}\dY^{(2)}_{j\rangle}\nn\\
	& \quad\quad
	+ \epsilon_{ab\langle i}\bigg[
	\frac{1}{3}\dI_{j\rangle a}\dZ^{(3)}_b
	- \dI_{j\rangle a}^{(3)}\dZ^{}_b
	+ \frac{4}{9}\dJ^{}_{j\rangle a}\dW^{(3)}_b
	- \frac{4}{9}\dJ^{}_{j\rangle a}\dY^{(2)}_b
	+ \frac{8}{9}\dJ^{(1)}_{j\rangle a}\dY^{(1)}_b\bigg]\Bigg\}
	\nn\\ &+ \calO \left(\frac{1}{c^9}\right)\,.
\end{align}
Here, for instance, $\dW$ denotes the monopole moment associated with the moment $\dW_L$, and $\dY_i$ is the dipole moment corresponding to $\dY_L$. Notice that the difference between the canonical and source moments starts at the relatively high 2.5PN order.

We need also the relations for the higher-order moments, notably for the control of the full waveform, which contain similarly some correction terms starting at the 2.5PN order. Here we present the most up-to-date results \cite{BFIS08, FMBI12}:
\begin{subequations}\label{cansourceMS}
	\begin{align}  
		\dM_{ijk} &= \dI_{ijk} + \frac{4G}{c^5}\biggl[\dW^{(2)}\dI_{ijk}-\dW^{(1)}\dI_{ijk}^{(1)}+3\,\dI_{\langle
			ij}\dY_{k\rangle }^{(1)}\biggr] +
		\calO\left(\frac{1}{c^7}\right)\label{M3} \,,\\ 
		\dM_{ijkl} &= \dI_{ijkl} + \frac{4 G}{c^5} \bigg[- \dW^{(1)} \dI_{ijkl}^{(1)} +
		\dW^{(2)} \dI_{ijkl} + 4 \dY_{\langle i}^{(1)} \dI_{jkl \rangle} \bigg] +
		\calO\left(\frac{1}{c^7}\right)\,,\\
		\dS_{ij} &= \dJ_{ij} + \frac{2G}{c^5}\biggl[\epsilon_{ab\langle
			i}\biggl(-\dI_{j\rangle b}^{(3)}\dW_{a}-2\dI_{j\rangle b}\dY_{a}^{(2)}
		+\dI_{j\rangle b}^{(1)}\dY_{a}^{(1)}\biggr) \nn\\&\qquad\qquad +3\dJ_{\langle i}\dY_{j\rangle
		}^{(1)}-2\dJ_{ij}^{(1)}\dW^{(1)}\biggr] +
		\calO\left(\frac{1}{c^7}\right)\,,\\
		\dS_{ijk} &= \dJ_{ijk} + \frac{4G}{c^5} \bigg[ \epsilon_{ab
			\langle i} \bigg(- \frac{1}{3} \dI_{jk \rangle a}^{(1)} \dY_b^{(1)} +
		\dI_{jk \rangle a} \dY_b^{(2)} + \dI_{j\underline{a}}^{(3)} \dW_{k \rangle
			b} \bigg) \nn\\&\qquad\qquad - \dW^{(1)} \dJ_{ijk}^{(1)} +
		\frac{8}{3} \dY_{\langle i}^{(1)} \dJ_{jk \rangle} \bigg] +
		\calO\left(\frac{1}{c^7}\right)\,,
	\end{align}
\end{subequations}
For all the other moments $\dM_L$, $\dS_L$ we can just say that they agree with their source counterparts $\dI_L$, $\dJ_L$ up to 2.5PN order,
\begin{subequations}\label{MLSL}
	\begin{align}
		\dM_L = \dI_L +
		\calO\left(\frac{1}{c^5}\right)\,,\qquad
		\dS_L = \dJ_L +
		\calO\left(\frac{1}{c^5}\right)\,.
\end{align}\end{subequations}
Finally, combining the previous formulas \eqref{cansourceMS} together with all the results of Sect.~\ref{sec:radcanonical} we control the full waveform in terms of the basic source and gauge moments up to order 3.5PN, with the dominant mode controlled to 4PN.

With all the formulas in Sects.~\ref{sec:radcanonical} and \ref{sec:cansource} we have related the radiative moments $\{\dU_L, \dV_L\}$ parametrizing the waveform at future null infinity \eqref{hijTT} to the six types of source multipole moments $\{\dI_L, \dJ_L, \dW_L, \dX_L,\dY_L, \dZ_L\}$. The general expression of the source moments in terms of the source parameters (actually, the PN expansion of the matter $+$ gravitation pseudo-tensor) has been obtained in Eqs. \eqref{sourcemoments} and \eqref{gaugemoments}. Hence at this stage the problem of the wave generation by an arbitrary post-Newtonian source is solved. What is missing is the explicit implementation of the formalism for an actual model of the source. We come to grips with the important physical case of compact binary systems in the next section.

\section{Compact binary systems}
\label{sec:compactbinary}

As reviewed in Sect.~\ref{sec:intro} the appropriate theoretical description of the early inspiral of compact binaries is by two structureless point-particles, characterized solely by their masses $m_1$ and $m_2$ and possibly their spins. Furthermore, when approaching the merger some extra parameters describing the internal structure of the bodies (and defined within the effective field theory) become necessary and in particular characterize the tidal interaction between the two objects (see Sect.~\ref{sec:intstructure}). Here we compute the equations of motion and radiation of compact binary systems within this model for the compact objects.


\subsection{Regularization of the field of point particles}
\label{sec:reg}

One aim is to compute the metric (and its gradient needed in the equations of motion) at some high post-Newtonian order for a system of two point-like particles. \emph{A priori} one is not allowed to use directly some metric expressions like in Eqs. \eqref{gmunu3PN}, which have been derived under the assumption of a continuous (smooth) matter distribution. Applying them to a system of point particles, we find that the metric becomes divergent at the location of the particles, i.e., when $\mathbf{x}\to\bm{y}_1(t)$ or $\bm{y}_2(t)$, where $\bm{y}_\text{a}(t)$ denote the two trajectories. Consequently, we must supplement the calculation by a prescription for how to remove the infinite self-field of point particles. At this stage different choices for a ``self-field'' regularization are possible: 
\begin{enumerate}
\item The Hadamard self-field regularization has long been used and proved to be very convenient for doing practical computations (in particular, by computer). However it suffers from the important drawback of yielding some physical ambiguity parameters, which cannot be determined within this regularization, starting essentially at the 3PN order;
\item The dimensional self-field regularization is an extremely powerful regularization which is free of ambiguities (at least up to the 4PN level). In particular it has been used to uniquely fix the values of the ambiguity parameters coming from Hadamard's regularization. However, in high post-Newtonian approximations, dimensional regularization cannot be implemented for an arbitrary space dimension $d\in\mathbb{C}$, but has to be applied in the limit $\varepsilon=d-3\to 0$, in which limit it is in fact connected in a precise way to the Hadamard regularization.
\end{enumerate}
We expect the results to be unique and independent of the employed regularization, in agreement with the physical intuition. This can be justified from the effacing principle of general relativity \citep{Dhouches} -- namely that the internal structure of the compact bodies makes a contribution only at the formal 5PN approximation. However note that all PN results derived below have been obtained using dimensional regularization (or have been shown to be equivalent to those of dimensional regularization).

In principle the use of a self-field regularization, be it dimensional or based on Hadamard's partie finie, signals a somewhat unsatisfactory situation on the physical point of view, because, ideally, we would like to perform a complete calculation valid for extended bodies, taking into account the details of the internal structure of the bodies (energy density, pressure, internal velocity field, etc.). By considering the limit where the radii of the objects tend to zero, one should recover the same result as obtained by means of the point-mass regularization. This would demonstrate the suitability of the regularization. This program was undertaken at the 2PN order by \cite{Kop85, GKop86} who derived the equations of motion of two extended fluid balls, and obtained equations of motion depending only on the two masses $m_1$ and $m_2$ of the compact bodies.\footnote{See some comments on the latter works by \cite{D300}, pp. 168-169.} At the 3PN or even 4PN order we expect that the extended-body program should give exactly the same result as given by dimensional regularization. Ideally, this should also be confirmed by independent and more physical methods like those of \cite{ThH85, Kop88, DSX91}.

Another body of work, in a sense more physical than the formal use of a self-field regularization, is the one of \cite{IFA00, IFA01, itoh1, itoh2, itoh3}, where the equations of motion are derived in harmonic coordinates by means of a particular variant of the famous ``surface-integral'' method (i.e., \emph{\`a la} \citealt{EIH}). The aim is to describe extended relativistic compact binary systems in the strong-field point particle limit \citep{F87}. This approach is interesting because it is based on the physical notion of extended compact bodies in general relativity, and by-passes the problem of the self-field regularization. The end result of \cite{itoh1, itoh2} is in agreement with the harmonic coordinates 3PN equations of motion \citep{BF00, BFeom, BDE04}.

\subsubsection{Hadamard's regularization}
\label{sec:had}

In many practical computations it is useful to employ the Hadamard regularization \citep{Hadamard, Schwartz, Sellier}, in particular it recovers in a simple way the $d\to 3$ limit of dimensional regularization in the absence of poles. See \cite{BFreg, JaraSLRR} for reviews; \cite{BFreg} proposed an ``extended'' Hadamard regularization based on a generalization of the theory of \cite{Schwartz} distributions with generalized singular functions or pseudo-functions.
	
With Hadamard's regularization the PN iteration of the metric and equations of motion for a system of compact objects (point-particles) proceeds along the line of Sect.~\ref{sec:3PNmetric}, i.e. computing the hierarchy of post-Newtonian potentials in three dimensions, Eqs. \eqref{pot1PN} and \eqref{pot3PN}. At each step one has typically to solve a Poisson equation of the type $\Delta P = F$, where the function $F(\mathbf{x})$ is smooth ($C^\infty$) on ${\mathbb{R}}^3$ \emph{except} for the two points $\bm{y}_\text{a}$ at which it becomes singular (UV type divergence; for simplicity we consider only two point particles). We assume that $F$ admits a power-like singular expansion around 1 of the general structure (with $p\in\mathbb{Z}$):\footnote{The function $F(\mathbf{x})$ depends also on (coordinate) time $t$, through for instance its dependence on the velocities $\bm{v}_1(t)$ and $\bm{v}_2(t)$, but the time $t$ is purely ``spectator'' in the regularization process, and thus will not be indicated. See the footnote \ref{fnote:landau} for the definition of the Landau symbol $o$ for remainders.}
\begin{equation}\label{Fexp}
	\forall \mathcal{N}\in {\mathbb{N}}, \qquad F(\mathbf{x}) =
	\!\!\!\!\!  \sum_{p_0\leqslant p\leqslant \mathcal{N}} \!\!\!  r_1^p
	\, \mathop{f}_{1}{}_{\!\!p}(\mathbf{n}_1)+o(r_1^{\mathcal{N}})\,,
\end{equation}
and similarly for the other point 2. Here $r_1=|\mathbf{x}-\bm{y}_1|\to 0$, and the coefficients ${}_1f_p$ of the various powers of $r_1$ depend on the unit direction $\mathbf{n}_1=(\mathbf{x}-\bm{y}_1)/r_1$ of approach to the singular point. The powers $p$ of $r_1$ are bounded from below ($p_0\leqslant p$ with $p_0\in\mathbb{Z}$). The coefficients ${}_1f_p$ (and ${}_2f_p$) for which $p\leqslant -1$ are referred to as the \emph{singular} coefficients of $F$. If both $F$ and $G$ admit expansions of the type \eqref{Fexp} so does the ordinary product $FG$, as well as the ordinary gradient $\partial_iF$. We define the Hadamard \emph{partie finie} of $F$ at the location of the point 1 where it is singular as
\begin{equation}
	(F)_1= \int \frac{\dd\Omega_1}{
		4\pi}\,\mathop{f}_{1}{}_{\!\!0}(\mathbf{n}_1)\,,
	\label{hadPF}
\end{equation}
where $\dd\Omega_1= \dd\Omega (\mathbf{n}_1)$ denotes the solid angle element centered on $\bm{y}_1$ and of direction $\mathbf{n}_1$. Notice that because of the angular integration in Eq.~\eqref{hadPF}, the Hadamard \emph{partie finie} is ``non-distributive'' in the sense that
\begin{equation}
	(FG)_1\not= (F)_1(G)_1 \quad \mathrm{in\ general}\,.
	\label{nondistr}
\end{equation}
The non-distributivity of Hadamard's partie finie is the main source of the appearance of ambiguity parameters with this regularization.

The second notion of Hadamard \emph{partie finie} ($\pf$) concerns that of the integral $\int \dd^3\mathbf{x} \, F$, which is generically divergent at the location of the two singular points $\bm{y}_1$ and $\bm{y}_2$ (we assume that the integral converges at infinity). It is defined by removing the singular parts of the expansion near the singularities:
\begin{align}\label{pfint}
	\Pf \int \dd^3\mathbf{x} \, F(\mathbf{x}) &= \lim_{s \to 0} \,
	\biggl\{\int_{\mathcal{S}(s)} \!\!\! \dd^3\mathbf{x} \, F +
	4\pi\sum_{p+3\leqslant -1}{\frac{s^{p+3}}{p+3}} \left( \frac{F}{r_1^p}
	\right)_1 \nn\\& \qquad\qquad + 4 \pi \ln \left(\frac{s}{s_1}\right) \left(r_1^3
	F\right)_1 + 1\leftrightarrow 2\biggr\}\,.
\end{align}
The first term integrates over a domain $\mathcal{S}(s)$ defined as ${\mathbb{R}}^3$ from which the two spherical balls $r_1<s$ and $r_2<s$ of radius $s$ and centered on the two singularities, i.e. $\mathcal{B}_1(s)$ and $\mathcal{B}_2(s)$, are excised: $\mathcal{S}(s)\equiv {\mathbb{R}}^3\setminus \mathcal{B}_1(s)\cup \mathcal{B}_2(s)$. The other terms, where the value of a function at point 1 takes the meaning \eqref{hadPF}, are precisely such that they cancel out the divergent part of the first term in the limit where $s\to 0$ (the symbol $1\leftrightarrow 2$ means the same terms but corresponding to point 2). The Hadamard partie-finie integral depends on two strictly positive constants $s_1$ and $s_2$, associated with the logarithms present in Eq.~\eqref{pfint}. The appearance of these logarithms and arbitrary constants signals in fact the ambiguousness and incompleteness of the regularization. 

The Hadamard partie-finie integral in the form given by \eqref{pfint} is rather difficult to evaluate, because of the integration over the complicated volume $\mathcal{S}(s)$. Fortunately, there exist several alternative expressions of the Hadamard partie finie, much better suited for practical computations. One is based on a double analytic continuation, with two complex parameters $\alpha$, $\beta\in {\mathbb C}$, 
\begin{equation}\label{pfint2}
	\Pf \int \dd^3\mathbf{x} \, F = \mathop{\mathrm{FP}}_{\alpha \to 0 \atop
		\beta \to 0} \int \dd^3 \mathbf{x}
	\left(\frac{r_1}{s_1} \right)^\alpha \!\left(\frac{r_2}{s_2} \right)^\beta F \,,
\end{equation}
where the constants $s_1$ and $s_2$ are the same as in the definition \eqref{pfint}. The operation $\mathrm{FP}$ means taking the finite parts in the Laurent expansions when $\alpha\to 0$ and $\beta\to 0$ sucessively; furthermore one can apply the two limits in whatever order. See \cite{BFreg} for other expressions of the partie-finie integral \eqref{pfint}.

Within the Hadamard regularization the definitions \eqref{pfint} or \eqref{pfint2} are applied to Poisson integrals, solving the Poisson equation $\Delta P = F$. In this case we consider the partie-finie integral\footnote{To high post-Newtonian order, the Poisson integral will also diverge at infinity. In addition to the UV regularization, one must also apply an IR regularization, which will be given in the context of Hadamard's regularization by Eq.~\eqref{genpoiss}.}
\begin{equation}\label{pfPoisson}
	P(\mathbf{x}') = - \frac{1}{4\pi} \Pf \int \frac{\dd^3\mathbf{x}}{\vert\mathbf{x}-\mathbf{x}'\vert} \,F(\mathbf{x}) \,,
\end{equation} 
where the field point $\mathbf{x}'$ is located outside the singular points $\bm{y}_1$ and $\bm{y}_2$, and is just ``spectator'' in the regularization process. However we want to evaluate the value of the regularized Poisson integral \eqref{pfPoisson} at the singular point $\bm{y}_1$. Clearly the limit will not be continuous at this point, because the Poisson integral admits a singular expansion when $\mathbf{x}'\to\bm{y}_1$. Because of the integration in \eqref{pfint2} this expansion will be more general than in Eq.~\eqref{Fexp}, typically involving a logarithm of $r'_1\equiv \vert\mathbf{x}'-\bm{y}_1\vert\to 0$. Therefore the Hadamard partie finie of $P$ at 1 in the sense of Eq.~\eqref{hadPF} is actually infinite since it \textit{a priori} contains $\ln r'_1=-\infty$. Nevertheless, in Hadamard's regularization we must keep the ``constant'' $\ln r'_1$ in all calculations; such drawback will be solved when we add the corrections due do the dimensional regularization in Sect.~\ref{sec:DReom}. Finally, the Hadamard partie finie at the point 1 in the sense of Eq.~\eqref{hadPF}, of the Poisson integral \eqref{pfPoisson} is given by \citep{BFreg}
\begin{equation}\label{pfPoisson1}
		(P)_1 = -\frac{1}{4\pi} \Pf \int \frac{\dd^3\mathbf{x}}{r_1}
		\,F(\mathbf{x})+\biggl[\ln\left(\frac{r_1'}{s_1}\right)-1\biggr]
		\bigl(r_1^2 F\bigr)_1 \,,
\end{equation}
with the ``infinite'' constant $r_1'=\vert\mathbf{x}'-\bm{y}_1\vert$ appearing in the second term. Actually the constant $s_1$ cancels out from the two terms in the right side of \eqref{pfPoisson1} so the partie finie depends only on $r'_1$ and $s_2$.

As we said the Hadamard regularization yields some ambiguous results for the computation of certain integrals at the 3PN order. This was noticed by \cite{JaraS98, JaraS99, JaraS00} in their computation of the equations of motion within the ADM-Hamiltonian formulation of general relativity. They showed that there are two and only two types of ambiguous terms in the 3PN Hamiltonian, which were then parametrized by two unknown numerical coefficients. Similarly \cite{BF00, BFeom} obtained the 3PN equations of motion in harmonic coordinates complete except for one and only one unknown numerical constant. More precisely they showed that the infinite ``constant'' $r'_1$ in Eq.~\eqref{pfPoisson1}, together with $r'_2$ for the other particle, are unphysical, in the sense that they can be removed by a coordinate transformation. However the constant $s_2$ which remains in the result \eqref{pfPoisson1}, and the $s_1$ for the other particle, are the source for the appearance of a single physical ambiguity parameter (they called $\lambda$) which was defined from the unknown ratios $r'_1/s_1$ and $r'_2/s_2$; see~(7.9) of \cite{BFeom}. Concerning the binary's radiation field, the same phenomenon occurs, with the appearance of some Hadamard regularization ambiguity parameters starting at the 3PN order in the binary's mass quadrupole moment \citep{BIJ02}, and therefore ambiguity parameters appearing in the 3PN flux and phase evolution as well \citep{BFIJ02}. All these problems, which concern UV type divergences associated with the model of point masses, have been resolved, and the ambiguity parameters uniquely fixed, thanks to the dimensional regularization.


\subsubsection{Dimensional regularization of the equations of motion}
\label{sec:DReom}

Dimensional regularization was invented as a means to preserve the gauge symmetry of perturbative quantum field theories \citep{tHooft, Bollini, Breitenlohner, Collins}. Our basic problem here is to respect the gauge symmetry associated with the diffeomorphism invariance of the classical general relativistic description of interacting point masses. Hence, we use dimensional regularization not merely as a trick to compute some particular integrals which would otherwise be divergent, but as a powerful tool for solving in a consistent way the Einstein field equations with singular point-mass sources, while preserving its crucial symmetries. In particular, the dimensional regularization is able to correctly keep track of the global Lorentz-Poincar\'e invariance of the gravitational field of isolated systems. And of course, dimensional regularization is also a crucial ingredient of the EFT approach to equations of motion and gravitational radiation \citep{GR06, GLPR16, FPRS19}.
	
The Einstein field equations in $D=d+1$ space-time dimensions take exactly the same form as in 4 dimensions, say, in harmonic coordinates $H^\alpha = 0$,
\begin{equation}\label{EFEd}
	\Box h^{\alpha\beta} = \frac{16\pi G^{(d)}}{c^4} \tau^{\alpha\beta} = \frac{16\pi G^{(d)}}{c^4} \vert g\vert\,T^{\alpha\beta} + \Lambda^{\alpha\beta}_\text{harm}\,.
\end{equation}
The differences are: (i) the box operator $\Box$ now denotes the flat space-time d'Alembertian operator in $D$ dimensions with signature $(-1,1,1,\cdots)$; (ii) the gravitational constant $G^{(d)}$ is related to the usual three-dimensional Newton's constant $G$ by
\begin{equation}
	G^{(d)} = G\,\ell_0^{d-3}\,,
	\label{G}
\end{equation}
where $\ell_0$ denotes an arbitrary length scale; (iii) the gravitational source term in harmonic coordinates acquires a $d$-dependent coefficient:
\begin{align}\label{Lambdad}
	\Lambda^{\alpha\beta}_\text{harm} = &- h^{\mu\nu} \partial^2_{\mu\nu}
	h^{\alpha\beta}+\partial_\mu h^{\alpha\nu} \partial_\nu h^{\beta\mu}
	+\frac{1}{2}g^{\alpha\beta}g_{\mu\nu}\partial_\lambda h^{\mu\tau}
	\partial_\tau h^{\nu\lambda} \nn
	\\ &-g^{\alpha\mu}g_{\nu\tau}\partial_\lambda h^{\beta\tau}
	\partial_\mu h^{\nu\lambda} -g^{\beta\mu}g_{\nu\tau}\partial_\lambda
	h^{\alpha\tau} \partial_\mu h^{\nu\lambda}
	+g_{\mu\nu}g^{\lambda\tau}\partial_\lambda h^{\alpha\mu}
	\partial_\tau h^{\beta\nu} \nn
	\\ &+\frac{1}{4}(2g^{\alpha\mu}g^{\beta\nu}-g^{\alpha\beta}g^{\mu\nu})
	\Bigl(g_{\lambda\tau}g_{\epsilon\pi}-
	\frac{1}{d-1}g_{\tau\epsilon}g_{\lambda\pi}\Bigr) \partial_\mu
	h^{\lambda\pi} \partial_\nu h^{\tau\epsilon}\,.
\end{align}
When $d=3$ we recover Eq.~\eqref{Lambdadef}. In the following we assume, as required in dimensional regularization, that the dimension of space is a complex number, $d\in\mathbb{C}$, and prove many results by invoking complex analytic continuation in $d$. We often pose $\varepsilon\equiv d-3\in\mathbb{C}$ and systematically consider the limit $\varepsilon\to 0$.
	
We start by writing in a general way the retarded solution of the flat scalar wave equation in $D$ space-time dimensions (thus, with $\mathbf{x}\in\mathbb{R}^d$),
\begin{equation}\label{waveeq}
		\Box h(\mathbf{x},t) = S(\mathbf{x},t)\,,
\end{equation}
where the source will represent a generic term in the matter tensor or the gravitational source term \eqref{Lambdad}. The retarded Green's function of the scalar wave equation, satisfying $\Box G_\text{ret}(\mathbf{x},t) = \delta(c t)\,\delta^{(d)}(\mathbf{x})$ where $\delta^{(d)}$ is the Dirac function in $d$ dimensions, reads in $(\mathbf{x},t)$ space\footnote{In the Fourier domain the retarded Green's function reads (see \citealt{Cardoso})
\begin{equation*}\label{green3}
	G_\mathrm{ret}(\mathbf{x},t)=-\frac{\theta(t)}{(2\pi)^{d/2}}
	\int_0^{+\infty}\dd k \left(\frac{k}{r}\right)^{\frac{d}{ 2}-1}\sin
	(c k t)\,J_{\frac{d}{2}-1}(k r)\,,
\end{equation*}
where $J_{\frac{d}{2}-1}(k r)$ is the usual Bessel function.}
\begin{equation}\label{Gret}
		G_\text{ret}(\mathbf{x},t) = -
		\frac{\tilde{k}}{4\pi}\,\frac{\theta(ct-r)}{r^{d-1}}
		\,\gamma_{\frac{1-d}{2}}\left(\frac{ct}{r}\right)\,,
\end{equation}
where $\theta(ct-r)$ is the usual Heaviside step function, and $\tilde{k}$ is a constant related to the usual Eulerian $\Gamma$-function by\footnote{We have $\lim_{d\rightarrow 3}\tilde{k}=1$. Notice that $\tilde{k}$ is closely linked to the volume $\Omega_{d-1}$ of the sphere with $d-1$ dimensions (i.e. embedded into Euclidean $d$-dimensional space):
\begin{align*}
	\tilde{k}\,\Omega_{d-1}=\frac{4\pi}{d-2}\,.
\end{align*}
}
\begin{equation}
	\tilde{k}=\frac{\Gamma\left(\frac{d-2}{2}\right)}
	{\pi^{\frac{d-2}{2}}}\,.
	\label{ktilde}
\end{equation}
The corresponding advanced Green's function $G_\text{adv}(\mathbf{x},t)$ is given by the same expression but with $\theta(-ct-r)$ in place of $\theta(ct-r)$. Notice that the Green's function is in fact a function of $t$ and $r=\vert\mathbf{x}\vert$ only. In Eq.~\eqref{Gret} we have introduced the function $\gamma_s(z)$ defined for any $s\in\mathbb{C}$ and $\vert z\vert\geqslant 1$ by
\begin{align}\label{gammas}
	\gamma_s(z) &= \frac{2\sqrt{\pi}}{\Gamma(s+1)\Gamma(-s-\frac{1}{2})}
	\,\big(z^2-1\bigr)^s\,,
\end{align}
where the normalisation is such that $\int_1^{+\infty} \dd z \,\gamma_s(z) = 1$. In the Green's function this expression gives back the usual Minkowskian time like interval since $z^2-1\propto c^2t^2-r^2$ for $z=\frac{ct}{r}$. With those notations the retarded solution of the wave equation \eqref{waveeq} is given by
\begin{align}\label{retsol}
 h = \Box^{-1}_{\mathrm{ret}} S = - \frac{\tilde{k}}{4\pi}
		\int_1^{+\infty} \dd z \,\gamma_{\frac{1-d}{2}}(z) \int
		\dd^d\mathbf{x}'
		\,\frac{S(\mathbf{x}',t-z\vert\mathbf{x}-\mathbf{x}'\vert/c)}
		{\vert\mathbf{x}-\mathbf{x}'\vert^{d-2}}\,.
\end{align}

With dimensional regularization the metric and equations of motion will be parametrized
by means of some retarded potentials $V$, $V_i$, $\hat{W}_{ij}$, $\cdots$, which are straightforward $d$-dimensional generalizations of the potentials used in three dimensions and which were defined in Sect.~\ref{sec:3PNmetric}. Those are obtained by post-Newtonian iteration of the $d$-dimensional field equations, starting from appropriate definitions of matter source densities generalizing Eqs. \eqref{sigmadef}:
\begin{equation}\label{sigmadefd}
\sigma =
\frac{2}{d-1}\frac{(d-2)T^{00}+T^{ii}}{c^2}\,,\qquad\sigma_i =
\frac{T^{0i}}{c}\,,\qquad \sigma_{ij} = T^{ij}\,.
\end{equation}
As a result all the expressions of Sect.~\ref{sec:3PNmetric} acquire some explicit $d$-dependent coefficients. For instance we find \citep{BDE04}
\begin{subequations}\label{pot1PNd}
\begin{align}
	V &= \Box^{-1}_{\mathrm{ret}} \left[-4 \pi G^{(d)}
			\sigma\right]\,,\\ 
			\hat{W}_{ij} &= \Box^{-1}_{\mathrm{ret}} \left[ -4 \pi G^{(d)}
			\left(\sigma_{ij}-\delta_{ij} \frac{\sigma_{kk}}{d-2}\right)-
			\frac{d-1}{2(d-2)} \partial_i V \partial_j V \right]\,,
\end{align}
\end{subequations}
where $\Box^{-1}_{\mathrm{ret}}$ is the retarded integral operator \eqref{retsol}. Here again, in higher approximations there will be also IR divergences and one should really employ the $d$-dimensional version of Eq.~\eqref{BoxRreg}.
	
As reviewed in Sect.~\ref{sec:had}, the generic functions $F(\mathbf{x})$ we have to deal with in 3 dimensions, are smooth on $\mathbb{R}^3$ except at $\bm{y}_1$ and $\bm{y}_2$, around which they admit singular Laurent-type expansions given by Eq.~\eqref{Fexp}. In $d$ spatial dimensions, there is an analogue of the function $F$, which results from the same PN iteration process performed in $d$ dimensions as we just outlined. Let us call this function $F^{(d)}(\mathbf{x})$, where $\mathbf{x}\in\mathbb{R}^d$. When $r_1\rightarrow 0$ the function $F^{(d)}$ admits a singular expansion which is more involved than in 3 dimensions:
\begin{equation}\label{Fexpd}
		F^{(d)}(\mathbf{x}) = \sum_{\substack{p_0 \leqslant p
				\leqslant \mathcal{N} \\ q_0 \leqslant q\leqslant q_1}} r_1^{p+q\varepsilon}
		\mathop{f}_1{}_{p,q}^{(\varepsilon)}(\mathbf{n}_1) +
		o(r_1^{\mathcal{N}})\,.
\end{equation}
The coefficients $\mathop{f}_1{}_{p,q}^{(\varepsilon)}(\mathbf{n}_1)$ depend on $\varepsilon = d-3$, and the powers of $r_1$ involve the relative integers $p$ and $q$ whose values are limited by some $p_0,\,q_0,\,q_1$ $\in\mathbb{Z}$ as indicated. Here we will be interested in functions $F^{(d)}(\mathbf{x})$ which have no poles as $\varepsilon \rightarrow 0$ (this will always be the case up to 4PN order). Therefore, we can deduce from the fact that $F^{(d)}(\mathbf{x})$ is continuous at $d=3$ the constraint
\begin{equation}\label{constr}
	\sum_{q=q_0}^{q_1}\mathop{f}_1{}_{p,q}^{(\varepsilon=0)}
	(\mathbf{n}_1) = \mathop{f}_1{}_{\!p}(\mathbf{n}_1)\,.
\end{equation}
	
For the problem of equations of motion, we have to deal with the regularization of Poisson integrals, or iterated Poisson integrals (and their gradients), of the generic function $F^{(d)}$. The Poisson integral of $F^{(d)}$, in $d$ dimensions, is given by the Green's function for the Laplace operator,\footnote{Namely, $u(\mathbf{x})\equiv\tilde{k}\,r^{2-d}$, satisfying $\Delta u=-4\pi\,\delta^{(d)}(\mathbf{x})$, where $\tilde{k}$ is defined in \eqref{ktilde}.} 
\begin{equation}\label{Pdx}
		P^{(d)}({\mathbf{x}}')= \Delta^{-1} \left[ F^{(d)}({\mathbf{x}})
		\right] \equiv -\frac{\tilde{k}}{4\pi} \int
		\frac{\dd^d{\mathbf{x}}}{\vert{\mathbf{x}}-{\mathbf{x}}'\vert^{d-2}}
		\,F^{(d)}({\mathbf{x}})\,.
\end{equation}
We want to evaluate the Poisson integral at the point ${\mathbf{x}}' = \bm{y}_1$ where it is singular; this is quite easy in dimensional regularization, because the nice properties of analytic continuation in $d$ allow simply to get $[P^{(d)} ({\mathbf{x}}')]_{\mathbf{x}' = {\bm{y}}_1}$ by replacing ${\mathbf{x}}'$ by $\bm{y}_1$ in the integrand of \eqref{Pdx}. So we simply have
\begin{equation}
	P^{(d)}({\bm{y}}_1)=-\frac{\tilde{k}}{4\pi}
	\int\frac{\dd^d{\mathbf{x}}}{r_1^{d-2}}\,F^{(d)}({\mathbf{x}})\,.
	\label{Pd}
\end{equation}

When using dimensional regularization, remind that it is not possible in general to obtain exact expressions valid in any dimension $d$, but only in the limit where $\varepsilon\to 0$. Thus in general we need first to control the singular case $\varepsilon=0$ by applying first the Hadamard regularization.\footnote{Actually we must employ a crude version of Hadamard's regularization [in particular avoiding the non-distributivity issue, Eq.~\eqref{nondistr}] coined the ``pure-Hadamard-Schwartz'' regularization \citep{BDE04}.} In particular this permits to obtain all the contributions which do not pose a problem. Then we add the specific corrections (typically in the form of poles $1/\varepsilon$) due to the use of dimensional regularization. Thus the main step of our strategy consists of computing, in the limit $\varepsilon\to 0$, the \emph{difference} between the $d$-dimensional result and its Hadamard 3-dimensional counterpart, which will constitute the correction to the Hadamard result. The point is that such difference can be computed solely from the vicinity of the singular points, as the regular parts cancel out in the difference for $\varepsilon=0$.

The Hadamard counterpart of the Poisson integral \eqref{Pd} evaluated at 1 was denoted $(P)_1$ and obtained in Eq.~\eqref{pfPoisson1}. Denoting the difference between the two regularizations by means of the script letter $\mathcal{D}$, we pose (for what concerns the point 1)
\begin{equation}
	\mathcal{D}P_1\equiv P^{(d)}({\bm{y}}_1)-(P)_1\,.
	\label{DP1def}
\end{equation}
That is, $\mathcal{D}P_1$ is what we shall have to \emph{add} to the Hadamard-regularization result in order to get the $d$-dimensional result. With this method, we only compute the first two terms of the Laurent expansion of $\mathcal{D}P_1$ when $\varepsilon \rightarrow 0$, say $\mathcal{D}P_1 = a_{-1} \, \varepsilon^{-1} + a_0 + \calO (\varepsilon)$. This is all the information we need to clear up the ambiguities with Hadamard's regularization. The result depends solely on the singular coefficients in the expansion of $F^{(d)}$ around the singularities, Eq.~\eqref{Fexpd}. It reads \citep{BDE04}
\begin{align}\label{DP1}
	\mathcal{D}P_1 &= - \frac{\tilde{k}}{4\pi}\Biggl\{\sum_{q_0\leqslant
		q\leqslant q_1}\left(\frac{1}{q\varepsilon}+\ln
	r_1'-1\right)\bigl<\mathop
	{f}_1{}_{-2,q}^{(\varepsilon)}\bigr>\\
	&+ \sum_{q_0\leqslant q\leqslant
		q_1}\left(\frac{1}{(q+1)\varepsilon}+\ln s_2\right)
	\sum_{\ell=0}^{+\infty}\frac{(-)^\ell}{\ell!}\partial_L
	\left(\frac{1}{r_{12}^{1+\varepsilon}}\right)\bigl<
	n_2^L\mathop{f}_2{}_{-\ell-3,q}^{(\varepsilon)}\bigr> + \calO(\varepsilon)\Biggr\}\,.\nn
\end{align}
where the angular average is defined in $d$ dimensions as
\begin{equation}\label{IntAngul}
	\bigl<\mathop{f}_1{}_{p,q}^{(\varepsilon)}\bigr>
	\equiv\int\dd\Omega_1^{(d-1)} \mathop{f}_1{}_{p,q}^{(\varepsilon)}(\mathbf{n}_1)\,,
\end{equation}
with $\dd\Omega_1^{(d-1)}$ the solid angle element around the direction $\mathbf{n}_1$; recall that $\Omega^{(d-1)}=2\pi^{\frac{d}{2}}/\Gamma\left(\frac{d}{2}\right)$ is the volume of the unit sphere with $d-1$ dimensions. The first term of \eqref{DP1} corresponds to the contribution of the singularity 1, while the second term concerns the other singularity 2, where we has to perform a multipolar re-expansion of the factor $r_1^{2-d}$ around the point $\mathbf{y}_2$, which reads in STF form\footnote{The expansion is STF because $\Delta r^{2-d}=0$ in $d$ dimensions (in the sense of functions). The Appendix~B of \cite{BDE04} gives a compendium of $d$-dimensional formulae on STF expansions.}
\begin{equation}
	r_1^{2-d}=\sum_{\ell=0}^{+\infty}\frac{(-)^\ell}{\ell!}\partial_L
	\left(\frac{1}{r_{12}^{1+\varepsilon}}\right)r_2^\ell n_2^L\ .
	\label{r1ofr2}
\end{equation}
See Eq.~(4.24) of \cite{BDE04} for the similar formula concerning the gradient of the Poisson potential -- evidently crucial for the problem of equations of motion.

Note that \eqref{DP1} depends on the two ``constants'' $\ln r_1'$ and $\ln s_2$ appearing in \eqref{pfPoisson1} [remind that $s_1$ disappears from \eqref{pfPoisson1}]. One can check that these $\ln r_1'$ and $\ln s_2$ will exactly cancel out the same constants present in the Hadamard calculation, so that the dimensionally regularized equations of motion are finally free of these constants. However the coefficients ${}_1f_{p,q}^{(\varepsilon)}$ (and $1\leftrightarrow 2$) in $d$ dimensions depend on the length scale $\ell_0$ introduced in Eq.~\eqref{G}. Taking this dependence into account one can verify that the constants in \eqref{DP1} appear only in the combinations $\ln (r_1'/\ell_0)$ and $\ln (s_2/\ell_0)$.

Next we outline the way we obtain, starting from the computation of the previous difference, the equations of motion in dimensional regularization. As we already alluded, the common ``core'' of both dimensional and Hadamard regularizations is the pure-Hadamard-Schwartz (pHS) regularization \citep{BDE04}. This regularization is a specific, minimal Hadamard-type regularization of integrals, based on the partie finie integral \eqref{pfint}, together with a minimal treatment of ``contact'' terms, in which the definition \eqref{pfint} is applied separately to each of the elementary potentials $V$, $V_i$, etc. (and gradients) which enter the PN metric. Furthermore, the regularization of a product of these potentials is assumed to be distributive, i.e.,  $(FG)_1= (F)_1(G)_1$ in the case where $F$ and $G$ are given by such elementary potentials; this is thus in contrast with Eq.~\eqref{nondistr}. The pHS regularization also assumes the use of standard \cite{Schwartz} distributional derivatives. The interest of the pHS regularization is that the dimensional regularization is equal to it plus the ``difference''; see Eq.~\eqref{a1DimReg}. 

Therefore, summing up all contributions, the final result for the acceleration of body 1, regularized with dimensional regularization (DR), reads
\begin{equation}
	{\bm{a}}_1^\mathrm{DR} = {\bm{a}}_1^\mathrm{pHS} +
	\mathcal{D}{\bm{a}}_1\,,
	\label{a1DimReg}
\end{equation}
where $\mathcal{D}{\bm{a}}_1 \sim \sum \mathcal{D}P_1$ is the sum of all the individual differences of Poisson or Poisson-like integrals as computed in Eq.~\eqref{DP1} (together with the difference for gradients), and where ${\bm{a}}_1^\mathrm{pHS}$ is the acceleration obtained with the pHS ``core'' regularization described above. As obtained in \eqref{DP1}, the difference makes a contribution only when a term generates a pole $\sim 1/\varepsilon$, in which case the dimensional regularization adds an extra contribution, made of the pole and the finite part associated with the pole -- we consistently neglect all terms $\calO(\varepsilon)$. One must then be especially wary of combinations of terms whose pole parts finally cancel but whose dimensionally regularized finite parts generally do not; these must be evaluated with care. The poles $\sim 1/\varepsilon$ arise at order 3PN in the equations of motion in harmonic coordinates.\footnote{The situation in harmonic coordinates is to be contrasted with the calculation in ADM-type coordinates, where it was shown that the pole parts directly cancel out in the total Hamiltonian at 3PN order \citep{DJSdim}.} Then we have

\begin{theorem}
The pole part $\propto 1/\varepsilon$ of the dimensional-regularized acceleration \eqref{a1DimReg} can be re-absorbed (i.e. renormalized) into some shifts of the two ``bare'' world-lines: $\bm{y}_1 \rightarrow \bm{y}_1+\bm{\xi}_1$ and $\bm{y}_2 \rightarrow \bm{y}_2+\bm{\xi}_2$ (with $\bm{\xi}_{a} \sim 1/\varepsilon$ say), so that the result, expressed in terms of the ``dressed'' quantities, is finite:
\begin{equation}\label{a1renorm}
	{\bm{a}}_1 = \lim_{\varepsilon\rightarrow 0}
	\Bigl[{\bm{a}}_1^\mathrm{DR} + \delta_{\bm{\xi}}{\bm{a}}_1\Bigr]\,,
\end{equation}
with $\delta_{\bm{\xi}}{\bm{a}}_1$ denoting the contribution of the shifts. Furthermore the renormalized acceleration \eqref{a1renorm} is equivalent to the Hadamard-regularized acceleration if and only if the ambiguous parameters are uniquely fixed.
\label{th:poleDR}
\end{theorem}
All the previous discussion concerns the ultra-violet (UV) type divergences appearing in the equations of motion. However starting at the 4PN order the equations of motion also develop infra-red (IR) type divergences. Those too have to be treated with the dimensional regularization. We give more details on the DR of the IR divergences in Sect.~\ref{sec:IRreg}.


\subsubsection{Dimensional regularization of the radiation field}
\label{sec:DRrad}

We now address the similar problem concerning the binary's radiation field -- say, to 3PN or 4PN order beyond the Einstein quadrupole formalism. As for the equations of motion, when using the Hadamard regularization for the calculation of the binary's multipole moments, some ambiguity parameters arise \citep{BIJ02, BI04mult}. Again, those are fixed using the dimensional regularization.

Let us first review the multipole expansion in $d$ dimensions. To this end, we consider again the scalar wave equation \eqref{waveeq} in $D$-dimensions, but in which the source $S=\Box h$ has a \textit{compact} spatial support. The multipole expansion $\mathcal{M}(h)$ is the retarded homogeneous solution of the wave equation, i.e. $\Box \mathcal{M}(h) = 0$, valid outside the compact support of the source $S$. It reads
\begin{equation}\label{phimultG}
	\mathcal{M}(h)(\mathbf{x},t) = -\frac{1}{4\pi}
	\sum_{\ell=0}^{+\infty}\frac{(-)^\ell}{\ell!}
	\,\partial_L\biggl[\frac{\tilde{k}}{r^{d-2}}\int_1^{+\infty}\dd y\,\gamma_{\frac{1-d}{2}}(y)\,\mathcal{F}_L(t-y r/c)
	\biggr]\,,
\end{equation}
where $\gamma_s(z)$ is part of the Green's function \eqref{Gret}--\eqref{gammas}, and the multipole-moment functions $\mathcal{F}_L$ are STF (in $d$-dimensions) and related to the compact-support source term by
\begin{align}\label{FLu}
	\mathcal{F}_L(u) = \int
	\dd^d\mathbf{x}
	\,\hat{x}_L \,S_{[\ell]}(\mathbf{x},u)\,.
\end{align}
We denote $\hat{x}_L$ the STF product of spatial vectors in $d$ dimensions, and employ the convenient $\ell$-dependent weighted time average 
\begin{subequations}\label{Sldef}
\begin{align}
	S_{[\ell]}(\mathbf{x},u) &\equiv \int_{-1}^1 \dd z
	\,\delta_\ell^{(\varepsilon)} (z)
	\,S(\mathbf{x},u+z r/c)\,,\\
	\text{where}\quad\delta_\ell^{(\varepsilon)} (z) &=	\frac{\Gamma\left(\ell+\frac{3}{2}+\frac{\varepsilon}{2}\right)}{
		\Gamma\left(\frac{1}{2}\right)\Gamma
		\left(\ell+1+\frac{\varepsilon}{2}\right)}
	\,(1-z^2)^{\ell+\frac{\varepsilon}{2}}\,,
\end{align}
\end{subequations}
satisfies the normalization condition $\int_{-1}^{1} \dd z\,\delta_\ell^{(\varepsilon)}(z) = 1$, and is a simple $d$-dimensional generalization of Eq.~\eqref{deltal}. The formal PN expansion of Eq.~\eqref{Sldef}, generalizing the 3 dimensional result \eqref{intdeltaexp}, reads
\begin{align}\label{PNform}
		S_{[\ell]}(\mathbf{x},u) &= \sum_{k=0}^{+\infty}\frac{1}{2^{2k}k!}\frac{\Gamma\left(\frac{d}{2}+\ell\right)}{\Gamma\left(\frac{d}{2}+\ell+k\right)} \left(\frac{r}{c}\frac{\partial}{\partial t}\right)^{2k}S(\mathbf{x},u)\,.
\end{align}

In the second step we want to generalize the formalism of Sect.~\ref{sec:sourcemoments}, i.e. solving for the multipole expansion of a source term with \textit{non-compact spatial support}, given by the stress-energy matter plus gravitation pseudo tensor $\tau\ab$ of the Einstein field equations in harmonic coordinates \eqref{EFEd}. We obtain 
\begin{align}\label{multSTFd}
		\mathcal{M}(h^{\alpha\beta}) &= u\ab - \frac{4G}{c^4}
		\sum_{\ell=0}^{+\infty}\frac{(-)^\ell}{\ell!}
	\,\partial_L\biggl[\frac{\tilde{k}}{r^{d-2}}\int_1^{+\infty}\dd y\,\gamma_{\frac{1-d}{2}}(y)\,\mathcal{F}\ab_L(t-y r/c)
\biggr]\,.
\end{align}
The first term $u\ab$ takes the same form as the corresponding piece in 3 dimensions, given by Eq.~\eqref{defuab}. In the second term the multipole-moment function $\mathcal{F}_L\ab(u)$ is given by the integral extending over the PN expansion of the pseudo-tensor $\overline{\tau}\ab$ \citep{BDEI05dr,HFB21}:
\begin{align}\label{calFd}
	\mathcal{F}_L\ab(u)=\mathop{\mathrm{FP}}_{B=0}\int
	\dd^d\mathbf{x}\,\widetilde{r}^B\,
	\hat{x}_L\,\overline{\tau}_{[\ell]}\ab(\mathbf{x},u)\,.
\end{align}
Indeed, we can follow exactly the same procedure as in 3 dimensions (see Sect.~\ref{sec:sourcemoments}), which is appropriate for the non-compact support source term $\overline{\tau}\ab$. In particular, we introduced the regularization factor $\widetilde{r}^B$ [see Eq.~\eqref{regfactor}] and the finite part when $B\to 0$ in both terms of Eq.~\eqref{multSTFd}. One may wonder why it is necessary to introduce the FP when $B\to 0$ since we have at our disposal the dimensional regularization parameter $d=3+\varepsilon$ which should play the role of IR regularization. Actually, applying the FP process ``on the top'' of dimensional regularization changes the nature of the limit $B\to 0$. While this limit is singular in 3 dimensions, it turns out to be finite in any generic $d$ dimensions. We shall see an example in Sect.~\ref{sec:IRreg} where it is necessary to compute first the finite limit when $B\to 0$, and only then to apply the standard dimensional regularization, of course keeping consistently all poles $\propto 1/\varepsilon$. We call this variant of dimensional regularization the ``$B\varepsilon$'' regularization.

With the result \eqref{calFd} in hand, we can derive the expressions of the irreducible source moments in $d$ dimensions, generalizing those computed in Theorem \ref{th:sourcemoments}. Extending the definitions \eqref{Sigma} we pose
\begin{align}\label{Sigmad}
	\overline{\Sigma} = \frac{2}{d-1}\,
	\frac{(d-2)\overline{\tau}^{00}+\overline{\tau}^{ii}}{c^2}\,,\qquad
	\overline{\Sigma}^i = \frac{\overline{\tau}^{i0}}{c}\,,\qquad
	\overline{\Sigma}^{ij} = \overline{\tau}^{ij}\,,
\end{align}
where we recall that $\overline{\tau}^{\alpha\beta}$ is the PN expansion of the pseudo-tensor $\tau^{\alpha\beta}$, and obtain the mass-type moment as \citep{BDEI05dr}
\begin{align}\label{ILd}
	\dI_L &= \frac{d-1}{2(d-2)}\mathop{\mathrm{FP}}_{B=0} \int
	\dd^d\mathbf{x}\,\widetilde{r}^B
	\biggl\{\hat{x}^L \overline{\Sigma}_{[\ell]}-\frac{4(d+2\ell-2)}
	{c^2(d+\ell-2)(d+2\ell)}\,\hat{x}^{iL}\,
	\dot{\overline{\Sigma}}^{i}_{[\ell+1]}\nn\\
	&\qquad\qquad\qquad~~ +\frac{2(d+2\ell-2)}
	{c^4(d+\ell-1)(d+\ell-2)(d+2\ell+2)}
	\,\hat{x}^{ijL}\,
	\ddot{\overline{\Sigma}}^{ij}_{[\ell+2]}\nn\\
	&\qquad\qquad\qquad~ -
	\frac{4(d-3)(d+2\ell-2)}{c^2(d-1)(d+\ell-2)(d+2\ell)}
	B \,\hat{x}^{iL}\,\frac{x^j}{r^2}
	\,\overline{\Sigma}^{ij}_{[\ell+1]}
	\biggr\}\,.
\end{align}
At Newtonian
order, this expression reduces to the standard result $\dI_L = \int \dd^d\mathbf{x}\,\rho\,\hat{x}_L+\calO(1/c^{2})$ with $\rho=T^{00}/c^2$ denoting the usual Newtonian density. The post-Newtonian corrections can be computed explicitly using Eq.~\eqref{PNform}. The last term in \eqref{ILd} does not exist in 3 dimensions and plays no role in practical calculations of the mass moment up to 4PN order. 

The mass moment is just STF in all its indices so that the symmetry is given by a symmetric Young tableau (with the multi-index $L=i_1\cdots i_\ell$)
\begin{align}\label{ILyoung}
	\dI_L=\ytableausetup{boxsize=1.25em,textmode}
	\ytableaushort{{\scriptsize $i_\ell$} {...} {\scriptsize $i_1$}}\,.
\end{align}
In $d$ dimensions, since we do not dispose of the usual notion of Levi-Civita symbol, it is more tricky to define the proper generalization of the current-type moment. One has \citep{HFB21}
\begin{align}\label{JiLd}
\dJ_{i|L} = \FP \,\underset{i|L}{\text{Sym}} \int
	\dd^d \mathbf{x} 
	\,\widetilde{r}^{B} \biggl\{-2 \hat{x}^L \overline{\Sigma}^{i}_{[\ell]} + \frac{2 (2\ell +d-2)}{c^2(\ell+d-1)(2\ell+d)}
	\hat{x}^{aL} \dot{\overline{\Sigma}}^{ia}_{[\ell+1]} \biggr\}\,,
\end{align}
where we introduce the specific notation that allows reconstructing the symmetries of the object $\dJ_{i|L}\equiv\dJ_{i|i_1\cdots i_\ell}$:
\begin{equation}\label{SymJkji}
	\underset{i|L}{\text{Sym}} \equiv \mathcal{A}_{i i_{\ell}} \underset{iL}{\text{TF}}\; \underset{L}{\text{STF}}\,,
\end{equation}
where STF is the standard symmetric-trace-free projection with respect to the indices $L$, where TF means removing the traces (but not symmetrizing) of the resulting object with respect to the indices $iL$, and $\mathcal{A}_{i i_{\ell}}$ means the anti-symmetrization with respect to the pair of indices $i i_{\ell}$ (with the factor $\frac{1}{2}$ included). Thus, $J_{i\vert L}$ is trace-free, STF with respect to $L-1=i_1\cdots i_{\ell-1}$ and anti-symmetric with respect to the pair $i_\ell i$. Finally, besides $\dI_L$ and $\dJ_{i|L}$, there exists in $d$ dimensions an additional type of irreducible multipole moment, denoted $\dK_{ij\vert L}\equiv\dK_{ij|i_1\cdots i_\ell}$ and given by Eq.~(2.34) of \cite{HFB21}. The symmetries of the moments $\dJ_{i|L}$ and $\dK_{ij\vert L}$ are given by the mixed Young tableaux
\begin{align}\label{JLKLyoung}
	\dJ_{i\vert L}=
	\ytableaushort{{\scriptsize $i_\ell$} {\scriptsize
			$\,i_{\scalebox{0.6}{\text{$\ell\! -\! 1$}}}$} {...} {\scriptsize 
			$i_1$}, {\scriptsize $i$}}~\,,\qquad K_{ij\vert L} = \ytableaushort{{\scriptsize $i_\ell$} {\scriptsize
			$\,i_{\scalebox{0.6}{\text{$\ell\!- \! 1$}}}$} {\scriptsize
			$i_{\scalebox{0.6}{\text{$\ell\!- \! 2$}}}$} {...} {\scriptsize $i_1$},
		{\scriptsize $j$} 
		{\scriptsize $i$}}~\,,
\end{align}
with the convention that the indices are symmetrized over lines \textit{before} being antisymmetrized over columns. However, the tensor $\dK_{ij|L}$ actually vanishes in $3$ dimensions. This can be checked by counting the number of its independent components, which turns out to be proportional to $d-3$. In fact, the tensor $\dK_{ij|L}$ for $\ell=2$ has exactly the same trace-free property, symmetries, and number of independent components as the Weyl tensor. It plays no role with dimensional regularization, which is always performed in the limit $d\to 3$, and shall henceforth be ignored. 

The ordinary STF mass-type moment in 3 dimensions is simply recovered as the limit $\dI_L^{(d=3)} = \lim_{d\to 3} \dI_L$ (after the UV type renormalization). In particular, we observe that this limit removes the last term in Eq.~\eqref{ILd}. For the ordinary STF current-type moment in 3 dimensions we have
\begin{align}\label{lim3d}
	\lim_{d\to 3} \dJ_{i\vert L} = \epsilon_{ii_{\ell}a} \,\dJ^{(d=3)}_{aL-1} \quad\Longleftrightarrow\quad \dJ^{(d=3)}_{L} = \frac{1}{2}
	\,\epsilon_{ab(i_\ell} \,\lim_{d\to 3}\, \dJ_{\underline{a}|\underline{b} L-1)} \,,
\end{align}
where $\epsilon_{abi}$ is the usual Levi-Civita symbol in 3 dimensions and underlined indices are excluded from symmetrization denoted by parenthesis. Notice that $\dJ^{(d=3)}_L$, as recovered from \eqref{lim3d}, not only is symmetric in its indices $L$ but is also automatically trace-free. 

As in the problem of equations of motion, the strategy -- since the moments \eqref{ILd} and \eqref{JiLd} cannot be computed in closed form for any $d$ -- is to use the expressions of the $d$-dimensional moments to evaluate the difference between the dimensional-regularization result and the previous Hadamard (or pHS) one \citep{BDEI04, BDEI05dr}. The ambiguities arise from the integration regions near the particles, that give rise to poles $\propto 1/\varepsilon$, corresponding to logarithmic ultra-violet (UV) divergences in 3 dimensions.\footnote{For the moment we neglect IR divergences, which will be discussed in Sect.~\ref{sec:IRreg}.} Since the difference is evaluated neglecting $\calO(\varepsilon)$, the compact-support terms in the moments (proportional to the matter source densities $\sigma$, $\sigma_i$ and $\sigma_{ij}$) do not contribute. We are therefore left with evaluating the difference linked with the computation of the \emph{non-compact} terms in the moments \eqref{ILd} and \eqref{JiLd} near the singularities that produce poles in $d$ dimensions.

A generic non-compact support term in 3 dimensions will be given by a function $F(\mathbf{x})$ smooth except at the source points $\bm{y}_1$ and $\bm{y}_2$ around which it admits a power-like singular expansion of the type \eqref{Fexp}. The generic contribution in the multipole moments in 3 dimensions is defined by the Hadamard partie finie prescription, given either by \eqref{pfint} or \eqref{pfint2}, say  
\begin{equation}\label{intH}
H = \Pf \int \dd^3\mathbf{x} \, F(\mathbf{x}) \,,
\end{equation}
which is just a function of time. In $d$ dimensions the corresponding function $F^{(d)}(\mathbf{x})$ admits a generalized singular expansion of the type \eqref{Fexpd}; since $\lim_{d\to 3}F^{(d)}(\mathbf{x})=F(\mathbf{x})$ the singular coefficients in the expansion satisfy the constraint \eqref{constr}. In practice, the various coefficients ${}_1f_{p,q}^{(\varepsilon)}$ in the expansion are computed by specializing the general expressions of the non-linear retarded potentials $V$, $\hat{W}_{ij}$, etc. [see Eqs. \eqref{pot1PNd}], to point particles in $d$ dimensions. Now the analogue of the term \eqref{intH} in $d$ dimensions reads
\begin{equation}\label{intHd}
	H^{(d)} = \int \dd^d\mathbf{x}\,F^{(d)}(\mathbf{x})\,.
\end{equation}
Given the results of the two regularizations we compute the difference -- which will have to be added to the 3-dimensional result -- namely
\begin{equation}\label{diffH}
	\mathcal{D}H \equiv H^{(d)} - H\,,
\end{equation}
and as in Eq.~\eqref{DP1} we compute this difference keeping the pole part $\propto\varepsilon^{-1}$ and the finite part, but neglecting terms of order $\calO(\varepsilon)$. To this order the difference depends only on the UV behaviour of the integrands, and can therefore be computed ``locally'', i.e., in the vicinity of the particles, when $r_1 \rightarrow 0$ and $r_2 \rightarrow 0$. The outcome is \citep{BDEI05dr}
\begin{equation}\label{DHres}
	\mathcal{D}H =
	\frac{1}{\varepsilon}\!\sum_{q_0\leqslant
		q\leqslant q_1}
	\biggl[\frac{1}{q+1}+\varepsilon \ln s_1\biggr]
	\bigl<\mathop{f}_1{}_{-3,q}^{(\varepsilon)}\bigr> +
	1\leftrightarrow 2
	+ \calO(\varepsilon)\,,
\end{equation}
where the spherical average in $d$ dimensions is defined by \eqref{IntAngul}. This result depends on the two Hadamard regularization scales $s_1$ and $s_2$, and on the scale $\ell_0$ belonging to dimensional regularization and defined by \eqref{G}; taking into account the dependence of ${}_1f_{p,q}^{(\varepsilon)}$ upon $\ell_0$, the Hadamard scales enter \eqref{DHres} only through the combinations $\ln(s_1/\ell_0)$ and $\ln(s_2/\ell_0)$.

Having controlled the difference \eqref{DHres}, the dimensional regularization of the multipole moments is obtained as the sum of the pHS part and of that difference. Consider the case of the mass quadrupole moment $\dI_{ij}$ since this is that moments which requests the highest approximation. Denoting $\mathcal{D}\dI_{ij}\sim\sum\mathcal{D}H$ the sum of all the differences we obtain 
\begin{equation}\label{IijDR}
	\dI_{ij}^{\mathrm{DR}} = \dI_{ij}^{\mathrm{pHS}} +
	\mathcal{D}\dI_{ij}\,,
\end{equation}
which now no longer depends on the Hadamard scales $s_1$ and $s_2$; however it contains poles $\propto 1/\varepsilon$ coming from the differences \eqref{DHres}. At this stage all elements are in place to prove \citep{BDEI04, BDEI05dr}:

\begin{theorem}
The pole part $\propto 1/\varepsilon$ of the dimensional-regularized quadrupole moment \eqref{IijDR} can be renormalized into the same shifts of the world-lines as for the particle's accelerations [Eq.~\eqref{a1renorm} in Theorem \ref{th:poleDR}]:
\begin{equation}\label{Iijrenorm}
	\dI_{ij} = \lim_{\varepsilon\rightarrow 0}
	\Bigl[\dI_{ij}^{\mathrm{DR}} + \delta_{\bm{\xi}}\dI_{ij}\Bigr]\,,
\end{equation}
with $\delta_{\bm{\xi}}\dI_{ij}$ denoting the contribution of the shifts. 
\label{th:poleDRquad}
\end{theorem}
To 3PN order it has been checked that the finite renormalized quadrupole moment \eqref{Iijrenorm} is physically equivalent to the Hadamard-regularized one \citep{BIJ02,BI04mult} if and only if the ambiguity parameters in the Hadamard calculation take uniquely determined values. Of course the procedure is then extended to 4PN order.


\subsubsection{Infra-red divergences and radiative modes}
\label{sec:IRreg}

Besides the UV divergences described in the previous Sects.~\ref{sec:DReom} and \ref{sec:DRrad}, and which arises at 3PN order, there are also IR divergences, due to the behaviour of spatial integrals at infinity, which start to appear at 4PN order \citep{DJS14,GLPR16,BBBFMc,FS4PN}. The IR divergences are present in both the multipole moments of the radiation field and in the local equations of motion (more precisely the Fokker Lagrangian introduced in Sect.~\ref{sec:Fokker}).  

Consider a non-compact support term $F(\mathbf{x})$ which typically contributes to the binary's multipole moments or the Lagrangian, and is obtained by direct post-Newtonian iteration of the metric in 3 dimensions. The divergences occur from the expansion of $F$ when $r\to +\infty$, which is of the form (for any $\mathcal{N}\in\mathbb{N}$)
\begin{equation}\label{Fdev0}
	F(\mathbf{x}) = \sum_{p_0 \leqslant p
		\leqslant \mathcal{N}} \frac{1}{r^p}\,\varphi_p(\mathbf{n})
	+ o\left(\frac{1}{r^\mathcal{N}}\right)\,.
\end{equation}
The coefficients $\varphi_p$ depend on the unit direction $\mathbf{n}=\mathbf{x}/r$ and on $p$, with minimal value $p_0\in\mathbb{Z}$ corresponding to growing divergent behaviour with the distance. In $d$ dimensions the same PN iteration yields $F^{(d)}(\mathbf{x})$, which admits a more general expansion of the form
\begin{equation}\label{Fdevddim}
	F^{(d)}(\mathbf{x}) = \sum_{\substack{p_0 \leqslant p
			\leqslant \mathcal{N} \\ q_0 \leqslant q\leqslant q_1}}  \,\frac{1}{r^{p + q\varepsilon}}\,\varphi^{(\varepsilon)}_{p,q}(\mathbf{n})
	+ o\left(\frac{1}{r^\mathcal{N}}\right)\,.
\end{equation}
Comparing \eqref{Fdev0} and \eqref{Fdevddim} in the limit $\varepsilon\to 0$ we have
\begin{equation}\label{relcoeff}
	\sum_{q=q_0}^{q_1}
	\varphi^{(\varepsilon=0)}_{p,q}(\mathbf{n}) = \varphi_{p}(\mathbf{n})\,.
\end{equation}
With such behaviour in hand we want to regularize the three-dimensional divergent integral $K^{\text{(div)}} = \int \dd^3 \mathbf{x}\,F(\mathbf{x})$. Since we are dealing with the IR bound at infinity, we can consider the integration domain $r=\vert\mathbf{x}\vert>\mathcal{R}$, where $\mathcal{R}$ is some sufficiently large radius, thus $K_\mathcal{R}^{\text{(div)}} = \int_{r > \mathcal{R}}\dd^3 \mathbf{x}\,F(\mathbf{x})$. 

The Hadamard regularization is a good starting point, and we employ the partie finie integral, in the form based on analytic continuation with regularization factor $\widetilde{r}^B$, like in Eqs. \eqref{sourcemoments}. Thus we define
\begin{equation}\label{KcalR}
	K_\mathcal{R} = \FP \int_{r>\mathcal{R}} \dd^{3}\mathbf{x}
	\,\widetilde{r}^B F(\mathbf{x})\,.
\end{equation}
A straightforward calculation, plugging \eqref{Fdev0} into \eqref{KcalR}, yields the Hadamard-regularized version of the integral as
\begin{equation}\label{KHR}
K_\mathcal{R} = - \sum_{p\not=
		3}\frac{\mathcal{R}^{3-p}}{3-p}\,\bigl<\varphi_{p}\bigr>_3
	- \ln\left(\frac{\mathcal{R}}{r_0}\right)\,\bigl<\varphi_{3}\bigr>_3\,,
\end{equation}
where $\bigl<\varphi_{p}\bigr>_3 \equiv\int\dd\Omega \,\varphi_{p}(\mathbf{n})$ is the angular average in 3 dimensions. As we see the crucial coefficient in the expansion \eqref{Fdev0} is that for $p=3$; it generates a logarithmic divergence of the original integral. On the other hand, the dimensional regularization prescription, to be considered in the sense of analytic continuation in $d\in\mathbb{C}$, reads
\begin{equation}\label{IDRdef}
	K^{(d)}_\mathcal{R} = \int_{r>\mathcal{R}}
	\dd^{d}\mathbf{x}\,F^{(d)}(\mathbf{x}) \,.
\end{equation}
Inserting the expansion \eqref{Fdevddim}, keeping only the pole $\propto 1/\varepsilon$ followed by the finite part $\propto \varepsilon^0$, and using also Eq.~\eqref{relcoeff}, we obtain
\begin{equation}\label{KDR}
K^{(d)}_\mathcal{R} = - \sum_{p\not=
	3}\frac{\mathcal{R}^{3-p}}{3-p}\,\bigl<\varphi_{p}\bigr>_3 
	+ \sum_{q_0\leqslant q\leqslant
		q_1} \left[\frac{1}{(q-1)\varepsilon} -
	\ln\mathcal{R}\right]\bigl<\varphi_{3,q}^{(\varepsilon)}\bigr>
	+ \calO\left(\varepsilon\right)\,,
\end{equation}
where the angular integration in the second term, because of the presence of the pole, is to be performed in $d$ dimensions following \eqref{IntAngul}. \textit{A priori} the result \eqref{KDR} could contain terms that diverge at infinity. These terms correspond to the value $q=1$ but do not appear up to the 4PN level. One can check that the result \eqref{KDR} comes out the same with the $B\varepsilon$ regularization considered below.

Finally, exactly like in the UV case (Sects.~\ref{sec:DReom} and \ref{sec:DRrad}), we add the difference between the two prescriptions, i.e.
\begin{equation}\label{diffK}
\mathcal{D}K \equiv
K^{(d)}_\mathcal{R} - K_\mathcal{R}\,.
\end{equation}
The first term in \eqref{KHR} and \eqref{KDR} is the same and thus cancels out in the difference, so we obtain 
\begin{equation}\label{resdiff}
	\mathcal{D}K = \sum_{q_0\leqslant q\leqslant
		q_1} \left[\frac{1}{(q-1)\varepsilon} + \ln r_0\right]\bigl<\varphi_{3,q}^{(\varepsilon)}\bigr>
	+ \calO\left(\varepsilon\right)\,.
\end{equation}
As expected, the scale $\mathcal{R}$ has disappeared from the difference. Keeping track of the dependence over $\ell_0$ of the singular coefficients in \eqref{Fdevddim}, the remaining logarithm $\ln r_0$ is actually $\ln(r_0/\ell_0)$. The poles $\propto 1/\varepsilon$ appearing in \eqref{resdiff} can be qualified as ``IR poles''; they appear in both the expression of the (source) multipole moments and in the Fokker Lagrangian for the conservative equations of motion.\footnote{Unlike in the EFT approach \citep{Portorevue,PR17} one does not distinguish between IR poles, say $\sim 1/\varepsilon_\text{IR}$, and UV poles, $\sim 1/\varepsilon_\text{UV}$; here $\varepsilon$ is simply related to the space dimension by $\varepsilon=d-3$.} 

We now review that the poles in the ``difference'' \eqref{resdiff} are dropped when we add the ``radiative modes'' associated with tail (and alike non-linear effects) propagation in the far zone. To see this we start again with the generic d'Alembertian equation in $d$ dimensions, Eq.~\eqref{waveeq}, which we now write as $\Box h = N$ to emphasize that the source $N(\mathbf{x},t)$ will be a non-linear term in the vacuum Einstein field equations -- say a typical term in Eq.~\eqref{Nab}. Furthermore since we are in vacuum outside the source, this term will be in the form of a multipole expansion, for instance a quadratic product of two linear homogeneous retarded waves \eqref{phimultG}. The retarded solution is given by \eqref{retsol} which we recopy here: 
\begin{align}\label{retsol2}
	h = - \frac{\tilde{k}}{4\pi}
	\int_1^{+\infty} \dd z \,\gamma_{\frac{1-d}{2}}(z) \int
	\dd^d\mathbf{x}'
	\,\frac{N(\mathbf{x}',t-z\vert\mathbf{x}-\mathbf{x}'\vert/c)}
	{\vert\mathbf{x}-\mathbf{x}'\vert^{d-2}}\,.
\end{align}
We split the spatial integration over the source point $\mathbf{x}'$ in \eqref{retsol2} into an inner domain corresponding to $r'<r$ (where $r'\equiv\vert\mathbf{x}'\vert$ and $r\equiv\vert\mathbf{x}\vert$), and the outer domain for which $r'>r$. For the computation of the radiation modes in the conservative Lagrangian and equations of motion, the relevant integration to be considered is that of the outer domain $r'>r$, i.e.
\begin{align}\label{retsol>}
	h^{>} = - \frac{\tilde{k}}{4\pi}
	\int_1^{+\infty} \!\!\dd z \,\gamma_{\frac{1-d}{2}}(z) \int_{r'>r}
	\dd^d\mathbf{x}'
	\,\frac{N(\mathbf{x}',t-z\vert\mathbf{x}-\mathbf{x}'\vert/c)}
	{\vert\mathbf{x}-\mathbf{x}'\vert^{d-2}}\,.
\end{align}
Within the integration domain $r'>r$ we are entitled to replace the integrand by its formal expansion when $r\to 0$. Computing the Taylor expansion when $r\to 0$, and transforming it to STF form, we obtain 
\begin{align}\label{multexp}
	\frac{N(\mathbf{x}',t-z\vert\mathbf{x}-\mathbf{x}'\vert/c)}{\vert\mathbf{x}-\mathbf{x}'\vert^{d-2}} =\sum_{p=0}^{+\infty} \frac{(-)^p}{p!}  \sum_{j=0}^{+\infty}
	\Delta^{-j} \hat{x}_P \,\hat{\partial}'_P\Delta'^j\!\left[\frac{N(\mathbf{y},t-z r'/c)}{r'^{d-2}}\right]_{\mathbf{y}=\mathbf{x}'}\,,
\end{align}
where $\hat{\partial}'_P$ is the STF product of $p$ partial derivatives, $\partial'_P=\partial'_{i_1}\cdots\partial'_{i_p}$ with $\partial'_i=\partial/\partial x'^i$, and on the last factor $\mathbf{y}=\mathbf{x}'$ is set after differentiation. We also employ the useful short-hand notation 
%
\begin{equation}\label{Deltaj}
	\Delta^{-j} \hat{x}_P =
	\frac{\Gamma(p+\frac{d}{2})}{\Gamma(p+j+\frac{d}{2})} \,\frac{r^{2j}
		\hat{x}_P}{2^{2j} j!}\,,
\end{equation}
such a notation being motivated by the fact that $\Delta(\Delta^{-j} \hat{x}_P)=\Delta^{-j+1} \hat{x}_P$ in $d$ dimensions. Inserting \eqref{multexp} into \eqref{retsol>} we obtain
\begin{align}\label{varphiexp0}
h^{>} &= - \frac{\tilde{k}}{4\pi}
	\sum_{p=0}^{+\infty} \frac{(-)^p}{p!} \sum_{j=0}^{+\infty} \frac{\Delta^{-j}
		\hat{x}_P}{c^{2j}} \!\int_1^{+\infty} \!\!\!\!\dd z \,\gamma_{\frac{1-d}{2}}(z)\nn\\&\qquad\qquad\qquad\qquad\times\int_{r'>r}
	\dd^d\mathbf{x}'\,\hat{\partial}'_P\!\left[\frac{N^{(2j)}(\mathbf{y},t
		- z r'/c)}{{r'}^{d-2}}\right]_{\mathbf{y}=\mathbf{x}'}\,.
\end{align}
Note that the iterated Laplace operator $\Delta'^j$ in \eqref{multexp} has transformed into $2j$ time derivatives of the source as indicated with the superscript $(2j)$.

Next we consider the above general formula in the case where the source term has a definite multipolarity $\ell$, i.e. is of the type $N(\mathbf{x},t) = \hat{n}_L \,N(r,t)$. Clearly, after summing up all the multipolar pieces, while assuming the convergence of the multipolar series, there is no restriction on the generality of the solution. We shall denote the corresponding solution by $h^{>}_L$ since it depends on the multi-index $L=i_1\cdots i_\ell$. In that case we can explicitly perform the angular integration in $d$ dimensions [see Eq. (B23b) of \citealt{BDE04}] and obtain, after simplification by means of a series of integrations by parts over the variable $z$,
\begin{equation}\label{varphiexp2}
	h^{>}_L = \frac{-1}{d+2\ell-2}
	\sum_{j=0}^{+\infty} \frac{\Delta^{-j}
		\hat{x}_L}{c^{2j}}  \int_1^{+\infty} \!\!\!\!\dd z
	\,\gamma_{\frac{1-d}{2}-\ell}(z) \int_r^{+\infty} \!\!\!\!\dd
	r'\,{r'}^{-\ell+1} N^{(2j)}\Bigl(r',t-\frac{z r'}{c}\Bigr)\,.
\end{equation}
Note that the index of the $\gamma$-function is $\frac{1-d}{2}-\ell$ instead of $\frac{1-d}{2}$ previously. Now it was shown by \cite{BBBFMc} that the radiative modes needed for insertion into the Fokker Lagrangian describing the conservative dynamics, are given by the outer domain integral \eqref{varphiexp2}, but in which the radial integral from $r$ to $+\infty$ is replaced by the integral from 0 to $+\infty$. Finally the general form of the radiative modes is
\begin{equation}\label{radmodeform}
	h^\text{rad}_L(\mathbf{x},t) = \sum_{j=0}^{+\infty} \frac{\Delta^{-j}
		\hat{x}_L}{c^{2j}} \,\mathcal{N}^{(2j)}(t)\,,	
\end{equation}
where the function $\mathcal{N}(t)$ is related to the source term of the (vacuum) Einstein field equations by
\begin{equation}\label{radmode}
	\mathcal{N}(t) = \frac{-1}{d+2\ell-2} \int_1^{+\infty} \!\!\!\!\dd z
	\,\gamma_{\frac{1-d}{2}-\ell}(z) \FP\int_0^{+\infty} \!\!\!\!\dd
	r' \,\widetilde{r}'^B{r'}^{-\ell+1} N\Bigl(r',t-\frac{z r'}{c}\Bigr)\,.
\end{equation}
From the above form one can directly check that it is an homogeneous solution in $d$ dimensions, i.e. $\Box h^\text{rad}_L = 0$. Furthermore, that solution is manifestly regular when $r\to 0$, so it must be identified with an homogeneous \textit{anti-symmetric} solution of the wave equation in $d$ dimensions, of the type half-retarded minus half-advanced. As such this homogeneous solution can be directly imported into the near zone -- via the matching equation, see Eq.~\eqref{matchingeq} -- where it contributes to the conservative dynamics described by the Fokker Lagrangian (or the EFT action).

Important in this approach, the ``$B\varepsilon$'' regularization scheme \citep{BBBFMc, MBBF17, LHBF22} is used in Eq.~\eqref{radmode}. Thus the regularization factor $\widetilde{r}'^B=(r'/r_0)^B$ is inserted first in order to treat the divergence of the multipole expansion, see Eq.~\eqref{ungen}. As already mentioned after Eq.~\eqref{calFd}, this procedure is used even in $d$ dimensions, the reason being that for the radiative modes obtained in Eq.~\eqref{radmode}, the direct limit $\varepsilon\to 0$ is ill defined and must be ``protected'' by the finite part regularization when $B\to 0$. Following the $B\varepsilon$ regularization one considers \textit{first} the finite part at $B=0$ in a generic $d$ dimension, and, \textit{second}, the limit $\varepsilon\to 0$ is applied, keeping all the occuring poles as usual. With this method the limit $B\to 0$ is proved to be finite (no poles $\propto 1/B$), so the method actually reduces to the usual dimensional regularization.

We apply the above formulas to a quadratic tail interaction, made of the interaction between the constant mass $\dM$ and a varying multipole moment $\dM_L$ or $\dS_L$. In this case the generic form of the source term is
\begin{equation}\label{Ntail}
	N(r,t) = r^{-k-2\varepsilon} \,\int_1^{+\infty} \dd y
	\,y^p\,\gamma_{-1-\frac{\varepsilon}{2}}(y)\,\dF(t-y r)\,,
\end{equation}
where $k, p \in\mathbb{N}$ and the function $\dF(t)$ stands for some time derivative of either $\dM_L(t)$ or $\dS_L(t)$ (and multiplied by $\dM$). The non-locality of the source term \eqref{Ntail} comes directly from that of the multipolar homogeneous solution \eqref{phimultG} in $d$ dimensions. In the quadratic case \eqref{Ntail}, after a change of integration variable and repeated integrations by parts \citep{BBBFMc}, the radiative mode \eqref{radmode} becomes
\begin{equation}\label{radmodequad}
	\mathcal{N}(t) = \FP\left\{ D_\ell^{p,k}(B,\varepsilon)\,\int_0^{+\infty} \dd
	\tau\,\widetilde{\tau}^B \tau^{-2\varepsilon}\,\dF^{(\ell+k-1)}(t-\tau)\right\}\,,
\end{equation}
where $\widetilde{\tau}^B=(c\tau/r_0)^B$ and one has succeeded in factorizing out a pure numerical coefficient in front given by
\begin{align}\label{coeffD}
	& D_\ell^{p,k}(B,\varepsilon) = \frac{(-)^{\ell+k}}{2\ell+1+\varepsilon}
	\,\frac{\Gamma(2\varepsilon-B)}{\Gamma(\ell+k-1+2\varepsilon-B)}\nn\\&\qquad\times\int_1^{+\infty} \dd y
	\,y^p\,\gamma_{-1-\frac{\varepsilon}{2}}(y)\int_1^{+\infty} \dd z
	\,(y+z)^{\ell+k-2+2\varepsilon-B}\,\gamma_{-\ell-1-\frac{\varepsilon}{2}}(z)\,.
\end{align}
We are ultimately interested in the limit $\varepsilon\to 0$, and the integral in \eqref{radmodequad} is well defined in this limit, i.e. convergent at the bound $\tau=0$. Since $\dF$ is a time derivative of a multipole moment, we can assume it to be zero in the remote past so there is no problem with the bound $\tau=+\infty$ of the integral. However the coefficient \eqref{coeffD} in front of the integral will develop a pole when $\varepsilon\to 0$. In contrast with the IR pole in Eq.~\eqref{resdiff}, such a pole will be a UV pole. The coefficient \eqref{coeffD} has been computed in the limit $\varepsilon\to 0$ in the Appendix D of \cite{BBBFMc}; as it turns out it is ill-defined for some values of $\ell$, $p$ or $k$, thus necessitating a ``protection'' by the $B$-process which reduces in this case to a mere finite limit $B\to 0$. In conclusion, the UV type pole generated by the coefficient \eqref{coeffD} cancels out (without need of a further renormalization) the IR pole coming from the potential modes in Eq.~\eqref{resdiff}. 


\subsection{Newtonian-like equations of motion}
\label{sec:EOM}


\subsubsection{The Fokker action for point particles}
\label{sec:Fokker}

The main methods able to derive the equations of motion of compact binary systems to high post-Newtonian order, namely ADM Hamiltonian, Fokker Lagrangian and EFT, have been reviewed in Sect.~\ref{sec:PNeom}. Here we focus on the powerful Fokker Lagrangian method. The starting point is the action $S=S_g+S_m$ given by \eqref{EH}, but in $D=d+1$ space-time dimensions, and where the gravitational action $S_g$ is written in the \cite{LL} form and we add the standard gauge-fixing term appropriate to harmonic coordinates,\footnote{Where $\Gamma^{\alpha}_{\mu\nu}=\frac{1}{2}g^{\alpha\rho}(\partial_\mu g_{\nu\rho}+\partial_\nu g_{\mu\rho}-\partial_\rho g_{\mu\nu})$ is the usual Christoffel symbol and we define $\Gamma^{\alpha}\equiv g^{\mu\nu}\Gamma^{\alpha}_{\mu\nu}$.}
\begin{equation}\label{Sg}
	S_{g} = \frac{c^{3}}{16\pi G^{(d)}} \int \dd^{d+1}x \, \sqrt{-g} \left[
	g^{\alpha\beta} \left( \Gamma^{\mu}_{\alpha\nu}
	\Gamma^{\nu}_{\beta\mu} - \Gamma^{\mu}_{\alpha\beta}
	\Gamma^{\nu}_{\mu\nu} \right) -\frac{1}{2}
	g_{\alpha\beta}\Gamma^{\alpha}\Gamma^{\beta} \right] \,.
\end{equation}
The matter action $S_{m}$ takes the form appropriate for $N$ point-particles -- we neglect the spins and the body's internal structure:
\begin{equation}\label{Sm}
	S_{m} = - \sum_{\text{a}=1}^N m_\text{a} c \int \dd t \sqrt{-(g_{\alpha\beta})_\text{a}
		\,v_{\text{a}}^{\alpha}\,v_{\text{a}}^{\beta}}\,.
\end{equation}
Here $m_\text{a}$ is the mass of the particles, $v_\text{a}^\alpha=\dd y_\text{a}^\alpha/\dd t=(c,\bm{v}_\text{a})$ is the coordinate velocity, $y_\text{a}^\alpha=(c t,\bm{y}_\text{a})$ the ordinary trajectory, and $(g_{\alpha\beta})_\text{a}$ is the metric evaluated at the location of the particle $\text{a}$. A self-field regularization will be necessary to define the object $(g_{\alpha\beta})_\text{a}$ and we shall follow the dimensional regularization scheme. A closed-form expression for the gauge-fixed gravitational action can be written
with the help of the gothic metric $\mathfrak{g}^{\alpha\beta}\equiv\sqrt{-g}g^{\alpha\beta}$ and its inverse $\mathfrak{g}_{\alpha\beta}=g_{\alpha\beta}/\sqrt{-g}$ as
\begin{align}\label{Sgothic}
	S_{g} &= \frac{c^{3}}{32\pi G^{(d)}} \int \dd^{d+1} x
	\biggl[-\frac{1}{2}\Bigl(\mathfrak{g}_{\alpha\mu}\mathfrak{g}_{\beta\nu}
	- \frac{1}{d-1}\,\mathfrak{g}_{\alpha\beta}\mathfrak{g}_{\mu\nu}\Bigr)
	\mathfrak{g}^{\rho\sigma}\partial_{\rho}
	\mathfrak{g}^{\alpha\beta}\partial_{\sigma}
	\mathfrak{g}^{\mu\nu}\nn\\&~\qquad\qquad\qquad\qquad\qquad
	+\mathfrak{g}_{\alpha\beta}\Bigl(\partial_{\mu}\mathfrak{g}^{\alpha\nu}
	\partial_{\nu}\mathfrak{g}^{\beta\mu} -
	\partial_{\mu}\mathfrak{g}^{\alpha\mu}\partial_{\nu}
	\mathfrak{g}^{\beta\nu}\Bigr)\biggr]\,.
\end{align}
The relaxed Einstein field equations following from the variation of the gauge-fixed action are of the form
\begin{equation}\label{relaxedEFE}
		\Box h^{\alpha\beta} = \frac{16\pi G^{(d)}}{c^4}
		\vert g\vert \,T^{\alpha\beta} +
		\Sigma^{\alpha\beta}\,,
\end{equation}
where $h^{\alpha\beta} = \mathfrak{g}^{\alpha\beta} - \eta^{\alpha\beta}$ denotes the gothic metric deviation, the particle stress-energy tensor takes the standard form
\begin{equation}\label{Talphabeta}
	T^{\alpha\beta} = \sum_{\text{a}=1}^N \frac{m_\text{a} \,v_\text{a}^\alpha v_\text{a}^\beta}{\sqrt{-(g_{\mu\nu})_\text{a} \frac{v_\text{a}^\mu v_\text{a}^\nu}{c^2}}}\,\frac{\delta^{(d)}(\mathbf{x} - \bm{y}_\text{a})}{\sqrt{-g}} \,, 
\end{equation}
and the source term differs from the one in harmonic coordinates by terms depending on the ``harmonicity'' $H^\alpha \equiv \partial_\mu h^{\alpha\mu} = - \sqrt{-g}\,\Gamma^\alpha$:
\begin{equation}\label{Lambda}
	\Sigma^{\alpha\beta} = \Lambda_\text{harm}^{\alpha\beta} - H^\alpha H^\beta - H^\mu
	\partial_\mu h^{\alpha\beta} -
	\frac{1}{2}\mathfrak{g}^{\alpha\beta}\mathfrak{g}_{\mu\nu} H^\mu
	H^\nu +
	2\mathfrak{g}_{\mu\nu}\mathfrak{g}^{\rho(\alpha}\partial_\rho
	h^{\beta)\mu}H^\nu\,,
\end{equation}
where $\Lambda_\text{harm}^{\alpha\beta}$ is given in $d$ dimensions by Eq.~\eqref{Lambdad}.

The \cite{Fokker} action is obtained by inserting into the action \eqref{Sm}--\eqref{Sgothic} an explicit PN solution of the field equations, given as a functional of the particle's trajectories, i.e., an explicit post-Newtonian metric
\begin{equation}\label{PNmetric}
  \overline{g}_{\alpha\beta}(\mathbf{x}, t) \equiv \overline{g}_{\alpha\beta}[\mathbf{x}; \bm{y}_\text{b}(t), \bm{v}_\text{b}(t), \cdots]\,,
\end{equation}
where the overbar recalls the post-Newtonian character of this solution. Note that the source term \eqref{Sigma} contains all required harmonicities $H^\alpha$, which are not assumed to be zero in the PN iteration of the metric \eqref{PNmetric}. Hence the ellipsis in \eqref{PNmetric} indicate also the dependence on accelerations $\bm{a}_\text{b}(t)$, derivative of accelerations $\bm{b}_\text{b}(t)$, and so on, due to the fact that the equations of motion are ``off-shell'' at this stage and no replacement of accelerations is performed. Thus, the Fokker generalized action, which is a functional of the particles only, but depending not only on positions and velocities but also on accelerations and their derivatives, is given by
\begin{align}\label{SF}
	&S_\text{F}\left[\bm{y}_\text{b}(t), \bm{v}_\text{b}(t), \cdots\right] =
	\int\dd t\!\int \dd^{d}\mathbf{x} \,\overline{\mathcal{L}}_g\left[\mathbf{x};
	\bm{y}_\text{b}(t), \bm{v}_\text{b}(t), \cdots\right] \nn\\ &\qquad\qquad -
	\sum_\text{a} m_\text{a} c \int \dd t \sqrt{-\overline{g}_{\alpha\beta}[\bm{y}_\text{a}(t);
		\bm{y}_\text{b}(t), \bm{v}_\text{b}(t), \cdots]
		\,v_\text{a}^{\alpha}v_\text{a}^{\beta}}\,,
\end{align}
where $\overline{\mathcal{L}}_g$ is the Lagrangian density of the gravitational action \eqref{Sgothic} computed with the metric \eqref{PNmetric}. As is well known, it is always possible to eliminate from a generalized PN action a contribution that is quadratic in the accelerations by absorbing it into a ``double-zero'' term which does not contribute to the dynamics \citep{DS85}. The argument can be extended to any term polynomial in the accelerations (and their derivatives). The post-Newtonian equations of motion of the particles are obtained as the generalized Lagrange equations
\begin{align}\label{FokkerEOM}
	\frac{\delta S_\text{F}}{\delta \bm{y}_\text{b}} \equiv \frac{\partial
		L_\text{F}}{\partial \bm{y}_\text{b}} - \frac{\dd}{\dd
		t}\left(\frac{\partial L_\text{F}}{\partial \bm{v}_\text{b}}\right) + \frac{\dd^2}{\dd
		t^2}\left(\frac{\partial L_\text{F}}{\partial \bm{a}_\text{b}}\right) +
	\cdots = 0\,,
\end{align}
where $L_\text{F}$ is the corresponding Lagrangian (such that $S_\text{F}=\int\dd t\,L_\text{F}$). Once they have been constructed, the equations \eqref{FokkerEOM} can be order reduced by replacing iteratively all the higher-order accelerations by their expressions coming from the PN equations of motion themselves.\footnote{The classical Fokker action is equivalent, in the ``tree-level'' approximation, to the effective action used by the effective field theory \citep{FSrevue, Portorevue}.}

By definition of the Fokker action, varying it with respect to the post-Newtonian metric \eqref{PNmetric} yields back the relaxed field equations. Hence (coming back to the gothic variable):
\begin{equation}\label{deltaSFokker} 
	\frac{\delta S_\text{F}}{\delta\overline{h}_{\alpha\beta}} = \frac{c^4}{64\pi G^{(d)}}\biggl[\Box
	\overline{h}^{\alpha\beta} - \overline{\Sigma}^{\alpha\beta} - \frac{16\pi G}{c^4}\,\vert\overline{g}\vert\,\overline{T}^{\alpha\beta}\biggr] = 0\,.
\end{equation}
Now let us suppose that we have constructed the post-Newtonian metric up to some small post-Newtonian remainder, say $\overline{h} = \overline{h}_n + r_n$ (skipping space-time indices for a while), where $\overline{h}_n$ is explicitly known while the small remainder $r_n$ is unknown. Evaluating the Fokker action for this solution we get\footnote{Such writing is a bit sketchy; see Eq. (4.7) of \cite{BBBFMa} for more details.}
\begin{equation}\label{expSFokker1} 
	S_\text{F}\bigl[\overline{h}\bigr] = S_\text{F}\bigl[\overline{h}_n\bigr] + r_n \frac{\delta S_\text{F}}{\delta\overline{h}}\bigl[\overline{h}_n\bigr] + \calO(r_n^2)\,.
\end{equation}
Actually, because the variation of the action with respect to $\overline{h}$ is zero, Eq.~\eqref{FokkerEOM}, when evaluated for the approximate solution $\overline{h}_n$, the correction term in \eqref{expSFokker1} will acquire an extra small factor $r_n$. More precisely, taking also into account the factor $c^{4}$ in front of \eqref{deltaSFokker},
\begin{equation}\label{expSFokker2} 
	\frac{\delta S_\text{F}}{\delta\overline{h}}\bigl[\overline{h}_n\bigr] = \calO(c^4 \,r_n)\quad\Longrightarrow\quad S_\text{F}\bigl[\overline{h}\bigr] = S_\text{F}\bigl[\overline{h}_n\bigr] + \calO(c^4\,r_n^2)\,.
\end{equation}
If the unknown remainder is of order $r_n=\calO(c^{-n-3})$, which means that we control the post-Newtonian metric up to order $c^{-n-2}$ included, we see that the Fokker action will be controlled up to order
\begin{equation}\label{expSFokker3}
	S_\text{F}\bigl[\overline{h}\bigr] = S_\text{F}\bigl[\overline{h}_n\bigr] + \calO(c^{-2n-2})\,,
\end{equation}
which means that it is controlled up to $n$PN order included. Thus in order to compute the Fokker action to order $n$PN it suffices to insert the metric to order $c^{-n-2}$, which represents for high PN orders an important improvement with respect to the naive expectection $c^{-2n}$. However the previous reasoning neglects the tensorial nature of the metric. The more precise analysis by \cite{BBBFMa} concludes on the same result provided that the components of the metric are decomposed according to 
\begin{equation}
	\label{components}
	\overline{h}{}\ab\longrightarrow \left\{\begin{array}{l}
		\displaystyle \overline{h} = \bigl(\overline{h}{}^{00ii}, \overline{h}{}^{0i}, \overline{h}{}^{ij}\bigr)\,,\\[0.3cm]
		\displaystyle \overline{h}{}^{00ii} = \frac{2}{d-1}\Bigl[(d-2)\overline{h}{}^{00} + \overline{h}{}^{ii}\Bigr]\,.\end{array}\right.
\end{equation}
That is, in order to obtain the Fokker action accurate to $n$PN order, it suffices to insert (each component of) the previous $\overline{h}$ developed up to order $c^{-n-2}$ included. For instance, in order to obtain the Fokker action (and therefore the full dynamics) correct at 4PN order one needs only to insert $\overline{h}$ up to the relatively low level to $c^{-6}$, i.e. (since we are interested in the conservative dynamics) $\overline{h}{}^{00ii}$ and $\overline{h}{}^{ij}$ to level $c^{-6}$, and $\overline{h}{}^{0i}$ to level $c^{-5}$. The above technique has been referred to as the ``method $n+2$''.

As said in Sect.~\ref{sec:DRrad} the Fokker Lagrangian contains some IR poles coming from the behaviour when $r\to +\infty$ of the gravitational part of the Lagrangian. Thus, one computes the expansion $r\to +\infty$ of the various potentials parametrizing the metric in $d$ dimensions, see e.g. \eqref{pot1PNd}, and then apply the formula \eqref{resdiff}. The potentials are those needed at the 4PN order (say) following the previous method ``$n+2$''. In the language of EFT \citep{PR17} the previous calculation corresponds to ``potential mode'' contribution. 

However, there is also to take into account the contribution coming from the conservative part of the 4PN tail effect in $d$ dimensions, which corresponds in the EFT language to the ``radiation'' contribution and has been reviewed in Sect.~\ref{sec:IRreg}. The radiation mode is an imprint of tails in the local conservative dynamics of the source at the 4PN order \citep{BD88}. It contains a tail-induced modification of the dissipative radiation reaction force at the relative 1.5PN order, see Eq.~\eqref{Vreactail}. The conservative piece modifies the Fokker action of equivalently the EFT at the 4PN order \citep{FStail, DJS14, GLPR16}.

Applying Eqs. \eqref{radmodequad}--\eqref{coeffD} to the case of the quadratic interaction between the constant mass $\dM$ and the varying quadrupole moment $\dM_{ij}$, one obtains at the leading 4PN order \citep{BBBFMc}
\begin{subequations}\label{h2harmhom}
	\begin{align}
		\overline{h}{}^{00ii}_\text{rad} &= \frac{8 G^2 \dM}{15 c^{10}} \,x^{ij}
		\int_0^{+\infty}\!\!\!\dd\tau\left[\ln\left(\frac{c\sqrt{\bar{q}}\,\tau}{2\ell_0}
		\right) - \frac{1}{2\varepsilon} +
		\frac{61}{60}\right]\dM^{(7)}_{ij}(t-\tau)  \,,\\[0.2cm]
		\overline{h}{}^{0i}_\text{rad} &= - \frac{8 G^2 \dM}{3 c^{9}} \,x^j
		\int_0^{+\infty}\!\!\!\dd\tau\left[\ln\left(\frac{c\sqrt{\bar{q}}\,\tau}{2\ell_0}
		\right) - \frac{1}{2\varepsilon} +
		\frac{107}{120}\right]\dM^{(6)}_{ij}(t-\tau) \,,\\[0.2cm]
		\overline{h}{}^{ij}_\text{rad} &= \frac{8 G^2 \dM}{c^{8}}
		\int_0^{+\infty}\!\!\!\dd\tau\left[\ln\left(\frac{c\sqrt{\bar{q}}\,\tau}{2\ell_0}
		\right) - \frac{1}{2\varepsilon} +
		\frac{4}{5}\right]\dM^{(5)}_{ij}(t-\tau) \,.
	\end{align}
\end{subequations}
We do not indicate the 5PN remainder terms nor the neglected $\calO(\varepsilon)$ terms; $G$ is Newton's constant in Eq.~\eqref{G} and we pose $\bar{q}=4\pi\de^{\gamma_\text{E}}$ with $\gamma_\text{E}$ denoting the Euler constant.\footnote{Euler's constant (discovered in 1731) is sometimes called the Euler-Mascheroni constant. Mascheroni's contribution was to have computed (in 1790) the constant up to 19 decimal places, while Euler had previously computed it to 15 places. Nowadays, one credits the important mathematical contribution, rather than the record of the most numerical digits computed, and it is appropriate to name the constant after Euler alone \citep{EulerConstant}.\label{fnote:EulerConst}} The tail terms in \eqref{h2harmhom} yield a non-local modification of the equations of motion at 4PN order, and consequently a modification of the Fokker action. One finds
\begin{align}\label{StailDR0}
	S_\text{F}^\text{rad} = \frac{
		2G^2\dM}{5 c^8} \int_{-\infty}^{+\infty}\!\!\!\dd t
	\,\dM_{ij}^{(3)}(t) \int_0^{+\infty} \!\!\!\dd\tau \left[
	\ln\left(\frac{c\sqrt{\bar{q}}\,\tau}{2\ell_0}\right) -
	\frac{1}{2\varepsilon} + \frac{41}{60} \right]\dM_{ij}^{(4)}(t-\tau)\,.
\end{align}
An important point to notice is that \eqref{StailDR0} can be rewritten in a manifestly time-symmetric way; thus the procedure automatically selects the ``conservative'' part of the tail at 4PN order -- the dissipative part giving no contribution to the action. Hence we have in fact
\begin{align}\label{StailDR1}
	S_\text{F}^\text{rad} &= \frac{G^2\dM}{5 c^8} \int_{-\infty}^{+\infty}\dd t
	\,\dM_{ij}^{(3)}(t) \\&\quad\times\int_0^{+\infty} \dd\tau \left[
	\ln\left(\frac{c\sqrt{\bar{q}}\,\tau}{2\ell_0}\right) -
	\frac{1}{2\varepsilon} + \frac{41}{60} \right]\left(\dM_{ij}^{(4)}(t-\tau) -
	\dM_{ij}^{(4)}(t+\tau)\right) \,,\nn
\end{align}
which can elegantly be rewritten by means of the Hadamard partie finie (Pf) integral, defined as the one-dimensional version of Eq.~\eqref{pfint}, as \citep{FStail,DJS14,GLPR16,BBBFMc}
\begin{equation}\label{StailDR2}
	S_\text{F}^\text{rad} = \frac{G^2\dM}{5 c^8} \,\mathop{\text{Pf}}_{\tau_0}
	\,\int\!\!\!\int \frac{\dd t\,\dd t'}{\vert t-t'\vert}
	\,\dM_{ij}^{(3)}(t)\,\dM_{ij}^{(3)}(t')\,,
\end{equation}
with $\tau_0 = \frac{2\ell_0}{c \sqrt{\bar{q}}}\,\de^{\frac{1}{2\varepsilon}-\frac{41}{60}}$ playing the role of the Hadamard cut-off scale.\footnote{For any regular function $f(t)$ tending to zero sufficiently rapidly when $t\to\pm\infty$ we have
\begin{align*}
	\mathop{\text{Pf}}_{\tau_0} \int_{-\infty}^{+\infty} \dd
	t'\,\frac{f(t')}{\vert t-t'\vert} \equiv \int_0^{+\infty} \dd\tau
	\ln\left(\frac{\tau}{\tau_0}\right)\left[f^{(1)}(t-\tau) -
	f^{(1)}(t+\tau)\right]\,.
\end{align*}
\label{fnote:hadamardPF}}
Since the action \eqref{StailDR2} is non local its variation involves functional derivatives rather than ordinary derivatives. Varying with respect to the particle world-lines $y_\text{a}^i(t)$ -- for a system of point particles with mass $m_\text{a}$ -- one obtains
\begin{equation}\label{verStail}
	\frac{\delta S_\text{F}^\text{rad}}{\delta y_\text{a}^i(t)} =
	-\frac{4G^2\dM}{5c^8} \,m_\text{a} y_\text{a}^j(t)\,\mathop{\text{Pf}}_{\tau_{0}}
\,\int_{-\infty}^{+\infty} \frac{\dd t'}{\vert t-t'\vert}
	\dM_{ij}^{(6)}(t') \,,
\end{equation}
which coincides with the conservative part of the known 4PN tail contribution in the equations of motion \citep{BD88}.


\subsubsection{The 3.5PN acceleration and 3PN energy}
\label{sec:PNacc}

The acceleration of one of the particles, say the particle 1, is presented up to the 3.5PN order, together with the corresponding binary's energy (which is conserved in the absence of radiation reaction). Besides the Hamiltonian and Lagrangian methods reviewed in Sect.~\ref{sec:PNeom}, a ``direct'' post-Newtonian attack of the problem is possible, reducing the metric of an extended regular source, as given in harmonic coordinates to 3.5PN order by Eqs. \eqref{gmunu3PN}, to the case where the matter tensor is made of delta functions, and then curing the self-field divergences by means of the dimensional regularization (see Sect.~\ref{sec:DReom}). In this approach the equations of motion (in the case of spinless particles) are just the geodesic equations provided to 3.5PN order in Eqs. \eqref{dPdtF}--\eqref{PiFi}; the metric therein is the regularized metric generated by the system of particles themselves. The radiation reaction effects at 2.5PN and 3.5PN orders in harmonic coordinates are known from \cite{IW93, IW95, JaraS97, PW02, KFS03, NB05, itoh3}. 

Though the successive post-Newtonian approximations are really a consequence of general relativity, the final equations of motion must be interpreted in a Newtonian-like fashion. That is, once a convenient general-relativistic (Cartesian) coordinate system is chosen, we should express the results in terms of the \emph{coordinate} positions, velocities, and accelerations of the bodies, and view the trajectories of the particles as taking place in the absolute Euclidean space of Newton. But because the equations of motion are actually relativistic, they must stay manifestly invariant -- in harmonic coordinates -- when we perform a global post-Newtonian-expanded Lorentz transformation. Furthermore they must possess the correct ``perturbative'' limit, given by the geodesics of the (post-Newtonian-expanded) Schwarzschild metric, when one of the masses tends to zero, and be conservative, i.e., to admit a Lagrangian or Hamiltonian formulation, when the gravitational radiation reaction is turned off.

We denote by $r_{12}=\vert\bm{y}_1(t)-\bm{y}_2(t)\vert$ the harmonic-coordinate distance between the two particles, with $\bm{y}_1=(y_1^i)$ and $\bm{y}_2=(y_2^i)$, by $\bm{n}_{12}=(\bm{y}_1-\bm{y}_2)/r_{12}$ the corresponding unit direction, and by $\bm{v}_1=\dd \bm{y}_1/\dd t$ and $\bm{a}_1=\dd\bm{v}_1/\dd t$ the coordinate velocity and acceleration of the particle 1 (and \emph{idem} for 2). Sometimes we pose $\bm{v}_{12}=\bm{v}_1-\bm{v}_2$ for the relative velocity. The usual Euclidean scalar product of vectors is denoted with parenthesis, e.g.,  $(n_{12}v_1)=\bm{n}_{12}\cdot\bm{ v}_1$ and $(v_1v_2)=\bm{v}_1\cdot\bm{ v}_2$. The equations of the body 2 are obtained by exchanging all the particle labels $1\leftrightarrow 2$ (remembering that $\bm{n}_{12}$ and $\bm{v}_{12}$ change sign in this operation):
\begin{align}
	\bm{a}_1 &= -\frac{G m_2}{r_{12}^2}\bm{n}_{12} \nn\\&
	+\frac{1}{c^2} \Bigg\{ \left[ \frac{5 G^2 m_1 m_2}{r_{12}^3} +
	\frac{4 G^2 m_2^2}{r_{12}^3} + \frac{G m_2}{r_{12}^2} \left(
	\frac{3}{2} (n_{12}v_2)^2 - v_1^2 + 4 (v_1v_2) - 2 v_2^2 \right)
	\right] \bm{n}_{12} \nn\\&  \qquad \quad \: + \frac{G
		m_2}{r_{12}^2} \left( 4 (n_{12}v_1) - 3 (n_{12}v_2) \right)
	\bm{v}_{12} \Bigg\} \nn\\& 
	+ \frac{1}{c^4} \Bigg\{ \bigg[ \! -
	\frac{57 G^3 m_1^2 m_2}{4 r_{12}^4} - \frac{69 G^3 m_1 m_2^2}{2
		r_{12}^4} - \frac{9 G^3 m_2^3}{r_{12}^4} \nn\\& 
	\qquad \quad \: + \frac{G m_2}{r_{12}^2} \bigg( \!\! - \frac{15}{8}
	(n_{12}v_2)^4 + \frac{3}{2} (n_{12}v_2)^2 v_1^2 - 6 (n_{12}v_2)^2
	(v_1v_2) - 2 (v_1v_2)^2 \nn\\&  \qquad \quad \quad + \frac{9}{2} (n_{12}v_2)^2 v_2^2
	+ 4 (v_1v_2) v_2^2 - 2
	v_2^4 \bigg) \nn\\&  \quad \: + \frac{G^2 m_1
		m_2}{r_{12}^3} \left( \frac{39}{2} (n_{12}v_1)^2 - 39
	(n_{12}v_1) (n_{12}v_2) + \frac{17}{2} (n_{12}v_2)^2 -
	\frac{15}{4} v_1^2 - \frac{5}{2} (v_1v_2) + \frac{5}{4} v_2^2
	\right) \nn\\& \qquad \quad \: + \frac{G^2
		m_2^2}{r_{12}^3} \left( 2 (n_{12}v_1)^2 - 4 (n_{12}v_1)
	(n_{12}v_2) - 6 (n_{12}v_2)^2 - 8 (v_1v_2) + 4 v_2^2 \right)
	\bigg] \bm{n}_{12} \nn\\& \qquad \quad \: + \bigg[ \frac{G^2
		m_2^2}{r_{12}^3} \left( - 2 (n_{12}v_1) - 2 (n_{12}v_2) \right)
	+ \frac{G^2 m_1 m_2}{r_{12}^3} \left( \!\! - \frac{63}{4}
	(n_{12}v_1) + \frac{55}{4} (n_{12}v_2) \right) \nn\\&
	\qquad \qquad ~ + \frac{G m_2}{r_{12}^2} \bigg( \!\! - 6
	(n_{12}v_1) (n_{12}v_2)^2 + \frac{9}{2} (n_{12}v_2)^3 +
	(n_{12}v_2) v_1^2 - 4 (n_{12}v_1) (v_1v_2) \nn\\& \qquad \quad \qquad \,\:\: +\, 4 (n_{12}v_2) (v_1v_2) + 4
	(n_{12}v_1) v_2^2 - 5 (n_{12}v_2) v_2^2 \bigg) \bigg] \bm{v}_{12}
	\Bigg\} \nn\\& 
	+ \frac{1}{c^5} \Bigg\{ \left[ \frac{208 G^3
		m_1 m_2^2}{15 r_{12}^4} (n_{12}v_{12}) - \frac{24 G^3 m_1^2
		m_2}{5 r_{12}^4} (n_{12}v_{12}) + \frac{12 G^2 m_1 m_2}{5
		r_{12}^3} (n_{12}v_{12}) v_{12}^2 \right] \bm{n}_{12} \nn
	\\ & \qquad \quad \;\, + \left[ \frac{8 G^3 m_1^2 m_2}{5 r_{12}^4} -
	\frac{32\,G^3 m_1 m_2^2}{5 r_{12}^4} - \frac{4 G^2 m_1 m_2}{5
		r_{12}^3} v_{12}^2 \right] \bm{v}_{12} \Bigg\} \nn\\& 
	+ \frac{1}{c^6} \Bigg\{ \bigg[ \frac{G m_2}{r_{12}^2} \bigg(
	\frac{35}{16} (n_{12}v_2)^6 - \frac{15}{8} (n_{12}v_2)^4 v_1^2 +
	\frac{15}{2} (n_{12}v_2)^4 (v_1v_2) + 3 (n_{12}v_2)^2 (v_1v_2)^2
	\nn\\& \qquad \qquad \quad \; - \frac{15}{2}
	(n_{12}v_2)^4 v_2^2 + \frac{3}{2} (n_{12}v_2)^2 v_1^2 v_2^2 - 12
	(n_{12}v_2)^2 (v_1v_2) v_2^2 - 2 (v_1v_2)^2 v_2^2 \nn\\&
	\qquad \qquad \quad \; + \frac{15}{2} (n_{12}v_2)^2 v_2^4 + 4
	(v_1v_2) v_2^4 - 2 v_2^6 \bigg) \nn\\& \qquad \quad \: +
	\frac{G^2 m_1 m_2}{r_{12}^3} \bigg( \!\! - \frac{171}{8}
	(n_{12}v_1)^4 + \frac{171}{2} (n_{12}v_1)^3 (n_{12}v_2) -
	\frac{723}{4} (n_{12}v_1)^2 (n_{12}v_2)^2 \nn\\& \qquad \qquad \quad ~\: + \frac{383}{2} (n_{12}v_1) (n_{12}v_2)^3
	- \frac{455}{8} (n_{12}v_2)^4 + \frac{229}{4} (n_{12}v_1)^2 v_1^2
	\nn\\& \qquad \qquad \quad ~\: - \frac{205}{2}
	(n_{12}v_1) (n_{12}v_2) v_1^2 + \frac{191}{4} (n_{12}v_2)^2 v_1^2
	- \frac{91}{8} v_1^4 - \frac{229}{2} (n_{12}v_1)^2 (v_1v_2)
	\nn\\& \qquad \qquad \quad ~\: + 244 (n_{12}v_1)
	(n_{12}v_2) (v_1v_2) - \frac{225}{2} (n_{12}v_2)^2 (v_1v_2) +
	\frac{91}{2} v_1^2 (v_1v_2) \nn\\& \qquad \qquad
	\quad ~\: - \frac{177}{4} (v_1v_2)^2 + \frac{229}{4}
	(n_{12}v_1)^2 v_2^2 - \frac{283}{2} (n_{12}v_1) (n_{12}v_2) v_2^2
	\nn\\& \qquad \qquad \quad ~\: + \frac{259}{4}
	(n_{12}v_2)^2 v_2^2 - \frac{91}{4} v_1^2 v_2^2 + 43 (v_1v_2) v_2^2
	- \frac{81}{8} v_2^4 \bigg) \nn\\& \qquad \quad \: +
	\frac{G^2 m_2^2}{r_{12}^3} \bigg( \!\! - 6 (n_{12}v_1)^2
	(n_{12}v_2)^2 + 12 (n_{12}v_1) (n_{12}v_2)^3 + 6 (n_{12}v_2)^4
	\nn\\& \qquad \qquad \quad \quad \,\: +\, 4 (n_{12}v_1)
	(n_{12}v_2) (v_1v_2) + 12 (n_{12}v_2)^2 (v_1v_2) + 4 (v_1v_2)^2
	\nn\\& \qquad \qquad \quad \quad \,\: -\, 4 (n_{12}v_1)
	(n_{12}v_2) v_2^2 - 12 (n_{12}v_2)^2 v_2^2 - 8 (v_1v_2) v_2^2 + 4
	v_2^4 \bigg) \nn\\& \qquad \quad \: + \frac{G^3
		m_2^3}{r_{12}^4} \left( \!\! - (n_{12}v_1)^2 + 2 (n_{12}v_1)
	(n_{12}v_2) + \frac{43}{2} (n_{12}v_2)^2 + 18 (v_1v_2) - 9 v_2^2
	\right) \nn\\& \qquad \quad \: + \frac{G^3 m_1
		m_2^2}{r_{12}^4} \bigg( \frac{415}{8} (n_{12}v_1)^2 -
	\frac{375}{4} (n_{12}v_1) (n_{12}v_2) + \frac{1113}{8}
	(n_{12}v_2)^2 - \frac{615}{64} (n_{12}v_{12})^2 \pi^2 \nn
	\\ & \qquad \qquad \quad ~\,\, + 18 v_1^2 + \frac{123}{64}
	\pi^2 v_{12}^2 + 33 (v_1v_2) - \frac{33}{2} v_2^2 \bigg) \nn
	\\ & \qquad \quad \: + \frac{G^3 m_1^2 m_2}{r_{12}^4} \bigg( \!\!
	- \frac{45887}{168} (n_{12}v_1)^2 + \frac{24025}{42} (n_{12}v_1)
	(n_{12}v_2) - \frac{10469}{42} (n_{12}v_2)^2 + \frac{48197}{840}
	v_1^2 \nn\\& \qquad \qquad \quad ~\: -
	\frac{36227}{420} (v_1v_2) + \frac{36227}{840} v_2^2 + 110
	(n_{12}v_{12})^2 \ln \left( \frac{r_{12}}{r'_0} \right) - 22
	v_{12}^2 \ln \left( \frac{r_{12}}{r'_0} \right) \bigg) \nn
	\\ & \qquad \quad \: + \frac{16 G^4 m_2^4}{r_{12}^5} + \frac{G^4
		m_1^2 m_2^2}{r_{12}^5} \left( 175 - \frac{41}{16} \pi^2 \right)
	+ \frac{G^4 m_1^3 m_2}{r_{12}^5} \left( \!\! - \frac{3187}{1260} +
	\frac{44}{3} \ln \left( \frac{r_{12}}{r'_0} \right) \right)
	\nn\\&\qquad \quad \: + \frac{G^4 m_1 m_2^3}{r_{12}^5}
	\left( \frac{110741}{630} - \frac{41}{16} \pi^2 - \frac{44}{3} \ln
	\left( \frac{r_{12}}{r'_0} \right) \right) \bigg] \bm{n}_{12}
	\nn\\& \qquad \quad \: + \bigg[ \frac{G m_2}{r_{12}^2}
	\bigg( \frac{15}{2} (n_{12}v_1) (n_{12}v_2)^4 - \frac{45}{8}
	(n_{12}v_2)^5 - \frac{3}{2} (n_{12}v_2)^3 v_1^2 \nn\\& \qquad \qquad \quad + 6 (n_{12}v_1)
	(n_{12}v_2)^2 (v_1v_2) -\, 6 (n_{12}v_2)^3 (v_1v_2) - 2 (n_{12}v_2) (v_1v_2)^2 \nn\\& \qquad \qquad \quad -
	12 (n_{12}v_1)(n_{12}v_2)^2 v_2^2 + 12 (n_{12}v_2)^3 v_2^2
	+ (n_{12}v_2)
	v_1^2 v_2^2 - 4 (n_{12}v_1) (v_1v_2) v_2^2 \nn\\& \qquad \qquad \quad + 8 (n_{12}v_2)
	(v_1v_2) v_2^2 + 4 (n_{12}v_1) v_2^4 -\, 7 (n_{12}v_2) v_2^4 \bigg) \nn\\&
	\qquad \qquad ~ + \frac{G^2 m_2^2}{r_{12}^3} \bigg( \!\! - 2
	(n_{12}v_1)^2 (n_{12}v_2) + 8 (n_{12}v_1) (n_{12}v_2)^2 + 2
	(n_{12}v_2)^3 + 2 (n_{12}v_1) (v_1v_2) \nn\\& \qquad
	\qquad \quad + 4 (n_{12}v_2) (v_1v_2) - 2
	(n_{12}v_1) v_2^2 - 4 (n_{12}v_2) v_2^2 \bigg) \nn\\&
	\qquad \qquad ~ + \frac{G^2 m_1 m_2}{r_{12}^3} \bigg( \!\! -
	\frac{243}{4} (n_{12}v_1)^3 + \frac{565}{4} (n_{12}v_1)^2
	(n_{12}v_2) - \frac{269}{4} (n_{12}v_1) (n_{12}v_2)^2 \nn
	\\ & \qquad \qquad \quad \; - \frac{95}{12}
	(n_{12}v_2)^3 + \frac{207}{8} (n_{12}v_1) v_1^2 - \frac{137}{8}
	(n_{12}v_2) v_1^2 - 36 (n_{12}v_1) (v_1v_2) \nn\\& \qquad
	\qquad \quad \; + \frac{27}{4} (n_{12}v_2) (v_1v_2)
	+ \frac{81}{8} (n_{12}v_1) v_2^2 + \frac{83}{8} (n_{12}v_2) v_2^2
	\bigg) \nn\\& \qquad \qquad ~ + \frac{G^3 m_2^3}{r_{12}^4}
	\left( 4 (n_{12}v_1) + 5 (n_{12}v_2) \right) \nn\\& \qquad
	\qquad ~ + \frac{G^3 m_1 m_2^2}{r_{12}^4} \left( \!\! -
	\frac{307}{8} (n_{12}v_1) + \frac{479}{8} (n_{12}v_2) +
	\frac{123}{32} (n_{12}v_{12}) \pi^2 \right) \nn\\& \qquad
	\qquad ~ + \frac{G^3 m_1^2 m_2}{r_{12}^4} \left( \frac{31397}{420}
	(n_{12}v_1) - \frac{36227}{420} (n_{12}v_2) - 44 (n_{12}v_{12})
	\ln \left( \frac{r_{12}}{r'_0} \right) \right) \bigg] \bm{v}_{12}
	\Bigg\} \nn\\& 
	+ \frac{1}{c^7} \Bigg\{ \bigg[ \frac{G^4 m_1^3m_2}{r_{12}^5} \left(
	\frac{3992}{105} (n_{12} v_1) - \frac{4328}{105} (n_{12} v_2)
	\right) \nn\\& \qquad \quad \; + \frac{G^4 m_1^2
		m_2^2}{r_{12}^5} \left( \!\! - \frac{13576}{105} (n_{12} v_1) +
	\frac{2872}{21} (n_{12} v_2) \right) - \frac{3172}{21} \frac{G^4 m_1
		m_2^3}{r_{12}^6} (n_{12} v_{12}) \nn\\& \qquad \quad \; +
	\frac{G^3 m_1^2m_2}{r_{12}^4} \bigg( 48 (n_{12} v_1)^3 -
	\frac{696}{5} (n_{12} v_1)^2 (n_{12} v_2) + \frac{744}{5} (n_{12}
	v_1) (n_{12} v_2)^2 \nn\\&
	\qquad \qquad \quad- \frac{288}{5} (n_{12} v_2)^3 - \frac{4888}{105} (n_{12} v_1)
	v_1^2 + \frac{5056}{105} (n_{12} v_2) v_1^2 + \frac{2056}{21}
	(n_{12} v_1) (v_1 v_2) \nn\\& \qquad \qquad \quad - \frac{2224}{21} (n_{12} v_2) (v_1 v_2) - \frac{1028}{21}
	(n_{12} v_1) v_2^2 + \frac{5812}{105} (n_{12} v_2) v_2^2 \bigg)
	\nn\\& \qquad \quad \; + \frac{G^3 m_1 m_2^2}{r_{12}^4}
	\bigg( \!\!  - \frac{582}{5} (n_{12} v_1)^3 + \frac{1746}{5} (n_{12}
	v_1)^2 (n_{12} v_2) - \frac{1954}{5} (n_{12} v_1) (n_{12} v_2)^2
	\nn\\& \qquad \qquad \quad ~ + 158 (n_{12} v_2)^3
	+\frac{3568}{105} (n_{12} v_{12}) v_1^2 - \frac{2864}{35} (n_{12}
	v_1) (v_1 v_2) \nn\\& \qquad \qquad \quad ~ +
	\frac{10048}{105} (n_{12} v_2) (v_1 v_2) +\frac{1432}{35} (n_{12}
	v_1) v_2^2 - \frac{5752}{105} (n_{12} v_2) v_2^2 \bigg) \nn
	\\ & \qquad \quad \; + \frac{G^2 m_1 m_2}{r_{12}^3} \bigg( \!\! - 56
	(n_{12} v_{12})^5 + 60 (n_{12} v_{1})^3 v_{12}^2 - 180 (n_{12}
	v_{1})^2 (n_{12} v_{2}) v_{12}^2 \nn\\& \qquad \qquad
	\quad ~\; + 174 (n_{12} v_{1}) (n_{12} v_{2})^2 v_{12}^2 - 54
	(n_{12} v_{2})^3 v_{12}^2 - \frac{246}{35} (n_{12} v_{12}) v_1^4
	\nn\\& \qquad \qquad \quad ~\; + \frac{1068}{35}
	(n_{12} v_1) v_1^2 (v_1 v_2) - \frac{984}{35} (n_{12} v_2) v_1^2
	(v_1 v_2) - \frac{1068}{35} (n_{12} v_1) (v_1 v_2)^2 \nn\\&
	\qquad \qquad \qquad \qquad ~\; + \frac{180}{7} (n_{12} v_2) (v_1
	v_2)^2 - \frac{534}{35} (n_{12} v_1) v_1^2 v_2^2 + \frac{90}{7}
	(n_{12} v_2) v_1^2 v_2^2 \nn\\& \qquad \qquad \quad 
	~\; + \frac{984}{35} (n_{12} v_1) (v_1 v_2) v_2^2 - \frac{732}{35}
	(n_{12} v_2) (v_1 v_2) v_2^2 - \frac{204}{35} (n_{12} v_1) v_2^4
	\nn\\& \qquad \qquad \quad ~\; + \frac{24}{7}
	(n_{12} v_2) v_2^4 \bigg) \bigg] \bm{n}_{12} \nn\\& \qquad ~\,
	+ \bigg[ \! - \frac{184}{21} \frac{G^4 m_1^3 m_2}{r_{12}^5} +
	\frac{6224}{105} \frac{G^4 m_1^2 m_2^2}{r_{12}^6} + \frac{6388}{105}
	\frac{G^4 m_1 m_2^3}{r_{12}^6} \nn\\& \qquad \qquad \; +
	\frac{G^3 m_1^2 m_2}{r_{12}^4} \bigg( \frac{52}{15} (n_{12} v_1)^2 -
	\frac{56}{15} (n_{12} v_1) (n_{12} v_2) - \frac{44}{15} (n_{12}
	v_2)^2 \nn
	\\ & \qquad \qquad\ \quad - \frac{132}{35} v_1^2 + \frac{152}{35} (v_1 v_2) - \frac{48}{35} v_2^2 \bigg)
	\nn\\& \qquad \qquad \; + \frac{G^3 m_1 m_2^2}{r_{12}^4}
	\bigg( \frac{454}{15} (n_{12} v_1)^2 - \frac{372}{5} (n_{12} v_1)
	(n_{12} v_2) + \frac{854}{15} (n_{12} v_2)^2 - \frac{152}{21} v_1^2
	\nn\\& \qquad \qquad\ \quad + \frac{2864}{105}
	(v_1 v_2) - \frac{1768}{105} v_2^2 \bigg) \nn\\& \qquad
	\qquad \; + \frac{G^2 m_1 m_2}{r_{12}^3} \bigg( 60 (n_{12} v_{12})^4
	- \frac{348}{5} (n_{12} v_1)^2 v_{12}^2 + \frac{684}{5} (n_{12} v_1)
	(n_{12} v_2) v_{12}^2 \nn\\& \qquad \qquad \quad - 66 (n_{12} v_2)^2 v_{12}^2 + \frac{334}{35} v_1^4 -
	\frac{1336}{35} v_1^2 (v_1 v_2) + \frac{1308}{35} (v_1 v_2)^2 \nn\\& \qquad \qquad 
	+
	\frac{654}{35} v_1^2 v_2^2 - \frac{1252}{35} (v_1 v_2) v_2^2 + \frac{292}{35}
	v_2^4 \bigg) \bigg] \bm{v}_{12} \Bigg\} + \calO \left(
	\frac{1}{c^8} \right)\,.
	\label{acc3PN}
\end{align}
The harmonic-coordinates equations of motion depend on an arbitrary length scale $r'_0$ associated with the logarithms present at the 3PN order. This scale is merely linked with the choice of coordinates -- it can be referred to as a ``gauge constant''. Initially two gauge scales $r'_1$ and $r'_2$ were introduced by \cite{BFeom}; however it is useful to choose $r'_1=r'_2=r'_0$ for a reason given in Sect.~VI of \cite{MHLMFB20}. One can show that the logarithms appearing in Eq.~\eqref{acc3PN}, together with the constant $r'_0$ therein, can be removed by a coordinate transformation, or equivalently at that order by a linear gauge transformation. Therefore, the arbitrariness in the choice of the constant $r'_0$ does not affect the physics, because physical results must be gauge invariant. Indeed we shall verify that $r'_0$ cancels out in the final results.\footnote{Notice the dependence upon the irrational number $\pi^2$. Technically, the $\pi^2$ terms arise from non-linear interactions involving some integrals in 3 dimensions such as
\begin{align*}
		\frac{1}{\pi}\int\frac{\dd^3\mathbf{x}}{r_1^2r_2^2} =
		\frac{\pi^2}{r_{12}}\,.
\end{align*}
}

When retaining the ``even'' relativistic corrections at the 1PN, 2PN and 3PN orders, and neglecting the ``odd'' radiation reaction terms at the 2.5PN and 3.5PN orders, we find that the equations of motion admit a conserved energy (and a Lagrangian as well); that energy can be straightforwardly obtained by guess-work starting from Eq.~\eqref{acc3PN}. When taking into account the radiation reaction, the energy will enter the left-hand side of the ``flux-balance'' equation, with the gravitational-wave energy flux in the right-hand side. Several equivalent choices can be made at this stage. The ``canonical'' choice is to present the flux into manifestly positive definite form, such as Eq.~\eqref{fluxE} at Newtonian order, or, more generally, Eqs. \eqref{FluxFG}. Then we have to transfer to the left-hand side of the balance equation some odd-parity terms in the form of total time derivatives. Such time derivative represent the analogue of the \cite{Schott} terms in electromagnetism. Thus we include a contribution to the energy $E$ at the 2.5PN order, coming from the decision to express the flux into canonical form, i.e., the familiar Einstein quadrupole formula at 2.5PN order. We obtain 
\begin{align}
  \dE &= \frac{ m_1 v_1^2}{2} - \frac{G m_1 m_2}{2 r_{12}} \nn
  \\ & 
+ \frac{1}{c^2} \left\{ \frac{G^2 m_1^2 m_2}{2 r_{12}^2} +
  \frac{3 m_1 v_1^4}{8} + \frac{G m_1 m_2}{r_{12}} \left( -
  \frac{1}{4} (n_{12}v_1) (n_{12}v_2) + \frac{3}{2} v_1^2 -
  \frac{7}{4} (v_1v_2) \right) \right\} \nn\\& 
+ \frac{1}{c^4}
  \Bigg\{ - \frac{G^3 m_1^3 m_2}{2 r_{12}^3} - \frac{19 G^3 m_1^2
    m_2^2}{8 r_{12}^3} + \frac{5 m_1 v_1^6}{16} \nn\\& \qquad
  ~\, + \frac{G m_1 m_2}{r_{12}} \bigg( \frac{3}{8} (n_{12}v_1)^3
  (n_{12}v_2) + \frac{3}{16} (n_{12}v_1)^2 (n_{12}v_2)^2 - \frac{9}{8}
  (n_{12}v_1) (n_{12}v_2) v_1^2 \nn\\& \qquad \qquad
  \quad \: - \frac{13}{8} (n_{12}v_2)^2 v_1^2 + \frac{21}{8} v_1^4 +
  \frac{13}{8} (n_{12}v_1)^2 (v_1v_2) + \frac{3}{4} (n_{12}v_1)
  (n_{12}v_2) (v_1v_2) \nn\\& \qquad \qquad \quad \: -
  \frac{55}{8} v_1^2 (v_1v_2) + \frac{17}{8} (v_1v_2)^2 +
  \frac{31}{16} v_1^2 v_2^2 \bigg) \nn\\& \qquad ~\, +
  \frac{G^2 m_1^2 m_2}{r_{12}^2} \left( \frac{29}{4} (n_{12}v_1)^2 -
  \frac{13}{4} (n_{12}v_1) (n_{12}v_2) + \frac{1}{2}(n_{12}v_2)^2 -
  \frac{3}{2} v_1^2 + \frac{7}{4} v_2^2 \right) \Bigg\} \nn\\&
+ \frac{1}{c^5} \Bigg\{ \frac{4 G^2 m_1^2 m_2}{5 r_{12}^2} (n_{12}v_1)
\left( v_{12}^2 -\frac{2 G (m_1-m_2)}{r_{12}} \right)\Bigg\} \nn\\&
+ \frac{1}{c^6} \Bigg\{ \frac{35 m_1 v_1^8}{128} \nn\\&
  \qquad ~\, + \frac{G m_1 m_2}{r_{12}} \bigg( \!\! - \frac{5}{16}
  (n_{12}v_1)^5 (n_{12}v_2) - \frac{5}{16} (n_{12}v_1)^4 (n_{12}v_2)^2
  - \frac{5}{32} (n_{12}v_1)^3 (n_{12}v_2)^3 \nn\\& \qquad \qquad \quad + \frac{19}{16} (n_{12}v_1)^3 (n_{12}v_2) v_1^2
  + \frac{15}{16} (n_{12}v_1)^2 (n_{12}v_2)^2 v_1^2 + \frac{3}{4}
  (n_{12}v_1) (n_{12}v_2)^3 v_1^2 \nn\\& \qquad \qquad
  \quad + \frac{19}{16} (n_{12}v_2)^4 v_1^2 - \frac{21}{16}
  (n_{12}v_1) (n_{12}v_2) v_1^4 - 2 (n_{12}v_2)^2 v_1^4 \nn\\&
  \qquad \qquad \quad + \frac{55}{16} v_1^6 - \frac{19}{16}
  (n_{12}v_1)^4 (v_1v_2) - (n_{12}v_1)^3 (n_{12}v_2) (v_1v_2)
  \nn\\& \qquad \qquad \quad - \frac{15}{32}
  (n_{12}v_1)^2 (n_{12}v_2)^2 (v_1v_2) + \frac{45}{16} (n_{12}v_1)^2
  v_1^2 (v_1v_2) \nn\\& \qquad \qquad \quad +
  \frac{5}{4} (n_{12}v_1) (n_{12}v_2) v_1^2 (v_1v_2) +
  \frac{11}{4}(n_{12}v_2)^2 v_1^2 (v_1v_2) - \frac{139}{16} v_1^4
  (v_1v_2) \nn\\& \qquad \qquad \quad - \frac{3}{4}
  (n_{12}v_1)^2 (v_1v_2)^2 + \frac{5}{16} (n_{12}v_1) (n_{12}v_2)
  (v_1v_2)^2 + \frac{41}{8} v_1^2 (v_1v_2)^2 + \frac{1}{16} (v_1v_2)^3
  \nn\\& \qquad \qquad \quad - \frac{45}{16}
  (n_{12}v_1)^2 v_1^2 v_2^2 - \frac{23}{32} (n_{12}v_1) (n_{12}v_2)
  v_1^2 v_2^2 + \frac{79}{16} v_1^4 v_2^2 - \frac{161}{32} v_1^2
  (v_1v_2) v_2^2 \bigg) \nn\\& \qquad ~\, + \frac{G^2 m_1^2
    m_2}{r_{12}^2} \bigg( - \frac{49}{8} (n_{12}v_1)^4 +
  \frac{75}{8} (n_{12}v_1)^3 (n_{12}v_2) - \frac{187}{8} (n_{12}v_1)^2
  (n_{12}v_2)^2 \nn\\& \qquad \qquad \quad ~ +
  \frac{247}{24} (n_{12}v_1) (n_{12}v_2)^3 + \frac{49}{8}
  (n_{12}v_1)^2 v_1^2 + \frac{81}{8} (n_{12}v_1) (n_{12}v_2) v_1^2
  \nn\\& \qquad \qquad \quad ~ - \frac{21}{4}
  (n_{12}v_2)^2 v_1^2 + \frac{11}{2} v_1^4 - \frac{15}{2}
  (n_{12}v_1)^2 (v_1v_2) - \frac{3}{2} (n_{12}v_1) (n_{12}v_2)
  (v_1v_2) \nn\\& \qquad \qquad \quad ~ + \frac{21}{4}
  (n_{12}v_2)^2 (v_1v_2) - 27 v_1^2 (v_1v_2) + \frac{55}{2} (v_1v_2)^2
  + \frac{49}{4} (n_{12}v_1)^2 v_2^2 \nn\\& \qquad
  \qquad \quad ~ - \frac{27}{2} (n_{12}v_1) (n_{12}v_2) v_2^2 +
  \frac{3}{4} (n_{12}v_2)^2 v_2^2 + \frac{55}{4} v_1^2 v_2^2 - 28
  (v_1v_2) v_2^2 + \frac{135}{16} v_2^4 \bigg) \nn\\& \qquad
  ~\, + \frac{3 G^4 m_1^4 m_2}{8 r_{12}^4} + \frac{G^4 m_1^3
    m_2^2}{r_{12}^4} \left( \frac{9707}{420} - \frac{22}{3} \ln \left(
  \frac{r_{12}}{r'_0} \right) \! \right) \nn\\& \qquad ~\, +
  \frac{G^3 m_1^2 m_2^2}{r_{12}^3} \bigg( \frac{547}{12} (n_{12}v_1)^2
  - \frac{3115}{48} (n_{12}v_1) (n_{12}v_2) - \frac{123}{64}
  (n_{12}v_1)(n_{12}v_{12}) \pi^2 \nn
  \\ & \qquad \qquad \quad ~ - \frac{575}{18} v_1^2 + \frac{41}{64} \pi^2 (v_1v_{12})
  + \frac{4429}{144} (v_1v_2) \bigg) \nn\\& \qquad ~\, +
  \frac{G^3 m_1^3 m_2}{r_{12}^3} \bigg( \!\! - \frac{44627}{840}
  (n_{12}v_1)^2 + \frac{32027}{840} (n_{12}v_1) (n_{12}v_2) +
  \frac{3}{2} (n_{12}v_2)^2 \nn\\&
  \qquad \qquad \quad ~ + \frac{24187}{2520} v_1^2 - \frac{27967}{2520} (v_1v_2) +
  \frac{5}{4} v_2^2 + 22 (n_{12}v_1)(n_{12}v_{12}) \ln
  \!\left(\!\frac{r_{12}}{r'_0}\!\right)\! \nn\\&
  \qquad \qquad \quad ~ - \frac{22}{3} (v_1v_{12})
  \ln \!\left(\!\frac{r_{12}}{r'_0}\!\right) \! \bigg) \!\! \Bigg\}
  + 1 \leftrightarrow 2 +
  \calO\left(\frac{1}{c^7} \right)\,.
  \label{en3PN}
\end{align}
To the terms given above, we must add the same terms but corresponding to the relabelling $1\leftrightarrow 2$. The time derivative of this energy, as computed by means of the 3PN equations of motion themselves (i.e., by order-reducing all the accelerations), is purely equal to the 2.5PN effect, and agrees with the standard Einstein quadrupole formula \eqref{fluxE}:
\begin{equation}
	\frac{\dd \dE}{\dd t} = -\frac{G}{5 c^5} \frac{\dd^3
		\mathrm{Q}_{ij}}{\dd t^3} \frac{\dd^3 \mathrm{Q}_{ij}}{\dd t^3} +
	\calO\left( \frac{1}{c^7} \right)\,,
	\label{dEdtquad}
\end{equation}
where the Newtonian trace-free quadrupole moment reads $\mathrm{Q}_{ij}=m_1 (y_1^i y_1^j-\frac{1}{3}\delta^{ij} \bm{y}_1^2)+1 \leftrightarrow 2$. We refer to \cite{IW93, IW95, GII97} for discussions of the energy balance equation to the next 3.5PN and 4.5PN orders. See also Eq.~\eqref{balance15PN} for the energy balance equation at 4PN order for general fluid systems.


\subsubsection{The 4PN Lagrangian and Hamiltonian}
\label{sec:Lag4PN}

The conservative part of the equations of motion in harmonic coordinates \eqref{acc3PN} is derivable from a \emph{generalized} Lagrangian, depending not only on the positions and velocities of the bodies, but also on their accelerations: $\bm{a}_1=\dd \bm{v}_1/\dd t$ and $\bm{a}_2=\dd \bm{v}_2/\dd t$, the accelerations occuring already from the 2PN order \citep{DD81b}. A general result by \cite{MS} states that $N$-body equations of motion cannot be derived from an ordinary Lagrangian beyond the 1PN level, when the gauge conditions preserve the manifest Lorentz invariance. One can always arrange for the dependence of the Lagrangian on accelerations to be linear, at the price of adding some ``multi-zero'' terms to the Lagrangian, which do not modify the equations of motion \citep{DS85}.\footnote{The method is a particular application of a general algorithm to eliminate higher-derivative terms in a Lagrangian \citep{DS91}.} The Lagrangian of two particles in harmonic coordinates at 4PN order is really the Fokker Lagrangian $L\equiv L_\text{F}$ of Sect.~\ref{sec:Fokker}. It has been derived using the traditional method by \cite{BBBFMa,BBBFMb,BBBFMc,BBFM17}, and by \cite{GLPR16,FS4PN,FPRS19} within the EFT approach. The long explicit results are presented in the form of the sequence of PN coefficients,
\begin{equation}\label{notationL}
	L = L_\text{N} +
	\frac{1}{c^2}L_\text{1PN} +
	\frac{1}{c^4}L_\text{2PN} +
	\frac{1}{c^6}L_\text{3PN} +
	\frac{1}{c^8}L_\text{4PN} + \calO\left(\frac{1}{c^{10}}\right)\,.
\end{equation}
For convenience we do not write the irrelevant rest-mass contribution. Up to 3PN order the results read
\begin{subequations}\label{L3PN}
\begin{align}
			L_\text{N} &= \frac{m_1 v_1^2}{2} + \frac{G m_1 m_2}{2 r_{12}} +  1
			\leftrightarrow 2\,,\\
			L_\text{1PN} &= \frac{m_1
				v_1^4}{8} - \frac{G^2 m_1^2 m_2}{2 r_{12}^2} \nn\\& + \frac{G m_1 m_2}{r_{12}} \left( -
			\frac{1}{4} (n_{12}v_1) (n_{12}v_2) + \frac{3}{2} v_1^2 - \frac{7}{4}
			(v_1v_2) \right) + 1 \leftrightarrow 2\,,\\
			L_\text{2PN} &= \frac{m_1
				v_1^6}{16} + \frac{G^3 m_1^3 m_2}{2 r_{12}^3} + \frac{19 G^3 m_1^2
				m_2^2}{8 r_{12}^3} \nn\\&  + \frac{G^2 m_1^2
				m_2}{r_{12}^2} \left( \frac{7}{2} (n_{12}v_1)^2 - \frac{7}{2}
			(n_{12}v_1) (n_{12}v_2) + \frac{1}{2}(n_{12}v_2)^2 \right.\nn\\&\qquad\left.+ \frac{1}{4} v_1^2
			- \frac{7}{4} (v_1v_2) + \frac{7}{4} v_2^2 \right) \nn\\&
			+ \frac{G m_1 m_2}{r_{12}} \bigg( \frac{3}{16}
			(n_{12}v_1)^2 (n_{12}v_2)^2 - \frac{7}{8} (n_{12}v_2)^2 v_1^2 +
			\frac{7}{8} v_1^4 \nn\\&\qquad+ \frac{3}{4} (n_{12}v_1) (n_{12}v_2) (v_1v_2) - 2 v_1^2 (v_1v_2) +
			\frac{1}{8} (v_1v_2)^2 + \frac{15}{16} v_1^2 v_2^2 \bigg) \nn\\&  + G m_1 m_2 \left( -
			\frac{7}{4} (a_1 v_2) (n_{12}v_2) - \frac{1}{8} (n_{12} a_1)
			(n_{12}v_2)^2 + \frac{7}{8} (n_{12} a_1) v_2^2 \right) \nn\\&+ 1
			\leftrightarrow 2\,,\\
			L_\text{3PN} &= \frac{5m_1 v_1^8}{128} \nn\\&+ \frac{G^2 m_1^2 m_2}{r_{12}^2} \bigg( \frac{13}{18}
			(n_{12}v_1)^4 + \frac{83}{18} (n_{12}v_1)^3 (n_{12}v_2) - \frac{35}{6}
			(n_{12}v_1)^2 (n_{12}v_2)^2 \nn\\&\qquad - \frac{245}{24} (n_{12}v_1)^2 v_1^2 + \frac{179}{12} (n_{12}v_1) (n_{12}v_2) v_1^2 -
			\frac{235}{24} (n_{12}v_2)^2 v_1^2 + \frac{373}{48} v_1^4 \nn\\&\qquad+
			\frac{529}{24} (n_{12}v_1)^2 (v_1v_2) - \frac{97}{6}
			(n_{12}v_1) (n_{12}v_2) (v_1v_2) - \frac{719}{24} v_1^2 (v_1v_2) \nn\\&\qquad+
			\frac{463}{24} (v_1v_2)^2 - \frac{7}{24} (n_{12}v_1)^2 v_2^2 - \frac{1}{2} (n_{12}v_1) (n_{12}v_2) v_2^2 + \frac{1}{4}
			(n_{12}v_2)^2 v_2^2 \nn\\&\qquad+ \frac{463}{48} v_1^2 v_2^2 - \frac{19}{2}
			(v_1v_2) v_2^2 + \frac{45}{16} v_2^4 \bigg) \nn\\& + G m_1 m_2
			\bigg(\frac{3}{8} (a_1 v_2) (n_{12}v_1) (n_{12}v_2)^2 + \frac{5}{12}
			(a_1 v_2) (n_{12}v_2)^3 \nn\\&\qquad+ \frac{1}{8} (n_{12} a_1) (n_{12}v_1)
			(n_{12}v_2)^3 + \frac{1}{16} (n_{12} a_1) (n_{12}v_2)^4
			+ \frac{11}{4} (a_1 v_1) (n_{12}v_2) v_1^2 \nn\\&\qquad- (a_1 v_2) (n_{12}v_2)
			v_1^2 - 2 (a_1 v_1) (n_{12}v_2) (v_1v_2) + \frac{1}{4}
			(a_1 v_2) (n_{12}v_2) (v_1v_2) \nn\\&\qquad + \frac{3}{8} (n_{12}
			a_1) (n_{12}v_2)^2 (v_1v_2) - \frac{5}{8} (n_{12} a_1) (n_{12}v_1)^2
			v_2^2 + \frac{15}{8} (a_1 v_1) (n_{12}v_2) v_2^2 \nn\\&\qquad -
			\frac{15}{8} (a_1 v_2) (n_{12}v_2) v_2^2 - \frac{1}{2} (n_{12} a_1)
			(n_{12}v_1) (n_{12}v_2) v_2^2 \nn\\&\qquad - \frac{5}{16} (n_{12}
			a_1) (n_{12}v_2)^2 v_2^2 \bigg)  \nn\\& + \frac{G^2 m_1^2 m_2}{r_{12}} \bigg( - \frac{235}{24} (a_2 v_1)
			(n_{12}v_1) - \frac{29}{24} (n_{12} a_2) (n_{12}v_1)^2 \nn\\&\qquad-
			\frac{235}{24} (a_1 v_2) (n_{12}v_2) - \frac{17}{6}
			(n_{12} a_1) (n_{12}v_2)^2 + \frac{185}{16} (n_{12} a_1) v_1^2 \nn\\&\qquad-
			\frac{235}{48} (n_{12} a_2) v_1^2  - \frac{185}{8}
			(n_{12} a_1) (v_1v_2) + \frac{20}{3} (n_{12} a_1) v_2^2 \bigg)
			\nn\\& + \frac{G m_1 m_2}{r_{12}} \bigg( - \frac{5}{32}
			(n_{12}v_1)^3 (n_{12}v_2)^3 + \frac{1}{8} (n_{12}v_1) (n_{12}v_2)^3
			v_1^2 + \frac{5}{8} (n_{12}v_2)^4 v_1^2 \nn\\&\qquad - \frac{11}{16}
			(n_{12}v_1) (n_{12}v_2) v_1^4 + \frac{1}{4} (n_{12}v_2)^2 v_1^4 +
			\frac{11}{16} v_1^6 \nn\\&\qquad - \frac{15}{32} (n_{12}v_1)^2
			(n_{12}v_2)^2 (v_1v_2) + (n_{12}v_1) (n_{12}v_2) v_1^2 (v_1v_2)
			\nn\\&\qquad + \frac{3}{8} (n_{12}v_2)^2 v_1^2 (v_1v_2) -
			\frac{13}{16} v_1^4 (v_1v_2) + \frac{5}{16} (n_{12}v_1) (n_{12}v_2)
			(v_1v_2)^2 \nn\\&\qquad + \frac{1}{16} (v_1v_2)^3 - \frac{5}{8}
			(n_{12}v_1)^2 v_1^2 v_2^2 - \frac{23}{32} (n_{12}v_1) (n_{12}v_2)
			v_1^2 v_2^2 + \frac{1}{16} v_1^4 v_2^2 \nn\\&\qquad - \frac{1}{32}
			v_1^2 (v_1v_2) v_2^2 \bigg) \nn\\& - \frac{3 G^4 m_1^4 m_2}{8
				r_{12}^4} + \frac{G^4 m_1^3 m_2^2}{r_{12}^4} \left( -
			\frac{9707}{420} + \frac{22}{3} \ln \left(\frac{r_{12}}{r'_0} \right)
			\right) \nn\\& + \frac{G^3 m_1^2 m_2^2}{r_{12}^3} \bigg(
			\frac{383}{24} (n_{12}v_1)^2 - \frac{889}{48} (n_{12}v_1) (n_{12}v_2)
			\nn\\&\qquad - \frac{123}{64} (n_{12}v_1)(n_{12}v_{12}) \pi^2 -
			\frac{305}{72} v_1^2 + \frac{41}{64} \pi^2 (v_1v_{12}) +
			\frac{439}{144} (v_1v_2) \bigg) \nn\\& + \frac{G^3 m_1^3
				m_2}{r_{12}^3} \bigg( - \frac{8243}{210} (n_{12}v_1)^2 +
			\frac{15541}{420} (n_{12}v_1) (n_{12}v_2) + \frac{3}{2} (n_{12}v_2)^2
			\nn\\&\qquad+ \frac{15611}{1260} v_1^2  - \frac{17501}{1260}
			(v_1v_2) + \frac{5}{4} v_2^2 + 22 (n_{12}v_1)(n_{12}v_{12}) \ln \left(
			\frac{r_{12}}{r'_0} \right) \nn\\&\qquad - \frac{22}{3} (v_1v_{12})
			\ln \left( \frac{r_{12}}{r'_0} \right) \bigg) + 1 \leftrightarrow 2\,.
\end{align}
\end{subequations}
where the logarithms in harmonic coordinates contain the gauge constant $r'_0$, and the accelerations start appearing at 2PN order [recall the notation, e.g. $(a_1v_2)=\bm{a}_1\cdot\bm{v}_2$]. We refer to \cite{ABF01} for the expressions of the ten conserved quantities corresponding to the integrals of energy [given in Eq.~\eqref{en3PN}], linear and angular momenta, and center-of-mass position at 3PN order. Note that while it is strictly forbidden to replace the accelerations by the equations of motion in the Lagrangian, this can and \emph{should} be done in the final expressions of the conserved integrals derived from the Lagrangian.

Now we exhibit a transformation of the particles' dynamical variables -- or \emph{contact} transformation -- which transforms the harmonic-coordinates 3PN Lagrangian \eqref{L3PN} into a new Lagrangian, valid in some ADM or ADM-like coordinate system, and such that the associated Hamiltonian coincides with the 3PN Hamiltonian obtained by \cite{JaraS98, JaraS99}. In ADM coordinates the Lagrangian will be ordinary, depending only on the positions and velocities of the bodies. Let this contact transformation be $\bm{Y}_1(t)=\bm{y}_1(t)+\delta \bm{y}_1(t)$ and $1\leftrightarrow 2$, where $\bm{Y}_1$ and $\bm{y}_1$ denote the trajectories in ADM and harmonic coordinates, respectively. For this transformation to be able to remove the accelerations in the Lagrangian \eqref{L3PN} at 3PN order, we determine it to be necessarily of the form \citep{ABF01}
\begin{equation}
	\delta \bm{y}_1=\frac{1}{m_1} \left[\frac{\partial
		L}{\partial \bm{a}_1}+ \frac{\partial F}{\partial
		\bm{v}_1}+\frac{1}{c^6}\bm{X}_1\right]+
	\calO\left(\frac{1}{c^8}\right) \,,
	\label{contacttransf}
\end{equation}
and \textit{idem} for $1\leftrightarrow 2$, where $F$ is a freely adjustable function of the positions and velocities, made of 2PN and 3PN terms, and where $\bm{X}_1$ represents a special correction term, that is purely of order 3PN. The point is that once the function $F$ is specified there is a unique determination of the correction term $\bm{X}_1$ for the contact transformation to work \citep{ABF01}. Thus, the freedom we have is entirely encoded into the function $F$, and the work then consists in showing that there exists a unique choice of $F$ for which the Lagrangian \eqref{L3PN} is physically equivalent, via the contact transformation \eqref{contacttransf}, to the ADM Hamiltonian of \cite{JaraS98, JaraS99}. An interesting point is that not only the transformation must remove all the accelerations, but it should also cancel out all the logarithms $\ln(r_{12}/r'_0)$, because there are no logarithms in ADM coordinates. As a result, $F$ involves the logarithmic terms\footnote{Hence the logarithms at 3PN order can be removed by the coordinate transformation $x'^\alpha = x^\alpha + \eta^\alpha(x)$ associated with the vector (with $r_\text{a} = \vert\mathbf{ x}-\bm{y}_\text{a}(t)\vert$ and $\partial^\alpha=\eta^{\alpha\mu}\partial_\mu$):
\begin{align*}
	\eta^\alpha=-\frac{22}{3}\frac{G^2 m_1 m_2}{c^6}\,\partial^\alpha\!\left(\frac{G m_2}{r_1}+ \frac{G m_1}{r_2}\right)\ln\left(\frac{r_{12}}{r'_0}\right)\,.
	\label{gaugeeta}
\end{align*}
\label{fnote:log3PN}}
\begin{equation}
	F = \frac{22}{3}\frac{G^3 m_1 m_2}{c^6r_{12}^2}
	\Bigl[m_1^2(n_{12}v_1) -
	m_2^2(n_{12}v_2)\Bigr]\ln\left(\frac{r_{12}}{r'_0}\right)+\cdots\,,
	\label{Flog}
\end{equation}
together with other non-logarithmic terms (indicated by dots) that are entirely specified by the isometry of the harmonic and ADM descriptions of the motion. The function $F$ and correction term $\bm{X}_1$ are given by Eqs.~(4.9)--(4.10) of \cite{ABF01}. For these choices the ADM Lagrangian becomes
\begin{equation}
	L^\mathrm{ADM} = L+\frac{\delta L}{\delta
		y_1^i}\delta y_1^i+\frac{\delta L}{\delta
		y_2^i}\delta
	y_2^i+\frac{\dd F}{\dd t}+\calO\left(\frac{1}{c^8}\right)\,,
	\label{LADMform}
\end{equation}
and the ADM Hamiltonian is deduced by an ordinary Legendre transformation. The procedure can easily be generalized to 4PN order.

Up to 3PN order all terms were ``instantaneous'', as they depend on the system at current time only, see Eqs. \eqref{L3PN}. However at 4PN order a new feature arises: the non-locality (in time) due to infinite-range temporal correlations of gravitational wave tails \citep{BD88} and computed in the Fokker Lagrangian in Sect.~\ref{sec:Fokker}. Accordingly we split the 4PN coefficient as
\begin{equation}\label{notation4PN}
	L_\text{4PN} = L^\text{inst}_\text{4PN} + L^\text{tail}_\text{4PN}\,,
\end{equation}
where the tail term reads explicitly
\begin{align}\label{Ltail4PN}
	L^\text{tail}_\text{4PN} = \frac{G^2\dM}{5c^8}
	\,\dM_{ij}^{(3)}(t) \int_0^{+\infty} \! \dd\tau \,
	\ln{\left( \frac{c\tau}{2 r_{12}} \right)} \left[ \dM_{ij}^{(4)}(t-\tau)
	- \dM_{ij}^{(4)}(t+\tau) \right]\,. 
\end{align}
There is, though, an important difference with Eqs. \eqref{StailDR1}--\eqref{StailDR2}, is that the Hadamard cut-off scale $c\tau_0$ has been replaced by (two times) the physical separation between the two bodies, $r_{12}=\vert\bm{y}_1-\bm{y}_2\vert$. Therefore when varying the tail term \eqref{Ltail4PN} [remind that this is a functional variation] one has also to take into account the variation of the ``constant'' $r_{12}(t)$. See Eq.~(4.3) of \cite{BBFM17} for the result of this variation. Note also that this variation concerns the conservative part of the acceleration; in addition one has to take into account the dissipative part. See Eq.~\eqref{acctailCM} for the complete result in the center-of-mass frame, including both conservative and dissipative contributions. 

As for the instantaneous part of the 4PN term, we split it into non-linear contributions corresponding to increasing powers of $G$:
\begin{equation}\label{notationLG}
	L^\text{inst}_\text{4PN} = L^{(0)}_\text{4PN} + G\,L^{(1)}_\text{4PN} + G^2 L^{(2)}_\text{4PN} + G^3 L^{(3)}_\text{4PN}+ G^4 L^{(4)}_\text{4PN}+ G^5 L^{(5)}_\text{4PN} \,,
\end{equation}
where each of the terms reads explicitly
\begin{subequations}\label{resultL4PN}
\begin{align}
L_\text{4PN}^{(0)}&= \frac{7}{256} m_1 v_1^{10}		
 + 1 \leftrightarrow 2\,,\\
L_\text{4PN}^{(1)}&= m_1 m_2 \left(\frac{13}{64} (a_2 v_1) (n_{12} v_1)^5
	+ \frac{5}{128} (a_2 n_{12}) (n_{12} v_1)^6
	-  \frac{13}{64} (n_{12} v_1)^5 (a_2 v_2)\right.\nn\\&\quad + \frac{11}{64} (a_2 v_1) (n_{12} v_1)^4 (n_{12} v_2)
	+ \frac{5}{64} (a_2 n_{12}) (n_{12} v_1)^5 (n_{12} v_2)\nn\\&\quad + \frac{5}{32} (a_2 v_1) (n_{12} v_1)^3 (n_{12} v_2)^2
	-  \frac{5}{32} (a_1 n_{12}) (n_{12} v_1)^3 (n_{12} v_2)^3\nn\\&\quad -  \frac{1}{16} (a_2 v_1) (n_{12} v_1)^3 (v_1 v_2)
	+ \frac{11}{64} (a_2 n_{12}) (n_{12} v_1)^4 (v_1 v_2)\nn\\&\quad + \frac{5}{16} (a_2 n_{12}) (n_{12} v_1)^3 (n_{12} v_2) (v_1 v_2)
	+ \frac{5}{16} (n_{12} v_1)^2 (a_1 v_2) (n_{12} v_2) (v_1 v_2)\nn\\&\quad -  \frac{3}{16} (a_2 v_1) (n_{12} v_1) (v_1 v_2)^2
	+ \frac{1}{16} (a_1 v_1) (n_{12} v_2) (v_1 v_2)^2\nn\\&\quad + \frac{5}{16} (a_1 n_{12}) (n_{12} v_1) (n_{12} v_2) (v_1 v_2)^2
	-  \frac{77}{96} (a_2 v_1) (n_{12} v_1)^3 v_1^{2}
	\nn\\&\quad-  \frac{27}{128} (a_2 n_{12}) (n_{12} v_1)^4 v_1^{2} + \frac{77}{96} (n_{12} v_1)^3 (a_2 v_2) v_1^{2}
	\nn\\&\quad-  \frac{13}{32} (a_2 v_1) (n_{12} v_1)^2 (n_{12} v_2) v_1^{2}
	-  \frac{11}{32} (a_2 n_{12}) (n_{12} v_1)^3 (n_{12} v_2) v_1^{2}\nn\\&\quad -  \frac{7}{32} (a_2 v_1) (n_{12} v_1) (n_{12} v_2)^2 v_1^{2}
	-  \frac{27}{64} (a_2 n_{12}) (n_{12} v_1)^2 (n_{12} v_2)^2 v_1^{2}
	\nn\\&\quad-  \frac{19}{32} (a_1 v_1) (n_{12} v_2)^3 v_1^{2}-  \frac{3}{16} (a_2 v_1) (n_{12} v_1) (v_1 v_2) v_1^{2}
	\nn\\&\quad-  \frac{13}{32} (a_2 n_{12}) (n_{12} v_1)^2 (v_1 v_2) v_1^{2}+ \frac{33}{16} (n_{12} v_1) (a_2 v_2) (v_1 v_2) v_1^{2}
	\nn\\&\quad-  \frac{7}{16} (a_2 n_{12}) (n_{12} v_1) (n_{12} v_2) (v_1 v_2) v_1^{2}+ \frac{1}{16} (a_1 v_2) (n_{12} v_2) (v_1 v_2) v_1^{2}
	\nn\\&\quad+ \frac{123}{64} (a_2 v_1) (n_{12} v_1) v_1^{4}
	+ \frac{53}{128} (a_2 n_{12}) (n_{12} v_1)^2 v_1^{4}-  \frac{123}{64} (n_{12} v_1) (a_2 v_2) v_1^{4}
	\nn\\&\quad+ \frac{49}{64} (a_2 v_1) (n_{12} v_2) v_1^{4}
	+ \frac{31}{64} (a_2 n_{12}) (n_{12} v_1) (n_{12} v_2) v_1^{4}\nn\\&\quad + \frac{49}{64} (a_2 n_{12}) (v_1 v_2) v_1^{4}
	-  \frac{75}{128} (a_2 n_{12}) v_1^{6}
	+ \frac{17}{96} (n_{12} v_1)^3 (a_1 v_2) v_2^{2}\nn\\&\quad -  \frac{23}{32} (a_1 v_1) (n_{12} v_1)^2 (n_{12} v_2) v_2^{2}
	+ \frac{15}{32} (a_1 n_{12}) (n_{12} v_1)^3 (n_{12} v_2) v_2^{2}\nn\\&\quad + \frac{17}{32} (a_1 n_{12}) (n_{12} v_1)^2 (v_1 v_2) v_2^{2}
	+ \frac{33}{32} (a_2 v_1) (n_{12} v_1) v_1^{2} v_2^{2}
	\nn\\&\quad\left.+ \frac{93}{32} (a_1 v_1) (n_{12} v_2) v_1^{2} v_2^{2}-  \frac{23}{32} (a_1 n_{12}) (n_{12} v_1) (n_{12} v_2) v_1^{2} v_2^{2}\right)\nn\\&
+ \frac{m_1 m_2}{r_{12}} \left(\frac{35}{128} (n_{12} v_1)^5 (n_{12} v_2)^3
	-  \frac{35}{256} (n_{12} v_1)^4 (n_{12} v_2)^4\right.\nn\\&\quad -  \frac{15}{128} (n_{12} v_1)^4 (n_{12} v_2)^2 (v_1 v_2)
	+ \frac{15}{32} (n_{12} v_1)^3 (n_{12} v_2)^3 (v_1 v_2)\nn\\&\quad -  \frac{15}{32} (n_{12} v_1)^3 (n_{12} v_2) (v_1 v_2)^2
	+ \frac{5}{32} (n_{12} v_1)^2 (v_1 v_2)^3
	-  \frac{3}{16} (n_{12} v_1) (n_{12} v_2) (v_1 v_2)^3\nn\\&\quad + \frac{1}{32} (v_1 v_2)^4
	-  \frac{5}{32} (n_{12} v_1)^3 (n_{12} v_2)^3 v_1^{2}
	-  \frac{5}{16} (n_{12} v_1) (n_{12} v_2)^3 (v_1 v_2) v_1^{2}\nn\\&\quad + \frac{9}{32} (n_{12} v_1) (n_{12} v_2) (v_1 v_2)^2 v_1^{2}
	-  \frac{15}{32} (n_{12} v_2)^2 (v_1 v_2)^2 v_1^{2}
	+ \frac{1}{32} (v_1 v_2)^3 v_1^{2}\nn\\&\quad + \frac{57}{128} (n_{12} v_1) (n_{12} v_2)^3 v_1^{4}
	-  \frac{15}{128} (n_{12} v_2)^4 v_1^{4}
	+ \frac{39}{128} (n_{12} v_2)^2 (v_1 v_2) v_1^{4}
	+ \frac{3}{4} (v_1 v_2)^2 v_1^{4}\nn\\&\quad -  \frac{11}{32} (n_{12} v_2)^2 v_1^{6}
	-  \frac{5}{4} (v_1 v_2) v_1^{6}
	+ \frac{75}{128} v_1^{8}
	-  \frac{75}{128} (n_{12} v_1)^5 (n_{12} v_2) v_2^{2}\nn\\&\quad -  \frac{53}{128} (n_{12} v_1)^4 (v_1 v_2) v_2^{2}
	+ \frac{99}{64} (n_{12} v_1)^3 (n_{12} v_2) v_1^{2} v_2^{2}
	-  \frac{21}{64} (n_{12} v_1)^2 (n_{12} v_2)^2 v_1^{2} v_2^{2}\nn\\&\quad + \frac{11}{64} (n_{12} v_1)^2 (v_1 v_2) v_1^{2} v_2^{2}
	+ \frac{35}{32} (n_{12} v_1) (n_{12} v_2) (v_1 v_2) v_1^{2} v_2^{2}
	-  \frac{1}{32} (v_1 v_2)^2 v_1^{2} v_2^{2}\nn\\&\quad -  \frac{185}{128} (n_{12} v_1) (n_{12} v_2) v_1^{4} v_2^{2}
	+ \frac{23}{64} (n_{12} v_2)^2 v_1^{4} v_2^{2}
	-  \frac{99}{128} (v_1 v_2) v_1^{4} v_2^{2}
	+ \frac{5}{8} v_1^{6} v_2^{2}\nn\\&\quad\left. + \frac{3}{256} v_1^{4} v_2^{4}\right)
	+ 1 \leftrightarrow 2\,,\\
L_\text{4PN}^{(2)}&= \frac{m_1^2 m_2}{r_{12}} \left(\frac{2099}{288} (a_1 v_1) (n_{12} v_1)^3
			+ \frac{3341}{480} (a_2 v_1) (n_{12} v_1)^3
			+ \frac{59}{180} (n_{12} v_1)^3 (a_2 v_2)\right.\nn\\&\quad + \frac{2197}{240} (a_1 n_{12}) (n_{12} v_1)^3 (n_{12} v_2)
			+ \frac{6661}{720} (a_2 n_{12}) (n_{12} v_1)^3 (n_{12} v_2)\nn\\&\quad + \frac{10223}{480} (n_{12} v_1)^2 (a_1 v_2) (n_{12} v_2)
			+ \frac{3059}{96} (a_1 v_1) (n_{12} v_1) (n_{12} v_2)^2\nn\\&\quad -  \frac{3781}{160} (n_{12} v_1) (a_1 v_2) (n_{12} v_2)^2
			+ \frac{1337}{240} (a_1 n_{12}) (n_{12} v_1) (n_{12} v_2)^3
			+ \frac{4621}{480} (a_1 v_2) (n_{12} v_2)^3\nn\\&\quad + \frac{1133}{960} (a_1 n_{12}) (n_{12} v_2)^4
			+ \frac{3613}{48} (a_1 v_1) (n_{12} v_1) (v_1 v_2)
			+ \frac{4529}{240} (a_2 v_1) (n_{12} v_1) (v_1 v_2)\nn\\&\quad + \frac{2099}{96} (a_1 n_{12}) (n_{12} v_1)^2 (v_1 v_2)
			-  \frac{23}{60} (a_1 n_{12}) (n_{12} v_1) (n_{12} v_2) (v_1 v_2)\nn\\&\quad + \frac{6499}{240} (a_1 v_2) (n_{12} v_2) (v_1 v_2)
			+ \frac{247}{30} (a_1 n_{12}) (n_{12} v_2)^2 (v_1 v_2)
			+ \frac{2503}{96} (a_1 n_{12}) (v_1 v_2)^2\nn\\&\quad + \frac{7193}{240} (a_2 v_1) (n_{12} v_1) v_1^{2}
			+ \frac{3021}{320} (a_2 n_{12}) (n_{12} v_1)^2 v_1^{2}
			+ \frac{4723}{96} (n_{12} v_1) (a_1 v_2) v_1^{2}\nn\\&\quad + \frac{3679}{480} (n_{12} v_1) (a_2 v_2) v_1^{2}
			+ \frac{13549}{480} (a_1 v_1) (n_{12} v_2) v_1^{2}
			+ \frac{8849}{480} (a_2 v_1) (n_{12} v_2) v_1^{2}\nn\\&\quad + \frac{2063}{96} (a_1 n_{12}) (n_{12} v_1) (n_{12} v_2) v_1^{2}
			+ \frac{166}{15} (a_2 n_{12}) (n_{12} v_1) (n_{12} v_2) v_1^{2}\nn\\&\quad + \frac{4621}{320} (a_2 n_{12}) (n_{12} v_2)^2 v_1^{2}
			+ \frac{8849}{480} (a_2 n_{12}) (v_1 v_2) v_1^{2}
			+ \frac{6943}{384} (a_1 n_{12}) v_1^{4}\nn\\&\quad + \frac{2293}{160} (n_{12} v_1) (a_1 v_2) v_2^{2}
			+ \frac{3733}{160} (a_1 v_1) (n_{12} v_2) v_2^{2}
			+ \frac{7}{5} (a_1 n_{12}) (n_{12} v_1) (n_{12} v_2) v_2^{2}\nn\\&\quad -  \frac{3733}{160} (a_1 v_2) (n_{12} v_2) v_2^{2}
			-  \frac{139}{20} (a_1 n_{12}) (n_{12} v_2)^2 v_2^{2}
			-  \frac{5593}{240} (a_1 n_{12}) (v_1 v_2) v_2^{2}\nn\\&\quad\left. + \frac{3613}{192} (a_1 n_{12}) v_1^{2} v_2^{2}
			+ \frac{2931}{320} (a_1 n_{12}) v_2^{4}\right) \nn\\&
			 + \frac{m_1^2 m_2}{r_{12}^{2}} \left(- \frac{4027}{800} (n_{12} v_1)^6 + \frac{3227}{800} (n_{12} v_1)^5 (n_{12} v_2)
			-  \frac{6301}{240} (n_{12} v_1)^4 (n_{12} v_2)^2 \right.\nn\\&\quad + \frac{6661}{240} (n_{12} v_1)^3 (n_{12} v_2)^3 
			-  \frac{2221}{64} (n_{12} v_1)^4 (v_1 v_2)
			+ \frac{25267}{720} (n_{12} v_1)^3 (n_{12} v_2) (v_1 v_2)\nn\\&\quad -  \frac{6661}{480} (n_{12} v_1)^2 (n_{12} v_2)^2 (v_1 v_2)
			-  \frac{23401}{480} (n_{12} v_1)^2 (v_1 v_2)^2\nn\\&\quad + \frac{4529}{240} (n_{12} v_1) (n_{12} v_2) (v_1 v_2)^2
			-  \frac{8369}{480} (v_1 v_2)^3
			+ \frac{17393}{2880} (n_{12} v_1)^4 v_1^{2}\nn\\&\quad -  \frac{26237}{720} (n_{12} v_1)^3 (n_{12} v_2) v_1^{2}
			+ \frac{4561}{160} (n_{12} v_1)^2 (n_{12} v_2)^2 v_1^{2}
			+ \frac{691}{240} (n_{12} v_1) (n_{12} v_2)^3 v_1^{2}\nn\\&\quad + \frac{4621}{240} (n_{12} v_2)^4 v_1^{2}
			-  \frac{14987}{960} (n_{12} v_1)^2 (v_1 v_2) v_1^{2}
			+ \frac{2601}{160} (n_{12} v_1) (n_{12} v_2) (v_1 v_2) v_1^{2}\nn\\&\quad + \frac{14649}{320} (n_{12} v_2)^2 (v_1 v_2) v_1^{2}
			+ \frac{97}{5} (v_1 v_2)^2 v_1^{2}
			-  \frac{4879}{320} (n_{12} v_1)^2 v_1^{4}\nn\\&\quad + \frac{5399}{192} (n_{12} v_1) (n_{12} v_2) v_1^{4}
			+ \frac{83}{15} (n_{12} v_2)^2 v_1^{4}
			+ \frac{749}{128} (v_1 v_2) v_1^{4}
			+ \frac{20389}{1920} v_1^{6}\nn\\&\quad + \frac{107}{180} (n_{12} v_1)^4 v_2^{2}
			-  \frac{1823}{240} (n_{12} v_1)^3 (n_{12} v_2) v_2^{2}
			-  \frac{1}{2} (n_{12} v_1)^2 (n_{12} v_2)^2 v_2^{2}\nn\\&\quad + \frac{1021}{120} (n_{12} v_1)^2 (v_1 v_2) v_2^{2}
			+ \frac{1}{2} (n_{12} v_1) (n_{12} v_2) (v_1 v_2) v_2^{2}
			+ \frac{67}{4} (v_1 v_2)^2 v_2^{2}\nn\\&\quad -  \frac{21709}{960} (n_{12} v_1)^2 v_1^{2} v_2^{2}
			-  \frac{1873}{480} (n_{12} v_1) (n_{12} v_2) v_1^{2} v_2^{2}
			-  \frac{4621}{320} (n_{12} v_2)^2 v_1^{2} v_2^{2}\nn\\&\quad -  \frac{42017}{960} (v_1 v_2) v_1^{2} v_2^{2}
			+ \frac{11119}{960} v_1^{4} v_2^{2}
			-  \frac{21}{8} (n_{12} v_1)^2 v_2^{4}
			-  \frac{3}{8} (n_{12} v_1) (n_{12} v_2) v_2^{4}\nn\\&\quad\left. + \frac{3}{16} (n_{12} v_2)^2 v_2^{4}
			-  \frac{105}{8} (v_1 v_2) v_2^{4}
			+ \frac{105}{16} v_1^{2} v_2^{4}
			+ \frac{115}{32} v_2^{6}\right)
			\nn\\&+ 1 \leftrightarrow
			2\,,\\
L_\text{4PN}^{(3)}&= \frac{m_1^2 m_2^2}{r_{12}^{2}} \left(- \frac{1099}{144} (a_2 v_1) (n_{12} v_1)
			+ \frac{41}{64} \pi^2 (a_2 v_1) (n_{12} v_1)
			+ \frac{2005}{96} (a_2 n_{12}) (n_{12} v_1)^2 \right. \nn\\&\quad -  \frac{123}{128} \pi^2 (a_2 n_{12}) (n_{12} v_1)^2
			+ \frac{225233}{1800} (a_1 v_1) (n_{12} v_2)
			-  \frac{43}{64} \pi^2 (a_1 v_1) (n_{12} v_2)\nn\\&\quad -  \frac{477941}{3600} (a_1 n_{12}) (v_1 v_2)
			+ \frac{21}{16} \pi^2 (a_1 n_{12}) (v_1 v_2)
			+ \frac{477941}{7200} (a_1 n_{12}) v_1^{2}\nn\\&\quad\left. -  \frac{21}{32} \pi^2 (a_1 n_{12}) v_1^{2}\right) \nn\\&
			 + \frac{m_1^2 m_2^2}{r_{12}^{3}} \left(- \frac{173617}{2880} (n_{12} v_1)^4
			-  \frac{2155}{1024} \pi^2 (n_{12} v_1)^4\right.\nn\\&\quad + \frac{173587}{720} (n_{12} v_1)^3 (n_{12} v_2)
			+ \frac{2155}{256} \pi^2 (n_{12} v_1)^3 (n_{12} v_2)
			-  \frac{85871}{480} (n_{12} v_1)^2 (n_{12} v_2)^2\nn\\&\quad -  \frac{6465}{1024} \pi^2 (n_{12} v_1)^2 (n_{12} v_2)^2
			+ \frac{5651}{300} (n_{12} v_1)^2 (v_1 v_2)
			-  \frac{939}{256} \pi^2 (n_{12} v_1)^2 (v_1 v_2)\nn\\&\quad -  \frac{6851}{300} (n_{12} v_1) (n_{12} v_2) (v_1 v_2)
			+ \frac{939}{256} \pi^2 (n_{12} v_1) (n_{12} v_2) (v_1 v_2)
			+ \frac{49139}{720} (v_1 v_2)^2\nn\\&\quad -  \frac{195}{512} \pi^2 (v_1 v_2)^2
			-  \frac{3677}{1200} (n_{12} v_1)^2 v_1^{2}
			+ \frac{447}{512} \pi^2 (n_{12} v_1)^2 v_1^{2}
			-  \frac{222679}{1200} (n_{12} v_1) (n_{12} v_2) v_1^{2}\nn\\&\quad -  \frac{189}{256} \pi^2 (n_{12} v_1) (n_{12} v_2) v_1^{2}
			+ \frac{153079}{800} (n_{12} v_2)^2 v_1^{2}
			-  \frac{69}{512} \pi^2 (n_{12} v_2)^2 v_1^{2}\nn\\&\quad -  \frac{61733}{900} (v_1 v_2) v_1^{2}
			-  \frac{55}{256} \pi^2 (v_1 v_2) v_1^{2}
			+ \frac{10337}{320} v_1^{4}
			+ \frac{133}{1024} \pi^2 v_1^{4}
			-  \frac{116123}{3600} v_1^{2} v_2^{2}\nn\\&\quad\left. + \frac{477}{1024} \pi^2 v_1^{2} v_2^{2}\right) \nn\\&
			 +  \frac{m_1^3 m_2}{r_{12}^{2}} \left(\frac{44023}{720} (a_1 v_1) (n_{12} v_1)
			+ \frac{562}{9} (a_2 v_1) (n_{12} v_1)\right.\nn\\&\quad + 44 \ln\Big(\frac{r_{12}}{r'_0}\Big) (n_{12} v_1) (a_1 v_2)
			- 44 \ln\Big(\frac{r_{12}}{r'_0}\Big) (n_{12} v_1) (a_2 v_2)
			+ \frac{110}{3} \ln\Big(\frac{r_{12}}{r'_0}\Big) (a_1 v_1) (n_{12} v_2)\nn\\&\quad + \frac{6397}{75} (a_1 n_{12}) (n_{12} v_1) (n_{12} v_2)
			+ \frac{198097}{4200} (a_2 n_{12}) (n_{12} v_1) (n_{12} v_2)\nn\\&\quad + 22 \ln\Big(\frac{r_{12}}{r'_0}\Big) (a_2 n_{12}) (n_{12} v_1) (n_{12} v_2)
			+ \frac{14377}{280} (a_1 v_2) (n_{12} v_2)\nn\\&\quad -  \frac{110}{3} \ln\Big(\frac{r_{12}}{r'_0}\Big) (a_1 v_2) (n_{12} v_2)
			+ \frac{44023}{720} (a_1 n_{12}) (v_1 v_2)
			- 44 \ln\Big(\frac{r_{12}}{r'_0}\Big) (a_1 n_{12}) (v_1 v_2)\nn\\&\quad\left. + 44 \ln\Big(\frac{r_{12}}{r'_0}\Big) (a_1 n_{12}) v_1^{2}
			+ \frac{14377}{560} (a_2 n_{12}) v_1^{2}
			+ \frac{937}{1440} (a_1 n_{12}) v_2^{2}\right)\nn\\&\quad + \frac{m_1^3 m_2}{r_{12}^{3}} \left(- \frac{30313}{360} (n_{12} v_1)^4
			+ \frac{64001}{720} (n_{12} v_1)^3 (n_{12} v_2)
			-  \frac{185917}{1680} (n_{12} v_1)^2 (n_{12} v_2)^2\right.\nn\\&\quad - 55 \ln\Big(\frac{r_{12}}{r'_0}\Big) (n_{12} v_1)^2 (n_{12} v_2)^2
			+ \frac{179617}{1680} (n_{12} v_1) (n_{12} v_2)^3
			+ 55 \ln\Big(\frac{r_{12}}{r'_0}\Big) (n_{12} v_1) (n_{12} v_2)^3\nn\\&\quad -  \frac{94667}{400} (n_{12} v_1)^2 (v_1 v_2)
			+ \frac{338099}{1400} (n_{12} v_1) (n_{12} v_2) (v_1 v_2)\nn\\&\quad + 22 \ln\Big(\frac{r_{12}}{r'_0}\Big) (n_{12} v_1) (n_{12} v_2) (v_1 v_2)
			-  \frac{214897}{8400} (n_{12} v_2)^2 (v_1 v_2)\nn\\&\quad - 11 \ln\Big(\frac{r_{12}}{r'_0}\Big) (n_{12} v_2)^2 (v_1 v_2)
			-  \frac{737}{18} (v_1 v_2)^2
			+ \frac{903589}{16800} (n_{12} v_1)^2 v_1^{2}\nn\\&\quad - 55 \ln\Big(\frac{r_{12}}{r'_0}\Big) (n_{12} v_1)^2 v_1^{2}
			-  \frac{96287}{1120} (n_{12} v_1) (n_{12} v_2) v_1^{2}
			+ 55 \ln\Big(\frac{r_{12}}{r'_0}\Big) (n_{12} v_1) (n_{12} v_2) v_1^{2}\nn\\&\quad + \frac{426731}{4200} (n_{12} v_2)^2 v_1^{2}
			+ 11 \ln\Big(\frac{r_{12}}{r'_0}\Big) (n_{12} v_2)^2 v_1^{2}
			+ \frac{202687}{3360} (v_1 v_2) v_1^{2}\nn\\&\quad -  \frac{55}{3} \ln\Big(\frac{r_{12}}{r'_0}\Big) (v_1 v_2) v_1^{2}
			+ \frac{22769}{2016} v_1^{4}
			+ \frac{55}{3} \ln\Big(\frac{r_{12}}{r'_0}\Big) v_1^{4}
			-  \frac{177}{8} (n_{12} v_1)^2 v_2^{2}\nn\\&\quad + 66 \ln\Big(\frac{r_{12}}{r'_0}\Big) (n_{12} v_1)^2 v_2^{2}
			-  \frac{120397}{4200} (n_{12} v_1) (n_{12} v_2) v_2^{2}
			- 88 \ln\Big(\frac{r_{12}}{r'_0}\Big) (n_{12} v_1) (n_{12} v_2) v_2^{2}\nn\\&\quad + \frac{7}{4} (n_{12} v_2)^2 v_2^{2}
			-  \frac{43}{2} (v_1 v_2) v_2^{2}
			+ 22 \ln\Big(\frac{r_{12}}{r'_0}\Big) (v_1 v_2) v_2^{2}
			-  \frac{8357}{560} v_1^{2} v_2^{2}
			- 22 \ln\Big(\frac{r_{12}}{r'_0}\Big) v_1^{2} v_2^{2}\nn\\&\quad\left. + \frac{91}{16} v_2^{4}\right)
			+ 1 \leftrightarrow
			2\,,\\
L_\text{4PN}^{(4)}&= \frac{m_1^4 m_2}{r_{12}^4} \left(\frac{282629}{900} (n_{12} v_1)^2
			-  \frac{880}{3} \ln\Big(\frac{r_{12}}{r'_0}\Big) (n_{12} v_1)^2
			-  \frac{283979}{900} (n_{12} v_1) (n_{12} v_2)\right.\nn\\&\quad + \frac{880}{3} \ln\Big(\frac{r_{12}}{r'_0}\Big) (n_{12} v_1) (n_{12} v_2)
			+ \frac{9}{4} (n_{12} v_2)^2
			+ \frac{208529}{3600} (v_1 v_2)
			-  \frac{220}{3} \ln\Big(\frac{r_{12}}{r'_0}\Big) (v_1 v_2)\nn\\&\quad\left. -  \frac{211229}{3600} v_1^{2}
			+ \frac{220}{3} \ln\Big(\frac{r_{12}}{r'_0}\Big) v_1^{2}
			+ \frac{15}{16} v_2^{2}\right) \nn\\&
			 + \frac{m_1^3 m_2^2}{r_{12}^4} \left(- \frac{1268557}{50400} (n_{12} v_1)^2  + \frac{659}{96} \pi^2 (n_{12} v_1)^2
			-  \frac{286}{3} \ln\Big(\frac{r_{12}}{r'_0}\Big) (n_{12} v_1)^2
			\right.\nn\\&\quad+ \frac{11530469}{25200} (n_{12} v_1) (n_{12} v_2) -  \frac{1715}{48} \pi^2 (n_{12} v_1) (n_{12} v_2)
			+ 108 \ln\Big(\frac{r_{12}}{r'_0}\Big) (n_{12} v_1) (n_{12} v_2)\nn\\&\quad -  \frac{2233689}{5600} (n_{12} v_2)^2
			+ \frac{2771}{96} \pi^2 (n_{12} v_2)^2
			- \frac{82}{3} \ln\Big(\frac{r_{12}}{r'_0}\Big) (n_{12} v_2)^2\nn\\&\quad -  \frac{959797}{8400} (v_1 v_2)
			+ \frac{103}{16} \pi^2 (v_1 v_2)
			-  \frac{202}{3} \ln\Big(\frac{r_{12}}{r'_0}\Big) (v_1 v_2)\nn\\&\quad + \frac{858533}{50400} v_1^{2}
			-  \frac{15}{32} \pi^2 v_1^{2}
			+ \frac{121}{3} \ln\Big(\frac{r_{12}}{r'_0}\Big) v_1^{2}
			+ \frac{5482669}{50400} v_2^{2}
			\nn\\&\quad \left.-  \frac{191}{32} \pi^2 v_2^{2}
			+ \frac{70}{3} \ln\Big(\frac{r_{12}}{r'_0}\Big) v_2^{2}\right) + 1 \leftrightarrow
			2\,,\\
L_\text{4PN}^{(5)}&= \frac{3}{8} \frac{m_1^5 m_2}{r_{12}^5}
			+ \frac{m_1^3 m_2^3}{r_{12}^5} \left(\frac{597771}{5600}
			-  \frac{71}{32} \pi^2
			-  \frac{110}{3} \ln\Big(\frac{r_{12}}{r'_0}\Big)\right)
			\nn\\&+ \frac{m_1^4 m_2^2}{r_{12}^5} \left(\frac{1734977}{25200}
			+ \frac{105}{32} \pi^2 -  \frac{290}{3} \ln\Big(\frac{r_{12}}{r'_0}\Big)\right) + 1 \leftrightarrow
			2\,.
\end{align}
\end{subequations}

By a contact transformation at 4PN order, extending Eq.~\eqref{contacttransf}, we can transform the 4PN harmonic-coordinates Lagrangian into a new Lagrangian, valid in some ADM or ADM-like coordinate system, and such that the associated Hamiltonian coincides with the 4PN Hamiltonian obtained with the ADM Hamiltonian formalism of \cite{JaraS13, JaraS15, DJS14, DJS15eob}. Denoting by $\bm{P}_1$ and $\bm{P}_2$ the canonical momenta in ADM coordinates, and by $R_{12}\bm{N}_{12} = \bm{Y}_1-\bm{Y}_2$ the relative separation, the 4PN ADM Hamiltonian is [following the same presentation as in \eqref{notationL} and \eqref{notationLG}]
\begin{subequations}\label{H3PN}
\begin{align}
H_\text{N} &= \frac{P_1^{2}}{2 m_1}-\frac{G m_1 m_2}{2
	R_{12}} + 1 \leftrightarrow 2\,,\\
H_\text{1PN} &= -\frac{P_1^{4}}{8 m_1^3} + \frac{G^2 m_1^2 m_2}{2 R_{12}^2}\nn\\&
+\frac{G m_1
	m_2}{R_{12}} \left(-\frac{3 P_1^{2}}{2
		m_1^2}+\frac{(N_{12}P_1) (N_{12}P_2)}{4 m_1
		m_2}+\frac{7 (P_1P_2)}{4 m_1
		m_2}\right) + 1 \leftrightarrow 2\,,\\
H_\text{2PN} &= \frac{P_1^{6}}{16
	m_1^5} + \frac{G^3 m_1 m_2}{R_{12}^3} \left(-\frac{m_1^2}{4}-\frac{5
		m_1 m_2}{8}\right)\nn\\&+\frac{G^2 m1
	m_2}{R_{12}^2} \left(\frac{5 m_2 P_1^{2}}{2
		m_1^2}+\frac{19 P_1^{2}}{4 m_1}-\frac{3
		(N_{12}P_1) (N_{12}P_2)}{2 m_2}-\frac{27 (P_1P_2)}{4
		m_2}\right)\nn\\&+\frac{G m_1 m_2}{R_{12}}
	\left(\frac{5 P_1^{4}}{8 m_1^4}-\frac{3 (N_{12}P_1)^2
		(N_{12}P_2)^2}{16 m_1^2 m_2^2}-\frac{3 (N_{12}P_1)
		(N_{12}P_2) (P_1P_2)}{4 m_1^2 m_2^2}\right.\nn\\&\left.\quad+\frac{5
		(N_{12}P_2)^2 P_1^{2}}{8 m_1^2 m_2^2}-\frac{11
		P_1^{2} P_2^{2}}{16 m_1^2
		m_2^2}-\frac{(P_1P_2)^2}{8 m_1^2
		m_2^2}\right)+ 1 \leftrightarrow 2\,,\\
H_\text{3PN} &= -\frac{5 P_1^{8}}{128
	m_1^7} + \frac{G^4 m_1 m_2}{R_{12}^4} \left(\frac{m_1^3}{8}+\frac{227
		m_1^2 m_2}{24}-\frac{21}{32} \pi ^2 m_1^2
	m_2\right)\nn\\&+\frac{G^3 m_1 m_2}{R_{12}^3}
	\left(-\frac{25 m_2^2 P_1^{2}}{8 m_1^2}-\frac{43
		m_2 (N_{12}P_1)^2}{16 m_1}-\frac{3 \pi ^2 m_2
		(N_{12}P_1)^2}{64 m_1}\right.\nn\\&\left.\quad+\frac{21 m_1 (N_{12}P_1)
		(N_{12}P_2)}{8 m_2}-\frac{473 m_2 P_1^{2}}{48
		m_1}+\frac{\pi ^2 m_2 P_1^{2}}{64
		m_1}+\frac{77 m_1 (P_1P_2)}{8 m_2}\right.\nn\\&\left.\quad+\frac{5
		(N_{12}P_1)^2}{4}+\frac{119 (N_{12}P_1)
		(N_{12}P_2)}{16}+\frac{3}{64} \pi ^2 (N_{12}P_1)
	(N_{12}P_2)\right.\nn\\&\left.\quad-\frac{425 P_1^{2}}{48}+\frac{143
		(P_1P_2)}{16}-\frac{\pi ^2
		(P_1P_2)}{64}\right)\nn\\&+\frac{G^2 m_1
	m_2}{R_{12}^2} \left(-\frac{27 m_2 P_1^{4}}{16
		m_1^4}+\frac{5 (N_{12}P_1)^4}{12 m_1^3}+\frac{17
		(N_{12}P_1)^2 P_1^{2}}{16
		m_1^3}+\frac{P_1^{4}}{16 m_1^3}\right.\nn\\&\left.\quad-\frac{3
		(N_{12}P_1)^3 (N_{12}P_2)}{2 m_1^2 m_2}+\frac{10
		(N_{12}P_1)^2 (N_{12}P_2)^2}{3 m_1^2 m_2}-\frac{11
		(N_{12}P_1)^2 (P_1P_2)}{8 m_1^2 m_2}\right.\nn\\&\left.\quad-\frac{15
		(N_{12}P_1) (N_{12}P_2) P_1^{2}}{8 m_1^2
		m_2}+\frac{125 (N_{12}P_1) (N_{12}P_2) (P_1P_2)}{12
		m_1^2 m_2}-\frac{193 (N_{12}P_2)^2 P_1^{2}}{48
		m_1^2 m_2}\right.\nn\\&\left.\quad-\frac{115 P_1^{2} (P_1P_2)}{16
		m_1^2 m_2}+\frac{371 P_1^{2} P_2^{2}}{48
		m_1^2 m_2}+\frac{25 (P_1P_2)^2}{48 m_1^2
		m_2}-\frac{55 (N_{12}P_2)^2 P_1^{2}}{12 m_1
		m_2^2}\right)\nn\\&+\frac{G m_1 m_2}{R_{12}}
	\left(-\frac{7 P_1^{6}}{16 m_1^6}+\frac{3
		(N_{12}P_1)^2 (N_{12}P_2)^2 P_1^{2}}{16 m_1^4
		m_2^2}-\frac{5 (N_{12}P_1)^2 P_1^{2} P_2^{2}}{16
		m_1^4 m_2^2}\right.\nn\\&\left.\quad+\frac{3 (N_{12}P_1) (N_{12}P_2)
		P_1^{2} (P_1P_2)}{4 m_1^4 m_2^2}-\frac{5
		(N_{12}P_2)^2 P_1^{4}}{16 m_1^4
		m_2^2}+\frac{P_1^{4} P_2^{2}}{2 m_1^4
		m_2^2}\right.\nn\\&\left.\quad+\frac{P_1^{2} (P_1P_2)^2}{8 m_1^4
		m_2^2}+\frac{5 (N_{12}P_1)^3 (N_{12}P_2)^3}{32 m_1^3
		m_2^3}+\frac{15 (N_{12}P_1)^2 (N_{12}P_2)^2
		(P_1P_2)}{32 m_1^3 m_2^3}\right.\nn\\&\left.\quad-\frac{9 (N_{12}P_1)
		(N_{12}P_2)^3 P_1^{2}}{16 m_1^3 m_2^3}+\frac{7
		(N_{12}P_1) (N_{12}P_2) P_1^{2} P_2^{2}}{32 m_1^3
		m_2^3}\right.\nn\\&\left.\quad-\frac{5 (N_{12}P_1) (N_{12}P_2) (P_1P_2)^2}{16
		m_1^3 m_2^3}+\frac{(N_{12}P_2)^2 P_1^{2}
		(P_1P_2)}{16 m_1^3 m_2^3}\right.\nn\\&\left.\quad+\frac{P_1^{2}
		(P_1P_2) P_2^{2}}{32 m_1^3
		m_2^3}-\frac{(P_1P_2)^3}{16 m_1^3
		m_2^3}\right) + 1 \leftrightarrow 2\,.
\end{align}
\end{subequations}
and, for the instantaneous (local-in-time) part at 4PN order,
\begin{subequations}\label{resultH4PN}
\begin{align}
H_\text{4PN}^{(0)} &= \frac{7 P_1^{10}}{256 m_1^9}
			+ 1 \leftrightarrow 2\,,\\
H_\text{4PN}^{(1)} &= \frac{m_1 m_2}{R_{12}} \Biggl\{ 
\frac{45 P_1^{8}}{128 m_1^8}\nn\\&\quad+\frac{1}{m_1^6
	m_2^2}\left(-\frac{9}{64}
	(N_{12}P_1)^2 (N_{12}P_2)^2 P_1^{4}+\frac{15}{64}
	(N_{12}P_1)^2 P_1^{4} P_2^{2}\right.\nn\\&\quad\quad\left.-\frac{9}{16} (N_{12}P_1)
	(N_{12}P_2) P_1^{4} (P_1P_2)+\frac{15 (N_{12}P_2)^2
		P_1^{6}}{64}\right.\nn\\&\quad\quad\left.-\frac{21 P_1^{6} P_2^{2}}{64}-\frac{3
		P_1^{4} (P_1P_2)^2}{32}\right)\nn\\&\quad+\frac{1}{m_1^5
		m_2^3}\left(-\frac{35}{256} (N_{12}P_1)^5
	(N_{12}P_2)^3-\frac{85}{256} (N_{12}P_1)^4 (N_{12}P_2)^2
	(P_1P_2)\right.\nn\\&\quad\quad\left.+\frac{25}{128} (N_{12}P_1)^3 (N_{12}P_2)^3
	P_1^{2}-\frac{7}{128} (N_{12}P_1)^3 (N_{12}P_2) P_1^{2}
	P_2^{2}\right.\nn\\&\quad\quad\left.+\frac{25}{64} (N_{12}P_1)^3 (N_{12}P_2)
	(P_1P_2)^2-\frac{45}{128} (N_{12}P_1)^2 (N_{12}P_2)^2
	P_1^{2} (P_1P_2)\right.\nn\\&\quad\quad\left.+\frac{7}{128} (N_{12}P_1)^2 P_1^{2}
	(P_1P_2) P_2^{2}-\frac{3 (N_{12}P_1)^2
		(P_1P_2)^3}{64}\right.\nn\\&\quad\quad\left.+\frac{33}{256} (N_{12}P_1) (N_{12}P_2)^3
	P_1^{4}-\frac{25}{256} (N_{12}P_1) (N_{12}P_2)
	P_1^{4} P_2^{2}\right.\nn\\&\quad\quad\left.+\frac{7}{64} (N_{12}P_1) (N_{12}P_2)
	P_1^{2} (P_1P_2)^2-\frac{1}{256} (N_{12}P_2)^2
	P_1^{4} (P_1P_2)\right.\nn\\&\quad\quad\left.-\frac{7}{256} P_1^{4} (P_1P_2)
	P_2^{2}+\frac{3 P_1^{2} (P_1P_2)^3}{64}\right)\nn\\&\quad+\frac{1}{m_1^4 m_2^4}\left(-\frac{5}{32} (N_{12}P_1)^2 (N_{12}P_2)^4
	P_1^{2}+\frac{21}{64} (N_{12}P_1)^2 (N_{12}P_2)^2
	P_1^{2} P_2^{2}\right.\nn\\&\quad\quad\left.-\frac{1}{2} (N_{12}P_1) (N_{12}P_2)^3
	P_1^{2} (P_1P_2)+\frac{1}{16} (N_{12}P_1) (N_{12}P_2)
	P_1^{2} (P_1P_2) P_2^{2}\right.\nn\\&\quad\quad\left.+\frac{7 (N_{12}P_2)^4
		P_1^{4}}{32}-\frac{3}{16} (N_{12}P_2)^2 P_1^{4}
	P_2^{2}+\frac{1}{8} (N_{12}P_2)^2 P_1^{2}
	(P_1P_2)^2\right.\nn\\&\quad\quad\left.-\frac{7 P_1^{4}
		P_2^{4}}{128}-\frac{1}{32} P_1^{2} (P_1P_2)^2
	P_2^{2}\right)\nn\\&\quad+\frac{1}{m_1^3 m_2^5}\left(\frac{55}{256}
	(N_{12}P_1) (N_{12}P_2)^5 P_1^{2}-\frac{23}{256}
	(N_{12}P_2)^4 P_1^{2} (P_1P_2)\right)
\Biggr\} \nn\\&+ 1 \leftrightarrow 2\,,\\
H_\text{4PN}^{(2)} &= \frac{m_1 m_2}{R_{12}^2} \Biggl\{ 
\frac{13 m_2 P_1^{6}}{8 m_1^6}\nn\\&\quad+\frac{1}{m_1^5}\left(\frac{369
		(N_{12}P_1)^6}{160}-\frac{889 (N_{12}P_1)^4
		P_1^{2}}{192}+\frac{49 (N_{12}P_1)^2
		P_1^{4}}{16}-\frac{63
		P_1^{6}}{64}\right)\nn\\&\quad+\frac{1}{m_1^4 m_2}\left(-\frac{549 (N_{12}P_1)^5
		(N_{12}P_2)}{128}-\frac{7 (N_{12}P_1)^4
		(N_{12}P_2)^2}{4}\right.\nn\\&\quad\quad\left.+\frac{1547 (N_{12}P_1)^4
		(P_1P_2)}{256}+\frac{67}{16} (N_{12}P_1)^3 (N_{12}P_2)
	P_1^{2}\right.\nn\\&\quad\quad\left.-\frac{7}{2} (N_{12}P_1)^3 (N_{12}P_2)
	(P_1P_2)+\frac{7}{6} (N_{12}P_1)^2 (N_{12}P_2)^2
	P_1^{2}\right.\nn\\&\quad\quad\left.-\frac{851}{128} (N_{12}P_1)^2 P_1^{2}
	(P_1P_2)+\frac{49}{48} (N_{12}P_1)^2 P_1^{2}
	P_2^{2}+\frac{21 (N_{12}P_1)^2
		(P_1P_2)^2}{16}\right.\nn\\&\quad\quad\left.-\frac{167}{128} (N_{12}P_1) (N_{12}P_2)
	P_1^{4}-\frac{133}{24} (N_{12}P_1) (N_{12}P_2) P_1^{2}
	(P_1P_2)\right.\nn\\&\quad\quad\left.+\frac{197 (N_{12}P_2)^2 P_1^{4}}{96}+\frac{1099
		P_1^{4} (P_1P_2)}{256}-\frac{173 P_1^{4}
		P_2^{2}}{48}-\frac{77 P_1^{2}
		(P_1P_2)^2}{96}\right)\nn\\&\quad+\frac{1}{m_1^3
		m_2^2}\left(\frac{3263
		(N_{12}P_1)^4 (N_{12}P_2)^2}{1280}-\frac{3571}{320}
	(N_{12}P_1)^3 (N_{12}P_2) (P_1P_2)\right.\nn\\&\quad\quad\left.+\frac{1067}{480}
	(N_{12}P_1)^2 (N_{12}P_2)^2 P_1^{2}-\frac{1999
		(N_{12}P_1)^2 P_1^{2} P_2^{2}}{3840}\right.\nn\\&\quad\quad\left.+\frac{4349
		(N_{12}P_1)^2 (P_1P_2)^2}{1280}+\frac{233}{96} (N_{12}P_1)
	(N_{12}P_2)^3 P_1^{2}\right.\nn\\&\quad\quad\left.+\frac{3073}{480} (N_{12}P_1)
	(N_{12}P_2) P_1^{2} (P_1P_2)-\frac{4567 (N_{12}P_2)^2
		P_1^{4}}{3840}\right.\nn\\&\quad\quad\left.+\frac{1}{384} (N_{12}P_2)^2 P_1^{2}
	(P_1P_2)+\frac{2081 P_1^{4} P_2^{2}}{3840}-\frac{3461
		P_1^{2} (P_1P_2)^2}{3840}\right)\nn\\&\quad+\frac{1}{m_1^2
		m_2^3}\left(-\frac{13 (N_{12}P_1)^3
		(N_{12}P_2)^3}{8}-\frac{19}{384} (N_{12}P_1)^2 (N_{12}P_2)^2
	(P_1P_2)\right.\nn\\&\quad\quad\left.+\frac{191}{192} (N_{12}P_1) (N_{12}P_2)^3
	P_1^{2}-\frac{47}{32} (N_{12}P_1) (N_{12}P_2) P_1^{2}
	P_2^{2}\right.\nn\\&\quad\quad\left.+\frac{11}{192} (N_{12}P_1) (N_{12}P_2)
	(P_1P_2)^2+\frac{1673 (N_{12}P_2)^4
		P_1^{2}}{1920}\right.\nn\\&\quad\quad\left.-\frac{5}{384} (N_{12}P_2)^2 P_1^{2}
	(P_1P_2)-\frac{185 P_1^{2} (P_1P_2)
		P_2^{2}}{384}+\frac{77 (P_1P_2)^3}{96}\right)\nn\\&\quad+\frac{7 (N_{12}P_2)^4 P_1^{2}}{4 m_1
	m_2^4}
\Biggr\} + 1 \leftrightarrow 2\,,\\
H_\text{4PN}^{(3)} &= \frac{m_1 m_2}{R_{12}^3} \Biggl\{ 
\frac{105 m_2^2 P_1^{4}}{32 m_1^4}\nn\\&\quad+\frac{m_2}{m_1^3}
	\left(\frac{375 \pi ^2 (N_{12}P_1)^4}{8192}-\frac{23533
		(N_{12}P_1)^4}{1280}+\frac{63347 (N_{12}P_1)^2
		P_1^{2}}{1600}\right.\nn\\&\quad\quad\left.-\frac{1059 \pi ^2 (N_{12}P_1)^2
		P_1^{2}}{1024}-\frac{211189 P_1^{4}}{19200}+\frac{2749
		\pi ^2 P_1^{4}}{8192}\right)\nn\\&\quad+\frac{1}{m_1^2}\left(\frac{5027
		(N_{12}P_1)^4}{384}+\frac{2369 (N_{12}P_1)^3
		(N_{12}P_2)}{60}\right.\nn\\&\quad\quad\left.+\frac{35655 \pi ^2 (N_{12}P_1)^3
		(N_{12}P_2)}{16384}-\frac{22993 (N_{12}P_1)^2
		P_1^{2}}{960}\right.\nn\\&\quad\quad\left.-\frac{1646983 (N_{12}P_1)^2
		(P_1P_2)}{19200}+\frac{56955 \pi ^2 (N_{12}P_1)^2
		(P_1P_2)}{16384}\right.\nn\\&\quad\quad\left.-\frac{391711 (N_{12}P_1) (N_{12}P_2)
		P_1^{2}}{6400}+\frac{43101 \pi ^2 (N_{12}P_1) (N_{12}P_2)
		P_1^{2}}{16384}\right.\nn\\&\quad\quad\left.+\frac{26627 (N_{12}P_2)^2
		P_1^{2}}{1600}-\frac{6695 P_1^{4}}{1152}+\frac{1243717
		P_1^{2} (P_1P_2)}{14400}-\frac{40483 \pi ^2 P_1^{2}
		(P_1P_2)}{16384}\right)\nn\\&\quad+\frac{1}{m_1
		m_2}\left(-\frac{3191 (N_{12}P_1)^3
		(N_{12}P_2)}{640}-\frac{30383 (N_{12}P_1)^2
		(N_{12}P_2)^2}{960}\right.\nn\\&\quad\quad\left.-\frac{36405 \pi ^2 (N_{12}P_1)^2
		(N_{12}P_2)^2}{16384}+\frac{8777 (N_{12}P_1)^2
		(P_1P_2)}{384}\right.\nn\\&\quad\quad\left.+\frac{28561 (N_{12}P_1) (N_{12}P_2)
		P_1^{2}}{1920}+\frac{248991 (N_{12}P_1) (N_{12}P_2)
		(P_1P_2)}{6400}\right.\nn\\&\quad\quad\left.-\frac{6153 \pi ^2 (N_{12}P_1) (N_{12}P_2)
		(P_1P_2)}{2048}+\frac{1411429 (N_{12}P_2)^2
		P_1^{2}}{19200}\right.\nn\\&\quad\quad\left.-\frac{1059}{512} \pi ^2 (N_{12}P_2)^2
	P_1^{2}+\frac{752969 P_1^{2}
		(P_1P_2)}{28800}-\frac{2492417 P_1^{2}
		P_2^{2}}{57600}\right.\nn\\&\quad\quad\left.+\frac{13723 \pi ^2 P_1^{2}
		P_2^{2}}{16384}-\frac{1918349
		(P_1P_2)^2}{57600}+\frac{10631 \pi ^2
		(P_1P_2)^2}{8192}\right)\nn\\&\quad+\frac{1}{m_2^2}\left(-\frac{16481}{960} (N_{12}P_1)^2
	(N_{12}P_2)^2-\frac{103957 (N_{12}P_1) (N_{12}P_2)
		(P_1P_2)}{2400}\right.\nn\\&\quad\quad\left.+\frac{94433 (N_{12}P_2)^2
		P_1^{2}}{4800}-\frac{118261 P_1^{2}
		P_2^{2}}{4800}+\frac{791 (P_1P_2)^2}{400}\right)
\Biggr\} + 1 \leftrightarrow 2\,,\\
H_\text{4PN}^{(4)} &= \frac{m_1 m_2}{R_{12}^4} \Biggl\{ 
\frac{105 m_2^3 P_1^{2}}{32 m_1^2}\nn\\&\quad+\frac{m_1^2}{m_2}
	\left(-\frac{7 (N_{12}P_1) (N_{12}P_2)}{2}-\frac{91
		(P_1P_2)}{8}\right)\nn\\&\quad+\frac{m_2^2}{m_1}
	\left(-\frac{28691 \pi ^2 (N_{12}P_1)^2}{24576}+\frac{3200179
		(N_{12}P_1)^2}{57600}+\frac{282361
		P_1^{2}}{19200}-\frac{21837 \pi ^2
		P_1^{2}}{8192}\right)\nn\\&\quad+m_1 \left(-\frac{9841
	(N_{12}P_1)^2}{1600}-\frac{2712013 (N_{12}P_1)
	(N_{12}P_2)}{19200}+\frac{176033 \pi ^2
	(P_1P_2)}{24576}\right.\nn\\&\quad\quad\left.+\frac{63641 \pi ^2 (N_{12}P_1)
	(N_{12}P_2)}{24576}+\frac{64861 P_1^{2}}{4800}-\frac{2864917
	(P_1P_2)}{57600}\right)\nn\\&\quad+m_2 \left(\frac{21745 \pi ^2
	(N_{12}P_1)^2}{16384}+\frac{698723
	(N_{12}P_1)^2}{19200}+\frac{1937033
	P_1^{2}}{57600}\right.\nn\\&\quad\quad\left.-\frac{199177 \pi ^2 P_1^{2}}{49152}\right)
\Biggr\} + 1 \leftrightarrow 2\,,\\
H_\text{4PN}^{(5)} &= \frac{m_1 m_2}{R_{12}^5} \Biggl\{ 
-\frac{m_1^4}{16}+\left(\frac{6237 \pi
	^2}{1024}-\frac{169799}{2400}\right) m_1^3
m_2\nn\\&\quad+\left(\frac{44825 \pi
	^2}{6144}-\frac{609427}{7200}\right) m_1^2 m_2^2
\Biggr\} + 1 \leftrightarrow 2\,.
\end{align}
\end{subequations}
The non-local tail term in the Hamiltonian is just the opposite of the tail term in the Lagrangian: $H^\text{tail}_\text{4PN}=-L^\text{tail}_\text{4PN}$ \citep{DJS14,BBBFMa}.

Arguably, the result given by the ADM-Hamiltonian formalism looks simpler than the harmonic-coordinate counterpart. Indeed, the Lagrangian is ordinary -- no accelerations -- and there are no logarithms nor associated gauge constant $r'_0$. Of course, one is free to describe the binary motion in whatever coordinates one likes, and the two formalisms, harmonic and ADM describe rigorously the same physics. On the other hand, the higher complexity of the harmonic-coordinates formalism enables one to perform more tests of the computations, notably by inquiring about the fate of the constant $r'_0$, which \emph{must} disappear from physical quantities such as the center-of-mass energy and the total gravitational-wave flux. On the other hand, the ADM-Hamiltonian formalism currently provides a limited description of the radiation field, compared to what will be done using harmonic coordinates in Sect.~\ref{sec:GW}. Therefore it is important to derive the equations of motion in harmonic coordinates for consistency with the calculation of the radiation field.

We have reviewed in Sect.~\ref{sec:PNeom} the efforts made to extend the equations of motion beyond the current state-of-the-art 4PN order.


\subsubsection{Equations of motion in the frame of the center of mass}
\label{sec:eomCM}

We translate the origin of coordinates to the binary's center-of-mass (CM) by imposing the vanishing of the binary's mass dipole moment: $\dI_i = 0$ in the notation of Sect.~\ref{sec:linvac}. Actually the dipole moment is computed as the center-of-mass conserved integral $\bm{\dG}\equiv\bm{\dI}$ associated with the boost symmetry of the equations of motion \citep{DJSpoinc,ABF01,BI03CM,BF19,BBFM17}. The condition $\bm{\dG}=\bm{0}$ is solved iteratively with usual order reduction of accelerations using the CM equations of motion. This results in the relationship between the individual positions in the center-of-mass frame $\bm{y}_1$ and $\bm{y}_2$, and the relative position $\bm{x}\equiv \bm{y}_{1}-\bm{y}_{2}$ and velocity $\bm{v}\equiv \bm{v}_{1}-\bm{v}_{2} = \dd \bm{x}/\dd t$ (formerly denoted $\bm{y}_{12}$ and $\bm{v}_{12}$). We also use the orbital separation $r\equiv\vert\bm{x}\vert$, together with $\bm{n}=\bm{x}/r$ and $\dot{r}=(nv)=\bm{n}\cdot\bm{v}$. Mass parameters are: The total mass $m=m_1+m_2$ (to be distinguished from the conserved ADM mass $\dM$); the relative mass difference $\Delta=(m_1-m_2)/m$; the reduced mass $\mu=m_1m_2/m$; the symmetric mass ratio
\begin{equation}
	\nu\equiv \frac{\mu}{m}\equiv \frac{m_1m_2}{(m_1+m_2)^2}\,.
	\label{nu}
\end{equation}
The usefulness of this ratio lies in its range of variation: $0<\nu\leqslant 1/4$, with $\nu=1/4$ in the case of equal masses, and $\nu\to 0$ in the test-mass limit for one of the bodies. Thus $\nu$ is numerically rather small and may be viewed as a small expansion parameter. We also pose $X_1=m_1/m$ and $X_2=m_2/m$ so that $\Delta=X_1-X_2$ and $\nu=X_1X_2$. The general form of the individual positions of the particles in the CM frame is
\begin{subequations}\label{CMrel}\begin{align}
		\bm{y}_1 &= \Bigl[X_2+\nu\,\Delta\,\mathcal{P}\Bigr] \bm{x}
		+\nu\,\Delta\,\mathcal{Q}\,\bm{v}\,,\\ \bm{y}_2 &= \Bigl[-X_1+\nu\,\Delta\,\mathcal{P}\Bigr]
		\bm{x} +\nu\,\Delta\,\mathcal{Q}\,\bm{v}\,,
\end{align}\end{subequations}
where all post-Newtonian corrections, beyond Newtonian order, are proportional to the mass ratio $\nu$ and the mass difference $\Delta$. And, of course, we have $\bm{y}_1=-\bm{y}_2$ for equal masses. Up to the 4PN order the two dimensionless coefficients $\mathcal{P}$ and $\mathcal{Q}$ read [with presentation similar to \eqref{notationL} and \eqref{notationLG}; the PN coefficients not indicated are zero]
\begin{subequations}\label{Pres}
	\begin{align}
		\mathcal{P}_\text{1PN} &= \frac{v^2}{2} -\frac{G m}{2\,r}\,,\\
		\mathcal{P}_\text{2PN} &= \frac{3\,v^4}{8} - \frac{3\,\nu\,v^4}{2}
		+ \frac{G m}{r}\,\left( -\frac{\dot{r}^2}{8} +
		\frac{3\,\dot{r}^2\,\nu}{4} + \frac{19\,v^2}{8} +
		\frac{3\,\nu\,v^2}{2} \right)\nn\\ &
		+\frac{G^2m^2}{r^2}\left(\frac{7}{4} - \frac{\nu}{2} \right)\,,\\
		\mathcal{P}_\text{3PN} &= \frac{5\,v^6}{16} -
		\frac{11\,\nu\,v^6}{4} + 6\,\nu^2\,v^6 \nn\\ & 
		+\frac{G m}{r}\left( \frac{\dot{r}^4}{16} -
		\frac{5\,\dot{r}^4\,\nu}{8} + \frac{21\,\dot{r}^4\,\nu^2}{16} -
		\frac{5\,\dot{r}^2\,v^2}{16} + \frac{21\,\dot{r}^2\,\nu\,v^2}{16}
		\right.\nn\\ & \quad\quad -\left.
		\frac{11\,\dot{r}^2\,\nu^2\,v^2}{2} + \frac{53\,v^4}{16} -
		7\,\nu\,v^4 - \frac{15\,\nu^2\,v^4}{2} \right) \nn\\ & 
		+\frac{G^2m^2}{r^2}\left( -\frac{7\,\dot{r}^2}{3} +
		\frac{73\,\dot{r}^2\,\nu}{8} + 4\,\dot{r}^2\,\nu^2 +
		\frac{101\,v^2}{12} - \frac{33\,\nu\,v^2}{8} + 3\,\nu^2\,v^2 \right)
		\nn\\ &  + \frac{G^3m^3}{r^3}\left( -\frac{14351}{1260} +
		\frac{\nu}{8} - \frac{\nu^2}{2} + \frac{22}{3}\,\ln
		\Big(\frac{r}{r'_0}\Big) \right)\,,\\ 
		\mathcal{P}_\text{3.5PN} &= \frac{G^2m^2}{r^2}\dot{r}\left( -\frac{8\,\dot{r}^2}{5} -
		\frac{2\,v^2}{15} + \frac{68\,v^2\,\nu}{35}\right)
		+ \frac{G^3m^3}{r^3}\dot{r}\left( \frac{52}{15} - \frac{116}{35} \nu\right)
		\,,\\
		\mathcal{P}^{(0)}_\text{4PN} &= \left(\frac{35}{128}
		-  \frac{125}{32} \nu
		+ \frac{145}{8} \nu^2
		-  \frac{55}{2} \nu^3\right) v^{8}
		\,,\\
		\mathcal{P}^{(1)}_\text{4PN} &= \frac{m}{r} \left(- \frac{5}{128} \dot{r}^6
		+ \frac{35}{64} \nu \dot{r}^6
		-  \frac{125}{64} \nu^2 \dot{r}^6
		+ \frac{55}{32} \nu^3 \dot{r}^6
		+ \frac{27}{128} \dot{r}^4 v^{2}
		-  \frac{115}{64} \nu \dot{r}^4 v^{2}
		\right.\nn\\
		& \quad\quad+ \frac{517}{64} \nu^2 \dot{r}^4 v^{2} -  \frac{213}{16} \nu^3 \dot{r}^4 v^{2}
		-  \frac{53}{128} \dot{r}^2 v^{4}
		+ \frac{3}{2} \nu \dot{r}^2 v^{4}
		-  \frac{95}{8} \nu^2 \dot{r}^2 v^{4}
		\nn\\
		&\quad\quad \left. + 36 \nu^3 \dot{r}^2 v^{4}
		+ \frac{515}{128} v^{6}
		-  \frac{749}{32} \nu v^{6} + \frac{91}{4} \nu^2 v^{6}
		+ 42 \nu^3 v^{6}\right)
		\,,\\
		\mathcal{P}^{(2)}_\text{4PN} &= \frac{m^2}{r^2} \left(\frac{1133}{960} \dot{r}^4
		-  \frac{1007}{48} \nu \dot{r}^4
		+ \frac{169}{24} \nu^2 \dot{r}^4
		+ 9 \nu^3 \dot{r}^4
		-  \frac{31}{5} \dot{r}^2 v^{2}
		+ 26 \nu \dot{r}^2 v^{2}
		\right.\nn\\
		&\left.\quad\quad -  \frac{541}{8} \nu^2 \dot{r}^2 v^{2}
		-  \frac{83}{2} \nu^3 \dot{r}^2 v^{2}
		+ \frac{5631}{320} v^{4}
		-  \frac{139}{4} \nu v^{4}
		+ \frac{71}{4} \nu^2 v^{4}
		-  \frac{45}{2} \nu^3 v^{4}\right)
		\,,\\
		\mathcal{P}^{(3)}_\text{4PN} &= \frac{m^3}{r^3} \left(- \frac{185497}{8400} \dot{r}^2
		-  \frac{64347}{1120} \nu \dot{r}^2
		-  \frac{123}{128} \pi^2 \nu \dot{r}^2
		+ \frac{495}{16} \nu^2 \dot{r}^2
		+ \frac{55}{4} \nu^3 \dot{r}^2\right.\nn\\
		&\quad\quad - 11 \ln\Big(\frac{r}{r'_{0}}\Big) \dot{r}^2
		+ 44 \nu \ln\Big(\frac{r}{r'_{0}}\Big) \dot{r}^2
		+ \frac{2737}{1440} v^{2}
		-  \frac{87181}{3360} \nu v^{2}
		+ \frac{123}{128} \pi^2 \nu v^{2}
		\nn\\
		&\quad\quad\left. -  \frac{117}{8} \nu^2 v^{2} + 5 \nu^3 v^{2}
		- 11 \nu \ln\Big(\frac{r}{r'_{0}}\Big) v^{2}
		+ 22 \ln\Big(\frac{r}{r'_{0}}\Big) v^{2}\right)
		\,,\\
		\mathcal{P}^{(4)}_\text{4PN} &= \frac{m^4}{r^4} \left(\frac{215279}{3600}
		+ \frac{22043}{720} \nu
		-  \frac{11}{2} \pi^2 \nu
		-  \frac{3}{2} \nu^2
		-  \frac{1}{2} \nu^3
		\right.\nn\\
		&\quad\quad\left. -  \frac{220}{3} \ln\Big(\frac{r}{r'_{0}}\Big)
		+ 60 \nu \ln\Big(\frac{r}{r'_{0}}\Big)\right)
		\,.
	\end{align}
\end{subequations}
and
\begin{subequations}\label{Qres}
	\begin{align}
		\mathcal{Q}_\text{2PN} &= -\frac{7\,G m\,\dot{r}}{4}\,,\\
		\mathcal{Q}_\text{2.5PN} &= \frac{4\,G m\,v^2}{5}
		-\frac{8\,G^2m^2}{5\,r} \,,\\ 
		\mathcal{Q}_\text{3PN} &= G m\,\dot{r}\left( \frac{5\,\dot{r}^2}{12} -
		\frac{19\,\dot{r}^2\,\nu}{24} - \frac{15\,v^2}{8} +
		\frac{21\,\nu\,v^2}{4} \right) \nn\\ &  +
		\frac{G^2m^2\,\dot{r}}{r}\left( -\frac{235}{24}- \frac{21\,\nu}{4}
		\right) \,,\\
		\mathcal{Q}_\text{3.5PN} &= G m\,v^4\left( \frac{6}{7} -
		\frac{22\,\nu}{7}\right) +
		\frac{G^2m^2}{r}\left( \frac{44}{15}\dot{r}^2 + \frac{6}{35}v^2 + \frac{68}{35}\dot{r}^2\nu + \frac{132}{35}v^2\nu
		\right)\nn\\ &  +
		\frac{G^3m^3}{r^2}\left( - \frac{172}{105} - \frac{64}{35}\nu \right)\,,\\ 
		\mathcal{Q}^{(1)}_\text{4PN} &= m \left(- \frac{13}{64} \dot{r}^5
		+ \frac{25}{32} \nu \dot{r}^5
		-  \frac{17}{32} \nu^2 \dot{r}^5
		+ \frac{77}{96} \dot{r}^3 v^{2}
		-  \frac{187}{48} \nu \dot{r}^3 v^{2}
		+ \frac{19}{4} \nu^2 \dot{r}^3 v^{2}
		\right.\nn\\
		&\left.\quad\quad-  \frac{123}{64} \dot{r} v^{4} + \frac{199}{16} \nu \dot{r} v^{4}
		- 21 \nu^2 \dot{r} v^{4}\right)
		\,,\\
		\mathcal{Q}^{(2)}_\text{4PN} &= \frac{m^2}{r} \left(\frac{4621}{480} \dot{r}^3
		+ \frac{113}{24} \nu \dot{r}^3
		+ \frac{7}{12} \nu^2 \dot{r}^3
		-  \frac{3733}{160} \dot{r} v^{2}
		+ \frac{95}{4} \nu \dot{r} v^{2}
		+ 28 \nu^2 \dot{r} v^{2}\right)
		\,,\\
		\mathcal{Q}^{(3)}_\text{4PN} &= \frac{m^3}{r^2} \left(\frac{14377}{280}
		+ \frac{71509}{5040} \nu
		-  \frac{41}{64} \pi^2 \nu
		-  \frac{49}{4} \nu^2
		+ \frac{22}{3} \nu \ln\Big(\frac{r}{r'_{0}}\Big)
		\right.\nn\\
		&\left.\quad\quad -  \frac{110}{3} \ln\Big(\frac{r}{r'_{0}}\Big)
		-  \frac{44}{3} \nu \ln\Big(\frac{r}{r'_{0}}\Big)\right) \dot{r}
		\,.
\end{align}\end{subequations}
The CM velocities are obtained by differentiating the above expressions -- with order reduction of accelerations. There are no tail terms, but the formulas contain logarithmic terms depending on the gauge constant $r'_0$ not affecting physical results.

To proceed in the CM frame we consider first the relative acceleration $\bm{a}= \bm{a}_1-\bm{a}_2$ and then make the replacements \eqref{CMrel}. The general form of the order-reduced acceleration (considering first the instantaneous, local part of the acceleration) is
\begin{equation}\label{dvdt}
	\bm{a}_\text{inst} = -\frac{G m}{r^2}\Big[\big(1+\mathcal{A}\big)\,\bm{n} +
	\mathcal{B}\,\bm{v} \Big] \,,
\end{equation}
where the various PN coefficients up to 4PN order are given by \citep{LW90, BI03CM, BBFM17}
\begin{subequations}\label{calAcoeff}
	\begin{align}
		\mathcal{A}_\text{1PN} &= -\frac{3\,\dot{r}^2\,\nu}{2} + v^2 +
		3\,\nu\,v^2-\frac{G m}{r}\left(4 +2\,\nu \right) \,,\\
		\mathcal{A}_\text{2PN} &= \frac{15\,\dot{r}^4\,\nu}{8} -
		\frac{45\,\dot{r}^4\,\nu^2}{8} - \frac{9\,\dot{r}^2\,\nu\,v^2}{2} +
		6\,\dot{r}^2\,\nu^2\,v^2 + 3\,\nu\,v^4 -
		4\,\nu^2\,v^4 \nn\\ & + \frac{G m}{r}\left(
		-2\,\dot{r}^2 - 25\,\dot{r}^2\,\nu - 2\,\dot{r}^2\,\nu^2 -
		\frac{13\,\nu\,v^2}{2} + 2\,\nu^2\,v^2 \right) \nn\\
		& + \frac{G^2m^2}{r^2}\,\left( 9 + \frac{87\,\nu}{4}
		\right)\,,\\
		\mathcal{A}_\text{2.5PN} &= \frac{8
			G\,m\,\nu}{5r}\dot{r}\left[-\frac{17}{3} \frac{G m}{r} - 3
		v^2\right]\,,\\
		\mathcal{A}_\text{3PN} &= -\frac{35\,\dot{r}^6\,\nu}{16} +
		\frac{175\,\dot{r}^6\,\nu^2}{16} -
		\frac{175\,\dot{r}^6\,\nu^3}{16}+\frac{15\,\dot{r}^4\,\nu\,v^2}{2}
		\nn\\& - \frac{135\,\dot{r}^4\,\nu^2\,v^2}{4} +
		\frac{255\,\dot{r}^4\,\nu^3\,v^2}{8} -
		\frac{15\,\dot{r}^2\,\nu\,v^4}{2} +
		\frac{237\,\dot{r}^2\,\nu^2\,v^4}{8} \nn\\ &
		-\frac{45\,\dot{r}^2\,\nu^3\,v^4}{2} + \frac{11\,\nu\,v^6}{4} -
		\frac{49\,\nu^2\,v^6}{4} + 13\,\nu^3\,v^6 \nn\\ & +
		\frac{G m}{r}\left( 79\,\dot{r}^4\,\nu -
		\frac{69\,\dot{r}^4\,\nu^2}{2} - 30\,\dot{r}^4\,\nu^3 -
		121\,\dot{r}^2\,\nu\,v^2 + 16\,\dot{r}^2\,\nu^2\,v^2
		\right.\nn\\&\quad\quad\quad \left.+
		20\,\dot{r}^2\,\nu^3\,v^2+\frac{75\,\nu\,v^4}{4} + 8\,\nu^2\,v^4 -
		10\,\nu^3\,v^4 \right) \nn\\ & +
		\frac{G^2m^2}{r^2}\,\left( \dot{r}^2 +
		\frac{32573\,\dot{r}^2\,\nu}{168} + \frac{11\,\dot{r}^2\,\nu^2}{8} -
		7\,\dot{r}^2\,\nu^3 + \frac{615\,\dot{r}^2\,\nu\,\pi^2}{64} \right.\nn\\&\quad\quad\quad
		\left. -
		\frac{26987\,\nu\,v^2}{840} + \nu^3\,v^2 - \frac{123\,\nu\,\pi^2\,v^2}{64} \right.\nn\\&\quad\quad\quad
		\left.-
		110\,\dot{r}^2\,\nu\,\ln \Big(\frac{r}{r'_0}\Big) + 22\,\nu\,v^2\,\ln
		\Big(\frac{r}{r'_0}\Big) \right)\nn\\&
		+\frac{G^3m^3}{r^3}\left( -16 - \frac{437\,\nu}{4} -
		\frac{71\,\nu^2}{2} + \frac{41\,\nu\,{\pi }^2}{16} \right)\,,\\
		\mathcal{A}_\text{3.5PN} &= \frac{G m \nu}{r} \dot{r}\biggl[\frac{G^2
			m^2}{r^2} \, \left( \frac{3956}{35} + \frac{184}{5} \nu \right) +
		\frac{G m \,v^2}{r} \, \left( \frac{692}{35} - \frac{724}{15} \nu
		\right)\nn\\ & \quad\quad + v^4 \, \left(
		\frac{366}{35} + 12 \nu \right) + \frac{G m \,\dot{r}^2}{r} \, \left(
		\frac{294}{5} + \frac{376}{5} \nu \right) \nn\\ &
		\quad\quad  - v^2 \dot{r}^2 \, \left( 114 + 12 \nu \right) + 112
		\dot{r}^4 \biggr]\,,\\
		\mathcal{A}^{(0)}_\text{4PN} &= \left(\frac{315}{128} \nu -  \frac{2205}{128} \nu^2 +
		\frac{2205}{64} \nu^3 -  \frac{2205}{128} \nu^4\right)
		\dot{r}^8 \nn\\&\quad+ \left(- \frac{175}{16} \nu + \frac{595}{8}
		\nu^2 -  \frac{2415}{16} \nu^3  + \frac{735}{8} \nu^4\right) \dot{r}^6 v^{2} \nn\\&\quad+
		\left(\frac{135}{8} \nu -  \frac{1875}{16} \nu^2 + \frac{4035}{16} \nu^3 -
		\frac{1335}{8} \nu^4\right) \dot{r}^4 v^{4} \nn\\&\quad+ \left(- \frac{21}{2} \nu +
		\frac{1191}{16} \nu^2 -  \frac{327}{2} \nu^3 + 99 \nu^4\right) \dot{r}^2 v^{6} \nn\\&\quad
		+
		\left(\frac{21}{8} \nu -  \frac{175}{8} \nu^2 + 61 \nu^3  - 54 \nu^4\right)
		v^{8}
		\,,\\
		\mathcal{A}^{(1)}_\text{4PN} &= \frac{m}{r} \biggl(\frac{2973}{40} \nu \dot{r}^6
		+ 407 \nu^2 \dot{r}^6+ \frac{181}{2} \nu^3 \dot{r}^6 - 86 \nu^4 \dot{r}^6 
		+ \frac{1497}{32} \nu \dot{r}^4 v^{2}
		\nn\\&\quad\quad\quad -  \frac{1627}{2} \nu^2 \dot{r}^4 v^{2} - 81 \nu^3 \dot{r}^4 v^{2} + 228 \nu^4 \dot{r}^4 v^{2} 
		-  \frac{2583}{16} \nu \dot{r}^2 v^{4} \nn\\&\quad\quad\quad+ \frac{1009}{2} \nu^2 \dot{r}^2 v^{4} 
		+ 47 \nu^3 \dot{r}^2 v^{4} - 104 \nu^4 \dot{r}^2 v^{4} + \frac{1067}{32} \nu v^{6} \nn\\&\quad\quad\quad - 58 \nu^2 v^{6} - 44 \nu^3 v^{6} 
		+ 58 \nu^4 v^{6}\biggr)
		\,,\\
		\mathcal{A}^{(2)}_\text{4PN} &= \frac{m^2}{r^2} \left(\frac{2094751}{960} \nu \dot{r}^4
		+ \frac{45255}{1024} \pi^2 \nu \dot{r}^4
		+ \frac{326101}{96} \nu^2 \dot{r}^4
		-  \frac{4305}{128} \pi^2 \nu^2 \dot{r}^4
		\right.\nn\\&\quad\quad\quad\left.-  \frac{1959}{32} \nu^3 \dot{r}^4 - 126 \nu^4 \dot{r}^4
		- 1155 \nu^2 \ln\Big(\frac{r}{r'_{0}}\Big) \dot{r}^4
		+ 385 \nu \ln\Big(\frac{r}{r'_{0}}\Big) \dot{r}^4
		\right.\nn\\&\quad\quad\quad\left.
		-  \frac{1636681}{1120} \nu \dot{r}^2 v^{2}  -  \frac{12585}{512} \pi^2 \nu \dot{r}^2 v^{2} -  \frac{255461}{112} \nu^2 \dot{r}^2 v^{2}
		\right.\nn\\&\quad\quad\quad\left.+ \frac{3075}{128} \pi^2 \nu^2 \dot{r}^2 v^{2}
		-  \frac{309}{4} \nu^3 \dot{r}^2 v^{2}
		+ 63 \nu^4 \dot{r}^2 v^{2} \right.\nn\\&\quad\quad\quad\left. - 605 \nu \ln\Big(\frac{r}{r'_{0}}\Big) \dot{r}^2 v^{2}
		+ 825 \nu^2 \ln\Big(\frac{r}{r'_{0}}\Big) \dot{r}^2 v^{2}
		+ \frac{1096941}{11200} \nu v^{4}
		\right.\nn\\&\quad\quad\quad\left.+ \frac{1155}{1024} \pi^2 \nu v^{4}
		+ \frac{7263}{70} \nu^2 v^{4}
		-  \frac{123}{64} \pi^2 \nu^2 v^{4}
		+ \frac{145}{2} \nu^3 v^{4}
		- 16 \nu^4 v^{4} \right.\nn\\&\quad\quad\quad\left.+ 88 \nu \ln\Big(\frac{r}{r'_{0}}\Big) v^{4} - 66 \nu^2 \ln\Big(\frac{r}{r'_{0}}\Big) v^{4}
		\right) \,,\\
		\mathcal{A}^{(3)}_\text{4PN} &= \frac{m^3}{r^3} \left(-2 \dot{r}^2
		+ \frac{1297943}{8400} \nu \dot{r}^2
		-  \frac{2969}{16} \pi^2 \nu \dot{r}^2
		+ \frac{1255151}{840} \nu^2 \dot{r}^2
		\right.\nn\\
		&\quad\quad\quad\left.+ \frac{7095}{32} \pi^2 \nu^2 \dot{r}^2
		- 17 \nu^3 \dot{r}^2 - 24 \nu^4 \dot{r}^2
		\right.\nn\\
		&\quad\quad\quad\left. +2232 \nu \ln\Big(\frac{r}{r'_{0}}\Big) \dot{r}^2
		- 3364 \nu^2 \ln\Big(\frac{r}{r'_{0}}\Big) \dot{r}^2\right.\nn\\
		&\quad\quad\quad\left. 
		+ \frac{1237279}{25200} \nu v^{2}
		+ \frac{3835}{96} \pi^2 \nu v^{2}
		-  \frac{693947}{2520} \nu^2 v^{2} -  \frac{229}{8} \pi^2 \nu^2 v^{2}
		\right.\nn\\
		&\quad\quad\quad\left. + \frac{19}{2} \nu^3 v^{2}
		- \frac{1336}{3} \nu \ln\Big(\frac{r}{r'_{0}}\Big) v^{2}
		+  \frac{1616}{3} \nu^2 \ln\Big(\frac{r}{r'_{0}}\Big) v^{2}\right) \,,\\
		\mathcal{A}^{(4)}_\text{4PN} &= \frac{m^4}{r^4} \left(25
		+ \frac{6625537}{12600} \nu
		-  \frac{4543}{96} \pi^2 \nu
		+ \frac{477763}{720} \nu^2
		+ \frac{3}{4} \pi^2 \nu^2
		\right. \nn\\
		&\quad\quad\quad \left.+ \frac{334}{3} \nu \ln\Big(\frac{r}{r'_{0}}\Big)  - \frac{514}{3} \nu^2 \ln\Big(\frac{r}{r'_{0}}\Big)
		\right)\,.
\end{align}\end{subequations}
\begin{subequations}\label{calBcoeff}
	\begin{align}
		\mathcal{B}_\text{1PN} &= -4\,\dot{r} + 2\,\dot{r}\,\nu\,,\\
		\mathcal{B}_\text{2PN} &= \frac{9\,\dot{r}^3\,\nu}{2} +
		3\,\dot{r}^3\,\nu^2 -\frac{15\,\dot{r}\,\nu\,v^2}{2} -
		2\,\dot{r}\,\nu^2\,v^2\nn\\ &  +
		\frac{G m}{r}\left( 2\,\dot{r} + \frac{41\,\dot{r}\,\nu}{2} +
		4\,\dot{r}\,\nu^2 \right)\,,\\
		\mathcal{B}_\text{2.5PN} &= \frac{8 G\,m\,\nu}{5r}\left[ 3\frac{G
			m}{r} + v^2\right]\,,\\
		\mathcal{B}_\text{3PN} &= -\frac{45\,\dot{r}^5\,\nu}{8} +
		15\,\dot{r}^5\,\nu^2 + \frac{15\,\dot{r}^5\,\nu^3}{4} +
		12\,\dot{r}^3\,\nu\,v^2 \nn\\&
		- \frac{111\,\dot{r}^3\,\nu^2\,v^2}{4}
		-12\,\dot{r}^3\,\nu^3\,v^2 -\frac{65\,\dot{r}\,\nu\,v^4}{8} +
		19\,\dot{r}\,\nu^2\,v^4 + 6\,\dot{r}\,\nu^3\,v^4
		\nn\\& + \frac{G m}{r}\left(
		\frac{329\,\dot{r}^3\,\nu}{6} + \frac{59\,\dot{r}^3\,\nu^2}{2} +
		18\,\dot{r}^3\,\nu^3 - 15\,\dot{r}\,\nu\,v^2 - 27\,\dot{r}\,\nu^2\,v^2
		- 10\,\dot{r}\,\nu^3\,v^2 \right) \nn\\&
		+ \frac{G^2m^2}{r^2}\,\left( -4\,\dot{r} -
		\frac{18169\,\dot{r}\,\nu}{840} + 25\,\dot{r}\,\nu^2 +
		8\,\dot{r}\,\nu^3 - \frac{123\,\dot{r}\,\nu\,\pi^2}{32} 
		\right.\nn\\&\quad\quad\quad\left.+ 44\,\dot{r}\,\nu\,\ln \Big(\frac{r}{r'_0}\Big)
		\right)\,,\\
		\mathcal{B}_\text{3.5PN} &= \frac{G m \nu}{r}\left[\frac{G^2
			m^2}{r^2} \, \left( - \frac{1060}{21} - \frac{104}{5}\nu \right) +
		\frac{G m v^2}{r} \, \left( \frac{164}{21} + \frac{148}{5} \nu
		\right) \right. \nn\\ & \quad\quad + v^4\, \left(
		- \frac{626}{35} - \frac{12}{5} \nu \right) + \frac{G m \dot{r}^2}{r} \,
		\left( - \frac{82}{3} - \frac{848}{15} \nu \right) \nn\\ &
		\quad\quad\left. + v^2 \dot{r}^2 \left( \frac{678}{5} +
		\frac{12}{5} \nu \right) - 120 \dot{r}^4 \right]\,,\\
		\mathcal{B}^{(0)}_\text{4PN} &= \left(\frac{105}{16} \nu
		-  \frac{245}{8} \nu^2
		+ \frac{385}{16} \nu^3
		+ \frac{35}{8} \nu^4\right) \dot{r}^7
		\nn\\
		&\quad + \left(- \frac{165}{8} \nu
		+ \frac{1665}{16} \nu^2
		-  \frac{1725}{16} \nu^3
		-  \frac{105}{4} \nu^4\right) \dot{r}^5 v^{2}\nn\\
		&\quad + \left(\frac{45}{2} \nu
		-  \frac{1869}{16} \nu^2
		+ 129 \nu^3
		+ 54 \nu^4\right) \dot{r}^3 v^{4}
		\nn\\
		&\quad + \left(- \frac{157}{16} \nu
		+ 54 \nu^2
		- 69 \nu^3
		- 24 \nu^4\right) \dot{r} v^{6}\,,\\
		\mathcal{B}^{(1)}_\text{4PN} &= \frac{m}{r} \left(- \frac{54319}{160} \nu \dot{r}^5
		-  \frac{901}{8} \nu^2 \dot{r}^5
		+ 60 \nu^3 \dot{r}^5
		+ 30 \nu^4 \dot{r}^5
		+ \frac{25943}{48} \nu \dot{r}^3 v^{2}
		\right.\nn\\&\quad\quad\quad\left.+ \frac{1199}{12} \nu^2 \dot{r}^3 v^{2}  -  \frac{349}{2} \nu^3 \dot{r}^3 v^{2}
		- 98 \nu^4 \dot{r}^3 v^{2}
		-  \frac{5725}{32} \nu \dot{r} v^{4}
		\right.\nn\\&\quad\quad\quad\left. -  \frac{389}{8} \nu^2 \dot{r} v^{4}
		+ 118 \nu^3 \dot{r} v^{4}
		+ 44 \nu^4 \dot{r} v^{4}\right)
		\,,\\
		\mathcal{B}^{(2)}_\text{4PN} &= \frac{m^2}{r^2} \left(- \frac{9130111}{3360} \nu \dot{r}^3
		-  \frac{4695}{256} \pi^2 \nu \dot{r}^3
		-  \frac{184613}{112} \nu^2 \dot{r}^3
		+ \frac{1845}{64} \pi^2 \nu^2 \dot{r}^3
		\right.\nn\\&\quad\quad\quad\left. + \frac{209}{2} \nu^3 \dot{r}^3  + 74 \nu^4 \dot{r}^3
		+ 440 \nu \ln\Big(\frac{r}{r'_{0}}\Big) \dot{r}^3
		+ 550 \nu^2 \ln\Big(\frac{r}{r'_{0}}\Big) \dot{r}^3
		\right.\nn\\
		&\quad\quad\quad\left. + \frac{8692601}{5600} \nu \dot{r} v^{2}
		+ \frac{1455}{256} \pi^2 \nu \dot{r} v^{2}
		+ \frac{58557}{70} \nu^2 \dot{r} v^{2}
		-  \frac{123}{8} \pi^2 \nu^2 \dot{r} v^{2}
		\right.\nn\\
		& \quad\quad\quad\left.- 70 \nu^3 \dot{r} v^{2}
		- 34 \nu^4 \dot{r} v^{2} - 154 \nu \ln\Big(\frac{r}{r'_{0}}\Big) \dot{r} v^{2}
		- 264 \nu^2 \ln\Big(\frac{r}{r'_{0}}\Big) \dot{r} v^{2}
		\right) \,,\\
		\mathcal{B}^{(3)}_\text{4PN} &= \frac{m^3}{r^3} \left(2
		-  \frac{619267}{525} \nu
		+ \frac{791}{16} \pi^2 \nu
		-  \frac{28406}{45} \nu^2
		-  \frac{2201}{32} \pi^2 \nu^2
		\right.\nn\\
		& \quad\quad\quad\left.+ 66 \nu^3
		+ 16 \nu^4 - \frac{1484}{3} \nu \ln\Big(\frac{r}{r'_{0}}\Big)
		+ 1268 \nu^2 \ln\Big(\frac{r}{r'_{0}}\Big)
		\right) \dot{r}\,.
\end{align}\end{subequations}
The tail part of the acceleration comes from the variation of the tail part in the Lagrangian \eqref{Ltail4PN}, taking into account the variation of the scale $r=r_{12}$ entering into the logarithmic kernel, to which one has to add the dissipative contribution. The result is 
\begin{align}\label{acctailCM}
	a^{i}_{\text{tail}} =& -\frac{8G^2m}{5c^8} \,x^a \int_0^{+\infty}
	\dd\tau \ln\left(\frac{c\tau}{2r}\right) \dQ_{ia}^{(7)}(t-\tau)\nn\\ & +\frac{8G^2m}{5c^8}
	\,x^a\left[\left(\dQ_{ia}^{(3)}\ln r\right)^{(3)}-\dQ_{ia}^{(6)}\ln
	r\right] -\frac{2G^2}{5c^8\nu}
	\frac{n^i}{r}\Bigl(\dQ_{ab}^{(3)}\Bigr)^2\,,
\end{align}
where to this order the ADM mass $\dM$ reduces to the rest mass $m$ and the quadrupole moment is the Newtonian one $\dQ_{ab} = m\nu \,x^{\langle a}x^{b\rangle}$.\footnote{We give for reference [in particular the last term in \eqref{acctailCM} tends to zero when $\nu\to 0$]
	\begin{align*}
		\dQ_{ab}^{(3)} &= \frac{2G m^2\nu}{r^2}\biggl[-4 n^{\langle
			a}v^{b\rangle} + 3\dot{r}\,n^{\langle
			a}n^{b\rangle}\biggr]\,,\\ \Bigl(\dQ_{ab}^{(3)}\Bigr)^2 &=
		\frac{8G^2 m^4\nu^2}{r^4}\biggl(4 v^2 -
		\frac{11}{3}\dot{r}^2\biggr)\,.
\end{align*}}
The result \eqref{acctailCM} is made of both conservative and dissipative parts according to the split (see the footnote \ref{fnote:hadamardPF} for the definition of the Hadamard Pf)
\begin{subequations}\label{acctailCMconsdiss}
\begin{align}\label{acctailCMcons}
	a^{i}_{\text{tail}}\Big|_{\text{cons}} &= -\frac{4G^2m}{5c^8}
	\,x^a \!\mathop{\text{Pf}}_{2r/c}
	\int_{-\infty}^{+\infty} \frac{\dd t'}{\vert t-t'\vert}
	\,\dQ_{ia}^{(6)}(t')\nn\\ & +\frac{8G^2m}{5c^8}
	\,x^a\left[\left(\dQ_{ia}^{(3)}\ln r\right)^{(3)}-\dQ_{ia}^{(6)}\ln
	r\right] -\frac{2G^2}{5c^8\nu}
	\frac{n^i}{r}\Bigl(\dQ_{ab}^{(3)}\Bigr)^2\,,\\a^{i}_{\text{tail}}\Big|_{\text{diss}} =& -\frac{4G^2m}{5c^8} \,x^a \int_0^{+\infty}
	\dd\tau \ln\left(\frac{c\tau}{2r}\right) \left[\dQ_{ia}^{(7)}(t-\tau) + \dQ_{ia}^{(7)}(t+\tau)\right]\,.\label{acctailCMdiss}
\end{align}
\end{subequations}

The logarithms in \eqref{calAcoeff}--\eqref{calBcoeff}, together with the constant $r'_0$ therein, can be removed at 3PN order by a simple coordinate transformation (see the footnote \ref{fnote:log3PN}), while staying within the class of harmonic coordinates. The modification of the equations of motion affects the 3PN coefficients; we denote the new coefficients as $\mathcal{A}^\text{MH}_\mathrm{3PN}$ and $\mathcal{B}^\text{MH}_\mathrm{3PN}$ where MH stands for the ``modified harmonic'' coordinate system, as opposed to the SH (``standard harmonic'') coefficients $\mathcal{A}_\mathrm{3PN}$ and $\mathcal{B}_\mathrm{3PN}$ containing logarithms. Limiting ourselves to 3PN order, we have \citep{MW03, ABIQ08} 
\begin{subequations}\label{ABcoeffMH}
	\begin{align}
		\mathcal{A}^\text{MH}_\mathrm{3PN} &= -\frac{35
			\dot{r}^6 \nu}{16} + \frac{175 \dot{r}^6 \nu^2}{16} - \frac{175
			\dot{r}^6 \nu^3}{16} + \frac{15 \dot{r}^4 \nu v^2}{2} - \frac{135
			\dot{r}^4 \nu^2 v^2}{4} + \frac{255 \dot{r}^4 \nu^3 v^2}{8} \nn\\& -
		\frac{15 \dot{r}^2 \nu v^4}{2} + \frac{237
			\dot{r}^2 \nu^2 v^4}{8} - \frac{45 \dot{r}^2 \nu^3 v^4}{2} +
		\frac{11 \nu v^6}{4} - \frac{49 \nu^2 v^6}{4} + 13 \nu^3 v^6
		\nn \\ & + \frac{G m}{r} \bigg( 79 \dot{r}^4 \nu -
		\frac{69 \dot{r}^4 \nu^2}{2} - 30 \dot{r}^4 \nu^3 - 121 \dot{r}^2
		\nu v^2 + 16 \dot{r}^2 \nu^2 v^2 + 20 \dot{r}^2 \nu^3 v^2 \nn\\& \qquad \quad + \frac{75
			\nu v^4}{4} + 8 \nu^2 v^4
		- 10 \nu^3 v^4 \bigg) \nn \\ & + \frac{G^2 m^2}{r^2}
		\bigg( \dot{r}^2 + \frac{22717 \dot{r}^2 \nu}{168} + \frac{11
			\dot{r}^2 \nu^2}{8} - 7 \dot{r}^2 \nu^3 + \frac{615 \dot{r}^2 \nu
			\pi^2}{64} - \frac{20827 \nu v^2}{840} \nn\\& \qquad \quad + \nu^3 v^2 - \frac{123 \nu \pi^2 v^2}{64} \bigg)
		\nn \\ & + \frac{G^3 m^3}{r^3} \left( -16 - \frac{1399
			\nu}{12} - \frac{71 \nu^2}{2} + \frac{41 \nu {\pi }^2}{16} \right)\,,
		\label{AcoeffMH}\\
		\mathcal{B}^\text{MH}_\mathrm{3PN} &= -
		\frac{45 \dot{r}^5 \nu}{8} + 15 \dot{r}^5 \nu^2 + \frac{15 \dot{r}^5
			\nu^3}{4} + 12 \dot{r}^3 \nu v^2 - \frac{111 \dot{r}^3 \nu^2
			v^2}{4} - 12 \dot{r}^3 \nu^3 v^2 \nn\\& - \frac{65 \dot{r} \nu v^4}{8} + 19 \dot{r} \nu^2 v^4 + 6 \dot{r} \nu^3 v^4
		\nn \\ & + \frac{G m}{r}\left( \frac{329 \dot{r}^3
			\nu}{6} + \frac{59 \dot{r}^3 \nu^2}{2} + 18 \dot{r}^3 \nu^3 - 15
		\dot{r} \nu v^2 - 27 \dot{r} \nu^2 v^2 - 10 \dot{r} \nu^3 v^2
		\right) \nn \\ & + \frac{G^2 m^2}{r^2} \left( -4
		\dot{r} - \frac{5849 \dot{r} \nu}{840} + 25 \dot{r} \nu^2 + 8
		\dot{r} \nu^3 - \frac{123 \dot{r} \nu \pi^2}{32} \right) \,.
		\label{BcoeffMH}
\end{align}\end{subequations}
In the following we shall use the MH coordinate system when constructing the generalized quasi-Keplerian representation of the 3PN motion in Sect.~\ref{sec:QK}. We now come back the the SH coordinate system.

The conservative part of the CM equations of motion derive from a Lagrangian, which can be constructed as follows. We start from the general-frame Lagrangian \eqref{L3PN}--\eqref{resultL4PN}, which is a functional of $\bm{y}_\text{a}$, $\bm{v}_\text{a}$ and $\bm{a}_\text{a}$, and admits the CM integral $\bm{\dG}[\bm{y}_\text{a}, \bm{v}_\text{a}]$ given at 4PN order by Eqs.~(B1)--(B3) of \cite{BBFM17}. We perform the change of variables $(\bm{y}_1, \bm{y}_2)\longrightarrow (\bm{x}, \bm{\dG})$, where $\bm{x} = \bm{y}_1 - \bm{y}_2$. Since $\bm{\dG} = m_1\,\bm{y}_1 + m_2\,\bm{y}_2 + \calO(c^{-2})$ to Newtonian order, we find $\bm{y}_1 = X_2\bm{x}+\frac{1}{m}\bm{\dG} + \calO(c^{-2})$ and $\bm{v}_1 = X_2\bm{v}+\frac{1}{m}\frac{\dd \bm{\dG}}{\dd t} + \calO(c^{-2})$ (and $1\leftrightarrow 2$). Proceeding iteratively, it is easy to see that the old variables $\bm{y}_\text{a}$ are obtained as functionals of the new variables $(\bm{x}, \bm{\dG})$ and their derivatives, up to some differentiation order depending on the looked for PN order. In the process, we do not perform any order reduction of accelerations. Thus we get
\begin{equation}
	\bm{y}_\text{a} = \bm{y}_\text{a}\bigl[\bm{x}, \bm{v}, \bm{a}, \cdots; \bm{\dG},
	\hbox{$\frac{\dd \bm{\dG}}{\dd t}$}, \hbox{$\frac{\dd^2 \bm{\dG}}{\dd t^2}$},
	\cdots\bigr]\,,
\end{equation} 
which we plug into the original general-frame Lagrangian, still without order reduction of the accelerations. This yields an equivalent Lagrangian, which is now of the ``doubly generalized'' type
\begin{equation}\label{Ldouble}
	L = L\bigl[\bm{x}, \bm{v}, \bm{a}, \cdots; \bm{\dG}, \hbox{$\frac{\dd
			\bm{\dG}}{\dd t}$}, \hbox{$\frac{\dd^2 \bm{\dG}}{\dd t^2}$}, \cdots\bigr]\,.
\end{equation} 
The ensuing equations of motion read $\frac{\delta L}{\delta \bm{x}} = 0$ as usual, together with $\frac{\delta L}{\delta \bm{\dG}} = 0$. Now the latter equation is necessarily equivalent to the conservation law for the CM integral, hence we have
\begin{equation}\label{d2G}
	\frac{\delta L}{\delta \bm{\dG}} = 0 \quad\Longleftrightarrow\quad \frac{\dd^2\bm{\dG}}{\dd t^2} = 0\,.
\end{equation} 
As a result, we can choose $\bm{\dG}=0$ as a solution of these equations. The CM equations of motion are then given by $\frac{\delta L}{\delta \bm{x}} = 0$ in which we pose $\bm{\dG}=0$ [these equations are nothing but the CM equations of motion \eqref{dvdt} with \eqref{calAcoeff}--\eqref{calBcoeff}] and the CM Lagrangian is obtained by setting $\bm{\dG}=0$ in Eq.~\eqref{Ldouble}.
Defining as usual the reduced CM Lagrangian as $\mathcal{L}=L/\mu$, we explicitly get \citep{BI03CM,BBFM17}
\begin{subequations}\label{calL}\begin{align}
		\mathcal{L}_\text{N} &= \frac{v^2}{2} + \frac{G m}{r}\,,\\
		\mathcal{L}_\text{1PN} &= \frac{v^4}{8} - \frac{3\,\nu\,v^4}{8} +
		\frac{G m}{r}\,\left( \frac{\dot{r}^2\,\nu}{2} + \frac{3\,v^2}{2} +
		\frac{\nu\,v^2}{2} \right)-\frac{G^2m^2}{2\,r^2}\,,\\ 
		\mathcal{L}_\text{2PN} &= \frac{v^6}{16} - \frac{7\,\nu\,v^6}{16} +
		\frac{13\,\nu^2\,v^6}{16} \nn\\ & + \frac{G m}{r}\,\left(
		\frac{3\,\dot{r}^4\,\nu^2}{8} - \frac{\dot{r}^2\,(an)\,\nu\,r}{8} +
		\frac{\dot{r}^2\,\nu\,v^2}{4} - \frac{5\,\dot{r}^2\,\nu^2\,v^2}{4} +
		\frac{7\,(an)\,\nu\,r\,v^2}{8} \right.\nn\\ &\quad\quad~
		+ \left.\frac{7\,v^4}{8} - \frac{5\,\nu\,v^4}{4} -
		\frac{9\,\nu^2\,v^4}{8} - \frac{7\,\dot{r}\,\nu\,r\,(av)}{4} \right)
		\nn\\ & +\frac{G^2m^2}{r^2}\,\left( \frac{\dot{r}^2}{2} +
		\frac{41\,\dot{r}^2\,\nu}{8} + \frac{3\,\dot{r}^2\,\nu^2}{2} +
		\frac{7\,v^2}{4} - \frac{27\,\nu\,v^2}{8} + \frac{\nu^2\,v^2}{2}
		\right) \nn\\ &\quad~ +\frac{G^3m^3}{r^3}\,\left( \frac{1}{2} +
		\frac{15\,\nu}{4} \right)\,,\\ 
		\mathcal{L}_\text{3PN} &= \frac{5\,v^8}{128} -
		\frac{59\,\nu\,v^8}{128} + \frac{119\,\nu^2\,v^8}{64} -
		\frac{323\,\nu^3\,v^8}{128} \nn\\ & + \frac{G
			m}{r}\,\left( \frac{5\,\dot{r}^6\,\nu^3}{16} +
		\frac{\dot{r}^4\,(an)\,\nu\,r}{16} -
		\frac{5\,\dot{r}^4\,(an)\,\nu^2\,r}{16} -
		\frac{3\,\dot{r}^4\,\nu\,v^2}{16}
		\right.\nn\\ &\quad\quad~\left.+
		\frac{7\,\dot{r}^4\,\nu^2\,v^2}{4} -
		\frac{33\,\dot{r}^4\,\nu^3\,v^2}{16} -
		\frac{3\,\dot{r}^2\,(an)\,\nu\,r\,v^2}{16} -
		\frac{\dot{r}^2\,(an)\,\nu^2\,r\,v^2}{16}
		\right.\nn\\ &\quad\quad~\left.+
		\frac{5\,\dot{r}^2\,\nu\,v^4}{8} - 3\,\dot{r}^2\,\nu^2\,v^4
		+\frac{75\,\dot{r}^2\,\nu^3\,v^4}{16} + \frac{7\,(an)\,\nu\,r\,v^4}{8}
		\right.\nn\\ &\quad\quad~\left.-
		\frac{7\,(an)\,\nu^2\,r\,v^4}{2} + \frac{11\,v^6}{16} -
		\frac{55\,\nu\,v^6}{16} + \frac{5\,\nu^2\,v^6}{2}
		\right.\nn\\ &\quad\quad~ +\left.
		\frac{65\,\nu^3\,v^6}{16} + \frac{5\,\dot{r}^3\,\nu\,r\,(av)}{12} -
		\frac{13\,\dot{r}^3\,\nu^2\,r\,(av)}{8}
		\right.\nn\\ &\quad\quad~\left.-
		\frac{37\,\dot{r}\,\nu\,r\,v^2\,(av)}{8} +
		\frac{35\,\dot{r}\,\nu^2\,r\,v^2\,(av)}{4} \right) \nn\\ &
		+ \frac{G^2m^2}{r^2}\,\left( -\frac{109\,\dot{r}^4\,\nu}{144} -
		\frac{259\,\dot{r}^4\,\nu^2}{36} + 2\,\dot{r}^4\,\nu^3 -
		\frac{17\,\dot{r}^2\,(an)\,\nu\,r}{6}
		\right.\nn\\ &\quad\quad~ +\left.
		\frac{97\,\dot{r}^2\,(an)\,\nu^2\,r}{12} +\frac{\dot{r}^2\,v^2}{4} -
		\frac{41\,\dot{r}^2\,\nu\,v^2}{6} -
		\frac{2287\,\dot{r}^2\,\nu^2\,v^2}{48}
		\right.\nn\\ &\quad\quad~ -\left.
		\frac{27\,\dot{r}^2\,\nu^3\,v^2}{4} + \frac{203\,(an)\,\nu\,r\,v^2}{12}
		+ \frac{149\,(an)\,\nu^2\,r\,v^2}{6}
		\right.\nn\\ &\quad\quad~ +\left. \frac{45\,v^4}{16} +
		\frac{53\,\nu\,v^4}{24} + \frac{617\,\nu^2\,v^4}{24} -
		\frac{9\,\nu^3\,v^4}{4} \right.\nn\\ &\quad\quad~
		-\left. \frac{235\,\dot{r}\,\nu\,r\,(av)}{24} +
		\frac{235\,\dot{r}\,\nu^2\,r\,(av)}{6} \right) \nn\\ & +
		\frac{G^3m^3}{r^3}\,\left( \frac{3\,\dot{r}^2}{2} -
		\frac{12041\,\dot{r}^2\,\nu}{420} + \frac{37\,\dot{r}^2\,\nu^2}{4} +
		\frac{7\,\dot{r}^2\,\nu^3}{2} - \frac{123\,\dot{r}^2\,\nu\,{\pi
			}^2}{64} \right.\nn\\ &\quad\quad~
		+\left. \frac{5\,v^2}{4} + \frac{387\,\nu\,v^2}{70} -
		\frac{7\,\nu^2\,v^2}{4} + \frac{\nu^3\,v^2}{2} + \frac{41\,\nu\,{\pi
			}^2\,v^2}{64} \right.\nn\\ &\quad\quad~ \left. +
		22\,\dot{r}^2\,\nu\,\ln \Big(\frac{r}{r'_0}\Big) -
		\frac{22\,\nu\,v^2}{3}\ln \Big(\frac{r}{r'_0}\Big)
		\right)\nn\\ & + \frac{G^4m^4}{r^4}\,\left( -\frac{3}{8}-
		\frac{18469\,\nu}{840} + \frac{22\,\nu}{3}\ln \Big(\frac{r}{r'_0}\Big)
		\right) \,,\\
		\mathcal{L}^{(0)}_\text{4PN} &=\frac{7}{256} v^{10}
		-  \frac{121}{256} \nu v^{10}
		+ \frac{785}{256} \nu^2 v^{10}
		-  \frac{1127}{128} \nu^3 v^{10}
		+ \frac{2415}{256} \nu^4 v^{10} \,,\\
		\mathcal{L}^{(1)}_\text{4PN} &= \frac{m}{r} 
		\left(\frac{23}{20} \nu^2 (av) r \dot{r}^5
		-  \frac{5}{128} \nu \dot{r}^8
		+ \frac{35}{128} \nu^2 \dot{r}^8
		-  \frac{35}{64} \nu^3 \dot{r}^8
		+ \frac{35}{128} \nu^4 \dot{r}^8
		\right.\nn\\
		&\left. \quad\quad+ \frac{7}{4} \nu (av) r \dot{r}^3 v^{2}+ \frac{361}{24} \nu^3 (av) r \dot{r}^3 v^{2}
		+ \frac{19}{32} \nu (an) r \dot{r}^4 v^{2}
		\right.\nn\\
		&\left. \quad\quad+ \frac{85}{32} \nu^3 (an) r \dot{r}^4 v^{2}
		-  \frac{5}{16} \nu \dot{r}^6 v^{2}
		-  \frac{31}{16} \nu^2 \dot{r}^6 v^{2} + \frac{45}{32} \nu^3 \dot{r}^6 v^{2}
		\right.\nn\\
		&\left. \quad\quad-  \frac{85}{32} \nu^4 \dot{r}^6 v^{2}
		+ \frac{341}{4} \nu^2 (av) r \dot{r} v^{4}
		+ \frac{245}{32} \nu^2 (an) r \dot{r}^2 v^{4}
		-  \frac{17}{64} \nu \dot{r}^4 v^{4}
		\right.\nn\\
		&\left. \quad\quad-  \frac{11}{8} \nu^2 \dot{r}^4 v^{4} -  \frac{193}{16} \nu^3 \dot{r}^4 v^{4}
		+ \frac{693}{64} \nu^4 \dot{r}^4 v^{4}
		+ \frac{217}{96} \nu (an) r v^{6}
		\right.\nn\\
		&\left. \quad\quad+ \frac{2261}{96} \nu^3 (an) r v^{6}
		-  \frac{11}{48} \nu \dot{r}^2 v^{6}
		-  \frac{29}{4} \nu^2 \dot{r}^2 v^{6}+ \frac{1021}{96} \nu^3 \dot{r}^2 v^{6}
		\right.\nn\\
		&\left. \quad\quad-  \frac{665}{32} \nu^4 \dot{r}^2 v^{6}
		+ \frac{75}{128} v^{8}
		-  \frac{1595}{384} \nu v^{8}
		+ \frac{2917}{128} \nu^2 v^{8}
		+ \frac{493}{192} \nu^3 v^{8}\right.\nn\\
		&\quad\quad\left. -  \frac{2261}{128} \nu^4 v^{8}\right) \,,\\
		\mathcal{L}^{(2)}_\text{4PN} &= \frac{m^2}{r^2} 
		\left(\frac{5407}{288} \nu (av) r \dot{r}^3
		-  \frac{5531}{3200} \nu \dot{r}^6
		-  \frac{487}{40} \nu^2 \dot{r}^6
		+ \frac{13}{5} \nu^3 \dot{r}^6
		+ \frac{3}{2} \nu^4 \dot{r}^6
		\right.\nn\\
		&\left. \quad\quad+ \frac{11497}{48} \nu^2 (av) r \dot{r} v^{2} + \frac{469}{4} \nu^3 (av) r \dot{r} v^{2}
		+ \frac{517}{48} \nu^2 (an) r \dot{r}^2 v^{2}
		\right.\nn\\
		&\left. \quad\quad-  \frac{61183}{5760} \nu \dot{r}^4 v^{2}
		+ \frac{3079}{96} \nu^2 \dot{r}^4 v^{2}
		-  \frac{161}{8} \nu^3 \dot{r}^4 v^{2} -  \frac{27}{2} \nu^4 \dot{r}^4 v^{2}
		\right.\nn\\
		&\left. \quad\quad+ \frac{14627}{384} \nu (an) r v^{4}
		+ \frac{3}{16} \dot{r}^2 v^{4}
		-  \frac{12091}{640} \nu \dot{r}^2 v^{4}
		-  \frac{22045}{192} \nu^2 \dot{r}^2 v^{4}
		\right.\nn\\
		&\left. \quad\quad+ \frac{255}{64} \nu^3 \dot{r}^2 v^{4} + \frac{569}{16} \nu^4 \dot{r}^2 v^{4}
		+ \frac{115}{32} v^{6}
		+ \frac{593}{120} \nu v^{6}
		+ \frac{9467}{192} \nu^2 v^{6}
		\right.\nn\\
		&\left. \quad\quad+ \frac{599}{64} \nu^3 v^{6}
		+ \frac{195}{16} \nu^4 v^{6}\right)
		\,,\\
		\mathcal{L}^{(3)}_\text{4PN} &= \frac{m^3}{r^3} \left(\frac{4937}{1260} 
		\nu^2 (av) r \dot{r}
		-  \frac{41}{32} \pi^2 \nu^2 (av) r \dot{r}
		+ \frac{44}{3} \nu^2 (av) r \ln\Big(\frac{r}{r'_{0}}\Big) \dot{r}
		\right.\nn\\
		&\left. \quad\quad+ \frac{22}{3} \nu (av) r \ln\Big(\frac{r}{r'_{0}}\Big) \dot{r}
		-  \frac{246373}{2240} \nu \dot{r}^4 -  \frac{2155}{1024} \pi^2 \nu \dot{r}^4
		-  \frac{210733}{2016} \nu^2 \dot{r}^4
		\right.\nn\\
		&\left. \quad\quad+ \frac{205}{128} \pi^2 \nu^2 \dot{r}^4
		+ \frac{367}{32} \nu^3 \dot{r}^4
		+ \frac{29}{4} \nu^4 \dot{r}^4 -  \frac{55}{3} \nu \ln\Big(\frac{r}{r'_{0}}\Big) \dot{r}^4
		\right.\nn\\
		&\left. \quad\quad+ 55 \nu^2 \ln\Big(\frac{r}{r'_{0}}\Big) \dot{r}^4
		+ \frac{229319}{6300} \nu (an) r v^{2}
		-  \frac{21}{32} \pi^2 \nu (an) r v^{2}
		\right.\nn\\
		&\left. \quad\quad + \frac{49}{4} \nu^3 (an) r v^{2} + 44 \nu (an) r \ln\Big(\frac{r}{r'_{0}}\Big) v^{2}
		+ \frac{44}{3} \nu^2 (an) r \ln\Big(\frac{r}{r'_{0}}\Big) v^{2}
		\right.\nn\\
		&\left. \quad\quad+ \frac{7}{4} \dot{r}^2 v^{2}
		+ \frac{516319}{4200} \nu \dot{r}^2 v^{2}
		+ \frac{447}{512} \pi^2 \nu \dot{r}^2 v^{2} + \frac{53099}{560} \nu^2 \dot{r}^2 v^{2}
		\right.\nn\\
		&\left. \quad\quad+ \frac{123}{64} \pi^2 \nu^2 \dot{r}^2 v^{2}
		-  \frac{1003}{16} \nu^3 \dot{r}^2 v^{2}
		-  \frac{47}{2} \nu^4 \dot{r}^2 v^{2} - 44 \nu \ln\Big(\frac{r}{r'_{0}}\Big) \dot{r}^2 v^{2}
		\right.\nn\\
		&\left. \quad\quad - 110 \nu^2 \ln\Big(\frac{r}{r'_{0}}\Big) \dot{r}^2 v^{2}
		+ \frac{91}{16} v^{4}
		-  \frac{166703}{20160} \nu v^{4} + \frac{133}{1024} \pi^2 \nu v^{4}
		\right.\nn\\
		&\left. \quad\quad+ \frac{10601}{3360} \nu^2 v^{4}
		-  \frac{123}{128} \pi^2 \nu^2 v^{4}
		+ \frac{567}{32} \nu^3 v^{4}
		-  \frac{15}{4} \nu^4 v^{4}
		\right.\nn\\
		&\left. \quad\quad+ \frac{55}{3} \nu \ln\Big(\frac{r}{r'_{0}}\Big) v^{4} + \frac{77}{3} \nu^2 \ln\Big(\frac{r}{r'_{0}}\Big) v^{4}\right) \,,\\
		\mathcal{L}^{(4)}_\text{4PN} &=\frac{m^4}{r^4} \left(\frac{9}{4} \dot{r}^2
		-  \frac{245971}{4200} \nu \dot{r}^2
		+ \frac{2771}{96} \pi^2 \nu \dot{r}^2
		-  \frac{8089}{140} \nu^2 \dot{r}^2
		- 44 \pi^2 \nu^2 \dot{r}^2
		\right.\nn\\
		&\left. \quad\quad+ \frac{185}{8} \nu^3 \dot{r}^2
		+ \frac{15}{2} \nu^4 \dot{r}^2 - 339 \nu \ln\Big(\frac{r}{r'_{0}}\Big) \dot{r}^2
		+ \frac{1616}{3} \nu^2 \ln\Big(\frac{r}{r'_{0}}\Big) \dot{r}^2\right.\nn\\
		&\quad\quad \left.
		+ \frac{15}{16} v^{2}
		+ \frac{2039993}{50400} \nu v^{2}
		-  \frac{191}{32} \pi^2 \nu v^{2}
		-  \frac{52907}{1008} \nu^2 v^{2}
		+ 11 \pi^2 \nu^2 v^{2}\right.\nn\\
		&\quad\quad\left. -  \frac{1}{8} \nu^3 v^{2}
		+ \frac{1}{2} \nu^4 v^{2}		
		+ \frac{301}{3} \nu \ln\Big(\frac{r}{r'_{0}}\Big) v^{2}
		- \frac{371}{3} \nu^2 \ln\Big(\frac{r}{r'_{0}}\Big) v^{2}\right)
		\,,\\
		\mathcal{L}^{(5)}_\text{4PN} &=\frac{m^5}{r^5} \left(\frac{3}{8}
		+ \frac{1697177}{25200} \nu
		+ \frac{105}{32} \pi^2 \nu
		+ \frac{55111}{720} \nu^2
		- 11 \pi^2 \nu^2
		\right.\nn\\
		&\quad\quad\left. 
		- \frac{290}{3} \nu \ln\Big(\frac{r}{r'_{0}}\Big)
		+ 120 \nu^2 \ln\Big(\frac{r}{r'_{0}}\Big)\right)
		\,.
\end{align}\end{subequations}
Remind that one has still to add the conservative 4PN tail term given by
\begin{align}\label{LtailCM}
	\mathcal{L}_\text{tail} = \frac{G^2}{5c^8\nu}
	\,\dQ_{ab}^{(3)}(t)\!\mathop{\text{Pf}}_{2r/c}
	\int_{-\infty}^{+\infty} \frac{\dd t'}{\vert t-t'\vert}
	\dQ_{ab}^{(3)}(t')\,,
\end{align}
whose functional variation yields Eq.~\eqref{acctailCMcons}. 

The above CM Lagrangian in harmonic coordinates depends on accelerations starting at 2PN order, through $(an)\equiv\bm{a}\cdot\bm{n}$ and $(av)\equiv\bm{a}\cdot\bm{v}$. Furthermore, it depends on the logarithm of $r/r'_0$. Both the accelerations and the logarithms can be removed by applying a contact transformation, and one can prove that the harmonic-coordinates CM Lagrangian is equivalent to the ADM-coordinates CM Hamiltonian (which has no accelerations nor logarithms). In the CM frame the conjugate variables are the relative separation $\bm{X}=R\,\bm{N}=\bm{Y}_1-\bm{Y}_2$ and the conjugate momentum (per unit reduced mass) $\bm{P}$ such that $\mu\bm{P}=\bm{P}_1=-\bm{P}_2$ ($\bm{P}_1$ and $\bm{P}_2$ are defined in Sect.~\ref{sec:Lag4PN}). Posing $P^2\equiv\bm{P}^2$ and $P_R\equiv\bm{N}\cdot\bm{P}$ and $\mathcal{H}=H/\mu$, we have \citep{DJS14, DJS15eob, JaraSLRR, BlumMMS20a}
\begin{subequations}\label{HADMcm}
	\begin{align}
	\mathcal{H}_\text{N} &= \frac{P^2}{2} -\frac{G m}{R}\,,\\ 
	\mathcal{H}_\text{1PN} &=
	\left(- \frac{1}{8} +
	\frac{3\,\nu}{8}\right)P^4 + \frac{G m}{R}\left( - \frac{{P_R}^2\,\nu}{2} -
	\frac{3\,P^2}{2} - \frac{\nu\,P^2}{2} \right)+\frac{G^2m^2}{2R^2} \,,\\ 
	\mathcal{H}_\text{2PN} &=
	\left(\frac{1}{16} -
	\frac{5\nu}{16} + \frac{5\nu^2}{16}\right)P^6
	\nn\\&\quad~+ \frac{G m}{R}\left( - \frac{3\,{P_R}^4\,\nu^2}{8}
	- \frac{{P_R}^2\,P^2\,\nu^2}{4} + \frac{5\,P^4}{8} -
	\frac{5\,\nu\,P^4}{2} - \frac{3\,\nu^2\,P^4}{8}
	\right)\nn\\&\quad~ + \frac{G^2m^2}{R^2}\,\left(
	\frac{3\,{P_R}^2\,\nu}{2} + \frac{5\,P^2}{2} + 4\,\nu\,P^2 \right)
	\nn\\&\quad~+\frac{G^3m^3}{R^3}\left( -\frac{1}{4} -
	\frac{3\,\nu}{4} \right)\,, \\ 
	\mathcal{H}_\text{3PN} &=
	\left(-\frac{5}{128} + \frac{35\nu}{128} -
	\frac{35\nu^2}{64} + \frac{35\nu^3}{128}\right)P^8
	\nn\\&\quad~ + \frac{G m}{R}\left(
	-\frac{5\,{P_R}^6\,\nu^3}{16} + \frac{3\,{P_R}^4\,P^2\,\nu^2}{16} -
	\frac{3\,{P_R}^4\,P^2\,\nu^3}{16} + \frac{{P_R}^2\,P^4\,\nu^2}{8}
	\right.\nn\\&\quad\qquad\quad~ \left. -
	\frac{3\,{P_R}^2\,P^4\,\nu^3}{16}-\frac{7\,P^6}{16} +
	\frac{21\,\nu\,P^6}{8} - \frac{53\,\nu^2\,P^6}{16} -
	\frac{5\,\nu^3\,P^6}{16} \right) \nn\\&\quad~ +
	\frac{G^2m^2}{R^2}\,\left( \frac{5\,{P_R}^4\,\nu}{12} +
	\frac{43\,{P_R}^4\,\nu^2}{12} + \frac{17\,{P_R}^2\,P^2\,\nu}{16}
	\right.\nn\\&\quad\qquad\quad~ \left.+
	\frac{15\,{P_R}^2\,P^2\,\nu^2}{8} - \frac{27\,P^4}{16} +
	\frac{17\,\nu\,P^4}{2} + \frac{109\,\nu^2\,P^4}{16} \right)
	\nn\\&\quad~ + \frac{G^3m^3}{R^3}\,\left(
	-\frac{85\,{P_R}^2\,\nu}{16} - \frac{7\,{P_R}^2\,\nu^2}{4} -
	\frac{25\,P^2}{8} - \frac{335\,\nu\,P^2}{48}
	\right.\nn\\&\quad\qquad\quad~ \left.- \frac{23\,\nu^2\,P^2}{8}
	- \frac{3\,{P_R}^2\,\nu\,\pi^2}{64} + \frac{\nu\,P^2\,\pi^2}{64}
	\right)\nn\\&\quad~ + \frac{G^4m^4}{R^4}\, \left( \frac{1}{8} +
	\frac{109\,\nu}{12} - \frac{21\,\nu\,\pi^2}{32} \right) \,,\\
\mathcal{H}^{(0)}_\text{4PN} &= \left(
		\frac{7}{256}
		-\frac{63}{256}\nu
		+\frac{189}{256}\nu^2
		-\frac{105}{128}\nu^3
		+\frac{63}{256}\nu^4
		\right)P^{10} \,,\\ 
\mathcal{H}^{(1)}_\text{4PN} &=
		\frac{m}{R}\Biggl\{
		\frac{45}{128} P^8
		-\frac{45}{16} P^8\nu
		+\left(
		\frac{423}{64} P^8
		-\frac{3}{32} P_R^2 P^6
		-\frac{9}{64} P_R^4 P^4
		\right)\nu^2
		\nn\\ &
		\quad + \left(
		-\frac{1013}{256} P^8
		+\frac{23}{64} P_R^2 P^6
		+\frac{69}{128} P_R^4 P^4
		-\frac{5}{64} P_R^6 P^2
		+\frac{35}{256} P_R^8
		\right)\nu^3
		\nn\\&
		\quad + \left(
		-\frac{35}{128} P^8
		-\frac{5}{32} P_R^2 P^6
		-\frac{9}{64} P_R^4 P^4
		-\frac{5}{32} P_R^6 P^2
		-\frac{35}{128} P_R^8
		\right)\nu^4
		\Biggr\}\,,\\
\mathcal{H}^{(2)}_\text{4PN} &=
		\frac{m^2}{R^2}\Biggl\{
		\frac{13}{8} P^6
		+ \left(
		-\frac{791}{64}P^6
		+\frac{49}{16} P_R^2 P^4
		-\frac{889}{192} P_R^4 P^2
		+\frac{369}{160} P_R^6
		\right)\nu
		\nn\\&
		\quad + \left(
		\frac{4857}{256} P^6
		-\frac{545}{64} P_R^2 P^4
		+\frac{9475}{768} P_R^4 P^2
		-\frac{1151}{128} P_R^6
		\right)\nu^2
		\nn\\&
		\quad + \left(
		\frac{2335}{256} P^6
		+\frac{1135}{256} P_R^2 P^4
		-\frac{1649}{768} P_R^4 P^2
		+\frac{10353}{1280} P_R^6
		\right)\nu^3
		\Biggr\}\,,\\ 
\mathcal{H}^{(3)}_\text{4PN} &=
		\frac{m^3}{R^3}\Biggl\{ \frac{105}{32} P^4
		\nn\\&\quad+ \left[ \left(\frac{2749}{8192}\pi^2-\frac{589189}{19200}\right) P^4
		+ \left(\frac{63347}{1600} - \frac{1059}{1024}\pi^2\right) P_R^2 P^2 
		\right.  \nn \\& \quad\left. 
		+ \left(\frac{375}{8192}\pi^2-\frac{23533}{1280}\right) P_R^4 \right]\nu
		+ \bigg[ \left(\frac{18491}{16384}\pi^2 - \frac{1189789}{28800}\right) P^4
		\nn\\&\quad
		- \left(\frac{127}{3} + \frac{4035}{2048}\pi^2\right) P_R^2 P^2
		+ \left(\frac{57563}{1920} - \frac{38655}{16384}\pi^2 \right) P_R^4
		\bigg]\nu^2
		\nn\\&\quad
		+ \bigg(
		-\frac{553}{128} P^4
		-\frac{225}{64} P_R^2 P^2
		-\frac{381}{128} P_R^4
		\bigg)\nu^3
		\Biggr\}\,,\\
\mathcal{H}^{(4)}_\text{4PN} &=
		\frac{m^4}{R^4}\Biggl\{
		\frac{105}{32}P^2
		\\&\quad + \left[  \left(\frac{185761}{19200} - \frac{21837}{8192}\pi^2\right) P^2
		+ \left(\frac{3401779}{57600} - \frac{28691}{24576}\pi^2\right) P_R^2 \right]\nu
		\nn\\[1ex]&\quad
		+ \left[ \left(\frac{672811}{19200} - \frac{158177}{49152}\pi^2\right) P^2
		+ \left(-\frac{21827}{3840} + \frac{110099}{49152}\pi^2\right) P_R^2 \right]\nu^2
		\Biggr\}\,,\nn\\
\mathcal{H}^{(5)}_\text{4PN} &=
		\frac{m^5}{R^5}\Biggl\{
		-\frac{1}{16}
		+ \left({-\frac{169199}{2400} + \frac{6237}{1024}\pi^2}\right) \, \nu
		+ \left(-\frac{1256}{45} + \frac{7403}{3072}\pi^2\right)\,\nu^2
		\Biggr\}\,,
	\end{align}
\end{subequations}
while to this order the tail part is the opposite of that in the Lagrangian: $\mathcal{H}_\text{tail} = - \mathcal{L}_\text{tail}$, where $\mathcal{L}_\text{tail}$ is given by \eqref{LtailCM} with $r\rightarrow R$.


\subsubsection{Equations of motion and energy for quasi-circular orbits}
\label{sec:eomcirc}

Most inspiralling compact binaries will have been circularized by the time they become visible by the detectors LIGO and Virgo; see Sect.~\ref{sec:quadform} for a discussion. In the case of orbits that are circular -- apart from the gradual radiation-reaction inspiral -- the complicated equations of motion simplify drastically, since we have $\dot{r}=(nv)=\calO(1/c^5)$ and, for instance, $\ddot{r}=\calO(1/c^{10})$. In the case of (quasi-)circular orbits it is convenient to display the successive post-Newtonian corrections employing the useful post-Newtonian parameter
\begin{equation}
	\gamma \equiv \frac{G m}{r c^2} =
	\calO\left(\frac{1}{c^2}\right)\,.
	\label{gammadef}
\end{equation}
The relation between center-of-mass variables and the relative ones still reads \eqref{CMrel} but with the simplified coefficients
\begin{subequations}\label{PQcirc}
	\begin{align}
	\mathcal{P} &= \frac{G^2m^2}{r^2 c^4}\Biggl\{ 3+ \left(-\frac{7211}{1260} + \frac{22}{3}
	\ln \left( \frac{r}{r'_0}\right) - \nu\right)\gamma \nn\\ & \quad + 
	\left(\frac{31669}{600} - \frac{154}{3}
	\ln \left( \frac{r}{r'_0}\right) \right.\nn\\ & \left. \qquad\quad + \left[ - \frac{13}{180} - \frac{135}{32}\pi^2 + 60
	\ln \left( \frac{r}{r'_0}\right) \right]\nu\right)\gamma^2\Biggr\}
	+ \calO\left(\frac{1}{c^{9}}\right)\,,\\
	\mathcal{Q} &= \frac{G^2m^2}{r c^5}\Biggl\{ - \frac{4}{5} + \left(-\frac{316}{105} - \frac{2}{5} \nu\right)\gamma \Biggr\}
	+ \calO\left(\frac{1}{c^{9}}\right)\,,
\end{align}
\end{subequations}

The relative acceleration $\bm{a}\equiv \bm{a}_1-\bm{a}_2$ of two bodies moving on a circular orbit at the 4PN order is then given by (in harmonic gauge)
\begin{equation}
  \bm{a} = -\Omega^2 \bm{x}-
  \frac{32}{5}\frac{G^3m^3\nu}{c^5r^4}\left[1+ \gamma
    \left(-\frac{743}{336} - \frac{11}{4}\nu\right) + 4\pi\gamma^{3/2}\right]\bm{v} +
  \calO\left(\frac{1}{c^{9}}\right)\,,
  \label{aieom}
\end{equation}
where $\bm{x}\equiv \bm{y}_1-\bm{y}_2$ is the relative separation (in harmonic coordinates) and $\Omega$ denotes the angular frequency of the quasi-circular motion. The second term in Eq.~\eqref{aieom}, opposite to the velocity $\bm{v}\equiv \bm{v}_1-\bm{v}_2$, represents the radiation reaction force up to 4PN order, which comes from the 2.5PN and 3.5PN coefficients in Eqs. \eqref{calAcoeff}--\eqref{calBcoeff}, plus the dissipative part of the 4PN tail contribution, given by Eq.~\eqref{acctailCMdiss} and which produces the $4\pi$ term in \eqref{aieom}. The calculation of the next 4.5PN dissipative coefficient has been carried out using the EFT by \cite{LPY23}. The results at 2.5PN, 3.5PN and 4.5PN orders are in agreement with the flux-balance method \citep{IW93,IW95,GII97}. The radiation-reaction force is responsible for the secular decrease of the separation $r$ and increase of the orbital frequency $\Omega$: 
\begin{subequations}\label{rOmdot}
\begin{align}
\dot{r} &= - \frac{64}{5} \frac{G^3 m^3\nu}{r^3 c^5} \left[1+ \gamma
  \left(-\frac{1751}{336} -
  \frac{7}{4}\nu\right) + 4\pi \gamma^{3/2} +
  \calO\left(\frac{1}{c^{4}}\right)\right]\,,\\ \dot{\Omega} &= \frac{96}{5}
\,\frac{G m \nu}{r^3}\,\gamma^{5/2}\left[1+ \gamma
  \left(-\frac{2591}{336} - \frac{11}{12}\nu\right) + 4\pi \gamma^{3/2} +
  \calO\left(\frac{1}{c^{4}}\right)\right]\,,
\end{align}
\end{subequations}
which can be recovered and extended to very high order (namely 2.5PN $+$ 4.5PN) using the phase and frequency evolution described in Sect.~\ref{sec:orbevol}.

Concerning conservative effects, the main content of the 4PN equations of motion \eqref{aieom} is the relation between the frequency $\Omega$ and the orbital separation $r$ or equivalently the parameter \eqref{gammadef}. This is given by the following generalized version of ``Kepler's third law'':\footnote{In the limit of quasi-circular orbits (in the adiabatic approximation) the tail integral in the equation of motion becomes local and we have 
\begin{align*}
\left(\mathop{\text{Pf}}_{2r/c}
\int_{-\infty}^{+\infty} \frac{\dd t'}{\vert t-t'\vert}
\,\dQ_{ij}^{(n)}(t')\right)\bigg|_\text{circ} = - 2 \dQ_{ij}^{(n)}\biggl[ \ln\left(\frac{2r\Omega}{c}\right) + \gamma_\text{E}\biggr]\,.
\end{align*}
}
\begin{align}\label{keplerlaw}
	\Omega^2 &= \frac{G m}{r^3} \Bigg\{ 1+(-3+\nu) \gamma + \left( 6 +
	\frac{41}{4}\nu + \nu^2 \right) \gamma^2 \nn\\ &  +
	\biggl( -10 + \left[- \frac{75707}{840} + \frac{41}{64} \pi^2 + 22
	\ln \left( \frac{r}{r'_0}\right) \right]\nu + \frac{19}{2}\nu^2 +
	\nu^3 \biggr) \gamma^3 \nn\\ & +
	\biggl(15 + \left[\frac{19644217}{33600} + \frac{163}{1024} \pi^2
	+ \frac{256}{5} \gamma_\text{E} + \frac{128}{5} \ln (16 \gamma ) 
	- 290 \ln\Big(\frac{r}{r'_{0}}\Big) \right] \nu \nn\\
	&\qquad  + \left[\frac{44329}{336} 
	-  \frac{1907}{64} \pi^2 + \frac{1168}{3} \ln\Big(\frac{r}{r'_{0}}\Big) \right] \nu^2 + \frac{51}{4} \nu^3 
	+ \nu^4 \biggr)\gamma^4 \Biggr\} \nn\\ &+
	\calO\left(\frac{1}{c^{10}}\right)\,,
\end{align}
where $r'_0$ is the UV type length scale in Eqs. \eqref{calAcoeff}--\eqref{calBcoeff}, and $\gamma_\text{E}$ is Euler's constant. As for the circular energy, it is inferred from the CM and then circular-orbit reduction of the general frame expression. We have
\begin{align}\label{Ecircgam}
	\dE &= -\frac{\mu c^2 \gamma}{2} \Biggl\{ 1 + \left( - \frac{7}{4} +
	\frac{\nu}{4} \right) \gamma + \left( - \frac{7}{8} + \frac{49}{8}
	\nu + \frac{\nu^2}{8} \right) \gamma^2 \nn\\ & 
	+ \left(-\frac{235}{64} + \left[\frac{46031}{2240} - \frac{123}{64}
	\pi^2 + \frac{22}{3} \ln \left( \frac{r}{r_0'} \right) \right] \nu
	+ \frac{27}{32} \nu^2 + \frac{5}{64} \nu^3 \right) \gamma^3
	\nn\\ & + \left(- \frac{649}{128} + \left[- \frac{3357833}{28800} + \frac{384}{5} \gamma_\text{E} + \frac{192}{5}\ln(16\gamma) + \frac{14935}{1024}\pi^2 - 93 \ln\Big(\frac{r}{r'_{0}}\Big)\right]\nu \right. \nn\\
	&\left.  \qquad+ \left[\frac{83959}{8064} -  \frac{957}{128} \pi^2 +  \frac{371}{3} \ln\Big(\frac{r}{r'_{0}}\Big)\right]\nu^2 + \frac{69}{64} \nu^3 + \frac{7}{128} \nu^4
	\right)\gamma^4 \Biggr\} \nn\\ &+ \calO \left(
	\frac{1}{c^{10}} \right)\,.
\end{align}
This expression is that of a physical observable $\dE$; however, it depends on the choice of a coordinate system, as it involves the post-Newtonian parameter $\gamma$ defined from the harmonic-coordinate separation $r$. But the \emph{numerical} value of $\dE$ should not depend on the choice of a coordinate system, so $\dE$ must admit a frame-invariant expression, the same in all coordinate systems. To find it we re-express $\dE$ with the help of the following frequency-related parameter $x$, instead of the post-Newtonian parameter $\gamma$ (with $m=m_1+m_2$):\footnote{This parameter is an invariant in a large class of coordinate systems, namely all those for which the metric becomes asymptotically Minkowskian far away from the matter source, i.e., $g_{\alpha\beta}\to\text{diag}(-1,1,1,1)$.}
\begin{equation}
  x \equiv \left(\frac{G m \Omega}{c^3}\right)^{2/3} =
  \calO\left(\frac{1}{c^2}\right)\,.
  \label{xdef}
\end{equation}
We readily obtain from Eq.~\eqref{keplerlaw} the expression of $\gamma$ in terms of $x$ as
\begin{align}\label{gammax}
	\gamma &= x \,\Biggl\{1  + \left(1 -  \frac{\nu}{3} \right) x + \left(1 -  \frac{65}{12} \nu\right) x^2 \nn\\
	& + \left(1 +  \left[- \frac{2203}{2520} -  \frac{41}{192} \pi^2  +  \frac{22}{3} \ln\left(\frac{x \,r'_{0}c^2}{G m}\right)\right]\nu + \frac{229}{36} \nu^2 + \frac{\nu^3}{81} \right) x^3 \nn\\
	& + \left(1 + \left[- \frac{2067859}{33600} - \frac{256}{15}\gamma_\text{E} -  \frac{5411}{3072} \pi^2 - 38 \ln\left(\frac{x \,r'_{0}c^2}{G m}\right) - \frac{128}{15} \ln(16 x)\right) \right]\nu \nn\\
	&\qquad\left. + \left[\frac{153613}{15120} + \frac{6049}{576} \pi^2 + \frac{992}{9} \ln\left(\frac{x \,r'_{0}c^2}{G m}\right) \right]\nu^2 -  \frac{1261}{324} \nu^3 + \frac{\nu^4}{243} \right) x^4 \nn\\& + \calO\left(\frac{1}{c^{10}}\right)\Biggr\}\,,
\end{align}
that we substitute back into Eq.~\eqref{Ecircgam}, making all appropriate post-Newtonian re-expansions. As a result, we gladly discover that the gauge constant $r'_0$ has cancelled out. Therefore, the gauge-invariant result is \citep{DJSinv, BF00, DJSdim, BDE04, LBW12, LBB12, BiniD13, DJS14, BBBFMb, BBBFMc, FS4PN, FPRS19}\footnote{Note the exact (Schwarzschild) test-mass limit $\nu\to 0$ of the energy function:
\begin{align*}
	\dE^\text{Schw} &= \mu c^2\left\{\frac{1-2x}{\sqrt{1-3x}}-1 \right\} = - \frac{\mu c^2 x}{2}  \sum_{p=0}^{+\infty}\frac{(2p-1)!!(1-2p)}{(p+1)!}\left(\frac{3x}{2}\right)^p \,.
\end{align*}
}
\begin{align}\label{Ecirc}
	\dE &= - \frac{\mu c^2 x}{2} \Biggl\{ 1 + \biggl( - \frac{3}{4} -
	\frac{\nu}{12} \biggr) x + \biggl( - \frac{27}{8} +
	\frac{19}{8} \nu - \frac{\nu^2}{24} \biggr) x^2 \nn\\
	&\qquad + \biggl( - \frac{675}{64} + \biggl[
	\frac{34445}{576} - \frac{205}{96} \pi^2 \biggr] \nu -
	\frac{155}{96} \nu^2 - \frac{35}{5184} \nu^3 \biggr) x^3
	\nn\\ 
	&\qquad + \biggl( - \frac{3969}{128} +
	\biggl[-\frac{123671}{5760}+\frac{9037}{1536}\pi^2 +
	\frac{896}{15}\gamma_\text{E}+ \frac{448}{15} \ln(16
	x)\biggr]\nu\nn\\
	& \qquad\qquad +
	\biggl[-\frac{498449}{3456}+\frac{3157}{576}\pi^2\biggr] \nu^2
	+\frac{301}{1728}\nu^3 + \frac{77}{31104}\nu^4\biggr) x^4 \nn\\&\qquad + \calO\left(\frac{1}{c^{10}}\right)
	\Biggr\} \,.
\end{align}
For quasi-circular orbits one can check that there are no terms of order $\propto x^{9/2}$ in \eqref{Ecirc}, so the result is actually valid up to the 4.5PN order. We shall discuss in Sect.~\ref{sec:spins} how the effects of the spins of the two black holes affect the latter formula. We can write also a similar expression for the angular momentum,
\begin{align}\label{Jcirc}
	\dJ &= \frac{G \mu m}{c x^{1/2}} \Biggl\{ 1 + \left(
	\frac{3}{2} + \frac{\nu}{6} \right) x + \left( \frac{27}{8} -
	\frac{19}{8} \nu + \frac{\nu^2}{24} \right) x^2 \nn
	\\ &\qquad + \left( \frac{135}{16} + \biggl[ -
	\frac{6889}{144} + \frac{41}{24} \pi^2 \biggr] \nu +
	\frac{31}{24} \nu^2 + \frac{7}{1296} \nu^3 \right) x^3 \nn
	\\ &\qquad + \left( \frac{2835}{128} +
	\left[\frac{98869}{5760}-\frac{6455}{1536}\pi^2 -
	\frac{128}{3}\gamma_\text{E} -
	\frac{64}{3}\ln(16x)\right]\nu\right.\nn\\ &
	\qquad\qquad \left.+
	\left[\frac{356035}{3456}-\frac{2255}{576}\pi^2\right]\nu^2
	-\frac{215}{1728}\nu^3 - \frac{55}{31104}\nu^4\right) x^4 \nn\\&\qquad + \calO\left(\frac{1}{c^{10}}\right)\Biggr\}
	\,,
\end{align}
For circular orbits the energy $\dE$ and angular momentum $\dJ$ are known to be linked together by the so-called ``thermodynamic'' relation
\begin{equation}\label{thermo}
	\frac{\partial \dE}{\partial \Omega} = \Omega \, \frac{\partial
		\dJ}{\partial \Omega}\,,
\end{equation}
which is actually just one aspect of the ``first law of binary black hole mechanics'' that we shall discuss in more details in Sect.~\ref{sec:firstlaw}.

We observe in Eqs. \eqref{Ecirc} and \eqref{Jcirc} the occurence of logarithmic terms $\propto\ln x$ starting at the 4PN order. These terms are due to tail effects occurring in the near zone. They contribute to the Fokker action \textit{via} the generalization of Eq.~\eqref{Ltail4PN} to include higher mass and current multipole moments \citep{FS20}. The 4PN, 5PN and 6PN coefficients of the logarithms in the circular energy function have been explicitly computed using a mix between the traditional approach \citep{BDLW10b, D10sf, LBW12} and the EFT method \citep{FS20, BFLS20}:
\begin{align}\label{Elogtail}
	\dE_\text{tail}^\text{log} &= -\frac{\mu c^2x}{2}\Biggl\{\frac{448}{15}\nu x^4\ln x+\left(-\frac{4988}{35}-\frac{656}{5}\nu\right)\nu x^5\ln x \nn\\&\quad + \left(-\frac{1967284}{8505}+\frac{914782}{945}\nu+\frac{32384}{135}\nu^2\right)\nu x^6\ln x + \calO\left(\frac{1}{c^{14}}\right)\Biggr\}\,.
\end{align}
The 7PN coefficient is also known, however in this case the iterated ``tail-of-tail-of-tail'' process is also relevant, and the complete 7PN result is not yet known. Furthermore it was shown how to compute, by means of renormalization group (RG) techniques \citep{GRoss10,GRR12}, the \textit{leading} logarithmic terms to generic power $n$, which turn out to appear at orders $(3n + 1)$PN, with $n=1$ corresponding to the 4PN logarithm in \eqref{Elogtail}, then $n=2$ being the first logarithm \textit{square} at 7PN order, and so on. We have, up to any power $n$ \citep{BFLS20}: 
\begin{subequations}\label{EJlogn}
\begin{align}
		\label{Elogn}
		\dE_\text{tail}^\text{leading-$(\log)^n$} &= -\frac{\mu c^2 x}{2} \Biggl\{ \frac{64 \nu}{15} \sum_{n = 1}^{+\infty} \frac{6n+1}{n!}\bigl(4\beta_2^{(m)}\bigr)^{n-1} x^{3n+1}(\ln x)^n\Biggr\}\,,\\
		\label{Jlogn}
		\dJ_\text{tail}^\text{leading-$(\log)^n$} &= \frac{G\mu m}{c x^{1/2}} \Biggl\{ - \frac{64 \nu}{15} \sum_{n = 1}^{+\infty} \frac{3n+2}{n!}\bigl(4\beta_2^{(m)}\bigr)^{n-1} x^{3n+1}(\ln x)^n\Biggr\}\,,
\end{align}
\end{subequations}
where $\beta^{(m)}_2 = -\frac{214}{105}$ is the $\beta$-function coefficient associated with the logarithmic renormalization of the mass-type quadrupole moment ($\ell=2$), and given for an arbitrary multipolarity by Eq.~\eqref{betamell}. For the linear log-term ($n=1$) one recovers the 4PN coefficient in \eqref{Elogtail}, while further comparison with high-order state-of-the-art self-force calculations by \cite{KOW15} reveals perfect match with the first terms up to $n\leqslant 7$ in the infinite series \eqref{Elogn}. Indeed, Eqs. \eqref{EJlogn} are first order effects in the test-mass limit $\nu\to 0$. Note that the comparison necessitates the first law of compact binary dynamics, reviewed in Sect.~\ref{sec:firstlaw}. The gravitational self-force (GSF) results are expressed in terms of the \cite{Det08} redshift variable. The first law permits to convey all the information from the redshift variable to the binary's energy function. The leading logarithms to any power $n$ in the redshift are given by Eq.~\eqref{u1tlogs}. All of this shows the great consistency between EFT and RG techniques, the traditional PN method and high-accuracy GSF calculations.

We end this section by giving also the periastron advance in the case of circular orbits to 4PN order \citep{DJS15eob,BBBFMb}:
\begin{align}\label{Ktotal}
	K &= 1 + 3 x + \left(\frac{27}{2} - 7\nu\right) x^2 +
	\left(\frac{135}{2}
	+\left[-\frac{649}{4}+\frac{123}{32}\pi^2\right]\nu+ 7\nu^2\right) x^3
	\nn\\& + \left(\frac{2835}{8}
	+\left[-\frac{275941}{360}+\frac{48007}{3072}\pi^2 -
	\frac{1256}{15}\ln x - \frac{592}{15}\ln 2 - \frac{1458}{5}\ln 3 \right.\right.\nn\\&\left.\left.\quad\quad -
	\frac{2512}{15}\gamma_\text{E}\right]\nu +
	\left[\frac{5861}{12}-\frac{451}{32}\pi^2\right]\nu^2 -
	\frac{98}{27}\nu^3\right) x^4 + \calO\left(\frac{1}{c^{10}}\right)\,,
\end{align}
The GSF contribution therein is generally described by the function $\rho(x)$ such that $K^{-2} = 1 - 6 x + \nu\rho(x) + \calO(\nu^2)$, i.e.
\begin{align}\label{rho}
	\rho &= 14 x^2 + \left(\frac{397}{2}-\frac{123}{16}\pi^2\right) x^3
	\nn\\& + \left(-\frac{215729}{180} + \frac{58265}{1536}\pi^2 +
	\frac{1184}{15}\ln 2 + \frac{2916}{5}\ln 3 +
	\frac{5024}{15}\gamma_\text{E} + \frac{2512}{15}\ln x\right) x^4 \nn\\& + \calO\left(\frac{1}{c^{10}}\right)\,.
\end{align}
The 4PN coefficient is of the type $\rho_\text{4PN}=a_\text{4PN}+b_\text{4PN}\ln x$ and in particular, the numerical value of $a_\text{4PN}$ (together with $a_\text{3PN}$ and $a_\text{2PN}$) is in perfect agreement with the GSF calculations of \cite{vdM16}. A comparison between the PN prediction for the periastron advance of circular orbits and numerical calculations based on self-force theory in the small mass ratio limit is done by \cite{Letal11}.


\subsubsection{The 2.5PN metric in the near zone}
\label{sec:metric}

The near-zone metric is given by Eqs. \eqref{gmunu3PN} for general post-Newtonian matter sources. For point-particles binaries all the potentials $V$, $V_i$, $\cdots$ parametrizing the metric must be computed and iterated for delta-function sources. This entails UV-type divergencies. Up to the 2.5PN order it is sufficient to cure the divergences by means of the Hadamard self-field regularization (Sect.~\ref{sec:had}), equivalent to this order to the dimensional regularization (Sect.~\ref{sec:DReom}). 

The post-Newtonian iteration of the metric in closed analytic form is greatly helped -- and indeed is made possible at all -- by the existence of the \cite{Fock} elementary solution of the Poisson equation
\begin{equation}\label{Poissong}
	\Delta g^\text{Fock} = \frac{1}{r_1 r_2} \,,  
\end{equation}
which takes the interesting closed analytic form 
\begin{subequations}\label{gFock}
	\begin{align}
		g^\text{Fock} &= \ln S\,, \\ \text{where}~~S &\equiv r_1+r_2+r_{12} \,,
	\end{align} 
\end{subequations}
with $r_\text{a}=\vert\mathbf{x}-\bm{y}_\text{a}\vert$ and $r_{12}=\vert\bm{y}_1-\bm{y}_2\vert$. Thus $S$ is the Euclidean perimeter of the triangle made of the two source points $\bm{y}_\text{a}$ and the field point $\bm{x}$. An important point is that the Fock function solves the Poisson equation \eqref{Poissong} in the sense of distribution theory. The generalization to $d$ dimensions can be found in the Appendix C of \cite{BDE04}.

The Fock function is the starting point of a hierarchy of elementary kernel functions, typically solving Poisson equations of the type $\Delta \varphi = r_1^{2n-1}r_2^{2m-1}$, with $(n,m) \in \mathbb{N}^2$, and their iterations. Such kernels play a crucial role in the post-Newtonian iteration to high order. For definiteness we limit ourselves to the following equations:
\begin{subequations}\label{kernels_def}
	\begin{align}
		\Delta g &= \frac{1}{r_1r_2}\,, \\ \Delta f &= g\,, \qquad\quad \Delta f_{12} = \frac{r_1}{2r_2}\,, \qquad\quad \Delta f_{21} = \frac{r_2}{2r_1}\,, \\ \Delta h &= f\,, \qquad\quad \Delta h_{12} = \frac{r_1^3}{24r_2}\,, \qquad\quad \Delta h_{21} = \frac{r_2^3}{24r_1}\,,\\\Delta k &= \frac{r_1 r_2}{4}\,, \qquad \Delta k_{12} = f_{21}\,, \qquad\qquad \Delta k_{21} = f_{12}\,,
	\end{align}
\end{subequations}
where the numerical coefficients are chosen for convenience. The general structure of the solutions is made of two parts: an homogeneous regular solution of the Laplace or iterated Laplace operator multiplied by $\ln S$; and a specific polynomial of $r_1$, $r_2$ and $r_{12}$ \citep{BDI95, JaraS98, BFeom, JaraS15}. If we seek for particular solutions, we are free to add a global homogeneous solution, for example adding a numerical constant to the Fock function. However the final solution should contain the proper homogeneous function which is to be selected by the matching procedure of Sect.~\ref{sec:matchingeq}, see \cite{MHLMFB20} for details. Particular solution of the above equations are 
\begin{subequations}\label{kernels_part}
	\begin{align}
		g &= \ln S + \frac{197}{810}\,, \\
		f &= \frac{1}{12}\Biggl[\bigl(r_1^2+r_2^2-r_{12}^2\bigr)\Bigl(\ln S - \frac{73}{810}\Bigr) + r_1r_{12}+r_2r_{12}-r_1r_2\Biggr]\,,\\
		h &= \frac{1}{320}\Biggl[\Bigl(r_1^4+r_2^4-r_{12}^4-2r_{12}^2(r_1^2+r_2^2)+\frac{2}{3}r_1^2r_2^2\Bigr)\Bigl(\ln S - \frac{37}{81}\Bigr) + \frac{5}{3}r_{12}(r_1^3+r_2^3)\nn\\
		&\quad\quad  + r_1r_2(r_{12}^2-r_1^2-r_2^2)+r_{12}(r_1+r_2)(r_1r_2-r_{12}) + \frac{4}{9}r_1^2r_2^2 \Biggr]\,,\\
		k &=\frac{1}{120}\Biggl[ \Bigl(r_{12}^4-3r_1^4-3r_2^4+6r_1^2r_2^2+2r_{12}^2(r_1^2+r_2^2)\Bigr)\ln S  + \frac{21}{10}(r_1^4+r_2^4) \nn\\ &\quad\quad  -\frac{r_{12}^4}{30} + 3r_{12}(r_1^3+r_2^3) +  (r_1^2+r_2^2)\Bigl(3r_1r_2-\frac{31}{15}r_{12}^2\Bigr) \nn\\ &\quad\quad +r_1r_2r_{12}^2 - \frac{21}{5}r_1^2r_2^2-r_{12}(r_1+r_2)\bigl(r_{12}^2-3r_1r_2\bigr)\Biggr]\,,
	\end{align}
\end{subequations}
together with $f_{12}$, $h_{12}$ and $k_{12}$ obtained by exchanging the field point $\mathbf{x}$ with the source point $\bm{y}_1$, i.e. $f_{12} = f\vert_{\mathbf{x}\longleftrightarrow\bm{y}_1}$, $h_{12} = h\vert_{\mathbf{x}\longleftrightarrow\bm{y}_1}$ and $k_{12} = k\vert_{\mathbf{x}\longleftrightarrow\bm{y}_1}$, and similarly $f_{21}$, $h_{21}$ and $k_{21}$ obtained by exchanging $\mathbf{x}$ and $\bm{y}_2$. In addition to satisfying Eqs. \eqref{kernels_def} the kernels also satisfy 
\begin{equation}\label{extra}
	\Delta_1f_{12} = g\,,\qquad  \Delta_1h_{12} = f_{12}\,,\qquad  \Delta_1 k = f_{21}\,,\qquad  \Delta_1 k_{21} = f\,,
\end{equation}
where $\Delta_\text{a}$ is the Laplacian with respect to the source point $\bm{y}_\text{a}$, together with the equations obtained from $1\leftrightarrow 2$. Note that these equations are true for the specific homogeneous solutions chosen in Eqs. \eqref{kernels_part}. More formulas can be found in the Appendix B of \cite{JaraS15}.

To obtain the metric at the 2.5PN order, we need $g$ and also the solutions of more difficult elementary Poisson equations. Namely we meet
\begin{subequations}\label{eqK1H1}
\begin{align}
\Delta K_1 &= 2\,\partial_{i}\partial_{j}\left(\frac{1}{r_2}\right)
\partial_{i}\partial_{j}\ln r_1 \,, \label{eqK1} \\ \Delta H_1 &=
2\,\partial_{i}\partial_{j}\left(\frac{1}{r_1}\right)
\mathop{\partial}_{1}{}_{\!\!i}\!\mathop{\partial}_{2}{}_{\!\!j}g \,,
\label{eqH1}
\end{align}
\end{subequations}
with ${}_\text{a}\partial_i$ denoting the partial derivatives with respect to the source point $y_\text{a}^i$ (and as usual $\partial_i$ being the partial derivative with respect to the field point $x^i$). It is quite remarkable that the solutions of the latter equations are known in closed analytic form. By combining several earlier results from \cite{Carmeli65, OO73a, S87}, one can write these solutions as \citep{BDI95, BFP98}
\begin{subequations}\label{K1H1}
\begin{align}
K_1 &= \left(\frac{1}{2}\Delta-\Delta_1\right) \left[\frac{\ln r_1}{r_2}\right] + \frac{1}{2}\Delta_2\left[\frac{\ln r_{12}}{r_2}\right] + \frac{r_2}{2r_{12}^2r_1^2}+\frac{1}{ r_{12}^2r_2}
\,, \label{K1} \\ H_1 &= \frac{1}{2}\Delta_1\left[\frac{g}{ r_1}+\frac{\ln
    r_1}{ r_{12}}- \Delta_1\left(\frac{r_1+r_{12}}{ 2}g
  \right)\right] - \frac{r_2}{ 2r_1^2r_{12}^2}\nn\\ &\quad +
\partial_i\!\mathop{\partial}_{2}{}_{\!\!i}\left[\frac{\ln r_{12} }{
    r_1}+ \frac{\ln r_1}{ 2r_{12}}\right] +
\frac{1}{2}\Delta_2\left[\frac{\ln r_{12}}{ r_{1}}\right] \,, \label{H1}
\end{align}
\end{subequations}
or, equivalently, in expanded form,
\begin{subequations}\label{KHexpl}
	\begin{align}
		K_1 &= -\frac{1}{r_2^3} + \frac{1}{r_2r_{12}^2}-\frac{1}{r_1^2r_2}+
		\frac{r_2}{2r_1^2r_{12}^2}+\frac{r_{12}^2}{2r_1^2r_2^3}
		+\frac{r_1^2}{2r_2^3r_{12}^2} \,, \\
		H_1 &= -\frac{1}{2r_1^3}-\frac{1}{4r_{12}^3} - \frac{1}{4r_1^2r_{12}}-
		\frac{r_2}{2 r_1^2r_{12}^2}+\frac{r_2}{2r_1^3r_{12}}
		+\frac{3r_2^2}{4r_1^2r_{12}^3}\nn\\&\quad+\frac{r_2^2}{2r_1^3r_{12}^2}-
		\frac{r_2^3}{2r_1^3r_{12}^3} \,.
\end{align}\end{subequations}
Again, those solutions are valid in the sense of distribution theory. Furthermore they tend to zero when $r\to\infty$.

We report the complete expression of the 2.5PN metric in harmonic coordinates, valid at any field point in the near zone. Posing $g_{\alpha\beta}=\eta_{\alpha\beta}+k_{\alpha\beta}$ -- to distinguish it from the gothic metric deviation $\sqrt{-g}g^{\alpha\beta}=\eta^{\alpha\beta}+h^{\alpha\beta}$ -- we have \citep{BFP98}
\begin{subequations}\label{metricNZ}
\begin{align}
k_{00} &= \frac{2 G m_1}{c^2 r_1} \nn \\&
+ \frac{1}{c^4} \left[\frac{G
    m_1}{r_1} \left(-(n_1v_1)^2+4 v_1^2 \right) -2 \frac{G^2
    m_1^2}{r_1^2}\right. \nn\\& \qquad \quad+
  \left. G^2 m_1 m_2 \left(-\frac{2}{r_1 r_2}- \frac{r_1}{2 r_{12}^3}+
  \frac{r_1^2}{2 r_2 r_{12}^3}- \frac{5}{2 r_2 r_{12}}\right)\right]\nn \\&
    +\frac{4 G^2 m_1 m_2}{3 c^5 r_{12}^2} (n_{12}v_{12}) \nn \\&
+ \frac{1}{c^6} \left[ \frac{G m_1}{r_1} \left(\frac{3}{4} (n_1v_1)^4-3
  (n_1v_1)^2 v_1^2+ 4 v_1^4\right)+ \frac{G^2 m_1^2}{r_1^2} \left(3
  (n_1v_1)^2-v_1^2\right) +2 \frac{G^3 m_1^3}{r_1^3}\right.  \nn
  \\ &\qquad \quad + G^2 m_1 m_2 \left(v_1^2 \left(\frac{3 r_1^3}{8 r_{12}^5}-
  \frac{3 r_1^2 r_2}{8 r_{12}^5}- \frac{3 r_1 r_2^2}{8 r_{12}^5}+
  \frac{3 r_2^3}{8 r_{12}^5}-\frac{37 r_1}{8 r_{12}^3}+
  \frac{r_1^2}{r_2 r_{12}^3}+\frac{3 r_2}{8 r_{12}^3}
  \right. \right. \nn\\& \left.  \qquad \qquad \quad + \frac{2
    r_2^2}{r_1 r_{12}^3}+\frac{6}{r_1 r_{12}}- \frac{5}{r_2
    r_{12}}-\frac{8r_{12}}{r_1r_2S} +\frac{16}{r_{12}
    S}\right)\nn\\& \qquad \quad+ (v_1v_2) \left(\frac{8}{r_1
    r_2}-\frac{3 r_1^3}{4 r_{12}^5}+ \frac{3 r_1^2 r_2}{4 r_{12}^5}+
  \frac{13 r_1}{4 r_{12}^3}-\frac{2 r_1^2}{r_2 r_{12}^3}- \frac{6}{r_1
    r_{12}}- \frac{16}{r_1 S}- \frac{12}{r_{12} S}\right) \nn
  \\& \qquad \quad+ (n_{12}v_1)^2 \left(-\frac{15 r_1^3}{8 r_{12}^5}+
  \frac{15 r_1^2 r_2}{8 r_{12}^5}+ \frac{15 r_1 r_2^2}{8 r_{12}^5}-
  \frac{15 r_2^3}{8 r_{12}^5}+ \frac{57 r_1}{8 r_{12}^3}- \frac{3
    r_1^2}{4 r_2 r_{12}^3}- \frac{33 r_2}{8 r_{12}^3}
  \right. \nn\\& \left.  \qquad \qquad \quad+
  \frac{7}{4 r_2 r_{12}} - \frac{16}{S^2}-\frac{16}{r_{12} S}\right)
  \nn\\& \quad \qquad + (n_{12}v_1) (n_{12}v_2) \left(\frac{15
    r_1^3}{4 r_{12}^5}- \frac{15 r_1^2 r_2}{4 r_{12}^5}- \frac{9
    r_1}{4 r_{12}^3}+\frac{12}{S^2}+ \frac{12}{r_{12} S}\right)
  \nn\\& \quad \qquad+ (n_1v_1)^2 \left(\frac{2}{r_1
    r_2}-\frac{r_1}{4 r_{12}^3}- \frac{3 r_2^2}{4 r_1 r_{12}^3}+
  \frac{7}{4 r_1 r_{12}}-\frac{8}{S^2}- \frac{8}{r_1 S}\right)
  \nn\\& \quad \qquad+ (n_1v_1) (n_1v_2)
  \left(\frac{r_1}{r_{12}^3}+\frac{16}{S^2}+ \frac{16}{r_1 S}\right)-
  (n_1v_2)^2 \left(\frac{8}{S^2}+\frac{8}{r_1 S}\right) \nn\\&
  \qquad \quad+ (n_{12}v_1) (n_1v_1) \left(-\frac{3 r_1^2}{r_{12}^4}+
  \frac{3 r_2^2}{2 r_{12}^4}+ \frac{3}{2
    r_{12}^2}+\frac{16}{S^2}\right)+ \frac{16 (n_1v_2) (n_2v_1)}{S^2}
  \nn\\& \qquad \quad + (n_{12}v_2) (n_1v_1) \left(\frac{3
    r_1^2}{r_{12}^4}- \frac{3 r_2^2}{2 r_{12}^4}+ \frac{13}{2
    r_{12}^2}- \frac{40}{S^2}\right)- \frac{12(n_1v_1) (n_2v_2)}{S^2}
  \nn\\&  \qquad \quad + (n_{12}v_1) (n_1v_2)
  \left(\frac{3 r_1^2}{2 r_{12}^4}+ \frac{4}{r_{12}^2}+ \frac{16}{S^2}
  \right)\nn\\& \left. \qquad \quad+ (n_{12}v_2) (n_1v_2) \left(-\frac{3 r_1^2}{2 r_{12}^4}-
  \frac{3}{r_{12}^2}+\frac{16}{S^2}\right) \right) \nn\\&\quad + G^3
  m_1^2 m_2 \left(\frac{4}{r_1^3}+\frac{1}{2 r_2^3}+\frac{9}{2 r_1^2
    r_2}- \frac{r_1^3}{4 r_{12}^6}+\frac{3 r_1^4}{16 r_2 r_{12}^6}-
  \frac{r_1^2 r_2}{8 r_{12}^6}+ \frac{r_1 r_2^2}{4 r_{12}^6}-
  \frac{r_2^3}{16 r_{12}^6}\right. \nn\\& \qquad\quad + \frac{5 r_1}{4 r_{12}^4}
  -\frac{23 r_1^2}{8 r_2 r_{12}^4}+
  \frac{43 r_2}{8 r_{12}^4}- \frac{5 r_2^2}{2 r_1
    r_{12}^4}-\frac{3}{r_{12}^3}+ \frac{3 r_1}{r_2
    r_{12}^3}+\frac{r_2}{r_1 r_{12}^3}- \frac{5 r_2^2}{r_1^2
    r_{12}^3}\nn\\& \qquad\quad + \frac{4 r_2^3}{r_1^3 r_{12}^3}+\frac{3}{2 r_1 r_{12}^2} - \frac{r_1^2}{4 r_2^3
    r_{12}^2} + \frac{3}{16 r_2 r_{12}^2}+ \frac{15 r_2}{4 r_1^2
    r_{12}^2}-\frac{4 r_2^2}{r_1^3 r_{12}^2}+ \frac{5}{r_1^2
    r_{12}}\nn\\& \qquad\quad\left. \left. +\frac{5}{r_1 r_2 r_{12}}- \frac{4 r_2}{r_1^3 r_{12}}-
  \frac{r_{12}^2}{4 r_1^2 r_2^3}\right)\right] \nn\\&
  +\frac{1}{c^7} \left[ G^2 m_1 m_2 \left((n_{12}v_{12})^2 (n_1v_1)
  \left(-\frac{8 r_1^3}{r_{12}^5}- \frac{16 r_1}{r_{12}^3}\right)+
  (n_{12}v_{12})^2 (n_1v_2) \left(\frac{8 r_1^3}{r_{12}^5}+ \frac{5
    r_1}{r_{12}^3}\right) \right. \right.  \nn\\& \qquad
  \quad + (n_{12}v_{12})^3 \left(-\frac{7 r_1^4}{2 r_{12}^6}+
  \frac{7 r_1^2 r_2^2}{2 r_{12}^6}- \frac{11
    r_1^2}{r_{12}^4}-\frac{37}{4 r_{12}^2}\right) \nn\\&
  \qquad \quad + (n_{12}v_1) (n_{12}v_{12})^2 \left(\frac{20
    r_1^2}{r_{12}^4}- \frac{11}{2 r_{12}^2}\right)- 4 (n_{12}v_{12})
  (n_1v_1)^2 \frac{r_1^2}{r_{12}^4} \nn\\& \qquad \quad
  +4 (n_{12}v_{12}) (n_1v_1) (n_1v_2)\frac{r_1^2}{r_{12}^4}+
  (n_{12}v_1)^2 (n_1v_1)\frac{r_1}{r_{12}^3} \nn\\& \qquad
  \quad+ 22 (n_{12}v_1) (n_{12}v_{12})
  (n_1v_1)\frac{r_1}{r_{12}^3}- (n_{12}v_1)^2
  (n_1v_2)\frac{r_1}{r_{12}^3} \nn\\& \qquad \quad+ 4
  (n_{12}v_1) (n_{12}v_{12}) (n_1v_2)\frac{r_1}{r_{12}^3}+ 11
  (n_{12}v_1)^2 (n_{12}v_{12})\frac{1}{2 r_{12}^2} \nn\\&
  \qquad \quad+ (n_1v_2) v_{12}^2 \left(-\frac{8 r_1^3}{5
    r_{12}^5}- \frac{2 r_1}{3 r_{12}^3}\right)+ (n_1v_1) v_{12}^2
  \left(\frac{8 r_1^3}{5 r_{12}^5}+ \frac{11 r_1}{3 r_{12}^3}\right)
  \nn\\& \qquad \quad - (n_{12}v_1) v_{12}^2
  \left(\frac{4 r_1^2}{r_{12}^4}+ \frac{5}{2 r_{12}^2}\right)-
  (n_{12}v_{12}) v_1^2 \left(\frac{12 r_1^2}{r_{12}^4}+ \frac{5}{2
    r_{12}^2}\right) \nn\\& \qquad \quad+
  (n_{12}v_{12}) v_{12}^2 \left(\frac{3 r_1^4}{2 r_{12}^6}- \frac{3
    r_1^2 r_2^2}{2 r_{12}^6}+ \frac{7 r_1^2}{r_{12}^4}+\frac{27}{4
    r_{12}^2}\right) \nn\\& \qquad \quad- 29 (n_1v_1)
  v_1^2\frac{r_1}{3 r_{12}^3}+ (n_1v_2) v_1^2\frac{r_1}{r_{12}^3}+ 5
  (n_{12}v_1) v_1^2\frac{1}{r_{12}^2} \nn\\& \qquad \quad
  + (n_{12}v_{12}) (v_1v_2) \left(\frac{12 r_1^2}{r_{12}^4}+
  \frac{3}{r_{12}^2}\right)+ 8 (n_1v_1) (v_1v_2)\frac{r_1}{r_{12}^3}
  \nn\\& \left. \qquad \quad + 2 (n_1v_2)
  (v_1v_2)\frac{r_1}{3 r_{12}^3}- 5 (n_{12}v_1)
  (v_1v_2)\frac{1}{r_{12}^2}\right) \nn\\&\quad + G^3 m_1^2 m_2
  \left((n_1v_{12}) \left(-\frac{8 r_1^3}{15 r_{12}^6}+ \frac{8 r_1
    r_2^2}{15 r_{12}^6}- \frac{16 r_1}{3 r_{12}^4}+\frac{8}{r_{12}^3}-
  \frac{8 r_2^2}{r_1^2 r_{12}^3}+ \frac{8}{r_1^2 r_{12}}\right)
  \right.  \nn\\& \qquad \quad +(n_{12}v_1) \left(-\frac{4
    r_1^2}{3 r_{12}^5}+ \frac{4 r_2^2}{3 r_{12}^5}+\frac{20}{3
    r_{12}^3}\right)+ 8 (n_1v_1)\frac{r_1}{3 r_{12}^4} \nn\\&
  \qquad \quad + (n_{12}v_{12}) \left(\frac{3}{r_1^3}-\frac{4 r_1^2}{3
    r_{12}^5}+ \frac{68 r_2^2}{15 r_{12}^5}+ \frac{3
    r_1}{r_{12}^4}-\frac{6 r_2^2}{r_1 r_{12}^4}+ \frac{3 r_2^4}{r_1^3
    r_{12}^4}- \frac{76}{3 r_{12}^3} \right.  \nn\\&
  \left. \left. \left. \quad \qquad \qquad +\frac{2}{r_1
    r_{12}^2}- \frac{6 r_2^2}{r_1^3 r_{12}^2}\right)\right) \right] \nn\\& +
1 \leftrightarrow 2 + \calO\left(\frac{1}{c^{8}}\right)
\,, \label{metricNZ00} \\ 
k_{0i} &= - \frac{4 G m_1}{c^3 r_1} v_1^i\nn\\&
+ \frac{1}{c^5} \left[ n_1^i
  \biggl(-\frac{G^2 m_1^2}{r_1^2} (n_1v_1)+ \frac{G^2 m_1 m_2}{S^2}
  \Bigl(-16 (n_{12}v_1)+ 12 (n_{12}v_2)
  \right. \nn\\&  
   \qquad \qquad \quad -16 (n_2v_1) +12 (n_2v_2) \Bigr)
  \biggr) \nn\\& \qquad \quad + n_{12}^i G^2 m_1 m_2 \left(-6
  (n_{12}v_{12}) \frac{r_1}{r_{12}^3}- 4 (n_1v_1)\frac{1}{r_{12}^2}+12
  (n_1v_1)\frac{1}{S^2} \right.  \nn\\& \left. \qquad \qquad
  \quad - 16 (n_1v_2)\frac{1}{S^2}+ 4
  (n_{12}v_1)\frac{1}{S} \left(\frac{1}{S}+ \frac{1}{r_{12}}\right)
  \right) \nn\\& \qquad \quad + v_1^i \left(\frac{G m_1}{r_1}
  \left(2 (n_1v_1)^2-4 v_1^2\right)+ \frac{G^2 m_1^2}{r_1^2}+ G^2 m_1
  m_2 \left(\frac{3 r_1}{r_{12}^3}-\frac{2 r_2}{r_{12}^3}\right)
  \right. \nn\\& \left. \left.  \qquad \qquad \quad +
  G^2 m_1 m_2 \left(-\frac{r_2^2}{r_1 r_{12}^3} -\frac{3}{r_1 r_{12}}+
  \frac{8}{r_2 r_{12}}-\frac{4}{r_{12} S}\right)\right)\right]
\nn\\&
+ \frac{1}{c^6} \left[ n_{12}^i \left(G^2 m_1 m_2
  \left(-10 (n_{12}v_{12})^2 \frac{r_1^2}{r_{12}^4}- 12 (n_{12}v_{12})
  (n_1v_1) \frac{r_1}{r_{12}^3}\right. \right.\right.\nn\\&\left.\left.\qquad  + 2 v_{12}^2 \frac{r_1^2}{r_{12}^4}-4
  \frac{v_1^2}{r_{12}^2}\right) + G^3 m_1^2 m_2 \left(\frac{2 r_1^2}{3 r_{12}^5}-
  \frac{2 r_2^2}{3 r_{12}^5}- \frac{2}{r_{12}^3}\right)\right)
  \nn\\& \qquad + v_1^i \frac{G^2 m_1 m_2}{r_{12}^3}
  \left(\frac{16 (n_1v_{12}) r_1}{3}- 4 (n_{12}v_2) r_{12} \right)
  \nn\\& \qquad \left. + v_{12}^i \frac{G^2 m_1 m_2}{r_{12}^2}
  \left(-2 (n_{12}v_1)+ 6 (n_{12}v_{12}) \frac{r_1^2}{r_{12}^2}\right)
  \right] \nn\\& + 1 \leftrightarrow 2 +
\calO\left(\frac{1}{c^{7}}\right) \,, \label{metricNZ0i}\\
k_{ij} &= \frac{2 G m_1}{c^2 r_1} \delta^{ij}\nn\\&
+ \frac{1}{c^4} \left[
  \delta_{ij} \left( -\frac{G m_1}{r_1} (n_1v_1)^2+\frac{G^2
    m_1^2}{r_1^2} \right. \right.  \nn\\& \left. \qquad \qquad
  + G^2 m_1 m_2 \left(\frac{2}{r_1 r_2}-\frac{r_1}{2
    r_{12}^3}+ \frac{r_1^2}{2 r_2 r_{12}^3}-\frac{5}{2 r_1 r_{12}}+
  \frac{4}{r_{12} S}\right)\right) \nn\\& \qquad \quad
  + 4 \frac{G m_1}{r_1} v_1^i v_1^j+ \frac{G^2 m_1^2}{r_1^2} n_1^i
  n_1^j - 4 G^2 m_1 m_2 n_{12}^i n_{12}^j \left(\frac{1}{S^2}+
  \frac{1}{r_{12} S}\right) \nn\\& \left.  \qquad \quad
  + \frac{4 G^2 m_1 m_2}{S^2} \left(n_1^{(i} n_2^{j)}+ 2n_1^{(i}
  n_{12}^{j)} \right)\right] \nn\\&
 + \frac{G^2 m_1
  m_2}{c^5r_{12}^2} \left(-\frac{2}{3}(n_{12}v_{12})\delta^{ij}- 6
(n_{12}v_{12}) n_{12}^i n_{12}^j+ 8 n_{12}^{(i} v_{12}^{j)}\right) \nn\\& + 1
\leftrightarrow 2 + \calO\left(\frac{1}{c^{6}}\right)
\,.\label{metricNZij}
\end{align}
\end{subequations}
To higher order one needs the higher kernels $f$, $h$, $\cdots$ given by \eqref{kernels_part} but also the solution of elementary equations still more intricate than \eqref{eqK1H1}; the 3PN metric valid in closed form all over the near zone is not currently known.

Finally let us display the 2.5PN metric computed at the location of the particle 1 for instance; to this order the Hadamard self-field regularization, Eq.~\eqref{hadPF}, gives the same result as dimensional regularization. We get
\begin{subequations}\label{metricNZ1}
\begin{align}
(k_{00})_1 &= \frac{2 G m_2}{c^2r_{12}} \nn\\& +\frac{G m_2}{c^4 r_{12}}
  \left(4 v_2^2-(n_{12}v_2)^2-3\frac{G m_1}{r_{12}} - 2 \frac{G
    m_2}{r_{12}} \right) \nn\\&+\frac{8G^2 m_1 m_2}{3c^5
    r_{12}^2}(n_{12}v_{12}) \nn\\&+ \frac{1}{c^6}\biggl[\frac{G m_2}{r_{12}} \left(
  \frac{3}{4} (n_{12}v_2)^4- 3 (n_{12}v_2)^2 v_2^2 + 4 v_2^4 \right)
  \nn\\&\qquad + \frac{G^2 m_1 m_2}{r_{12}^2} \left(-\frac{87}{4}
  (n_{12}v_1)^2 + \frac{47}{2}(n_{12}v_1) (n_{12}v_2)- \frac{55}{4}
  (n_{12}v_2)^2 \right.\nn\\& \qquad\quad\left.+ \frac{23}{4} v_1^2 - \frac{39}{2} (v_1v_2) \right) + \frac{47}{4} \frac{G^2 m_1 m_2}{r_{12}^2} v_2^2\nn\\&\qquad +
  \frac{G m_2}{r_{12}} \left(\frac{G m_2}{r_{12}} \left[3
    (n_{12}v_2)^2 - v_2^2 \right] -\frac{G^2 m_1^2}{r_{12}^2} +
  \frac{17}{2} \frac{G^2 m_1 m_2}{r_{12}^2} + 2 \frac{G^2
    m_2^2}{r_{12}^2} \right)\biggr] \nn\\&+ \frac{G^2 m_1 m_2}{c^7
    r_{12}^2} \biggl[ -20 (n_{12}v_1)^3 + 40 (n_{12}v_1)^2 (n_{12}v_2)
  - 36 (n_{12}v_1) (n_{12}v_2)^2 + 16 (n_{12}v_2)^3 \nn
  \\ & \qquad +\frac{296}{15} (n_{12}v_1) v_1^2 - \frac{116}{15}
  (n_{12}v_2) v_1^2 - \frac{104}{5} (n_{12}v_1) (v_1v_2) +
  \frac{232}{15} (n_{12}v_2) (v_1v_2) \nn\\& \qquad +
  \frac{56}{15} (n_{12}v_1) v_2^2 - \frac{52}{5} (n_{12}v_2) v_2^2
  +\frac{G m_1}{r_{12}} \left(-\frac{64}{5} (n_{12}v_1)+ \frac{104}{5}
  (n_{12}v_2) \right) \nn\\& \qquad   +\frac{G
    m_2}{r_{12}} \left(-\frac{144}{5}(n_{12}v_1)+
  \frac{392}{15}(n_{12}v_2)\right) \biggr] +
  \calO\left(\frac{1}{c^{8}}\right) \,, \label{metricNZ001}\\
(k_{0i})_1 &= - \frac{4 G m_2}{c^3 r_{12}} v_2^i \nn\\& +\frac{G m_2}{c^5
    r_{12}}\left[n_{12}^i \left(\frac{G m_1}{r_{12}} \left(10
    (n_{12}v_1) + 2 (n_{12}v_2)\right) -\frac{G m_2}{r_{12}}
    (n_{12}v_2)\right) \right. \nn\\& \left. \qquad + 4 \frac{G m_1}{r_{12}} v_1^i +v_2^i \left(2
  (n_{12}v_2)^2 - 4 v_2^2- 2 \frac{G m_1}{r_{12}} + \frac{G
    m_2}{r_{12}} \right) \right]\nn\\&+\frac{G^2 m_1
    m_2}{c^6 r_{12}^2} \biggl[n_{12}^i \biggl(10 (n_{12}v_1)^2 - 8
  (n_{12}v_1) (n_{12}v_2) - 2 (n_{12}v_2)^2 - 6 v_1^2 
  \nn\\& \qquad + 4 (v_1v_2) + 2 v_2^2-
  \frac{8}{3} \frac{G m_1}{r_{12}n} + \frac{4}{3} \frac{G m_2}{r_{12}}
  \biggr) \nn\\& \qquad -8 (n_{12}v_1) v_1^i
  +v_2^i \left(\frac{20}{3} (n_{12}v_1) + \frac{4}{3}
  (n_{12}v_2)\right) \biggr] + \calO\left(\frac{1}{c^{7}}\right)
  \,, \label{metricNZ0i1} \\
(k_{ij})_1 &= \frac{2 G m_2}{c^2 r_{12}} \delta_{ij} \nn\\& +\frac{G m_2}{c^4 r_{12}}\left[ \delta_{ij} \left(-(n_{12}v_2)^2 + \frac{G m_1}{r_{12}} +
  \frac{G m_2}{r_{12}} \right) \right. \nn\\&\qquad \left.+n_{12}^{i}n_{12}^{j} \left(-8 \frac{G m_1}{r_{12}} + \frac{G
    m_2}{r_{12}}\right) + 4 v_2^{i}v_2^{j}\right] \nn\\& + \frac{G^2 m_1
    m_2}{c^5 r_{12}^2} \left[ - \frac{4}{3}\delta_{ij} (n_{12}v_{12})  - 12
  n_{12}^{i}n_{12}^{j}(n_{12}v_{12})+ 16 n_{12}^{(i} v_{12}^{j)} \right] \nn\\& +
  \calO\left(\frac{1}{c^{6}}\right) \,. \label{metricNZij1}
\end{align}
\end{subequations}
The metric regularized at the location of the particles has been computed to 3PN order by \cite{BDLW10a}. At that order it becomes crucial to make full use of dimensional regularization, and poles $\propto 1/\varepsilon$ appear in the near-zone metric. The metric regularized at the location of the particles is needed when we compute the redshift observable \eqref{z1expr} in Sects.~\ref{sec:firstlaw} and \ref{sec:SF}.


\subsection{Conservative dynamics of compact binaries}
\label{sec:conserv}


\subsubsection{The innermost circular orbit}
\label{sec:ICO}

Having in hand the total relativistic energy for circular orbits (including the rest-mass contribution) at 4PN order, thus $\dM = m + \dE$ where $\dE(x)$ is given by Eq.~\eqref{Ecirc},\footnote{In all of Sect.~\ref{sec:conserv} we pose $G=1=c$.} we define the innermost circular orbit (ICO) as the minimum, when it exists, of this energy function \citep{B02ico}. Notice that the ICO is not defined as a point of dynamical general-relativistic unstability. Hence, we prefer to call this point the ICO rather than, strictly speaking, an innermost stable circular orbit or ISCO. A study of the dynamical stability of circular binary orbits in the post-Newtonian approximation is reported in Sect.~\ref{sec:stab}.

The previous definition of the ICO is motivated by the comparison with some results of numerical relativity. Indeed we shall confront the prediction of the standard (Taylor-based) post-Newtonian approximation with numerical computations of the energy of binary black holes under the assumptions of conformal flatness for the spatial metric and of exactly circular orbits \citep{GGB1, GGB2, CPf04, CCGPf06}. The latter restriction is implemented by requiring the existence of an ``helical'' Killing vector (HKV), which is time-like inside the light cylinder associated with the circular motion, and space-like outside. The HKV will be defined in Eq.~\eqref{killing} below. In these numerical approaches there are no gravitational waves; the field is periodic in time, and the gravitational potentials tend to zero at spatial infinity, within a restricted model equivalent to solving five out of the ten Einstein field equations -- the \cite{IN80, WM80} approximation. Considering an evolutionary sequence of equilibrium configurations the circular-orbit energy $\dE(\Omega)$ and the ICO of binary black holes are obtained numerically (see also \citealt{BGM99, GGTMB01, LGG05} for related calculations of the ICO in the case of binary neutron stars and strange quark stars; see also \citealt{MW03}).

Since the numerical calculations of \cite{GGB2, CPf04} have been performed in the case of two \emph{corotating} black holes, which are essentially spinning with the orbital angular velocity, for the comparison we must include within our PN formalism the effects of spins appropriate to two corotating Kerr black holes. The total relativistic masses of the two Kerr black holes (with $\text{a}=1,2$ labelling the black holes) are given by
\begin{equation}
  m_\text{a}^2=\mu_\text{a}^2+\frac{S_\text{a}^2}{4\mu_\text{a}^2}\,.
  \label{massformula}
\end{equation}
We assume the validity of the \cite{Chr70, ChrR71} mass formula for each Kerr black holes; i.e., we neglect the influence of the companion. Here $S_\text{a}$ is the spin, related to the usual Kerr parameter by $S_\text{a}=m_\text{a} a_\text{a}$, and $\mu_\text{a}\equiv m_\text{a}^\text{irr}$ is the irreducible mass, not to be confused with the reduced mass, and given by $16\pi\mu_\text{a}^2=A_\text{a}$ (with $A_\text{a}$ the black hole's surface area). The angular velocity of the black hole, defined by the angular velocity of the outgoing photons located at the light-like horizon, is
\begin{equation}
  \omega_\text{a} = \frac{\partial m_\text{a}}{\partial
    S_\text{a}}{\bigg|}_{\mu_\text{a}} = \frac{S_\text{a}}{4m_\text{a}
    \mu_\text{a}^2}\,.
  \label{omegan}
\end{equation}
We shall give in Eq.~\eqref{dmn} a more general formulation of the ``internal structure'' of the black holes. Combining Eqs. \eqref{massformula}--\eqref{omegan} we obtain $m_\text{a}$ and $S_\text{a}$ as functions of $\mu_\text{a}$ and $\omega_\text{a}$,
\begin{subequations}\label{mnSn}
  \begin{align}
    m_\text{a} &= \frac{\mu_\text{a}}{\sqrt{1-4\mu_\text{a}^2
        \,\omega_\text{a}^2}}\,,\\ S_\text{a} &=
    \frac{4\mu_\text{a}^3\omega_\text{a}}{\sqrt{1-4\mu_\text{a}^2
        \,\omega_\text{a}^2}}\,.
  \end{align}
\end{subequations}
In the limit of slow rotation we get
\begin{subequations}\label{slowomn}
\begin{align}
  S_\text{a} &= I_\text{a}\,\omega_\text{a} +
  \calO\left(\omega_\text{a}^3\right)\,,\label{defIA}
  \\ m_\text{a} &= \mu_\text{a} +\frac{1}{2}
  I_\text{a}\,\omega_\text{a}^2 +
  \calO\left(\omega_\text{a}^4\right)\,,\label{spinkin}
\end{align}
\end{subequations}
where $I_\text{a}=4 \mu_\text{a}^3$ is the moment of inertia of the black hole. We see that the total mass-energy $m_\text{a}$ involves the irreducible mass augmented by the usual kinetic energy of the spin.

We now need the relation between the rotation frequency $\omega_\text{a}$ of each of the corotating black holes and the orbital frequency $\Omega$ of the binary system. Indeed $\Omega$ is the basic variable describing each equilibrium configuration calculated numerically by \cite{GGB2, CPf04}, with the irreducible masses held constant along the numerical evolutionary sequences. Here we report the result of an investigation of the condition for corotation based on the first law of mechanics for spinning black holes \citep{BBL13}, which concludes that the corotation condition (at 2PN order) is
\begin{equation}\label{corot}
\omega_\text{a} = \Omega \left\{ 1 - \nu \, x + \nu \left( -
\frac{3}{2} + \frac{\nu}{3} \right) x^2 + \calO(x^3) \right\}
\,,
\end{equation}
where $x$ denotes the post-Newtonian parameter \eqref{xdef} and $\nu$ is the symmetric mass ratio \eqref{nu}. This condition is issued from the general relation given in Eq.~\eqref{corotcond}. Interestingly, notice that $\omega_1=\omega_2$ up to the rather high 2PN order. In the Newtonian limit $x \to 0$ or in the test-particle limit $\nu \to 0$ we simply have $\omega_\text{a} = \Omega$, in agreement with physical intuition.

To take into account the spin effects our first task is to replace all the masses entering the energy function \eqref{Ecirc} by their equivalent expressions in terms of $\omega_\text{a}$ and the irreducible masses $\mu_\text{a}$, and then to replace $\omega_\text{a}$ in terms of $\Omega$ according to the corotation prescription \eqref{corot}.\footnote{Note that this is an iterative process because the masses in Eq.~\eqref{corot} are themselves to be replaced by their expression in terms of the irreducible masses.} It is clear that the leading contribution is that of the spin kinetic energy given in Eq.~\eqref{spinkin}, and it comes from the replacement of the rest mass-energy $m=m_1+m_2$ in $\dM=m+\dE$. From Eq.~\eqref{spinkin} this effect is of order $\Omega^2$ in the case of corotating binaries, which means by comparison with Eq.~\eqref{Ecirc} that it is equivalent to an ``orbital'' effect at the 2PN order (i.e., $\propto x^2$). Higher-order corrections in Eq.~\eqref{spinkin}, which behave at least like $\Omega^4$, will correspond to the orbital 5PN order at least and are negligible for the present purpose, limited to 4PN order. In addition there will be a subdominant contribution, of the order of $\Omega^{8/3}$ equivalent to 3PN order, which comes from the replacement of the masses into the Newtonian part, proportional to $x\propto \Omega^{2/3}$, of the energy $\dE$; see Eq.~\eqref{Ecirc}. With the 4PN accuracy we need only to replace the masses that enter into the 1PN corrections in $\dE$; in higher order terms the masses can be considered to be the irreducible ones.

Our second task is to include the specific relativistic effects due to the spins, namely the spin-orbit (SO) interaction and the spin-spin (SS) one. In the case of spins $S_1$ and $S_2$ aligned parallel to the orbital angular momentum (and right-handed with respect to the sense of motion) the SO energy to ``next-to-leading order'' reads
\begin{align}\label{ESO1PN}
	\dE_\mathrm{SO} &= -m \nu\, (m\Omega)^{5/3}  \left(\frac{m_1^2}{m^2}\left[\frac{4}{3}+\Bigl(4-\frac{31}{18}\nu\Bigr)(m\Omega)^{2/3}\right] \right.\nn\\&\qquad\qquad\qquad\qquad\quad\left.+\nu\left[1+\Bigl(\frac{3}{2}-\frac{5}{3}\nu\Bigr)(m\Omega)^{2/3}\right]\right) \frac{S_1}{m_1^2}
	+ 1\leftrightarrow 2\,.
\end{align}
We shall review in Sect.~\ref{sec:spins} the most up-to-date results for the spin-orbit energy and related quantities; here we are simply employing the formula \eqref{ESO}. We immediately infer from Eq.~\eqref{ESO1PN} that in the case of corotating black holes the SO effect is equivalent dominantly to a 3PN orbital effect and thus must be retained with the present accuracy. Since our goaled precision is 4PN we have to retain also the next-to-leading correction in \eqref{ESO1PN}. With this approximation, the masses in Eq.~\eqref{ESO1PN} can be replaced by the irreducible ones. As for the SS interaction (still in the case of aligned spins) it is given by
\begin{equation}\label{ESSN}
  \dE_\mathrm{SS} = m \nu\, (m\Omega)^2
  \frac{S_1\,S_2}{m_1^2\,m_2^2}\,.
\end{equation}
The SS effect can be neglected here because it is of order 5PN for corotating systems. Summing up all the spin contributions to 4PN order we find that the supplementary energy due to the corotating spins is\footnote{This formula extends the works of \cite{B02ico, BBL13} to the next 4PN order. Note that \cite{B02ico} had initially assumed that the corotation condition was given by the leading-order result $\omega_\text{a} = \Omega$; but as was shown by \cite{BBL13} the 1PN correction in Eq.~\eqref{corot} modifies the 3PN terms in \eqref{Ecorot} with respect to the initial result of \cite{B02ico}.} 
\begin{align}\label{Ecorot}
  \Delta \dE^\mathrm{corot} &= \mu \,x_\mu \biggl[\bigl( 2 - 6 \eta \bigr)
    x_\mu^2 + \eta \bigl( - 10 + 25 \eta \bigr) x_\mu^3 \nn\\& \qquad\qquad + \eta \left( - 21 + \frac{224}{3} \eta - \frac{455}{12} \eta^2 \right) x_\mu^4 +
    \calO\bigl(x_\mu^5\bigr)\biggr] \,.
\end{align}
The total mass $\mu = \mu_1 + \mu_2$, the symmetric mass ratio $\eta = \mu_1 \mu_2 / \mu^2$, and the dimensionless invariant post-Newtonian parameter $x_\mu = (\mu\Omega)^{2/3}$ are now expressed in terms of the irreducible ``bare'' masses $\mu_\text{a}$, rather than the ``dressed'' masses $m_\text{a}$. The complete 4PN energy of the corotating binary is finally given by the sum of Eqs. \eqref{Ecirc} and \eqref{Ecorot}, in which all the masses are now understood as being the irreducible masses, which must be assumed to stay constant when the binary evolves for the comparison with the numerical calculation.
\begin{figure}[htbp]
\centering
  \includegraphics[width=0.6\textwidth]{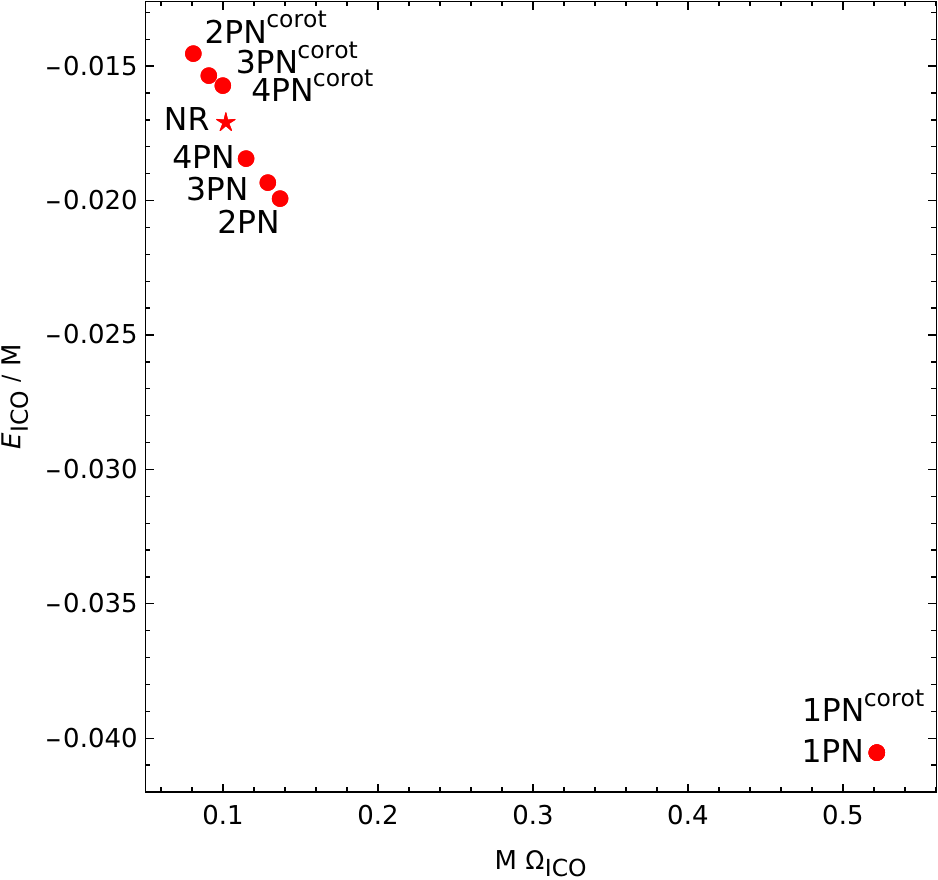}
  \includegraphics[width=0.6\textwidth]{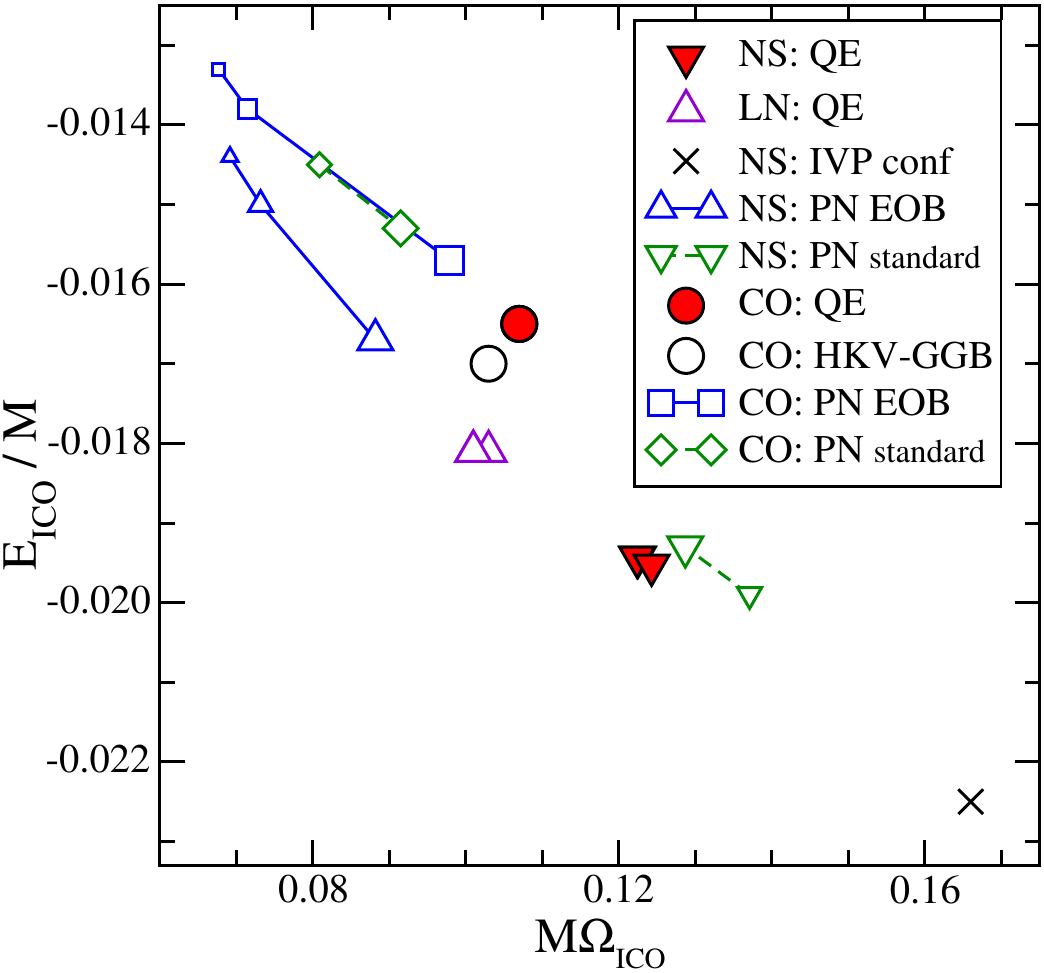}
\caption{The binding energy $\dE_{\mathrm{ICO}}$ versus $\Omega_{\mathrm{ICO}}$ in the equal-mass case ($\nu=1/4$). \emph{Top panel:} Comparison with the numerical relativity result of \cite{GGB1, GGB2} valid in the corotating case (marked by a star). Points indicated by $n\mathrm{PN}$ are computed from the minimum of Eq.~\eqref{Ecirc}, and correspond to irrotational binaries. Points denoted by $n\mathrm{PN}^\mathrm{corot}$ include the contribution \eqref{Ecorot} to the energy, and describe corotational binaries. \emph{Bottom panel:} Numerical relativity results of \cite{CPf04, CCGPf06} for quasi-equilibrium (QE) configurations and various boundary conditions for the lapse function, in the non-spinning (NS), leading-order non spinning (LN) and corotating (CO) cases. The point from \cite{GGB1, GGB2} (HKV-GGB) is also reported, together with IVP, the initial value approach with effective potential of \cite{Cook94, PfTC00}, as well as standard PN predictions from the left panel and non-standard (EOB) ones. 
}
\label{fig:ICO}
\end{figure}

Figure~\ref{fig:ICO} shows the post-Newtonian results for $\dE_{\mathrm{ICO}}$ in the case of irrotational and corotational binaries and make comparisons with numerical results. Since $\Delta \dE^\mathrm{corot}$, given by Eq.~\eqref{Ecorot}, is at least of order 2PN, the result for $\mathrm{1PN}^\mathrm{corot}$ is the same as for 1PN in the irrotational case; then, obviously, $\mathrm{2PN}^\mathrm{corot}$ takes into account only the leading 2PN corotation effect, i.e., the spin kinetic energy given by Eq.~\eqref{spinkin}, while $\mathrm{3PN}^\mathrm{corot}$ and $\mathrm{4PN}^\mathrm{corot}$ involve also the corotational SO coupling \eqref{Ecorot}. As we can see the 3PN and 4PN points, and even the 2PN ones, are in good agreement with the numerical values.\footnote{As a caveat recall that the numerical works of \cite{GGB1, GGB2, CPf04, CCGPf06} assume that the spatial metric is conformally flat, which is incompatible with the post-Newtonian approximation starting from the 2PN order (see \citealt{FJS04} for a discussion).} The fact that the 2PN, 3PN and 4PN values are so close to each other is a good sign of the convergence of the (Taylor-expanded) series. The agreement found in Fig.~\ref{fig:ICO} is remarquable, all the more if we remember the previous situation, since the first estimations of the ICO in post-Newtonian theory \citep{KWWisco} and numerical relativity \citep{Cook94, PfTC00, Baum00} disagreed with each other, and do not match with the present 3PN/4PN results. The convergence of the standard (Taylor-expanded) post-Newtonian approximation is further discussed by \cite{B11}.


\subsubsection{Dynamical stability of circular orbits}
\label{sec:stab}

In this section, we investigate the problem of the stability of circular orbits against dynamical perturbations. We propose to use two different methods, one based on a linear perturbation at the level of the center-of-mass equations of motion in harmonic coordinates, the other one consisting of perturbing the center-of-mass Hamiltonian equations in ADM coordinates. We shall find a gauge invariant criterion for the stability of circular orbits \citep{BI03CM}. We shall check that our two methods agree on the result.

We deal first with the perturbation of the equations of motion, following \cite{KWWisco}, see their Sect.~III.A. We introduce polar coordinates $(r,\varphi)$ in the orbital plane and pose $u\equiv \dot{r}$ and $\Omega\equiv \dot{\varphi}$. In this section we restrict ourselves to the 3PN order. Then the equations of motion are purely instantaneous, as we neglect the non-local tail term \eqref{acctailCM}. Then Eq.~\eqref{dvdt} yields the system of equations
\begin{subequations}\label{systeq}\begin{align}
\dot{r} &= u\,,\label{systeqa}\\ \dot{u} &= -\frac{G
  m}{r^2}\Big[1+\mathcal{
    A}+\mathcal{B}u\Big]+r\Omega^2\,,\label{systeqb}\\ \dot{\Omega} &=
-\Omega\left[\frac{G m}{r^2}\mathcal{B}+\frac{2
    u}{r}\right]\,,\label{systeqc}
\end{align}\end{subequations}
where $\mathcal{A}$ and $\mathcal{B}$ are provided by Eqs. \eqref{calAcoeff}--\eqref{calBcoeff}, but limited to 3PN order, as functions of $r$, $u$ and $\Omega$ (through $v^2=u^2+r^2\Omega^2$).

In the case of an orbit that is circular apart from the adiabatic inspiral at the 2.5PN order (we neglect the radiation-reaction damping effects), we have $\dot{r}_0=\dot{u}_0=\dot{\Omega}_0=0$ hence $u_0=0$. In this section we shall indicate quantities associated with the circular orbit, which constitutes the zero-th approximation in our perturbation scheme, using the subscript $0$. Hence Eq.~\eqref{systeqb} gives the angular velocity $\Omega_0$ of the circular orbit as
\begin{equation}\label{Om0harm}
\Omega_0^2 = \frac{G m}{r_0^3}\big(1+\mathcal{A}_0\big)\,.
\end{equation}
Solving iteratively this relation at the 3PN order using the equations of motion, we obtain $\Omega_0$ as a function of the circular-orbit radius $r_0$ in standard harmonic coordinates; the result agrees with the 3PN restriction of Eq.~\eqref{keplerlaw}.\footnote{Of course one should distinguish $r_0$ from the regularization constant $r'_0$ entering the logarithms at the 3PN and 4PN orders in Eq.~\eqref{keplerlaw}, and also from $r_0$ in Eq.~\eqref{regfactor}.}

We now investigate the linear perturbation around the circular orbit defined by the constants $r_0$, $u_0=0$ and $\Omega_0$. We pose
\begin{subequations}\label{pertruOm}\begin{align}
r &= r_0 + \delta r\,,\\ u &= \delta u\,,\\ \Omega &= \Omega_0 +
\delta \Omega\,,
\end{align}\end{subequations}
where $\delta r$, $\delta u$ and $\delta\Omega$ denote the linear perturbations of the circular orbit. Then a system of linear equations readily follows:
\begin{subequations}\label{perteq}\begin{align}
\dot{\delta r} &= \delta u\,,\\ \dot{\delta u} &= \alpha_0\, \delta r
+ \beta_0\, \delta \Omega\,,\\ \dot{\delta \Omega} &= \gamma_0\,
\delta u\,,
\end{align}\end{subequations}
where the coefficients, which solely depend on the unperturbed circular orbit (hence the added subscript $0$), read as
\begin{subequations}\label{abc0}\begin{align}
\alpha_0 &= 3 \Omega_0^2 - \frac{G m}{r_0^2}\left(\frac{\partial
  \mathcal{A}}{\partial r}\right)_0\,,\\ \beta_0 &= 2 r_0 \Omega_0 -
\frac{G m}{r_0^2}\left(\frac{\partial \mathcal{A}}{\partial
  \Omega}\right)_0\,,\\ \gamma_0 &= -\Omega_0 \left[\frac{2}{r_0} +
  \frac{G m}{r_0^2}\left(\frac{\partial \mathcal{B}}{\partial
    u}\right)_0\right]\,.
\end{align}\end{subequations}
In obtaining these equations we use the fact that $\mathcal{A}$ is a function of the square $u^2$ through $v^2=u^2+r^2\Omega^2$, so that $\partial \mathcal{A}/\partial u$ is proportional to $u$ and thus vanishes in the unperturbed configuration (because $u=\delta u$). On the other hand, since the radiation reaction is neglected, $\mathcal{B}$ is also proportional to $u$ [see Eq.~\eqref{calBcoeff}], so only $\partial \mathcal{B}/\partial u$ can contribute at the zero-th perturbative order. Now by examining the fate of perturbations that are proportional to some $\de^{\di\sigma t}$, we arrive at the condition for the frequency $\sigma$ of the perturbation to be real, and hence for stable circular orbits to exist, as being \citep{KWWisco}
\begin{equation}\label{crit}
\hat{C}_0 \equiv -\alpha_0 - \beta_0\, \gamma_0 ~> 0\,.
\end{equation}
Substituting into this $\mathcal{A}$ and $\mathcal{B}$ at the 3PN order we then arrive at the orbital-stability criterion
\begin{align}\label{criterion}
	\hat{C}_0 &=
	\frac{G m}{r_0^3}\biggl\{1+\gamma_0\bigl(-9+\nu\bigr)
	+\gamma_0^2\left(30
	+\frac{65}{4}\nu+\nu^2\right)\\ & +\gamma_0^3
	\left(-70+\left[-\frac{29927}{840}-\frac{451}{64}\pi^2+22\ln
	\Big(\frac{r_0}{r'_0}\Big) \right]\nu + \frac{19}{2}\nu^2+\nu^3\right)
	+\calO\left(\frac{1}{c^8}\right)\biggr\}\,,\nn
\end{align}
where we denote $\gamma_0=\frac{G m}{r_0 c^2}$ and we recall that $r_0$ is the radius of the circular orbit in harmonic coordinates.

Our second method is to use the Hamiltonian equations associated with the 3PN CM Hamiltonian in ADM coordinates, 
given by the 3PN restriction of Eq.~\eqref{HADMcm}. We introduce the polar coordinates $(R,\Psi)$ in the orbital plane -- we assume that the orbital plane is equatorial, given by $\Theta=\frac{\pi}{2}$ in the spherical coordinate system $(R,\Theta,\Psi)$ -- and make the substitution
\begin{equation}\label{P2}
  P^2={P_R}^2+\frac{P_\Psi^2}{R^2}\,.
\end{equation}
This yields a reduced Hamiltonian that is a function of $R$, $P_R$ and $P_\Psi$, and describes the motion in polar coordinates in the orbital plane, hence $\mathcal{H}=\mathcal{H}[R,P_R,P_\Psi]$. The Hamiltonian equations then read
\begin{subequations}\label{Hameq}\begin{align}
\dot{R} &= \frac{\partial \mathcal{H}}{\partial
  P_R}\,, \qquad\quad\dot{\Psi} = \frac{\partial
  \mathcal{H}}{\partial P_\Psi}\,,\\ \dot{P_R} &=
-\frac{\partial \mathcal{H}}{\partial R}\,, \qquad\quad\dot{P_\Psi} = 0\,.
\end{align}\end{subequations}
Evidently the constant $P_\Psi$ is nothing but the conserved angular-momentum integral. For circular orbits we have $R=R_0$ (a constant) and $P_R=0$, so
\begin{equation}\label{AM0}
\frac{\partial \mathcal{H}}{\partial R}\big[R_0,0,P_\Psi^0\big] = 0\,,
\end{equation}
which gives the angular momentum $P_\Psi^0$ of the orbit as a function of $R_0$, and
\begin{equation}\label{OmADM}
\Omega_0 \equiv \left(\frac{\dd\Psi}{\dd t}\right)_0 = \frac{\partial
  \mathcal{H}}{\partial P_\Psi}\big[R_0,0,P_\Psi^0\big]\,,
\end{equation}
which yields the angular frequency of the circular orbit $\Omega_0$, which is evidently the same numerical quantity as in Eq.~\eqref{Om0harm}, but is here expressed in terms of the separation $R_0$ in ADM coordinates. The last equation, which is equivalent to $R=\text{const}=R_0$, is
\begin{equation}\label{lasteq}
\frac{\partial \mathcal{H}}{\partial P_R}\big[R_0,0,P_\Psi^0\big] =
0\,.
\end{equation}
It is automatically verified because $\mathcal{H}$ is a quadratic function of $P_R$ and hence $\partial \mathcal{H}/\partial P_R$ is zero for circular orbits.

We consider now a perturbation of the circular orbit defined by
\begin{subequations}\label{pertADM}\begin{align}
P_R &= \delta P_R\,,\\ P_\Psi &= P_\Psi^0 + \delta P_\Psi\,,\\ R &=
R_0 + \delta R\,,\\ \Omega &= \Omega_0 + \delta \Omega\,.
\end{align}\end{subequations}
The Hamiltonian equations \eqref{Hameq}, worked out at the linearized order, read as
\begin{subequations}\label{eqpertADM}\begin{align}
\dot{\delta P_R} &= -\pi_0\, \delta R - \rho_0\, \delta
P_\Psi\,,\\ \dot{\delta P_\Psi} &= 0\,,\\ \dot{\delta R} &= \sigma_0\,
\delta P_R\,,\\ \delta \Omega &= \rho_0\, \delta R + \tau_0\, \delta
P_\Psi\,,
\end{align}\end{subequations}
where the coefficients, which depend on the unperturbed orbit, are given by
\begin{subequations}\label{coeffs0}\begin{align}
\pi_0&=\frac{\partial^2 \mathcal{H}}{\partial
  R^2}\big[R_0,0,P_\Psi^0\big]\,,\\ \rho_0&=\frac{\partial^2
  \mathcal{H}}{\partial R\, \partial
  P_\Psi}\big[R_0,0,P_\Psi^0\big]\,,\\ \sigma_0&=\frac{\partial^2
  \mathcal{H}}{\partial
  {P_R}^2}\big[R_0,0,P_\Psi^0\big]\,,\\ \tau_0&=\frac{\partial^2
  \mathcal{H}}{\partial {P_\Psi}^2}\big[R_0,0,P_\Psi^0\big]\,.
\end{align}\end{subequations}
By looking to solutions proportional to some $\de^{\di\sigma t}$ one obtains some real frequencies, and therefore one finds stable circular orbits, if and only if
\begin{equation}\label{critADM}
\hat{C}_0 \equiv  \pi_0\, \sigma_0 ~> 0\,.
\end{equation}
Using explicitly the Hamiltonian \eqref{HADMcm} we readily obtain to 3PN order
\begin{align}\label{criterionADM}
\hat{C}_0 &= \frac{G m}{R_0^3}\biggl\{1+\Gamma_0\bigl(-9+\nu\bigr)
+\Gamma_0^2\left(\frac{117}{4}
+\frac{43}{8}\nu+\nu^2\right)\\ &\qquad\quad
+\Gamma_0^3\left(-61+\left[\frac{4777}{48}
  -\frac{325}{64}\pi^2\right]\nu-\frac{31}{8}\nu^2+\nu^3\right)
+\calO\left(\frac{1}{c^8}\right)\biggr\}\,,\nn
\end{align}
where $\Gamma_0=\frac{G m}{R_0 c^2}$. This result does not look the same as our previous result \eqref{criterion}, but this is simply due to the fact that it depends on the ADM radial separation $R_0$ instead of the harmonic one $r_0$. 

Fortunately we have derived in Sect.~\ref{sec:Lag4PN} the material needed to connect $R_0$ to $r_0$ with the 3PN accuracy. Indeed, with Eqs. \eqref{contacttransf} [together with~(4.9)--(4.10) of \citealt{ABF01}] we have the relation valid for general orbits in an arbitrary frame between the separation vectors in both coordinate systems. Specializing that relation to circular orbits we find
\begin{align}\label{R0r0}
R_0 = r_0\biggl\{1 &+\gamma_0^2\left(-\frac{1}{4}-\frac{29}{8}\nu\right) \\& +\gamma_0^3\left(\left[\frac{3163}{1680} +\frac{21}{32}\pi^2
  -\frac{22}{3}\ln\Big(\frac{r_0}{r'_0}\Big)\right]\nu
+\frac{3}{8}\nu^2\right) +
\calO\left(\frac{1}{c^8}\right)\biggr\}\,.\nn
\end{align}
Note that the difference between $R_0$ and $r_0$ starts at 2PN order. That relation easily permits to perfectly reconcile both expressions \eqref{criterion} and \eqref{criterionADM}.

Finally let us give to $\hat{C}_0$ an invariant meaning by expressing it with the help of the orbital frequency $\Omega_0$ of the circular orbit, or, more conveniently, of the frequency-related parameter $x_0\equiv(G m\,\Omega_0/c^3)^{2/3}$ -- cf. Eq.~\eqref{xdef}. This allows us to write the criterion for stability as $C_0 > 0$, where $C_0=\frac{G^2m^2}{x_0^3}\hat{C}_0$ admits the gauge-invariant form \citep{BI03CM}
\begin{equation}\label{C0}
C_0 = 1-6\,x_0 + 14\,\nu\,x_0^2 +
\left(\left[\frac{397}{2}-\frac{123}{16}\pi^2\right]\nu
-14\nu^2\right)\,x_0^3 + \calO\left(x_0^4\,\right)\,.
\end{equation} 
This form is more interesting than the coordinate-dependent expressions \eqref{criterion} or \eqref{criterionADM}, not only because of its invariant meaning, but also because as we see the 1PN term yields exactly the Schwarzschild result that the innermost stable circular orbit or ISCO of a test particle (i.e., in the limit $\nu\to 0$) is located at $x_{\mathrm{ISCO}}=\frac{1}{6}$. Thus we find that, at the 1PN order, but for \emph{any} mass ratio $\nu$,
\begin{equation}\label{C01PN}
x_{\mathrm{ISCO}}^\text{1PN} = \frac{1}{6}\,.
\end{equation}
One could have expected that some deviations of the order of $\nu$ already occur at the 1PN order, but it turns out that only from the 2PN order does one find the occurrence of some non-Schwarzschildean corrections proportional to $\nu$. At the 2PN order we obtain
\begin{equation}\label{C02PN}
x_{\mathrm{ISCO}}^{\mathrm{2PN}} =
\frac{3}{14\nu}\biggl(1-\sqrt{1-\frac{14\nu}{9}}\,\biggr)\,.
\end{equation}
For equal masses this gives $x_{\mathrm{ISCO}}^{\mathrm{2PN}}\simeq 0.187$. Notice also that the effect of the finite mass corrections is to increase the frequency of the ISCO with respect to the Schwarzschild result, i.e., to make it more \emph{inward}:\footnote{This tendency is in agreement with numerical and analytical self-force calculations \citep{BarackS09, LBB12}.}
\begin{equation}\label{C02PN0}
x_{\mathrm{ISCO}}^\text{2PN}=\frac{1}{6}\left[1+\frac{7}{18}\nu
  +\calO(\nu^2)\right]\,.
\end{equation}
Finally, at the 3PN order, for equal masses $\nu=\frac{1}{4}$, we find that according to this criterion all the circular orbits are stable. More generally, we find that at the 3PN order all orbits are stable when the mass ratio $\nu$ is larger than some critical value $\nu_c \simeq 0.183$.

The stability criterion \eqref{C0} has been compared by \cite{F11a} to various other stability criteria and shown to perform very well. It has also been generalized to spinning black hole binaries by \cite{F11b}. Note that this criterion is based on the physical requirement that a stable perturbation should have a real frequency. It gives an innermost stable circular orbit, when it exists, which differs from the innermost circular orbit or ICO defined in Sect.~\ref{sec:ICO}; see \cite{Schaferorleans} for a discussion on the difference between an ISCO and the ICO in the PN context. Note also that the criterion \eqref{C0} is based on systematic post-Newtonian expansions, without resorting for instance to Pad\'e approximants. Nevertheless, it performs better than other criteria based on various resummation techniques, as discussed by \cite{F11a}.


\subsubsection{The first law of binary point-particle mechanics}
\label{sec:firstlaw}

In this section we review an interesting relation for binary systems of point particles modelling black hole binaries and moving on circular orbits, known as the ``first law of point-particle mechanics''. This law was obtained using post-Newtonian methods by \cite{LBW12}, but is actually a particular case of a more general law, valid for systems of black holes and extended fluid balls, derived by \cite{FUS02}.

Before tackling the problem it is necessary to make more precise the notion of circular orbits. These are obtained from the \emph{conservative} part of the dynamics, neglecting the dissipative radiation-reaction force responsible for the gravitational-wave inspiral. In post-Newtonian theory this means neglecting the radiation-reaction force at 2.5PN and 3.5PN orders, i.e., considering only the conservative dynamics at the even-parity 1PN, 2PN and 3PN orders. We have seen in Sects.~\ref{sec:radreac} and \ref{sec:4PNradreac} that this clean separation between conservative even-parity and dissipative odd-parity terms breaks at 4PN order, because of a contribution originating from gravitational-wave tails in the radiation-reaction force. The dissipative part of the 4PN tail term has been given in Eq.~\eqref{acctailCMdiss}. In fact at any high order 4PN, 4.5PN, 5PN, etc. there will be a mixture of conservative and dissipative contributions; here we assume that at any order we can neglect the radiation-reaction dissipation effects.

Consider a system of two compact objects moving on circular orbits. We examine first the case of non-spinning objects. With exactly circular orbits the geometry admits a \emph{helical} Killing vector (HKV) field $K^\alpha$, satisfying the Killing equation $\nabla^{(\alpha} K^{\beta)}=0$. Imposing the existence of the HKV is the rigorous way to implement the notion of circular orbits. A Killing vector is only defined up to an overall constant factor. The helical Killing vector $K^\alpha$ extends out to a large distance where the geometry is essentially flat. There,
\begin{equation}\label{killing}
  K\ua\partial\la = \partial_t + \Omega\,\partial_\varphi\,,
\end{equation}
in any natural coordinate system which respects the helical symmetry \citep{SBD08}. We let this equality define the overall constant factor, thereby specifying the Killing vector field uniquely. In Eq.~\eqref{killing} $\Omega$ denotes the angular frequency of the binary's circular motion.

An observer moving with one of the particles (say the particle 1), while orbiting around the other particle, would detect no change in the local geometry. Thus the four-velocity $u_1^\alpha$ of that particle is tangent to the Killing vector $K^\alpha$ evaluated at the location of the particle, which we denote by $K_1^\alpha$. A physical quantity is then defined as the constant of proportionality, say $u_1^T$, between these two vectors, namely
\begin{equation}\label{utdef}
   u_1\ua = u_1^T \,K_1\ua\,.
\end{equation}
The four-velocity of the particle is normalized by $(g_{\mu\nu})_1 u_1^\mu u_1^\nu = -1$, where $(g_{\mu\nu})_1$ denotes the metric at the location of the particle. For a self-gravitating compact binary system, the metric at point 1 is generated by the two particles and has to be regularized according to the self-field regularization discussed in Sect.~\ref{sec:reg}. It will in fact be sometimes more convenient to work with the inverse of $u_1^T$, denoted $z_1\equiv 1/u_1^T$. From Eq.~\eqref{utdef} we get
\begin{equation}\label{z1def}
   z_1 = - (u_1 K_1)\,,
\end{equation}
where $(u_1 K_1) = (g_{\mu\nu})_1 u_1^\mu K_1^\nu$ denotes the usual space-time dot product. Thus we can regard $z_1$ as the Killing energy of the particle that is associated with the HKV field $K^\alpha$. The quantity $z_1$ represents also the redshift of light rays emitted from the particle and received on the helical symmetry axis perpendicular to the orbital plane at large distances from it. In the following we shall refer to $z_1$ as the \emph{redshift observable}; it has been introduced by \cite{Det08}.

If we choose a coordinate system such that Eq.~\eqref{killing} is satisfied everywhere, then in particular $K_1^t=1$, thus $u_1^T$ simply agrees with $u_1^t$, the $t$-component of the four-velocity of the particle. The Killing vector on the particle is then $K_1\ua=u_1\ua/u_1^t$, and simply reduces to the particle's ordinary coordinate velocity: $K_1\ua = v_1\ua$ where $v_1\ua=\dd y_1\ua/\dd t$ and $y_1\ua(t)=(t, \bm{y}_1(t))$ denotes the particle's trajectory in that coordinate system. The redshift observable is then
\begin{equation}\label{z1expr}
    z_1 = \frac{1}{u_1^T} = \sqrt{- (g_{\mu\nu})_1 v_1^\mu
      v_1^\nu} \,.
\end{equation}
It is important to note that for circular orbits this quantity does not depend upon the choice of coordinates; in a perturbative approach in which the perturbative parameter is the particles' mass ratio $\nu\ll 1$, it does not depend upon the choice of perturbative gauge with respect to the background metric. We shall be interested in the \emph{invariant} scalar function $z_1(\Omega)$, where $\Omega$ is the angular frequency of the circular orbit defined by Eq.~\eqref{killing}.

We have obtained in Sect.~\ref{sec:eomcirc} the expressions of the post-Newtonian binding energy $\dE$ and angular momentum $\dJ$ for point-particle binaries on circular orbits. We shall now show that there are some differential and algebraic relations linking $\dE$ and $\dJ$ to the redshift observables $z_1$ and $z_2$ associated with the two individual particles. Here we prefer to introduce instead of $\dE$ the total relativistic (ADM) mass of the binary system
\begin{equation}\label{PNmass}
\dM = m + \dE \, ,
\end{equation}
where $m$ is the sum of the two post-Newtonian individual masses $m_1$ and $m_2$ -- those which have been used up to now, for instance in Eq.~\eqref{acc3PN}. Note that in the spinning case such post-Newtonian masses acquire some spin contributions given by Eqs. \eqref{massformula}--\eqref{slowomn}.

For point particles without spins, the ADM mass $\dM$, angular momentum $\dJ$, and redshifts $z_\text{a}$, are functions of three independent variables, namely the orbital frequency $\Omega$ that is imposed by the existence of the HKV, and the individual masses $m_\text{a}$. For spinning point particles, we have also the two spins $S_\text{a}$ which are necessarily aligned with the orbital angular momentum, and furthermore should satisfy a corotation condition. We first recall that the ADM quantities obey the ``thermodynamical'' relation already met in Eq.~\eqref{thermo},
\begin{equation}\label{partialOmega}
\frac{\partial \dM}{\partial \Omega} = \Omega \, \frac{\partial
  \dJ}{\partial \Omega}\,.
\end{equation}
Such relation is commonly used in post-Newtonian theory, see e.g., \cite{DJSinv, B02ico,BFLS20}; it states that the gravitational-wave energy and angular momentum \emph{fluxes} are strictly proportional for circular orbits, with $\Omega$ being the coefficient of proportionality. The relation is also used in numerical computations of the binary evolution based on a sequence of quasi-equilibrium configurations \citep{GGB1, GGB2, CPf04, CCGPf06}, as reviewed in Sect.~\ref{sec:ICO}.

The first law is a generalization of Eq.~\eqref{partialOmega}, describing the changes in the ADM quantities not only when the orbital frequency $\Omega$ varies with fixed masses, but also when the individual masses $m_\text{a}$ of the particles vary with fixed orbital frequency. That is, one compares together different conservative dynamics with different masses but the same orbital frequency. This situation is answered by the differential equations
\begin{equation}\label{partialmn}
\frac{\partial \dM}{\partial m_\text{a}} - \Omega \frac{\partial
  \dJ}{\partial m_\text{a}} = z_\text{a} \quad (\text{a}=1,2) \,.
\end{equation}
Finally the relations \eqref{partialOmega}--\eqref{partialmn} can be summarized by the following.

\begin{theorem} \citep{FUS02, LBW12}
The changes in the ADM mass and angular momentum of a binary system made of point particles on a circular orbit, in response to infinitesimal variations of the individual masses of the point particles, are related together by the first law of binary point-particle mechanics
\begin{equation}\label{firstlaw}
\delta \dM - \Omega \, \delta \dJ = \sum_\text{a}
z_\text{a} \, \delta m_\text{a} \,.
\end{equation}
\label{th:firstlaw}
\end{theorem}
This law was initially proved in a very general way by \cite{FUS02} for systems of black holes and extended bodies under some arbitrary Killing symmetry. The particular form given in Eq.~\eqref{firstlaw} is a specialization to the case of point particle binaries with helical Killing vector. It has been derived directly in this form by \cite{LBW12} up to high post-Newtonian order, viz. 3PN and the logarithmic terms at 4PN and 5PN orders. Furthermore, it was proved by \cite{BL17} that the first law still holds when taking into account the non-local tail term at 4PN order. 

An interesting by-product of the first law \eqref{firstlaw} is the simple relation
\begin{equation}\label{firstintegral}
	\dM - 2 \Omega \dJ = \sum_\text{a} z_\text{a} m_\text{a}\,,
\end{equation}
which can be seen as a first integral of the differential relation \eqref{firstlaw}. The existence of such a simple algebraic relation between the quantities $z_\text{a}$ belonging to each particles, and the globally defined quantities $\dM$ and $\dJ$, is remarkable.

The law \eqref{firstlaw} has been generalized by \cite{LeT15} for binary systems of point masses moving on generic eccentric orbits, using a natural orbit-averaged definition of the redshift observable \citep{BarackS11}. An application of the first law is the calculation of the exact gravitational self-force contributions in the binding energy and angular momentum for circular orbits, allowing a coordinate-invariant comparison to numerical relativity \citep{LBB12}.

In the case of an asymptotically flat vacuum spacetime with two black holes on circular orbits, we have \citep{FUS02}
\begin{equation}\label{BBHmechanics}
	\delta \dM - \Omega \, \delta \dJ = \sum_\text{a}
	\frac{\kappa_\text{a}}{8\pi} \, \delta A_\text{a} \,,
\end{equation}
which can be viewed as a generalization of the celebrated first law of black hole mechanics $\delta \dM - \omega_\text{H} \, \delta \dJ = \frac{\kappa}{8\pi} \delta A$, which holds for any non-singular, asymptotically flat perturbation of a stationary and axisymmetric black hole of mass $\dM$, intrinsic angular momentum (or spin) $\dJ \equiv \dM a$, surface area $A$, uniform surface gravity $\kappa$, and angular frequency $\omega_\text{H}$ on the horizon \citep{BarCH73, Wald73}. In the binary black hole case, the role played by the horizon angular velocity $\omega_\text{H}$ of a single rotating black hole is taken by the orbital frequency $\Omega$ of the binary. 

In order to interpret the first law \eqref{firstlaw} (and the redshift factor therein) for actual binary black holes, and to reconcile it with \eqref{BBHmechanics}, we have to remember that the assumption of helical Killing symmetry requires that the two black holes are in co-rotation, and must therefore have non-zero spins. For rotating black holes, the post-Newtonian mass $m_\text{a}$ is given by the total mass of the black hole, including the effect of the spin, see Eq.~\eqref{massformula}. We therefore need the generalization of the first law \eqref{firstlaw} applicable to systems of point particles with spins (moving on circular orbits). The result will be valid through linear order in the spin of each particle, but will also hold for the quadratic coupling between different spins (interaction spin terms $S_1\times S_2$ in the language of Sect.~\ref{sec:spins}). To be consistent with the HKV symmetry, we must assume that the two spins $S_\text{a}$ are aligned or anti-aligned with the orbital angular momentum. We introduce the total (ADM-like) angular momentum $\dJ$ which is related to the orbital angular momentum $\dL$ by $\dJ = \dL + \sum_\text{a}S_\text{a}$ for aligned or anti-aligned spins. The first law now reads
\begin{equation}\label{firstlawspin}
\delta \dM - \Omega \, \delta \dJ = \sum_\text{a} \Bigl[
  z_\text{a} \, \delta m_\text{a} + \bigl( \Omega_\text{a} - \Omega
  \bigr) \delta S_\text{a} \Bigr] \,,
\end{equation}
where $\Omega_\text{a}=\vert\bm{\Omega}_\text{a}\vert$ denotes the \emph{precession} frequency of the spins. This law has been derived by \cite{BBL13} using the canonical Hamiltonian formalism. The spin variables in \eqref{firstlawspin} are the canonical spins $\bm{S}_\text{a}$, that are easily seen to obey, from the algebra satisfied by the canonical variables, the usual Newtonian-looking precession equations $\dd\bm{S}_\text{a}/\dd t=\bm{\Omega}_\text{a}\times\bm{S}_\text{a}$. These variables are identical to the ``constant-in-magnitude'' spins which will be defined and extensively used in Sect.~\ref{sec:spins}. Similarly to Eq.~\eqref{firstintegral}, we have also a first integral associated with the variational law \eqref{firstlawspin}:
\begin{equation}\label{firstintegralspin}
\dM - 2 \Omega \dJ = \sum_\text{a} \Bigl[ z_\text{a}
  m_\text{a} + 2 (\Omega_\text{a} - \Omega) S_\text{a} \Bigr] \,.
\end{equation}

Notice that the relation \eqref{firstlawspin} has been derived for point particles and arbitrary aligned spins. We want now to derive the analogous relation for binary black holes. The key difference is that black holes are extended finite-size objects while point particles have by definition no spatial extension. For point particle binaries the spins can have arbitrary magnitude and still be compatible with the HKV. In this case the law \eqref{firstlawspin} would describe also (super-extremal) naked singularities. For black hole binaries the HKV constraints the rotational state of each black hole and the binary system must be corotating.

Let us derive, in a heuristic way, the analogue of the first law \eqref{firstlawspin} for general bodies, by introducing some ``constitutive relations'' $m_\text{a}(\mu_\text{a},S_\text{a},\cdots)$ specifying the energy content of the bodies, i.e., the relations linking their masses $m_\text{a}$ to the spins $S_\text{a}$ and to some irreducible masses $\mu_\text{a}$. More precisely, we define for each spinning particle the analogue of an irreducible mass $\mu_\text{a}\equiv m_\text{a}^\text{irr}$ via the variational relation $\delta m_\text{a} = c_\text{a} \, \delta \mu_\text{a} + \omega_\text{a} \, \delta S_\text{a}$, in which the ``response coefficient'' $c_\text{a}$ of the body and its proper rotation frequency $\omega_\text{a}$ are associated with the internal structure:
\begin{subequations}\label{dmn}
\begin{align} 
c_\text{a} &\equiv \frac{\partial m_\text{a}}{\partial
  \mu_\text{a}}{\bigg|}_{S_\text{a}}\,,\label{dmn1}\\ \omega_\text{a}
&\equiv \frac{\partial m_\text{a}}{\partial
  S_\text{a}}{\bigg|}_{\mu_\text{a}}\,.\label{dmn2}
\end{align}\end{subequations}
For instance, using the Christodoulou mass formula \eqref{massformula} for Kerr black holes, we obtain the rotation frequency $\omega_\text{a}$ given by Eq.~\eqref{omegan}. On the other hand, the response coefficient $c_\text{a}$ differs from 1 only because of spin effects, and we can check that $c_\text{a} = 1 + \calO(S_\text{a}^2)$.

Within the latter heuristic model a condition for the corotation of black hole binaries has been proposed by \cite{BBL13} as
\begin{equation}\label{corotcond}
	z_\text{a} \, \omega_\text{a} + \Omega_\text{a} = \Omega \,.
\end{equation}
This condition determines the value of the proper frequency $\omega_\text{a}$ of each black hole appropriate to the corotation state. When expanded to 2PN order the condition \eqref{corotcond} leads to Eq.~\eqref{corot} that we have already used in Sect.~\ref{sec:ICO}. With Eq.~\eqref{corotcond} imposed, the first law \eqref{firstlawspin} simplifies considerably:
\begin{equation}\label{firstlawcorot}
\delta \dM - \Omega \, \delta \dJ = \sum_\text{a}
c_\text{a} z_\text{a} \, \delta \mu_\text{a} \,.
\end{equation}
This is almost identical to the first law for non-spinning binaries given by Eq.~\eqref{firstlaw}; indeed it simply differs from it by the substitutions $c_\text{a} \rightarrow 1$ and $\mu_\text{a} \rightarrow m_\text{a}$. Since the irreducible mass $\mu_\text{a}$ of a rotating black hole is the spin-independent part of its total mass $m_\text{a}$, this observation suggests that corotating binaries are very similar to non-spinning binaries, at least from the perspective of the first law. Finally we can easily reconcile the first law \eqref{firstlawcorot} for corotating systems with the known first law of binary black hole mechanics given by \eqref{BBHmechanics}. Indeed, since the surface area is related to the irreducible mass by $\mu_\text{a}^2 = A_\text{a}/(16\pi)$, it suffices to make the formal identification in Eq.~\eqref{firstlawcorot}:
\begin{equation}\label{correspondence}
	c_\text{a} z_\text{a} \equiv 4
\mu_\text{a} \kappa_\text{a}\,,
\end{equation}
where $\kappa_\text{a}$ denotes the constant surface gravity. In particular, for two black holes on circular orbits, but far enough apart so that they can be viewed -- in first approximation -- as isolated, the surface gravity $\kappa_\text{a}\to (4\mu_\text{a})^{-1}$, while for the point particles we have both $z_\text{a}\to 1$ and $c_\text{a}\to 1$. The nice correspondence \eqref{correspondence} shows that the heuristic model based on the constitutive relations \eqref{dmn} is able to capture the physics of corotating black hole binary systems.


\subsubsection{Post-Newtonian approximation versus gravitational self-force}
\label{sec:SF}

The high-accuracy predictions from GR we have drawn up to now are well suited to describe the inspiralling phase of compact binaries, when the post-Newtonian parameter \eqref{epsPN} is small, independently of the mass ratio $q\equiv m_1/m_2$ between the compact bodies. In this section we investigate how well does the post-Newtonian expansion, compared with another very important approximation scheme in general relativity: The gravitational \emph{self-force} approach, based on black-hole perturbation theory, which gives an accurate description of \emph{extreme mass ratio} binaries having $q\ll 1$ or equivalently $\nu\ll 1$, even in the strong field regime. It is thus expected to provide templates for extreme mass ratio inspirals (EMRI) anticipated to be present in the bandwidth of space-based detectors. 

The gravitational self-force analysis was pioneered by \cite{dWB60, MiSaTa, QuWa, DW03, GW08}; see \cite{PoissonLR, Detweilerorleans, Barackorleans, BP18} for reviews. The expansion at leading order in the mass ratio reaches impressive PN equivalent levels \citep{Fuj14PN,Fuj22PN,KOW15}. The state-of-the-art is currently the second order in the mass ratio, for both the binding energy \citep{Poundetal20} and the energy flux and waveform \citep{WarburtonPound21,WPWMDL23}.

Consider a system of two (non-spinning) compact objects with mass ratio $q=m_1/m_2 \ll 1$; we shall call the smaller mass $m_1$ the ``particle'', and the larger mass $m_2$ the ``black hole''. The orbit of the particle around the black hole is supposed to be exactly circular as we neglect the radiation-reaction effects. With circular orbits and no dissipation, we are considering the conservative part of the dynamics, and the geometry admits the HKV field \eqref{killing}. Note that in self-force theory there is a clean split between the dissipative and conservative parts of the dynamics (see e.g., \citealt{Ba09}). This split is particularly transparent for an exact circular orbit, since the radial component (along $r$) is the only non-vanishing component of the conservative self-force, while the dissipative part of the self-force are the components along $t$ and $\varphi$.

\begin{figure}[htb]
\centerline{\includegraphics[width=\textwidth]{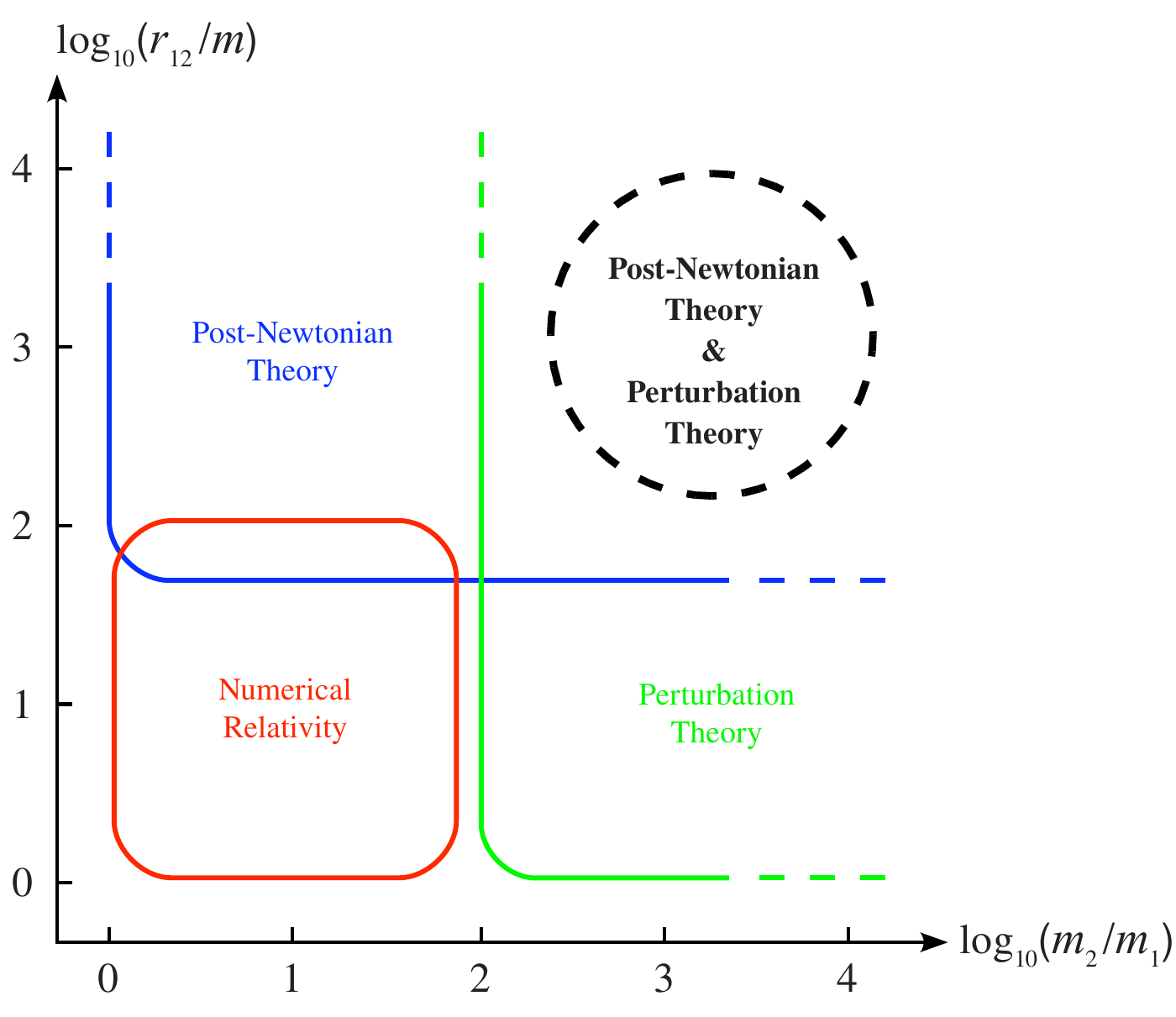}}
\caption{Different analytical approximation schemes and numerical techniques to study black hole binaries, depending on the mass ratio $q=m_1/m_2$ and the post-Newtonian parameter $\epsilon^2 \sim v^2/c^2 \sim G m/(c^2r_{12})$. Post-Newtonian theory and perturbative self-force analysis can be compared in the post-Newtonian regime ($\epsilon\ll 1$ thus $r_{12}\gg G m/c^2$) of an extreme mass ratio ($m_1 \ll m_2$) binary. Image courtesy A.~Le Tiec.}
\label{fig:PN-SF}
\end{figure}

The problem of the comparison between the post-Newtonian and perturbative self-force analyses in their common domain of validity, that of the slow-motion and weak-field regime of an extreme mass ratio binary, is illustrated in Fig.~\ref{fig:PN-SF}. This problem has been tackled by \cite{Det08}, who computed numerically within the self-force (SF) approach the redshift observable $u_1^T$ associated with the particle, and compared it with the 2PN prediction extracted from the post-Newtonian results of \cite{BFP98}. This comparison proved to be successful, and therefore was later systematically implemented and extended to higher PN orders by \cite{BDLW10a, BDLW10b} (see also \citealt{BDS10,BarackS11} for related comparisons). In this section we review the latter works which have demonstrated an excellent agreement between the analytical post-Newtonian result derived through 3PN order, with inclusion of specific logarithmic terms at 4PN and 5PN orders, and the exact numerical SF result.

For the PN-SF comparison, we require two physical quantities which are precisely defined in the context of each of the approximation schemes. The orbital frequency $\Omega$ of the circular orbit as measured by a distant observer is one such quantity and has been introduced in Eq.~\eqref{killing}; the second quantity is the \cite{Det08} redshift observable $u_1^T$ (or equivalently $z_1=1/u_1^T$) associated with the smaller mass $m_1\ll m_2$ and defined by Eqs. \eqref{utdef} or \eqref{z1def}. The truly coordinate and perturbative-gauge independent properties of $\Omega$ and the redshift observable $u_1^T$ play a crucial role in this comparison. In the perturbative self-force approach we use Schwarzschild coordinates for the background, and we refer to ``gauge invariance'' as a property which holds within the restricted class of gauges for which \eqref{killing} is a helical Killing vector. In all other respects, the gauge choice is arbitrary. In the post-Newtonian approach we generally work with harmonic coordinates and compute the explicit expression \eqref{z1expr} of the redshift observable.

The main difficulty in the post-Newtonian calculation is the control to high PN order of the near-zone metric $(g_{\mu\nu})_1$ entering the definition of the redshift observable \eqref{z1expr}, and which has to be regularized at the location of the particle by means of dimensional regularization (see Sects.~\ref{sec:DReom} and \ref{sec:DRrad}). Up to 2.5PN order the Hadamard regularization is sufficient and the regularized metric has been provided in Eqs. \eqref{metricNZ1}. However at 3PN order dimensional regularization plays a crucial role as there are poles $\propto 1/\varepsilon$ which develop in the near zone metric. As shown by \cite{BDLW10a} the poles are cancelled in fine when the redshift observable is expressed in terms of the gauge invariant post-Newtonian parameter \eqref{xdef}. At the next 4PN order we do not know the near zone metric and thus a direct computation of the redshift using Eq.~\eqref{z1expr} is not possible. Nevertheless we can infer \textit{indirectly} the 4PN coefficient in the redshift from the known 4PN coefficient in the energy function for circular orbits \eqref{Ecirc} by using the first law \eqref{firstlaw}. Indeed, as found by \cite{LBW12,LBB12} the first law is fully satisfied if and only if there is a precise relationship between the post-Newtonian coefficients in the redshift on the one hand, and those in the conserved energy and angular momentum on the other hand. Finally the complete result for the redshift at 4PN  order is
\begin{align}\label{z1x}
	z_1 &= 1 + \left( - \frac{3}{2}X_2 + \frac{\nu}{2} \right) x \nn \\
	& + \left( - \frac{9}{8}X_2 +\left[ -\frac{3}{8} -\frac{X_2}{4}\right]\nu + \frac{5}{24} \nu^2 \right) x^2 \nn \\
	& + \left( - \frac{27}{16}X_2 +\left[- \frac{27}{16} + \frac{19}{8}X_2\right]\nu +\left[- \frac{19}{16} - \frac{X_2}{16}\right]\nu^2 + \frac{\nu^3}{16} \right) x^3 \nn \\
 & + \left( - \frac{405}{128}X_2 + \left[ -\frac{675}{128} + \left(\frac{6889}{192} - \frac{41}{32} \pi^2\right)X_2 \right] \nu + \left[ -\frac{6889}{1152} + \frac{41}{192} \pi^2 - \frac{93}{64} X_2 \right]\nu^2 \right. \nn\\ &\qquad\quad \left. + \left[ \frac{217}{192} - \frac{7}{864} X_2 \right]\nu^3 + \frac{91}{10368}\nu^4 \right) x^4 \nn \\
  & + \left( - \frac{1701}{256}X_2 + \left[- \frac{3969}{256} + \left( - \frac{24689}{1920} + \frac{128}{5}\gamma_\text{E} + \frac{1291}{512}\pi^2 + \frac{64}{5}\ln(16x) \right)X_2\right]\nu \right. \nn\\ &\qquad\quad \left. + \left[ \frac{24463}{11520} + \frac{64}{15}\gamma_\text{E} + \frac{1291}{3072}\pi^2 + \frac{32}{15}\ln(16x) + \left( - \frac{71207}{768} + \frac{451}{128}\pi^2 \right)X_2\right]\nu^2 \right. \nn\\ &\qquad\quad \left. + \left[\frac{356035}{6912} - \frac{2255}{1152}\pi^2 + \frac{43}{288} X_2\right]\nu^3 + \left[ - \frac{473}{3456} + \frac{55}{20736} X_2\right]\nu^4 \right. \nn\\ &\qquad\quad \left. - \frac{187}{62208}\nu^5 \right) x^5 + \calO\left(x^6\right) \,, 
\end{align}
where $x$ is defined by \eqref{xdef}, $\nu$ is the mass ratio \eqref{nu}, and we recall that $X_2 \equiv m_2/m$ (with 2 being the ``black hole''). The redshift of the other particle is deduced by changing $X_2 \rightarrow X_1$.

The post-Newtonian result \eqref{z1x} is valid for any mass ratio, and for comparison purpose with the SF calculation we now investigate the small mass ratio regime $q\ll 1$. We introduce a PN parameter appropriate to the small mass limit of the ``particle'',
\begin{equation}\label{y}
    y \equiv \left( \frac{G\,m_2\,\Omega}{c^3} \right)^{2/3} = x
    \,\bigl(1+q\bigr)^{-2/3}\,.
\end{equation}
We express the symmetric mass ratio in terms of the asymmetric one: $\nu = q (1+q)^{-2}$, together with $X_2=(q+1)^{-1}$. Then Eq.~\eqref{z1x} [or, rather, its inverse $u_1^T=1/z_1$], expanded through first order in $q$, which means including the linear self-force level, reads
\begin{equation}\label{utexp}
    u_1^T = u^T_\mathrm{Schw} + q \, u^T_\mathrm{SF} + \calO(q^2) \,.
\end{equation}
The Schwarzschildean result is known in closed form as
\begin{equation}\label{utSchw}
   u^T_\mathrm{Schw} = \frac{1}{\sqrt{1 - 3 y }} \,,
\end{equation}
and for the SF contribution one obtains\footnote{Since there are logarithms in this expansion we use the Landau $o$-symbol for remainders; see the footnote \ref{fnote:landau}.}
\begin{align}\label{utSF}
	u^T_\mathrm{SF}(y) &= - y - 2 y^2 - 5 y^3 + \left( - \frac{121}{3} +
	\frac{41}{32} \pi^2 \right) y^4 \nn\\&+ \left( - \frac{1157}{15} + \frac{677}{512}\pi^2 - \frac{128}{5}\gamma_\text{E} - \frac{64}{5}\ln(16y)\right) y^5 + o(y^5)\,.
\end{align}
The analytic coefficients in \eqref{utSF} were determined up to 2PN order by \cite{Det08} using the regularized post-Newtonian metric of \cite{BFP98}. The 3PN term was computed by \cite{BDLW10a} making full use of dimensional regularization. The comparison of the post-Newtonian expansion \eqref{utSF} with the numerical SF data has confirmed with high precision the analytic determination of the 3PN coefficient. Notice that such agreement provides an independent check of the dimensional regularization procedure at the basis of the PN expansion (see Sects.~\ref{sec:DReom}--\ref{sec:DRrad}). It is remarkable that such procedure gives results equivalent to those obtained with the subtraction of the singular field in the SF approach \citep{DW03}. The comparison with SF data also revealed the presence of logarithmic contributions at the 4PN and 5PN orders, which have then been computed analytically \citep{BDLW10a, BDLW10b, D10sf}. Furthermore the PN-SF comparison has permitted to measure the 4PN and 5PN coefficients, with high significant digits \citep{LBW12}. The 4PN coefficient in \eqref{utSF} was computed analytically by \cite{BiniD13}.

For reference let us give the higher-order 2SF and 3SF corrections at the 4PN order, i.e. the next two terms in \eqref{utexp}, which are straightforward to compute from Eq.~\eqref{z1x}:
\begin{subequations}\label{ut2SF3SF}
	\begin{align}
		u^T_\mathrm{2SF}(y) &= y + 3 y^2 + \frac{97}{8} y^3 + \left( \frac{725}{12} -
		\frac{41}{64} \pi^2 \right) y^4 \nn\\&+ \left( \frac{674801}{1920} - \frac{9491}{1024}\pi^2 - \frac{64}{5}\gamma_\text{E} - \frac{32}{5}\ln(16y)\right) y^5 + o(y^5)\,,\\ 
		u^T_\mathrm{3SF}(y) &= -\frac{28}{27}y - \frac{67}{18} y^2 - \frac{97}{6} y^3 + \left( - \frac{2099}{27} +
		\frac{41}{72} \pi^2 \right) y^4 \nn\\&+ \left( - \frac{73417}{135} + \frac{32123}{2304}\pi^2 + \frac{256}{45}\gamma_\text{E} + \frac{128}{45}\ln(16y)\right) y^5 + o(y^5)\,.
	\end{align}
\end{subequations}
We report also the \textit{leading} logarithmic contributions in the redshift variable at any generic power $n$. These have been computed in the energy and angular momentum using renormalization group techniques, see Eqs. \eqref{EJlogn}. We have (at linear order in the mass ratio $q$)
\begin{align}\label{u1tlogs}
		u^T_\mathrm{SF}(y)\Big|_\text{leading-$(\log)^n$} &= - \frac{64}{5} \sum_{n = 1}^{+\infty} \frac{1}{n!}\bigl(4\beta_2^{(m)}\bigr)^{n-1} y^{3n+2}(\ln y)^n\,,
\end{align}
where we recall that $\beta^{(m)}_2 = -\frac{214}{105}$ is the RG coefficient associated with the mass quadrupole moment. We recover the 4PN logarithm in \eqref{utSF}, followed by a logarithm square at 7PN order, a logarithm cube at 10PN order, and so on. The agreement with the high-order analytical (linear in mass ratio) self force calculation of \cite{KOW15} is perfect. This also highlights the predictive power of perturbative self-force calculations, which is obviously due to the fact (illustrated in Fig.~\ref{fig:PN-SF}) that perturbation theory is legitimate in the strong field regime of the coalescence of black hole binary systems, which is inaccessible to the post-Newtonian method.

The accuracy of the numerical computation of the self-force has been drastically improved by \cite{SFW14}. The PN coefficients of the redshift observable were obtained to very high 10.5PN order both numerically and analytically, for a subset of coefficients that are either rational or made of the product of $\pi$ with a rational. The analytical values of the coefficients up to 6PN order have also been obtained from a perturbative self-force calculation by \cite{BiniD14a, BiniD14b}. Note that the works of \cite{SFW14} and \cite{BiniD13, BiniD14a, BiniD14b} use alternative techniques (developed by \citealt{MST96a, MST96b, MT97}) with which to represent metric perturbation solutions for black hole space-times.

An interesting feature of the post-Newtonian expansion of the redshift at high order is the appearance of half-integral PN coefficients (i.e., of the type $\frac{p}{2}$PN where $p$ is an \emph{odd} integer) in the conservative dynamics of binary point particles, i.e. moving on exactly circular orbits. This is interesting because any instantaneous (non-tail or ``local'') term at any half-integral PN order in the redshift will be zero for circular orbits, as can be shown by a simple dimensional argument \citep{BFW14a}. Therefore half-integral coefficients can appear only due to truly hereditary (tail or ``non-local'') integrals. Using standard post-Newtonian methods it has been proved by \cite{BFW14a, BFW14b} that the dominant half-integral PN term in the redshift observable \eqref{utSF} occurs at the 5.5PN order (confirming the earlier finding of \citealt{SFW14}) and originates from the non-linear ``tail-of-tail'' integrals investigated in Sect.~\ref{sec:gravtails}. The results for the 5.5PN coefficient, and also for the next-to-leading 6.5PN and 7.5PN ones due to higher-order tails-of-tails and computed by \cite{BFW14a, BFW14b}, fully agree with the SF computations. We emphasize that that the latter results (contrary to various analytical and numerical SF calculations by \citealt{SFW14, BiniD13, BiniD14a, BiniD14b, KOW15}) are achieved with the ``traditional'' PN approach, which is completely general, i.e. is not tuned to a particular type of source but is applicable to any extended PN source; see the formalism of Sect.~\ref{sec:PNsource}. 

To conclude, the consistency of this cross-cultural comparison between the analytical post-Newtonian and the perturbative self-force approaches confirms the soundness of both approximations in describing the dynamics of compact binaries. Furthermore this interplay between PN and SF efforts is important for the synthesis of template waveforms of EMRIs to be analysed by space-based gravitational-wave detectors, and has also an impact on efforts of numerical relativity in the case of comparable masses.


\subsection{Gravitational waves from compact binaries}
\label{sec:GW}

We pointed out that the 4PN equations of motion are merely ``1.5PN'' as regards the radiative aspects of the problem, because the radiation reaction force starts at the 2.5PN order. Because of this difference between conservative and dissipative (radiation reaction) effects, we stop developing high order PN corrections in the radiation reaction force, and instead apply the gravitational wave-generation formalism described in Sect.~\ref{sec:PNsource}. In this approach, the work done by the radiation reaction force is computed as a flux at future null infinity. The total energy flux $\mathcal{F}\equiv(\dd\dE/\dd t)^\text{GW}$ drives the evolution of the energy in the system through the flux-balance equation
\begin{equation}
  \frac{\dd \dE}{\dd t}=-\mathcal{F}\,.
  \label{baleq}
\end{equation}
Therefore, consistently with the result \eqref{Ecirc} we obtained for the 4.5PN binary's center-of-mass energy $\dE$, we must now control the total gravitational-wave flux $\mathcal{F}$ at the same relative 4.5PN order, which means, beyond the Einstein quadrupole formula \eqref{fluxE}.

Because the orbit of inspiralling compact binaries is circular, the energy balance equation is sufficient, and there is no need to invoke the angular momentum balance equation.\footnote{By contrast, remind the computation of the evolution of the orbital period $\dot{P}$ and eccentricity $\dot{e}$ of the binary pulsar, see Eqs. \eqref{balanceEJ}--\eqref{peters}.} Furthermore the time average over one orbital period is here irrelevant, and the angular momentum flux $\mathcal{G}\equiv(\dd\dJ/\dd t)^\text{GW}$ is related to the energy flux by $\mathcal{F}=\Omega\,\mathcal{G}$, which is a consequence of \eqref{partialOmega}.

We keep in mind that we shall use the flux-balance equation \eqref{baleq} at very high order, where one is missing a complete proof of it, following from first principles in GR. Nevertheless, Eq.~\eqref{baleq} has been verified by radiation-reaction calculations for extended post-Newtonian fluids at 1.5PN order \citep{B97, BF19}, and by explicit computations for compact binaries at the 3.5PN \citep{JaraS97, PW02, KFS03, NB05, itoh3} and 4.5PN \citep{LPY23} levels, fully consistent with the balance equations \citep{IW93, IW95, GII97}.

Obtaining the energy flux $\mathcal{F}$ can be divided into two equally important tasks: computing the \emph{source} multipole moments $\dI_L$ and $\dJ_L$ of the compact binary system with due account of self-field and IR regularizations; and controlling the tails and related non-linear effects occurring in the relation between the binary's source moments and the \emph{radiative} ones $\dU_L$ and $\dV_L$ observed at future null infinity (cf. the general formalism of Sect.~\ref{sec:PNsource}).


\subsubsection{The binary's multipole moments for circular orbits}
\label{sec:binarymoments}

The general expressions of the source multipole moments given by Eqs. \eqref{sourcemoments} in Theorem \ref{th:sourcemoments} are first worked out explicitly for general fluid systems. For this computation one uses the formula \eqref{intdeltaexp}, and we insert the metric coefficients (in harmonic coordinates) expressed by means of the retarded-type elementary potentials.\footnote{See Eqs. \eqref{pot1PN}--\eqref{pot3PN} and the generalization at 4PN order given in the Appendix A of \cite{MHLMFB20}.} Then we specialize each of the numerous terms to the case of point-particle binaries by inserting the standard expression \eqref{Talphabeta} for the matter stress-energy tensor made out of Dirac delta-functions. 

The most important input for the computation of the waveform and flux is the mass quadrupole moment $\dI_{ij}$, since this moment necessitates the full PN precision. A first computation of the quadrupole moment at the 4PN order was performed by \cite{MHLMFB20}, using dimensional regularization for the UV divergences but the Hadamard regularization for the IR ones. However, in order to be consistent with the 4PN equations of motion \citep{BBBFMa,BBBFMb,BBBFMc,BBFM17}, it should be computed with dimensional regularization also for the IR divergences. It was thus completed by \cite{LHBF22}, and the expression of the source quadrupole moment, obtained with full dimensional regularization,\footnote{More precisely: the ``$B\varepsilon$'' variant, see Sect.~\ref{sec:DRrad}.} has the expected feature of exhibiting poles in $\varepsilon = d-3$. Those poles are crucial to cancel the divergences linked with the $d$ dimensional computation of the radiative moment, i.e. $\dU_{ij}$, as performed by \cite{LBHF22}. Indeed, the source moments are not observables \emph{per se}, but the radiative moments are; therefore, only the latter have to be finite in the $\varepsilon \rightarrow 0$ limit. However, for the sake of computational simplicity, it was deemed useful to introduce the notion of a ``renormalized'' source quadrupole moment $\dI_{ij}$, defined by Eq.~(6.2) of \cite{LBHF22}. This renormalized quantity is nothing but the sum of the $d$-dimensional source quadrupole and the corrections, in the form of a ``difference'' (see Sect.~\ref{sec:DRrad}), arising notably from the relation between radiative and canonical moments, involving the non-linear tails-of-tails, tails-of-memory, etc. listed in Sect.~\ref{sec:radcanonical}. The renormalized quadrupole, now truly called $\dI_{ij}$, has no poles in $\varepsilon$ and when injected into the three dimensional MPM algorithm, yields the correct radiative quadrupole moment $\dU_{ij}$, by definition. Last but not least, up to 3PN order, the corrections due to the IR dimensional regularization exactly cancel the ones due to the renormalization, as proven by \cite{LBHF22}. Thus, up to 3PN order, the renormalized source quadrupole coincides with the one computed with Hadamard regularization in the IR, which is why those subtleties did not hit previous computations of the flux. However, this equivalence is no more valid at 4PN order, thus the crucial need for the implementation of dimensional regularization and the previous ``renormalization''. 

The mass quadrupole moment complete to order 4PN, for non-spinning compact binaries on quasi circular orbits, is
\begin{equation}\label{Iijcirc}
	\dI_{ij} = \mu \left\{
	A \, x_{\langle i}x_{j \rangle}
	+B \, \frac{r^2}{c^2}v_{\langle i}v_{j \rangle}
	+ \frac{G^2 m^2\nu}{c^5r}\,C\,x_{\langle i}v_{j \rangle}\right\}
	+ \calO\left(\frac{1}{c^{9}}\right)\,,
\end{equation}
where $\bm{x}=\bm{y}_1-\bm{y}_2=(x_i)$ and $\bm{v}=\bm{v}_1-\bm{v}_2=(v_i)$ are the orbital separation and relative velocity. Introducing the post-Newtonian parameter \eqref{gammadef}, the 4PN coefficients are given by\footnote{To 3PN order the mass quadrupole was computed by \cite{BIJ02, BI04mult, BDEI05dr}; to 2PN order it was computed by \cite{BDI95,WW96,LMRY20}; to 1PN order by \cite{WagW76, BS89}.}
\begin{subequations}\label{Iijrenorm_AB}
	\begin{align}
		A &= 1
		+ \gamma \biggl(- \frac{1}{42}
		-  \frac{13}{14} \nu \biggr)
		+ \gamma^2 \biggl(- \frac{461}{1512}
		-  \frac{18395}{1512} \nu
		-  \frac{241}{1512} \nu^2\biggr)
		\nn\\
		& + \gamma^3 \biggl(\frac{395899}{13200}
		-  \frac{428}{105} \ln\biggl(\frac{r}{r_{0}{}} \biggr)
		+ \biggl[\frac{3304319}{166320}
		-  \frac{44}{3} \ln\biggl(\frac{r}{r'_{0}}\biggr) \biggr]\nu
		\nn\\& \qquad\quad + \frac{162539}{16632} \nu^2 + \frac{2351}{33264} \nu^3
		\biggr)
		\nn\\
		& + \gamma^4 \biggl (- \frac{1067041075909}{12713500800}
		+ \frac{31886}{2205} \ln\biggl(\frac{r}{r_{0}{}} \biggr)\nn\\
		& \qquad\quad
		+ \biggl[-\frac{85244498897}{470870400}
		-  \frac{2783}{1792} \pi^2 -\frac{128}{7}\gamma_\text{E} -\frac{64}{7}\ln(16\gamma)\nn\\& \qquad\qquad\quad -  \frac{10886}{735} \ln\biggl(\frac{r}{r_{0}{}} \biggr)
		+ \frac{8495}{63} \ln\biggl(\frac{r}{r'_{0}} \biggr)\biggr] \nu\nn\\
		& \qquad\quad
		+ \biggl[\frac{171906563}{4484480} + \frac{44909}{2688} \pi^2-  \frac{4897}{21}
		\ln\biggl(\frac{r}{r'_{0}} \biggr)\biggr]\nu^2 \nn\\& \qquad\qquad\quad - \frac{22063949}{5189184} \nu^3 + \frac{71131}{314496} \nu^4
		\biggl)\,, \\
		B &= \frac{11}{21}
		-  \frac{11}{7} \nu
		+ \gamma \biggl(\frac{1607}{378}
		-  \frac{1681}{378} \nu
		+ \frac{229}{378} \nu^2\biggr) \nn\\
		& + \gamma^2 \biggl(- \frac{357761}{19800}
		+ \frac{428}{105} \ln\biggl(\frac{r}{r_{0}{}} \biggr)
		-  \frac{92339}{5544} \nu
		+ \frac{35759}{924} \nu^2
		+ \frac{457}{5544} \nu^3 \biggr)  \nn\\
		& + \gamma^3 \biggl(\frac{23006898527}{1589187600} -  \frac{4922}{2205}
		\ln\biggl(\frac{r}{r_{0}{}} \biggr)\nn\\
		& \qquad\quad + \biggl[\frac{8431514969}{529729200}
		+ \frac{143}{192} \pi^2-\frac{64}{7}\gamma_\text{E} -\frac{32}{7}\ln(16\gamma)
		\nn\\& \qquad\qquad\quad -  \frac{1266}{49} \ln\biggl(\frac{r}{r_{0}{}} \biggr)
		- \frac{968}{63} \ln\biggl(\frac{r}{r'_{0}} \biggr)\biggr] \nu  \nn\\
		& \qquad\quad + 
		\biggl[\frac{351838141}{5045040} 
		-  \frac{41}{24} \pi^2
		+ \frac{968}{21} \ln\biggl(\frac{r}{r'_{0}} \biggr)\biggr] \nu^2
		\nn\\& \qquad\qquad\quad -  \frac{1774615}{81081} \nu^3
		-  \frac{3053}{432432} \nu^4 \biggl)\,, \\
		C &= \frac{48}{7} + \gamma \left(-\frac{4096}{315} - \frac{24512}{945}\nu \right)-\frac{32}{7}\pi\,\gamma^{3/2}\,.
	\end{align}
\end{subequations}
The quadrupole depends on both the scale $r_0$ associated with the Finite Part regularization for the IR, see Eq.~\eqref{regfactor}, and the one $r'_0$ associated with the UV regularization and present in the equations of motion, see Eq.~\eqref{acc3PN}. The coefficients $A$ and $B$ represent the conservative part of the quadrupole moment, while $C$ is due to the radiation reaction dissipative effects. As shown by \cite{LHBF22}, the source quadrupole moment contains a non-local tail integral. The 4PN logarithms $\ln(16\gamma)$ in $A$ and $B$ are due to the conservative part of this non-local tail term, while the last term $\propto\pi$ in $C$ corresponds to the dissipative part of the tail term.

Besides the 4PN mass quadrupole, we need also (for the 4PN flux) the mass octupole moment $\dI_{ijk}$ and current quadrupole moment $\dJ_{ij}$, both of them at the 3PN order. The renormalized versions of these moments coincide at 3PN order with the ones computed with Hadamard regularization for the IR. The octupole mass moment reads \citep{FBI15}
\begin{align}\label{Iijk}
	\dI_{ijk} &= - \mu \Delta\left\{D\,x_{\langle i}x_{j}x_{k\rangle} +
	E\,\frac{r}{c}\,v_{\langle i}x_{j}x_{k\rangle} +
	F\,\frac{r^2}{c^2}\,v_{\langle i}v_{j}x_{k\rangle} +
	G\,\frac{r^3}{c^3}\,v_{\langle i}v_{j}v_{k\rangle} \right\} \nn\\&+ \calO\left(\frac{1}{c^{7}}\right)\,,
\end{align}
where $\Delta = (m_1-m_2)/m$, and the coefficients are
\begin{subequations}\label{Iijk_DEFG}
	\begin{align}
		D &= 1 -\gamma \nu + \gamma^2 \left( - \frac{139}{330} -
		\frac{11923}{660}\nu - \frac{29}{110}\nu^2\right) \nn \\
		& +
		\gamma^3 \left( \frac{1229440}{63063} + \frac{610499}{20020}\nu +
		\frac{319823}{17160} \nu^2 - \frac{101}{2340} \nu^3 \right.\nn\\&\qquad\qquad\left. 
		- \frac{26}{7} \ln
		\Bigl(\frac{r}{r_0}\Bigr) - 22 \nu \ln \Bigl(\frac{r}{r'_{0}}\Bigr)\right)\,,\\
		E &= \frac{196}{15}\gamma^2 \nu \,,\\
		F &= 1 - 2\nu +
		\gamma \left(\frac{1066}{165} - \frac{1433}{330}\nu + \frac{21}{55}\nu^2\right) \nn\\& + \gamma^2 \left( - \frac{1130201}{48510} - \frac{989}{33} \nu + \frac{20359}{330} \nu^2 - \frac{37}{198} \nu^3 + \frac{52}{7} \ln \Bigl(\frac{r}{r_0}\Bigr)\right)\,,\\
		G &= 0\,.
	\end{align}
\end{subequations}
The 3PN current quadrupole moment is \citep{HFB21}
\begin{equation}\label{J2circ}
	\dJ_{ij} = - \mu \Delta\left\{ H \,L^{\langle i} x^{j\rangle} + K \,\frac{G m}{c^3}\,L^{\langle i} v^{j\rangle} \right\} + \calO\left(\frac{1}{c^7}\right)\,,
\end{equation}
where we denote $L^i \equiv \epsilon_{ijk}x^jv^k$, and where
\begin{subequations}\label{J2circ_HK}
	\begin{align}
		H =\ & 1 +\gamma \left(\frac{67}{28}-\frac{2}{7}\nu \right)+\gamma^2\left(\frac{13}{9} -\frac{4651}{252}\nu -\frac{\nu^2}{168} \right) \nn\\
		& + \gamma^3\left(\frac{2301023}{415800} - \frac{214}{105}\ln\left(\frac{r}{r_0}\right) + \left[ - \frac{243853}{9240} + \frac{123}{128}\pi^2 - 22 \ln\left(\frac{r}{r'_0}\right)\right]\nu \right.\nn\\&\qquad\qquad\left. + \frac{44995}{5544}\nu^2 + \frac{599}{16632}\nu^3\right)\,,\\
		K =\ & \frac{188}{35} \nu \,\gamma\,.
\end{align}\end{subequations}

The list of required source moments for the 4PN flux and 3.5PN waveform continues with the 2PN mass $2^4$-pole and current $2^3$-pole (octupole) moments, and so on. These moments are relatively easy to compute as no regularization subtleties arise. Since the first non-local feature appears at 4PN order, these higher-order source moments are instantaneous. In particular, they cannot contribute to the 4.5PN term in the quasi-circular flux. Here we give the most updated moments \citep{BFIS08, H23}\footnote{The STF projection $\langle\rangle$ applies only on ``alive'' indices $ijkl\cdots$ but not on the summed up indices $a$ and $b$.}
\begin{subequations}\label{highermoments}\begin{align}
		\dI_{ijkl} &= \mu\left\{ x_{\langle ijkl\rangle}\left[1 -
		3\nu + \gamma \left(\frac{3}{ 110} - \frac{25}{ 22}\nu + \frac{69}{
			22}\nu^2\right)\right.\right.\nn\\ &\quad
		+\left.\left.\gamma^2\left(-\frac{126901}{200200}-\frac{58101}{2600}\nu
		+\frac{204153}{2860}\nu^2+\frac{1149}{1144}\nu^3\right)
		\right]\right.\nn\\ & +\left.\right. \frac{r}{c}\,x_{\langle
			ijk}v_{l\rangle} \gamma^2\left(\frac{976}{55}\nu-\frac{3104}{55}\nu^2\right)\nn\\ &\left.+\frac{r^2}{ c^2}\,x_{\langle
			ij}v_{kl\rangle}\left[\frac{78}{55} ( 1 - 5\nu + 5\nu^2
		)\right.\right. \nn
		\\ & \quad +\left.\left.\gamma\,\left(\frac{30583}{3575}-\frac{107039}{3575}\nu
		+\frac{8792}{715}\nu^2-\frac{639}{715}\nu^3\right)
		\right]\right. \nn\\&\left.+\frac{71}{715}\,\frac{r^4}{
			c^4}\,v_{\langle
			ijkl\rangle}\left(1-7\nu+14\nu^2-7\nu^3\right)\right\}
		+\calO\left(\frac{1}{c^6}\right)\label{I4}\,,\\
		\dJ_{ijk} &= \mu \left\{ \epsilon_{ab\langle i}
		x_{jk\rangle a} v_b \left[1 - 3\nu + \gamma \left(\frac{181}{ 90} -
		\frac{109}{ 18}\nu + \frac{13}{ 18}\nu^2\right)\right. \right.\nn
		\\ &\quad +\left.\left. \gamma^2\left(\frac{1469}{3960}-\frac{5681}{264}\nu
		+\frac{48403}{660}\nu^2-\frac{559}{3960}\nu^3\right)\right]\right.\nn
		\\ &+\left.\frac{r}{ c}\,\epsilon_{ab\langle i}x_{ja} v_{k\rangle
			b}\gamma^2\left(\frac{472}{45}\nu-\frac{496}{15}\nu^2\right)\right.\nn
		\\ &+\left.\frac{r^2}{ c^2}\,\epsilon_{ab\langle i}x_a v_{jk\rangle
			b}\left[\frac{7}{45}\left(1 - 5\nu +
		5\nu^2\right) \right.\right.\nn\\&\quad\left.\left.+\gamma\left(\frac{1621}{990}-\frac{4879}{990}\nu+\frac{1084}{495}\nu^2
		-\frac{259}{990}\nu^3\right)\right]\right\} +\calO\left(\frac{1}{c^6}\right)\,.\label{J3}\\
		\dI_{ijklm}&=-\mu \Delta \left\{ x_{\langle
			ijklm\rangle}\left[1-2\nu+\gamma\left(\frac{2}{ 39}-\frac{47}{
			39}\nu+\frac{28}{ 13}\nu^2 \right)\right.\right.\nn\\&\left.\left. \qquad +\gamma^2\left(-\frac{716}{819}-\frac{15595}{546}\nu+\frac{32245}{546}\nu^2 +\frac{631}{819}\nu^3 \right) \right]\right.\nn
		\\ &+\left. \frac{r^2 }{ c^2} x_{\langle
			ijk}v_{lm\rangle}\left[\frac{70}{39}\left(1-4\nu+3\nu^2\right) \right.\right.\nn\\&\left.\left. \qquad + \gamma\left(\frac{1238}{117}-\frac{986}{39}\nu+\frac{290}{39}\nu^2 -\frac{70}{117}\nu^3 \right) \right]	\right.\nn\\ &+\left. \frac{85}{273}\frac{r^4}{c^4} \,x_{\langle
			i}v_{jklm\rangle}\bigl(1-6\nu+10\nu^2-4\nu^3\bigr)\right\}
		 +\calO\left(\frac{1}{c^5}\right)\label{I5}\,,\\
		\dJ_{ijkl} &= -\mu \Delta \left\{ \epsilon_{ab\langle
			i}x_{jkl\rangle a} v_b \left[1-2\nu +\gamma \left(\frac{20}{
			11}-\frac{155}{ 44}\nu+\frac{5}{ 11}\nu^2\right) \right.\right.\nn\\&\left.\left. \qquad +\gamma^2 \left(-\frac{1103}{5005}-\frac{141333}{5720}\nu+\frac{69783}{1144}\nu^2 - \frac{83}{572}\nu^3\right)
		\right]\right.\nn\\&~~\left.  + \frac{r^2}{
			c^2}\,\epsilon_{ab\langle i}x_{ja} v_{kl\rangle
			b}\left[\frac{4}{11}\bigl(1-4\nu+3\nu^2\bigr) \right.\right.\nn\\&\left.\left. \qquad + \gamma\left( \frac{2583}{715}-\frac{21887}{2860}\nu+\frac{1411}{572}\nu^2 - \frac{28}{143}\nu^3 \right)\right]\right\}
		+\calO\left(\frac{1}{c^5}\right)\,.\label{J4}\\
		\dI_{ijklmn}&=\mu\left\{ x_{\langle
			ijklmn\rangle}\left[1-5 \nu+5 \nu^2
		+\gamma\,\left(\frac{1}{14}-\frac{3}{2}\nu+6\nu^2
		-\frac{11}{2}\nu^3\right)\right]\right.\nn
		\\ &+\left.\frac{15}{7}\,\frac{r^2}{ c^2}\, x_{\langle
			ijkl}v_{mn\rangle}\left(1-7\nu+14\nu^2-7\nu^3\right) \right\} +
		\calO\left(\frac{1}{c^4}\right)\label{I6}\,,\\
		\dJ_{ijklm} &= \mu \biggl\{ \epsilon_{ab\langle
			i}x_{jklm\rangle a}v_b\left[ 1-5\nu + 5\nu^2 \right.\nn\\&\quad\left. +
		\gamma\left(\frac{1549}{910}-\frac{1081}{130}\nu+\frac{107}{13}\nu^2
		-\frac{29}{26}\nu^3\right)\right] \nn
		\\ & +\frac{54}{91}\,\frac{r^2}{ c^2}\epsilon_{ab\langle
			i}x_{jka}v_{lm\rangle b} \left(1-7\nu+14\nu^2-7\nu^3\right) \biggr\}
		+ \calO\left(\frac{1}{c^4}\right)\,.\label{J5}\\
		\dI_{ijklmno} &= -\mu \Delta \biggl\{x_{\langle
			ijklmno\rangle} \left[1-4 \nu + 3 \nu^2 + \gamma\left( \frac{3}{34}-\frac{26}{17}\nu+\frac{83}{17}\nu^2
		-\frac{57}{17}\nu^3\right)\right] \nn\\&\qquad+ \frac{42}{17}\frac{r^2}{c^2}x_{\langle
			ijklm} v_{no\rangle} \bigl(1-6\nu+10\nu^2-4\nu^3\bigr)\biggr\}+
		\calO\left(\frac{1}{c^3}\right)\label{I7}\,,\\
		\dJ_{ijklmn} &= -\mu \Delta \biggl\{\epsilon_{ab\langle i} x_{jklmn\rangle a}v_{b} \biggl[ 1-4\nu + 3\nu^2 \nn\\&\qquad + \gamma \left(\frac{13}{8}-\frac{25}{4}\nu+\frac{55}{12}\nu^2
		-\frac{2}{3}\nu^3\right)\biggr] \nn\\& + \frac{5}{6}\frac{r^2}{c^2}\epsilon_{ab\langle i} x_{jkla}v_{mn\rangle b}\bigl(1-6\nu+10\nu^2-4\nu^3\bigr)\biggr\} +
		\calO\left(\frac{1}{c^3}\right)\,.\label{J6}
\end{align}\end{subequations}
All the other higher-order moments are required at the Newtonian order (there are no $1/c$ corrections in $\dI_{ijklmnop}$ and $\dJ_{ijklmno}$), at which order they are trivial to compute, with result ($\forall\,\ell\in\mathbb{N}$)
\begin{subequations}\label{IJnewtonian}\begin{align}
		\dI_L &= \mu \sigma_\ell(\nu)\,x_{\langle L\rangle}+
		\calO\left(\frac{1}{c}\right)\label{Il}\,,\\ 
		\dJ_{L-1} &= \mu \sigma_\ell(\nu)\,\epsilon_{ab\langle
			i_{l-1}} x_{L-2\rangle a}v_b+
		\calO\left(\frac{1}{c}\right)\,.\label{Jl}
\end{align}\end{subequations}
Here we use the notation $\sigma_\ell(\nu)\equiv X_2^{\ell-1}+(-)^\ell X_1^{\ell-1}$, with $X_1=\frac{m_1}{m}$ and $X_2=\frac{m_2}{m}$ such that $X_1+X_2=1$, $X_1-X_2=\Delta$ and $X_1 X_2=\nu$. More explicit expressions are ($k\in\mathbb{N}$):
\begin{subequations}\label{sigmaell}\begin{align}
		\sigma_{2k}(\nu)&=\sum_{p=0}^{k-1}(-)^p\frac{2k-1}{2k-1-p}\left({2k-1-p\atop
			p}\right)\nu^p\,,\\\sigma_{2k+1}(\nu)&=-\Delta\sum_{p=0}^{k-1}(-)^p
		\left({2k-1-p\atop p}\right)\nu^p\,,
\end{align}\end{subequations}
where $\left({n\atop p}\right)$ is the usual binomial coefficient.

Evidently the results \eqref{Iijcirc}--\eqref{IJnewtonian} permit the control only of that part of the total energy flux which is generated by the source multipole moments $\dI_L$ and $\dJ_L$; from \eqref{FluxF} we have up to 4.5PN order
\begin{align}\label{flux45PNinst}
	\mathcal{F}\Big|_{\dI_L,\,\dJ_L} &= \frac{G}{c^5} \left\{ \frac{1}{5}
	\bigl(\dI^{(3)}_{ij}\bigr)^2 + \frac{1}{c^2}
	\left[ \frac{1}{189} \bigl(\dI^{(4)}_{ijk}\bigr)^2 +
	\frac{16}{45} \bigl(\dJ^{(3)}_{ij}\bigr)^2\right] \right. \\
	& \qquad
	\left. +
	\frac{1}{c^4} \left[ \frac{1}{9072}
	\bigl(\dI^{(5)}_{ijkl}\bigr)^2 +
	\frac{1}{84} \bigl(\dJ^{(4)}_{ijk}\bigr)^2\right] \right.
	\nn \\
	& \qquad
	\left. + \frac{1}{c^6} \left[ \frac{1}{594000}
	\bigl(\dI^{(6)}_{ijklm}\bigr)^2 + \frac{4}{14175}
	\bigl(\dJ^{(5)}_{ijkl}\bigr)^2\right] \right.
	\nn \\
	& \qquad
	\left. + \frac{1}{c^8} \left[ \frac{1}{52123500}
	\bigl(\dI^{(7)}_{ijklmn}\bigr)^2 + \frac{1}{213840}
	\bigl(\dJ^{(6)}_{ijklm}\bigr)^2\right] +
	\calO\left(\frac{1}{c^{10}}\right) \right\}\,.\nn
\end{align}
The time derivatives of these source moments are computed by means of the quasi circular-orbit equations of motion given by Eq.~\eqref{aieom} together with \eqref{keplerlaw}. Next we need to add to the flux the contributions coming from the relations between radiative and canonical moments in Sect.~\ref{sec:radcanonical}, and between canonical and source ones in Sect.~\ref{sec:cansource}. 

When computing the tails we shall also require, in addition to the above source moments $\dI_L$ and $\dJ_L$, the binary's mass monopole $\dM$ or ADM mass. In a realistic model where the binary system has been formed as a close compact binary at a finite instant in the past, this mass is equal to the sum of the rest masses $m=m_1+m_2$, plus the total (constant) binary's mass-energy $\dE/c^2$ given by Eq.~\eqref{Ecircgam}. To 3PN order the mass reads
\begin{equation}
	\dM = m \left[ 1 - \frac{\nu}{2}\gamma + \frac{\nu}{8}
	\left(7-\nu\right)\gamma^2 + \frac{\nu}{16}
	\bigl(7-49\nu-\nu^2\bigr)\gamma^2 +
	\calO\left(\frac{1}{c^8}\right)\right]\,.
	\label{ADM}
\end{equation}
This is required to compute the leading tail at 4.5PN order (i.e. 3PN relative order). Notice that 3PN order in $\dM$ corresponds to 2PN order in $\dE$.


\subsubsection{Post-adiabatic evaluation of hereditary integrals}
\label{sec:PA}

The most important problem we face is the computation of the \emph{hereditary} tail and related non-linear integrals occuring at the 4PN order, for instance the dominant 1.5PN quadrupole tail \eqref{Uij15PN}. To compute the tail integrals, one needs to specify the orbit's behaviour of the compact binary system in the remote past, as the effect is not localized in time, but integrates over the whole past history of the source. In the case of quasi-circular orbits, and up to 3.5PN precision in the moments, an adiabatic approximation (considering the orbital elements $r$ and $\Omega$ to be constant in time) is sufficient, and the integrals can be computed following the method of \cite{BS93, ABIQ04}. However, Eqs. \eqref{rOmdot} show that this adiabatic approximation is no longer valid at the relative 2.5PN precision. As the tail enters at 1.5PN order in the moments, the first ``post-adiabatic'' correction will affect the moments at the 4PN order, thus we need to properly evaluate it in order to consistently derive the 4PN flux and $(2,2)$ GW mode.

As shown by \cite{ABIQ04}, the tail integrals on quasi-circular orbits reduce to elementary integrals of the type
\begin{align}\label{Ian}
	\mathcal{I}_{a,n}(U) = \int_0^{+\infty} \dd\tau \, 
	\bigl[x(U-\tau)\bigr]^a \,\de^{-\di n \,\phi(U-\tau)} \ln \left(\frac{\tau}{\tau_0}\right) \,,
\end{align}
where $U$ is the current time, $U-\tau$ is any time in the past like in \eqref{Uij15PN}, and the orbital phase $\phi$, the orbital frequency $\Omega=\dot{\phi}$ and the PN parameter $x$ defined by \eqref{xdef} are integrated over the binary's past evolution. In \eqref{Ian}, $n$ is a non-zero integer (the case $n=0$ does not appear at 4PN order), $a$ can be any power ($a=5$ for the leading tail), and the constant $\tau_0$ denotes either $2r_0/c$ or $2b_0/c$, see for instance Eq.~\eqref{Uij4PN}. The strategy is to notice that the integral \eqref{Ian} involves a fast oscillating exponential, which is derivable in the form of an asymptotic series which is nothing but the post-adiabatic approximation. In the application to 4PN order we need only the first post-adiabatic correction, relevant to the dominant 1.5PN tail term \eqref{Uij15PN}; the other tail integrals like those appearing in the tails-of-memory \eqref{Uij4PN} can be computed in the adiabatic limit.

Remark first that the difference between the current phase $\phi(U)$ and the phase $\phi(U-\tau)$ in the past is of the order of the inverse of the radiation reaction scale, \emph{i.e.}, $\phi(U) - \phi(U-\tau)$ scales as $\calO(c^5)$. To describe the radiation-reaction scale, we introduce a dimensionless adiabatic parameter at current time $U$, 
\begin{align}
	\xi(U) \equiv \frac{\dot{\Omega}(U)}{\Omega^2(U)} = \calO\left(\frac{1}{c^5}\right)\,,
\end{align}	
and compute the integral \eqref{Ian} in the post-adiabatic limit where $\xi(U)\to 0$. To this end, let us pose
\begin{align}\label{lambdadef}
	\phi(U) - \phi(U-\tau) = \frac{\lambda}{\xi(U)}\,,
\end{align}
and change the integration variable from $\tau$ to $\lambda$, so that 
\begin{align}\label{Ian2}
	\mathcal{I}_{a,n} =  \frac{Gm}{c^3}\,\frac{\de^{-\di n \phi(U)}}{\xi(U)}\,\int_0^{+\infty} \dd \lambda\, 
	\bigl[x(U-\tau(\lambda))\bigr]^{a-3/2}  \,\ln \Bigl(\frac{\tau(\lambda)}{\tau_0}\Bigr) \,\de^{\di n \lambda/\xi(U)}\,.
\end{align}
Here $\tau(\lambda)$ is obtained by inverting Eq.~\eqref{lambdadef} for $\tau$ as a function of $\lambda$. The main point is that, in the limit $\xi(U)\to 0$, since the imaginary exponential in the integrand of \eqref{Ian2} oscillates very rapidly, the integral is a sum of alternatively positive and negative contributions which essentially sum up to zero. However there are no oscillations when $\lambda=0$, therefore the integral is essentially given by the contribution in the neighbourhood of the bound at $\lambda=0$. In other words we are entitled to compute the integral by replacing the integrand by its formal expansion series when $\lambda\to 0$, which gives a formal (asymptotic) series in powers of $\xi(U)$. In the limit $\tau\to 0$ we can expand Eq.~\eqref{lambdadef} as
\begin{align}\label{lambdadefsol}
	\lambda = \xi \sum_{m=0}^{+\infty} \frac{(-)^m}{(m+1)!} \,\Omega^{(m)} \tau^{n+1}\,,
\end{align}
where we now pose $\xi=\xi(U)$ and $\Omega^{(m)}=(\dd^m\Omega/\dd U^m) (U)$ for the quantities at the \emph{current} time $U$. Since $\Omega^{(1)}\equiv\dot{\Omega}=\xi \Omega^2$ the expansion \eqref{lambdadefsol} clearly represents the post-adiabatic approximation in powers of $\xi$ and arbitrary high time derivatives of $\xi$ denoted $\xi^{(m)}$. Note that each time derivative of $\xi$ adds a radiation reaction scale, hence we have $\xi^{(m)} = \calO(\xi^{m+1}) = \calO(c^{-5m-5})$. To obtain the expansion of \eqref{Ian2} we need to invert the series \eqref{lambdadefsol} and obtain $\tau$ as a power series in $\lambda$. This is given by the Lagrange inversion theorem as
\begin{align}
	\tau = \sum_{p=0}^{+\infty} \frac{1}{(p+1)!}\,f_p\left(\frac{\lambda}{\xi}\right)^{p+1}\,,
\end{align}
where the general coefficients $f_p$ read
\begin{align}\label{taudefsol}
	f_p &= \left(\frac{\dd}{\dd\tau}\right)^p\biggl[\left(\frac{\tau}{\phi(U)-\phi(U-\tau)}\right)^{p+1}\biggr]_{\tau=0} \nn\\& = \left(\frac{\dd}{\dd\tau}\right)^p\biggl[\biggl(
	\sum_{m=0}^{+\infty} \frac{(-)^m}{(m+1)!} \,\Omega^{(m)} \tau^{m}
	\biggr)^{\!-p-1\,}\biggr]_{\tau=0}\,,
\end{align}
where $\tau=0$ is applied after the $p$ differentiations with respect to $\tau$. For instance, up to second order we obtain
\begin{align}\label{tau2PA}
	\tau = \frac{\lambda}{\xi\Omega}\biggl[1 + \frac{\lambda}{2} + \biggl(1 - \frac{\dot{\xi}}{\xi^2 \Omega}\biggr)\frac{\lambda^2}{6} + \calO\left(\lambda^3\right)\biggr]\,.
\end{align}

The previous formulas show that the method can be straightforwardly extended to any order. But to compute the tail term at 4PN order, we need only the relative correction of order 2.5PN, which corresponds to the first post-adiabatic approximation, hence just $\tau = \frac{\lambda}{\xi\,\Omega}[1 + \frac{\lambda}{2} + \calO(\lambda^2)]$. To this order the integral \eqref{Ian2} becomes \citep{BFHLT23b}
\begin{align}\label{Ian3}
	\mathcal{I}_{a,n} &=  \frac{Gm}{c^3}\,\frac{x^{a-3/2}\,\de^{-\di n \phi}}{\xi}\\&\qquad\quad\times\int_0^{+\infty} \dd \lambda\, 
	\biggl[ \biggl( 1+ \Bigl(1-\frac{2a}{3}\Bigr)\lambda \biggr)\ln\left(\frac{\lambda}{\xi\Omega\tau_0}\right) + \frac{\lambda}{2} \biggr] \,\de^{\di n \lambda/\xi}  \,,\nn
\end{align}
modulo $\calO(\lambda^2)$ terms. The remaining integral is computed as follows. We transform the complex exponential into a real one by performing the change of variable $\lambda=\di\, \sigma$ for $n> 0$, and $\lambda=-\di\, \sigma$ for $n< 0$. The integration now takes place along the imaginary axis, which can be remedied by resorting to the Cauchy theorem on the closed contour made of the three following pieces, to be considered in the limit $R\to +\infty$: (i) the path from $\sigma=0$ to $\sigma=R$ on the real axis, (ii)~the oriented quarter of circle of radius $|\sigma|=R$ from $\arg \sigma = 0$ to $\arg \sigma=-\pi/2$ (or $\arg \sigma=\pi/2$ if $n < 0$) and (iii)~the segment of the imaginary axis going from $\sigma=-\di R$ ($\sigma=\di R$ if $n<0$) to $\sigma=0$. This leads to
\begin{align}\label{Ian4}
	\mathcal{I}_{a,n} &=  \frac{Gm}{c^3}\,\frac{x^{a-3/2}\,\de^{-\di n \phi}}{\di\,\xi}\\&\times\int_0^{+\infty} \dd \sigma\, 
	\Biggl[ \biggl( - \text{s}(n)+ \frac{2a-3}{3}\,\di\,\sigma \biggr)\left(\ln\left(\frac{\sigma}{\xi\Omega\tau_0}\right) +\di \frac{\pi}{2} \,\text{s}(n)\right) + \frac{\di \,\sigma}{2}\Biggr] \,\de^{-\vert n\vert \sigma/\xi}  \,,\nn
\end{align}
modulo $\calO(\sigma^2)$ terms, where $\text{s}(n)$ is the sign function. In this form, $\mathcal{I}_{a,n}$ can be integrated, as it boils down to elementary integrals. At the first post-adiabatic order the result is
\begin{align}
	\mathcal{I}_{a,n} &= \frac{Gm}{c^3}\,\frac{x^{a-3/2}\,\de^{-\di n \phi}}{\di \,n}\biggl\{ \bigg( 1 + \frac{2a-3}{3}\,\frac{\xi}{\di \,n}\bigg)\bigg[ \ln\bigl(\vert n \vert \Omega \tau_0\bigr) + \gamma_\text{E} - \di \,\frac{\pi}{2} \,\text{s}(n) \bigg] \nn\\& \qquad\qquad\qquad\qquad\qquad - \frac{4a-9}{6}\,\frac{\xi}{\di\,n} + \calO\left(\xi^2\right)\biggr\}\,.
\end{align}
Recall that all quantities are evaluated at current time $U$ and that the adiabatic parameter $\xi = \dot{\Omega}/\Omega^2$ is easily computed with Eqs. \eqref{rOmdot}. With this result at hand, we are able to derive the tail integral \eqref{Uij15PN} with 2.5PN relative precision, which impacts the computation and final results at the 4PN order.

For completeness we give the formulas needed to handle any approximation in the post-adiabatic expansion for tail like integrals:
\begin{subequations}\label{formulegen}
	\begin{align}
		\int_0^{+\infty} \dd \sigma\, \sigma^j \,\de^{-\vert n\vert \sigma/\xi}
		&
		= j! \left(\frac{\xi}{\vert n \vert}\right)^{j+1}\,, \\ 
		\int_0^{+\infty} \dd \sigma\, \sigma^j \, \ln\left(\frac{\sigma}{\xi\Omega\tau_0}\right) \,\de^{-\vert n\vert \sigma/\xi}	&
		= j! \left(\frac{\xi}{\vert n \vert}\right)^{j+1} \Big[ H_j - \gamma_\text{E} - \ln \bigl(\vert n \vert \Omega \tau_0\bigr) \Big]\,,
	\end{align}
\end{subequations}
where $H_j=\sum_{k=1}^j \frac{1}{k}$ denotes the harmonic number. For instance the case $j=1$ yields the standard integral for adiabatic tails (coming back to $\lambda=\pm\di\sigma$)
\begin{align}
	\int^{+\infty}_0 \dd\lambda\,\ln
	\!\left(\frac{\lambda}{\xi\Omega\tau_0}\right)\, \de^{-\di n\lambda/\xi} &=
	\frac{\di\,\xi}{n} \left(\ln\bigl(\vert n\vert\Omega\tau_0\bigr)
	+\gamma_\text{E}+\di\frac{\pi}{2}\,\text{s}(n)\right)\,.
\end{align}
For iterated tails, i.e. tails-of-tails at 3PN order and tails-of-tails-of-tails at 4.5PN order (see Sect.~\ref{sec:radcanonical}), the adiabatic approximation is sufficient and the required formulas are available from \cite{GR}:
\begin{subequations}\label{intlog23}
	\begin{align}
		\int^{+\infty}_0
		\dd\lambda\,\ln^2\!\left(\frac{\lambda}{\xi\Omega\tau_0}\right) \,
		\de^{-\di n\lambda/\xi} &= - \frac{\di\,\xi}{n} \left[
		\left(\ln\bigl(\vert n\vert\Omega\tau_0\bigr)
		+\gamma_\text{E}+\di\frac{\pi}{2}\,\text{s}(n)\right)^2 +
		\frac{\pi^2}{6}\right]\,,\\
		 \int^{+\infty}_0
		\dd\lambda\,\ln^3\!\left(\frac{\lambda}{\xi\Omega\tau_0}\right) \,
		\de^{-\di n\lambda/\xi} &= \frac{\di\,\xi}{n} \left[
		\left(\ln\bigl(\vert n\vert\Omega\tau_0\bigr)
		+\gamma_\text{E}+\di\frac{\pi}{2}\,\text{s}(n)\right)^3 \right.\label{intlog3}\\&~~\left. +
		\frac{\pi^2}{2}\left(\ln\bigl(\vert n\vert\Omega\tau_0\bigr)
		+\gamma_\text{E}+\di\frac{\pi}{2}\,\text{s}(n)\right)
		+2\zeta(3)\right]\,.\nn
	\end{align}
\end{subequations}
Note the presence in the latter formula of the Ap\'ery constant $\zeta(3)$, where $\zeta$ is the Riemann function, but this constant disappears from final results.

Next we deal with hereditary memory type terms, such as the first term of Eq.~\eqref{Uij25PN} -- leading memory at 2.5PN order -- or the first term of Eq.~\eqref{Uij4PN} -- tail-of-memory at 4PN order. As was shown in Sect.~\ref{sec:memory} the memory terms enter only in the mass-type radiative moments $\dU_L$. In the case of quasi-circular orbits, they reduce to a sum of terms of the form
\begin{align}\label{mem_Jan}
	\mathcal{J}_{a,n}(U) = \int_0^{+\infty} \dd\tau \, 
	\bigl[x(U-\tau)\bigr]^a \,\de^{-\di n \,\phi(U-\tau)} \,.
\end{align}
Besides the absence of the logarithm, the main difference with the tail integral \eqref{Ian} is the possibility of having $n = 0$, which corresponds to persistent or direct-current (``DC'') terms, in addition to oscillatory alternative-current (``AC'') terms having $n \neq 0$. As we remind below, the memory effect properly speaking is associated with the DC terms.

We focus first on AC terms, having $n \neq 0$. As they only involve a simple, logarithmic-free, integration, such terms will enter in the flux as instantaneous contributions. In particular, since they arise at the half-integer approximations 2.5PN and 3.5PN, they do not contribute to the flux for quasi-circular orbits. The evaluation of the integrals \eqref{mem_Jan} is thus only required for the derivation of the modes. Concerning the quadrupole moment, since the memory effect enters at 2.5PN order [see Eq.~\eqref{Uij25PN}], it is thus required at relative 1.5PN order for the mode $(2,2)$ and thus, one can safely compute it in the adiabatic approximation. As proved by \cite{ABIQ04}, for circular orbits this is equivalent to taking the frequency $x$ constant together with a linear phase (appropriate for the exact circular orbit), and keeping only the contribution of the integral due to the bound $\tau=0$. One finds
\begin{equation}\label{DCmem}
	\mathcal{J}_{a,n}  =  x^a \,\de^{-\di n \,\phi} \int_0 \dd\tau \,\de^{\di n \Omega \tau} + \calO\left(\xi\right) = - \frac{x^a \,\de^{-\di n \phi}}{\di n \Omega} + \calO\left(\xi\right) \,.
\end{equation}

The DC terms obviously do not contribute to the flux, and they only contribute to the modes which have $m=0$, for example $(\ell,m) = (2,0)$ for the mass quadrupole. The absence of the fast oscillating exponential in $\mathcal{J}_{a,0}$ -- in contrast to \eqref{Ian} -- generates an inverse power of the adiabatic parameter $\xi$, thus degrading the precision by the radiation-reaction scale 2.5PN. This is the memory effect: as it starts at 2.5PN order in the waveform, it finally enters the $(2,0)$ mode at Newtonian order. Several methods are possible to evaluate $\mathcal{J}_{a,0}$ \citep{WW91,ABIQ04,F09}. Here we rely on the change of integration variables from $\tau$ to $x'=x(U-\tau)$: 
\begin{align}\label{Ja0}
	\mathcal{J}_{a,0} 
	= \int_0^{+\infty} \dd\tau \, 
	\bigl[x(U-\tau)\bigr]^a   
	= \int_0^{x(U)} \dd x'\, \frac{x'^a}{\dot{x}(U-\tau)}\,,
\end{align}
reading $\dot{x}(U-\tau)$ as a function of $x'$ from Eqs. \eqref{rOmdot}, and supposing that $a > 4$, as is the case in practical computations. To 1.5PN relative order, we have
\begin{align}\label{Ja0res}
	\mathcal{J}_{a,0} 
	= \frac{5 G m}{64c^3\nu}\,\frac{x^{a-4}}{a-4}\left[1 + \frac{a-4}{a-3} \left(\frac{743}{336}+\frac{11}{4}\nu\right)x -8\pi \frac{a-4}{2a-5} x^{3/2} + \calO\left(x^2\right)\right] \,.
\end{align}
It is interesting to observe that the tail effect in the flux directly influences the DC memory in \eqref{Ja0res}. Finally we look at the interesting case of the tails-of-memory; strictly speaking this is the term in the first line of \eqref{Uij4PN}, namely
\begin{equation}\label{ToMKij}
	\mathcal{K}_{ij} = \frac{8G^2\dM}{7c^8}\int_0^{+\infty} \!\dd\rho\,  \dM_{a \langle i}^{(4)}(u-\rho) \int_0^{+\infty} \!\dd \tau\,  \dM_{j \rangle a}^{(4)}(u-\rho-\tau)  \ln\left(\frac{\tau}{\tau_0}\right)\,, 
\end{equation}
where $c \,\tau_0 = 2 r_0\,\de^\frac{1613}{270}$. This expression \citep{TB23} agrees with the tail-of-memory directly computed from the radiative quadrupole moment at infinity \citep{F09,F11}, see Sect.~\ref{sec:memory}. We first perform the tail-like integral over $\tau$; as we need to evaluate it at relative Newtonian order only, we can safely use the adiabatic approximation. Next we perform the integral over $\tau$, which is found to be a simple DC memory integral of the type \eqref{Ja0} with $a=13/2$. Hence we find
\begin{equation}\label{ToMKijres}
	\mathcal{K}_{ij} =
	\frac{128\pi}{7}\frac{\nu^2 c^5}{G } \,\ell_{\langle i}\ell_{j \rangle} \int_0^{+\infty} \!\dd\rho\,\bigl[x(U-\rho)\bigr]^{13/2}
	= \frac{4\pi}{7} m \nu \,c^2\,x^{5/2} \,\ell_{\langle i}\ell_{j \rangle}\,,
\end{equation}
where $\ell^i=L^i/|\bm{L}|$ is the constant unit direction of the Newtonian angular momentum, orthogonal to the orbital plane. Interestingly, the tail-of-memory effect \eqref{ToMKijres} gives a contribution to the $(2,0)$ mode which exactly cancels the 1.5PN correction to the ordinary memory effect, i.e. the term $\propto\pi$ in \eqref{Ja0res}. Hence there is no tail term at 1.5PN order in Eq.~\eqref{Hmem20} below.


\subsubsection{The 4.5PN energy flux for circular orbits}
\label{sec:gravflux}

The flux \eqref{flux45PNinst} due to source multipole moments at 4PN order can now be completed by all contributions coming from the non-linear interactions between the moments. The flux for quasi-circular orbits will finally be expressed as a function of the frequency of gravitational waves. However we have to take into account an important effect, which is that the GW frequency is modified with respect to the orbital frequency because of the  propagation of tails in the wave zone. Indeed the phase $\psi$ of the gravitational wave [actually half the phase of the leading (2,2) mode], differs from the orbital phase $\phi$ by a logarithmic, tail-induced phase modulation, as
\begin{equation}\label{phasemod}
	\psi = \phi - \frac{2 G \dM \Omega}{c^3} \ln\biggl(\frac{\Omega}{\Omega_0}\biggr)\,,
\end{equation}
where $\dM$ is the constant ADM mass (i.e. $\dM=m+\dE/c^2$), and $\Omega_0$ is related to the length scale $b_0$ parametrizing the gauge transformation between radiative and harmonic coordinates, see Eq.~\eqref{Uu}, by $c\,\Omega_0^{-1} = 4b_0 \,\de^{\gamma_\text{E}-11/12}$. The phase modulation \eqref{phasemod} was determined by \cite{Wi93, BS93}, and argued by \cite{BIWW96, ABIQ04} to affect the waveform and flux at the 4PN order. Indeed, this phase modulation is responsible for a shift of the GW frequency, i.e. $\Omega_\text{GW} = \dd\psi/\dd t$, with respect to the orbital one, i.e. $\Omega = \dd\phi/\dd t$. Therefore the GW frequency, directly measurable from the waveform at infinity, reads
\begin{equation}
	\Omega_\text{GW} = \Omega - \frac{2 G \dM \dot{\Omega}}{c^3}\left[ \ln\biggl(\frac{\Omega}{\Omega_0}\biggr) + 1\right]\,.
\end{equation}
Using Eqs. \eqref{rOmdot} for the frequency chirp at the dominant order, and replacing $\dM$ by $m$ at that order, yields
\begin{equation}\label{OmOmGW}
	\Omega_\text{GW} = \Omega\,\bigg\{ 1 - \frac{192}{5} \,\nu x^4 \bigg[\ln\biggl(\frac{\Omega}{\Omega_0}\biggl) + 1 \bigg]  + \calO\left(\frac{1}{c^{10}}\right)\bigg\}\,,
\end{equation}
showing that this is a 4PN effect (i.e. $\propto x^4$), hence the fact that it did not affect previous computations of the phase \citep{BIWW96,ABIQ04}. Therefore, once the flux has been obtained at 4PN order in terms of $x=(\frac{Gm\Omega}{c^3})^{2/3}$, we must reexpress it in terms of the measurable GW frequency using the inverse of \eqref{OmOmGW}, i.e.
\begin{equation}\label{xxGW}
	x = x_\text{GW}\,\biggl\{ 1 + \frac{192}{5} \,\nu \, x_\text{GW}^4 \biggl[\ln\biggl(\frac{x_\text{GW}}{x_0}\biggr) + \frac{2}{3} \biggr]  + \calO\left(\frac{1}{c^{10}}\right)  \biggr\}\,,
\end{equation}
where $x_\text{GW}=(\frac{Gm\Omega_\text{GW}}{c^3})^{2/3}$ and we pose $x_0=(\frac{Gm\Omega_0}{c^3})^{2/3}$. With this procedure, once the flux is entirely expressed in terms of $x_\text{GW}$ (after required reexpansion at the 4PN order), we find that the constant~$b_0$ cancels out as expected. 

At this stage, it is appropriate to rename $x_\text{GW}$ as simply ``$x$'' in order to be consistent with the calculation of the binary's equations of motion, and notably the binary's energy $\dE$ for circular orbits. Indeed the energy was expressed in Eq.~\eqref{Ecirc} in terms of the orbital frequency $x$, which represents exactly the quantity which would be measured far away from the binary detector\footnote{For instance by analyzing EM radio pulses sent out to infinity by an observer equipped with a flashlight and sitting on one of the binary's particles \citep{Det05}.} and actually agrees with the measurable GW frequency $x_\text{GW}$ obtained in \eqref{OmOmGW}. In particular we shall consistently use the balance equation \eqref{baleq}, where both $\dE(x)$ and $\mathcal{F}(x)$ are expressed in terms of the same frequency $x$, which is the orbital frequency measured far away in the left-hand side, and the GW frequency in the right-hand side. 

At long last, adding all pieces together, i.e. Eq.~\eqref{flux45PNinst} plus contributions from tails, iterated tails and tails-of-memory in Sect.~\ref{sec:radcanonical}, we obtain the quasi-circular 4.5PN energy flux as \citep{BFHLT23a, BFHLT23b} -- remind our rename of $x_\text{GW}$ as $x$:
\begin{align}\label{Fluxx}
	\mathcal{F} &= \frac{32c^5}{5G}\nu^2 x^5 \Biggl\{
	1 
	+ \biggl(-\frac{1247}{336} - \frac{35}{12}\nu \biggr) x 
	+ 4\pi x^{3/2}
	\nn\\
	&\quad 
	+ \biggl(-\frac{44711}{9072} +\frac{9271}{504}\nu + \frac{65}{18} \nu^2\biggr) x^2 
	+ \biggl(-\frac{8191}{672}-\frac{583}{24}\nu\biggr)\pi x^{5/2}
	\nn\\
	&\quad 
	+ \Biggl[\frac{6643739519}{69854400}+ \frac{16}{3}\pi^2-\frac{1712}{105}\gamma_\text{E} - \frac{856}{105} \ln (16\,x) 
	+ \biggl(-\frac{134543}{7776} + \frac{41}{48}\pi^2 \biggr)\nu 
	\nn\\
	&\quad \qquad - \frac{94403}{3024}\nu^2 
	- \frac{775}{324}\nu^3 \Biggr] x^3 
	\nn\\
	&\quad
	+ \biggl(-\frac{16285}{504} + \frac{214745}{1728}\nu +\frac{193385}{3024}\nu^2\biggr)\pi x^{7/2} 
	\nn\\
	&\quad
	+ \Biggl[ -\frac{323105549467}{3178375200} + \frac{232597}{4410}\gamma_\text{E} - \frac{1369}{126} \pi^2 + \frac{39931}{294}\ln 2 - \frac{47385}{1568}\ln 3 
	\nn\\
	&\quad \qquad + \frac{232597}{8820}\ln x   
	\nn\\
	&\quad \qquad
	+ \biggl( -\frac{1452202403629}{1466942400} + \frac{41478}{245}\gamma_\text{E} - \frac{267127}{4608}\pi^2 + \frac{479062}{2205}\ln 2 + \frac{47385}{392}\ln 3  
	\nn\\
	&\quad \qquad\qquad + \frac{20739}{245}\ln x \biggr)\nu
	\nn\\
	&\quad \qquad
	+ \biggl( \frac{1607125}{6804} - \frac{3157}{384}\pi^2 \biggr)\nu^2 + \frac{6875}{504}\nu^3 + \frac{5}{6}\nu^4 \Biggr] x^4
	\nn\\ 
	&\quad 
	+ \Biggl[ \frac{265978667519}{745113600} - \frac{6848}{105}\gamma_\text{E} - \frac{3424}{105} \ln (16 \,x)
	+ \biggl( \frac{2062241}{22176} + \frac{41}{12}\pi^2 \biggr)\nu
	\nn\\ 
	&\quad \qquad
	- \frac{133112905}{290304}\nu^2 - \frac{3719141}{38016}\nu^3 \Biggr] \pi x^{9/2} 
	+ \calO\left(\frac{1}{c^{10}}\right)\Biggr\}\,.
\end{align}
A technical check is that all arbitrary constants ($r_0$, $r'_0$ and $b_0$) which played a role in intermediate steps of the calculation finally cancel out. An important physical check is that the test-mass limit $\nu\to 0$ perfectly agrees with the result of linear black-hole perturbation theory by \cite{TNaka94, Sasa94, TSasa94, TTS96, Fuj14PN, Fuj22PN}. Since BH perturbations have recently been extended numerically to second order in the mass ratio $\nu$ \citep{WPWMDL23, ANP22a, ANP22b}, it would be interesting to verify the consistency of this numerical result with the PN prediction \eqref{Fluxx}. Note also that in the case of black holes, the contributions due to the absorption by the BH horizons are not included in the PN calculation, and should be added separately. The BH absorption is a 4PN effect for Schwarzschild black holes \citep{PS95}, and a 2.5PN effect for spinning ones \citep{TMT97, Alvi01, Porto08, Chatz12, Saketh22}. In the latter case, the flux has to be augmented by direct spin contributions, see Sect.~\ref{sec:fluxSO}. Note also that the flux has been confirmed by different groups up to 2PN order \citep{WW96, LMRY20}, and that the 4.5PN piece is in agreement with independent work by \cite{MNagar17}. 


\subsubsection{The 4.5PN phase and frequency evolution}
\label{sec:orbevol}

We shall now deduce the laws of variation with time of the orbital frequency and phase of an inspiralling compact binary from the energy balance equation \eqref{baleq}. The center-of-mass energy $\dE$ is given by Eq.~\eqref{Ecirc} and the total flux $\mathcal{F}$ by Eq.~\eqref{Fluxx}. For convenience we adopt the dimensionless time
variable
\begin{equation}\label{tauPN}
\tau \equiv \frac{\nu c^3}{5Gm}\bigl(t_0-t\bigr)\,,
\end{equation}
where $t$ is the coordinate time in the asymptotic radiative coordinate system [formerly denoted $U$, see Eq.~\eqref{hijTT}], and $t_0$ an integration constant. We transform the balance equation into an ordinary differential equation for the parameter $x$, which is immediately integrated. Then we use the freedom to redefine the integration constant $t_0$ as $t_0 + \zeta \,\frac{G m}{c^3}$ where $\zeta$ is any constant, which amounts to the replacement of $\tau$ by $\tau [1 + \frac{\zeta \nu}{5\tau}]$. Although $t_0$ is not uniquely defined, it might be formally interpreted as the instant of coalescence, when $x\to+\infty$, and then it satisfies $t_0-t= \calO(c^5)$ in the PN regime. Thus we see that $\tau^{-1} = \calO(c^{-8})$ is a small 4PN quantity. Using this fact, we conveniently adjust $\zeta$ so as to simplify as much as possible the result:
\begin{align}\label{xtau}
	x &= \frac{\tau^{-1/4}}{4}\Biggl\{ 1 + \biggl( \frac{743}{4032} +
	\frac{11}{48}\nu\biggr)\tau^{-1/4} - \frac{1}{5}\pi\,\tau^{-3/8}
	\nn \\ & \quad + \biggl( \frac{19583}{254016} +
	\frac{24401}{193536} \nu + \frac{31}{288} \nu^2 \biggr)
	\tau^{-1/2} + \biggl(-\frac{11891}{53760} +
	\frac{109}{1920}\nu\biggr) \pi \,\tau^{-5/8} 
	\nn \\ & 
	\quad + \Biggl[-\frac{10052469856691}{6008596070400} +
	\frac{1}{6}\pi^2 + \frac{107}{420}\gamma_\text{E} -
	\frac{107}{3360} \ln\biggl(\frac{\tau}{256}\biggr)
	\nn \\ &  \quad \qquad +
	\biggl(\frac{3147553127}{780337152} - \frac{451}{3072}\pi^2 \biggr)
	\nu - \frac{15211}{442368}\nu^2 +
	\frac{25565}{331776}\nu^3\Biggr] \tau^{-3/4} 
	\nn \\ &
	\quad + \biggl(-\frac{113868647}{433520640} -
	\frac{31821}{143360}\nu + \frac{294941}{3870720}\nu^2\biggr) \pi
	\tau^{-7/8} 
	\nn \\ & 
	\quad + \Biggl[-\frac{2518977598355703073}{3779358859513036800} + \frac{9203}{215040}\gamma_\text{E} + \frac{9049}{258048}\pi^2 \nn\\ & \quad \qquad + \frac{14873}{1128960}\ln 2	+ \frac{47385}{1605632}\ln 3 - \frac{9203}{3440640}\ln\tau
	\nn \\ & \quad \qquad +
	\Biggl(  \frac{718143266031997}{576825222758400} + \frac{244493}{1128960}\gamma_\text{E} - \frac{65577}{1835008}\pi^2
	 \nn\\ & \quad \quad \qquad + \frac{15761}{47040}\ln 2 - \frac{47385}{401408}\ln 3 - \frac{244493}{18063360}\ln\tau\Biggr) \nu 
	\nn \\ & \quad \qquad + \Biggl(  - \frac{1502014727}{8323596288} + \frac{2255}{393216}\pi^2 \Biggr)\nu^2 - \frac{258479}{33030144} \nu^3 + \frac{1195}{262144} \nu^4 \Biggr] \tau^{-1}\ln\tau
	\nn \\ & 
	\quad + \Biggl[-\frac{9965202491753717}{5768252227584000} + \frac{107}{600}\gamma_\text{E} + \frac{23}{600}\pi^2 - \frac{107}{4800}\ln\biggl(\frac{\tau}{256}\biggr) 
	\nn \\ & \quad \qquad +
	\biggl( \frac{8248609881163}{2746786775040} - \frac{3157}{30720}\pi^2 \biggr)
	\nu - \frac{3590973803}{20808990720}\nu^2 
	\nn \\ & \quad \qquad -
	\frac{520159}{1634992128}\nu^3 \Biggr] \pi \,\tau^{-9/8} + \calO\left(\frac{1}{c^{10}}\right)\Biggr\}\,.
\end{align}
We have $\dot{\psi}=\Omega$, where we recall that $\Omega\equiv\Omega_\text{GW}$ is the wave frequency, from which we determine the explicit expression of the time-domain GW half-phase $\psi$ in terms of $x=(Gm\Omega/c^3)^{2/3}$ as \citep{BFHLT23a, BFHLT23b}\footnote{This procedure for computing analytically the orbital phase corresponds to what is called the ``Taylor T2 approximant''. We refer to \cite{Boyle08, BIO09} for the definition and usefulness of several types of approximants for computing (in general numerically) the orbital phase.}
\begin{align}\label{phix}
	\psi &= \psi_0 - \frac{x^{-5/2}}{32\nu}\Biggl\{ 1 + \biggl(
	\frac{3715}{1008} + \frac{55}{12}\nu\biggr)x - 10\pi x^{3/2}
	\nn \\ & \quad + \biggl( \frac{15293365}{1016064} +
	\frac{27145}{1008} \nu + \frac{3085}{144} \nu^2 \biggr) x^2 +
	\biggl(\frac{38645}{1344} - \frac{65}{16}\nu\biggr) \pi x^{5/2} \ln x
	\nn \\ & \quad +
	\Biggl[\frac{12348611926451}{18776862720} - \frac{160}{3}\pi^2 -
	\frac{1712}{21}\gamma_\text{E} - \frac{856}{21} \ln (16\,x)
	\nn \\ & \quad \qquad +
	\biggl(-\frac{15737765635}{12192768} + \frac{2255}{48}\pi^2
	\biggr)\nu + \frac{76055}{6912}\nu^2 -
	\frac{127825}{5184}\nu^3\Biggr] x^3 
	\nn \\ & \quad +
	\biggl(\frac{77096675}{2032128} + \frac{378515}{12096}\nu -
	\frac{74045}{6048}\nu^2\biggr) \pi x^{7/2}
	\nn \\ & \quad +
	\Biggl[ \frac{2550713843998885153}{2214468081745920} - \frac{9203}{126}\gamma_\text{E} - \frac{45245}{756}\pi^2 - \frac{252755}{2646}\ln 2  
	\nn\\ & \quad \qquad - \frac{78975}{1568}\ln 3 - \frac{9203}{252}\ln x 
	\nn \\ & \quad \qquad +
	\biggl(-\frac{680712846248317}{337983528960} - \frac{488986}{1323}\gamma_\text{E} + \frac{109295}{1792}\pi^2 - \frac{1245514}{1323}\ln 2 
	\nn \\ & \quad \qquad \qquad + \frac{78975}{392}\ln 3 - \frac{244493}{1323}\ln x \biggr)\nu 
	\nn \\ & \quad \qquad +
	\biggl( \frac{7510073635}{24385536} - \frac{11275}{1152}\pi^2 \biggr)\nu^2 +
	\frac{1292395}{96768}\nu^3 - \frac{5975}{768}\nu^4\Biggr] x^{4} 
	\nn \\ & \quad +
	\Biggl[ - \frac{93098188434443}{150214901760} + \frac{1712}{21}\gamma_\text{E} + \frac{80}{3}\pi^2 +  \frac{856}{21}\ln (16 x)
	\nn \\ & \quad \qquad +
	\biggl( \frac{1492917260735}{1072963584} - \frac{2255}{48}\pi^2\biggr)\nu
	- \frac{45293335}{1016064}\nu^2 - \frac{10323755}{1596672}\nu^3 \Biggr] \pi x^{9/2} 
	\nn \\ & \quad \quad 
	+ \calO\left(\frac{1}{c^{10}}\right)\Biggr\}\,,
\end{align}
where the integration constant $\psi_0$ is determined by initial conditions, e.g. when the wave frequency enters the detector's band. The results \eqref{xtau}--\eqref{phix} give the unique prediction of Einstein's general relativity for the GW frequency and phase chirp of non-spinning circular compact binaries up to 4.5PN precision. Note that the contributions of the quadrupole moments of compact objects which are induced by tidal effects, are expected to come into play only at the 5PN order, see Sect.~\ref{sec:intstructure}. On the other hand, most of the frameworks for data analysis rely on the stationary phase approximation \citep{TFP00}, for which the phase of the dominant quadrupole mode reads 
\begin{align}\label{phiSPA}
	\psi_\text{SPA} &= 2\pi F\,T_0 + \Psi_0 \nn\\
	& \nn
	+\frac{3\,v^{-5}}{128\nu}\Biggl\{
	1
	+ \biggl(\frac{3715}{756} + \frac{55}{9}\nu\biggr)v^2
	- 16\pi v^3 
	\\
	& \quad\nn
	+ \biggl(\frac{15293365}{508032} +	\frac{27145}{504} \nu + \frac{3085}{72} \nu^2 \biggr) v^4 
	+ \biggl(\frac{38645}{252} - \frac{65}{3}\nu\biggr) \pi v^5 \ln v
	\\
	& \quad\nn
	+ \Biggl[\frac{11583231236531}{4694215680}- \frac{640}{3}\pi^2- \frac{6848}{21}\gamma_\text{E}- \frac{6848}{21}\ln(4v) \\ & \quad\qquad + \biggl(-\frac{15737765635}{3048192}+\frac{2255}{12}\pi^2 \biggr)\nu 
	+ \frac{76055}{1728}\nu^2- \frac{127825}{1296}\nu^3\Biggr]v^6
	\nn\\
	& \quad\nn
	+ \Biggl[\frac{77096675}{254016} + \frac{378515}{1512}\nu-\frac{74045}{756}\nu^2\Biggr] \pi v^7
	\\
	& \quad\nn
	+ \Biggl[ - \frac{2550713843998885153}{276808510218240}+ \frac{90490}{189}\pi^2 + \frac{36812}{63} \gamma_\text{E} + \frac{1011020}{1323}\ln 2 \nn\\ & \quad\qquad + \frac{78975}{196} \ln 3 + \frac{18406}{63} \ln v 
	\nn\\
	& \quad\qquad\nn
	+\biggl(\frac{680712846248317}{42247941120} - \frac{109295}{224}\pi^2 + \frac{3911888}{1323} \gamma_\text{E} + \frac{9964112}{1323}\ln 2 \nn\\ & \quad\qquad\qquad - \frac{78975}{49} \ln 3 + \frac{1955944}{1323} \ln v \biggr)\nu
	\nn\\
	& \quad\qquad\nn
	+ \biggl( - \frac{7510073635}{3048192} + \frac{11275}{144} \pi^2\biggr)\nu^2 -\frac{1292395}{12096}\nu^3+\frac{5975}{96}\nu^4 \Biggr] v^8 \ln v
	\\
	& 
	\quad\nn
	+ \Biggl[\frac{105344279473163}{18776862720}- \frac{640}{3}\pi^2- \frac{13696}{21}\gamma_\text{E}- \frac{13696}{21}\ln(4v) \\
	&
	\quad\qquad 
	+ \biggl(-\frac{1492917260735}{134120448}+\frac{2255}{6}\pi^2 \biggr)\nu
	+ \frac{45293335}{127008}\nu^2 \nn\\& \quad\qquad\qquad + \frac{10323755}{199584}\nu^3\Biggr]\pi v^9
	+ \calO\left(\frac{1}{c^{10}}\right) \Biggr\}\,,
\end{align}
where $v\equiv \left(\frac{\pi G m F}{c^3}\right)^{1/3}$ with $F$ being the Fourier frequency, and where $T_0$ and $\Psi_0$ are two integration constants. Again we have adjusted $T_0$ in order to simplify the result. See e.g. \cite{PH20} for application to gravitational-wave data analysis.

As a rough estimate of the relative importance of the various PN terms, we provide in Table \ref{tab:Ncycle} numerical estimates for the number of accumulated GW cycles in the frequency band of current and future detectors. Such an estimate is only indicative, because a full treatment would require the knowledge of the detector's noise, and a complete simulation of the parameter estimation using matched filtering techniques \citep{CF94, PW95, KKS95, AISS05}. The number of cycles between entry and exit frequencies $f_\text{entry}$ and $f_\text{exit}$ (where $f\equiv\Omega/\pi$) is
\begin{equation}
	\mathcal{N}_\text{cycle}\equiv\frac{\psi_{\text{exit}}
		-\psi_{\text{entry}}}{\pi}\,.
	\label{GWcycle}
\end{equation}
As we see from Table \ref{tab:Ncycle}, the PN expansion seems to converge well. For all the typical compact binaries, we find that the 4PN and 4.5PN orders amount to about a tenth of a cycle (less than 1 radian). This suggests that systematic errors due to the PN modeling may be dominated by statistical errors and negligible for LISA. However, this should be confirmed by detailed investigations along the lines of \cite{Owen23}. 
\begin{table}[ht]
	\caption{Contribution of each PN order to the total number of accumulated cycles \eqref{GWcycle} inside the detector's frequency band, for typical (non-spinning) quasi-circular compact binaries observed by current and future detectors. We have approximated the frequency bands of LIGO/Virgo, Einstein Telescope (ET) and LISA with step functions, respectively between $[f_\text{entry},f_\text{exit}]=[30\,\text{Hz},10^3\,\text{Hz}]$, $[1\,\text{Hz},10^4\,\text{Hz}]$ and $[10^{-4}\,\text{Hz},10^{-1}\,\text{Hz}]$. When the merger occurs within the frequency band of the detector, the exit frequency is taken to be the Schwarzschild ISCO, $f_\text{exit}=c^3/(6^{3/2}\pi G m)$.}
	\label{tab:Ncycle}
	\centering
	\begin{tabular}{|l||c|c||c|c||c|c|}
		\hline
		Detector & \multicolumn{2}{c||}{LIGO/Virgo} & \multicolumn{2}{c||}{ET}& \multicolumn{2}{c|}{LISA} \\
		\hline
		Masses ($M_\odot$) & $1.4 \times 1.4$ & $ 10\times 10$ &  $1.4 \times 1.4$ & $500 \times 500$ & $10^5 \times 10^5$ & $10^7\times 10^7$ \\
		\hline
		PN order &\multicolumn{6}{|c|}{cumulative number of cycles $\mathcal{N}_\text{cycle}$} \\
		\hline
		\hline
		Newtonian & $2\,562.599$ & $95.502$ & $744\,401.36$ & $37.90$ & $28\,095.39$ & $9.534$ \\
		\hline
		1PN & $143.453$ & $17.879$ & $4\,433.85$ & $9.60$ & $618.31$ & $3.386$ \\
		\hline
		1.5PN & $-94.817$ & $-20.797$ & $-1\,005.78$ & $-12.63$ & $-265.70$ & $-5.181$ \\
		\hline
		2PN & $5.811$ & $2.124$ & $23.94$ &  $1.44$ & $11.35$ & $0.677$ \\
		\hline
		2.5PN & $-8.105$ & $-4.604$ & $-17.01$ & $-3.42$ & $-12.47$ & $-1.821$ \\
		\hline
		3PN & $1.858$ & $1.731$ & $2.69$ & $1.43$ & $2.59$ & $0.876$ \\
		\hline	
		3.5PN & $-0.627$ & $-0.689$ & $-0.93$ & $-0.59$ & $-0.91$ & $-0.383$ \\
		\hline 
		4PN & $-0.107$ & $-0.064$ & $-0.12$ & $-0.04$ & $-0.12$ & $-0.013$ \\
		\hline	
		4.5PN & $0.098$ & $0.118$ & $0.14$ & $0.10$ & $0.14$ & $0.065$ \\
		\hline
	\end{tabular}
\end{table}


\subsubsection{Spherical modes for data analysis and numerical relativity}
\label{sec:sphharm}

Besides the chirp described by Eqs. \eqref{xtau}--\eqref{phix}, it is also important to compute the wave amplitude, in view of the data analysis of LISA \citep{HM1, HM2}. For instance it has been shown that using the full waveform information for LISA will yield substantial improvements (with respect to the ``restricted post-Newtonian waveform'') of the angular resolution and the estimation of the luminosity distance of super-massive black hole binaries \citep{AISS07, AISSV07, TS08}. We also require the wave amplitude for high-accuracy comparisons with numerical relativity (see e.g. \citealt{Boyle07,BTB11,BAPS20}). 

The two polarization waveforms defined by Eq.~\eqref{hpc}\footnote{For the choice of polarization vectors we adopt the following: $\bm{P}$ and $\bm{Q}$ lie along the major and minor axis, respectively, of the projection onto the plane of the sky of the binary's orbit. Thus $\bm{P}$ is oriented toward the orbit's \emph{ascending node} -- namely the point $\mathcal{N}$ at which the orbit intersects the plane of the sky and the bodies are moving \emph{toward} the observer located in the direction $\bm{N}$. The ascending node for the particle 1 is chosen for the origin of the phase $\psi$. The inclination angle $i$ is between the direction of the detector $\bm{N}$ as seen from the binary's center-of-mass, and the normal to the orbital plane (supposed to be right-handed with respect to the sense of motion, so that $0\leqslant i\leqslant \pi$).\label{fnote:polar}} are decomposed, at leading order in the distance $R$ to the binary (in radiative coordinates), onto the basis of spin-weighted spherical harmonics following Eq.~\eqref{spinw}. The spherical modes are the coefficients in this decomposition and can be evaluated either from the angular integration \eqref{decomp}, or from the relations \eqref{inv}--\eqref{UV} giving them directly in terms of the radiative moments $\dU_L$ and $\dV_L$. 

A useful fact to remember is that for \textit{non-spinning} binaries, the mode $h^{\ell m}$ is entirely given by the \emph{mass} multipole moment $\dU_L$ when $\ell+m$ is even, and by the \emph{current} one $\dV_L$ when $\ell+m$ is odd. This is valid in general for non-spinning binaries, regardless of the orbit being quasi-circular or elliptical. The important point is only that the motion of the two particles must be \emph{planar}, i.e., takes place in a fixed plane. This is the case if the particles are non-spinning, but this will also be the case if, more generally, the spins are aligned or anti-aligned with the orbital angular momentum, since there is no orbital precession in this case. Thus, for any ``planar'' binaries, Eq.~\eqref{inv} splits into (see \citealt{FMBI12} for a proof)
\begin{subequations}\label{hlmUV}\begin{align}
		h^{\ell m} &= - \frac{G}{\sqrt{2} R c^{\ell +2}} \,\dU^{\ell m}
		\qquad \text{(when } \ell+m \text{ is even)} \,,\\ h^{\ell m} &=
		\frac{G}{\sqrt{2} R c^{\ell +3}} \,\di\,\dV^{\ell m}
		\qquad~ \text{(when } \ell+m \text{ is odd)} \,.
	\end{align} 
\end{subequations}

Let us factorize out in all the modes an overall coefficient including the appropriate phase factor $\de^{-\di m \psi}$, where we recall that $\psi$ denotes the gravitational-wave half phase \eqref{phix}, and such that the dominant mode with $(\ell,m)=(2,2)$ conventionally starts with one at the Newtonian order. We thus pose
\begin{equation}\label{modedef}
	h^{\ell m} = \frac{2 G \mu x}{R c^2}
	\,\sqrt{\frac{16\pi}{5}}\, H^{\ell m}\,\de^{-\di m \psi}
	\,.
\end{equation}
The dominant mode is $H^{22}$, due to the mass quadrupole radiation, and is primarily important for numerical relativity comparisons. This mode has been computed to 4PN order by \cite{BFHLT23a,BFHLT23b}, extending previous calculations to 3PN order by \cite{K07,BFIS08} and to 3.5PN by \cite{FMBI12}. We have
\begin{align}\label{h22}
	H^{22} &= 1 + x\biggl(-\frac{107}{42}+\frac{55}{42}\nu\biggr)  
	+2 \pi x^{3/2} 
	\nn\\
	& + x^2\biggl(-\frac{2173}{1512}-\frac{1069}{216}\nu+\frac{2047}{1512}\nu^2\biggr)
	+  x^{5/2}\left[-\frac{107 \pi }{21} +\left(\frac{34 \pi}{21}-24 \,\di\right)\nu\right] \nn \\
	& + x^3\Biggl[\frac{27027409}{646800}-\frac{856}{105}\,\gamma_\text{E} +\frac{428\,\di\,\pi }{105}+\frac{2 \pi ^2}{3} + \biggl(-\frac{278185}{33264}+\frac{41 \pi^2}{96}\biggr) \nu 
	\nn\\
	&\quad\quad -\frac{20261}{2772}\nu^2+\frac{114635}{99792}\nu^3-\frac{428}{105} \ln (16 x)\Biggr]
	\nn \\
	& +  x^{7/2}\Biggl[ -\frac{2173\pi}{756} + \biggl(
	-\frac{2495\pi}{378}+\frac{14333\,\di}{162} \biggr)\nu + \biggl(
	\frac{40\pi}{27}-\frac{4066\,\di}{945} \biggr)\nu^2 \Biggr]  
	\nn \\
	& + x^4\Biggl[- \frac{846557506853}{12713500800} + \frac{45796}{2205}\gamma_\text{E} - \frac{22898}{2205}\di \pi - \frac{107}{63}\pi^2 + \frac{22898}{2205}\ln(16x) \nn \\
	&\quad\quad  +\biggl(- \frac{336005827477}{4237833600} + \frac{15284}{441}\gamma_\text{E} - \frac{219314}{2205}\di \pi - \frac{9755}{32256}\pi^2 + \frac{7642}{441}\ln(16x)\biggr)\nu \nn \\
	&\quad\quad +\biggl( \frac{256450291}{7413120} - \frac{1025}{1008}\pi^2 \biggr)\nu^2 - \frac{81579187}{15567552}\nu^3 + \frac{26251249}{31135104}\nu^4 \Biggr]   \nn \\
	& + \calO\left(\frac{1}{c^{9}}\right)\,.
\end{align}
This result is in perfect agreement with linear black-hole perturbation theory in the limit when $\nu\to 0$ \citep{TSasa94}. 

We notice that the mode \eqref{h22} contains some imaginary parts. These can clearly be viewed as corrections to the phase rather than corrections in the amplitude. However, as corrections to the phase they are to be compared to the leading phase evolution $\psi(x)$ given by \eqref{phix}. Thus we see that their real order of magnitude is a factor $x^{5/2}=\calO(c^{-5})$ smaller than what is read from the amplitude. Since the imaginary parts in \eqref{h22} start at 2.5PN order, they correspond to very small 5PN modulations in the phase, and therefore can be neglected (the same conclusion also applies to the other modes below).

The mass octupole modes $H^{33}$ and $H^{31}$ were computed to 3.5PN order by \cite{FBI15}, and the current quadrupole mode $H^{21}$ to 3.5PN order by \cite{HFB21}. Note that 3.5PN refers here to the precision of the waveform; for that one needs to control the mass octupole and current quadrupole radiative moments ($\dU_{ijk}$ and $\dV_{ij}$) to 3PN order. Therefore these modes are comparatively easier than the mass quadrupole mode \eqref{h22} which required the full 4PN precision in the mass quadrupole moment $\dU_{ij}$. All the modes have been completed up to the 3.5PN level in the waveform by \cite{H23} -- the overlap agrees with previous calculations \citep{BFIS08}. We have\footnote{We assume $m\geqslant 0$; the modes having $m<0$ are deduced using $H^{\ell,-m}=(-)^\ell\overline{H}^{\ell m}$.} 
\begin{align}
	H^{21} &=\frac{1}{3} \di \,\Delta \bigg[x^{1/2}+x^{3/2} \left(-\frac{17}{28}+\frac{5
		\nu }{7}\right)+x^2 \left(\pi +\di \left(-\frac{1}{2}-2 \ln 2\right)\right)
	\nn \\ & +x^{5/2} \left(-\frac{43}{126}-\frac{509 \nu }{126}+\frac{79
		\nu^2}{168}\right)+x^3 \bigg(-\frac{17 \pi }{28}+\frac{3 \pi \nu }{14}
	\nn \\ & \qquad +\di \left(\frac{17}{56}+\nu \left(-\frac{353}{28}-\frac{3 \ln
		2}{7}\right)+\frac{17 \ln 2}{14}\right)\bigg) \nn \\
	& + x^{7/2}\biggl(\frac{15223771}{1455300}+\frac{\pi^2}{6}-\frac{214}{105}\gamma_\text{E}-\frac{107}{105}\ln(4x)-\ln 2-2(\ln 2)^2\nn \\ & \qquad +\nu\biggl(-\frac{102119}{2376}+\frac{205}{128}\pi^2\biggr)-\frac{4211}{8316}\nu^2+\frac{2263}{8316}\nu^3+\di \pi\biggl(\frac{109}{210}-2\ln 2\biggr)\biggr)\biggr]\nn\\&+
	\calO\left(\frac{1}{c^8}\right)\,,\\
	H^{33} &=-\frac{3}{4}\di\sqrt{\frac{15}{14}} \,\Delta \bigg[x^{1/2}+x^{3/2} (-4+2 \nu
	)+x^2 \left(3 \pi +\di \left(-\frac{21}{5}+6 \ln \left(3/2\right)\right)\right)
	\nn \\ &+x^{5/2} \left(\frac{123}{110}-\frac{1838 \nu }{165}+\frac{887
		\nu ^2}{330}\right)+x^3 \bigg(-12 \pi +\frac{9 \pi \nu }{2} \nn \\ &\qquad +\di
	\left(\frac{84}{5}-24 \ln \left(3/2\right)+\nu \left(-\frac{48103}{1215}+9 \ln
	\left(3/2\right)\right)\right)\bigg)\bigg]\nn \\ & + x^{7/2}\biggl(\frac{19388147}{280280} +
	\frac{492}{35} \ln \left(3/2\right) -18\ln^2 (3/2) -
	\frac{78}{7}\gamma_\text{E} + \frac{3}{2} \pi^2 \nn\\& \qquad + 6 \di \pi
	\left(-\frac{41}{35} + 3 \ln (3/2) \right) \nn \\ 
	&\qquad   + \frac{\nu}{8} \left(- \frac{7055}{429} + \frac{41}{8} \pi^2 \right) -
	\frac{318841}{17160} \nu^2 + \frac{8237}{2860} \nu^3 - \frac{39}{7}
	\ln (16x) \biggr)\bigg]\nn\\&+
	\calO\left(\frac{1}{c^8}\right)\,,\\ 
	H^{32} &=\frac{1}{3}
	\sqrt{\frac{5}{7}} \bigg[x (1-3 \nu )+x^2 \left(-\frac{193}{90}+\frac{145 \nu
	}{18}-\frac{73 \nu ^2}{18}\right)\nn\\& +x^{5/2} \left(2 \pi -6 \pi \nu +\di
	\left(-3+\frac{66 \nu }{5}\right)\right) \nn \\ &+x^3
	\left(-\frac{1451}{3960}-\frac{17387 \nu }{3960}+\frac{5557
		\nu^2}{220}-\frac{5341 \nu^3}{1320}\right)+ x^{7/2}\left(\frac{193}{30}\di-\frac{193}{45}\pi  \nn \right. \\
	& \qquad \left. + \nu \left(-\frac{258929}{5400}\di+\frac{136}{9}\pi\right) + \nu^2\left( \frac{33751}{450}\di-\frac{46}{9}\pi \right)\right)\bigg] \nn\\&+
	\calO\left(\frac{1}{c^8}\right)\,,\\  
	H^{31} &=\frac{\di \,\Delta}{12\sqrt{14}} \bigg[x^{1/2}+x^{3/2} \left(-\frac{8}{3}-\frac{2 \nu
	}{3}\right)+x^2 \left(\pi +\di \left(-\frac{7}{5}-2 \ln 2\right)\right)
	\nn \\ &+x^{5/2} \left(\frac{607}{198}-\frac{136 \nu }{99}-\frac{247
		\nu^2}{198}\right)+x^3 \bigg(-\frac{8 \pi }{3}-\frac{7 \pi \nu }{6} \nn
	\\ &\qquad +\di \left(\frac{56}{15}+\frac{16 \ln 2}{3}+\nu \left(-\frac{1}{15}+\frac{7
		\ln 2}{3}\right)\right)\bigg)\bigg] \nn \\ &+ x^{7/2}\biggl( \frac{10753397}{1513512} - 2 \ln 2
	\left( \frac{212}{105} + \ln 2\right) - \frac{26}{21} \gamma_\text{E} +
	\frac{\pi^2}{6} -2 \di \pi \left( \frac{41}{105} + \ln 2 \right)
	\nn \\ & \qquad + \frac{\nu}{8} \bigg(- \frac{1738843}{19305}
	+ \frac{41}{8} \pi^2 \bigg) + \frac{327059}{30888} \nu^2 -
	\frac{17525}{15444} \nu^3 - \frac{13}{21} \ln x \biggr)\bigg]\nn\\&+
	\calO\left(\frac{1}{c^8}\right)\,,\\ 
	H^{30} &=-\frac{2}{5} \di \sqrt{\frac{6}{7}} \nu \left[ x^{5/2} +x^{7/2}\left( -\frac{5017}{1296} -\frac{25}{108}\nu  \right)\right] +
	\calO\left(\frac{1}{c^8}\right)\,,\\ 
	H^{44} &=-\frac{8}{9} \sqrt{\frac{5}{7}} \bigg[x (1-3 \nu )+x^2
	\left(-\frac{593}{110}+\frac{1273 \nu }{66}-\frac{175 \nu^2}{22}\right)
	\nn \\ &+x^{5/2} \left(4 \pi -12 \pi \nu +\di \left(-\frac{42}{5}+\nu
	\left(\frac{1193}{40}-24 \ln 2\right)+8 \ln 2\right)\right) \nn \\ &+x^3
	\left(\frac{1068671}{200200}-\frac{1088119 \nu }{28600}+\frac{146879 \nu
		^2}{2340}-\frac{226097 \nu^3}{17160}\right) \nn \\
	& + x^{7/2}\left( \frac{12453}{275}\di -\frac{1186}{55}\pi-\frac{2372}{55}\di \ln 2 + \nu\left( -\frac{31525499}{140800}\di+\frac{2480}{33}\pi +\frac{4960}{33}\di \ln 2 \right) \nn \right. \\
	& \qquad \left. + \nu^2\left( \frac{4096237}{21120}\di-\frac{284}{11}\pi -\frac{568}{11}\di \ln 2 \right)\right)\bigg] +
	\calO\left(\frac{1}{c^8}\right)\,,\\  
	H^{43} &=-\frac{9 \di
		\,\Delta}{4 \sqrt{70}} \bigg[x^{3/2} (1-2 \nu )+x^{5/2}
	\left(-\frac{39}{11}+\frac{1267 \nu }{132}-\frac{131 \nu^2}{33}\right)
	\nn \\ &+x^3 \left(3 \pi -6 \pi \nu +\di \left(-\frac{32}{5}+\nu
	\left(\frac{16301}{810}-12 \ln \left(3/2\right)\right)+6 \ln
	\left(3/2\right)\right)\right) \nn \\
	& + x^{7/2}\left( \frac{7206}{5005}-\frac{82869}{5720}\nu+\frac{104839}{3432}\nu^2-\frac{2987}{572}\nu^3 \right)\bigg]+
	\calO\left(\frac{1}{c^8}\right)\,,\\
	H^{42} &=\frac{1}{63}
	\sqrt{5} \bigg[x (1-3 \nu )+x^2 \left(-\frac{437}{110}+\frac{805 \nu
	}{66}-\frac{19 \nu^2}{22}\right)\nn\\&+x^{5/2} \bigg(2 \pi -6 \pi \nu 
	+\di \left(-\frac{21}{5}+\frac{84 \nu }{5}\right)\bigg) \nn\\& +x^3
	\left(\frac{1038039}{200200}-\frac{606751 \nu }{28600}+\frac{400453 \nu
		^2}{25740}+\frac{25783 \nu^3}{17160}\right)\nn \\
	& + x^{7/2}\left( \frac{9177}{550}\di - \frac{437}{55}\pi +\nu\left(-\frac{83029}{880}\di + \frac{772}{33}\pi \right)+\nu^2\left(\frac{93081}{1100}\di + \frac{14}{11}\pi \right) \right)\bigg]\nn\\&+
	\calO\left(\frac{1}{c^8}\right)\,,\\  
	H^{41} &=\frac{\di \,\Delta}{84\sqrt{10}} \bigg[x^{3/2} (1-2 \nu )+x^{5/2} \left(-\frac{101}{33}+\frac{337
		\nu }{44}-\frac{83 \nu^2}{33}\right) \nn \\ & +x^3 \left(\pi -2 \pi \nu
	+\di \left(-\frac{32}{15}-2 \ln 2+\nu \left(\frac{1661}{30}+4 \ln
	2\right)\right)\right)\nn \\
	&+ x^{7/2}\left(\frac{42982}{15015}-\frac{513989}{51480}\nu+\frac{196957}{10296}\nu^2-\frac{1195}{572}\nu^3 \right)\bigg]+
	\calO\left(\frac{1}{c^8}\right)\,,\\ 
	H^{55} &=\frac{625 \di
		\,\Delta}{96 \sqrt{66}} \bigg[x^{3/2} (1-2 \nu )+x^{5/2}
	\left(-\frac{263}{39}+\frac{688 \nu }{39}-\frac{256 \nu^2}{39}\right)
	\nn \\ &+x^3 \left(5 \pi -10 \pi \nu +\di \left(-\frac{181}{14}+\nu
	\left(\frac{105834}{3125}-20 \ln \left(5/2\right)\right)+10 \ln
	\left(5/2\right)\right)\right)\nn \\
	&+ x^{7/2}\left(\frac{9185}{819}-\frac{188765}{3276}\nu+\frac{54428}{819}\nu^2-\frac{10567}{819}\nu^3 \right)\bigg]+
	\calO\left(\frac{1}{c^8}\right)\,,\\  
	H^{54} &=-\frac{32}{9\sqrt{165}} \bigg[x^2 \left(1-5 \nu +5 \nu^2\right)+x^3
	\left(-\frac{4451}{910}+\frac{3619 \nu }{130}-\frac{521 \nu ^2}{13}+\frac{339
		\nu^3}{26}\right) \nn \\
	&+ x^{7/2}\left( -\frac{52}{5}\di +4\pi +8\di \ln 2 +\nu\left(\frac{3351011}{53760}\di -20\pi -40\di \ln 2\right) \nn \right. \\
	&\qquad \left. +\nu^2\left(-\frac{10923}{128}\di +20\pi +40\di \ln 2\right) \right)\bigg]+
	\calO\left(\frac{1}{c^8}\right)\,,\\ 
	H^{53} &=-\frac{9}{32}\di\sqrt{\frac{3}{110}} \,\Delta \bigg[x^{3/2}
	(1-2 \nu )+x^{5/2} \left(-\frac{69}{13}+\frac{464 \nu }{39}-\frac{88 \nu
		^2}{39}\right)\nn \\ & +x^3 \left(3 \pi -6 \pi \nu +\di
	\left(-\frac{543}{70}+\nu \left(\frac{83702}{3645}-12 \ln
	\left(3/2\right)\right)+6 \ln \left(3/2\right)\right)\right) \nn \\
	& + x^{7/2}\left(\frac{12463}{1365}-\frac{56969}{1820}\nu+\frac{2172}{91}\nu^2-\frac{365}{273}\nu^3 \right)\bigg]+
	\calO\left(\frac{1}{c^8}\right)\,,\\  
	H^{52} &=\frac{2}{27\sqrt{55}} \bigg[x^2 \left(1-5 \nu +5 \nu^2\right)+x^3
	\left(-\frac{3911}{910}+\frac{3079 \nu }{130}-\frac{413 \nu ^2}{13}+\frac{231
		\nu^3}{26}\right)\nn \\
	& + x^{7/2}\left( -\frac{26}{5}\di+2\pi+\nu\left( \frac{16237}{336}\di-10\pi \right)+\nu^2\left(-\frac{1861}{20}\di +10\pi  \right) \right)\bigg]\nn\\&+
	\calO\left(\frac{1}{c^8}\right)\,,\\ 
	H^{51} &=\frac{\di \,\Delta}{288 \sqrt{385}} \bigg[x^{3/2} (1-2 \nu
	)+x^{5/2} \left(-\frac{179}{39}+\frac{352 \nu }{39}-\frac{4 \nu
		^2}{39}\right)\nn \\ & +x^3 \left(\pi -2 \pi \nu +\di
	\left(-\frac{181}{70}-2 \ln 2+\nu \left(\frac{626}{5}+4 \ln
	2\right)\right)\right) \nn \\
	& +x^{7/2}\left(\frac{5023}{585}-\frac{49447}{2340}\nu+\frac{68}{9}\nu^2+\frac{287}{117}\nu^3 \right)\bigg]+
	\calO\left(\frac{1}{c^8}\right)\,,\\ 
	H^{50} &= \frac{4117\di \nu}{7560\sqrt{462}}x^{7/2}\left[1-\frac{21588}{4117}\nu \right]+
	\calO\left(\frac{1}{c^8}\right)\,,\\ 
	H^{66}&=\frac{54}{5 \sqrt{143}} \bigg[x^2 \left(1-5 \nu +5 \nu^2\right)+x^3
	\left(-\frac{113}{14}+\frac{91 \nu }{2}-64 \nu^2+\frac{39 \nu
		^3}{2}\right)\nn \\
	& +x^{7/2}\left(-\frac{249}{14}\di+6\pi+12\di \ln 3 +\nu\left(\frac{21787499}{217728}\di-30\pi-60\di \ln 3 \right) \nn \right. \\
	&\qquad \left. +\nu^2\left(-\frac{323903}{2592}\di+30\pi+60\di \ln 3 \right) \right)\bigg]+
	\calO\left(\frac{1}{c^8}\right)\,,\\  
	H^{65} &=\frac{3125\di \,\Delta}{504 \sqrt{429}} \bigg[x^{5/2}\left(1-4 \nu +3 \nu^2 \right) \nn\\& + x^{7/2}\left(-\frac{149}{24} +\frac{349}{12}\nu-\frac{409}{12}\nu^2+\frac{29}{3}\nu^3 \right)\bigg]+
	\calO\left(\frac{1}{c^8}\right)\,,\\  
	H^{64}
	&=-\frac{128}{495} \sqrt{\frac{2}{39}} \bigg[x^2 \left(1-5 \nu +5
	\nu^2\right)+x^3 \left(-\frac{93}{14}+\frac{71 \nu }{2}-44 \nu^2+\frac{19 \nu
		^3}{2}\right)\nn \\
	& +x^{7/2}\left(-\frac{83}{7}\di+4\pi+8\di \ln 2 +\nu\left(\frac{3432215}{49152}\di-20\pi-40\di \ln 2 \right) \nn \right. \\
	& \qquad\left. +\nu^2\left(-\frac{382365}{4096}\di+20\pi+40\di \ln 2 \right) \right)\bigg]+
	\calO\left(\frac{1}{c^8}\right)\,,\\  
	H^{63}&=-\frac{81\di\,\Delta}{616 \sqrt{65}} \bigg[x^{5/2}\left(1-4 \nu +3 \nu^2\right) \nn\\&+x^{7/2}\left(-\frac{133}{24}+\frac{301}{12}\nu-\frac{329}{12}\nu^2+7\nu^3 \right) \bigg]+
	\calO\left(\frac{1}{c^8}\right)\,,\\  
	H^{62} &=\frac{2}{297\sqrt{65}} \bigg[x^2 \left(1-5 \nu +5 \nu^2\right)+x^3
	\left(-\frac{81}{14}+\frac{59 \nu }{2}-32 \nu^2+\frac{7 \nu
		^3}{2}\right)\nn \\ 
	& + x^{7/2}\left(-\frac{83}{14}\di +2\pi+\nu\left(\frac{799789}{13440}\di -10\pi \right)+\nu^2\left(-\frac{19193}{160}\di +10\pi \right) \right)\bigg]\nn\\&+
	\calO\left(\frac{1}{c^8}\right)\,,\\  
	H^{61}&=\frac{\di \,\Delta}{8316 \sqrt{26}} \bigg[x^{5/2}\left(1-4 \nu +3 \nu^2\right) \nn\\& + x^{7/2}\left(-\frac{125}{24}+\frac{277}{12}\nu -\frac{289}{12}\nu^2+\frac{17}{3}\nu^3\right)\bigg] +
	\calO\left(\frac{1}{c^8}\right)\,,\\ 
	H^{77} &=-\frac{16807\di \,\Delta}{1440} \sqrt{\frac{7}{858}} \bigg[x^{5/2}\left(1-4 \nu +3 \nu^2\right) \nn\\& + x^{7/2}\left(-\frac{319}{34}+\frac{2225}{51}\nu-\frac{2558}{51}\nu^2+\frac{230}{17}\nu^3 \right)\bigg] +
	\calO\left(\frac{1}{c^8}\right)\,,\\  
	H^{76} &=\frac{81}{35}
	\sqrt{\frac{3}{143}} x^3 \bigg[1-7 \nu +14 \nu^2-7 \nu^3\bigg]+
	\calO\left(\frac{1}{c^8}\right)\,,\\  
	H^{75} &=\frac{15625\di \,\Delta}{26208 \sqrt{66}} \bigg[x^{5/2}\left(1-4 \nu +3 \nu^2\right) \nn\\& +x^{7/2}\left( -\frac{271}{34}+\frac{1793}{51}\nu-\frac{1838}{51}\nu^2+\frac{134}{17}\nu^3 \right)\bigg] +
	\calO\left(\frac{1}{c^8}\right)\,,\\  
	H^{74} &=-\frac{128x^3}{1365}\sqrt{\frac{2}{33}} \bigg[1-7 \nu +14 \nu^2-7 \nu^3\bigg]+
	\calO\left(\frac{1}{c^8}\right)\,,\\  
	H^{73} &=-\frac{243\di \,\Delta}{160160}\sqrt{\frac{3}{2}} \bigg[x^{5/2}\left(1-4 \nu +3 \nu^2\right)\nn\\&+x^{7/2}\left(-\frac{239}{34} +\frac{1505}{51}\nu-\frac{1358}{51}\nu^2+\frac{70}{17}\nu^3\right)\bigg] +
	\calO\left(\frac{1}{c^8}\right)\,,\\  
	H^{72} &=\frac{x^3}{3003 \sqrt{3}} (1-7
	\nu +14 \nu^2-7 \nu^3) +
	\calO\left(\frac{1}{c^8}\right)\,,\\  
	H^{71} &=\frac{\di\,\Delta}{864864 \sqrt{2}} \bigg[x^{5/2}\left(1-4 \nu +3 \nu^2\right)\nn\\& +x^{7/2}\left(-\frac{223}{34}+\frac{1361}{51}\nu-\frac{1118}{51}\nu^2 +\frac{38}{17}\nu^3\right)\bigg] +
	\calO\left(\frac{1}{c^8}\right)\,,\\
	H^{70} &= \calO\left(\frac{1}{c^8}\right)\,.
\end{align}
With the 3.5PN approximation in the waveform all the modes with $\ell\geqslant 8$ can be considered as merely Newtonian. We give here the general Newtonian expressions of any mode with arbitrary $\ell$ and non-zero $m$ \citep{K07}:
\begin{subequations}
	\begin{align}\label{lmeven}
		H^{\ell m}\Big|_\text{leading} &= \frac{(-)^{(\ell-m+2)/2}}{2^{\ell+1} (\frac{\ell+m}{2})!
			(\frac{\ell-m}{2})!(2\ell-1)!!}
		\left(\frac{5(\ell+1)(\ell+2)(\ell+m)!(\ell-m)!}{\ell
			(\ell-1)(2\ell+1)}\right)^{1/2} \nn\\ &\qquad\times \sigma_\ell(\nu)
		\,(\di m)^\ell \,x^{\ell/2-1} \qquad\qquad \text{(for }
		\ell + m \text{ even)}\,,\\
		H^{\ell m}\Big|_\text{leading} &= \frac{(-)^{(\ell-m-1)/2}}{2^{\ell-1}
			(\frac{\ell+m-1}{2})!  (\frac{\ell-m-1}{2})! (2\ell+1)!!}
		\left(\frac{5(\ell+2)(2\ell+1)(\ell+m)!(\ell-m)!}{\ell (\ell-1)
			(\ell+1) }\right)^{1/2} \nn\\ &\qquad\times \sigma_{\ell+1}(\nu)
		\,\di \,(\di m)^\ell \,x^{(\ell-1)/2} \qquad \text{(for } \ell +
		m \text{ odd)}\,,
	\end{align}
\end{subequations}
where we employ the function of the mass ratio $\sigma_\ell(\nu)=X_2^{\ell-1}+(-)^\ell X_1^{\ell-1}$ (with $X_\text{a}=m_\text{a}/m$), which is also given by Eqs. \eqref{sigmaell}.

We have not yet written the modes $(\ell,0)$ with $\ell$ even. Indeed these modes with $m=0$ arise from the DC memory integrals in the mass-type radiative moments $\dU_L$; hence they have $\ell+m$ even, see \eqref{hlmUV}. From Eq.~\eqref{DCmem} we know that these integrals exhibit the memory effect, i.e. their formal post-Newtonian order is enhanced by the inverse of the adiabatic parameter $\xi=\calO(c^{-5})$. Therefore, starting with the explicit waveform at the 4PN order we can control the DC memory contributions only with the 1.5PN precision; the result is
\begin{subequations}\label{Hmem}
\begin{align}
	H^{20}& =-\frac{5}{14\sqrt{6}}\left[ 1+ x\left( -\frac{4075}{4032} + \frac{67\nu}{48}\right)  \right] + \calO\left(\frac{1}{c^4}\right)\,,\label{Hmem20}\\
	H^{40} &=-\frac{1}{504 \sqrt{2}}\left[ 1+x\left( -\frac{180101}{29568}+\frac{27227}{1056}\nu \right)  \right] + \calO\left(\frac{1}{c^4}\right)\,,\\ 
	H^{60} &=\frac{4195}{1419264\sqrt{273}}x\left[1-\frac{3612}{839}\nu \right]+\calO\left(\frac{1}{c^4}\right)\,,\\
	H^{80} &= \calO\left(\frac{1}{c^4}\right)\,.
\end{align}
\end{subequations}
However, the DC modes can also be computed in a general way, using the general formalism of Sect.~\ref{sec:memory}, in terms of the radiative moments, see Eqs. \eqref{ULmemres} and \eqref{hijTTmemfinal}. Using this formalism the DC memory terms have been computed to 3PN order by \cite{F09}; of course, the 1.5PN results \eqref{Hmem} agree.


\subsection{Eccentric compact binaries}
\label{sec:eccentric}

Inspiralling compact binaries are usually modelled as moving in quasi-circular orbits since gravitational radiation reaction circularizes the orbit towards the late stages of inspiral \citep{Pe64}, as discussed in Sect.~\ref{sec:quadform}. Nevertheless, there is an increased interest in inspiralling binaries moving in \emph{quasi-eccentric} orbits. Astrophysical scenarios currently exist which lead to binaries with non-zero eccentricity in the gravitational-wave detector bandwidth, both terrestrial and space-based. For instance, inner binaries of hierarchical triplets undergoing ZLK \citep{vonZeipel, Lidov62, Kozai62} oscillations could not only merge due to gravitational radiation reaction but a fraction of them should have non negligible eccentricities when they enter the sensitivity band of advanced ground based interferometers \citep{Wen03}. On the other hand the population of stellar mass binaries in globular clusters is expected to have a thermal distribution of eccentricities \citep{BenaLR}. In a study of the growth of intermediate black holes \citep{GMH04} in globular clusters it was found that the binaries have eccentricities between 0.1 and 0.2 in the LISA bandwidth. Though, supermassive black hole binaries are powerful gravitational-wave sources for LISA, it is not known if they would be in quasi-circular or quasi-eccentric orbits. If a ZLK mechanism is at work, these supermassive black hole binaries could be in highly eccentric orbits and merge within the Hubble time \citep{BLS02}. Sources of the kind discussed above provide the prime motivation for investigating higher post-Newtonian order modelling for quasi-eccentric binaries.


\subsubsection{Doubly periodic structure of the motion of eccentric binaries}
\label{sec:eccstruct} 

In Sect.~\ref{sec:eomCM} we have investigated the equations of motion of non-spinning compact binary systems in the frame of the center-of-mass for general orbits. We shall now (in this section and the next one) look for the explicit solution to those equations. In particular, let us discuss the general ``doubly-periodic'' structure of the post-Newtonian solution, closely following \cite{Dhouches,D83,DD85}.

The 4PN equations of motion admit, when neglecting the radiation reaction terms at 2.5PN and 3.5PN orders, ten first integrals of the motion corresponding to the conservation of energy, angular momentum, linear momentum, and center of mass position. When restricted to the frame of the center of mass, the equations admit four first integrals associated with the energy $\dE$ and the angular momentum vector $\mathbf{J}$.

The motion takes place in the plane orthogonal to $\mathbf{J}$. Denoting by $r=\vert\bm{x}\vert$ the binary's orbital separation in that plane, and by $\bm{v}=\bm{v}_1-\bm{v}_2$ the relative velocity, we find that $\dE$ and $\mathbf{J}$ are functions of $r$, $\dot{r}^2$, $v^2$ and $\bm{x}\times\bm{v}$. We adopt polar coordinates $(r, \phi)$ in the orbital plane, and express $\dE$ and the norm $\dJ=\vert\mathbf{J}\vert$, thanks to $v^2=\dot{r}^2+r^2\dot{\phi}^2$, as some explicit functions of $r$, $\dot{r}^2$ and $\dot{\phi}$. The latter functions can be inverted by means of a straightforward post-Newtonian iteration to give $\dot{r}^2$ and $\dot{\phi}$ in terms of $r$ and the constants of motion $\dE$ and $\dJ$. Hence,
\begin{subequations}\label{eom}
\begin{align} \dot{r}^2 &= \mathcal{R}\bigl[r;\dE,\dJ\bigr]\,,\label{eomr}\\ 
\dot{\phi} &= \mathcal{S}\bigl[r;\dE,\dJ\bigr]\,,\label{eomphi}
\end{align}\end{subequations}
where $\mathcal{R}$ and $\mathcal{S}$ denote certain polynomials in $1/r$, the degree of which depends on the post-Newtonian approximation in question; for instance it is seventh degree for both $\mathcal{R}$ and $\mathcal{S}$ at 3PN order \citep{MGS04}. At high PN orders one must use coordinate systems which avoid the presence of the logarithm of $r$. The various coefficients of the powers of $1/r$ are themselves polynomials in $\dE$ and $\dJ$, and also, of course, depend on the total mass $m$ and symmetric mass ratio $\nu$. In the case of bounded elliptic-like motion, one can prove \citep{D83} that the function $\mathcal{R}$ admits two real roots, say $r_p$ and $r_a$ such that $r_p\leqslant r_a$, which admit some non-zero finite Newtonian limits when $c\rightarrow\infty$, and represent respectively the radii of the orbit's periastron ($p$) and apastron ($a$). The other roots are complex and tend to zero when $c\rightarrow\infty$.

Let us consider a given binary's orbital configuration, fully specified by some values of the integrals of motion $\dE$ and $\dJ$ corresponding to quasi-elliptic motion.\footnote{The dependence on $\dE$ and $\dJ$ will no longer be indicated but is always understood as implicit in what follows.} The binary's orbital period, or time of return to the periastron, is obtained by integrating the radial motion as
\begin{equation}\label{P}
P = 2 \int_{r_p}^{r_a}\frac{\dd
  r}{\sqrt{\mathcal{R}\bigl[r\bigr]}}\,.
\end{equation}
We introduce the fractional angle (i.e., the angle divided by $2\pi$) of the advance of the periastron \emph{per} orbital revolution,
\begin{equation}\label{K}
K = \frac{1}{\pi}\int_{r_p}^{r_a}\dd r
\,\frac{\mathcal{S}\bigl[r\bigr]}{\sqrt{\mathcal{R}\bigl[r\bigr]}}\,,
\end{equation}
which is such that the precession of the periastron \emph{per} period is given by $\Delta\Phi=2\pi(K-1)$. As $K$ tends to one in the limit $c\rightarrow\infty$ (as is easily checked from the usual Newtonian solution), it is often convenient to pose $k\equiv K-1$, which will then entirely describe the \emph{relativistic precession}.

Let us then define the mean anomaly $\ell$ and the mean motion $n$ by
\begin{subequations}\label{elln}
  \begin{align} \ell &\equiv n\left(t-t_p\right)\,,\\ 
n &\equiv \frac{2\pi}{P}\,.
\end{align}\end{subequations}
Here $t_p$ denotes the instant of passage to the periastron. For a given value of the mean anomaly $\ell$, the orbital separation $r$ is obtained by inversion of the integral equation
\begin{equation}\label{ell}
\ell = n \int_{r_p}^{r} \frac{\dd
  s}{\sqrt{\mathcal{R}\left[s\right]}}\,.
\end{equation}
This defines the function $r(\ell)$ which is a periodic function in $\ell$ with period $2\pi$. The orbital phase $\phi$ is then obtained in terms of the mean anomaly $\ell$ by integrating the angular motion as
\begin{equation}\label{phi}
\phi = \phi_p + \frac{1}{n} \int_0^{\ell} \dd
l\,\mathcal{S}\left[r(l)\right]\,,
\end{equation}
where $\phi_p$ denotes the value of the phase at the instant $t_p$. We may define the origin of the orbital phase at the ascending node $\mathcal{N}$ with respect to some observer. In the particular case of a circular orbit, $r=\mathrm{const}$, the phase evolves linearly with time, $\dot{\phi}=\mathcal{S}\left[r\right]=\Omega$, where $\Omega$ is the orbital frequency of the circular orbit given by
\begin{equation}\label{Omega} 
\Omega = K n = (1+k)\,n\,.
\end{equation}
In the general case of a non-circular orbit it is convenient to keep that definition $\Omega=K n$ and to explicitly introduce the linearly growing part of the orbital phase \eqref{phi} by writing it in the form
\begin{align}\label{phidecomp}
\phi &= \phi_p + \Omega\,\left(t-t_p\right) +
W(\ell)\nn\\ &= \phi_p + K\,\ell + W(\ell)\,.
\end{align}
Here $W(\ell)$ denotes a certain function of the mean anomaly which is periodic in $\ell$ with period $2\pi$, hence periodic in time with period $P$. According to Eq.~\eqref{phi} this function is given in terms of the mean anomaly $\ell$ by
\begin{equation}\label{W}
W(\ell) = \frac{1}{n} \int_0^{\ell} \dd
l\,\bigl(\mathcal{S}\left[r(l)\right]-\Omega\bigr)\,.
\end{equation}
Finally, the decomposition \eqref{phidecomp} exhibits clearly the nature of the compact binary motion, which may be called doubly periodic in that the mean anomaly $\ell$ is periodic with period $2\pi$, and the periastron advance $K \ell$ is periodic with period $2\pi K$. Notice however that, though standard, the term ``doubly periodic'' is misleading since the motion in physical space is not periodic in general. The radial motion $r(t)$ is periodic with period $P$ while the angular motion $\phi(t)$ is periodic [modulo $2\pi$] with period $P/k$ where $k=K-1$. Only when the two periods are commensurable, i.e., when $k=1/N$ where $N\in\mathbb{N}$, is the motion periodic in physical space (with period $N P$).


\subsubsection{Quasi-Keplerian representation of the motion}
\label{sec:QK}

The quasi-Keplerian (QK) representation of the motion of compact binaries is an elegant formulation of the solution of the 1PN equations of motion parametrized by the eccentric anomaly $u$ (entering a specific generalization of Kepler's equation) and depending on various orbital elements, such as several types of eccentricities. It was introduced by \cite{DD85, DD86} to study the problem of binary pulsar timing data including relativistic corrections at the 1PN order, where notably the relativistic periastron precession complicates the simpler Keplerian solution.

In the QK representation the radial motion is given as
\begin{equation} \label{1PNQKradial}
r = a_r \left( 1 -e_r\,\cos u \right) + \calO\left(\frac{1}{c^4}\right)\,,
\end{equation}
which is the standard form parametrized by the eccentric anomaly $u$, with $a_r$ and $e_r$ denoting two constants representing the semi-major axis of the orbit and its eccentricity. However, these constants are labelled after the radial coordinate $r$ to remember that they enter (by definition) into the radial equation; in particular $e_r$ will differ from other kinds of eccentricities $e_t$ and $e_\phi$. The ``time'' eccentricity $e_t$ enters the Kepler equation which at the 1PN order also takes the usual form
\begin{equation} \label{1PNQKkepler}
\ell = u -e_t\,\sin u + \calO\left(\frac{1}{c^4}\right)\,,
\end{equation}
where the mean anomaly $\ell$, proportional to the time elapsed since the instant $t_p$ of passage at the periastron, and the mean motion $n$ are given by \eqref{elln}. The ``angular'' eccentricity $e_\phi$ enters the equation for the angular motion at 1PN order which is written in the form
\begin{equation}\label{1PNQKphi}
\frac{\phi - \phi_p}{K} = v +
\calO\left(\frac{1}{c^4}\right)\,,
\end{equation}
where we define\footnote{Comparing with Eqs. \eqref{phidecomp} we have also
\begin{align*}
	v = \ell + \frac{W(\ell)}{K} +
	\calO\left(\frac{1}{c^4}\right)\,.
\end{align*}
}
\begin{equation}\label{Vdef}
	v \equiv 2 \arctan \biggl[ \biggl( \frac{ 1 + e_{\phi}}{ 1 - e_{\phi}}
	\biggr)^{1/2} \tan \frac{u}{2} \biggr]\,.
\end{equation}
The constant $K$ is the advance of periastron \emph{per} orbital revolution defined by Eq.~\eqref{K}; it may be written as $K=\frac{\Phi}{2\pi}$ where $\Phi$ is the angle of return to the periastron.

Crucial to the QK formalism are the explicit expressions for the orbital elements $n$, $K$, $a_r$, $e_r$, $e_t$ and $e_\phi$ in terms of the conserved energy $\dE$ and angular momentum $\dJ$ of the orbit. For convenience we introduce two dimensionless parameters directly linked to $\dE$ and $\dJ$ by
\begin{subequations}\label{epsj}
\begin{align}
\varepsilon &\equiv -\frac{2 \dE}{\mu c^2}\,,\label{eps}\\ j &\equiv
-\frac{2 \dE \,h^2}{\mu^3}\,,\label{j}
\end{align}\end{subequations}
where $\mu=m\nu$ is the reduced mass with $m$ the total mass (recall that $\dE<0$ for bound orbits) and we have used the intermediate standard notation $h\equiv \frac{\dJ}{G m}$. The equations to follow will then appear as expansions in powers of the small post-Newtonian parameter $\varepsilon=\calO(1/c^{2})$,\footnote{Note that the post-Newtonian parameter $\varepsilon$ is precisely defined by Eq.~\eqref{eps}, and should not be confused with $\epsilon$ in Eq.~\eqref{epsPN} which is intended to represent a post-Newtonian \emph{estimate}.} with coefficients depending on $j$ and the dimensionless reduced mass ratio $\nu$; notice that the parameter $j$ is of Newtonian order: $j=\calO(1/c^0)$. We have \citep{DD85}
\begin{subequations}\label{paramnKae}\begin{align}
n &= \frac{\varepsilon^{3/2}\,c^3}{G\,m} \biggl\{
1+\frac{\varepsilon}{8}\, ( -15+\nu ) +
\calO\left(\frac{1}{c^4}\right) \biggr\} \,,\label{n1PN}\\
K &= 1+\frac{3\varepsilon}{j} + \calO\left(\frac{1}{c^4}\right)
\,,\label{K1PN}\\
a_r &= \frac{G\,m}{\varepsilon\,c^2}\biggl\{ 1+\frac{\varepsilon}{4} (
-7+\nu ) + \calO\left(\frac{1}{c^4}\right)\biggr\}\,, \\
e_r &= \sqrt{1 -j} + \frac{\varepsilon}{8\sqrt{1 -j}} \Bigl[24 -4
  \nu+5 j (-3+ \nu ) \Bigr] + \calO\left(\frac{1}{c^4}\right)\,,
\\
e_t &= \sqrt{1 -j} + \frac{\varepsilon}{8\sqrt{1 -j}} \Bigl[ -8+8\nu
+j ( 17-7\nu ) \Bigr] + \calO\left(\frac{1}{c^4}\right)\,, \\
e_\phi&= \sqrt{1 -j} + \frac{\varepsilon}{8\sqrt{1 -j}} \Bigl[ 24+ j
(-15+\nu )\Bigr] + \calO\left(\frac{1}{c^4}\right)\,.
\end{align}\end{subequations}
The dependence of such relations on the coordinate system will be discussed later. Notice the interesting point that there is no dependence of the mean motion $n$ and the radial semi-major axis $a_r$ on the angular momentum $\dJ$ up to the 1PN order; such dependence starts at 2PN order $\propto\varepsilon^2$, see e.g. \eqref{n3PN}.

The QK representation of the compact binary motion has been generalized at the 2PN order by \cite{DS88,SW93,Wex95}, at 3PN order by \cite{MGS04}, and at 4PN order by \cite{CTG22} with, however, an incomplete treatment of the 4PN tail term.\footnote{See also \cite{Trestini2024} for the QK representation in scalar-tensor theory at 2PN order.} The construction of a generalized QK representation at 3PN order exploits the fact that the radial equation given by Eq.~\eqref{eomr} is a \emph{polynomial} in $1/r$. However, this is true only in coordinate systems avoiding the appearance of terms with the logarithm $\ln r$; the presence of logarithms in the standard harmonic (SH) coordinates at 3PN order will obstruct the construction of the QK parametrization. Therefore \cite{MGS04} obtained it in the ADM coordinate system -- free of logarithms -- and also in the modified harmonic (MH) coordinate system, for which the coefficients in the equations of motion are given by Eqs. \eqref{ABcoeffMH}.

At the 3PN order the radial equation in ADM or MH coordinates is still given by Eq.~\eqref{1PNQKradial}. However, the Kepler equation \eqref{1PNQKkepler} and angular equation \eqref{1PNQKphi} acquire extra contributions. The equations become
\begin{subequations}\label{3PNQK}
\begin{align}
&r = a_r \left( 1 -e_r\,\cos u \right) + \calO\left(\frac{1}{c^8}\right)\,,\\
&\ell = u -e_t\,\sin u + f_t\,\sin v + g_t\,(v -u) + i_t\,\sin 2 v +
h_t\,\sin 3 v +
\calO\left(\frac{1}{c^8}\right)\,,\label{3PNQKkepler}\\ 
&\!\!\frac{\phi
  - \phi_p}{K} = v + f_\phi\,\sin 2v + g_\phi\, \sin 3v +
i_\phi\,\sin 4v + h_\phi\,\sin 5v +
\calO\left(\frac{1}{c^8}\right)\,,\label{3PNQKang}
\end{align}\end{subequations}
in which $v$ is exactly given by Eq.~\eqref{Vdef}. The new orbital elements $f_t$, $f_\phi$, $g_t$, $g_\phi$ are composed of 2PN and 3PN terms, while $i_t$, $i_\phi$, $h_t$, $h_\phi$ start only at 3PN order. All the orbital elements are now to be related, similarly to Eqs. \eqref{paramnKae}, to the constants $\varepsilon$ and $j$ with 3PN accuracy in a given coordinate system. 

Let us make clear that in different coordinate systems such as MH and ADM coordinates, the QK representation takes exactly the same form as given by Eqs. \eqref{3PNQK}. \emph{But}, the relations linking the various orbital elements $a_r$, $e_r$, $e_t$, $e_\phi$, $f_t$, $f_\phi$, $\cdots$ to $\dE$ and $\dJ$ or $\varepsilon$ and $j$, are different, with the notable exceptions of $n$ and $K$. Indeed, an important point related to the use of gauge invariant variables in the elliptical orbit case is that the functional forms of the mean motion $n$ and periastron advance $K$ in terms of the gauge invariant variables $\varepsilon$ and $j$ are identical in different coordinate systems like the MH and ADM coordinates \citep{DS88}. Their explicit expressions at 3PN order read
\begin{subequations}\label{nK}
\begin{align}
n &= \frac{\varepsilon^{3/2}\,c^3}{G\,m} \bigg\{
1+\frac{\varepsilon}{8}\, ( -15+\nu ) +\frac{\varepsilon^{2}}{128}
\biggl[ 555 +30\,\nu +11\,\nu^{2} + \frac{192}{j^{1/2}} ( -5+2\,\nu )
  \biggr] \nn\\& + \frac{\varepsilon^{3}}{3072} \biggl[ -29385
  -4995\,\nu-315\,\nu^{2}+135 \,\nu^{3} + \frac{5760}{j^{1/2}}
  \biggl(17 -9\,\nu+2\,\nu^2 \biggr)\nn\\&\qquad\quad +
  \frac{16}{j^{3/2}} \bigg( -10080+(13952-123\pi^2)\nu-1440\nu^2\bigg)
  \biggr] \nn\\& +
\calO\left(\frac{1}{c^8}\right)\bigg\} \,.\label{n3PN}\\
K &= 1+\frac{3\varepsilon}{j}+ \frac{\varepsilon^{2}}{4} \biggl [
  \frac{3}{j} ( -5+2\,\nu ) + \frac{15}{j^2} ( 7 -2\,\nu ) \biggr]
\nn\\& +\frac{\varepsilon^{3}}{128} \biggl[ \frac{24}{j} (
  5 -5\nu+ 4\nu^2) + \frac{1}{j^2} \biggl( -10080
  +(13952-123\pi^{2})\nu-1440\nu^{2} \biggr)\nn\\&\qquad\quad+
  \frac{5}{j^3} \biggl(7392+(-8000+ 123\pi^2)\nu + 336\nu^2 \biggr)
  \biggr]+
\calO\left(\frac{1}{c^8}\right)\,.\label{K3PN}
\end{align}\end{subequations}
Because of their gauge invariant meaning, it is natural to use $n$ and $K$ as two independent gauge-invariant variables in the general orbit case. Actually, instead of working with the mean motion $n$ it is often preferable to use the orbital frequency $\Omega$ which has been defined for general quasi-elliptic orbits in Eq.~\eqref{Omega}. Moreover we can pose
\begin{equation}\label{xeccentric}
x = \left(\frac{G\,m\,\Omega}{c^3}\right)^{2/3}\quad \text{(with $\Omega = K n$)} \,,
\end{equation} 
which constitutes the obvious generalization of the gauge invariant variable $x$ used in the circular orbit case. The use of $x$ as an independent parameter will thus facilitate the straightforward reading out and check of the circular orbit limit. The parameter $x$ is related to the energy and angular momentum variables $\varepsilon$ and $j$ up to 3PN order by
\begin{align}\label{xepsj}
x &= \varepsilon\Biggl\{ 1 +\varepsilon
\left[-\frac{5}{4}+\frac{1}{12}\nu +\frac{2}{j}\right] \nn\\&
+\varepsilon^2\left[\frac{5}{2}+\frac{5}{24}\nu +\frac{1}{18}\nu
  ^2+\frac{1}{j^{1/2}}(-5+2\nu )+\frac{1}{j}\left(-5+\frac{7}{6}\nu
  \right)+\frac{1}{j^2}\left(\frac{33}{2}-5\nu
  \right)\right]\nn\\&
+\varepsilon^3\left[-\frac{235}{48}-\frac{25}{24}\nu
  -\frac{25}{576}\nu^2+\frac{35}{1296}\nu^3
  +\frac{1}{j}\left(\frac{35}{4}-\frac{5}{3}\nu
  +\frac{25}{36}\nu^2\right)\right.\nn\\&\qquad\quad
  +\frac{1}{j^{1/2}}\left(\frac{145}{8}-\frac{235}{24}\nu
  +\frac{29}{12}\nu^2\right)+\frac{1}{j^{3/2}}\left(-45
  +\left(\frac{472}{9}-\frac{41}{96}\pi^2\right)\nu
  -5\nu^2\right)\nn\\&\qquad\quad
  +\frac{1}{j^2}\left(-\frac{565}{8}+\left(\frac{1903}{24}
  -\frac{41}{64}\pi^2\right)\nu
  -\frac{95}{12}\nu^2\right)\nn\\&\qquad\quad\left.
  +\frac{1}{j^3}\left(\frac{529}{3}
  +\left(-\frac{610}{3}+\frac{205}{64}\pi^2\right)\nu +\frac{35}{4}\nu
  ^2\right)\right]+ \calO\left(\frac{1}{c^8}\right)\Biggr\}\,.
\end{align}
Besides the very useful gauge-invariant quantities $n$, $K$ and $x$, the other orbital elements $a_r$, $e_r$, $e_t$, $e_\phi$, $f_t$, $g_t$, $i_t$, $h_t$, $f_\phi$, $g_\phi$, $i_\phi$, $h_\phi$ parametrizing Eqs. \eqref{1PNQKradial} and \eqref{3PNQK} are not gauge invariant; their expressions in terms of $\varepsilon$ and $j$ depend on the coordinate system in use. We refer to \cite{MGS04, ABIQ08} for the full expressions of all the orbital elements at 3PN order in both MH and ADM coordinate systems. Here, for future use, we only give the expression of the time eccentricity $e_t$ (squared) in MH coordinates:
\begin{align}\label{etMH}
e_t^2 &= 1-j+ \frac{\varepsilon}{4}\, \biggl[-8+8\nu +j ( 17-7\nu )
  \biggr] \\& + \frac{\varepsilon^{2}}{8} \biggl[12+72\nu +20\nu^2 +
  j( -112+47\nu -16\nu^2) +24j^{1/2}( 5-2\nu )
  \nn\\ &\qquad\quad+\frac{16}{j} (4 - 7 \nu ) +
  \frac{24}{j^{1/2}} \, ( -5 +2\,\nu ) \biggr]\nn\\ & +
\frac{\varepsilon^{3}}{6720} \biggl[
  23520-464800\nu+179760\nu^2+16800\nu^3 \nn\\&\qquad \quad + 525\,j \biggl
  (528-200\nu+77\nu^2 - 24\nu^3 \biggr) \nn\\&\qquad\quad +
  2520j^{1/2}( -265+193\,\nu -46\,\nu^{2} ) \nn\\&\qquad \quad + \frac{6}{j} \bigg(
  -73920+(260272-4305\pi^2)\nu-61040\nu^2\bigg)
  \nn\\&\qquad\quad + \frac{70}{j^{1/2}} \bigg(
  16380+(-19964+123\pi^2)\nu+3240\,\nu^{2} \bigg)
  \nn\\&\qquad\quad +\frac{70}{j^{3/2}} \bigg(
  -10080+(13952-123\pi^2)\nu-1440\nu^2 \bigg) \nn\\&\qquad\quad+
  \frac{8}{j^2} \bigg( 53760 +(-176024+4305\pi^2)\nu+15120\nu^2 \bigg)
  \biggr]+ \calO\left(\frac{1}{c^8}\right)\,.\nn
\end{align}
Again, with our notation \eqref{epsj}, this appears as a
post-Newtonian expansion in the small parameter $\varepsilon\to 0$
with fixed ``Newtonian'' parameter $j$.


\subsubsection{Averaged energy and angular momentum fluxes}
\label{sec:averflux}

The gravitational-wave energy and angular momentum fluxes from a system of two point masses in elliptic motion was first computed by \cite{PM63,Pe64} at Newtonian level. The 1PN and 1.5PN corrections to the fluxes were provided by \cite{WagW76, BS89, JS92, BS93, RS97} and used to study the associated secular evolution of orbital elements under gravitational radiation reaction using the QK representation of the binary's orbit at 1PN order \citep{DD85}. These results were extended to 2PN order by \cite{GI97, GI02} for the instantaneous terms (leaving aside the tails) using the generalized QK representation \citep{DS88, SW93, Wex95}; the energy flux and waveform were in agreement with those of \cite{WW96} obtained using a different method. \cite{ABIQ08tail, ABIQ08, ABIS09} have generalized the results at 3PN order, including all tails and related hereditary contributions \citep{ABIQ08tail, LY17}, by computing the averaged energy and angular momentum fluxes for quasi-elliptical orbits using the QK representation at 3PN order \citep{MGS04}, and deriving the secular evolution of the orbital elements under 3PN gravitational radiation reaction. The waveform of eccentric binaries has been derived up to the 2PN order in the Fourier domain by \cite{TessS10, TessS11}. The gravitational-wave modes, including the non-linear memory and tail contributions, have been derived at the 3PN order by \cite{EBFMIJ19, BMCFGI19}. The gravitational energy and angular momentum fluxes of eccentric orbits in the small mass ratio limit (case of EMRI systems) have been derived to high 9PN order by \cite{FEH16,MEHF20}. Furthermore the expansion in powers of the eccentricity has been extended to high order using the method of singular eccentricity factors, allowing the flux to be accurately determined when $e\to 1$ \citep{FEH16}.

The secular evolution of orbital elements under gravitational radiation reaction is in principle only the starting point for constructing templates for eccentric binary orbits. To go beyond the secular evolution one needs to include in the evolution of the orbital elements, besides the averaged contributions in the fluxes, the post-adiabatic terms rapidly oscillating at the orbital period. An analytic approach, based on an improved method of variation of constants, has been introduced by \cite{DGI04} at the leading post-adiabatic approximation, i.e. the level of 2.5PN radiation reaction.

The generalized QK representation of the motion discussed in Sect.~\ref{sec:QK} plays a crucial role in the procedure of averaging the energy and angular momentum fluxes $\mathcal{F}$ and $\mathcal{G}_i$ over one orbit.\footnote{Recall that the fluxes are defined in a general way, for any matter system, in terms of the radiative multipole moments by the expressions \eqref{FluxFG}.} Actually the averaging procedure applies to the ``instantaneous'' parts of the fluxes, while the ``hereditary'' parts are treated separately for technical reasons \citep{ABIQ08tail, LY17}. We pose $\mathcal{F}=\mathcal{F}_{\text{inst}}+\mathcal{F}_{\text{hered}}$ where the hereditary part of the energy flux is composed of tails and tail-of-tails (limiting ourselves to 3PN). For the angular momentum flux one needs also to include a contribution from the memory \citep{ABIS09, LY17}. We thus have to compute for the instantaneous part
\begin{equation}
\langle\,\mathcal{F}_{\text{inst}}\rangle = \frac{1}{P} \int^{P}_{0}
\dd t\,\mathcal{F}_{\text{inst}}=\frac{1}{2\,\pi} \int^{2\pi}_{0} \dd
u\,{\dd\ell\over \dd u}\,\mathcal{F}_{\text{inst}}\,, \label{avginteg}
\end{equation}
and similarly for the instantaneous part of the angular momentum flux $\mathcal{G}_i$.

Thanks to the QK representation, we can express $\mathcal{F}_{\text{inst}}$, which is initially a function of the natural variables $r$, $\dot{r}$ and $v^2$, as a function of the varying eccentric anomaly $u$, and depending on two constants: The frequency-related parameter $x$ defined by \eqref{xeccentric}, and the ``time'' eccentricity $e_t$ given by \eqref{etMH}. To do so one must select a particular coordinate system -- the MH coordinates for instance. The choice of $e_t$ rather than $e_r$ (say) is a matter of convenience; since $e_t$ appears in the Kepler-like equation \eqref{3PNQKkepler} at leading order, it will directly be dealt with when averaging over one orbit. We note that in the expression of the energy flux at the 3PN order there are some logarithmic terms of the type $\ln (r/r_0)$ even in MH coordinates. Indeed, as we have seen in Sect.~\ref{sec:eomCM}, the MH coordinates permit the removal of the logarithms $\ln (r/r'_0)$ in the equations of motion, where $r'_0$ is the UV scale in Eq.~\eqref{dvdt}; however there are still some logarithms $\ln (r/r_0)$ which involve the IR constant $r_0$ entering the definition of the multipole moments for general sources, see Theorem \ref{th:sourcemoments} and the regularization factor \eqref{regfactor}. As a result we find that the general structure of $\mathcal{F}_{\text{inst}}$ (and similarly for $\mathcal{G}_{\text{inst}}$, the norm of the angular momentum flux) consists of a finite sum of terms of the type
\begin{equation}\label{QKE-3PN} 
\mathcal{F}_{\text{inst}} = \frac{\dd
  u}{\dd\ell}\sum_{l}\frac{\alpha_l(x,e_t)+\beta_l(x,e_t)\sin u +
  \gamma_l(x,e_t)\ln(1-e_t \cos u)}{(1- e_t\cos u)^{l+1}}\,.
\end{equation}
The factor $\dd u/\dd\ell$ has been inserted to prepare for the orbital average \eqref{avginteg}. The coefficients $\alpha_l$, $\beta_l$ and $\gamma_l$ are straightforwardly computed using the QK parametrization as functions of $x$ and the time eccentricity $e_t$. The $\beta_l$'s correspond to 2.5PN radiation-reaction terms and will play no role, while the $\gamma_l$'s correspond to the logarithmic terms $\ln (r/r_0)$ arising at the 3PN order. For convenience the dependence on the constant $\ln r_0$ has been included into the coefficients $\alpha_l$'s. To compute the average we dispose of the following integration formulas ($l\in\mathbb{N}$)\footnote{The second of these formulas can alternatively be written with the standard Legendre polynomial $P_{l}$ as
\begin{align*}
\int_0^{2\pi}\frac{\dd u}{2\pi}\,\frac{1}{(1-e_t \cos
  u)^{l+1}}=\frac{1}{(1-e_t^2)^{\frac{l+1}{2}}} P_{l}
\biggl(\frac{1}{\sqrt{1-e_t^2}}\biggr)\,.
\end{align*}
}
\begin{subequations}
\begin{align}
\int_0^{2\pi}\frac{\dd u}{2\pi}\,\frac{\sin u }{(1-e_t \cos u)^{l+1}}
&= 0\,,\label{int1}\\
\int_0^{2\pi}\frac{\dd u}{2\pi}\,\frac{1}{(1-e_t \cos u)^{l+1}} &=
\frac{(-)^{l}}{l!}\biggl(\frac{\dd^{l}}{\dd
  z^{l}}\biggl[\frac{1}{\sqrt{z^2-e_t^2}}
  \biggr]\biggr){\bigg|}_{z=1}\,,\label{int2}\\
\int_0^{2\pi}\frac{\dd u}{2\pi}\,\frac{\ln(1-e_t \cos u)}{(1-e_t \cos
  u)^{l+1}} &= \frac{(-)^{l}}{l!}\biggl(\frac{\dd^{l}}{\dd
  z^{l}}\biggl[\frac{Z(z,e_t)}{\sqrt{z^2-e_t^2}}
  \biggr]\biggr){\bigg|}_{z=1}\,.\label{int3}
\end{align}
\end{subequations}
In the right-hand sides of Eqs. \eqref{int2} and \eqref{int3} we have to differentiate $l$ times with respect to the intermediate variable $z$ before applying $z=1$. The equation \eqref{int3}, necessary for dealing with the logarithmic terms, contains the non-trivial function
\begin{equation}\label{Zz}
Z(z,e_t)=\ln \biggl[\frac{\sqrt{1-e_t^2}+1}{2}\biggr]+2\ln\biggl[1 +
  \frac{\sqrt{1-e_t^2}-1}{z +\sqrt{z^2-e_t^2}}\biggr]\,.
\end{equation}
From Eq.~\eqref{int1} we see that there will be no radiation-reaction terms at 2.5PN order in the final result; the 2.5PN contribution is proportional to $\dot{r}$ and vanishes after averaging since it involves only odd functions of $u$.

Finally, after implementing all the above integrations, the averaged instantaneous energy flux in MH coordinates at the 3PN order is obtained as
\begin{equation}\label{Finst}
\langle\,\mathcal{F}_{\text{inst}}\rangle =\frac{32 c^5}{5
  G}\,\nu^2\,x^5\,\Bigl(\mathcal{I}_0+ x\,\mathcal{I}_1 +
x^2\,\mathcal{I}_2 + x^3\,\mathcal{I}_3 \Bigr)\,,
\end{equation}
where we recall that the post-Newtonian parameter $x$ is defined by \eqref{xeccentric}. The various instantaneous post-Newtonian pieces depend on the symmetric mass ratio $\nu$ and the time eccentricity $e_t$ in MH coordinates as \citep{ABIQ08}
\begin{subequations}\label{AvEMha}
\begin{align}
\label{AvEMh0} \mathcal{I}_0 &= \frac{1}{(1-e_t^2)^{7/2}}{\left(
1+\frac{73}{24}~e_t^2 + \frac{37}{96}~e_t^4\right)}\,,\\
\label{AvEMh1} \mathcal{I}_1 &=
\frac{1}{(1-e_t^2)^{9/2}}{\left\{-\frac{1247}{336} -\frac{35}{12} \nu
  +e_t^2\left(\frac{10475}{672}-\frac{1081}{36} \nu
  \right)\right.}\\&
  {\left.+e_t^4\left(\frac{10043}{384}-\frac{311}{12} \nu
  \right)+e_t^6\left(\frac{2179}{1792}-\frac{851}{576} \nu
  \right)\right\}}\,,\nn\\
 \mathcal{I}_2 &= \frac{1}{(1-e_t^2)^{11/2}}
{\left\{-\frac{203471}{9072}+\frac{12799}{504} \nu +\frac{65}{18} \nu
  ^2\right.}\nn\\&
{+e_t^2\left(-\frac{3807197}{18144}+\frac{116789}{2016} \nu
  +\frac{5935}{54} \nu ^2\right)}
{+e_t^4\left(-\frac{268447}{24192}-\frac{2465027}{8064} \nu
  +\frac{247805}{864} \nu ^2\right)}\nn\\&
{+e_t^6\left(\frac{1307105}{16128}-\frac{416945}{2688} \nu
  +\frac{185305}{1728} \nu ^2\right)}
{+e_t^8\left(\frac{86567}{64512}-\frac{9769}{4608} \nu
  +\frac{21275}{6912} \nu ^2\right)}\nn\\&
{+\sqrt{1-e_t^2}\left[\frac{35}{2}-7 \nu
    +e_t^2\left(\frac{6425}{48}-\frac{1285}{24} \nu
    \right)\right.}\nn\\&
  {\left.\left.+e_t^4\left(\frac{5065}{64}-\frac{1013}{32} \nu
    \right)+e_t^6\left(\frac{185}{96}-\frac{37}{48} \nu
    \right)\right]\right\}}\,,\label{AvEMh2}\\
 \mathcal{I}_3 &= \frac{1} {(1-e_t^2)^{13/2}}
{\left\{ \frac{2193295679}{9979200}+\left[\frac{8009293
    }{54432}-\frac{41 \pi ^2 }{64}\right]\nu-\frac{209063 }{3024}\nu
  ^2-\frac{775}{324} \nu ^3 \right.} \nn\\& {+e_t^2\left(
  \frac{20506331429}{19958400}+\left[\frac{649801883}{272160}+\frac{4879
      \pi ^2}{1536}\right]\nu-\frac{3008759 }{3024}\nu ^2-\frac{53696
  }{243}\nu ^3 \right)}\nn\\& {+e_t^4\left(
  -\frac{3611354071}{13305600}+\left[\frac{755536297
    }{136080}-\frac{29971 \pi ^2 }{1024}\right]\nu-\frac{179375
  }{576}\nu ^2-\frac{10816087 }{7776}\nu ^3 \right)}\nn\\&
{+e_t^6\left( \frac{4786812253}{26611200}+\left[\frac{1108811471
    }{1451520}-\frac{84501 \pi ^2 }{4096}\right]\nu+\frac{87787969
  }{48384}\nu ^2-\frac{983251 }{648}\nu ^3 \right)}\nn\\&
{+e_t^8\left(
  \frac{21505140101}{141926400}+\left[-\frac{32467919}{129024}
    -\frac{4059 \pi ^2 }{4096}\right]\nu+\frac{79938097 }{193536}\nu
  ^2-\frac{4586539 }{15552}\nu ^3 \right)}\nn\\&
{+e_t^{10}\left( -\frac{8977637}{11354112}+\frac{9287
  }{48384}\nu+\frac{8977 }{55296}\nu ^2-\frac{567617 }{124416}\nu ^3
  \right)}\nn\\& +\sqrt{1-e_t^2}\left[
  -\frac{14047483}{151200}+\left[-\frac{165761 }{1008}+\frac{287 \pi
      ^2 }{192}\right]\nu+\frac{455 }{12}\nu ^2\right. \nn\\&
  +e_t^2\left(
  \frac{36863231}{100800}+\left[-\frac{14935421}{6048}+\frac{52685 \pi
      ^2 }{4608}\right]\nu+\frac{43559 }{72}\nu ^2 \right)
  \nn\\&+e_t^4\left(
  \frac{759524951}{403200}+\left[-\frac{31082483 }{8064}+\frac{41533
      \pi ^2 }{6144}\right]\nu+\frac{303985 }{288}\nu ^2
  \right)\nn\\& +e_t^6\left(
  \frac{1399661203}{2419200}+\left[-\frac{40922933}{48384}+\frac{1517
      \pi ^2}{9216}\right]\nu+\frac{73357 }{288}\nu ^2
  \right)\nn\\& + \left.e_t^8\left( \frac{185}{48}-\frac{1073
  }{288}\nu+\frac{407 }{288}\nu ^2 \right)\right] \nn\\&
\left. + \left(\frac{1712}{105}+\frac{14552}{63}
e_t^2+\frac{553297}{1260} e_t^4\right.\right.\nn\\&\qquad\quad\left.\left.+\frac{187357}{1260}
e_t^6+\frac{10593}{2240} e_t^8\right) \ln\left[\frac{x}{z_0}
  \frac{1+\sqrt{1-e_t^2}}{2(1-e_t^2)}\right]\right\}\,.\label{AvEMh3}
\end{align}
\end{subequations}
We recall that $e_t$ corresponds to MH coordinates and is given by Eq.~\eqref{etMH}. On the contrary $x$ is gauge invariant -- since both $n$ and $K$ are gauge invariant. In the small mass ratio limit all the coefficients in \eqref{AvEMha} have been obtained independently by \cite{FEH16}.

The Newtonian coefficient in \eqref{AvEMha} is nothing but the ``enhancement'' function of eccentricity which enters in the orbital radiation decay of the binary pulsar, see Eq.~\eqref{Pdot}. It was computed by \cite{PM63} from the average of the usual quadrupole formula in the time domain, but also in the Fourier domain, using the Fourier decomposition of the Keplerian motion. We have $\mathcal{I}_0 \equiv f$ with
\begin{equation}\label{PM}
	f(e_t) = \sum_{p=1}^{+\infty} g(p,e_t) = \frac{
		1+\frac{73}{24} e_t^2 + \frac{37}{96} e_t^4}{(1-e_t^2)^{7/2}}\,,
\end{equation}
where $g(p,e_t)$ is directly given by the discrete Fourier coefficients of the Newtonian STF quadrupole moment $\hat{Q}_{ij}$ (normalized by $\mu a^2$) as
\begin{align}\label{gne}
	g(p,e_t) &= \frac{p^6}{16}\bigl|\mathop{\hat{Q}}_{p}{}_{ij}\big|^2\,,
\end{align}
and admits the following expression in terms of Bessel functions:  
\begin{align}\label{gneBessel}
	g(p,e_t) &= \frac{1}{2} p^2 \bigg\{ \left[-\frac{4}{e_t^3}-3 e_t+
	\frac{7}{e_t}\right] p J_p(p e_t) J_p'(p e_t) \nn\\ 
	&\qquad\quad +\left[\frac{1}{e_t^4}-\frac{1}{e_t^2}+\frac{1}{3}+
	\left(\frac{1}{e_t^4}-e_t^2-\frac{3}{e_t^2}+3\right) p^2\right] 
	\bigl(J_p(p e_t)\bigr)^2 \nn\\&\qquad\quad +
	\left[\left(e_t^2+\frac{1}{e_t^2}-2\right) p^2+\frac{1}{e_t^2}-1\right] 
	\bigl(J_p'(p e_t)\bigr)^2\bigg\}\,.
\end{align}

The last term in the 3PN coefficient of \eqref{AvEMha} is proportional to some logarithm which directly arises from the integration formula \eqref{int3}. Inside the logarithm we have posed $z_0 \equiv G m/(c^2 r_0)$, exhibiting the dependence upon the arbitrary length scale $r_0$ introduced in the formalism through Eq.~\eqref{regfactor}. Only after adding the hereditary contribution to the 3PN energy flux can we check the required cancellation of $r_0$. The hereditary part at 3PN order is of the form
\begin{equation}\label{Fhered}
\langle\,\mathcal{F}_{\text{hered}}\rangle =\frac{32 c^5}{5
  G}\,\nu^2\,x^5\,\Bigl( x^{3/2}\,\mathcal{K}_{3/2} +
x^{5/2}\,\mathcal{K}_{5/2} + x^3\,\mathcal{K}_{3} \Bigr)\,,
\end{equation}
where the 1.5PN and 2.5PN terms are due to tails, and the 3PN term is due to tails-of-tails. We have \citep{ABIQ08tail}
\begin{subequations}\label{hered2}\begin{align}
\mathcal{K}_{3/2} &= 4\pi\,\varphi(e_t)\,,\\ 
\mathcal{K}_{5/2} &=
-\frac{8191}{672}\,\pi\,\psi(e_t)
-\frac{583}{24}\nu\,\pi\,\zeta(e_t)\,,\\ 
\mathcal{K}_{3} &= -\frac{116761}{3675}\,\kappa(e_t) \nn\\& +
\left[\frac{16}{3} \,\pi^2 -\frac{1712}{105}\,\gamma_\text{E} -
  \frac{1712}{105}\ln\left(\frac{4x^{3/2}}{z_0}\right)\right]\,F(e_t)\,,
\label{hered2log}
\end{align}
\end{subequations}
where $\varphi(e_t)$, $\psi(e_t)$, $\zeta(e_t)$, $\kappa(e_t)$ and $F(e_t)$ denote certain ``enhancement'' functions of the eccentricity, with normalization such that they equal 1 for circular orbits, when $e_t=0$. The quadrupole moment \eqref{gne} also determines the function $F(e_t)$ in factor of the logarithm in the 3PN piece, which turns out to admit a closed analytic form:
\begin{align}\label{Fet}
	F(e_t) &= \frac{1}{4}\sum_{p=1}^{+\infty} p^2 g(p,e_t) \nn\\& = \frac{1}{(1-e_t^2)^{13/2}}
	\left[1+\frac{85}{6}e_t^2+\frac{5171}{192}e_t^4+\frac{1751}{192}e_t^6
	+\frac{297}{1024}e_t^8\right]\,.
\end{align}
This result is important for checking that the arbitrary constant $z_0$ disappears from the final result. Indeed this is immediately seen from comparing the last term in \eqref{AvEMh3} with \eqref{hered2log}.

\begin{figure}[ht]
  \centering
  \includegraphics[width=0.9\textwidth]{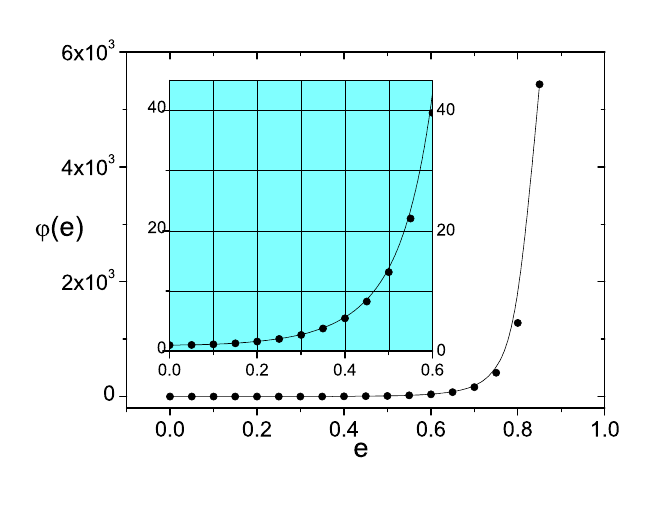}
	\caption{The enhancement factor $\varphi(e)$ as function of the eccentricity $e$. This function agrees with the numerical calculation of \cite{BS93} modulo a trivial rescaling with the Peters--Mathews function \eqref{AvEMh0}. The inset graph is a zoom of the function at a smaller scale. The dots represent the numerical computation and the solid line is a fit to the numerical points. In the circular orbit limit we have $\varphi(0)=1$.}
	\label{fig:phie}
\end{figure}
On the other hand, the four functions $\varphi(e_t)$, $\psi(e_t)$, $\zeta(e_t)$ and $\kappa(e_t)$ defined in Eqs. \eqref{hered2} do not admit analytic closed-form expressions. They have been derived by \cite{ABIQ08tail} (extending \citealt{BS93}) in the form of infinite series made out of quadratic products of Bessel functions, and computed numerically. For instance the numerical graph of the function $\varphi(e_t)$ is given in Fig.~\ref{fig:phie}. The leading expansions when $e_t\to 0$ has been computed:
\begin{subequations}\label{enhancexp}
	\begin{align}
		\label{phiexp} \varphi\left(e_t\right) &= 1+\frac{2335
		}{192}\,e_t^2+\calO\left(e_t^4\right)\,,\\
		\label{psiexp} \psi\left(e_t\right) &= 1-\frac{22988
		}{8191}\,\,e_t^2+\calO\left(e_t^4\right)\,,\\
		\label{zetaexp} \zeta\left(e_t\right) &= 1+
		\frac{1011565} {48972}  \,e_t^2+\calO\left(e_t^4\right)\,,\\
		\label{kappaexp} \kappa\left(e_t\right) &= 1+\left(\frac{62}{3}
		-\frac{4613840}{350283}\ln 2+\frac{24570945}{1868176}\ln
		3\right)\,e_t^2 +\calO\left(e_t^4\right)\,.
\end{align}\end{subequations}

Important improvements on the analytic determination of these functions have been made by \cite{FEH16}, using the method of ``singular eccentricity factors''.\footnote{See also the resummation of asymptotic enhancement factors \citep{LY17}.} Consider the function $\varphi(e_t)$ which enters the dominant 1.5PN tail term in Eq.~\eqref{Fhered}. It is obtained as
\begin{equation}\label{PMtail}
	\varphi(e_t) = \sum_{p=1}^{+\infty} \frac{p}{2} \,g(p,e_t)\,.
\end{equation}
The method of singular factors consists of determining a factor $(1-e_t^2)^{-q}$ blowing up when $e_t\to 1$, such that $(1-e_t^2)^q \varphi(e_t)$ admits a convergent expansion in powers of $e_t^2$ taking a finite value when $e_t=1$. So, the divergence of the function when $e_t\to 1$ is factorized out into some $(1-e_t^2)^{-q}$. \cite{FEH16} find $q=5$ in the case of the function \eqref{PMtail} and are able to determine
\begin{align}
	\label{vphiExpand}
	\varphi(e_t) = \frac{1}{(1-e_t^2)^5} \, 
	\bigg(1+\frac{1375}{192} e_t^2+\frac{3935}{768} e_t^4 
	+\frac{10007}{36864} e_t^6 
	+\cdots \bigg) \,,
\end{align}
with many more coefficients in the series shown in Eq.~(4.17) of \cite{FEH16}. As checked numerically the power series which is left after the factorization \eqref{vphiExpand} converges at $e_t=1$. Finally all the functions can be pushed all the way to $e_t=1$ which permits to extend analytically the validity of the flux in the regime of EMRIs, where $e_t$ is close to one. This is also useful for studying gravitational waves from binary black holes with moderately high eccentricities, such as those formed in globular clusters \citep{GMH04}. On the other hand one can retrieve from \eqref{vphiExpand} the expansion when $e_t\to 0$ to very high order,
\begin{align}
	\label{vphiExpand0}
	\varphi(e_t) = 1 + \frac{2335}{192}\,e_t^2 + \frac{42955}{756}\,e_t^4 + \frac{6204647}{36864}\,e_t^6 + \calO\left(e_t^8\right)\,.
\end{align}

The averaged energy flux was presented above using $x$ -- the gauge invariant variable \eqref{xeccentric} -- and the time eccentricity $e_t$ which however is gauge dependent. Of course it is possible to provide a fully gauge invariant formulation of the energy flux. The most natural choice is to express the result in terms of the conserved energy $E$ and angular momentum $\dJ$, or, rather, in terms of the pair of rescaled variables ($\varepsilon$, $j$) defined by Eqs. \eqref{epsj}. To this end it suffices to replace $e_t$ by its MH-coordinate expression \eqref{etMH} and to use Eq.~\eqref{xepsj} to re-express $x$ in terms of $\varepsilon$ and $j$. However, there are other possible choices for a couple of gauge invariant quantities. As we have seen the mean motion $n$ and the periastron precession $K$ are separately gauge invariant so we may define the pair of variables to be ($x$, $\iota$), where $x$ is given by \eqref{xeccentric} and we pose
\begin{equation}
\iota \equiv \frac{3x}{K-1}\,.
\end{equation}
Such choice would be motivated by the fact that $\iota$ reduces to the angular-momentum related variable $j$ in the limit $\varepsilon\rightarrow 0$. Note however that with the latter choices ($\varepsilon$, $j$) or ($x$, $\iota$) of gauge-invariant variables, the circular-orbit limit is not directly readable from the result; this is why we have preferred to present it in terms of the gauge dependent couple of variables ($x$, $e_t$).

As we are interested in the phasing of binaries moving in quasi-eccentric orbits in the adiabatic approximation, we require the orbital averages not only of the energy flux $\mathcal{F}$ but also of the angular momentum flux $\mathcal{G}_i$. Since the quasi-Keplerian orbit is planar, we only need to average the magnitude $\mathcal{G}$ of the angular momentum flux. The complete computation thus becomes a generalisation of the previous computation of the averaged energy flux requiring similar steps. The angular momentum flux is split into instantaneous $\mathcal{G}_{\text{inst}}$ and hereditary $\mathcal{G}_{\text{hered}}$ contributions; the instantaneous part is averaged using the QK representation in either MH or ADM coordinates; the hereditary part is evaluated separately and defined by means of several types of enhancement functions of the time eccentricity $e_t$; finally these are obtained numerically or analytically using the method leading to \eqref{vphiExpand}. At this stage we dispose of both the averaged energy and angular momentum fluxes $\langle\mathcal{F}\rangle$ and $\langle\mathcal{G}\rangle$.

The procedure to compute the secular evolution of the orbital elements under gravitational radiation-reaction is straightforward. Differentiating the orbital elements with respect to time, and using the heuristic balance equations, we equate the decreases of energy and angular momentum to the corresponding averaged fluxes $\langle\mathcal{F}\rangle$ and $\langle\mathcal{G}\rangle$ at 3PN order \citep{ABIS09}. This extends earlier analyses at previous orders: Newtonian \citep{Pe64} as we have reviewed in Sect.~\ref{sec:quadform}; 1PN \citep{BS89, JS92}; 1.5PN \citep{BS93, RS97} and 2PN \citep{GI97, DGI04}. Let us take the example of the mean motion $n$. From Eq.~\eqref{n3PN} together with the definitions \eqref{epsj} we know the function $n(\dE,\dJ)$ at 3PN order, where $\dE$ and $\dJ$ are the orbit's constant energy and angular momentum. Thus,
\begin{equation}\label{timeder}
\frac{\dd n}{\dd t} = \frac{\partial n}{\partial \dE}\,\frac{\dd \dE}{\dd
  t} + \frac{\partial n}{\partial \dJ}\,\frac{\dd
  \dJ}{\dd t}\,.
\end{equation}
The usual balance equations for energy and angular momentum
\begin{subequations}\label{balanceEJ3PN}
\begin{align}
\langle \frac{\dd \dE}{\dd t}\rangle &= - \langle \mathcal{F}\rangle
\,,\\ \langle \frac{\dd \dJ}{\dd t}\rangle &= - \langle \mathcal{G}\rangle
\,,
\end{align}\end{subequations}
have already been used at Newtonian order in Eqs. \eqref{balanceEJ}. Although heuristically assumed at 3PN order, they have been proved through 1.5PN order in Eq.~\eqref{balance15PN}. With the averaged fluxes known through 3PN order, we obtain the 3PN averaged evolution equation as
\begin{equation}\label{evoleq}
\langle\frac{\dd n}{\dd t}\rangle = - \frac{\partial n}{\partial
  \dE}\,\langle\mathcal{F}\rangle - \frac{\partial n}{\partial
  \dJ}\,\langle\mathcal{G}\rangle\,.
\end{equation}
We recall that this gives only the slow \emph{secular} evolution under gravitational radiation reaction for eccentric orbits. The complete evolution includes also, superimposed on the averaged adiabatic evolution, some fast but smaller post-adiabatic oscillations at the orbital time scale \citep{DGI04,KG06}.


\subsection{Spinning compact binaries}
\label{sec:spins}

The post-Newtonian templates have been developed so far for compact binary systems which can be described with high precision by point masses without spins. Here by spin, we mean the intrinsic (\emph{classical}) angular momentum $S$ of the individual compact body. Including the effects of spins is essential, as the astrophysical evidence indicates that stellar-mass black holes \citep{AK01, Stroh01, McClint06, Gou11, Nowak12} and supermassive black holes \citep{FM05, BrennR06, BrennR11} (see \citealt{Reyn13} for a review) can be generically close to maximally spinning. However the recent gravitational-wave observations have disfavored high spin magnitude \citep{GSM21}, although a fraction of binary black hole systems contain at least one black hole with a non-negligible spin \citep{ALT23}.

The spins affect the orbital dynamics of the binary, in particular leading to orbital plane precession if they are not aligned with the orbital angular momentum \citep{CF94, ACST94}, and thereby to strong modulations in the observed signal frequency and phase. An important effort has been undertaken to compute spin effects to high post-Newtonian order in the dynamics and radiation of compact binaries:
\begin{enumerate}
\item \emph{Dynamics}. The goal is to obtain the equations of motion and related conserved integrals of the motion, the equations of precession of the spins, and the post-Newtonian metric in the near zone. For this step we need a formulation of the conservative dynamics of particles with spins (either Lagrangian or Hamiltonian);
\item \emph{Radiation}. The mass and current radiative multipole moments, including all hereditary effects, are to be computed for spinning binaries. One then deduces the gravitational waveform and the fluxes, from which we compute the secular evolution of the orbital phase. This step requires plugging the previous \emph{dynamics} into the general wave generation formalism of Sect.~\ref{sec:PNsource}.
\end{enumerate}

We adopt a particular post-Newtonian counting for spin effects that actually refers to maximally spinning black holes. In this convention the two spin variables $S_\text{a}$ ($\text{a}=1,2$) have the dimension of an angular momentum multiplied by a factor $c$, and we pose
\begin{equation}\label{Sndef}
S_{\text{a}} = G m_{\text{a}}^{2} \chi_{\text{a}}\,,
\end{equation}
where $m_{\text{a}}$ is the mass of the compact body, and $\chi_{\text{a}}$ is the dimensionless spin parameter, which equals one for maximally spinning Kerr black holes. Thus the spins $S_{\text{a}}$ of the compact bodies can be considered as ``Newtonian'' quantities [there are no $c$'s in Eq.~\eqref{Sndef}], and all spin effects will carry (at least) an explicit $1/c$ factor with respect to non-spin effects. With this convention any post-Newtonian estimate is expected to be appropriate (i.e., numerically close to reality) in the case of maximal rotation. One should keep in mind that spin effects will be formally a factor $1/c$ smaller for non-maximally spinning objects such as neutron stars; thus in this case a given post-Newtonian computation will actually be a factor $1/c$ more accurate.

As usual we make a distinction between spin-orbit (SO) effects, which are linear in the spins, and spin-spin (SS) ones, which are quadratic. In the following sections we shall essentially focus on SO effects, as they play the most important role in gravitational wave detection and parameter estimation; we shall give a short account of SS effects in Sect.~\ref{sec:SS}.

The SO effects have been known at the leading level since the seminal works of \cite{Papa51spin, CPapa51, Tulc1, Tulc2, BOC75, BOC79, KWW93, K95}. With our post-Newtonian counting such leading level corresponds to the 1.5PN order. The SO terms have been computed to the next-to-leading level which corresponds to 2.5PN order by \cite{TOO01, FBB06, DJSspin, Le10so, Po10, HaS11} for the equations of motion or dynamics, and by \cite{BBF06, BBF11} for the gravitational radiation field. Note that \cite{TOO01, FBB06, DJSspin, HaS11} employ traditional post-Newtonian methods (both harmonic-coordinates and Hamiltonian), but that \cite{Le10so, Po10} are based on the effective field theory (EFT) approach. The next-to-next-to-leading SO level corresponding to 3.5PN order has been obtained by \cite{HaS11so, HaSS13} using the Hamiltonian method for the equations of motion, by \cite{LS15b} using the EFT, and by \cite{MBFB13, BMFB13} using the harmonic-coordinates method. Here we shall focus on the harmonic-coordinates approach \citep{MBFB13, BMFB13, BMB13, MBBB13} which is in fact well formulated using a Lagrangian, see Sect.~\ref{sec:lagspins}. With this approach the next-to-next-to-leading SO level was derived not only for the equations of motion including precession, but also for the radiation field (energy flux and orbital phasing) by \cite{BMB13, MBBB13}. An analytic solution for the SO precession effects will be presented in Sect.~\ref{sec:eomspins}. Concerning the radiation field the highest known SO level actually contains specific tail-induced contributions at 3PN and 4PN orders \citep{BBF11, MBBB13, CPY22}. To next-to-next-to-next-to-leading order the SO terms in the equations of motion have been derived by \cite{Antonelli2020a, Kim2023a, Mandal2023a}.


\subsubsection{Lagrangian formalism for spinning point particles}
\label{sec:lagspins}

Some necessary material for constructing a Lagrangian for a spinning point particle in curved spacetime is presented here. The formalism is issued from early works by \cite{HRegge74, BIsrael75} and has also been developed in the context of the EFT approach by \cite{Po06}. Variants and alternatives (most importantly the associated Hamiltonian formalism) have been developed by \cite{SHS08b, St11rev, BarauRB09}. The formalism yields for the equations of motion of spinning particles and the equations of precession of the spins the classic results known in general relativity \citep{Tulc1, Tulc2, Mathisson37, Mathisson37repub, Papa51spin, CPapa51, Traut58, Dixon79}.

Let us consider a single spinning point particle moving in a given curved background metric $g_{\alpha\beta}(x)$. The particle follows the worldline $y^\alpha(\tau)$, with tangent four-velocity $u^\alpha = \dd y^\alpha/\dd \tau$, where $\tau$ is a parameter along the representative worldline. In a first stage we do not require that the four-velocity be normalized; thus $\tau$ needs not be the proper time elapsed along the worldline. To describe the internal degrees of freedom associated with the particle's spin, we introduce a moving orthonormal tetrad $e_A^{\phantom{A}\alpha}(\tau)$ along the trajectory, which defines a ``body-fixed'' frame.\footnote{The tetrad is orthonormal in the sense that $g_{\alpha\beta}\,e_A^{\phantom{A}\alpha}e_B^{\phantom{B}\beta} = \eta_{AB}$, where $\eta_{AB} = \text{diag}(-1,1,1,1)$ denotes a Minkowski metric. The indices $AB\cdots=0,1,2,3$ are the internal Lorentz indices, while as usual $\alpha\beta\cdots\mu\nu\cdots=0,1,2,3$ are the space-time covariant indices. The inverse dual tetrad $e^A_{\phantom{A}\alpha}$, defined by $e_A^{\phantom{A}\beta}e^A_{\phantom{A}\alpha}=\delta^\beta_\alpha$, satisfies $\eta_{AB} \,e^A_{\phantom{A}\alpha}e^B_{\phantom{B}\beta} = g_{\alpha\beta}$. We have also the completeness relation $e_A^{\phantom{A}\beta}e^B_{\phantom{B}\beta}=\delta^B_A$.} The rotation tensor $\omega^{\alpha\beta}$ associated with the tetrad is defined by
\begin{equation}\label{rot} 
\frac{\dD e_A^{\phantom{A}\alpha}}{\dd \tau} = - \omega^{\alpha\beta}
\,e_{A\beta}\,,
\end{equation}
where $\dD/\dd \tau \equiv u^\beta \nabla_\beta$ is the covariant derivative with respect to the parameter $\tau$ along the worldline; equivalently, we have
\begin{equation}\label{rot2}
\omega^{\alpha\beta}=e^{A\alpha}\,\frac{\dD
  e_A^{\phantom{A}\beta}}{\dd \tau}\,.
\end{equation}
Because of the normalization of the tetrad the rotation tensor is antisymmetric: $\omega^{\alpha\beta}=-\omega^{\beta\alpha}$.

We look for an action principle for the spinning particle. Following
\cite{HRegge74, Po06} we require the following symmetries to hold:
\begin{enumerate}
\item The action is a covariant scalar, i.e., behaves as a scalar with respect to general space-time diffeomorphisms;
\item It is a global Lorentz scalar, i.e., stays invariant under an arbitrary change of the tetrad vectors: $e_A^{\phantom{A}\alpha}(\tau) \longrightarrow \Lambda^B_{\phantom{B}A}\,e_B^{\phantom{B}\alpha}(\tau)$ where $\Lambda^B_{\phantom{B}A}$ is a constant Lorentz matrix;
\item It is parametrization-invariant, i.e.,  its form is independent of the parameter $\tau$ used to follow the particle's worldline.\footnote{This condition can always be realized. Suppose that $\tau$ initially denotes the proper time so that $g_{\alpha\beta}u^\alpha u^\beta=-1$. For an arbitrary parametrization $\bar{\tau}$ we pose 
\begin{align*}
		\dd\tau = \dd\bar{\tau}\sqrt{-\bar{u}^2}\,,\qquad u^\alpha = \frac{\bar{u}^\alpha}{\sqrt{-\bar{u}^2}}\,,\qquad L = \frac{\bar{L}}{\sqrt{-\bar{u}^2}}\,,
\end{align*}
where $\bar{u}^\alpha=\dd y^\alpha/\dd\bar{\tau}$ and $\bar{u}^2 = g_{\alpha\beta}\bar{u}^\alpha\bar{u}^\beta$. Then the action $S=\int\dd\bar{\tau}\bar{L}$ will be in parametrization-invariant form.}
\end{enumerate}
In addition to these symmetries we need to specify the dynamical degrees of freedom: These are chosen to be the particle's position $y^\alpha$ and the tetrad $e_A^{\phantom{A}\alpha}$. Furthermore we restrict ourselves to a Lagrangian depending only on the four-velocity $u^\alpha$, the rotation tensor $\omega^{\alpha\beta}$, and the metric $g_{\alpha\beta}$. Thus, the postulated action is of the type
\begin{equation}\label{action} 
S\bigl[y^\alpha, e_A^{\phantom{A}\alpha}\bigr] =
\int_{-\infty}^{+\infty} \dd \tau
\,L\bigl(u^\alpha,\omega^{\alpha\beta},g_{\alpha\beta}\bigr) \,.
\end{equation}
These assumptions confine the formalism to a ``pole-dipole'' model and to terms linear in the spins. An important point is that such a model is \emph{universal} in the sense that it can be used for black holes as well as neutrons stars. Indeed, the internal structure of the spinning body appears only at the quadratic order in the spins, through the rotationally induced quadrupole moment.

As it is written in \eqref{action}, i.e., depending only on Lorentz scalars, $L$ is automatically a Lorentz scalar. By performing an infinitesimal coordinate transformation, one easily sees that the requirement that the Lagrangian be a covariant scalar specifies its dependence on the metric to be such that (see e.g. \citealt{BIsrael75})
\begin{equation}\label{scalar} 
2 \frac{\partial L}{\partial g_{\alpha\beta}} = p^\alpha u^\beta +
S^{\alpha}_{\phantom{\alpha}\gamma}\omega^{\beta\gamma}\,.
\end{equation}
We have defined the conjugate linear momentum $p^\alpha$ and the antisymmetric spin tensor $S^{\alpha\beta}$ by the partial derivatives
\begin{align}\label{conjugate}
p_\alpha \equiv \frac{\partial L}{\partial u^\alpha}{\Big|}_{\omega, g}
\,,\qquad S_{\alpha\beta} \equiv 2 \frac{\partial L}{\partial
  \omega^{\alpha\beta}}{\Big|}_{u, g} \,.
\end{align}
Note that the right-hand side of Eq.~\eqref{scalar} is necessarily symmetric by exchange of the indices $\alpha$ and $\beta$. Finally, imposing the invariance of the action \eqref{action} by reparametrization of the worldline, we find that the Lagrangian must be a homogeneous function of degree one in the velocity $u^\alpha$ and rotation tensor $\omega^{\alpha\beta}$. Applying Euler's theorem to the function $L(u^\alpha,\omega^{\alpha\beta})$ immediately gives
\begin{equation}\label{Euler} 
L = p_\alpha u^\alpha + \frac{1}{2} S_{\alpha\beta}
\omega^{\alpha\beta} \,, 
\end{equation}
where the functions $p_\alpha(u,\omega)$ and $S_{\alpha\beta}(u,\omega)$ must be parametrization invariant. Note that, at this stage, their explicit expressions are not known. They will be specified only when a spin supplementary condition is imposed, see \eqref{SSC}.

We now investigate the unconstrained variations of the action \eqref{action} with respect to the dynamical variables $e_A^{\phantom{A}\alpha}$, $y^\alpha$ and the metric. First, we vary it with respect to the tetrad $e_A^{\phantom{A}\alpha}$ while keeping the position $y^\alpha$ fixed. A worry is that we must have a way to distinguish intrinsic variations of the tetrad from variations which are induced by a change of the metric $g_{\alpha\beta}$. This is conveniently solved by decomposing the variation $\delta e_A^{\phantom{A}\beta}$ according to
\begin{equation}\label{variation} 
\delta e_A^{\phantom{A}\beta} = e_{A\alpha} \Bigl(
\delta\theta^{\alpha\beta} + \frac{1}{2} \delta g^{\alpha\beta}
\Bigr)\,,
\end{equation}
in which we have introduced the antisymmetric tensor $\delta\theta^{\alpha\beta}\equiv e^{A[\alpha}\delta e_A^{\phantom{A}\beta]}$, and where the corresponding symmetric part is simply given by the variation of the metric, i.e. $e^{A(\alpha}\delta e_A^{\phantom{A}\beta)}\equiv\frac{1}{2}\delta g^{\alpha\beta}$. Then we can consider the independent variations $\delta\theta^{\alpha\beta}$ and $\delta g^{\alpha\beta}$. Varying with respect to $\delta\theta^{\alpha\beta}$, but holding the metric fixed, gives the equation of spin precession which is found to be
\begin{equation}\label{prec0} 
\frac{\dD S_{\alpha\beta}}{\dd \tau} =
\omega_{\alpha}^{\phantom{\alpha}\gamma}S_{\beta\gamma} -
\omega_{\beta}^{\phantom{\beta}\gamma}S_{\alpha\gamma}\,,
\end{equation}
or, alternatively, using the fact that the right-hand side of \eqref{scalar} is symmetric,
\begin{equation}\label{precgen}
\frac{\dD S_{\alpha\beta}}{\dd \tau} = p_\alpha u_\beta - p_\beta
u_\alpha\,.
\end{equation}
We next vary with respect to the particle's position $y^\alpha$ while holding the tetrad $e_A^{\phantom{A}\alpha}$ fixed. Operationally, this means that we have to parallel-transport the tetrad along the displacement vector, i.e., to impose
\begin{equation}\label{parallel}
\delta y^\beta \nabla_\beta e_A^{\phantom{A}\alpha}=0\,.
\end{equation}
A simple way to derive the result is to use locally inertial coordinates, such that the Christoffel symbols $\Gamma_{\beta\gamma}^\alpha = 0$ along the particle's worldline $y^\alpha(\tau)$; then, Eq.~\eqref{parallel} gives $\delta e_A^{\phantom{A}\alpha}=\delta y^\beta \partial_\beta e_A^{\phantom{A}\alpha}=-\delta y^\beta \Gamma_{\beta\gamma}^\alpha e_A^{\phantom{A}\gamma}=0$. The variation leads then to the well-known \cite{Mathisson37, Mathisson37repub, Papa51spin, CPapa51} equation of motion
\begin{equation}\label{MathPapa} 
\frac{\dD p_\alpha}{\dd \tau} = -\frac{1}{2} u^\beta
R_{\alpha\beta\mu\nu} S^{\mu\nu}\,,
\end{equation}
which involves the famous coupling of the spin tensor to the Riemann curvature.\footnote{Our conventions for the Riemann tensor $R_{\alpha\beta\mu\nu}$ follow those of \cite{MTW}.} With a little more work, the equation of motion \eqref{MathPapa} can also be derived using an arbitrary coordinate system, making use of the parallel transport equation \eqref{parallel}. 

Finally, varying with respect to the metric while keeping $\delta\theta^{\alpha\beta}=0$, gives the stress-energy tensor of the spinning particle. We must again take into account the scalarity of the action, as imposed by Eq.~\eqref{scalar}. We obtain the standard pole-dipole result \citep{Tulc1, Tulc2, Mathisson37, Mathisson37repub, Papa51spin, CPapa51, Traut58,
  Dixon79}:
\begin{equation}\label{Tabspin}
T^{\alpha\beta} = \int_{-\infty}^{+\infty} \dd \tau
\,p^{(\alpha}\,u^{\beta)}\,\frac{\delta_{(4)} (x-y)}{\sqrt{-g}} -
\nabla_\gamma \int_{-\infty}^{+\infty} \dd \tau\,
S^{\gamma(\alpha}\,u^{\beta)}\,\frac{\delta_{(4)} (x-y)}{\sqrt{-g}}
\,,
\end{equation}
where $\delta_{(4)}(x-y)$ denotes the four-dimensional Dirac function. It can easily be checked that the covariant conservation law $\nabla_\beta T^{\alpha\beta}=0$ holds as a consequence of the equation of motion \eqref{MathPapa} and the equation of spin precession \eqref{precgen}.

Up to now we have considered unconstrained variations of the action \eqref{action}, describing the particle's internal degrees of freedom by the six independent components of the tetrad $e_A^{\phantom{A}\alpha}$ (namely a $4\times 4$ matrix subject to the 10 constraints $g_{\alpha\beta}\,e_A^{\phantom{A}\alpha}e_B^{\phantom{B}\beta} = \eta_{AB}$). To correctly account for the number of degrees of freedom associated with the spin, we must impose three \emph{supplementary spin conditions} (SSC). 

Several choices are possible for a sensible SSC. Notice that in the case of extended bodies the choice of a SSC corresponds to the choice of a central worldline inside the body with respect to which the spin angular momentum is defined (see \citealt{K95} for a discussion). Here we adopt the \cite{Tulc1, Tulc2} covariant SSC
%
\begin{equation}\label{SSC}
S^{\alpha\beta} p_\beta = 0 \,.
\end{equation}
As shown by \cite{HRegge74} in the flat space-time case, it is possible to specify the Lagrangian in our original action \eqref{action} in such a way that the constraints \eqref{SSC} are directly the consequence of the equations derived from that Lagrangian. Here, for simplicity's sake, we shall simply impose the constraints \eqref{SSC} in the space of solutions of the Euler-Lagrange equations. From Eq.~\eqref{SSC} we can introduce the covariant spin vector $S_{\mu}$ associated with the spin tensor by\footnote{The four-dimensional Levi-Civita \emph{tensor} is defined by $\varepsilon_{\alpha\beta\mu\nu} \equiv \sqrt{-g}\,\epsilon_{\alpha\beta\mu\nu}$ and $\varepsilon^{\alpha\beta\mu\nu} \equiv -\epsilon^{\alpha\beta\mu\nu}/\sqrt{-g}$; here $\epsilon_{\alpha\beta\mu\nu}=\epsilon^{\alpha\beta\mu\nu}$ denotes the completely anti-symmetric Levi-Civita \emph{symbol} such that $\epsilon_{0123}=\epsilon^{0123}=1$. For convenience we often pose $c=1$.}
\begin{equation}\label{defSalpha}
S^{\alpha\beta} \equiv
\frac{1}{m}\,\varepsilon^{\alpha\beta\mu\nu}p_{\mu}S_{\nu}\,,
\end{equation}
where we have defined the mass of the particle by $m^2 \equiv - g^{\mu\nu} p_\mu p_\nu$. By contracting Eq.~\eqref{precgen} with $p^\beta$ and using the equation of motion \eqref{MathPapa}, one obtains
\begin{equation}\label{pu} 
p_\alpha (pu) + m^2 u_\alpha = \frac{1}{2} u^\gamma
R^\beta_{\phantom{\beta} \gamma \mu\nu} S^{\mu\nu}S_{\alpha\beta}\,,
\end{equation}
where we denote $(pu)\equiv p_\mu u^\mu$. By further contracting Eq.~\eqref{pu} with $u^\alpha$ we obtain an explicit expression for $(pu)$, which can then be substituted back into \eqref{pu} to provide the relation linking the four-momentum $p_\alpha$ to the four-velocity $u_\alpha$. It can be checked using \eqref{SSC} and \eqref{pu} that the mass of the particle is constant along the particle's trajectory: $\dd m/\dd \tau=0$. Furthermore the four-dimensional magnitude $s$ of the spin defined by $s^2\equiv g^{\mu\nu}S_\mu S_\nu$ is also conserved: $\dd s/\dd \tau=0$.

Henceforth we shall restrict our attention to SO interactions, linear in the spins (see Sect.~\ref{sec:SS} for SS interactions). We shall also adopt for the parameter $\tau$ along the particle's worldline the proper time $\dd\tau\equiv\sqrt{-g_{\mu\nu}\dd y^\mu\dd y^\nu}$, so that $g_{\mu\nu} u^\mu u^\nu = -1$. Neglecting quadratic spin-spin (SS) and higher-order interactions, the linear momentum is simply proportional to the normalized four-velocity: $p_\alpha=m \,u_\alpha + \calO(S^2)$. Hence, from Eq.~\eqref{precgen} we deduce that $\dD S_{\alpha\beta} / \dd \tau = \calO(S^2)$. The equation for the spin covariant vector $S_\alpha$ then reduces at linear order to
\begin{equation}\label{transparallel}
  \frac{\dD S_{\alpha}}{\dd\tau} = \calO(S^2) \,.
\end{equation}
Thus the spin covector is parallel transported along the particle's trajectory at linear order in spin. For the spin covector $S_\alpha$ itself, we choose a four-vector which is purely spatial in the particle's instantaneous rest frame, where $u^\alpha= (1, \bm{0})$, hence the components of $S_\alpha$ are $(0, \bm{S})$ in that frame. Therefore, in any frame,
\begin{equation}\label{orthog}
S_{\alpha}u^{\alpha} = 0\,.
\end{equation}

In applications (viz. the construction of gravitational-wave templates for the compact binary inspiral) it is very useful to introduce new spin variables that are designed to have a conserved \emph{three-dimensional} Euclidean norm (numerically equal to $s$). Using conserved-norm spin vector variables is indeed the most natural choice when considering the dynamics of compact binaries reduced to the frame of the center of mass or to circular orbits \citep{BMFB13}. Indeed the evolution equations of such spin variables reduce, by construction, to ordinary precession equations, and these variables are secularly constant (see \citealt{W05}).

A standard, general procedure to define a (Euclidean) conserved-norm spin spatial vector consists of projecting the spin covector $S_\alpha$ onto an orthonormal tetrad $e_A^{\phantom{A}\alpha}$, which leads to the four scalar components ($A=0,1,2,3$)
\begin{equation}\label{tetradcomp}
S_A = e_A^{\phantom{A}\alpha} S_{\alpha}\,.
\end{equation}
If we choose for the time-like tetrad vector the four-velocity itself, $e_0^{\phantom{0}\alpha} = u^\alpha$,\footnote{Because of this choice, it is better to consider that the tetrad is not the same as the one we originally employed to construct the action \eqref{action}.} the time component tetrad projection $S_{0}$ vanishes because of the orthogonality condition \eqref{orthog}. We have seen that $S_{\alpha}S^{\alpha} = s^{2}$ is conserved along the trajectory; because of \eqref{orthog} we can rewrite this as $\gamma^{\alpha\beta}S_{\alpha}S_{\beta} = s^{2}$, in which we have introduced the projector $\gamma^{\alpha\beta}=g^{\alpha\beta}+u^\alpha u^\beta$ onto the spatial hypersurface orthogonal to $u^\alpha$. From the orthonormality of the tetrad and our choice $e_0^{\phantom{0}\alpha} = u^\alpha$, we have $\gamma^{\alpha\beta}=\delta^{ab}e_a^{\phantom{a}\alpha}e_b^{\phantom{b}\beta}$ in which $a, b =1,2,3$ refer to the spatial values of the tetrad indices, i.e., $A=(0,a)$ and $B=(0,b)$. Therefore the conservation law $\gamma^{\alpha\beta}S_{\alpha}S_{\beta} = s^{2}$ becomes
\begin{equation}\label{Euclidorthog}
\delta^{ab}S_{a}S_{b} = s^{2}\,,
\end{equation}
which is indeed the relation defining a Euclidean conserved-norm spin variable $S_a$.\footnote{Beware that here we employ the usual slight ambiguity in the notation when using the same carrier letter $S$ to denote the tetrad components \eqref{tetradcomp} and the original spin covector. Thus, $S_a$ should not be confused with the spatial components $S_i$ (with $i=1,2,3$) of the covariant vector $S_\alpha$.} However, note that the choice of the spin variable $S_a$ is still somewhat arbitrary, since a rotation of the tetrad vectors can freely be performed. We refer to \cite{DJSspin, BMFB13} for the definition of some ``canonical'' choice for the tetrad in order to fix this residual freedom. Such choice presents the advantage of providing a unique determination of the conserved-norm spin variable in a given gauge. This canonical choice will be the one adopted in all explicit results presented in Sect.~\ref{sec:fluxSO}.

The evolution equation \eqref{transparallel} for the original spin variable $S_\alpha$ now translates into an ordinary precession equation for the tetrad components $S_a$:
\begin{equation}\label{evolS}
\frac{\dd S_{a}}{\dd t} = \Omega_{a}^{\phantom{a}b} S_{b}\,,
\end{equation} 
where the precession tensor $\Omega_{ab}$ is related to the tetrad components $\omega_{AB}$ of the rotation tensor defined in \eqref{rot2} by $\Omega_{ab}=z\,\omega_{ab}$ where we pose $z\equiv\dd\tau/\dd t$, remembering the redshift variable \eqref{z1expr}. The antisymmetric character of the matrix $\Omega_{ab}$ guaranties that $S_a$ satisfies the Euclidean precession equation
\begin{equation}\label{prec}
\frac{\dd \bm{S}}{\dd t} = \bm{\Omega}\times\bm{S} \,,
\end{equation}
where we denote $\bm{S}=(S_a)$, and $\bm{\Omega}=(\Omega_a)$ with $\Omega_{a}=-\frac{1}{2}\epsilon_{abc}\,\Omega^{bc}$. As a consequence of \eqref{prec} the spin has a conserved Euclidean norm: $\bm{S}^2=s^2$. From now on we shall no longer make any distinction between the spatial tetrad indices $ab\cdots$ and the ordinary spatial indices $ij\cdots$ which are raised and lowered with the Kronecker metric. Explicit results for the equations of motion and gravitational wave templates will be given in Sects.~\ref{sec:eomspins} and \ref{sec:fluxSO} using the canonical choice for the conserved-norm spin variable $\bm{S}$.


\subsubsection{Equations of motion and precession for circular binaries}
\label{sec:eomspins}

The previous formalism can be generalized to self-gravitating systems consisting of two spinning point particles. The metric generated by the system of particles, interacting only through gravitation, is solution of the Einstein field equations \eqref{EinsteinG} with stress-energy tensor given by the sum of the individual stress-energy tensors \eqref{Tabspin} for each particles. The equations of motion of the particles are the Mathisson--Papapetrou equations \eqref{MathPapa} with the metric generated by the self-gravitating system evaluated at the location of the particles thanks to a regularization procedure (see Sect.~\ref{sec:reg}). The precession equations of each of the spins are given by
\begin{equation}\label{preceqn}
\frac{\dd \bm{S}_{\text{a}}}{\dd t} =
\bm{\Omega}_{\text{a}}\times\bm{S}_{\text{a}} \,,
\end{equation}
where $\text{a} = 1,2$ labels the particles. The spin variables $\bm{S}_\text{a}$ are the conserved-norm spins defined in Sect.~\ref{sec:lagspins}. In the following it is convenient to introduce two combinations of the individuals spins defined by\footnote{This convenient notation is adopted by \cite{K95}; the inverse formulas read (recall $X_\text{a}\equiv m_\text{a}/m$ and $\nu\equiv X_1X_2$)
\begin{align*}
\bm{S}_{1} = X_1 \bm{S} - \nu \bm{\Sigma}\,,\\
\bm{S}_{2} = X_2 \bm{S} + \nu \bm{\Sigma}\,.
\end{align*}
}
\begin{subequations}\label{defSSigma}
\begin{align}
\bm{S} &\equiv \bm{S}_{1} + \bm{S}_{2} \,,\\
\bm{\Sigma} &\equiv \frac{\bm{S}_{2}}{X_{2}} - \frac{\bm{S}_{1}}{X_{1}} \,.
\end{align}
\end{subequations}

We shall investigate the case where the binary's orbit is \emph{quasi-circular}, i.e., whose radius is constant apart from small perturbations induced by the spins (as usual we neglect the gravitational radiation damping effects). We denote by $\bm{x}=\bm{y}_1-\bm{y}_2$ and $\bm{v}=\dd\bm{x}/\dd t$ the relative position and velocity.\footnote{Note that the individual particle's positions $\bm{y}_\text{a}$ in the frame of the center-of-mass (defined by the cancellation of the center-of-mass integral: $\bm{\dG}=0$) are related to $\bm{x}$ and $\bm{v}$ by expressions similar to \eqref{CMrel}, but containing spin effects starting at the order 1.5PN, see Sect.~3 of \cite{BMFB13}.} We introduce an orthonormal moving triad $\{\bm{n}, \bm{\lambda}, \bm{\ell}\}$ defined by the unit separation vector $\bm{n}=\bm{x}/r$ (with $r=\vert\bm{x}\vert$) and the unit normal $\bm{\ell}$ to the instantaneous orbital plane given by $\bm{\ell}=\bm{n}\times\bm{v}/|\bm{n}\times\bm{v}|$; the orthonormal triad is then completed by $\bm{\lambda}=\bm{\ell}\times\bm{n}$. Those vectors are represented on Fig.~\ref{fig:geom}, which shows the geometry of the system. The orbital frequency $\Omega$ is defined for general orbits, not necessarily circular, by $\bm{v}=\dot{r}\bm{n}+r\Omega\bm{\lambda}$ where $\dot{r}=\bm{n}\cdot\bm{v}$ is the derivative of $r$ with respect to the coordinate time $t$. The general expression for the relative acceleration $\bm{a}\equiv\dd\bm{v}/\dd t$ decomposed in the moving basis $\{\bm{n},\bm{\lambda},\bm{\ell}\}$ is
\begin{equation}\label{acircspingen}
\bm{a} = \bigl(\ddot{r}-r\,\Omega^{2}\bigr)\bm{n}
+\bigl(r\,\dot{\Omega}+2\dot{r}\,\Omega\bigr)\bm{\lambda} +
r\,\varpi\,\Omega\,\bm{\ell} \,.
\end{equation}
Here we have introduced the \emph{orbital plane precession} $\varpi$ of the orbit defined by $\varpi\equiv-\bm{\lambda}\cdot \dd \bm{\ell}/\dd t$. Next we impose the restriction to quasi-circular precessing orbits which is defined by the conditions $\dot{r}=\dot{\Omega}=\calO(1/c^5)$ and $\ddot{r}=\calO(1/c^{10})$ so that $v^2 = r^2\Omega^2+\calO(1/c^{10})$; see Eqs. \eqref{rOmdot}. Then $\bm{\lambda}$ represents the direction of the velocity, and the precession frequency $\varpi$ is proportional to the variation of $\bm{\ell}$ in the direction of the velocity. In this way we find that the equations of the relative motion in the frame of the center-of-mass are
\begin{equation}\label{acircspin}
\bm{a} = - r\,\Omega^2\,\bm{n} + r\,\varpi\,\Omega\,\bm{\ell} +
\calO\left(\frac{1}{c^5}\right)\,.
\end{equation}
Since we neglect the radiation reaction damping there is no component of the acceleration along $\bm{\lambda}$. This equation represents the generalization of Eq.~\eqref{aieom} for spinning quasi-circular binaries with no radiation reaction. The orbital frequency $\Omega$ will contain spin effects in addition to the non-spin terms given by \eqref{keplerlaw}, while the precessional frequency $\varpi$ will entirely be due to spins. 

Here we report the results for the spin-orbit (SO) contributions into these quantities at the next-to-next-to-leading level corresponding to 3.5PN order \citep{MBFB13, BMFB13}. We project out the spins on the moving orthonormal basis, defining $\bm{S}= S_n \bm{n} + S_\lambda \bm{\lambda} + S_\ell \bm{\ell}$ and similarly for $\bm{\Sigma}$. We have
\begin{align}\label{omega2SO}
\Omega^2_\text{SO}&=\frac{\gamma^{3/2}}{m\,r^3}\biggl\{
-5S_\ell-3\Delta\Sigma _\ell +\gamma \left[\left(\frac{45}{2}
  -\frac{27}{2} \nu\right)S_\ell+\Delta \left(\frac{27}{2}
  -\frac{13}{2} \nu\right)\Sigma _\ell\right]\nn\\ &+\gamma^{2}
\left[\left(-\frac{495}{8} -\frac{561}{8} \nu -\frac{51}{8}
  \nu^2\right)S_\ell+ \Delta \left(-\frac{297}{8} -\frac{341}{8} \nu
  -\frac{21}{8} \nu^2\right)\Sigma _\ell\right]\biggr\} \nn\\& +
\calO\left(\frac{1}{c^8}\right) \,,
\end{align}
which has to be added to the non-spin terms \eqref{keplerlaw} up to 3.5PN order. We recall that the ordering post-Newtonian parameter is $\gamma = \frac{G m}{r c^2}$. The next-to-next-to-leading SO effects into the precessional frequency read
\begin{align}\label{varpiSO}
\varpi &= \frac{c^3 x^{3}}{G^2 m^3}\biggl\{ 7S_n+3\Delta\Sigma _n +x
\left[\left(-3 - 12 \nu\right)S_n+\Delta\left(-3 -\frac{11}{2}
  \nu\right)\Sigma _n\right]\nn\\ & +x^2 \left[\left(-\frac{3}{2} -
  \frac{59}{2} \nu + 9 \nu^2\right)S_n+\Delta\left(-\frac{3}{2} -
  \frac{77}{8} \nu + \frac{13}{3} \nu^2\right)\Sigma
  _n\right]\biggr\} \nn\\& +\calO\left(\frac{1}{c^8}\right)\,,
\end{align}
where this time the ordering post-Newtonian parameter is $x\equiv (\frac{G m \Omega}{c^3})^{2/3}$. In order to complete the evolution equations for quasi-circular orbits we need also the precession vectors $\bm{\Omega}_\text{a}$ of the two spins as defined by Eq.~\eqref{preceqn}. These are given by
\begin{align}\label{precessionSO}
\bm{\Omega}_1 &= \frac{c^3 x^{5/2}}{G\,m}\bm{\ell} \biggl\{
\frac{3}{4} + \frac{1}{2} \nu -\frac{3}{4}\Delta +x \left[\frac{9}{16}
  + \frac{5}{4} \nu -\frac{1}{24} \nu^2+\Delta\left(-\frac{9}{16} +
  \frac{5}{8} \nu\right)\right]\nn\\ &+x^2 \left[\frac{27}{32} +
  \frac{3}{16} \nu -\frac{105}{32} \nu^2 -\frac{1}{48}
  \nu^3+\Delta\left(-\frac{27}{32} + \frac{39}{8} \nu -\frac{5}{32}
  \nu^2\right)\right]\biggr\} \nn\\& +\calO\left(\frac{1}{c^8}\right)\,.
\end{align}
We obtain $\bm{\Omega}_2$ by exchanging the masses, $\Delta \rightarrow - \Delta$. At the linear SO level the precession vectors $\bm{\Omega}_\text{a}$ are independent of the spins.\footnote{Beware of our inevitably slightly confusing notation: $\Omega$ is the binary's \emph{orbital} frequency and $\Omega_\text{SO}$ refers to the spin-orbit terms therein; $\Omega_\text{a}$ is the \emph{precession} frequency of the $\text{a}$-th spin while $\varpi$ is the precession frequency of the orbital plane; and $\omega_\text{a}$ defined earlier in Eqs. \eqref{omegan} and \eqref{dmn} is the \emph{rotation} frequency of the $\text{a}$-th black hole. Such different notions nicely mix up in the first law of spinning binary black holes in Sect.~\ref{sec:firstlaw}; see Eq.~\eqref{firstlawspin} and the corotation condition \eqref{corotcond}.} Finally we give the SO contributions to next-to-next-to-leading order in the binary's conservative energy for circular orbits as \citep{MBFB13, BMFB13}
\begin{align}
	\label{ESO}
	\dE_\text{SO} &= -\frac{\nu c^2 x^{5/2}}{2G\,m} \biggl\{
	\frac{14}{3}S_\ell+2\Delta\Sigma_\ell \nn\\& \qquad + x \left[\left(11
	-\frac{61}{9} \nu\right)S_\ell+\Delta\left(3
	-\frac{10}{3} \nu\right)\Sigma _\ell\right]\nn\\ & \qquad +x^2
	\left[\left(\frac{135}{4} -\frac{367}{4} \nu + \frac{29}{12}
	\nu^2\right)S_\ell+\Delta\left(\frac{27}{4} -39 \nu +
	\frac{5}{4} \nu^2\right)\Sigma_\ell\right]\biggr\} \nn\\ & \qquad +
	\calO\left(\frac{1}{c^8}\right)\,.
\end{align}
We recall that the non-spin contributions in $\dE$ were given in Eq.~\eqref{Ecirc}.


\subsubsection{Orbital precession at the spin-orbit level}
\label{sec:orbprecession}

We now investigate an analytical solution for the dynamics of compact spinning binaries on quasi-circular orbits, including the effects of spin precession \citep{BBF11, MBBB13}. This solution will be valid whenever the radiation reaction effects can be neglected, and is restricted to the linear SO level.
\begin{figure}[htb]
\centerline{\includegraphics[width=0.9\textwidth]{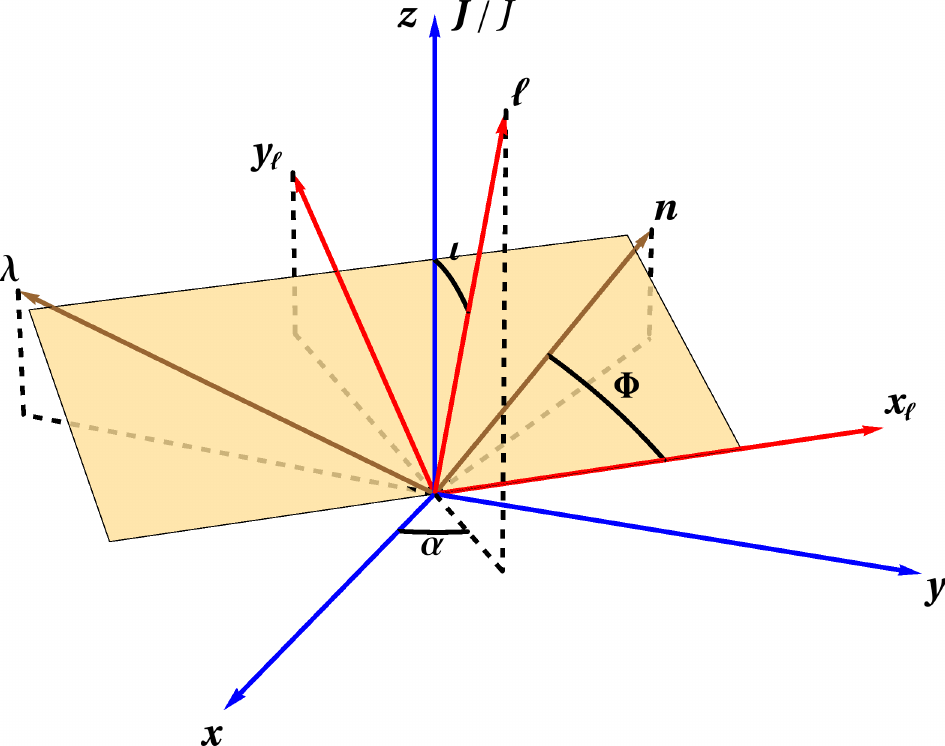}}
\caption{Geometric definitions for the precessional motion of spinning compact binaries. We show (i) the source frame defined by the fixed orthonormal basis $\{\bm{x},\bm{y},\bm{z}\}$; (ii) the instantaneous orbital plane which is described by the orthonormal basis $\{\bm{x}_\ell, \bm{y}_\ell, \bm{\ell}\}$; (iii) the moving triad $\{\bm{n},\bm{\lambda},\bm{\ell}\}$ and the associated three Euler angles $\alpha$, $\iota$ and $\Phi$; (iv) the direction of the total angular momentum $\mathbf{J}$ which coincides with the $z$--direction. Dashed lines show projections into the $x$--$y$ plane.}\label{fig:geom}
\end{figure}

In the following, we will extensively employ the total angular momentum of the system, that we denote by $\mathbf{J}$, and which is conserved when radiation-reaction effects are neglected,
\begin{equation}\label{Jconserved}
  \frac{\dd\mathbf{J}}{\dd t} = 0 \,.
\end{equation}
It is customary to decompose the conserved total angular momentum $\mathbf{J}$ as the sum of the orbital angular momentum $\bm{L}$ and of the two spins,\footnote{Recall from Eq.~\eqref{Sndef} that in our convention the spins have the dimension of an angular momentum times $c$.}
\begin{equation}\label{Jdecomp}
  \mathbf{J} = \bm{L} + \frac{\bm{S}}{c}\,.
\end{equation}
This split between $\bm{L}$ and $\bm{S}=\bm{S}_{1}+\bm{S}_{2}$ is specified by our choice of spin variables, here the conserved-norm spins defined in Sect.~\ref{sec:lagspins}. Although $\bm{L}$ is called the ``orbital'' angular momentum, it actually includes both non-spin and spin contributions. We refer to Eq.~(4.7) of \cite{BMFB13} for the expression of $\bm{L}$ at the next-to-next-to-leading SO level for quasi-circular orbits.

Our solution will consist of some explicit expressions for the moving triad $\{\bm{n}, \bm{\lambda}, \bm{\ell}\}$ at the SO level in the conservative dynamics for quasi-circular orbits. With the previous definitions of the orbital frequency $\Omega$ and the precessional frequency $\varpi$ we have the following system of equations for the time evolution of the triad vectors,
\begin{subequations}\label{precessionbasis}
\begin{align}
\frac{\dd \bm{n}}{\dd t} &= \Omega \,\bm{\lambda} \,, \\ \frac{\dd
  \bm{\lambda}}{\dd t} &= - \Omega\,\bm{n} + \varpi\,\bm{\ell} \,,
\\ \frac{\dd \bm{\ell}}{\dd t} &= - \varpi\,\bm{\lambda} \,.
\end{align}
\end{subequations}
Equivalently, introducing the orbital rotation vector $\bm{\Omega}\equiv\Omega\,\bm{\ell}$ and spin precession vector $\bm{\varpi}\equiv\varpi\,\bm{n}$, these equations can be elegantly written as
\begin{subequations}\label{precessionbasis2}
\begin{align}
\frac{\dd\bm{n}}{\dd t} &= \bm{\Omega}\times\bm{n}
\,,\\ \frac{\dd\bm{\lambda}}{\dd t} &= \bigl(\bm{\Omega}+
\bm{\varpi}\bigr)\times \bm{\lambda} \,, \\ \frac{\dd\bm{\ell}}{\dd t}
&= \bm{\varpi} \times\bm{\ell} \,.
\end{align}
\end{subequations}
Next we introduce a fixed (inertial) orthonormal basis $\{ \mathbf{x}, \mathbf{y}, \mathbf{z} \}$ as follows: $\mathbf{z}$ is defined to be the normalized value $\mathbf{J}/\dJ$ of the total angular momentum $\mathbf{J}$; $\mathbf{y}$ is orthogonal to the plane spanned by $\mathbf{z}$ and the direction $\bm{N}=\bm{X}/R$ of the detector as seen from the source (notation of Sect.~\ref{sec:radmult}) and is defined by $\mathbf{y} = \mathbf{z} \times \bm{N}/\vert\mathbf{z} \times \bm{N}\vert$; and $\mathbf{x}$ completes the triad -- see Fig.~\ref{fig:geom}. Then, we introduce the standard spherical coordinates $(\alpha, \iota)$ of the vector $\bm{\ell}$ measured in the inertial basis $\{\mathbf{x}, \mathbf{y}, \mathbf{z}\}$. Since $\iota$ is the angle between the total and orbital angular momenta, we have
\begin{equation} \label{siniota}
\sin \iota = \frac{\vert\mathbf{J} \times \bm{\ell}\vert}{\dJ} \,,
\end{equation}
where $\dJ\equiv\vert\mathbf{J}\vert$. The angles $(\alpha, \iota)$ are referred to as the \emph{precession angles}.

We now derive from the time evolution of our triad vectors those of the precession angles $(\alpha, \iota)$, and of an appropriate phase $\Phi$ that specifies the position of $\bm{n}$ with respect to some reference direction in the orbital plane denoted $\mathbf{x}_\ell$. Following \cite{ABFO08}, we pose
\begin{subequations}\label{xyell}
\begin{align}
\mathbf{x}_\ell &= \frac{\mathbf{J} \times \bm{\ell}}{\vert \mathbf{J} \times
  \bm{\ell} \vert}\,,\\\mathbf{y}_\ell &= \bm{\ell} \times
\mathbf{x}_\ell \,,
\end{align}
\end{subequations}
such that $\{\mathbf{x}_\ell, \mathbf{y}_\ell, \mathbf{\ell}\}$ forms an orthonormal basis. The motion takes place in the instantaneous orbital plane spanned by $\bm{n}$ and $\bm{\lambda}$, and the phase angle $\Phi$ is such that (see Fig.~\ref{fig:geom}):
\begin{subequations}\label{nlambdaexpr}
\begin{align}
\bm{n} &= \cos \Phi \, \mathbf{x}_\ell + \sin \Phi \,
\mathbf{y}_\ell\,, \label{nexpr} \\ \bm{\lambda} &= -\sin \Phi \,
\mathbf{x}_\ell + \cos \Phi \, \mathbf{y}_\ell \,,\label{lambdaexpr}
\end{align}
\end{subequations}
from which we deduce
\begin{equation}\label{expPhi}
e^{-\di \,\Phi} = \mathbf{x}_\ell\cdot\bigl(\bm{n}+\di
\bm{\lambda}\bigr) = \frac{\dJ_\lambda - \di\,\dJ_n
}{\sqrt{\dJ_n^2 + \dJ_\lambda^2}}\,.
\end{equation}
Combining Eqs. \eqref{expPhi} with \eqref{siniota} we also get
\begin{equation} \label{siniotaexpPhi}
\sin\iota\,\de^{-\di \,\Phi} = \frac{\dJ_\lambda -
  \di\,\dJ_n }{\dJ}\,.
\end{equation}
By identifying the right-hand sides of \eqref{precessionbasis} with the time-derivatives of the relations \eqref{nlambdaexpr} we obtain the following system of equations for the variations of $\alpha$, $\iota$ and $\Phi$,
\begin{subequations} 
\begin{align}
\label{alphadot}
\frac{\dd \alpha}{\dd t} &= \varpi\,\frac{\sin \Phi}{\sin \iota}\,,\\
\label{iotadot}
\frac{\dd \iota}{\dd t} &= \varpi\,\cos \Phi\,,\\
\label{Phidot}
\frac{\dd \Phi}{\dd t} &= \Omega - \varpi\,\frac{\sin \Phi}{\tan
  \iota}\,.
\end{align}
\end{subequations}

On the other hand, using the decompositon of the total angular momentum \eqref{Jdecomp} together with the fact that the components of $\bm{L}$ projected along $\bm{n}$ and $\bm{\lambda}$ are of the order $\calO(S)$, Eq.~(4.7) of \cite{BMFB13}, we deduce that $\sin\iota$ is itself a small quantity of order $\calO(S)$. Since we also have $\varpi=\calO(S)$, we conclude by direct integration of the sum of Eqs. \eqref{alphadot} and \eqref{Phidot} that
\begin{equation}\label{Phialpha}
\Phi + \alpha = \phi + \calO(S^2) \,,
\end{equation}
in which we have defined the ``carrier'' phase as 
\begin{equation}\label{carrier}
\phi \equiv \int \Omega \, \dd t = \Omega (t-t_0) + \phi_0 \,,
\end{equation}
with $\phi_0$ the value of the carrier phase at some arbitrary initial time $t_0$. An important point we have used when integrating \eqref{carrier} is that the orbital frequency $\Omega$ is constant at linear order in the spins. Indeed, from Eq.~\eqref{omega2SO} we see that only the components of the conserved-norm spin vectors along $\bm{\ell}$ can contribute to $\Omega$ at linear order. As we show in Eq.~\eqref{dtSell} below, these components are in fact constant at linear order in spins. Thus we can treat $\Omega$ as a constant for our purpose.

The combination $\Phi+\alpha$ being known by Eq.~\eqref{Phialpha}, we can further express the precession angles $\iota$ and $\alpha$ at linear order in spins in terms of the components $\dJ_n$ and $\dJ_\lambda$; from Eqs. \eqref{siniota} and \eqref{siniotaexpPhi},
\begin{subequations}\label{iotaalpha}
\begin{align}
\sin\iota &= \frac{\sqrt{\dJ_n^2 +
    \dJ_\lambda^2}}{L_\text{NS}} + \calO(S^2)\,,\\ \de^{\di
  \,\alpha} &= \frac{\dJ_\lambda - \di\,\dJ_n
}{\sqrt{\dJ_n^2+\dJ_\lambda^2}}\,\de^{\di \,\phi} +
\calO(S^2)\,,
\end{align}\end{subequations}
where we denote by $L_\text{NS}$ the norm of the non-spin (NS) part of the orbital angular momentum $\bm{L}$.

It remains to obtain the explicit time variation of the components of the two individual spins $S_n^\text{a}$, $S_\lambda^\text{a}$ and $S_\ell^\text{a}$ (with $\text{a}=1,2$). Using Eqs. \eqref{iotaalpha} together with the decomposition \eqref{Jdecomp} and the explicit expression of $\bm{L}$ (see \citealt{BMFB13}), we shall then be able to obtain the explicit time variation of the precession angles $(\alpha, \iota)$ and phase $\Phi$. Combining \eqref{preceqn} and \eqref{precessionbasis} we obtain
\begin{subequations}\label{spinevolv}
\begin{align}
\frac{\dd S_n^\text{a}}{\dd t} &= \bigl(\Omega - \Omega_\text{a}
\bigr) S_\lambda^\text{a}\,,\\ \frac{\dd S_\lambda^\text{a}}{\dd t} &=
- \bigl(\Omega - \Omega_\text{a} \bigr) S_n^\text{a} +
\varpi\,S_\ell^\text{a}\,,\\ \frac{\dd S_\ell^\text{a}}{\dd t} &= -
\varpi\,S_\lambda^\text{a}\,,
\end{align}
\end{subequations}
where $\Omega_\text{a}$ is the norm of the precession vector of the $\text{a}$-th spin as given by \eqref{precessionSO}, and the precession frequency $\varpi$ is explicitly given by \eqref{varpiSO}. At linear order in spins these equations translate into
\begin{subequations}\label{spinevolvSO}
\begin{align}
\frac{\dd S_n^\text{a}}{\dd t} &= \bigl(\Omega - \Omega_\text{a}
\bigr) S_\lambda^\text{a}\,,\\ \frac{\dd S_\lambda^\text{a}}{\dd t} &=
- \bigl(\Omega - \Omega_\text{a} \bigr) S_n^\text{a} +
\calO(S^{2})\,,\\ \frac{\dd S_\ell^\text{a}}{\dd t} &=
\calO(S^{2})\,.\label{dtSell}
\end{align}
\end{subequations}
We see that, as stated before, the spin components along $\bm{\ell}$ are constant, and so is the orbital frequency $\Omega$ given by \eqref{omega2SO}. At the linear SO level, the equations \eqref{spinevolvSO} can be decoupled and integrated as
\begin{subequations}\label{spinperp}
\begin{align}
S^\text{a}_n &= S_\perp^\text{a} \cos\psi_\text{a}
\,,\\ S^\text{a}_\lambda &= - S_\perp^\text{a} \sin\psi_\text{a}
\,,\\ S^\text{a}_\ell &= S_{\parallel}^\text{a} \,.
\end{align}
\end{subequations}
Here $S_\perp^\text{a}$ and $S_{\parallel}^\text{a}$ denote two quantities for each spins, that are constant up to terms $\calO(S^2)$. The phase of the projection perpendicular to the direction $\bm{\ell}$ of each of the spins is given by
\begin{equation}\label{psin}
\psi_\text{a} = (\Omega - \Omega_\text{a})(t-t_0)+\psi^0_\text{a}\,,
\end{equation}
where $\psi^0_\text{a}$ is the constant initial phase at the reference time $t_0$.

Finally we can give in an explicit way, to linear SO order, the moving triad $\{\bm{n}, \bm{\lambda}, \bm{\ell}\}$ in terms of the reference triad $\{\bm{n}_{0}, \bm{\lambda}_{0}, \bm{\ell}_{0}\}$ at the reference time $t_{0}$ in Eqs. \eqref{psin} and \eqref{carrier}. The best way to express the result is to introduce the complex null vector $\bm{m} \equiv\frac{1}{\sqrt{2}}(\bm{n}+\di\bm{\lambda})$ and its complex conjuguate $\overline{\bm{m}}$; the normalization is chosen so that $\bm{m}\cdot\overline{\bm{m}}=1$. We obtain
\begin{subequations}\label{solutionml}
\begin{align}
\bm{m} &= \de^{-\di(\phi-\phi_{0})} \bm{m}_{0} + \frac{\di}{\sqrt{2}}
\left( \sin \iota \, \de^{\di\alpha} - \sin\iota_{0} \, \de^{\di
  \alpha_{0}}\right) \de^{-\di\phi} \bm{\ell}_{0} + \calO(S^2)
\,,\label{solutionm} \\ \bm{\ell} &= \bm{\ell}_{0} + \left[
  \frac{\di}{\sqrt{2}} \left( \sin \iota \, \de^{-\di\alpha} -
  \sin\iota_{0} \, \de^{-\di \alpha_{0}}\right) \de^{\di\phi_{0}}
  \bm{m}_{0} + \mathrm{c.c.} \right] + \calO(S^2)
\,.\label{solutionl}
\end{align}
\end{subequations}
The precession effects in the dynamical solution for the evolution of the basis vectors $\{\bm{n}, \bm{\lambda}, \bm{\ell}\}$ are given by the second terms in these equations. They depend only in the combination $\sin\iota\,e^{\di \alpha}$ and its complex conjugate $\sin\iota\,e^{- \di \alpha}$, which follows from Eqs. \eqref{iotaalpha} and the known spin and non-spin contributions to the total angular momentum $\mathbf{J}$. One can check that precession effects in the above dynamical solution \eqref{solutionml} for the moving triad start at order $\calO(1/c^3)$.


\subsubsection{Spin-orbit effects in the gravitational-wave flux and orbital phase}
\label{sec:fluxSO}

Like in Sect.~\ref{sec:GW} our main task is to control up to high post-Newtonian order the mass and current radiative multipole moments $\dU_L$ and $\dV_L$ which parametrize the asymptotic waveform and gravitational fluxes far away from the source, cf. Eqs. \eqref{hijTT}--\eqref{FluxFG}. The radiative multipole moments are in turn related to the source multipole moments $\dI_L$ and $\dJ_L$ through complicated relationships involving tails and related effects, and described in Sects.~\ref{sec:radcanonical} and \ref{sec:cansource}.

The source moments have been expressed in Eqs. \eqref{sourcemoments} in terms of some source densities $\Sigma$, $\Sigma_i$ and $\Sigma_{ij}$ defined from the components of the post-Newtonian expansion of the pseudo-tensor, denoted $\overline{\tau}^{\alpha\beta}$. To lowest order the (PN expansion of the) pseudo-tensor reduces to the matter tensor $T^{\alpha\beta}$ which has compact support, and the source densities $\Sigma$, $\Sigma_i$, $\Sigma_{ij}$ reduce to the compact support quantities $\sigma$, $\sigma_i$ $\sigma_{ij}$ given by Eqs. \eqref{sigmadef}. Now, computing spin effects, the matter tensor $T^{\alpha\beta}$ has been found to be given by \eqref{Tabspin} in the framework of the pole-dipole approximation suitable for SO couplings (and sufficient also for SS interactions between different spins). Here, to give a flavor of the computation, we present the lowest order SO contributions to the general mass and current source multipole moments ($\forall\ell\in\mathbb{N}$):
\begin{subequations}\label{IJSO}
\begin{align}
\dI_L^\text{SO} &= \frac{2\ell \,\nu}{c^3(\ell+1)}\biggl\{
\ell\left[\sigma_\ell(\nu)(\bm{v}\times\bm{S})^{\langle
    i_\ell}-\sigma_{\ell+1}(\nu)(\bm{v}\times\bm{\Sigma})^{\langle
    i_\ell}\right] x^{L-1\rangle}\nn\\&\qquad\quad
-(\ell-1)\left[\sigma_\ell(\nu)(\bm{x}\times\bm{S})^{\langle
    i_\ell}-\sigma_{\ell+1}(\nu)(\bm{x}\times\bm{\Sigma})^{\langle
    i_\ell}\right]
v^{i_{\ell-1}}x^{L-2\rangle}\biggr\} \nn\\& +
\calO\left(\frac{1}{c^5}\right)\,,\\ 
\dJ_L^\text{SO} &= \frac{(\ell+1)\nu}{2
  c}\left[\sigma_{\ell-1}(\nu)S^{\langle
    i_\ell}-\sigma_{\ell}(\nu)\Sigma^{\langle i_\ell}\right]
x^{L-1\rangle}+ \calO\left(\frac{1}{c^3}\right)\,.
\end{align}\end{subequations}
Paralleling the similar expressions \eqref{IJnewtonian} for the Newtonian approximation to the source moments in the non-spin case, we posed $\sigma_\ell(\nu)\equiv X_2^{\ell-1}+(-)^\ell X_1^{\ell-1}$, see also Eqs. \eqref{sigmaell}. In Eqs. \eqref{IJSO} we employ the notation \eqref{defSSigma} for the two spins and the ordinary cross product $\times$ of Euclidean vectors. Thus, the dominant level of spins is at the 1.5PN order in the mass-type moments $\dI_L$, but only at the 0.5PN order in the current-type moments $\dJ_L$. It is then evident that the spin part of the current-type moments will always dominate over that of the mass-type moments. We refer to \cite{BBF06, BMB13} for higher order post-Newtonian expressions of the source moments. If we insert the expressions \eqref{IJSO} into tail integrals like \eqref{radtail}, we find that some spin contributions originate from tails starting at the 3PN order (\citealt{BBF11, MBBB13}; the equivalent calculation has been done with EFT techniques: \citealt{CPY22}).

We are left with the following high-order result for the SO contributions to the gravitational-wave energy flux, with 4PN level precision \citep{BBF06, BMB13, MBBB13, CPY22}:
%
\begin{align}\label{fluxSO}
\mathcal{F}_\text{SO} &= \frac{32\nu^2c^5x^{13/2}}{5G^2
  m^2}\left\{ -4S_\ell -\frac{5}{4}\Delta\Sigma_\ell \right.
\nn\\&\left.+ x \left[
  \left(-\frac{9}{2}+\frac{272}{9}\nu\right)S_\ell
  +\left(-\frac{13}{16}+\frac{43}{4}\nu\right)\Delta
    \Sigma_\ell\right]\right.\nn\\&\left.+ x^{3/2}
\left[ -16 \pi\,S_\ell -\frac{31\pi}{6}\,\Delta
    \Sigma_\ell\right]\right.  \nn\\ &+ x^2
\left[\left(\frac{476645}{6804}+\frac{6172}{189}\nu
  -\frac{2810}{27}\nu^2\right)S_\ell
  +\left(\frac{9535}{336}+\frac{1849}{126}\nu
  -\frac{1501}{36}\nu^2\right)\Delta\Sigma_\ell \right]
\nn\\ &\left.+ x^{5/2} \left[ \left( -\frac{3485 \pi}{96}
  + \frac{13879 \pi}{72}\nu \right) S_{\ell} + \left( -\frac{7163
    \pi}{672} + \frac{130583 \pi}{2016}\nu \right)\Delta \Sigma_{\ell}
  \right] \right.\nn\\&\left. + \calO\left(\frac{1}{c^6}\right)\right\}\,.
\end{align}
We recall that $S_\ell\equiv\bm{\ell}\cdot\mathbf{S}$ and $\Sigma_\ell\equiv\bm{\ell}\cdot\bm{\Sigma}$, with $\mathbf{S}$ and $\bm{\Sigma}$ denoting the combinations \eqref{defSSigma}, and the individual spins are the specific conserved-norm spins that have been constructed in Sect.~\ref{sec:lagspins}. The result \eqref{fluxSO} superposes to the non-spin contributions given by Eq.~\eqref{Fluxx}. Satisfyingly it is in complete agreement in the test-mass limit where $\nu\to 0$ with the result of black-hole perturbation theory on a Kerr background obtained by \cite{TSTS96}.

Finally we can compute the spin effects in the time evolution of the binary's orbital frequency $\Omega$. As in Sect.~\ref{sec:GW} we rely on Eq.~\eqref{baleq} balancing the total emitted energy flux $\mathcal{F}$ with the variation of the binary's center-of-mass energy $\dE$. The SO part of the energy $\dE$ have been provided in Eq.~\eqref{ESO}. Using $\dE$ and $\mathcal{F}$ expressed as functions of the orbital frequency $\Omega$ (through $x$) and of the spin variables (through $S_\ell$ and $\Sigma_\ell$), we transform the balance equation into
\begin{equation}\label{balanceOmega}
\dot{\Omega}_\text{SO} = - \left(\frac{\mathcal{F}}{\dd E/\dd
  \Omega}\right)_\text{SO}\,.
\end{equation}

However, in writing the latter equation it is important to justify that the spin quantities $S_\ell$ and $\Sigma_\ell$ are secularly constant, i.e., do not evolve on a gravitational radiation reaction time scale so we can neglect their variations when taking the time derivative of Eq.~\eqref{ESO}. Fortunately, this is the case of the conserved-norm spin variables, as proved by \cite{W05} up to relative 1PN order, i.e., considering radiation reaction effects up to 3.5PN order. Furthermore this can also be shown from the following structural general argument valid at linear order in spins \citep{BBF11, BMB13}. In the center-of-mass frame, the only vectors at our disposal, except for the spins, are $\bm{n}$ and $\bm{v}$. Recalling that the spin vectors are pseudovectors regarding parity transformation, we see that the only way SO contributions can enter scalars such as the energy $\dE$ or the flux $\mathcal{F}$ is through the mixed products $(n,v,S_\text{a})$, i.e., through the components $S_{\ell}^\text{a}$. Now, the same reasoning applies to the precession vectors $\bm{\Omega}_\text{a}$ in Eqs. \eqref{preceqn}: They must be pseudovectors, and, at linear order in spin, they must only depend on $\bm{n}$ and $\bm{v}$; so that they must be proportional to $\bm{\ell}$, as can be explicitly seen for instance in Eq.~\eqref{precessionSO}. Now, the time derivative of the components along $\bm{\ell}$ of the spins are given by $\dd S_{\ell}^\text{a}/\dd t=\mathbf{S}_\text{a}\cdot(\dd \bm{\ell}/\dd t+\bm{\ell}\times\bm{\Omega}_\text{a})$. The second term vanishes because $\bm{\Omega}_\text{a}\propto\bm{\ell}$, and since $\dd \bm{\ell}/\dd t = \calO(S)$, we obtain that $S_{\ell}^\text{a}$ is constant at \emph{linear} order in the spins. We have already met an instance of this important fact in Eq.~\eqref{dtSell}. This argument is valid at any post-Newtonian order and for general orbits, but is limited to spin-orbit terms; furthermore it does not specify any time scale for the variation, so it applies to short time scales such as the orbital and precessional periods, as well as to the long gravitational radiation reaction time scale (see also \citealt{Ger00} and references therein for related discussions).

We are then allowed to apply Eq.~\eqref{balanceOmega} with conserved-norm spin variables at the SO level. We thus obtain the secular evolution of $\Omega$ and from that we deduce by a further integration (following the Taylor approximant T2) the secular evolution of the carrier phase $\phi\equiv\int\Omega\,\dd t$:
\begin{align}\label{phaseSO}
\phi_\text{SO} &=-\frac{x^{-1}}{32\,G\,m^2\,\nu}\left\{
\frac{235}{6}S_\ell +\frac{125}{8}\Delta\Sigma_\ell
\right. \nn\\&\left.\quad+x \ln x \left[
  \left(-\frac{554345}{2016}-\frac{55}{8}\nu\right)S_\ell
  +\left(-\frac{41745}{448}+\frac{15}{8}\nu\right)\Delta
  \Sigma_\ell\right]\right.  \nn\\ &\left.\quad+ x^{3/2} \left[
  \frac{940\pi}{3}\,S_\ell +\frac{745\pi}{6}\,\Delta
  \Sigma_\ell\right]\right.  \nn\\ &\quad+ x^2 \left[
  \left(-\frac{8980424995}{6096384}+\frac{6586595}{6048}\nu
  -\frac{305}{288}\nu^2\right)S_\ell\right.
  \nn\\ &\qquad\qquad\quad\left.
  +\left(-\frac{170978035}{387072}
  +\frac{2876425}{5376}\nu+\frac{4735}{1152}\nu^2\right)\Delta
    \Sigma_\ell \right]\nn\\ &\quad\left. + x^{5/2} \left[
  \left( \frac{2388425 \pi}{3024} - \frac{9925 \pi}{36}\nu \right)
  S_{\ell} + \left( \frac{3237995 \pi}{12096} - \frac{258245
    \pi}{2016}\nu \right)\Delta \Sigma_{\ell} \right] \right.\nn\\&\quad\left. +
\calO\left(\frac{1}{c^6}\right)\right\}\,.
\end{align}
This expression, when added to the expression for the non-spin terms given by Eq.~\eqref{phix}, and considering also the SS terms, constitutes the main theoretical input needed for the construction of templates for spinning compact binaries. However, recall that in the case of precessional binaries, for which the spins are not aligned or anti-aligned with the orbital angular momentum, we must subtract to the carrier phase $\phi$ the precessional correction $\alpha$ arising from the precession of the orbital plane. Indeed the physical phase variable $\Phi$ which is defined in Fig.~\ref{fig:geom}, has been proved to be given by $\Phi=\phi-\alpha$ at linear order in spins, cf. Eq.~\eqref{Phialpha}. The precessional correction $\alpha$ can be computed at linear order in spins from the results of Sect.~\ref{sec:eomspins}.


\subsubsection{Spin-spin effects}
\label{sec:SS}

The SS effects are known at the leading level corresponding to 2PN order from \cite{BOC75, BOC79} in the equations of motion (see \citealt{K95, Po06, BuonFH13} for subsequent derivations), and from \cite{KWW93, K95} in the radiation field. Next-to-leading SS contributions are at 3PN order and have been obtained with the Hamiltonian \citep{SHS08a, SHS08b, SHS08c, HSS10, HaS11}, the EFT \citep{PoR06, PoR08a, PoR08b, Le10ss, LS15a} and the harmonic coordinates \citep{BFMP15} techniques. \cite{BFMP15, CPP21} obtained also the next-to-leading SS terms in the gravitational-wave flux. The next-to-next-to-next-to-leading order SS effects in the equations of motion have been obtained by \cite{Antonelli2020b, Kim2023b, Mandal2023b}.

With SS effects in a compact binary system one must make a distinction between the \emph{spin squared} terms, involving the coupling between the two same spins $S_1$ or $S_2$, and the \emph{interaction} terms, involving the coupling between the two different spins $S_1$ and $S_2$. The spin-squared terms $S_1^2$ and $S_2^2$ arise due to the effects on the dynamics of the quadrupole moments of the compact bodies that are induced by their spins, as shown by \cite{P97quad}. They have been computed through 2PN order in the fluxes and orbital phase by \cite{Ger99, Ger00, MVGer05}. The interaction terms $S_1\times S_2$ are simpler, as they can be computed with the pole-dipole formalism reviewed in Sect.~\ref{sec:lagspins}. The interaction terms $S_1\times S_2$ between different spins have been derived to next-to-next-to-leading 4PN order for the equations of motion by \cite{Le12ss, LS16} (EFT) and by \cite{HaS11ss} (Hamiltonian). 

The SS interactions (spin-squared terms $S_1^2$ and $S_2^2$) can be computed by representing each compact object as a spinning point particle whose internal structure and finite-size effects are parametrized by an effective quadrupole moment. The dynamics of such particles with higher order multipoles was investigated by \cite{Dixon64,Dixon73,Dixon79,BIsrael75} who constructed the effective stress-energy tensor and Lagrangian to encode information about the internal structure of the body. In the case of a spinning particle endowed with a quadrupolar structure the equations generalizing the Mathisson-Papapetrou equations \eqref{MathPapa} and \eqref{precgen} read (with $c=1$)
\begin{subequations}\label{Dixon}
\begin{align}
	\frac{\dD p_\alpha}{\dd \tau} &= -\frac{1}{2} u^\beta
	R_{\alpha\beta\mu\nu} S^{\mu\nu} - \frac{1}{3}\nabla_\lambda R_{\alpha\beta\mu\nu} J^{\lambda\beta\mu\nu}\,,\\
	\frac{\dD S^{\alpha\beta}}{\dd \tau} &= 2p^{[\alpha} u^{\beta]} +
	\frac{4}{3} R^{[\alpha}_{\phantom{\alpha}\gamma\mu\nu} J^{\beta]\gamma\mu\nu}\,,
\end{align}
\end{subequations}
where $J^{\alpha\beta\mu\nu}$ is the Dixon quadrupolar tensor, which is at this stage only constrained to have the same symmetry properties as $R_{\alpha\beta\mu\nu}$. We account for the correct number of degrees of freedom of the spin by imposing the same covariant SSC \eqref{SSC} as before. Similarly for the quadrupole tensor $J^{\alpha\beta\mu\nu}$, not all of its degrees of freedom are physical, and we restrict these by relating the quadrupole tensor to the four-velocity $u^\mu$ and the spin tensor $S^{\mu\nu}$. In order to describe a physical body with a spin and a non-vanishing spin-induced quadrupole moment one has \citep{Po06, PoR08b, SP10, St11rev, BuonFH13, BFMP15}
\begin{equation}
	J^{\alpha\beta\mu\nu} = \frac{3\kappa}{m}S^{\lambda[\alpha}u^{\beta]}S_{\lambda}^{\phantom{\lambda}\![\mu}u^{\nu]} \,,
	\label{eq:Jdef}
\end{equation}
where the mass parameter is defined by $m^2=-g^{\mu\nu}p_\mu p_\nu$ as in Sect.~\ref{sec:lagspins}, but is here constant only modulo higher-order $\calO(S^2)$ terms. The constant $\kappa$ characterizes the quadrupolar ``polarisability'' of the body; for a neutron star the numerical value of $\kappa$ depends on the equation of state of the nuclear matter.

As in Sect.~\ref{sec:lagspins} we apply the formalism to self-gravitating binary systems (strictly speaking the formalism is applied to spin-squared contributions $S_1^2$ and $S_2^2$; the pole-dipole formalism is sufficient for $S_1\times S_2$). We present the complete SS contributions, including all $S_1^2$, $S_2^2$ and $S_1\times S_2$ terms, to the conservative energy in the case of spins aligned with the orbital angular momentum and for circular orbits \citep{BFMP15}:
\begin{align}\label{ESS}
	\dE_{\rm SS} &= - \frac{\nu c^2 x^3}{2G^{2} m^{3}} 
	\biggl\{ 
	S_{\ell}^2 \bigl(-\kappa_{+}-2\bigr)  + 
	S_{\ell} \Sigma_{\ell} \Bigl(-\Delta  \kappa_{+}-
	2 \Delta +\kappa_{-}\Bigr) \nn \\
	& \quad\qquad + 
	\Sigma_{\ell}^2 \left( \left(\frac{\Delta  \kappa_{-}}{2}-
	\frac{\kappa_{+}}{2}\right)  + 
	\nu \bigl(\kappa_{+}+2\bigr) \right)  \nn \\
	& \quad + x \left[ S_{\ell}^2 
	\left( \left(-\frac{5 \Delta  \kappa_{-}}{3} -
	\frac{25 \kappa_{+}}{6} + \frac{50}{9}\right) + 
	\nu \left(\frac{5 \kappa_{+}}{6} + 
	\frac{5}{3}\right) \right) \right. \nn \\
	& \quad\qquad \left. + 
	S_{\ell} \Sigma_{\ell} \left( \left(-\frac{5 \Delta  \kappa_{+}}{2} +
	\frac{25 \Delta }{3} + \frac{5 \kappa_{-}}{2}\right) + 
	\nu \left(\frac{5 \Delta  \kappa_{+}}{6} + \frac{5 \Delta }{3} +
	\frac{35 \kappa_{-}}{6}\right) \right) \right. \nn \\
	& \quad\qquad \left. + \Sigma_{\ell}^2 
	\left( \left(\frac{5 \Delta  \kappa_{-}}{4} - \frac{5 \kappa_{+}}{4} +
	5\right)  + \nu \left(\frac{5 \Delta  \kappa_{-}}{4} + 
	\frac{5 \kappa_{+}}{4} - 10\right)  \right.\right. \nn \\
	& \quad\qquad
	\left.\left.  + \nu ^2 \left(-\frac{5 \kappa_{+}}{6}-
	\frac{5}{3}\right) \right) \right] + \calO\left(\frac{1}{c^4}\right)\biggr\} \,.
\end{align}
This expression is to be added to the non-spin contributions \eqref{Ecirc} and to the SO ones given in Eq.~\eqref{ESO}. Here $\kappa_{+}=\kappa_1+\kappa_2$ and $\kappa_{-}=\kappa_1-\kappa_2$ where $\kappa_\text{a}$ are the polarisability parameters of the two bodies. The expression \eqref{ESS} can be shown to be in agreement, in the test-mass limit for one body, with the energy of a test particle in circular equatorial orbits around a Kerr black hole \citep{BPT72}. 

The energy flux for SS interactions -- thus complementing the non-spin \eqref{Fluxx} and SO \eqref{fluxSO} terms -- in the case of aligned spins (no orbital precession) and circular orbits, reads \citep{BFMP15, CPP21}
\begin{align}\label{FSScirc}
	\mathcal{F}_{\rm SS} &= \frac{32 \nu^{2}c^{5}x^{7}}{5G^3m^4} \biggl\{  S_{\ell}^2
	\left(2 \kappa_{+}+4\right) + S_{\ell} \Sigma_{\ell} \left(2 \Delta
	\kappa_{+}+4 \Delta -2 \kappa_{-}\right)  \nn \\
	& \qquad\quad  + \Sigma_{\ell}^2
	\left( \left(-\Delta  \kappa_{-}+\kappa_{+}+\frac{1}{16}\right) + \nu
	\left(-2 \kappa_{+}-4\right) \right)   \nn \\
	& \quad + x \left[ S_{\ell}^2 \left(
	\left(\frac{41 \Delta  \kappa_{-}}{16}-\frac{271
		\kappa_{+}}{112}-\frac{5239}{504}\right) + \nu \left(-\frac{43
		\kappa_{+}}{4}-\frac{43}{2}\right) \right) \right. \nn \\
	& \qquad\quad + S_{\ell} \Sigma_{\ell}
	\left( \left(-\frac{279 \Delta  \kappa_{+}}{56}-
	\frac{817 \Delta}{56}+\frac{279 \kappa_{-}}{56}\right) 
	\right.\nn \\
	& \qquad\quad \left. + 
	\nu \left(-\frac{43 \Delta  \kappa_{+}}{4}-
	\frac{43 \Delta }{2}+\frac{\kappa_{-}}{2}\right) \right)  
	 \nn \\
	& \qquad\quad  + 
	\Sigma_{\ell}^2 \left( \left(\frac{279 \Delta  \kappa_{-}}{112}-
	\frac{279 \kappa_{+}}{112}-\frac{25}{8}\right)  + 
	\nu \left(\frac{45 \Delta  \kappa_{-}}{16}+
	\frac{243 \kappa_{+}}{112}+\frac{344}{21}\right) 
	\right. \nn \\
	& \qquad\quad \left.\left. + 
	\nu ^2 \left(\frac{43 \kappa_{+}}{4}+
	\frac{43}{2}\right) \right)  \right] + \calO\left(\frac{1}{c^3}\right) \biggr\} \,.
\end{align}
This result is in agreement, in the limit of a test particle orbiting a Kerr black hole, with the black hole perturbation calculation of \cite{TSTS96}. Using \eqref{FSScirc} together with the expression of the orbital energy \eqref{ESS}, we can solve the energy balance equation to obtain the phase evolution of the binary for circular orbits and aligned spins. This yields \citep{BFMP15}
\begin{align}
	\phi_\text{SS} &= -\frac{x^{-1/2}}{32G^2 m^4 \nu} 
	\biggl\{ S_{\ell}^2 
	\bigl(-25 \kappa_{+}-50\bigr)  + 
	S_{\ell} \Sigma_{\ell} \bigl(-25 \Delta  \kappa_{+}-50 \Delta +
	25 \kappa_{-}\bigr) \nn \\
	& \left. \qquad\quad + 
	\Sigma_{\ell}^2 \left( \left(\frac{25 \Delta  \kappa_{-}}{2}-
	\frac{25 \kappa_{+}}{2}-\frac{5}{16}\right)  + 
	\nu \left(25 \kappa_{+}+50\right) \right)  \right. \nn \\
	& \left. \quad + x \biggl[ S_{\ell}^2 
	\left( \left(\frac{2215 \Delta  \kappa_{-}}{48}+
	\frac{15635 \kappa_{+}}{84}-\frac{31075}{126}\right)  + 
	\nu \left(30 \kappa_{+}+60\right) \right)  \right. \nn \\
	& \left.\left. \qquad\quad + 
	S_{\ell} \Sigma_{\ell} \left( 
	\left(\frac{47035 \Delta  \kappa_{+}}{336}-
	\frac{9775 \Delta }{42}-
	\frac{47035 \kappa_{-}}{336}\right)  \right.\right.\right. \nn \\
	& \left.\left.\left. \qquad\qquad\quad  + 
	\nu \left(30 \Delta  \kappa_{+}+60 \Delta -
	\frac{2575 \kappa_{-}}{12}\right) \right)  \right.\right. \nn \\
	& \left.\left. \qquad\quad  + 
	\Sigma_{\ell}^2 \left( 
	\left(-\frac{47035 \Delta  \kappa_{-}}{672}+
	\frac{47035 \kappa_{+}}{672}-\frac{410825}{2688}\right)  
	\right.\right.\right. \nn \\
	& \left.\left.\left. \qquad\quad  + 
	\nu \left(-\frac{2935 \Delta  \kappa_{-}}{48}-
	\frac{4415 \kappa_{+}}{56}+\frac{23535}{112}\right) 
	\right.\right.\right. \nn \\
	& \left.\qquad\quad + 
	\nu ^2 \left(-30 \kappa_{+}-60\right) \right) 
	 \biggr] + \calO\left(\frac{1}{c^3}\right) \biggr\} \,.
\end{align}

Concerning the waveform the spin contributions for spin-aligned circular binaries have been included in the polarization modes at the 3.5PN order \citep{HMK22}; for non-circular binaries (with aligned or anti-aligned spins) they are known to the third post-Newtonian order \citep{HK23}. We mention also the calculations of the subdominant cubic spin-spin-spin (SSS) and even quartic SSSS effects \citep{HergtS08a, HergtS08b, LS15a, M15, Vaidya15}.

\phantomsection
\addcontentsline{toc}{section}{References}
\bibliographystyle{spbasic-FS}      
\bibliography{ListeRef_LRR_2024}

\begin{thebibliography}{604}
\expandafter\ifx\csname url\endcsname\relax
 \def\url#1{\burl{#1}}\fi
\expandafter\ifx\csname urlprefix\endcsname\relax\def\urlprefix{URL }\fi
\providecommand{\bibinfo}[2]{#2}
\providecommand{\eprint}[2][]{\url{#2}}
\providecommand{\doi}[1]{\urlstyle{rm}\url{https://doi.org/#1}}

\bibitem[{Abbott et~al.(2016{\natexlab{a}})}]{LIGOtestGR}
Abbott B, et~al. (2016{\natexlab{a}}) {Tests of general relativity with
  {GW150914}}. Phys Rev Lett 116:221101. \doi{10.1103/PhysRevLett.116.221101}.
  {\href{https://arxiv.org/abs/1602.03841}{{arXiv:1602.03841}}} {[gr-qc]}

\bibitem[{Abbott et~al.(2017{\natexlab{a}})}]{LIGOmultmess}
Abbott B, et~al. (2017{\natexlab{a}}) {Multi-messenger Observations of a Binary
  Neutron Star Merger}. Astrophys J Lett 848:L12.
  {\href{https://arxiv.org/abs/arXiv:1710.05833}{{arXiv:1710.05833}}} {[gr-qc]}

\bibitem[{Abbott et~al.(2019)Abbott, Abbott, Abbott, Acernese, Ackley, Adams,
  Adams, Addesso, Adhikari, Adya et~al.}]{LIGOtestGR2}
Abbott BP, Abbott R, Abbott T, Acernese F, Ackley K, Adams C, Adams T, Addesso
  P, Adhikari RX, Adya VB, et~al. (2019) {Tests of general relativity with
  {GW170817}}. Phys Rev Lett 123(1):011102.
  \doi{10.1103/PhysRevLett.123.011102}

\bibitem[{Abbott et~al.(2016{\natexlab{b}})}]{GW150914}
Abbott BP, et~al. (2016{\natexlab{b}}) {Observation of Gravitational Waves from
  a Binary Black Hole Merger}. Phys Rev Lett 116:061102.
  \doi{10.1103/PhysRevLett.116.061102}.
  {\href{https://arxiv.org/abs/1602.03837}{{arXiv:1602.03837}}} {[gr-qc]}

\bibitem[{Abbott et~al.(2017{\natexlab{b}})}]{GW170817}
Abbott BP, et~al. (2017{\natexlab{b}}) {GW170817: Observation of Gravitational
  Waves from a Binary Neutron Star Inspiral}. Phys Rev Lett 119:161101.
  \doi{10.1103/PhysRevLett.119.161101}.
  {\href{https://arxiv.org/abs/1710.05832}{{arXiv:1710.05832}}} {[gr-qc]}

\bibitem[{Abdelsalhin et~al.(2018)Abdelsalhin, Gualtieri, and Pani}]{AGP18}
Abdelsalhin T, Gualtieri L, Pani P (2018) {Post-{N}ewtonian spin-tidal
  couplings for compact binaries}. Phys Rev D 98.
  \doi{10.1103/PhysRevD.98.104046}

\bibitem[{Abramowicz and Klu{\'{z}}niak(2001)}]{AK01}
Abramowicz MA, Klu{\'{z}}niak W (2001) {A precise determination of black hole
  spin in {GRO J1655-40}}. Astron Astrophys 374:L19--L20.
  \doi{10.1051/0004-6361:20010791}.
  {\href{https://arxiv.org/abs/astro-ph/0105077}{{astro-ph/0105077}}}

\bibitem[{{Adamcewicz} et~al.(2024){Adamcewicz}, {Galaudage}, {Lasky}, and
  {Thrane}}]{ALT23}
{Adamcewicz} C, {Galaudage} S, {Lasky} PD, {Thrane} E (2024) {Which black hole
  is spinning? Probing the origin of black hole spin with gravitational waves}.
  Astrophys J Lett 964(1):L6. \doi{10.3847/2041-8213/ad2df2}.
  {\href{https://arxiv.org/abs/2311.05182}{{arXiv:2311.05182}}} {[astro-ph.HE]}

\bibitem[{Ajith et~al.(2005)Ajith, Iyer, Robinson, and Sathyaprakash}]{AIRS05}
Ajith P, Iyer BR, Robinson CAK, Sathyaprakash BS (2005) {New class of
  post-{Newtonian} approximants to the waveform templates of inspiralling
  compact binaries: Test mass in the {Schwarzschild} spacetime}. Phys Rev D
  71:044029. {\href{https://arxiv.org/abs/gr-qc/0412033}{{gr-qc/0412033}}}

\bibitem[{Ajith et~al.(2008)Ajith, Babak, Chen, Hewitson, Krishnan, Sintes,
  Whelan, Br{\"{u}}gmann, Diener, Dorband, Gonzalez, Hannam, Husa, Pollney,
  Rezzolla, Santamar{\'{\i}}a, Sperhake, and Thornburg}]{Ajith08}
Ajith P, Babak S, Chen Y, Hewitson M, Krishnan B, Sintes AM, Whelan JT,
  Br{\"{u}}gmann B, Diener P, Dorband N, Gonzalez J, Hannam M, Husa S, Pollney
  D, Rezzolla L, Santamar{\'{\i}}a L, Sperhake U, Thornburg J (2008) {Template
  bank for gravitational waveforms from coalescing binary black holes:
  Nonspinning binaries}. Phys Rev D 77:104017.
  \doi{10.1103/PhysRevD.77.104017}, {Erratum}: Phys Rev D, 79, 129901(E)
  (2009). {\href{https://arxiv.org/abs/0710.2335}{{arXiv:0710.2335}}} {[gr-qc]}

\bibitem[{Albertini et~al.(2022{\natexlab{a}})Albertini, Nagar, Pound,
  Warburton, Wardell, Durkan, and Miller}]{ANP22b}
Albertini A, Nagar A, Pound A, Warburton N, Wardell B, Durkan L, Miller J
  (2022{\natexlab{a}}) {Comparing second-order gravitational self-force and
  effective one body waveforms from inspiralling, quasicircular and nonspinning
  black hole binaries. {II}. The large-mass-ratio case}. Phys Rev D
  106(8):084062. \doi{10.1103/PhysRevD.106.084062}.
  {\href{https://arxiv.org/abs/2208.02055}{{arXiv:2208.02055}}} {[gr-qc]}

\bibitem[{Albertini et~al.(2022{\natexlab{b}})Albertini, Nagar, Pound,
  Warburton, Wardell, Durkan, and Miller}]{ANP22a}
Albertini A, Nagar A, Pound A, Warburton N, Wardell B, Durkan L, Miller J
  (2022{\natexlab{b}}) {Comparing second-order gravitational self-force,
  numerical relativity, and effective one body waveforms from inspiralling,
  quasicircular, and nonspinning black hole binaries}. Phys Rev D
  106(8):084061. \doi{10.1103/PhysRevD.106.084061}.
  {\href{https://arxiv.org/abs/2208.01049}{{arXiv:2208.01049}}} {[gr-qc]}

\bibitem[{Almeida et~al.(2021)Almeida, Foffa, and Sturani}]{AFS21a}
Almeida GL, Foffa S, Sturani R (2021) {Gravitational multipole
  renormalization}. Phys Rev D 104(8):084095.
  \doi{10.1103/PhysRevD.104.084095}.
  {\href{https://arxiv.org/abs/2107.02634}{{arXiv:2107.02634}}} {[gr-qc]}

\bibitem[{Almeida et~al.(2023)Almeida, Foffa, and Sturani}]{AFS23}
Almeida GL, Foffa S, Sturani R (2023) {Gravitational radiation contributions to
  the two-body scattering angle}. Phys Rev D 107(2):024020.
  \doi{10.1103/PhysRevD.107.024020}.
  {\href{https://arxiv.org/abs/2209.11594}{{arXiv:2209.11594}}} {[gr-qc]}

\bibitem[{Alvi(2001)}]{Alvi01}
Alvi K (2001) {Energy and angular momentum flow into a black hole in a binary}.
  Phys Rev D 64:104020. \doi{10.1103/PhysRevD.64.104020}.
  {\href{https://arxiv.org/abs/0107080}{{arXiv:0107080}}} {[gr-qc]}

\bibitem[{Anderson and DeCanio(1975)}]{AD75}
Anderson JL, DeCanio TC (1975) {Equations of hydrodynamics in general
  relativity in the slow motion approximation}. Gen Relativ Gravit 6:197--237.
  \doi{10.1007/BF00769986}

\bibitem[{Anderson et~al.(1982)Anderson, Kates, Kegeles, and Madonna}]{AKKM82}
Anderson JL, Kates RE, Kegeles LS, Madonna RG (1982) {Divergent integrals of
  post-{Newtonian} gravity: Nonanalytic terms in the near-zone expansion of a
  gravitationally radiating system found by matching}. Phys Rev D
  25:2038--2048. \doi{10.1103/PhysRevD.25.2038}

\bibitem[{Antonelli et~al.(2020{\natexlab{a}})Antonelli, Kavanagh, Khalil,
  Steinhoff, and Vines}]{Antonelli2020b}
Antonelli A, Kavanagh C, Khalil M, Steinhoff J, Vines J (2020{\natexlab{a}})
  {Gravitational spin-orbit and aligned spin1-spin2 couplings through
  third-subleading post-{Newtonian} orders}. Phys Rev D 102(12).
  \doi{10.1103/physrevd.102.124024}

\bibitem[{Antonelli et~al.(2020{\natexlab{b}})Antonelli, Kavanagh, Khalil,
  Steinhoff, and Vines}]{Antonelli2020a}
Antonelli A, Kavanagh C, Khalil M, Steinhoff J, Vines J (2020{\natexlab{b}})
  {Gravitational spin-orbit coupling through third-subleading post-{Newtonian}
  order: from first-order self-force to arbitrary mass ratios}. Phys Rev Lett
  125(1). \doi{10.1103/physrevlett.125.011103}

\bibitem[{Apostolatos et~al.(1994)Apostolatos, Cutler, Sussman, and
  Thorne}]{ACST94}
Apostolatos TA, Cutler C, Sussman GJ, Thorne KS (1994) {Spin induced orbital
  precession and its modulation of the gravitational wave forms from merging
  binaries}. Phys Rev D 49:6274--6297. \doi{10.1103/PhysRevD.49.6274}

\bibitem[{Arun et~al.(2004)Arun, Blanchet, Iyer, and Qusailah}]{ABIQ04}
Arun KG, Blanchet L, Iyer BR, Qusailah MSS (2004) {The 2.5PN gravitational wave
  polarizations from inspiralling compact binaries in circular orbits}. Class
  Quantum Grav 21:3771--3801. \doi{10.1088/0264-9381/21/15/010}, {Erratum}:
  Class. Quantum Grav., 22, 3115 (2005).
  {\href{https://arxiv.org/abs/gr-qc/0404185}{{gr-qc/0404185}}}

\bibitem[{Arun et~al.(2005)Arun, Iyer, Sathyaprakash, and
  Sundararajan}]{AISS05}
Arun KG, Iyer BR, Sathyaprakash BS, Sundararajan PA (2005) {Parameter
  estimation of inspiralling compact binaries using 3.5 post-{Newtonian}
  gravitational wave phasing: The nonspinning case}. Phys Rev D 71:084008.
  \doi{10.1103/PhysRevD.71.084008}.
  {\href{https://arxiv.org/abs/gr-qc/0411146}{{gr-qc/0411146}}}

\bibitem[{Arun et~al.(2006{\natexlab{a}})Arun, Iyer, Qusailah, and
  Sathyaprakash}]{AIQS06b}
Arun KG, Iyer BR, Qusailah MSS, Sathyaprakash BS (2006{\natexlab{a}}) {Probing
  the non-linear structure of general relativity with black hole binaries}.
  Phys Rev D 74:024006. \doi{10.1103/PhysRevD.74.024006}.
  {\href{https://arxiv.org/abs/gr-qc/0604067}{{gr-qc/0604067}}}

\bibitem[{Arun et~al.(2006{\natexlab{b}})Arun, Iyer, Qusailah, and
  Sathyaprakash}]{AIQS06a}
Arun KG, Iyer BR, Qusailah MSS, Sathyaprakash BS (2006{\natexlab{b}}) {Testing
  post-{Newtonian} theory with gravitational wave observations}. Class Quantum
  Grav 23:L37--L43. \doi{10.1088/0264-9381/23/9/L01}.
  {\href{https://arxiv.org/abs/gr-qc/0604018}{{arXiv:gr-qc/0604018}}}

\bibitem[{Arun et~al.(2007{\natexlab{a}})Arun, Iyer, Sathyaprakash, and
  Sinha}]{AISS07}
Arun KG, Iyer BR, Sathyaprakash BS, Sinha S (2007{\natexlab{a}}) {Higher
  harmonics increase LISA's mass reach for supermassive black holes}. Phys Rev
  D 75:124002. \doi{10.1103/PhysRevD.75.124002}.
  {\href{https://arxiv.org/abs/0704.1086}{{arXiv:0704.1086}}}

\bibitem[{Arun et~al.(2007{\natexlab{b}})Arun, Iyer, Sathyaprakash, Sinha, and
  Van Den~Broeck}]{AISSV07}
Arun KG, Iyer BR, Sathyaprakash BS, Sinha S, Van Den~Broeck C
  (2007{\natexlab{b}}) {Higher signal harmonics, LISA's angular resolution, and
  dark energy}. Phys Rev D 76:104016. \doi{10.1103/PhysRevD.76.104016}.
  {\href{https://arxiv.org/abs/0707.3920}{{arXiv:0707.3920}}}

\bibitem[{Arun et~al.(2008{\natexlab{a}})Arun, Blanchet, Iyer, and
  Qusailah}]{ABIQ08}
Arun KG, Blanchet L, Iyer BR, Qusailah MS (2008{\natexlab{a}}) {Inspiralling
  compact binaries in quasi-elliptical orbits: The complete 3PN energy flux}.
  Phys Rev D 77:064035. \doi{10.1103/PhysRevD.77.064035}.
  {\href{https://arxiv.org/abs/0711.0302}{{arXiv:0711.0302}}}

\bibitem[{Arun et~al.(2008{\natexlab{b}})Arun, Blanchet, Iyer, and
  Qusailah}]{ABIQ08tail}
Arun KG, Blanchet L, Iyer BR, Qusailah MS (2008{\natexlab{b}}) {Tail effects in
  the 3PN gravitational wave energy flux of compact binaries in
  quasi-elliptical orbits}. Phys Rev D 77:064034.
  \doi{10.1103/PhysRevD.77.064034}.
  {\href{https://arxiv.org/abs/0711.0250}{{arXiv:0711.0250}}}

\bibitem[{Arun et~al.(2009{\natexlab{a}})Arun, Blanchet, Iyer, and
  Sinha}]{ABIS09}
Arun KG, Blanchet L, Iyer BR, Sinha S (2009{\natexlab{a}}) {Third
  post-{Newtonian} angular momentum flux and the secular evolution of orbital
  elements for inspiralling compact binaries in quasi-elliptical orbits}. Phys
  Rev D 80:124018. \doi{10.1103/PhysRevD.80.124018}.
  {\href{https://arxiv.org/abs/0908.3854}{{arXiv:0908.3854}}}

\bibitem[{Arun et~al.(2009{\natexlab{b}})Arun, Buonanno, Faye, and
  Ochsner}]{ABFO08}
Arun KG, Buonanno A, Faye G, Ochsner E (2009{\natexlab{b}}) {Higher-order spin
  effects in the amplitude and phase of gravitational waveforms emitted by
  inspiraling compact binaries: Ready-to-use gravitational waveforms}. Phys Rev
  D 79:104023. \doi{10.1103/PhysRevD.79.104023}.
  {\href{https://arxiv.org/abs/0810.5336}{{arXiv:0810.5336}}}

\bibitem[{Ashtekar et~al.(2018)Ashtekar, Campiglia, and Laddha}]{ACL18}
Ashtekar A, Campiglia M, Laddha A (2018) {Null infinity, the {BMS} group and
  infrared issues}. Gen Relativ Gravit 50:1--23.
  \doi{10.1007/s10714-018-2464-3}

\bibitem[{Bailey and Israel(1975)}]{BIsrael75}
Bailey I, Israel W (1975) {Lagrangian dynamics of spinning particles and
  polarized media in general relativity}. Commun Math Phys 42:65.
  \doi{10.1007/BF01609434}

\bibitem[{Baker et~al.(2006{\natexlab{a}})Baker, Centrella, Choi, Koppitz, van
  Meter, and Miller}]{Bak06b}
Baker JG, Centrella J, Choi DI, Koppitz M, van Meter J, Miller MC
  (2006{\natexlab{a}}) {Getting a kick out of numerical relativity}. Astrophys
  J 653:L93--L96. \doi{10.1086/510448}.
  {\href{https://arxiv.org/abs/astro-ph/0603204}{{astro-ph/0603204}}}

\bibitem[{Baker et~al.(2006{\natexlab{b}})Baker, Centrella, Choi, Koppitz, and
  van Meter}]{Bak06}
Baker JG, Centrella J, Choi DI, Koppitz M, van Meter JR (2006{\natexlab{b}})
  {Gravitational-Wave Extraction from an Inspiraling Configuration of Merging
  Black Holes}. Phys Rev Lett 96:111102. \doi{10.1103/PhysRevLett.96.111102}.
  {\href{https://arxiv.org/abs/gr-qc/0511103}{{arXiv:gr-qc/0511103}}}

\bibitem[{{Banihashemi} and {Vines}(2020)}]{BV18}
{Banihashemi} B, {Vines} J (2020) {Gravitomagnetic tidal effects in
  gravitational waves from neutron star binaries}. Phys Rev D 101(6):064003.
  \doi{10.1103/PhysRevD.101.064003}.
  {\href{https://arxiv.org/abs/1805.07266}{{arXiv:1805.07266}}} {[gr-qc]}

\bibitem[{Barack(2009)}]{Ba09}
Barack L (2009) {Gravitational self-force in extreme mass-ratio inspirals}.
  Class Quantum Grav 26:213001. \doi{10.1088/0264-9381/26/21/213001}.
  {\href{https://arxiv.org/abs/0908.1664}{{arXiv:0908.1664}}} {[gr-qc]}

\bibitem[{Barack(2011)}]{Barackorleans}
Barack L (2011) Computational methods for the self-force in black hole
  spacetimes. In: Blanchet L, Spallicci A, Whiting B (eds) Mass and Motion in
  General Relativity. Fundamental Theories of Physics, vol 162. Springer,
  Dordrecht; New York, pp 327--366. \doi{10.1007/978-90-481-3015-3_12}

\bibitem[{Barack and Pound(2018)}]{BP18}
Barack L, Pound A (2018) {Self-force and radiation reaction in general
  relativity}. Rep Prog Phys 82(1):016904. \doi{10.1088/1361-6633/aae552}.
  {\href{https://arxiv.org/abs/1805.10385}{{arXiv:1805.10385}}} {[gr-qc]}

\bibitem[{Barack and Sago(2009)}]{BarackS09}
Barack L, Sago N (2009) {Gravitational self-force correction to the innermost
  stable circular orbit of a {Schwarzschild} black hole}. Phys Rev Lett
  102:191101. \doi{10.1103/PhysRevLett.102.191101}.
  {\href{https://arxiv.org/abs/0902.0573}{{arXiv:0902.0573}}}

\bibitem[{Barack and Sago(2011)}]{BarackS11}
Barack L, Sago N (2011) Beyond the geodesic approximation: conservative effects
  of the gravitational self-force in eccentric orbits around a schwarzschild
  black hole. Phys Rev D 83:084023. \doi{10.1103/PhysRevD.83.084023}.
  {\href{https://arxiv.org/abs/1101.3331}{{arXiv:1101.3331}}} {[gr-qc]}

\bibitem[{Barack et~al.(2010)Barack, Damour, and Sago}]{BDS10}
Barack L, Damour T, Sago N (2010) {Precession effect of the gravitational
  self-force in a {Schwarzschild} spacetime and the effective one-body
  formalism}. Phys Rev D 82(8):084036. \doi{10.1103/PhysRevD.82.084036}

\bibitem[{Barausse et~al.(2009)Barausse, Racine, and Buonanno}]{BarauRB09}
Barausse E, Racine E, Buonanno A (2009) {{Hamiltonian} of a spinning test
  particle in curved spacetime}. Phys Rev D 80:104025.
  \doi{10.1103/PhysRevD.80.104025}.
  {\href{https://arxiv.org/abs/0907.4745}{{arXiv:0907.4745}}} {[gr-qc]}

\bibitem[{{Bardeen} et~al.(1972){Bardeen}, {Press}, and {Teukolsky}}]{BPT72}
{Bardeen} JM, {Press} WH, {Teukolsky} SA (1972) {Rotating black holes: locally
  nonrotating frames, energy extraction, and scalar synchrotron radiation}.
  Astrophys J 178:347. \doi{10.1086/151796}

\bibitem[{Bardeen et~al.(1973)Bardeen, Carter, and Hawking}]{BarCH73}
Bardeen JM, Carter B, Hawking SW (1973) {The Four Laws of Black Hole
  Mechanics}. Commun Math Phys 31:161--170. \doi{10.1007/BF01645742}

\bibitem[{Barker and {O'Connell}(1975)}]{BOC75}
Barker BM, {O'Connell} RF (1975) {Gravitational two-body problem with arbitrary
  masses, spins, and quadrupole moments}. Phys Rev D 12:329--335.
  \doi{10.1103/PhysRevD.12.329}

\bibitem[{Barker and {O'Connell}(1979)}]{BOC79}
Barker BM, {O'Connell} RF (1979) {The Gravitational Interaction: Spin,
  Rotation, and Quantum Effects -- A Review}. Gen Relativ Gravit 11:149--175.
  \doi{10.1007/BF00756587}

\bibitem[{Baumgarte(2000)}]{Baum00}
Baumgarte TW (2000) {Innermost stable circular orbit of binary black holes}.
  Phys Rev D 62:024018. \doi{10.1103/PhysRevD.62.024018}

\bibitem[{Bekenstein(1973)}]{Bek73}
Bekenstein JD (1973) {Gravitational Radiation Recoil and Runaway Black Holes}.
  Astrophys J 183:657--664. \doi{10.1086/152255}

\bibitem[{Bel et~al.(1981)Bel, Damour, Deruelle, Ib{\'{a}}{\~{n}}ez, and
  Martin}]{BeDD81}
Bel L, Damour T, Deruelle N, Ib{\'{a}}{\~{n}}ez J, Martin J (1981)
  {Poincar{\'e}-Invariant Gravitational Field and Equations of Motion of two
  Pointlike Objects: The Postlinear Approximation of General Relativity}. Gen
  Relativ Gravit 13:963--1004. \doi{10.1007/BF00756073}

\bibitem[{Benacquista and Downing(2013)}]{BenaLR}
Benacquista MJ, Downing JMB (2013) {Relativistic Binaries in Globular
  Clusters}. Living Rev Relativ 16:4. \doi{10.12942/lrr-2013-4}.
  {\href{https://arxiv.org/abs/1110.4423}{{arXiv:1110.4423}}}

\bibitem[{Bern et~al.(2019{\natexlab{a}})Bern, Cheung, Roiban, Shen, Solon, and
  Zeng}]{Bern19b}
Bern Z, Cheung C, Roiban R, Shen CH, Solon MP, Zeng M (2019{\natexlab{a}})
  {Black hole binary dynamics from the double copy and effective theory}. J
  High Energy Phys 10:206. \doi{10.1007/JHEP10(2019)206}.
  {\href{https://arxiv.org/abs/1908.01493}{{arXiv:1908.01493}}} {[hep-th]}

\bibitem[{Bern et~al.(2019{\natexlab{b}})Bern, Cheung, Roiban, Shen, Solon, and
  Zeng}]{Bern19a}
Bern Z, Cheung C, Roiban R, Shen CH, Solon MP, Zeng M (2019{\natexlab{b}})
  {Scattering amplitudes and the conservative {Hamiltonian} for binary systems
  at third post-{Minkowskian} order}. Phys Rev Lett 122(20):201603.
  \doi{10.1103/PhysRevLett.122.201603}.
  {\href{https://arxiv.org/abs/1901.04424}{{arXiv:1901.04424}}} {[hep-th]}

\bibitem[{Bern et~al.(2021)Bern, Parra-Martinez, Roiban, Ruf, Shen, Solon, and
  Zeng}]{Bern21}
Bern Z, Parra-Martinez J, Roiban R, Ruf MS, Shen CH, Solon MP, Zeng M (2021)
  {Scattering amplitudes and conservative binary dynamics at $O(G^4)$}. Phys
  Rev Lett 126(17):171601. \doi{10.1103/PhysRevLett.126.171601}.
  {\href{https://arxiv.org/abs/2112.10750}{{arXiv:2112.10750}}}

\bibitem[{Bernard(2018)}]{BernardST1}
Bernard L (2018) {Dynamics of compact binary systems in scalar-tensor theories:
  Equations of motion to the third post-{Newtonian} order}. Phys Rev D
  98:044004. \doi{10.1103/PhysRevD.98.044004}.
  {\href{https://arxiv.org/abs/1802.10201}{{arXiv:1802.10201}}} {[gr-qc]}

\bibitem[{Bernard(2019)}]{BernardST2}
Bernard L (2019) {Dynamics of compact binary systems in scalar-tensor theories.
  {II}. {C}enter-of-mass and conserved quantities to 3{PN} order}. Phys Rev D
  99(4):044047. \doi{10.1103/PhysRevD.99.044047}.
  {\href{https://arxiv.org/abs/1812.04169}{{arXiv:1812.04169}}} {[gr-qc]}

\bibitem[{Bernard et~al.(2016)Bernard, Blanchet, Boh\'e, Faye, and
  Marsat}]{BBBFMa}
Bernard L, Blanchet L, Boh\'e A, Faye G, Marsat S (2016) {Fokker action of
  non-spinning compact binaries at the fourth post-{Newtonian} approximation}.
  Phys Rev D 93:084037. \doi{10.1103/PhysRevD.93.084037}.
  {\href{https://arxiv.org/abs/1512.02876}{{arXiv:1512.02876}}} {[gr-qc]}

\bibitem[{Bernard et~al.(2017{\natexlab{a}})Bernard, Blanchet, Boh\'e, Faye,
  and Marsat}]{BBBFMc}
Bernard L, Blanchet L, Boh\'e A, Faye G, Marsat S (2017{\natexlab{a}})
  {Dimensional regularization of the {IR} divergences in the {Fokker} action of
  point-particle binaries at the fourth post-{Newtonian} order}. Phys Rev D
  96:104043. \doi{10.1103/PhysRevD.96.104043}.
  {\href{https://arxiv.org/abs/1706.08480}{{arXiv:1706.08480}}} {[gr-qc]}

\bibitem[{Bernard et~al.(2017{\natexlab{b}})Bernard, Blanchet, Boh\'e, Faye,
  and Marsat}]{BBBFMb}
Bernard L, Blanchet L, Boh\'e A, Faye G, Marsat S (2017{\natexlab{b}}) {Energy
  and periastron advance of compact binaries on circular orbits at the fourth
  post-{Newtonian} order}. Phys Rev D 95:044026.
  \doi{10.1103/PhysRevD.95.044026}.
  {\href{https://arxiv.org/abs/1610.07934}{{arXiv:1610.07934}}} {[gr-qc]}

\bibitem[{Bernard et~al.(2018)Bernard, Blanchet, Faye, and Marchand}]{BBFM17}
Bernard L, Blanchet L, Faye G, Marchand T (2018) {Center-of-mass equations of
  motion and conserved integrals of compact binary systems at the fourth
  post-{Newtonian} order}. Phys Rev D 97:044037.
  \doi{10.1103/PhysRevD.97.044037}.
  {\href{https://arxiv.org/abs/1711.00283}{{arXiv:1711.00283}}} {[gr-qc]}

\bibitem[{Bernard et~al.(2022)Bernard, Blanchet, and Trestini}]{BBT22}
Bernard L, Blanchet L, Trestini D (2022) {Gravitational waves in scalar-tensor
  theory to one-and-a-half post-{Newtonian} order}. J Cosmol Astropart Phys
  2022(08):008. \doi{10.1088/1475-7516/2022/08/008}.
  {\href{https://arxiv.org/abs/2201.10924}{{arXiv:2201.10924}}} {[gr-qc]}

\bibitem[{Bernuzzi et~al.(2012)Bernuzzi, Thierfelder, and Br{\"u}gmann}]{BTB11}
Bernuzzi S, Thierfelder M, Br{\"u}gmann B (2012) {Accuracy of numerical
  relativity waveforms from binary neutron star mergers and their comparison
  with post-{Newtonian} waveforms}. Phys Rev D 85:104030.
  \doi{10.1103/PhysRevD.85.104030}.
  {\href{https://arxiv.org/abs/1109.3611}{{arXiv:1109.3611}}} {[gr-qc]}

\bibitem[{Bertotti and Pleba{\'{n}}ski(1960)}]{BertottiP60}
Bertotti B, Pleba{\'{n}}ski JF (1960) {Theory of gravitational perturbations in
  the fast motion approximation}. Ann Phys (NY) 11:169--200.
  \doi{10.1016/0003-4916(60)90132-9}

\bibitem[{Bini and Damour(2013)}]{BiniD13}
Bini D, Damour T (2013) {Analytical determination of the two-body gravitational
  interaction potential at the fourth post-{Newtonian} approximation}. Phys Rev
  D 87:121501. \doi{10.1103/PhysRevD.87.121501}.
  {\href{https://arxiv.org/abs/1305.4884}{{arXiv:1305.4884}}} {[gr-qc]}

\bibitem[{Bini and Damour(2014{\natexlab{a}})}]{BiniD14b}
Bini D, Damour T (2014{\natexlab{a}}) {Analytic determination of the
  eight-and-a-half post-{Newtonian} self-force contributions to the two-body
  gravitational interaction potential}. Phys Rev D 89:104047.
  \doi{10.1103/PhysRevD.89.104047}.
  {\href{https://arxiv.org/abs/1403.2366}{{arXiv:1403.2366}}} {[gr-qc]}

\bibitem[{Bini and Damour(2014{\natexlab{b}})}]{BiniD14a}
Bini D, Damour T (2014{\natexlab{b}}) {High-order post-{Newtonian}
  contributions to the two-body gravitational interaction potential from
  analytical gravitational self-force calculations}. Phys Rev D 89:064063.
  \doi{10.1103/PhysRevD.89.064063}.
  {\href{https://arxiv.org/abs/1312.2503}{{arXiv:1312.2503}}} {[gr-qc]}

\bibitem[{Bini et~al.(2012)Bini, Damour, and Faye}]{BiniDF12}
Bini D, Damour T, Faye G (2012) {Effective action approach to higher-order
  relativistic tidal interactions in binary systems and their effective one
  body description}. Phys Rev D 85:124034. \doi{10.1103/PhysRevD.85.124034}.
  {\href{https://arxiv.org/abs/1202.3565}{{arXiv:1202.3565}}} {[gr-qc]}

\bibitem[{{Bini} et~al.(2016){Bini}, {Damour}, and {Geralico}}]{BiniDG15}
{Bini} D, {Damour} T, {Geralico} A (2016) {Confirming and improving
  post-{Newtonian} and effective-one-body results from self-force computations
  along eccentric orbits around a {Schwarzschild} black hole}. Phys Rev D
  93:064023. \doi{10.1103/PhysRevD.93.064023}.
  {\href{https://arxiv.org/abs/1511.04533}{{arXiv:1511.04533}}} {[gr-qc]}

\bibitem[{Bini et~al.(2019)Bini, Damour, and Geralico}]{BiniDG19}
Bini D, Damour T, Geralico A (2019) {Novel approach to binary dynamics:
  application to the fifth post-{Newtonian} level}. Phys Rev Lett
  123(23):231104. \doi{10.1103/PhysRevLett.123.231104}.
  {\href{https://arxiv.org/abs/1909.02375}{{arXiv:1909.02375}}} {[gr-qc]}

\bibitem[{Bini et~al.(2020{\natexlab{a}})Bini, Damour, and
  Geralico}]{BiniDG20a}
Bini D, Damour T, Geralico A (2020{\natexlab{a}}) {Binary dynamics at the fifth
  and fifth-and-a-half post-{Newtonian} orders}. Phys Rev D 102(2):024062.
  \doi{10.1103/PhysRevD.102.024062}.
  {\href{https://arxiv.org/abs/2003.11891}{{arXiv:2003.11891}}} {[gr-qc]}

\bibitem[{Bini et~al.(2020{\natexlab{b}})Bini, Damour, and
  Geralico}]{BiniDG20b}
Bini D, Damour T, Geralico A (2020{\natexlab{b}}) {Sixth post-{N}ewtonian
  local-in-time dynamics of binary systems}. Phys Rev D 102(2):024061.
  \doi{10.1103/PhysRevD.102.024061}.
  {\href{https://arxiv.org/abs/2004.05407}{{arXiv:2004.05407}}} {[gr-qc]}

\bibitem[{Bini et~al.(2020{\natexlab{c}})Bini, Damour, and
  Geralico}]{BiniDG20c}
Bini D, Damour T, Geralico A (2020{\natexlab{c}}) {Sixth post-{Newtonian}
  nonlocal-in-time dynamics of binary systems}. Phys Rev D 102(8):084047.
  \doi{10.1103/PhysRevD.102.084047}.
  {\href{https://arxiv.org/abs/2007.11239}{{arXiv:2007.11239}}} {[gr-qc]}

\bibitem[{Bini et~al.(2021)Bini, Damour, Geralico, Laporta, and
  Mastrolia}]{BiniDGLM21}
Bini D, Damour T, Geralico A, Laporta S, Mastrolia P (2021) {Gravitational
  scattering at the seventh order in $G$: nonlocal contribution at the sixth
  post-{Newtonian} accuracy}. Phys Rev D 103(4):044038.
  \doi{10.1103/PhysRevD.103.044038}.
  {\href{https://arxiv.org/abs/2012.12918}{{arXiv:2012.12918}}} {[gr-qc]}

\bibitem[{Bini et~al.(2023)Bini, Damour, and Geralico}]{BDG23}
Bini D, Damour T, Geralico A (2023) {Comparing one-loop gravitational
  bremsstrahlung amplitudes to the multipolar-post-{Minkowskian} waveform}.
  Phys Rev D 108(12):124052. \doi{10.1103/PhysRevD.108.124052}.
  {\href{https://arxiv.org/abs/2309.14925}{{arXiv:2309.14925}}} {[gr-qc]}

\bibitem[{Bini et~al.(2024)Bini, Damour, {De Angelis}, Geralico, Herderschee,
  Roiban, and Teng}]{BDDGH24}
Bini D, Damour T, {De Angelis} S, Geralico A, Herderschee A, Roiban R, Teng F
  (2024) {Gravitational waveform: a tale of two formalisms}. arXiv e-prints
  {\href{https://arxiv.org/abs/2402.06604}{{arXiv:2402.06604}}} {[hep-th]}

\bibitem[{Binnington and Poisson(2009)}]{BinnP09}
Binnington T, Poisson E (2009) {Relativistic theory of tidal {Love} numbers}.
  Phys Rev D 80:084018. \doi{10.1103/PhysRevD.80.084018}.
  {\href{https://arxiv.org/abs/0906.1366}{{arXiv:0906.1366}}} {[gr-qc]}

\bibitem[{Blaes et~al.(2002)Blaes, Lee, and Socrates}]{BLS02}
Blaes O, Lee MH, Socrates A (2002) {The {Kozai} mechanism and the evolution of
  binary supermassive black holes}. Astrophys J 578:775--786.
  \doi{10.1086/342655}.
  {\href{https://arxiv.org/abs/astro-ph/0203370}{{astro-ph/0203370}}}

\bibitem[{Blanchet(1987)}]{B87}
Blanchet L (1987) {Radiative gravitational fields in general-relativity. {II}.
  {A}symptotic-behaviour at future null infinity}. Proc R Soc London, Ser A
  409:383--399. \doi{10.1098/rspa.1987.0022}

\bibitem[{Blanchet(1990)}]{B90mem}
Blanchet L (1990) {Contribution \`a l'\'etude du rayonnement gravitationnel
  \'emis par un syst\`eme isol\'e}. PhD thesis, Universit\'e Pierre et Marie
  Curie, Paris VI.
  \urlprefix\url{http://www2.iap.fr/users/blanchet/TheseHabilitation1990.pdf},
  see Chapter VI, pages 205-214

\bibitem[{Blanchet(1993)}]{B93}
Blanchet L (1993) {Time asymmetric structure of gravitational radiation}. Phys
  Rev D 47:4392--4420. \doi{10.1103/PhysRevD.47.4392}

\bibitem[{Blanchet(1995)}]{B95}
Blanchet L (1995) {Second-post-{Newtonian} generation of gravitational
  radiation}. Phys Rev D 51:2559--2583. \doi{10.1103/PhysRevD.51.2559}.
  {\href{https://arxiv.org/abs/gr-qc/9501030}{{gr-qc/9501030}}}

\bibitem[{Blanchet(1996)}]{B96}
Blanchet L (1996) {Energy losses by gravitational radiation in inspiralling
  compact binaries to five halves post-{Newtonian} order}. Phys Rev D
  54:1417--1438. \doi{10.1103/PhysRevD.54.1417}, {Erratum}: Phys Rev D, 71,
  129904(E) (2005).
  {\href{https://arxiv.org/abs/gr-qc/9603048}{{gr-qc/9603048}}}

\bibitem[{Blanchet(1997{\natexlab{a}})}]{Bhouches}
Blanchet L (1997{\natexlab{a}}) Gravitational radiation from relativistic
  sources. In: Marck JA, Lasota JP (eds) Relativistic Gravitation and
  Gravitational Radiation. Cambridge Contemporary Astrophysics. Cambridge
  University Press, Cambridge, pp 33--66.
  {\href{https://arxiv.org/abs/gr-qc/9609049}{{gr-qc/9609049}}}

\bibitem[{Blanchet(1997{\natexlab{b}})}]{B97}
Blanchet L (1997{\natexlab{b}}) {Gravitational radiation reaction and balance
  equations to post-{Newtonian} order}. Phys Rev D 55:714--732.
  \doi{10.1103/PhysRevD.55.714}.
  {\href{https://arxiv.org/abs/gr-qc/9609049}{{gr-qc/9609049}}}

\bibitem[{Blanchet(1998{\natexlab{a}})}]{B98tail}
Blanchet L (1998{\natexlab{a}}) {Gravitational-wave tails of tails}. Class
  Quantum Grav 15:113--141. \doi{10.1088/0264-9381/15/1/009}, {Erratum}: Class.
  Quantum Grav., 22, 3381 (2005).
  {\href{https://arxiv.org/abs/gr-qc/9710038}{{gr-qc/9710038}}}

\bibitem[{Blanchet(1998{\natexlab{b}})}]{B98mult}
Blanchet L (1998{\natexlab{b}}) {On the multipole expansion of the
  gravitational field}. Class Quantum Grav 15:1971--1999.
  \doi{10.1088/0264-9381/15/7/013}.
  {\href{https://arxiv.org/abs/gr-qc/9801101}{{gr-qc/9801101}}}

\bibitem[{Blanchet(1998{\natexlab{c}})}]{B98quad}
Blanchet L (1998{\natexlab{c}}) {Quadrupole-quadrupole gravitational waves}.
  Class Quantum Grav 15:89--111. \doi{10.1088/0264-9381/15/1/008}.
  {\href{https://arxiv.org/abs/gr-qc/9710037}{{gr-qc/9710037}}}

\bibitem[{Blanchet(2002)}]{B02ico}
Blanchet L (2002) {Innermost circular orbit of binary black holes at the third
  post-{Newtonian} approximation}. Phys Rev D 65:124009.
  \doi{10.1103/PhysRevD.65.124009}.
  {\href{https://arxiv.org/abs/gr-qc/0112056}{{gr-qc/0112056}}}

\bibitem[{Blanchet(2011{\natexlab{a}})}]{B11}
Blanchet L (2011{\natexlab{a}}) {On the accuracy of the post-{Newtonian}
  approximation}. In: Ciufolini I, Dominici D, Lusanna L (eds) 2001: A
  Relativistic Spacetime Odyssey. World Scientific, pp 411--430.
  \doi{10.1142/9789812791368_0022}.
  {\href{https://arxiv.org/abs/gr-qc/0207037}{{gr-qc/0207037}}}

\bibitem[{Blanchet(2011{\natexlab{b}})}]{Borleans}
Blanchet L (2011{\natexlab{b}}) {Post-{Newtonian} theory and the two-body
  problem}. In: Blanchet L, Spallicci A, Whiting B (eds) Mass and Motion in
  General Relativity. Fundamental Theories of Physics. Springer, Dordrecht; New
  York, pp 125--166. \doi{10.1007/978-90-481-3015-3_5}.
  {\href{https://arxiv.org/abs/0907.3596}{{arXiv:0907.3596}}} {[gr-qc]}

\bibitem[{Blanchet(2014)}]{BlanchetLR}
Blanchet L (2014) {Gravitational radiation from post-{N}ewtonian sources and
  inspiralling compact binaries}. Living Rev Relativ 17:2.
  \doi{10.12942/lrr-2014-2}.
  {\href{https://arxiv.org/abs/1310.1528}{{arXiv:1310.1528}}} {[gr-qc]}

\bibitem[{Blanchet and Damour(1984)}]{BD84}
Blanchet L, Damour T (1984) {Multipolar radiation reaction in general
  relativity}. Phys Lett A 104:82--86. \doi{10.1016/0375-9601(84)90967-8}

\bibitem[{Blanchet and Damour(1986)}]{BD86}
Blanchet L, Damour T (1986) {Radiative gravitational fields in general
  relativity {I}. {G}eneral structure of the field outside the source}. Philos
  Trans R Soc London, Ser A 320:379--430. \doi{10.1098/rsta.1986.0125}

\bibitem[{Blanchet and Damour(1988)}]{BD88}
Blanchet L, Damour T (1988) {Tail-transported temporal correlations in the
  dynamics of a gravitating system}. Phys Rev D 37:1410--1435.
  \doi{10.1103/PhysRevD.37.1410}

\bibitem[{Blanchet and Damour(1989)}]{BD89}
Blanchet L, Damour T (1989) {Post-{Newtonian} generation of gravitational
  waves}. Ann Inst Henri Poincare A 50:377--408

\bibitem[{Blanchet and Damour(1992)}]{BD92}
Blanchet L, Damour T (1992) {Hereditary effects in gravitational radiation}.
  Phys Rev D 46:4304--4319. \doi{10.1103/PhysRevD.46.4304}

\bibitem[{Blanchet and Faye(2000{\natexlab{a}})}]{BF00}
Blanchet L, Faye G (2000{\natexlab{a}}) {Equations of motion of point-particle
  binaries at the third post-{Newtonian} order}. Phys Lett A 271:58--64.
  \doi{10.1016/S0375-9601(00)00360-1}.
  {\href{https://arxiv.org/abs/gr-qc/0004009}{{gr-qc/0004009}}}

\bibitem[{Blanchet and Faye(2000{\natexlab{b}})}]{BFreg}
Blanchet L, Faye G (2000{\natexlab{b}}) {Hadamard regularization}. J Math Phys
  41:7675--7714. \doi{10.1063/1.1308506}.
  {\href{https://arxiv.org/abs/gr-qc/0004008}{{gr-qc/0004008}}}

\bibitem[{Blanchet and Faye(2001{\natexlab{a}})}]{BFeom}
Blanchet L, Faye G (2001{\natexlab{a}}) {General relativistic dynamics of
  compact binaries at the third post-{Newtonian} order}. Phys Rev D 63:062005.
  \doi{10.1103/PhysRevD.63.062005}.
  {\href{https://arxiv.org/abs/gr-qc/0007051}{{gr-qc/0007051}}}

\bibitem[{Blanchet and Faye(2001{\natexlab{b}})}]{BFregM}
Blanchet L, Faye G (2001{\natexlab{b}}) {Lorentzian regularization and the
  problem of point-like particles in general relativity}. J Math Phys
  42:4391--4418. \doi{10.1063/1.1384864}.
  {\href{https://arxiv.org/abs/gr-qc/0006100}{{gr-qc/0006100}}}

\bibitem[{Blanchet and Faye(2019)}]{BF19}
Blanchet L, Faye G (2019) {Flux-balance equations for linear momentum and
  center-of-mass position of self-gravitating post-{Newtonian} systems}. Class
  Quantum Grav 36(8):085003. \doi{10.1088/1361-6382/ab0d4f}

\bibitem[{Blanchet and Fokas(2018)}]{BFok18}
Blanchet L, Fokas A (2018) {Equations of motion of self-gravitating $N$-body
  systems in the first post-{Minkowskian} approximation}. Phys Rev D 98:084005.
  \doi{10.1103/PhysRevD.98.084005}.
  {\href{https://arxiv.org/abs/0812.4413}{{arXiv:0812.4413}}} {[gr-qc]}

\bibitem[{Blanchet and Iyer(2003)}]{BI03CM}
Blanchet L, Iyer BR (2003) {Third post-{Newtonian} dynamics of compact
  binaries: Equations of motion in the center-of-mass frame}. Class Quantum
  Grav 20:755. \doi{10.1088/0264-9381/20/4/309}.
  {\href{https://arxiv.org/abs/gr-qc/0209089}{{gr-qc/0209089}}}

\bibitem[{Blanchet and Iyer(2005)}]{BI04mult}
Blanchet L, Iyer BR (2005) {Hadamard regularization of the third
  post-{Newtonian} gravitational wave generation of two point masses}. Phys Rev
  D 71:024004. \doi{10.1103/PhysRevD.71.024004}.
  {\href{https://arxiv.org/abs/gr-qc/0409094}{{gr-qc/0409094}}}

\bibitem[{Blanchet and {Le Tiec}(2017)}]{BL17}
Blanchet L, {Le Tiec} A (2017) {First law of compact binary mechanics with
  gravitational-wave tails}. Class Quantum Grav 34:164001.
  \doi{10.1088/1361-6382/aa79d7}.
  {\href{https://arxiv.org/abs/1702.06839}{{1702.06839}}} {[gr-qc]}

\bibitem[{Blanchet and Sathyaprakash(1994)}]{BSat94}
Blanchet L, Sathyaprakash BS (1994) {Signal analysis of gravitational wave
  tails}. Class Quantum Grav 11:2807--2831. \doi{10.1088/0264-9381/11/11/020}

\bibitem[{Blanchet and Sathyaprakash(1995)}]{BSat95}
Blanchet L, Sathyaprakash BS (1995) {Detecting a Tail Effect in
  Gravitational-Wave Experiments}. Phys Rev Lett 74:1067--1070.
  \doi{10.1103/PhysRevLett.74.1067}

\bibitem[{Blanchet and Sch{\"{a}}fer(1989)}]{BS89}
Blanchet L, Sch{\"{a}}fer G (1989) {Higher order gravitational radiation losses
  in binary systems}. Mon Not R Astron Soc 239:845--867.
  \doi{10.1093/mnras/239.3.845}

\bibitem[{Blanchet and Sch{\"{a}}fer(1993)}]{BS93}
Blanchet L, Sch{\"{a}}fer G (1993) {Gravitational wave tails and binary star
  systems}. Class Quantum Grav 10:2699--2721. \doi{10.1088/0264-9381/10/12/026}

\bibitem[{Blanchet et~al.(1995{\natexlab{a}})Blanchet, Damour, and
  Iyer}]{BDI95}
Blanchet L, Damour T, Iyer BR (1995{\natexlab{a}}) {Gravitational waves from
  inspiralling compact binaries: Energy loss and wave form to second
  post-{Newtonian} order}. Phys Rev D 51:5360--5386.
  \doi{10.1103/PhysRevD.51.5360}.
  {\href{https://arxiv.org/abs/gr-qc/9501029}{{gr-qc/9501029}}}

\bibitem[{Blanchet et~al.(1995{\natexlab{b}})Blanchet, Damour, Iyer, Will, and
  Wiseman}]{BDIWW95}
Blanchet L, Damour T, Iyer BR, Will CM, Wiseman AG (1995{\natexlab{b}})
  {Gravitational-Radiation Damping of Compact Binary Systems to Second
  Post-{Newtonian} Order}. Phys Rev Lett 74:3515--3518.
  \doi{10.1103/PhysRevLett.74.3515}.
  {\href{https://arxiv.org/abs/gr-qc/9501027}{{gr-qc/9501027}}}

\bibitem[{Blanchet et~al.(1996)Blanchet, Iyer, Will, and Wiseman}]{BIWW96}
Blanchet L, Iyer BR, Will CM, Wiseman AG (1996) {Gravitational wave forms from
  inspiralling compact binaries to second-post-{Newtonian} order}. Class
  Quantum Grav 13:575--584. \doi{10.1088/0264-9381/13/4/002}.
  {\href{https://arxiv.org/abs/gr-qc/9602024}{{gr-qc/9602024}}}

\bibitem[{Blanchet et~al.(1998)Blanchet, Faye, and Ponsot}]{BFP98}
Blanchet L, Faye G, Ponsot B (1998) {Gravitational field and equations of
  motion of compact binaries to 5/2 post-{Newtonian} order}. Phys Rev D
  58:124002. \doi{10.1103/PhysRevD.58.124002}.
  {\href{https://arxiv.org/abs/gr-qc/9804079}{{gr-qc/9804079}}}

\bibitem[{Blanchet et~al.(2002{\natexlab{a}})Blanchet, Faye, Iyer, and
  Joguet}]{BFIJ02}
Blanchet L, Faye G, Iyer BR, Joguet B (2002{\natexlab{a}}) {Gravitational-wave
  inspiral of compact binary systems to 7/2 post-{Newtonian} order}. Phys Rev D
  65:061501(R). \doi{10.1103/PhysRevD.65.061501}, {Erratum}: Phys Rev D, 71,
  129902(E) (2005).
  {\href{https://arxiv.org/abs/gr-qc/0105099}{{gr-qc/0105099}}}

\bibitem[{Blanchet et~al.(2002{\natexlab{b}})Blanchet, Iyer, and
  Joguet}]{BIJ02}
Blanchet L, Iyer BR, Joguet B (2002{\natexlab{b}}) {Gravitational waves from
  inspiralling compact binaries: Energy flux to third post-{Newtonian} order}.
  Phys Rev D 65:064005. {Erratum}: Phys Rev D, 71, 129903(E) (2005).
  {\href{https://arxiv.org/abs/gr-qc/0105098}{{gr-qc/0105098}}}

\bibitem[{Blanchet et~al.(2004{\natexlab{a}})Blanchet, Damour, and
  Esposito-Far{\`{e}}se}]{BDE04}
Blanchet L, Damour T, Esposito-Far{\`{e}}se G (2004{\natexlab{a}}) {Dimensional
  regularization of the third post-{Newtonian} dynamics of point particles in
  harmonic coordinates}. Phys Rev D 69:124007.
  \doi{10.1103/PhysRevD.69.124007}.
  {\href{https://arxiv.org/abs/gr-qc/0311052}{{gr-qc/0311052}}}

\bibitem[{Blanchet et~al.(2004{\natexlab{b}})Blanchet, Damour,
  Esposito-Far{\`{e}}se, and Iyer}]{BDEI04}
Blanchet L, Damour T, Esposito-Far{\`{e}}se G, Iyer BR (2004{\natexlab{b}})
  {Gravitational radiation from inspiralling compact binaries completed at the
  third post-{Newtonian} order}. Phys Rev Lett 93:091101.
  \doi{10.1103/PhysRevLett.93.091101}.
  {\href{https://arxiv.org/abs/gr-qc/0406012}{{gr-qc/0406012}}}

\bibitem[{Blanchet et~al.(2005{\natexlab{a}})Blanchet, Damour,
  Esposito-Far{\`{e}}se, and Iyer}]{BDEI05dr}
Blanchet L, Damour T, Esposito-Far{\`{e}}se G, Iyer BR (2005{\natexlab{a}})
  {Dimensional regularization of the third post-{Newtonian} gravitational wave
  generation of two point masses}. Phys Rev D 71:124004.
  \doi{10.1103/PhysRevD.71.124004}.
  {\href{https://arxiv.org/abs/gr-qc/0503044}{{gr-qc/0503044}}}

\bibitem[{Blanchet et~al.(2005{\natexlab{b}})Blanchet, Damour, and
  Iyer}]{BDI04zeta}
Blanchet L, Damour T, Iyer BR (2005{\natexlab{b}}) {Surface-integral
  expressions for the multipole moments of post-{Newtonian} sources and the
  boosted {Schwarzschild} solution}. Class Quantum Grav 22:155.
  \doi{10.1088/0264-9381/22/1/011}.
  {\href{https://arxiv.org/abs/gr-qc/0410021}{{gr-qc/0410021}}}

\bibitem[{Blanchet et~al.(2005{\natexlab{c}})Blanchet, Faye, and
  Nissanke}]{BFN05}
Blanchet L, Faye G, Nissanke S (2005{\natexlab{c}}) {Structure of the
  post-{Newtonian} expansion in general relativity}. Phys Rev D 72:044024.
  \doi{10.1103/PhysRevD.72.044024}

\bibitem[{Blanchet et~al.(2005{\natexlab{d}})Blanchet, Qusailah, and
  Will}]{BQW05}
Blanchet L, Qusailah MS, Will CM (2005{\natexlab{d}}) {Gravitational recoil of
  inspiraling black-hole binaries to second post-{Newtonian} order}. Astrophys
  J 635:508. \doi{10.1086/497332}.
  {\href{https://arxiv.org/abs/astro-ph/0507692}{{astro-ph/0507692}}}

\bibitem[{Blanchet et~al.(2006)Blanchet, Buonanno, and Faye}]{BBF06}
Blanchet L, Buonanno A, Faye G (2006) {Higher-order spin effects in the
  dynamics of compact binaries {II}. {R}adiation field}. Phys Rev D 74:104034.
  \doi{10.1103/PhysRevD.74.104034}, {Erratum}: Phys Rev D, 75, 049903 (2007).
  {\href{https://arxiv.org/abs/gr-qc/0605140}{{gr-qc/0605140}}}

\bibitem[{Blanchet et~al.(2008)Blanchet, Faye, Iyer, and Sinha}]{BFIS08}
Blanchet L, Faye G, Iyer BR, Sinha S (2008) {The third post-{Newtonian}
  gravitational wave polarisations and associated spherical harmonic modes for
  inspiralling compact binaries in quasi-circular orbits}. Class Quantum Grav
  25:165003. \doi{10.1088/0264-9381/25/16/165003}.
  {\href{https://arxiv.org/abs/0802.1249}{{arXiv:0802.1249}}}

\bibitem[{Blanchet et~al.(2010{\natexlab{a}})Blanchet, Detweiler, Le~Tiec, and
  Whiting}]{BDLW10b}
Blanchet L, Detweiler S, Le~Tiec A, Whiting BF (2010{\natexlab{a}})
  {Higher-order Post-{Newtonian} fit of the gravitational self-force for
  circular orbits in the {Schwarzschild} geometry}. Phys Rev D 81:084033.
  \doi{10.1103/PhysRevD.81.084033}.
  {\href{https://arxiv.org/abs/1002.0726}{{arXiv:1002.0726}}} {[gr-qc]}

\bibitem[{Blanchet et~al.(2010{\natexlab{b}})Blanchet, Detweiler, Le~Tiec, and
  Whiting}]{BDLW10a}
Blanchet L, Detweiler S, Le~Tiec A, Whiting BF (2010{\natexlab{b}})
  {Post-{Newtonian} and numerical calculations of the gravitational self-force
  for circular orbits in the {Schwarzschild} geometry}. Phys Rev D 81:064004.
  \doi{10.1103/PhysRevD.81.064004}.
  {\href{https://arxiv.org/abs/0910.0207}{{arXiv:0910.0207}}} {[gr-qc]}

\bibitem[{Blanchet et~al.(2011)Blanchet, Buonanno, and Faye}]{BBF11}
Blanchet L, Buonanno A, Faye G (2011) {Tail-induced spin-orbit effect in the
  gravitational radiation of compact binaries}. Phys Rev D 84:064041.
  \doi{10.1103/PhysRevD.84.064041}.
  {\href{https://arxiv.org/abs/1104.5659}{{arXiv:1104.5659}}} {[gr-qc]}

\bibitem[{Blanchet et~al.(2013)Blanchet, Buonanno, and Le~Tiec}]{BBL13}
Blanchet L, Buonanno A, Le~Tiec A (2013) {First law of mechanics for black hole
  binaries with spins}. Phys Rev D 87:024030. \doi{10.1103/PhysRevD.87.024030}.
  {\href{https://arxiv.org/abs/1211.1060}{{arXiv:1211.1060}}} {[gr-qc]}

\bibitem[{Blanchet et~al.(2014{\natexlab{a}})Blanchet, Faye, and
  Whiting}]{BFW14a}
Blanchet L, Faye G, Whiting BF (2014{\natexlab{a}}) {Half-integral conservative
  post-{Newtonian} approximations in the redshift factor of black hole
  binaries}. Phys Rev D 89:064026. \doi{10.1103/PhysRevD.89.064026}.
  {\href{https://arxiv.org/abs/1312.2975}{{arXiv:1312.2975}}} {[gr-qc]}

\bibitem[{Blanchet et~al.(2014{\natexlab{b}})Blanchet, Faye, and
  Whiting}]{BFW14b}
Blanchet L, Faye G, Whiting BF (2014{\natexlab{b}}) {High-order half-integral
  conservative post-{Newtonian} coefficients in the redshift factor of black
  hole binaries}. Phys Rev D 90:044017. \doi{10.1103/PhysRevD.90.044017}.
  {\href{https://arxiv.org/abs/1405.5151}{{arXiv:1405.5151}}} {[gr-qc]}

\bibitem[{Blanchet et~al.(2020)Blanchet, Foffa, Larrouturou, and
  Sturani}]{BFLS20}
Blanchet L, Foffa S, Larrouturou F, Sturani R (2020) {Logarithmic tail
  contributions to the energy function of circular compact binaries}. Phys Rev
  D 101(8):084045. \doi{10.1103/PhysRevD.101.084045}.
  {\href{https://arxiv.org/abs/1912.12359}{{arXiv:1912.12359}}} {[gr-qc]}

\bibitem[{Blanchet et~al.(2021)Blanchet, Comp{\`e}re, Faye, Oliveri, and
  Seraj}]{BCFOS21}
Blanchet L, Comp{\`e}re G, Faye G, Oliveri R, Seraj A (2021) {Multipole
  expansion of gravitational waves: from harmonic to {Bondi} coordinates}. J
  High Energy Phys 2021(2):29. \doi{10.1007/JHEP02(2021)029}.
  {\href{https://arxiv.org/abs/2011.10000}{{arXiv:2011.10000}}} {[gr-qc]}

\bibitem[{Blanchet et~al.(2022)Blanchet, Faye, and Larrouturou}]{BFL22}
Blanchet L, Faye G, Larrouturou F (2022) {The quadrupole moment of compact
  binaries to the fourth post-{Newtonian} order: from source to canonical
  moment}. Class Quantum Grav 39(19):195003. \doi{10.1088/1361-6382/ac840c}

\bibitem[{Blanchet et~al.(2023{\natexlab{a}})Blanchet, Comp{\`e}re, Faye,
  Oliveri, and Seraj}]{BCFOS23}
Blanchet L, Comp{\`e}re G, Faye G, Oliveri R, Seraj A (2023{\natexlab{a}})
  {Multipole expansion of gravitational waves: memory effects and {Bondi}
  aspects}. J High Energy Phys 2023:123. \doi{10.1007/JHEP07(2023)123}.
  {\href{https://arxiv.org/abs/2303.07732}{{arXiv:2303.07732}}} {[gr-qc]}

\bibitem[{Blanchet et~al.(2023{\natexlab{b}})Blanchet, Faye, Henry,
  Larrouturou, and Trestini}]{BFHLT23b}
Blanchet L, Faye G, Henry Q, Larrouturou F, Trestini D (2023{\natexlab{b}})
  {Gravitational-wave flux and quadrupole modes from quasicircular nonspinning
  compact binaries to the fourth post-{Newtonian} order}. Phys Rev D
  108(6):064041. \doi{10.1103/PhysRevD.108.064041}.
  {\href{https://arxiv.org/abs/2304.11186}{{arXiv:2304.11186}}} {[gr-qc]}

\bibitem[{Blanchet et~al.(2023{\natexlab{c}})Blanchet, Faye, Henry,
  Larrouturou, and Trestini}]{BFHLT23a}
Blanchet L, Faye G, Henry Q, Larrouturou F, Trestini D (2023{\natexlab{c}})
  {Gravitational-Wave Phasing of Quasicircular Compact Binary Systems to the
  Fourth-and-a-Half Post-{Newtonian} Order}. Phys Rev Lett 131(12):121402.
  \doi{10.1103/PhysRevLett.131.121402}.
  {\href{https://arxiv.org/abs/2304.11185}{{arXiv:2304.11185}}} {[gr-qc]}

\bibitem[{Bl{\"u}mlein et~al.(2020{\natexlab{a}})Bl{\"u}mlein, Maier, Marquard,
  and Sch{\"a}fer}]{BlumMMS20a}
Bl{\"u}mlein J, Maier A, Marquard P, Sch{\"a}fer G (2020{\natexlab{a}}) {Fourth
  post-{Newtonian} {Hamiltonian} dynamics of two-body systems from an effective
  field theory approach}. Nucl Phys B 955:115041.
  \doi{10.1016/j.nuclphysb.2020.115041}.
  {\href{https://arxiv.org/abs/2003.01692}{{2003.01692}}}

\bibitem[{Bl{\"u}mlein et~al.(2020{\natexlab{b}})Bl{\"u}mlein, Maier, Marquard,
  and Sch{\"a}fer}]{BlumMMS20b}
Bl{\"u}mlein J, Maier A, Marquard P, Sch{\"a}fer G (2020{\natexlab{b}})
  {Testing binary dynamics in gravity at the sixth post-{Newtonian} level}.
  Phys Lett B 807:135496. \doi{10.1016/j.physletb.2020.135496}.
  {\href{https://arxiv.org/abs/2003.07145}{{arXiv:2003.07145}}} {[gr-qc]}

\bibitem[{Bl{\"u}mlein et~al.(2021)Bl{\"u}mlein, Maier, Marquard, and
  Sch{\"a}fer}]{BlumMMS21}
Bl{\"u}mlein J, Maier A, Marquard P, Sch{\"a}fer G (2021) {The fifth-order
  post-{Newtonian} {Hamiltonian} dynamics of two-body systems from an effective
  field theory approach: potential contributions}. Nucl Phys B 965:115352.
  \doi{10.1016/j.nuclphysb.2021.115352}.
  {\href{https://arxiv.org/abs/2010.13672}{{arXiv:2010.13672}}} {[gr-qc]}

\bibitem[{Bl{\"u}mlein et~al.(2022)Bl{\"u}mlein, Maier, Marquard, and
  Sch{\"a}fer}]{BlumMMS22a}
Bl{\"u}mlein J, Maier A, Marquard P, Sch{\"a}fer G (2022) {The fifth-order
  post-{Newtonian} {Hamiltonian} dynamics of two-body systems from an effective
  field theory approach}. Nucl Phys B 983:115900.
  \doi{10.1016/j.nuclphysb.2022.115900}.
  {\href{https://arxiv.org/abs/2110.13822}{{arXiv:2110.13822}}} {[gr-qc]}

\bibitem[{Boetzel et~al.(2019)Boetzel, Mishra, Faye, Gopakumar, and
  Iyer}]{BMCFGI19}
Boetzel Y, Mishra CK, Faye G, Gopakumar A, Iyer BR (2019) {Gravitational-wave
  amplitudes for compact binaries in eccentric orbits at the third
  post-{Newtonian} order: Tail contributions and postadiabatic corrections}.
  Phys Rev D 100(4):044018.
  {\href{https://arxiv.org/abs/1904.11814}{{arXiv:1904.11814}}} {[gr-qc]}

\bibitem[{Boh{\'{e}} et~al.(2013{\natexlab{a}})Boh{\'{e}}, Marsat, and
  Blanchet}]{BMB13}
Boh{\'{e}} A, Marsat S, Blanchet L (2013{\natexlab{a}})
  {Next-to-next-to-leading order spin-orbit effects in the gravitational wave
  flux and orbital phasing of compact binaries}. Class Quantum Grav 30:135009.
  {\href{https://arxiv.org/abs/1303.7412}{{arXiv:1303.7412}}}

\bibitem[{Boh{\'{e}} et~al.(2013{\natexlab{b}})Boh{\'{e}}, Marsat, Faye, and
  Blanchet}]{BMFB13}
Boh{\'{e}} A, Marsat S, Faye G, Blanchet L (2013{\natexlab{b}})
  {Next-to-next-to-leading order spin-orbit effects in the near-zone metric and
  precession equations of compact binary systems}. Class Quantum Grav
  30:075017. {\href{https://arxiv.org/abs/1212.5520}{{arXiv:1212.5520}}}

\bibitem[{Boh{\'{e}} et~al.(2015)Boh{\'{e}}, Faye, Marsat, and Porter}]{BFMP15}
Boh{\'{e}} A, Faye G, Marsat S, Porter EK (2015) {Quadratic-in-spin effects in
  the orbital dynamics and gravitational-wave energy flux of compact binaries
  at the 3PN order}. Class Quantum Grav 32:195010.
  \doi{10.1088/0264-9381/32/19/195010}.
  {\href{https://arxiv.org/abs/1501.01529}{{arXiv:1501.01529}}} {[gr-qc]}

\bibitem[{Bollini and Giambiagi(1972)}]{Bollini}
Bollini CG, Giambiagi JJ (1972) {Lowest order `divergent' graphs in
  $v$-dimensional space}. Phys Lett B 40:566--568.
  \doi{10.1016/0370-2693(72)90483-2}

\bibitem[{Bonazzola et~al.(1999)Bonazzola, Gourgoulhon, and Marck}]{BGM99}
Bonazzola S, Gourgoulhon E, Marck JA (1999) {Numerical models of irrotational
  binary neutron stars in general relativity}. Phys Rev Lett 82:892--895.
  \doi{10.1103/PhysRevLett.82.892}.
  {\href{https://arxiv.org/abs/gr-qc/9810072}{{arXiv:gr-qc/9810072}}} {[gr-qc]}

\bibitem[{Bondi et~al.(1962)Bondi, van~der Burg, and Metzner}]{BBM62}
Bondi H, van~der Burg MGJ, Metzner AWK (1962) {Gravitational Waves in General
  Relativity. {VII}. {W}aves from Axi-Symmetric Isolated Systems}. Proc R Soc
  London, Ser A 269:21--52. \doi{10.1098/rspa.1962.0161}

\bibitem[{Bonnor(1959)}]{Bo59}
Bonnor WB (1959) {Spherical gravitational waves}. Philos Trans R Soc London,
  Ser A 251:233--271. \doi{10.1098/rsta.1959.0003}

\bibitem[{Bonnor and Rotenberg(1961)}]{BoR61}
Bonnor WB, Rotenberg MA (1961) {Transport of momentum by gravitational waves:
  Linear approximation}. Proc R Soc London, Ser A 265:109--116.
  \doi{10.1098/rspa.1961.0226}

\bibitem[{Bonnor and Rotenberg(1966)}]{BoR66}
Bonnor WB, Rotenberg MA (1966) {Gravitational waves from isolated sources}.
  Proc R Soc London, Ser A 289:247--274. \doi{10.1098/rspa.1966.0010}

\bibitem[{Borhanian et~al.(2020)Borhanian, Arun, Pfeiffer, and
  Sathyaprakash}]{BAPS20}
Borhanian S, Arun KG, Pfeiffer HP, Sathyaprakash BS (2020) {Comparison of
  post-{Newtonian} mode amplitudes with numerical relativity simulations of
  binary black holes}. Class Quantum Grav 37(6):065006.
  \doi{10.1088/1361-6382/ab6a21}.
  {\href{https://arxiv.org/abs/1901.08516}{{arXiv:1901.08516}}} {[gr-qc]}

\bibitem[{Boyle et~al.(2007)Boyle, Brown, Kidder, Mrou{\'{e}}, Pfeiffer,
  Scheel, Cook, and Teukolsky}]{Boyle07}
Boyle M, Brown DA, Kidder LE, Mrou{\'{e}} AH, Pfeiffer HP, Scheel MA, Cook GB,
  Teukolsky SA (2007) {High-accuracy comparison of numerical relativity
  simulations with post-{Newtonian} expansions}. Phys Rev D 76:124038.
  \doi{10.1103/PhysRevD.76.124038}.
  {\href{https://arxiv.org/abs/0710.0158}{{arXiv:0710.0158}}} {[gr-qc]}

\bibitem[{Boyle et~al.(2008)Boyle, Buonanno, Kidder, Mrou{\'{e}}, Pan,
  Pfeiffer, and Scheel}]{Boyle08}
Boyle M, Buonanno A, Kidder LE, Mrou{\'{e}} AH, Pan Y, Pfeiffer HP, Scheel MA
  (2008) {High-accuracy numerical simulation of black-hole binaries:
  Computation of the gravitational-wave energy flux and comparisons with
  post-{Newtonian} approximants}. Phys Rev D 78:104020.
  \doi{10.1103/PhysRevD.78.104020}.
  {\href{https://arxiv.org/abs/0804.4184}{{arXiv:0804.4184}}} {[gr-qc]}

\bibitem[{Braginsky and Thorne(1987{\natexlab{a}})}]{BragTh76}
Braginsky VB, Thorne KS (1987{\natexlab{a}}) {Gravitational-wave bursts with
  memory and experimental prospects}. Nature 327:123--125.
  \doi{10.1038/327123a0}

\bibitem[{Braginsky and Thorne(1987{\natexlab{b}})}]{Braginsky87}
Braginsky VB, Thorne KS (1987{\natexlab{b}}) {Gravitational-wave bursts with
  memory and experimental prospects}. Nature 327(6118):123--125.
  \doi{10.1038/327123a0}

\bibitem[{Brandhuber et~al.(2023)Brandhuber, Brown, Chen, De~Angelis, Gowdy,
  and Travaglini}]{Brandhuber23}
Brandhuber A, Brown GR, Chen G, De~Angelis S, Gowdy J, Travaglini G (2023)
  {One-loop gravitational bremsstrahlung and waveforms from a heavy-mass
  effective field theory}. J High Energy Phys 2023(6).
  \doi{10.1007/jhep06(2023)048}

\bibitem[{Breitenlohner and Maison(1977)}]{Breitenlohner}
Breitenlohner P, Maison D (1977) {Dimensional renormalization and the action
  principle}. Commun Math Phys 52:11--38. \doi{10.1007/BF01609069}

\bibitem[{Brenneman and Reynolds(2006)}]{BrennR06}
Brenneman LW, Reynolds CS (2006) {Constraining Black Hole Spin via X-Ray
  Spectroscopy}. Astrophys J 652:1028--1043. \doi{10.1086/508146}.
  {\href{https://arxiv.org/abs/astro-ph/0608502}{{arXiv:astro-ph/0608502}}}

\bibitem[{Brenneman et~al.(2011)Brenneman, Reynolds, Nowak, Reis, Trippe,
  Fabian, Iwasawa, Lee, Miller, Mushotzky, Nandra, and Volonteri}]{BrennR11}
Brenneman LW, Reynolds CS, Nowak MA, Reis RC, Trippe M, Fabian AC, Iwasawa K,
  Lee JC, Miller JM, Mushotzky RF, Nandra K, Volonteri M (2011) {The Spin of
  the Supermassive Black Hole in NGC 3783}. Astrophys J 736:103.
  \doi{10.1088/0004-637X/736/2/103}.
  {\href{https://arxiv.org/abs/1104.1172}{{arXiv:1104.1172}}} {[astro-ph.HE]}

\bibitem[{Breuer and Rudolph(1981)}]{BR81}
Breuer R, Rudolph E (1981) {Radiation reaction and energy loss in the
  post-{Newtonian} approximation of general relativity}. Gen Relativ Gravit
  13:777. \doi{10.1007/BF00758216}

\bibitem[{Bruhat(1962)}]{YCB62}
Bruhat Y (1962) The {Cauchy} problem. In: Witten L (ed) Gravitation: An
  Introduction to Current Research. Wiley, New York; London, pp 130--168

\bibitem[{Buonanno and Damour(1999)}]{BuonD99}
Buonanno A, Damour T (1999) {Effective one-body approach to general
  relativistic two-body dynamics}. Phys Rev D 59:084006.
  \doi{10.1103/PhysRevD.59.084006}.
  {\href{https://arxiv.org/abs/gr-qc/9811091}{{arXiv:gr-qc/9811091}}}

\bibitem[{Buonanno and Damour(2000)}]{BuonD00}
Buonanno A, Damour T (2000) {Transition from inspiral to plunge in binary black
  hole coalescences}. Phys Rev D 62:064015. \doi{10.1103/PhysRevD.62.064015}.
  {\href{https://arxiv.org/abs/gr-qc/0001013}{{arXiv:gr-qc/0001013}}}

\bibitem[{Buonanno and Sathyaprakash(2015)}]{BuonSathya15}
Buonanno A, Sathyaprakash B (2015) Sources of gravitational waves: Theory and
  observations. In: Ashtekar A, Berger B, Isenberg J, MacCallum M (eds) General
  Relativity and Gravitation: A Centennial Perspective. Cambridge University
  Press, p 287. {\href{https://arxiv.org/abs/1410.7832}{{arXiv:1410.7832}}}
  {[gr-qc]}

\bibitem[{Buonanno et~al.(2003{\natexlab{a}})Buonanno, Chen, and
  Vallisneri}]{BCV03a}
Buonanno A, Chen Y, Vallisneri M (2003{\natexlab{a}}) {Detection template
  families for gravitational waves from the final stages of binary black-holes
  binaries: Nonspinning case}. Phys Rev D 67:024016.
  \doi{10.1103/PhysRevD.67.024016}.
  {\href{https://arxiv.org/abs/gr-qc/0205122}{{gr-qc/0205122}}}

\bibitem[{Buonanno et~al.(2003{\natexlab{b}})Buonanno, Chen, and
  Vallisneri}]{BCV03b}
Buonanno A, Chen Y, Vallisneri M (2003{\natexlab{b}}) {Detection template
  families for precessing binaries of spinning compact binaries: Adiabatic
  limit}. Phys Rev D 67:104025. \doi{10.1103/PhysRevD.67.104025}.
  {\href{https://arxiv.org/abs/gr-qc/0211087}{{gr-qc/0211087}}}

\bibitem[{Buonanno et~al.(2007)Buonanno, Cook, and Pretorius}]{BuonCP07}
Buonanno A, Cook GB, Pretorius F (2007) {Inspiral, merger, and ring-down of
  equal-mass black-hole binaries}. Phys Rev D 75:124018.
  \doi{10.1103/PhysRevD.75.124018}.
  {\href{https://arxiv.org/abs/gr-qc/0610122}{{gr-qc/0610122}}}

\bibitem[{Buonanno et~al.(2009{\natexlab{a}})Buonanno, Iyer, Pan, Ochsner, and
  Sathyaprakash}]{BIO09}
Buonanno A, Iyer BR, Pan Y, Ochsner E, Sathyaprakash BS (2009{\natexlab{a}})
  {Comparison of post-{Newtonian} templates for compact binary inspiral signals
  in gravitational-wave detectors}. Phys Rev D 80:084043.
  \doi{10.1103/PhysRevD.80.084043}.
  {\href{https://arxiv.org/abs/0907.0700}{{arXiv:0907.0700}}} {[gr-qc]}

\bibitem[{Buonanno et~al.(2009{\natexlab{b}})Buonanno, Pan, Pfeiffer, Scheel,
  Buchman, and Kidder}]{Buon09}
Buonanno A, Pan Y, Pfeiffer HP, Scheel MA, Buchman LT, Kidder LE
  (2009{\natexlab{b}}) {Effective-one-body waveforms calibrated to numerical
  relativity simulations: Coalescence of nonspinning, equal-mass black holes}.
  Phys Rev D 79:124028. \doi{10.1103/PhysRevD.79.124028}.
  {\href{https://arxiv.org/abs/0902.0790}{{arXiv:0902.0790}}} {[gr-qc]}

\bibitem[{Buonanno et~al.(2013)Buonanno, Faye, and Hinderer}]{BuonFH13}
Buonanno A, Faye G, Hinderer T (2013) {Spin effects on gravitational waves from
  inspiralling compact binaries at second post-{Newtonian} order}. Phys Rev D
  87:044009. \doi{10.1103/PhysRevD.87.044009}.
  {\href{https://arxiv.org/abs/1209.6349}{{arXiv:1209.6349}}}

\bibitem[{Burke(1971)}]{Bu71}
Burke WL (1971) {Gravitational radiation damping of slowly moving systems
  calculated using matched asymptotic expansions}. J Math Phys 12:401--418.
  \doi{10.1063/1.1665603}

\bibitem[{Burke and Thorne(1970)}]{BuTh70}
Burke WL, Thorne KS (1970) Gravitational radiation damping. In: Carmeli M,
  Fickler SI, Witten L (eds) Relativity. Plenum Press, New York; London, pp
  209--228

\bibitem[{Campanelli(2005)}]{Camp05}
Campanelli M (2005) {Understanding the fate of merging supermassive black
  holes}. Class Quantum Grav 22:S387. \doi{10.1088/0264-9381/22/10/034}.
  {\href{https://arxiv.org/abs/astro-ph/0411744}{{astro-ph/0411744}}}

\bibitem[{Campanelli et~al.(2006)Campanelli, Lousto, Marronetti, and
  Zlochower}]{Camp06}
Campanelli M, Lousto CO, Marronetti P, Zlochower Y (2006) {Accurate Evolutions
  of Orbiting Black-Hole Binaries without Excision}. Phys Rev Lett 96:111101.
  \doi{10.1103/PhysRevLett.96.111101}.
  {\href{https://arxiv.org/abs/gr-qc/0511048}{{arXiv:gr-qc/0511048}}}

\bibitem[{Campanelli et~al.(2007)Campanelli, Lousto, Zlochower, and
  Merritt}]{Camp07}
Campanelli M, Lousto CO, Zlochower Y, Merritt D (2007) {Large merger recoils
  and spin flips from generic black-hole binaries}. Astrophys J Lett
  659:L5--L8. \doi{10.1086/516712}.
  {\href{https://arxiv.org/abs/gr-qc/0701164}{{arXiv:gr-qc/0701164}}} {[gr-qc]}

\bibitem[{Campbell and Morgan(1971)}]{CM71}
Campbell WB, Morgan TA (1971) {Debye Potentials For Gravitational Field}.
  Physica 53(2):264. \doi{10.1016/0031-8914(71)90074-7}

\bibitem[{Campbell et~al.(1977)Campbell, Macek, and Morgan}]{CMM77}
Campbell WB, Macek J, Morgan TA (1977) {Relativistic time-dependent multipole
  analysis for scalar, electromagnetic, and gravitational fields}. Phys Rev D
  15:2156--2164. \doi{10.1103/PhysRevD.15.2156}

\bibitem[{Cardoso et~al.(2003)Cardoso, Dias, and Lemos}]{Cardoso}
Cardoso V, Dias OJC, Lemos JPS (2003) {Gravitational radiation in
  $D$-dimensional spacetimes}. Phys Rev D 67:064026.
  \doi{10.1103/PhysRevD.67.064026}

\bibitem[{Carmeli(1965)}]{Carmeli65}
Carmeli M (1965) {The equations of motion of slowly moving particles in the
  general theory of relativity}. Nuovo Cimento 37:842. \doi{10.1007/BF02773176}

\bibitem[{Caudill et~al.(2006)Caudill, Cook, Grigsby, and Pfeiffer}]{CCGPf06}
Caudill M, Cook GB, Grigsby JD, Pfeiffer HP (2006) {Circular orbits and spin in
  black-hole initial data}. Phys Rev D 74:064011.
  \doi{10.1103/PhysRevD.74.064011}.
  {\href{https://arxiv.org/abs/gr-qc/0605053}{{gr-qc/0605053}}}

\bibitem[{Chandrasekhar(1965)}]{C65}
Chandrasekhar S (1965) {The Post-{Newtonian} Equations of Hydrodynamics in
  General Relativity}. Astrophys J 142:1488--1540. \doi{10.1086/148432}

\bibitem[{Chandrasekhar and Esposito(1970)}]{CE70}
Chandrasekhar S, Esposito FP (1970) {The $2\frac{1}{2}$-Post-{Newtonian}
  Equations of Hydrodynamics and Radiation Reaction in General Relativity}.
  Astrophys J 160:153--179. \doi{10.1086/150414}

\bibitem[{Chandrasekhar and Nutku(1969)}]{CN69}
Chandrasekhar S, Nutku Y (1969) {The Second Post-{Newtonian} Equations of
  Hydrodynamics in General Relativity}. Astrophys J 158:55--79.
  \doi{10.1086/150171}

\bibitem[{Chatziioannou et~al.(2013)Chatziioannou, Poisson, and
  Yunes}]{Chatz12}
Chatziioannou K, Poisson E, Yunes N (2013) {Tidal heating and torquing of a
  {Kerr} black hole to next-to-leading order in the tidal coupling}. Phys Rev D
  87:044022. \doi{10.1103/PhysRevD.87.044022}.
  {\href{https://arxiv.org/abs/1211.1686}{{arXiv:1211.1686}}} {[gr-qc]}

\bibitem[{Chicone et~al.(2001)Chicone, Kopeikin, Mashhoon, and
  Retzloff}]{CKMR01}
Chicone C, Kopeikin SM, Mashhoon B, Retzloff DG (2001) {Delay equations and
  radiation damping}. Phys Lett A 285:17--26.
  \doi{10.1016/S0375-9601(01)00327-9}.
  {\href{https://arxiv.org/abs/gr-qc/0101122}{{gr-qc/0101122}}}

\bibitem[{Cho et~al.(2021)Cho, Pardo, and Porto}]{CPP21}
Cho G, Pardo B, Porto RA (2021) {Gravitational radiation from inspiralling
  compact objects: Spin-spin effects completed at the next-to-leading
  post-{Newtonian} order}. Phys Rev D 104(2):024037.
  \doi{10.1103/PhysRevD.104.024037}.
  {\href{https://arxiv.org/abs/2103.14612}{{arXiv:2103.14612}}} {[gr-qc]}

\bibitem[{Cho et~al.(2022{\natexlab{a}})Cho, Porto, and Yang}]{CPY22}
Cho G, Porto RA, Yang Z (2022{\natexlab{a}}) {Gravitational radiation from
  inspiralling compact objects: Spin effects to the fourth post-{Newtonian}
  order}. Phys Rev D 106(10):L101501. \doi{10.1103/PhysRevD.106.L101501}.
  {\href{https://arxiv.org/abs/2201.05138}{{arXiv:2201.05138}}} {[gr-qc]}

\bibitem[{Cho et~al.(2022{\natexlab{b}})Cho, Tanay, Gopakumar, and Lee}]{CTG22}
Cho G, Tanay S, Gopakumar A, Lee HM (2022{\natexlab{b}}) {Generalized
  quasi-{Keplerian} solution for eccentric, nonspinning compact binaries at
  4{PN} order and the associated inspiral-merger-ringdown waveform}. Phys Rev D
  105(6):064010. \doi{10.1103/PhysRevD.105.064010}.
  {\href{https://arxiv.org/abs/2110.09608}{{arXiv:2110.09608}}} {[gr-qc]}

\bibitem[{Christodoulou(1970)}]{Chr70}
Christodoulou D (1970) {Reversible and irreversible transformations in
  black-hole physics}. Phys Rev Lett 25:1596. \doi{10.1103/PhysRevLett.25.1596}

\bibitem[{Christodoulou(1991)}]{Chr91}
Christodoulou D (1991) {Nonlinear Nature of Gravitation and Gravitational-Wave
  Experiments}. Phys Rev Lett 67:1486--1489. \doi{10.1103/PhysRevLett.67.1486}

\bibitem[{Christodoulou and Ruffini(1971)}]{ChrR71}
Christodoulou D, Ruffini R (1971) {Reversible transformations of a charged
  black hole}. Phys Rev D 4:3552--3555. \doi{10.1103/PhysRevD.4.3552}

\bibitem[{Christodoulou and Schmidt(1979)}]{CS79}
Christodoulou D, Schmidt BG (1979) {Convergent and Asymptotic Iteration Methods
  in General Relativity}. Commun Math Phys 68:275--289.
  \doi{10.1007/BF01221128}

\bibitem[{Collins(1984)}]{Collins}
Collins JC (1984) Renormalization: An introduction to renormalization, the
  renormalization group, and the operator-product expansion. Cambridge
  University Press, Cambridge; New York

\bibitem[{Comp{\`e}re et~al.(2018)Comp{\`e}re, Oliveri, and Seraj}]{COS18}
Comp{\`e}re G, Oliveri R, Seraj A (2018) {Gravitational multipole moments from
  {Noether} charges}. J High Energy Phys 2018(5):54.
  \doi{10.1007/JHEP05(2018)054}.
  {\href{https://arxiv.org/abs/1711.08806}{{arXiv:1711.08806}}}

\bibitem[{Comp{\`e}re et~al.(2020)Comp{\`e}re, Oliveri, and Seraj}]{COS20}
Comp{\`e}re G, Oliveri R, Seraj A (2020) {The {Poincar{\'e}} and {BMS}
  flux-balance laws with application to binary systems}. J High Energy Phys
  2020(10):116. \doi{10.1007/JHEP10(2020)116}.
  {\href{https://arxiv.org/abs/1912.03164}{{arXiv:1912.03164}}}

\bibitem[{Cook(1994)}]{Cook94}
Cook GB (1994) {Three-dimensional initial data for the collision of two black
  holes. {II}. {Q}uasicircular orbits for equal-mass black holes}. Phys Rev D
  50:5025--5032. \doi{10.1103/PhysRevD.50.5025}

\bibitem[{Cook and Pfeiffer(2004)}]{CPf04}
Cook GB, Pfeiffer HP (2004) {Excision boundary conditions for black hole
  initial data}. Phys Rev D 70:104016. \doi{10.1103/PhysRevD.70.104016}

\bibitem[{Cooperstock and Booth(1969)}]{CB69}
Cooperstock FI, Booth DJ (1969) {Angular-Momentum Flux For Gravitational
  Radiation to Octupole Order}. Nuovo Cimento B 62(1):163--170.
  \doi{10.1007/BF02712475}

\bibitem[{Corinaldesi and Papapetrou(1951)}]{CPapa51}
Corinaldesi E, Papapetrou A (1951) {Spinning test-particles in general
  relativity. {II}}. Proc R Soc London, Ser A 209:259--268.
  \doi{10.1098/rspa.1951.0201}

\bibitem[{Crowley and Thorne(1977)}]{CTh77}
Crowley RJ, Thorne KS (1977) {Generation of gravitational waves. {II}.
  {P}ost-linear formalism revisited}. Astrophys J 215:624--635.
  \doi{10.1086/155397}

\bibitem[{Cutler and Flanagan(1994)}]{CF94}
Cutler C, Flanagan {\'{E}}{\'{E}} (1994) {Gravitational waves from merging
  compact binaries: How accurately can one extract the binary's parameters from
  the inspiral wave form?} Phys Rev D 49:2658--2697.
  \doi{10.1103/PhysRevD.49.2658}.
  {\href{https://arxiv.org/abs/gr-qc/9402014}{{arXiv:gr-qc/9402014}}}

\bibitem[{Cutler et~al.(1993{\natexlab{a}})Cutler, Apostolatos, Bildsten, Finn,
  Flanagan, Kennefick, Markovi{\'{c}}, Ori, Poisson, and Sussman}]{3mn}
Cutler C, Apostolatos TA, Bildsten L, Finn LS, Flanagan {\'{E}}{\'{E}},
  Kennefick D, Markovi{\'{c}} DM, Ori A, Poisson E, Sussman GJ
  (1993{\natexlab{a}}) {The Last Three Minutes: Issues in Gravitational-Wave
  Measurements of Coalescing Compact Binaries}. Phys Rev Lett 70:2984--2987.
  \doi{10.1103/PhysRevLett.70.2984}.
  {\href{https://arxiv.org/abs/astro-ph/9208005}{{astro-ph/9208005}}}

\bibitem[{Cutler et~al.(1993{\natexlab{b}})Cutler, Finn, Poisson, and
  Sussman}]{CFPS93}
Cutler C, Finn LS, Poisson E, Sussman GJ (1993{\natexlab{b}}) {Gravitational
  radiation from a particle in circular orbit around a black hole. {II}.
  {N}umerical results for the nonrotating case}. Phys Rev D 47:1511--1518.
  \doi{10.1103/PhysRevD.47.1511}

\bibitem[{D'Alembert(1743)}]{Dalembert}
D'Alembert J (1743) Trait\'e de Dynamique. David L'Aine, Paris

\bibitem[{Damour(1982)}]{D82}
Damour T (1982) {Probl{\`e}me des deux corps et freinage de rayonnement en
  relativit{\'e} g{\'e}n{\'e}rale}. C R Acad Sci Ser II 294:1355--1357

\bibitem[{Damour(1983{\natexlab{a}})}]{Dhouches}
Damour T (1983{\natexlab{a}}) Gravitational radiation and the motion of compact
  bodies. In: Deruelle N, Piran T (eds) Rayonnement Gravitationnel /
  Gravitational Radiation. North-Holland, Amsterdam, pp 59--144

\bibitem[{Damour(1983{\natexlab{b}})}]{D83}
Damour T (1983{\natexlab{b}}) {Gravitational radiation reaction in the binary
  pulsar and the quadrupole formula controvercy}. Phys Rev Lett 51:1019--1021.
  \doi{10.1103/PhysRevLett.51.1019}

\bibitem[{{Damour}(1986)}]{D86MG}
{Damour} T (1986) {Analytical calculations of gravitational radiation.} In:
  {Ruffini} R (ed) Fourth Marcel Grossmann Meeting on General Relativity. pp
  365--392

\bibitem[{Damour(1987{\natexlab{a}})}]{Dcargese}
Damour T (1987{\natexlab{a}}) An introduction to the theory of gravitational
  radiation. In: Carter B, Hartle JB (eds) Gravitation in Astrophysics:
  Carg\`{e}se 1986. NATO ASI Series B, vol 156. Plenum Press, New York, pp
  3--62

\bibitem[{Damour(1987{\natexlab{b}})}]{D300}
Damour T (1987{\natexlab{b}}) The problem of motion in {{Newtonian}} and
  {Einsteinian} gravity. In: Hawking SW, Israel W (eds) Three Hundred Years of
  Gravitation. Cambridge University Press, Cambridge; New York, pp 128--198

\bibitem[{Damour(2010)}]{D10sf}
Damour T (2010) {Gravitational self-force in a {Schwarzschild} background and
  the effective one-body formalism}. Phys Rev D 81:024017.
  \doi{10.1103/PhysRevD.81.024017}.
  {\href{https://arxiv.org/abs/0910.5533}{{arXiv:0910.5533}}} {[gr-qc]}

\bibitem[{Damour and Deruelle(1981{\natexlab{a}})}]{DD81b}
Damour T, Deruelle N (1981{\natexlab{a}}) {Lagrangien g{\'e}n{\'e}ralis{\'e} du
  syst{\`e}me de deux masses ponctuelles, {\`a} l'approximation
  post-post-newtonienne de la relativit{\'e} g{\'e}n{\'e}rale}. C R Acad Sci
  Ser II 293:537--540

\bibitem[{Damour and Deruelle(1981{\natexlab{b}})}]{DD81a}
Damour T, Deruelle N (1981{\natexlab{b}}) {Radiation reaction and angular
  momentum loss in small angle gravitational scattering}. Phys Lett A
  87:81--84. \doi{10.1016/0375-9601(81)90567-3}

\bibitem[{Damour and Deruelle(1985)}]{DD85}
Damour T, Deruelle N (1985) {General relativistic celestial mechanics of binary
  systems {I}. {T}he post-{Newtonian} motion}. Ann Inst Henri Poincare A
  43:107--132.
  \urlprefix\url{http://www.numdam.org/item/AIHPA_1985__43_1_107_0/}

\bibitem[{Damour and Deruelle(1986)}]{DD86}
Damour T, Deruelle N (1986) {General relativistic celestial mechanics of binary
  systems {II}. {T}he post-{Newtonian} timing formula}. Ann Inst Henri Poincare
  A 44:263--292.
  \urlprefix\url{http://www.numdam.org/item/AIHPA_1986__44_3_263_0/}

\bibitem[{Damour and Esposito-Far{\`{e}}se(1996)}]{Dgef96}
Damour T, Esposito-Far{\`{e}}se G (1996) {Testing gravity to second
  post-{Newtonian} order: A field-theory approach}. Phys Rev D 53:5541--5578.
  \doi{10.1103/PhysRevD.53.5541}.
  {\href{https://arxiv.org/abs/gr-qc/9506063}{{gr-qc/9506063}}}

\bibitem[{Damour and Gopakumar(2006)}]{DG06}
Damour T, Gopakumar A (2006) {Gravitational recoil during binary black hole
  coalescence using the effective one body approach}. Phys Rev D 73(12):124006.
  \doi{10.1103/PhysRevD.73.124006}.
  {\href{https://arxiv.org/abs/gr-qc/0602117}{{arXiv:gr-qc/0602117}}}

\bibitem[{Damour and Iyer(1991{\natexlab{a}})}]{DI91b}
Damour T, Iyer BR (1991{\natexlab{a}}) {Multipole analysis for electromagnetism
  and linearized gravity with irreducible Cartesian tensors}. Phys Rev D
  43:3259--3272. \doi{10.1103/PhysRevD.43.3259}

\bibitem[{Damour and Iyer(1991{\natexlab{b}})}]{DI91a}
Damour T, Iyer BR (1991{\natexlab{b}}) {Post-{Newtonian} generation of
  gravitational waves. {II}. {T}he spin moments}. Ann Inst Henri Poincare A
  54:115--164.
  \urlprefix\url{http://www.numdam.org/item?id=AIHPA_1991__54_2_115_0}

\bibitem[{Damour and Nagar(2009)}]{DN09tidal}
Damour T, Nagar A (2009) {Relativistic tidal properties of neutron stars}. Phys
  Rev D 80:084035. \doi{10.1103/PhysRevD.80.084035}.
  {\href{https://arxiv.org/abs/0906.0096}{{arXiv:0906.0096}}} {[gr-qc]}

\bibitem[{Damour and Nagar(2011)}]{DNorleans}
Damour T, Nagar A (2011) The effective one-body description of the two-body
  problem. In: Blanchet L, Spallicci A, Whiting B (eds) Mass and Motion in
  General Relativity. Fundamental Theories of Physics, vol 162. Springer,
  Dordrecht; New York, pp 211--252. \doi{10.1007/978-90-481-3015-3_7}

\bibitem[{Damour and Sch{\"{a}}fer(1985)}]{DS85}
Damour T, Sch{\"{a}}fer G (1985) {Lagrangians for $n$ Point Masses at the
  Second Post-{Newtonian} Approximation of General Relativity}. Gen Relativ
  Gravit 17:879--905. \doi{10.1007/BF00773685}

\bibitem[{Damour and Sch{\"{a}}fer(1988)}]{DS88}
Damour T, Sch{\"{a}}fer G (1988) {Higher-Order Relativistic Periastron Advances
  in Binary Pulsars}. Nuovo Cimento B 101:127--176. \doi{10.1007/BF02828697}

\bibitem[{Damour and Sch{\"a}fer(1991)}]{DS91}
Damour T, Sch{\"a}fer G (1991) {Redefinition of position variables and the
  reduction of higher-order {Lagrangians}}. J Math Phys 32(1):127--134.
  \doi{10.1063/1.529135}

\bibitem[{Damour and Schmidt(1990)}]{DS90}
Damour T, Schmidt BG (1990) {Reliability of perturbation theory in general
  relativity}. J Math Phys 31:2441--2458. \doi{10.1063/1.528850}

\bibitem[{Damour and Taylor(1991)}]{DT91}
Damour T, Taylor JH (1991) {On the Orbital Period Change of the Binary Pulsar
  PSR 1913+16}. Astrophys J 366:501--511. \doi{10.1086/169585}

\bibitem[{Damour et~al.(1991)Damour, Soffel, and Xu}]{DSX91}
Damour T, Soffel M, Xu C (1991) {General-relativistic celestial mechanics. {I}.
  {M}ethod and definition of reference systems}. Phys Rev D 43:3273--3307.
  \doi{10.1103/PhysRevD.43.3273}

\bibitem[{Damour et~al.(1998)Damour, Iyer, and Sathyaprakash}]{DIS98}
Damour T, Iyer BR, Sathyaprakash BS (1998) {Improved filters for gravitational
  waves from inspiraling compact binaries}. Phys Rev D 57:885--907.
  \doi{10.1103/PhysRevD.57.885}.
  {\href{https://arxiv.org/abs/gr-qc/9708034}{{gr-qc/9708034}}}

\bibitem[{Damour et~al.(2000{\natexlab{a}})Damour, Iyer, and
  Sathyaprakash}]{DIS00}
Damour T, Iyer BR, Sathyaprakash BS (2000{\natexlab{a}}) {Frequency-domain
  P-approximant filters for time-truncated inspiral gravitational wave signals
  from compact binaries}. Phys Rev D 62:084036.
  \doi{10.1103/PhysRevD.62.084036}.
  {\href{https://arxiv.org/abs/gr-qc/0001023}{{gr-qc/0001023}}}

\bibitem[{Damour et~al.(2000{\natexlab{b}})Damour, Jaranowski, and
  Sch{\"{a}}fer}]{DJSinv}
Damour T, Jaranowski P, Sch{\"{a}}fer G (2000{\natexlab{b}}) {Dynamical
  invariants for general relativistic two-body systems at the third
  post-{Newtonian} approximation}. Phys Rev D 62:044024.
  {\href{https://arxiv.org/abs/gr-qc/9912092}{{gr-qc/9912092}}}

\bibitem[{Damour et~al.(2000{\natexlab{c}})Damour, Jaranowski, and
  Sch{\"{a}}fer}]{DJSisco}
Damour T, Jaranowski P, Sch{\"{a}}fer G (2000{\natexlab{c}}) {On the
  determination of the last stable orbit for circular general relativistic
  binaries at the third post-{Newtonian} approximation}. Phys Rev D 62:084011.
  {\href{https://arxiv.org/abs/gr-qc/0005034}{{gr-qc/0005034}}}

\bibitem[{Damour et~al.(2000{\natexlab{d}})Damour, Jaranowski, and
  Sch{\"{a}}fer}]{DJSpoinc}
Damour T, Jaranowski P, Sch{\"{a}}fer G (2000{\natexlab{d}}) {Poincar\'e
  invariance in the {ADM} {Hamiltonian} approach to the general relativistic
  two-body problem}. Phys Rev D 62:021501(R). {Erratum}: Phys. Rev. D, 63,
  029903(E) (2000).
  {\href{https://arxiv.org/abs/gr-qc/0003051}{{gr-qc/0003051}}}

\bibitem[{Damour et~al.(2001{\natexlab{a}})Damour, Jaranowski, and
  Sch{\"{a}}fer}]{DJSdim}
Damour T, Jaranowski P, Sch{\"{a}}fer G (2001{\natexlab{a}}) {Dimensional
  regularization of the gravitational interaction of point masses}. Phys Lett B
  513:147--155. \doi{10.1016/S0370-2693(01)00642-6}.
  {\href{https://arxiv.org/abs/gr-qc/0105038}{{gr-qc/0105038}}}

\bibitem[{Damour et~al.(2001{\natexlab{b}})Damour, Jaranowski, and
  Sch{\"{a}}fer}]{DJSequiv}
Damour T, Jaranowski P, Sch{\"{a}}fer G (2001{\natexlab{b}}) {Equivalence
  between the {ADM}-{Hamiltonian} and the harmonic-coordinates approaches to
  the third post-{Newtonian} dynamics of compact binaries}. Phys Rev D
  63:044021. \doi{10.1103/PhysRevD.63.044021}, {Erratum}: Phys. Rev. D, 66,
  029901(E) (2002).
  {\href{https://arxiv.org/abs/gr-qc/0010040}{{gr-qc/0010040}}}

\bibitem[{Damour et~al.(2002)Damour, Iyer, and Sathyaprakash}]{DIS02}
Damour T, Iyer BR, Sathyaprakash BS (2002) {Comparison of search templates for
  gravitational waves from binary inspiral: 3.5PN update}. Phys Rev D
  66:027502. \doi{10.1103/PhysRevD.66.027502}.
  {\href{https://arxiv.org/abs/gr-qc/0207021}{{gr-qc/0207021}}}

\bibitem[{Damour et~al.(2003)Damour, Iyer, Jaranowski, and
  Sathyaprakash}]{DIJS03}
Damour T, Iyer BR, Jaranowski P, Sathyaprakash BS (2003) {Gravitational waves
  from black hole binary inspiral and merger: The span of third
  post-{Newtonian} effective-one-body templates}. Phys Rev D 67:064028.
  \doi{10.1103/PhysRevD.67.064028}.
  {\href{https://arxiv.org/abs/gr-qc/0211041}{{gr-qc/0211041}}}

\bibitem[{Damour et~al.(2004)Damour, Gopakumar, and Iyer}]{DGI04}
Damour T, Gopakumar A, Iyer BR (2004) {Phasing of gravitational waves from
  inspiralling eccentric binaries}. Phys Rev D 70:064028.
  \doi{10.1103/PhysRevD.70.064028}.
  {\href{https://arxiv.org/abs/gr-qc/0404128}{{gr-qc/0404128}}}

\bibitem[{Damour et~al.(2008)Damour, Jaranowski, and Sch{\"{a}}fer}]{DJSspin}
Damour T, Jaranowski P, Sch{\"{a}}fer G (2008) {{Hamiltonian} of two spinning
  compact bodies with next-to-leading order gravitational spin-orbit coupling}.
  Phys Rev D 77:064032. \doi{10.1103/PhysRevD.77.064032}.
  {\href{https://arxiv.org/abs/0711.1048}{{arXiv:0711.1048}}}

\bibitem[{Damour et~al.(2012)Damour, Nagar, and Villain}]{DNV12}
Damour T, Nagar A, Villain L (2012) {Measurability of the tidal polarizability
  of neutron stars in late-inspiral gravitational-wave signals}. Phys Rev D
  85:123007. \doi{10.1103/PhysRevD.85.123007}

\bibitem[{Damour et~al.(2014)Damour, Jaranowski, and Sch{\"{a}}fer}]{DJS14}
Damour T, Jaranowski P, Sch{\"{a}}fer G (2014) {Non-local-in-time action for
  the fourth post-{Newtonian} conservative dynamics of two-body systems}. Phys
  Rev D 89:064058. \doi{10.1103/PhysRevD.89.064058}.
  {\href{https://arxiv.org/abs/1401.4548}{{arXiv:1401.4548}}} {[gr-qc]}

\bibitem[{Damour et~al.(2015)Damour, Jaranowski, and Sch{\"{a}}fer}]{DJS15eob}
Damour T, Jaranowski P, Sch{\"{a}}fer G (2015) {Fourth post-{Newtonian}
  effective one-body dynamics}. Phys Rev D 91:084024.
  \doi{10.1103/PhysRevD.91.084024}.
  {\href{https://arxiv.org/abs/1502.07245}{{arXiv:1502.07245}}} {[gr-qc]}

\bibitem[{{de Andrade} et~al.(2001){de Andrade}, Blanchet, and Faye}]{ABF01}
{de Andrade} VC, Blanchet L, Faye G (2001) {Third post-{Newtonian} dynamics of
  compact binaries: Noetherian conserved quantities and equivalence between the
  harmonic-coordinate and {ADM}-{Hamiltonian} formalisms}. Class Quantum Grav
  18:753--778. \doi{10.1088/0264-9381/18/5/301}.
  {\href{https://arxiv.org/abs/gr-qc/0011063}{{gr-qc/0011063}}}

\bibitem[{{De Donder}(1921)}]{deDonder}
{De Donder} T (1921) {La gravifique einsteinienne}. Gauthier-Villars

\bibitem[{{De Witt} and Brehme(1960)}]{dWB60}
{De Witt} B, Brehme R (1960) Radiation damping in a gravitational field. Ann
  Phys (NY) 9:220--259

\bibitem[{Detweiler(2005)}]{Det05}
Detweiler S (2005) {Perspective on gravitational self-force analyses}. Class
  Quantum Grav 22:S681--S716. \doi{10.1088/0264-9381/22/15/006}.
  {\href{https://arxiv.org/abs/gr-qc/0501004}{{arXiv:gr-qc/0501004}}}

\bibitem[{Detweiler(2008)}]{Det08}
Detweiler S (2008) {Consequence of the gravitational self-force for circular
  orbits of the {Schwarzschild} geometry}. Phys Rev D 77:124026.
  \doi{10.1103/PhysRevD.77.124026}.
  {\href{https://arxiv.org/abs/0804.3529}{{arXiv:0804.3529}}}

\bibitem[{Detweiler(2011)}]{Detweilerorleans}
Detweiler S (2011) Elementary development of the gravitational self-force. In:
  Blanchet L, Spallicci A, Whiting B (eds) Mass and Motion in General
  Relativity. Fundamental Theories of Physics, vol 162. Springer, Dordrecht;
  New York, pp 271--307. \doi{10.1007/978-90-481-3015-3_10}

\bibitem[{Detweiler and Whiting(2003)}]{DW03}
Detweiler S, Whiting BF (2003) {Self-force via a Green's function
  decomposition}. Phys Rev D 67:024025. \doi{10.1103/PhysRevD.67.024025}.
  {\href{https://arxiv.org/abs/gr-qc/0202086}{{arXiv:gr-qc/0202086}}}

\bibitem[{Dixon(1964)}]{Dixon64}
Dixon WG (1964) A covariant multipole formalism for extended test bodies in
  general relativity. Il Nuovo Cimento 34:317--339. \doi{10.1007/BF02734579}

\bibitem[{Dixon(1973)}]{Dixon73}
Dixon WG (1973) The definition of multipole moments for extended bodies. Gen
  Relativ Gravit 4:199--209. \doi{10.1007/BF02412488}

\bibitem[{Dixon(1979)}]{Dixon79}
Dixon WG (1979) Extended bodies in general relativity: Their description and
  motion. In: Ehlers J (ed) Isolated Gravitating Systems in General Relativity
  (Sistemi gravitazionali isolati in relativit\`a generale). North-Holland,
  Amsterdam; New York, pp 156--219

\bibitem[{Dlapa et~al.(2022)Dlapa, K{\"a}lin, Liu, and Porto}]{DKLP22}
Dlapa C, K{\"a}lin G, Liu Z, Porto RA (2022) {Dynamics of binary systems to
  fourth Post-{Minkowskian} order from the effective field theory approach}.
  Phys Lett B 831:137203. \doi{10.1016/j.physletb.2022.137203}

\bibitem[{Droste(1917)}]{Droste17}
Droste J (1917) {The field of a single centre in Einstein's theory of
  gravitation, and the motion of a particle in that field}. K Nederl Akad
  Wetens Proc 19:197. Reprinted in Gen. Relativ. Gravit. 34, 1545--1563 (2002)

\bibitem[{Duez and Zlochower(2018)}]{DZ18}
Duez MD, Zlochower Y (2018) {Numerical relativity of compact binaries in the
  21st century}. Rep Prog Phys 82(1):016902. \doi{10.1088/1361-6633/aadb16}.
  {\href{https://arxiv.org/abs/1808.06011}{{arXiv:1808.06011}}}

\bibitem[{Dyson(1963)}]{Dyson}
Dyson F (1963) Gravitational machines. New-York: Benjamin Press

\bibitem[{Ebersold et~al.(2019)Ebersold, Boetzel, Faye, Mishra, Iyer, and
  Jetzer}]{EBFMIJ19}
Ebersold M, Boetzel Y, Faye G, Mishra CK, Iyer BR, Jetzer P (2019)
  {Gravitational-wave amplitudes for compact binaries in eccentric orbits at
  the third post-{Newtonian} order: Memory contributions}. Phys Rev D
  100(8):084043. {\href{https://arxiv.org/abs/1906.06263}{{arXiv:1906.06263}}}
  {[gr-qc]}

\bibitem[{Eder(1983)}]{Eder}
Eder E (1983) {Existence, uniqueness and iterative construction of motions of
  charged particles with retarded interactions}. Ann Inst Henri Poincare A
  39:1--27. \urlprefix\url{http://www.numdam.org/item?id=AIHPA_1983__39_1_1_0}

\bibitem[{Ehlers(1980)}]{Ehl80}
Ehlers J (1980) {Isolated systems in general relativity}. Ann NY Acad Sci
  336:279--294. \doi{10.1111/j.1749-6632.1980.tb15936.x}

\bibitem[{Ehlers et~al.(1976)Ehlers, Rosenblum, Goldberg, and
  Havas}]{Ehletal76}
Ehlers J, Rosenblum A, Goldberg JN, Havas P (1976) {Comments on gravitational
  radiation damping and energy loss in binary systems}. Astrophys J Lett
  208:L77--L81. \doi{10.1086/182236}

\bibitem[{Einstein(1915)}]{E15a}
Einstein A (1915) {Die {F}eldgleichungen der {G}ravitation}. Sitzungsber K
  Preuss Akad Wiss 1915:844--847.
  \urlprefix\url{http://echo.mpiwg-berlin.mpg.de/MPIWG:ZZB2HK6W}

\bibitem[{Einstein(1918)}]{E18}
Einstein A (1918) {\"{U}ber Gravitationswellen}. Sitzungsber K Preuss Akad Wiss
  1918:154--167. \urlprefix\url{http://echo.mpiwg-berlin.mpg.de/MPIWG:8HSP60BU}

\bibitem[{Einstein et~al.(1938)Einstein, Infeld, and Hoffmann}]{EIH}
Einstein A, Infeld L, Hoffmann B (1938) {The Gravitational Equations and the
  Problem of Motion}. Ann Math (2) 39:65--100. \doi{10.2307/1968714}

\bibitem[{Elkhidir et~al.(2023)Elkhidir, {O'Connell}, Sergola, and
  Vazquez-Holm}]{elkhidir2023}
Elkhidir A, {O'Connell} D, Sergola M, Vazquez-Holm IA (2023) {Radiation and
  reaction at one loop}. arXiv e-prints
  {\href{https://arxiv.org/abs/2303.06211}{{arXiv:2303.06211}}} {[hep-th]}

\bibitem[{Epstein(1978)}]{Epstein78}
Epstein R (1978) {The generation of gravitational radiation by escaping
  supernova neutrinos}. Astrophys J 223:1037--1045

\bibitem[{Epstein and Wagoner(1975)}]{EW75}
Epstein R, Wagoner RV (1975) {Post-{Newtonian} Generation of Gravitational
  Waves}. Astrophys J 197:717--723. \doi{10.1086/153561}

\bibitem[{Esposito and Harrison(1975)}]{EH75}
Esposito LW, Harrison ER (1975) {Properties of the {Hulse}-{Taylor} binary
  pulsar system}. Astrophys J Lett 196:L1--L2. \doi{10.1086/181729}

\bibitem[{Faber and Rasio(2012)}]{FR12}
Faber JA, Rasio FA (2012) {Binary Neutron Star Mergers}. Living Rev Relativ
  15:8. \doi{10.12942/lrr-2012-8}.
  {\href{https://arxiv.org/abs/1204.3858}{{arXiv:1204.3858}}} {[gr-qc]}

\bibitem[{Fabian and Miniutti(2009)}]{FM05}
Fabian AC, Miniutti G (2009) {The X-ray spectra of accreting {Kerr} black
  holes}. In: Wiltshire DL, Visser M, Scott SM (eds) The {Kerr} Spacetime:
  Rotating Black Holes in General Relativity. Cambridge University Press,
  Cambridge; New York, chap~9.
  {\href{https://arxiv.org/abs/astro-ph/0507409}{{arXiv:astro-ph/0507409}}}

\bibitem[{Favata(2009)}]{F09}
Favata M (2009) {Post-{Newtonian} corrections to the gravitational-wave memory
  for quasicircular, inspiralling compact binaries}. Phys Rev D 80:024002.
  \doi{10.1103/PhysRevD.80.024002}.
  {\href{https://arxiv.org/abs/0812.0069}{{arXiv:0812.0069}}}

\bibitem[{Favata(2011{\natexlab{a}})}]{F11b}
Favata M (2011{\natexlab{a}}) {Conservative corrections to the innermost stable
  circular orbit (ISCO) of a {Kerr} black hole: a new gauge-invariant
  post-{Newtonian} ISCO condition, and the ISCO shift due to test-particle spin
  and the gravitational self-force}. Phys Rev D 83:024028.
  \doi{10.1103/PhysRevD.83.024028}.
  {\href{https://arxiv.org/abs/1010.2553}{{arXiv:1010.2553}}}

\bibitem[{Favata(2011{\natexlab{b}})}]{F11a}
Favata M (2011{\natexlab{b}}) {Conservative self-force correction to the
  innermost stable circular orbit: comparison with multiple
  post-{Newtonian}-based methods}. Phys Rev D 83:024027.
  \doi{10.1103/PhysRevD.83.024027}.
  {\href{https://arxiv.org/abs/1008.4622}{{arXiv:1008.4622}}}

\bibitem[{Favata(2011{\natexlab{c}})}]{F11}
Favata M (2011{\natexlab{c}}) {The gravitational-wave memory from eccentric
  binaries}. Phys Rev D 84:124013. \doi{10.1103/PhysRevD.84.124013}.
  {\href{https://arxiv.org/abs/1108.3121}{{arXiv:1108.3121}}}

\bibitem[{Favata(2014)}]{F14}
Favata M (2014) {Systematic parameter errors in inspiraling neutron star
  binaries}. Phys Rev Lett 112:101101. \doi{10.1103/PhysRevLett.112.101101}.
  {\href{https://arxiv.org/abs/1310.8288}{{arXiv:1310.8288}}} {[gr-qc]}

\bibitem[{Favata et~al.(2004)Favata, Hughes, and Holz}]{FHH04}
Favata M, Hughes SA, Holz DE (2004) How black holes get their kicks:
  Gravitational radiation recoil revisited. Astrophys J Lett 607:L5.
  \doi{10.1086/421552}.
  {\href{https://arxiv.org/abs/astro-ph/0402056}{{astro-ph/0402056}}}

\bibitem[{Faye et~al.(2004)Faye, Jaranowski, and Sch{\"{a}}fer}]{FJS04}
Faye G, Jaranowski P, Sch{\"{a}}fer G (2004) {Skeleton approximate solution of
  the {Einstein} field equations for multiple black-hole systems}. Phys Rev D
  69:124029. \doi{10.1103/PhysRevD.69.124029}.
  {\href{https://arxiv.org/abs/gr-qc/0311018}{{gr-qc/0311018}}}

\bibitem[{Faye et~al.(2006)Faye, Blanchet, and Buonanno}]{FBB06}
Faye G, Blanchet L, Buonanno A (2006) {Higher-order spin effects in the
  dynamics of compact binaries {I}. {E}quations of motion}. Phys Rev D
  74:104033. \doi{10.1103/PhysRevD.74.104033}.
  {\href{https://arxiv.org/abs/gr-qc/0605139}{{gr-qc/0605139}}}

\bibitem[{Faye et~al.(2012)Faye, Marsat, Blanchet, and Iyer}]{FMBI12}
Faye G, Marsat S, Blanchet L, Iyer BR (2012) {The third and a half
  post-{Newtonian} gravitational wave quadrupole mode for quasi-circular
  inspiralling compact binaries}. Class Quantum Grav 29:175004.
  \doi{10.1088/0264-9381/29/17/175004}.
  {\href{https://arxiv.org/abs/1204.1043}{{arXiv:1204.1043}}}

\bibitem[{Faye et~al.(2015)Faye, Blanchet, and Iyer}]{FBI15}
Faye G, Blanchet L, Iyer BR (2015) {Non-linear multipole interactions and
  gravitational-wave octupole modes for inspiralling compact binaries to
  third-and-a-half post-{Newtonian} order}. Class Quantum Grav 32:045016.
  {\href{https://arxiv.org/abs/1409.3546}{{arXiv:1409.3546}}} {[gr-qc]}

\bibitem[{Finn and Chernoff(1993)}]{FCh93}
Finn LS, Chernoff DF (1993) {Observing binary inspiral in gravitational
  radiation: One interferometer}. Phys Rev D 47:2198--2219.
  \doi{10.1103/PhysRevD.47.2198}.
  {\href{https://arxiv.org/abs/gr-qc/9301003}{{arXiv:gr-qc/9301003}}}

\bibitem[{Fitchett(1983)}]{Fit83}
Fitchett MJ (1983) {The influence of gravitational wave momentum losses on the
  centre of mass motion of a {Newtonian} binary system}. Mon Not R Astron Soc
  203:1049--1062

\bibitem[{Fitchett and Detweiler(1984)}]{FitDet84}
Fitchett MJ, Detweiler S (1984) {Linear momentum and gravitational
  waves-Circular orbits around a Schwarzschild black hole}. Mon Not R Astron
  Soc 211:933. \doi{10.1093/mnras/211.4.933}

\bibitem[{Flanagan and Hinderer(2008)}]{FHind08}
Flanagan {\'{E}}, Hinderer T (2008) {Constraining neutron star tidal Love
  numbers with gravitational wave detectors}. Phys Rev D 77:021502.
  \doi{10.1103/PhysRevD.77.021502}.
  {\href{https://arxiv.org/abs/0709.1915}{{arXiv:0709.1915}}} {[astro-ph]}

\bibitem[{Fock(1939)}]{Fock39}
Fock VA (1939) {On motion of finite masses in general relativity}. J Phys
  (Moscow) 1(2):81--116

\bibitem[{Fock(1959)}]{Fock}
Fock VA (1959) Theory of space, time and gravitation. Pergamon, London

\bibitem[{Foffa and Sturani(2011)}]{FS3PN}
Foffa S, Sturani R (2011) {Effective field theory calculation of conservative
  binary dynamics at third post-{Newtonian} order}. Phys Rev D 84:044031.
  \doi{10.1103/PhysRevD.84.044031}.
  {\href{https://arxiv.org/abs/1104.1122}{{arXiv:1104.1122}}} {[gr-qc]}

\bibitem[{Foffa and Sturani(2012)}]{FS4PN}
Foffa S, Sturani R (2012) {The dynamics of the gravitational two-body problem
  in the post-{Newtonian} approximation at quadratic order in the {Newton}'s
  constant}. Phys Rev D 87:064011.
  {\href{https://arxiv.org/abs/1206.7087}{{arXiv:1206.7087}}} {[gr-qc]}

\bibitem[{Foffa and Sturani(2013)}]{FStail}
Foffa S, Sturani R (2013) {Tail terms in gravitational radiation reaction via
  effective field theory}. Phys Rev D 87:044056.
  \doi{10.1103/PhysRevD.87.044056}.
  {\href{https://arxiv.org/abs/1111.5488}{{arXiv:1111.5488}}} {[gr-qc]}

\bibitem[{Foffa and Sturani(2014)}]{FSrevue}
Foffa S, Sturani R (2014) {Effective field theory methods to model compact
  binaries}. Class Quantum Grav 31:043001. \doi{10.1088/0264-9381/31/4/043001}.
  {\href{https://arxiv.org/abs/1309.3474}{{arXiv:1309.3474}}} {[gr-qc]}

\bibitem[{Foffa and Sturani(2019)}]{FS19}
Foffa S, Sturani R (2019) {Conservative dynamics of binary systems to fourth
  post-{Newtonian} order in the {EFT} approach. {I}. Regularized {Lagrangian}}.
  Phys Rev D 100:024047. \doi{10.1103/PhysRevD.100.024047}.
  {\href{https://arxiv.org/abs/1903.05113}{{arXiv:1903.05113}}} {[gr-qc]}

\bibitem[{Foffa and Sturani(2020)}]{FS20}
Foffa S, Sturani R (2020) {Hereditary terms at next-to-leading order in
  two-body gravitational dynamics}. Phys Rev D 101(6):064033.
  \doi{10.1103/PhysRevD.101.064033}, [Erratum: Phys. Rev. D 103, 089901
  (2021)]. {\href{https://arxiv.org/abs/1907.02869}{{arXiv:1907.02869}}}
  {[gr-qc]}

\bibitem[{Foffa and Sturani(2021)}]{FS21}
Foffa S, Sturani R (2021) {Near and far zones in two-body dynamics: An
  effective field theory perspective}. Phys Rev D 104(2):024069.
  \doi{10.1103/PhysRevD.104.024069}.
  {\href{https://arxiv.org/abs/2103.03190}{{arXiv:2103.03190}}} {[gr-qc]}

\bibitem[{Foffa et~al.(2017)Foffa, Mastrolia, Sturani, and Sturm}]{FMSS17}
Foffa S, Mastrolia P, Sturani R, Sturm C (2017) {Effective field theory
  approach to the gravitational two-body dynamics at fourth post-{Newtonian}
  order and quintic in the {Newton} constant}. Phys Rev D 95:104009.
  \doi{10.1103/PhysRevD.95.104009}.
  {\href{https://arxiv.org/abs/1612.00482}{{arXiv:1612.00482}}} {[gr-qc]}

\bibitem[{Foffa et~al.(2019)Foffa, Porto, Rothstein, and Sturani}]{FPRS19}
Foffa S, Porto R, Rothstein I, Sturani R (2019) {Conservative dynamics of
  binary systems to fourth post-{Newtonian} order in the {EFT} approach. {II}.
  Renormalized {Lagrangian}}. Phys Rev D 100:024048.
  \doi{10.1103/PhysRevD.100.024048}.
  {\href{https://arxiv.org/abs/1903.05118}{{arXiv:1903.05118}}} {[gr-qc]}

\bibitem[{Foffa et~al.(2021)Foffa, Sturani, and Torres~Bobadilla}]{FST21}
Foffa S, Sturani R, Torres~Bobadilla WJ (2021) {Efficient resummation of high
  post-{Newtonian} contributions to the binding energy}. J High Energy Phys
  2021(02):165. \doi{10.1007/JHEP02(2021)165}.
  {\href{https://arxiv.org/abs/2010.13730}{{arXiv:2010.13730}}} {[gr-qc]}

\bibitem[{Fokker(1929)}]{Fokker}
Fokker AD (1929) {Ein invarianter Variationssatz f{\"u}r die Bewegung mehrerer
  elektrischer Massenteilchen}. Z Phys 58:386--393. \doi{10.1007/BF01340389}

\bibitem[{Forseth et~al.(2016)Forseth, Evans, and Hopper}]{FEH16}
Forseth E, Evans CR, Hopper S (2016) {Eccentric-orbit {EMRI} gravitational wave
  energy fluxes to 7{PN} order}. Phys Rev D 93:064058.
  \doi{10.1103/PhysRevD.93.064058}.
  {\href{https://arxiv.org/abs/1512.03051}{{arXiv:1512.03051}}}

\bibitem[{Friedman et~al.(2002)Friedman, Ury{\={u}}, and Shibata}]{FUS02}
Friedman JL, Ury{\={u}} K, Shibata M (2002) {Thermodynamics of binary black
  holes and neutron stars}. Phys Rev D 65:064035.
  \doi{10.1103/PhysRevD.65.064035}, {Erratum}: Phys. Rev. D, 70, 129904(E)
  (2004)

\bibitem[{Fujita(2012{\natexlab{a}})}]{Fuj14PN}
Fujita R (2012{\natexlab{a}}) {Gravitational Radiation for Extreme Mass Ratio
  Inspirals to the 14th Post-{Newtonian} Order}. Prog Theor Phys 127:583--590.
  \doi{10.1143/PTP.127.583}.
  {\href{https://arxiv.org/abs/1104.5615}{{arXiv:1104.5615}}} {[gr-qc]}

\bibitem[{Fujita(2012{\natexlab{b}})}]{Fuj22PN}
Fujita R (2012{\natexlab{b}}) {Gravitational Waves from a Particle in Circular
  Orbits around a {Schwarzschild} Black Hole to the 22nd Post-{Newtonian}
  Order}. Prog Theor Phys 128:971--992. \doi{10.1143/PTP.128.971}.
  {\href{https://arxiv.org/abs/1211.5535}{{arXiv:1211.5535}}} {[gr-qc]}

\bibitem[{Fujita and Iyer(2010)}]{FI10}
Fujita R, Iyer B (2010) {Spherical harmonic modes of 5.5 post-{N}ewtonian
  gravitational wave polarizations and associated factorized resummed waveforms
  for a particle in circular orbit around a {Schwarzschild} black hole}. Phys
  Rev D 82:044051. \doi{10.1103/PhysRevD.82.044051}.
  {\href{https://arxiv.org/abs/1005.2266}{{arXiv:1005.2266}}} {[gr-qc]}

\bibitem[{Futamase(1983)}]{F83}
Futamase T (1983) {Gravitational radiation reaction in the {Newtonian} limit}.
  Phys Rev D 28:2373--2381. \doi{10.1103/PhysRevD.28.2373}

\bibitem[{Futamase(1987)}]{F87}
Futamase T (1987) {Strong-field point-particle limit and the equations of
  motion in the binary pulsar}. Phys Rev D 36:321--329.
  \doi{10.1103/PhysRevD.36.321}

\bibitem[{Futamase and Itoh(2007)}]{ItohLR}
Futamase T, Itoh Y (2007) {The Post-{Newtonian} Approximation for Relativistic
  Compact Binaries}. Living Rev Relativ 10:2. \doi{10.12942/lrr-2007-2}

\bibitem[{Futamase and Schutz(1983)}]{FS83}
Futamase T, Schutz BF (1983) {{Newtonian} and post-{Newtonian} approximations
  are asymptotic to general relativity}. Phys Rev D 28:2363--2372.
  \doi{10.1103/PhysRevD.28.2363}

\bibitem[{Galley et~al.(2016)Galley, Leibovich, Porto, and Ross}]{GLPR16}
Galley CR, Leibovich AK, Porto RA, Ross A (2016) {Tail effect in gravitational
  radiation reaction: Time nonlocality and renormalization group evolution}.
  Phys Rev D 93:124010. \doi{10.1103/PhysRevD.93.124010}.
  {\href{https://arxiv.org/abs/1511.07379}{{arXiv:1511.07379}}} {[gr-qc]}

\bibitem[{{Gal'tsov} et~al.(1980){Gal'tsov}, Matiukhin, and Petukhov}]{Galtsov}
{Gal'tsov} DV, Matiukhin AA, Petukhov VI (1980) {Relativistic corrections to
  the gravitational radiation of a binary system and the fine structure of the
  spectrum}. Phys Lett A 77:387--390. \doi{10.1016/0375-9601(80)90728-8}

\bibitem[{Garc{\'\i}a-Bellido et~al.(2021)Garc{\'\i}a-Bellido, Siles, and
  Morales}]{GSM21}
Garc{\'\i}a-Bellido J, Siles JFN, Morales ER (2021) {Bayesian analysis of the
  spin distribution of {LIGO}/{Virgo} black holes}. Phys Dark Univ 31:100791.
  \doi{10.1016/j.dark.2021.100791}.
  {\href{https://arxiv.org/abs/2010.13811}{{arXiv:2010.13811}}}

\bibitem[{Georgoudis et~al.(2023)Georgoudis, Heissenberg, and
  Vazquez-Holm}]{Georgoudis2023}
Georgoudis A, Heissenberg C, Vazquez-Holm I (2023) {Inelastic exponentiation
  and classical gravitational scattering at one loop}. J High Energy Phys
  2023(6). \doi{10.1007/jhep06(2023)126}

\bibitem[{Georgoudis et~al.(2024{\natexlab{a}})Georgoudis, Heissenberg, and
  Russo}]{GHR24a}
Georgoudis A, Heissenberg C, Russo R (2024{\natexlab{a}}) {An eikonal-inspired
  approach to the gravitational scattering waveform}. J High Energy Phys
  2024(3):1--39. {\href{https://arxiv.org/abs/2312.07452}{{arXiv:2312.07452}}}
  {[hep-th]}

\bibitem[{Georgoudis et~al.(2024{\natexlab{b}})Georgoudis, Heissenberg, and
  Russo}]{GHR24b}
Georgoudis A, Heissenberg C, Russo R (2024{\natexlab{b}}) {Post-{Newtonian}
  multipoles from the next-to-leading post-{Minkowskian} gravitational
  waveform}. arXiv e-prints
  {\href{https://arxiv.org/abs/2402.06316}{{arXiv:2402.06316}}} {[hep-th]}

\bibitem[{Gergely(1999)}]{Ger99}
Gergely L (1999) {Spin-spin effects in radiating compact binaries}. Phys Rev D
  61:024035. \doi{10.1103/PhysRevD.61.024035}.
  {\href{https://arxiv.org/abs/gr-qc/9911082}{{gr-qc/9911082}}}

\bibitem[{Gergely(2000)}]{Ger00}
Gergely L (2000) {Second post-{Newtonian} radiative evolution of the relative
  orientations of angular momenta in spinning compact binaries}. Phys Rev D
  62:024007. \doi{10.1103/PhysRevD.62.024007}.
  {\href{https://arxiv.org/abs/gr-qc/0003037}{{gr-qc/0003037}}}

\bibitem[{Geroch(1970)}]{G70}
Geroch R (1970) {Multipole Moments. {II}. {C}urved Space}. J Math Phys
  11:2580--2588. \doi{10.1063/1.1665427}

\bibitem[{Geroch and Horowitz(1978)}]{GH78}
Geroch R, Horowitz GT (1978) {Asymptotically simple does not imply
  asymptotically Minkowskian}. Phys Rev Lett 40:203--206

\bibitem[{Goldberger and Ross(2010)}]{GRoss10}
Goldberger WD, Ross A (2010) {Gravitational radiative corrections from
  effective field theory}. Phys Rev D 81:124015.
  \doi{10.1103/PhysRevD.81.124015}.
  {\href{https://arxiv.org/abs/0912.4254}{{arXiv:0912.4254}}}

\bibitem[{Goldberger and Rothstein(2006)}]{GR06}
Goldberger WD, Rothstein IZ (2006) {Effective field theory of gravity for
  extended objects}. Phys Rev D 73:104029. \doi{10.1103/PhysRevD.73.104029}.
  {\href{https://arxiv.org/abs/hep-th/0409156}{{arXiv:hep-th/0409156}}}
  {[hep-th]}

\bibitem[{Goldberger et~al.(2014)Goldberger, Ross, and Rothstein}]{GRR12}
Goldberger WD, Ross A, Rothstein IZ (2014) {Black hole mass dynamics and
  renormalization group evolution}. Phys Rev D 89:124033.
  \doi{10.1103/PhysRevD.89.124033}.
  {\href{https://arxiv.org/abs/1211.6095}{{arXiv:1211.6095}}} {[hep-th]}

\bibitem[{Gonz{\'a}lez et~al.(2007)Gonz{\'a}lez, Sperhake, Br{\"u}gmann,
  Hannam, and Husa}]{GSBHH07}
Gonz{\'a}lez JA, Sperhake U, Br{\"u}gmann B, Hannam M, Husa S (2007) {Maximum
  kick from nonspinning black-hole binary inspiral}. Phys Rev Lett
  98(9):091101. \doi{10.1103/PhysRevLett.98.091101}.
  {\href{https://arxiv.org/abs/gr-qc/0610154}{{arXiv:gr-qc/0610154}}}

\bibitem[{Gonz{\'a}lez et~al.(2009)Gonz{\'a}lez, Sperhake, and
  Br{\"u}gmann}]{GSB09}
Gonz{\'a}lez JA, Sperhake U, Br{\"u}gmann B (2009) {Black-hole binary
  simulations: The mass ratio 10:1}. Phys Rev D 79(12):124006.
  \doi{10.1103/PhysRevD.79.124006}.
  {\href{https://arxiv.org/abs/0811.3952}{{arXiv:0811.3952}}} {[gr-qc]}

\bibitem[{Gopakumar and Iyer(1997)}]{GI97}
Gopakumar A, Iyer BR (1997) {Gravitational waves from inspiraling compact
  binaries: Angular momentum flux, evolution of the orbital elements and the
  waveform to the second post-{Newtonian} order}. Phys Rev D 56:7708--7731.
  \doi{10.1103/PhysRevD.56.7708}.
  {\href{https://arxiv.org/abs/gr-qc/9710075}{{arXiv:gr-qc/9710075}}}

\bibitem[{Gopakumar and Iyer(2002)}]{GI02}
Gopakumar A, Iyer BR (2002) {Second post-{Newtonian} gravitational wave
  polarizations for compact binaries in elliptical orbits}. Phys Rev D
  65:084011. \doi{10.1103/PhysRevD.65.084011}.
  {\href{https://arxiv.org/abs/gr-qc/0110100}{{arXiv:gr-qc/0110100}}}

\bibitem[{Gopakumar et~al.(1997)Gopakumar, Iyer, and Iyer}]{GII97}
Gopakumar A, Iyer BR, Iyer S (1997) {Second post-{Newtonian} gravitational
  radiation reaction for two-body systems: Nonspinning bodies}. Phys Rev D
  55:6030--6053. \doi{10.1103/PhysRevD.55.6030}.
  {\href{https://arxiv.org/abs/gr-qc/9703075}{{arXiv:gr-qc/9703075}}}

\bibitem[{Gou et~al.(2011)Gou, McClintock, Reid, Orosz, Steiner, Narayan,
  Xiang, Remillard, Arnaud, and Davis}]{Gou11}
Gou L, McClintock JE, Reid MJ, Orosz JA, Steiner JF, Narayan R, Xiang J,
  Remillard RA, Arnaud KA, Davis SW (2011) {The extreme spin of the black hole
  in Cygnus X-1}. Astrophys J 742:85. \doi{10.1088/0004-637X/742/2/85}.
  {\href{https://arxiv.org/abs/1106.3690}{{arXiv:1106.3690}}} {[astro-ph.HE]}

\bibitem[{Gourgoulhon et~al.(2001)Gourgoulhon, Grandcl{\'{e}}ment, Taniguchi,
  Marck, and Bonazzola}]{GGTMB01}
Gourgoulhon E, Grandcl{\'{e}}ment P, Taniguchi K, Marck JA, Bonazzola S (2001)
  {Quasi-equilibrium sequences of synchronized and irrotational binary neutron
  stars in general relativity}. Phys Rev D 63:064029.
  \doi{10.1103/PhysRevD.63.064029}.
  {\href{https://arxiv.org/abs/gr-qc/0007028}{{gr-qc/0007028}}}

\bibitem[{Gourgoulhon et~al.(2002)Gourgoulhon, Grandcl{\'{e}}ment, and
  Bonazzola}]{GGB1}
Gourgoulhon E, Grandcl{\'{e}}ment P, Bonazzola S (2002) {Binary black holes in
  circular orbits. I. A global spacetime approach}. Phys Rev D 65:044020.
  \doi{10.1103/PhysRevD.65.044020}.
  {\href{https://arxiv.org/abs/gr-qc/0106015}{{gr-qc/0106015}}}

\bibitem[{Gradshteyn and Ryzhik(1980)}]{GR}
Gradshteyn IS, Ryzhik IM (1980) Table of Integrals, Series and Products.
  Academic Press, San Diego; London

\bibitem[{Gralla and Wald(2008)}]{GW08}
Gralla SE, Wald RM (2008) {A rigorous derivation of gravitational self-force}.
  Class Quantum Grav 25:205009. \doi{10.1088/0264-9381/25/20/205009}.
  {\href{https://arxiv.org/abs/0806.3293}{{arXiv:0806.3293}}}

\bibitem[{Grandcl{\'{e}}ment and Novak(2009)}]{GN09}
Grandcl{\'{e}}ment P, Novak J (2009) {Spectral Methods for Numerical
  Relativity}. Living Rev Relativ 12:1. \doi{10.12942/lrr-2009-1}.
  {\href{https://arxiv.org/abs/0706.2286}{{arXiv:0706.2286}}} {[gr-qc]}

\bibitem[{Grandcl{\'{e}}ment et~al.(2002)Grandcl{\'{e}}ment, Gourgoulhon, and
  Bonazzola}]{GGB2}
Grandcl{\'{e}}ment P, Gourgoulhon E, Bonazzola S (2002) {Binary black holes in
  circular orbits. {II}. {N}umerical methods and first results}. Phys Rev D
  65:044021. \doi{10.1103/PhysRevD.65.044021}

\bibitem[{Grishchuk and Kopeikin(1986)}]{GKop86}
Grishchuk LP, Kopeikin SM (1986) Equations of motion for isolated bodies with
  relativistic corrections including the radiation-reaction force. In:
  Kovalevsky J, Brumberg VA (eds) Relativity in Celestial Mechanics and
  Astrometry: High Precision Dynamical Theories and Observational
  Verifications. Reidel, Dordrecht; Boston, pp 19--34

\bibitem[{Gultekin et~al.(2004)Gultekin, Miller, and Hamilton}]{GMH04}
Gultekin K, Miller MC, Hamilton DP (2004) {Growth of Intermediate-Mass Black
  Holes in Globular Clusters}. Astrophys J 616:221. \doi{10.1086/424809}.
  {\href{https://arxiv.org/abs/astro-ph/0402532}{{astro-ph/0402532}}}

\bibitem[{Hadamard(1932)}]{Hadamard}
Hadamard J (1932) Le probl\`{e}me de {Cauchy} et les \'{e}quations aux
  d\'{e}riv\'{e}es partielles lin\'{e}aires hyperboliques. Hermann, Paris

\bibitem[{Hansen(1974)}]{H74}
Hansen RO (1974) {Multipole moments of stationary space-times}. J Math Phys
  15:46--52. \doi{10.1063/1.1666501}

\bibitem[{Hanson and Regge(1974)}]{HRegge74}
Hanson AJ, Regge T (1974) {The Relativistic Spherical Top}. Ann Phys (NY)
  87:498--566. \doi{10.1016/0003-4916(74)90046-3}

\bibitem[{Hari~Dass and Soni(1982)}]{HariDass}
Hari~Dass ND, Soni V (1982) {Feynman graph derivation of the {Einstein}
  quadrupole formula}. J Phys A: Math Gen 15:473--492.
  \doi{10.1088/0305-4470/15/2/019}

\bibitem[{Hartung and Steinhoff(2011{\natexlab{a}})}]{HaS11}
Hartung J, Steinhoff J (2011{\natexlab{a}}) {Next-to-leading order spin-orbit
  and spin(a)-spin(b) {Hamiltonian}s for $n$ gravitating spinning compact
  objects}. Phys Rev D 83:044008. \doi{10.1103/PhysRevD.83.044008}.
  {\href{https://arxiv.org/abs/1011.1179}{{arXiv:1011.1179}}} {[gr-qc]}

\bibitem[{Hartung and Steinhoff(2011{\natexlab{b}})}]{HaS11so}
Hartung J, Steinhoff J (2011{\natexlab{b}}) {Next-to-next-to-leading order
  post-{Newtonian} spin-orbit {Hamiltonian} for self-gravitating binaries}. Ann
  Phys (Berlin) 523:783--790. \doi{10.1002/andp.201100094}.
  {\href{https://arxiv.org/abs/1104.3079}{{arXiv:1104.3079}}} {[gr-qc]}

\bibitem[{Hartung and Steinhoff(2011{\natexlab{c}})}]{HaS11ss}
Hartung J, Steinhoff J (2011{\natexlab{c}}) {Next-to-next-to-leading order
  post-{Newtonian} spin(1)-spin(2) {Hamiltonian} for self-gravitating
  binaries}. Ann Phys (Berlin) 523:919--924. \doi{10.1002/andp.201100163}.
  {\href{https://arxiv.org/abs/1107.4294}{{arXiv:1107.4294}}} {[gr-qc]}

\bibitem[{Hartung et~al.(2013)Hartung, Steinhoff, and Sch{\"{a}}fer}]{HaSS13}
Hartung J, Steinhoff J, Sch{\"{a}}fer G (2013) {Next-to-next-to-leading order
  post-{Newtonian} linear-in-spin binary {Hamiltonians}}. Ann Phys (Berlin)
  525:359--394. \doi{10.1002/andp.201200271}.
  {\href{https://arxiv.org/abs/1302.6723}{{arXiv:1302.6723}}} {[gr-qc]}

\bibitem[{Hellings and Moore(2003)}]{HM1}
Hellings RW, Moore TA (2003) {The information content of gravitational wave
  harmonics in compact binary inspiral}. Class Quantum Grav 20(10):S181.
  \doi{10.1088/0264-9381/20/10/321}

\bibitem[{Henry(2023)}]{H23}
Henry Q (2023) {Complete gravitational-waveform amplitude modes for
  quasicircular compact binaries to the 3.5PN order}. Phys Rev D 107(4):044057.
  \doi{10.1103/PhysRevD.107.044057}.
  {\href{https://arxiv.org/abs/2210.15602}{{arXiv:2210.15602}}} {[gr-qc]}

\bibitem[{Henry and Khalil(2023)}]{HK23}
Henry Q, Khalil M (2023) {Spin effects in gravitational waveforms and fluxes
  for binaries on eccentric orbits to the third post-{Newtonian} order}. Phys
  Rev D 108(10):104016. \doi{10.1103/PhysRevD.108.104016}.
  {\href{https://arxiv.org/abs/2308.13606}{{arXiv:2308.13606}}} {[gr-qc]}

\bibitem[{Henry et~al.(2020{\natexlab{a}})Henry, Faye, and Blanchet}]{HFB20a}
Henry Q, Faye G, Blanchet L (2020{\natexlab{a}}) {Tidal effects in the
  equations of motion of compact binary systems to next-to-next-to-leading
  post-{Newtonian} order}. Phys Rev D 101:064047.
  \doi{10.1103/PhysRevD.101.064047}.
  {\href{https://arxiv.org/abs/1912.01920}{{arXiv:1912.01920}}} {[gr-qc]}

\bibitem[{Henry et~al.(2020{\natexlab{b}})Henry, Faye, and Blanchet}]{HFB20b}
Henry Q, Faye G, Blanchet L (2020{\natexlab{b}}) {Tidal effects in the
  gravitational-wave phase evolution of compact binary systems to
  next-to-next-to-leading post-{Newtonian} order}. Phys Rev D 102:044033.
  \doi{10.1103/PhysRevD.102.044033}.
  {\href{https://arxiv.org/abs/2005.13367}{{arXiv:2005.13367}}} {[gr-qc]}

\bibitem[{Henry et~al.(2021)Henry, Faye, and Blanchet}]{HFB21}
Henry Q, Faye G, Blanchet L (2021) {The current-type quadrupole moment and
  gravitational-wave mode (\ensuremath{\ell}, m) = (2, 1) of compact binary
  systems at the third post-{N}ewtonian order}. Class Quantum Grav
  38(18):185004. \doi{10.1088/1361-6382/ac1850}.
  {\href{https://arxiv.org/abs/2105.10876}{{arXiv:2105.10876}}} {[gr-qc]}

\bibitem[{Henry et~al.(2022)Henry, Marsat, and Khalil}]{HMK22}
Henry Q, Marsat S, Khalil M (2022) {Spin contributions to the
  gravitational-waveform modes for spin-aligned binaries at the 3.5PN order}.
  Phys Rev D 106(12):124018. \doi{10.1103/PhysRevD.106.124018}.
  {\href{https://arxiv.org/abs/2209.00374}{{arXiv:2209.00374}}} {[gr-qc]}

\bibitem[{Herderschee et~al.(2023)Herderschee, Roiban, and
  Teng}]{Herderschee23}
Herderschee A, Roiban R, Teng F (2023) {The sub-leading scattering waveform
  from amplitudes}. J High Energy Phys 2023(6). \doi{10.1007/jhep06(2023)004}

\bibitem[{Hergt and Sch{\"{a}}fer(2008)}]{HergtS08b}
Hergt S, Sch{\"{a}}fer G (2008) {Higher-order-in-spin interaction
  {Hamiltonians} for binary black holes from Poincar\'e invariance}. Phys Rev D
  78:124004. \doi{10.1103/PhysRevD.78.124004}.
  {\href{https://arxiv.org/abs/0809.2208}{{arXiv:0809.2208}}} {[gr-qc]}

\bibitem[{Hergt and Sch{\"a}fer(2008)}]{HergtS08a}
Hergt S, Sch{\"a}fer G (2008) {{Higher-order-in-spin interaction {Hamiltonians}
  for binary black holes from source terms of {Kerr} geometry in approximate
  {ADM} coordinates}}. Phys Rev D 77:104001. \doi{10.1103/PhysRevD.77.104001}.
  {\href{https://arxiv.org/abs/0712.1515}{{arXiv:0712.1515}}} {[gr-qc]}

\bibitem[{Hergt et~al.(2010)Hergt, Steinhoff, and Sch{\"{a}}fer}]{HSS10}
Hergt S, Steinhoff J, Sch{\"{a}}fer G (2010) {The reduced {Hamiltonian} for
  next-to-leading-order spin-squared dynamics of general compact binaries}.
  Class Quantum Grav 27:135007. \doi{10.1088/0264-9381/27/13/135007}.
  {\href{https://arxiv.org/abs/1002.2093}{{arXiv:1002.2093}}} {[gr-qc]}

\bibitem[{Herrmann et~al.(2007)Herrmann, Hinder, Shoemaker, and
  Laguna}]{HHSL07}
Herrmann F, Hinder I, Shoemaker D, Laguna P (2007) {Unequal mass binary black
  hole plunges and gravitational recoil}. Class Quantum Grav 24(12):S33.
  \doi{10.1088/0264-9381/24/12/S04}.
  {\href{https://arxiv.org/abs/gr-qc/0601026}{{arXiv:gr-qc/0601026}}}

\bibitem[{Hinderer(2008)}]{Hind08}
Hinderer T (2008) {Tidal {Love} numbers of neutron stars}. Astrophys J
  677:1216. \doi{10.1086/533487}

\bibitem[{'t~Hooft and Veltman(1972)}]{tHooft}
't~Hooft G, Veltman M (1972) {Regularization and renormalization of gauge
  fields}. Nucl Phys B 44:139. \doi{10.1016/0550-3213(72)90279-9}

\bibitem[{Hopper et~al.(2016)Hopper, Kavanagh, and Ottewill}]{HKO16}
Hopper S, Kavanagh C, Ottewill AC (2016) {Analytic self-force calculations in
  the post-{Newtonian} regime: eccentric orbits on a {Schwarzschild}
  background}. Phys Rev D 93:044010. \doi{10.1103/PhysRevD.93.044010}.
  {\href{https://arxiv.org/abs/1512.01556}{{arXiv:1512.01556}}} {[gr-qc]}

\bibitem[{Hulse and Taylor(1975)}]{HulseTaylor}
Hulse RA, Taylor JH (1975) {Discovery of a pulsar in a binary system}.
  Astrophys J 195:L51--L53. \doi{10.1086/181708}

\bibitem[{Hunter and Rotenberg(1969)}]{HR69}
Hunter AJ, Rotenberg MA (1969) {The double-series approximation method in
  general relativity. I. Exact solution of the (24) approximation. {II}.
  {D}iscussion of `wave tails' in the (2s) approximation}. J Phys A: Math Gen
  2:34--49. \doi{10.1088/0305-4470/2/1/007}

\bibitem[{Isaacson(1968)}]{Isaacson}
Isaacson RA (1968) {Gravitational radiation in the limit of high frequency. II.
  Nonlinear terms and the effective stress tensor}. Phys Rev 166(5):1272

\bibitem[{Isaacson and Winicour(1968)}]{IW68}
Isaacson RA, Winicour J (1968) {Harmonic and Null Descriptions of Gravitational
  Radiation}. Phys Rev 168:1451--1456. \doi{10.1103/PhysRev.168.1451}

\bibitem[{Isenberg and Nester(1980)}]{IN80}
Isenberg J, Nester J (1980) Canonical gravity. In: Held A (ed) General
  Relativity and Gravitation. Vol. 1. One hundred years after the birth of
  Albert Einstein. Plenum Press, New York, p~23

\bibitem[{Ito and Ohtsuka(2019)}]{IO19}
Ito T, Ohtsuka K (2019) {The {Lidov}-{Kozai} oscillation and {Hugo von
  Zeipel}}. Monogr Environ Earth Planets 7:1.
  {\href{https://arxiv.org/abs/1911.03984}{{arXiv:1911.03984}}} {[astro-ph]}

\bibitem[{Itoh(2004)}]{itoh2}
Itoh Y (2004) {Equation of motion for relativistic compact binaries with the
  strong field point particle limit: Third post-{Newtonian} order}. Phys Rev D
  69:064018. \doi{10.1103/PhysRevD.69.064018}

\bibitem[{Itoh(2009)}]{itoh3}
Itoh Y (2009) {Third-and-a-half order post-{Newtonian} equations of motion for
  relativistic compact binaries using the strong field point particle limit}.
  Phys Rev D 80:124003. \doi{10.1103/PhysRevD.80.124003}.
  {\href{https://arxiv.org/abs/0911.4232}{{arXiv:0911.4232}}} {[gr-qc]}

\bibitem[{Itoh and Futamase(2003)}]{itoh1}
Itoh Y, Futamase T (2003) {New derivation of a third post-{Newtonian} equation
  of motion for relativistic compact binaries without ambiguity}. Phys Rev D
  68:121501(R). \doi{10.1103/PhysRevD.68.121501}.
  {\href{https://arxiv.org/abs/gr-qc/0310028}{{gr-qc/0310028}}}

\bibitem[{Itoh et~al.(2000)Itoh, Futamase, and Asada}]{IFA00}
Itoh Y, Futamase T, Asada H (2000) {Equation of motion for relativistic compact
  binaries with the strong field point particle limit: Formulation, the first
  post-{Newtonian} order, and multipole terms}. Phys Rev D 62:064002.
  \doi{10.1103/PhysRevD.62.064002}.
  {\href{https://arxiv.org/abs/gr-qc/9910052}{{gr-qc/9910052}}}

\bibitem[{Itoh et~al.(2001)Itoh, Futamase, and Asada}]{IFA01}
Itoh Y, Futamase T, Asada H (2001) {Equation of motion for relativistic compact
  binaries with the strong field point particle limit: The second and half
  post-{Newtonian} order}. Phys Rev D 63:064038.
  \doi{10.1103/PhysRevD.63.064038}.
  {\href{https://arxiv.org/abs/gr-qc/0101114}{{gr-qc/0101114}}}

\bibitem[{Iyer and Will(1993)}]{IW93}
Iyer BR, Will CM (1993) {Post-{Newtonian} gravitational radiation reaction for
  two-body systems}. Phys Rev Lett 70:113--116.
  \doi{10.1103/PhysRevLett.70.113}

\bibitem[{Iyer and Will(1995)}]{IW95}
Iyer BR, Will CM (1995) {Post-{Newtonian} gravitational radiation reaction for
  two-body systems: Nonspinning bodies}. Phys Rev D 52:6882--6893.
  \doi{10.1103/PhysRevD.52.6882}

\bibitem[{Jaranowski and Sch{\"{a}}fer(1997)}]{JaraS97}
Jaranowski P, Sch{\"{a}}fer G (1997) {Radiative 3.5 post-{Newtonian} {ADM}
  {Hamiltonian} for many-body point-mass systems}. Phys Rev D 55:4712--4722.
  \doi{10.1103/PhysRevD.55.4712}

\bibitem[{Jaranowski and Sch{\"{a}}fer(1998)}]{JaraS98}
Jaranowski P, Sch{\"{a}}fer G (1998) {Third post-{Newtonian} higher order {ADM}
  {Hamilton} dynamics for two-body point-mass systems}. Phys Rev D
  57:7274--7291. \doi{10.1103/PhysRevD.57.7274}.
  {\href{https://arxiv.org/abs/gr-qc/9712075}{{gr-qc/9712075}}}

\bibitem[{Jaranowski and Sch{\"{a}}fer(1999)}]{JaraS99}
Jaranowski P, Sch{\"{a}}fer G (1999) {Binary black-hole problem at the third
  post-{Newtonian} approximation in the orbital motion: Static part}. Phys Rev
  D 60:124003. \doi{10.1103/PhysRevD.60.124003}.
  {\href{https://arxiv.org/abs/gr-qc/9906092}{{gr-qc/9906092}}}

\bibitem[{Jaranowski and Sch{\"{a}}fer(2000)}]{JaraS00}
Jaranowski P, Sch{\"{a}}fer G (2000) {The binary black-hole dynamics at the
  third post-{Newtonian} order in the orbital motion}. Ann Phys (Berlin)
  9:378--383.
  \doi{10.1002/(SICI)1521-3889(200005)9:3/5<378::AID-ANDP378>3.0.CO;2-M}.
  {\href{https://arxiv.org/abs/gr-qc/0003054}{{gr-qc/0003054}}}

\bibitem[{Jaranowski and Sch{\"{a}}fer(2012)}]{JaraS12}
Jaranowski P, Sch{\"{a}}fer G (2012) {Towards the fourth post-{Newtonian}
  {Hamiltonian} for two-point-mass systems}. Phys Rev D 86:061503(R).
  \doi{10.1103/PhysRevD.86.061503}.
  {\href{https://arxiv.org/abs/1207.5448}{{arXiv:1207.5448}}} {[gr-qc]}

\bibitem[{Jaranowski and Sch{\"{a}}fer(2013)}]{JaraS13}
Jaranowski P, Sch{\"{a}}fer G (2013) {Dimensional regularization of local
  singularities in the 4th post-{Newtonian} two-point-mass {Hamiltonian}}. Phys
  Rev D 87:081503(R). \doi{10.1103/PhysRevD.87.081503}.
  {\href{https://arxiv.org/abs/1303.3225}{{arXiv:1303.3225}}} {[gr-qc]}

\bibitem[{Jaranowski and Sch{\"{a}}fer(2015)}]{JaraS15}
Jaranowski P, Sch{\"{a}}fer G (2015) {Derivation of the local-in-time fourth
  post-{Newtonian} {ADM} {Hamiltonian} for spinless compact binaries}. Phys Rev
  D 92:124043. \doi{10.1103/PhysRevD.92.124043}.
  {\href{https://arxiv.org/abs/1508.01016}{{arXiv:1508.01016}}} {[gr-qc]}

\bibitem[{Junker and Sch{\"{a}}fer(1992)}]{JS92}
Junker W, Sch{\"{a}}fer G (1992) {Binary systems: higher order gravitational
  radiation damping and wave emission}. Mon Not R Astron Soc 254:146--164

\bibitem[{Kavanagh et~al.(2015)Kavanagh, Ottewill, and Wardell}]{KOW15}
Kavanagh C, Ottewill AC, Wardell B (2015) {Analytical high-order
  post-{Newtonian} expansions for extreme mass ratio binaries}. Phys Rev D
  92:084025. \doi{10.1103/PhysRevD.92.084025}.
  {\href{https://arxiv.org/abs/1503.02334}{{arXiv:1503.02334}}} {[gr-qc]}

\bibitem[{Kennefick(2007)}]{Kennefick}
Kennefick D (2007) {Traveling at the speed of thought: {Einstein} and the quest
  for gravitational waves}. Princeton University Press

\bibitem[{Kerlick(1980{\natexlab{a}})}]{K80a}
Kerlick GD (1980{\natexlab{a}}) {Finite reduced hydrodynamic equations in the
  slow-motion approximation to general relativity. Part {I}. {F}irst
  post-{Newtonian} equations}. Gen Relativ Gravit 12:467--482.
  \doi{10.1007/BF00756177}

\bibitem[{Kerlick(1980{\natexlab{b}})}]{K80b}
Kerlick GD (1980{\natexlab{b}}) {Finite reduced hydrodynamic equations in the
  slow-motion approximation to general relativity. Part {II}. {R}adiation
  reaction and higher-order divergent terms}. Gen Relativ Gravit 12:521--543.
  \doi{10.1007/BF00756528}

\bibitem[{Kidder(1995)}]{K95}
Kidder LE (1995) {Coalescing binary systems of compact objects to
  (post)$^{5/2}$-{Newtonian} order. {V}. {S}pin effects}. Phys Rev D
  52:821--847. \doi{10.1103/PhysRevD.52.821}

\bibitem[{Kidder(2008)}]{K07}
Kidder LE (2008) {Using full information when computing modes of
  post-{Newtonian} waveforms from inspiralling compact binaries in circular
  orbits}. Phys Rev D 77:044016. \doi{10.1103/PhysRevD.77.044016}.
  {\href{https://arxiv.org/abs/0710.0614}{{arXiv:0710.0614}}}

\bibitem[{Kidder et~al.(1993{\natexlab{a}})Kidder, Will, and Wiseman}]{KWWisco}
Kidder LE, Will CM, Wiseman AG (1993{\natexlab{a}}) {Coalescing binary systems
  of compact objects to (post)$^{5/2}$-{Newtonian} order. {III}. {T}ransition
  from inspiral to plunge}. Phys Rev D 47:3281--3291.
  \doi{10.1103/PhysRevD.47.3281}

\bibitem[{Kidder et~al.(1993{\natexlab{b}})Kidder, Will, and Wiseman}]{KWW93}
Kidder LE, Will CM, Wiseman AG (1993{\natexlab{b}}) {Spin effects in the
  inspiral of coalescing compact binaries}. Phys Rev D 47:R4183--R4187.
  \doi{10.1103/PhysRevD.47.R4183}.
  {\href{https://arxiv.org/abs/gr-qc/9211025}{{gr-qc/9211025}}}

\bibitem[{Kim et~al.(2023{\natexlab{a}})Kim, Levi, and Yin}]{Kim2023b}
Kim JW, Levi M, Yin Z (2023{\natexlab{a}}) {{N3LO} quadratic-in-spin
  interactions for generic compact binaries}. J High Energy Phys 2023(3).
  \doi{10.1007/jhep03(2023)098}

\bibitem[{Kim et~al.(2023{\natexlab{b}})Kim, Levi, and Yin}]{Kim2023a}
Kim JW, Levi M, Yin Z (2023{\natexlab{b}}) {{N3LO} spin-orbit interaction via
  the {EFT} of spinning gravitating objects}. J High Energy Phys 2023(5).
  \doi{10.1007/jhep05(2023)184}

\bibitem[{Kochanek(1992)}]{Kochanek}
Kochanek CS (1992) {Coalescing binary neutron stars}. Astrophys J 398:234--247.
  \doi{10.1086/171851}

\bibitem[{Kol and Smolkin(2008)}]{KolS08}
Kol B, Smolkin M (2008) {Non-relativistic gravitation: From {Newton} to
  {Einstein} and back}. Class Quantum Grav 25:145011.
  \doi{10.1088/0264-9381/25/14/145011}.
  {\href{https://arxiv.org/abs/0712.4116}{{arXiv:0712.4116}}} {[hep-th]}

\bibitem[{K{\"{o}}nigsd{\"{o}}rffer and Gopakumar(2006)}]{KG06}
K{\"{o}}nigsd{\"{o}}rffer C, Gopakumar A (2006) {Phasing of gravitational waves
  from inspiralling eccentric binaries at the third-and-a-half post-{Newtonian}
  order}. Phys Rev D 73:124012. \doi{10.1103/PhysRevD.73.124012}.
  {\href{https://arxiv.org/abs/gr-qc/0603056}{{gr-qc/0603056}}}

\bibitem[{K{\"{o}}nigsd{\"{o}}rffer et~al.(2003)K{\"{o}}nigsd{\"{o}}rffer,
  Faye, and Sch{\"{a}}fer}]{KFS03}
K{\"{o}}nigsd{\"{o}}rffer C, Faye G, Sch{\"{a}}fer G (2003) {The binary
  black-hole dynamics at the third-and-a-half post-{Newtonian} order in the
  {ADM}-formalism}. Phys Rev D 68:044004. \doi{10.1103/PhysRevD.68.044004}.
  {\href{https://arxiv.org/abs/astro-ph/0305048}{{astro-ph/0305048}}}

\bibitem[{Kopeikin(1985)}]{Kop85}
Kopeikin SM (1985) {The equations of motion of extended bodies in
  general-relativity with conservative corrections and radiation damping taken
  into account}. Astron Zh 62:889--904

\bibitem[{Kopeikin(1988)}]{Kop88}
Kopeikin SM (1988) {Celestial Coordinate Reference Systems in Curved
  Spacetime}. Celest Mech 44:87. \doi{10.1007/BF01230709}

\bibitem[{Kopeikin et~al.(1999)Kopeikin, Sch{\"{a}}fer, Gwinn, and
  Eubanks}]{KSGE}
Kopeikin SM, Sch{\"{a}}fer G, Gwinn CR, Eubanks TM (1999) {Astrometric and
  timing effects of gravitational waves from localized sources}. Phys Rev D
  59:084023. \doi{10.1103/PhysRevD.59.084023}.
  {\href{https://arxiv.org/abs/gr-qc/9811003}{{gr-qc/9811003}}}

\bibitem[{Kozai(1962)}]{Kozai62}
Kozai Y (1962) {Secular perturbations of asteroids with high inclination and
  eccentricity}. Astron J 67:591--598. \doi{10.1086/108790}

\bibitem[{Kozameh and Quirega(2016)}]{KQ16}
Kozameh C, Quirega G (2016) {Center of mass and spin for isolated sources of
  gravitational radiation}. Phys Rev D 93:064050.
  \doi{10.1103/PhysRevD.93.064050}.
  {\href{https://arxiv.org/abs/1311.5854}{{arXiv:1311.5854}}} {[gr-qc]}

\bibitem[{Kozameh et~al.(2018)Kozameh, Nieva, and Quirega}]{KNQ18}
Kozameh C, Nieva J, Quirega G (2018) {Spin and center of mass comparison
  between the post-{Newtonian} approach and the asymptotic formulation}. Phys
  Rev D 98:064005. \doi{10.1103/PhysRevD.98.064005}.
  {\href{https://arxiv.org/abs/1711.11375}{{arXiv:1711.11375}}} {[gr-qc]}

\bibitem[{Kramer and Wex(2009)}]{KW09}
Kramer M, Wex N (2009) {The double pulsar system: A unique laboratory for
  gravity}. Class Quantum Grav 26(7):073001.
  \doi{10.1088/0264-9381/26/7/073001}

\bibitem[{Kramer et~al.(2021)Kramer, Stairs, Manchester, Wex, Deller, Coles,
  Ali, Burgay, Camilo, Cognard et~al.}]{Kramer21}
Kramer M, Stairs IH, Manchester RN, Wex N, Deller AT, Coles WA, Ali M, Burgay
  M, Camilo F, Cognard I, et~al. (2021) {Strong-field gravity tests with the
  double pulsar}. Phys Rev X 11(4):041050. \doi{10.1103/PhysRevX.11.041050}.
  {\href{https://arxiv.org/abs/2112.06795}{{arXiv:2112.06795}}} {[astro-ph]}

\bibitem[{Kr{\'{o}}lak et~al.(1995)Kr{\'{o}}lak, Kokkotas, and
  Sch{\"{a}}fer}]{KKS95}
Kr{\'{o}}lak A, Kokkotas KD, Sch{\"{a}}fer G (1995) {Estimation of the
  post-{Newtonian} parameters in the gravitational-wave emission of a
  coalescing binary}. Phys Rev D 52:2089--2111. \doi{10.1103/PhysRevD.52.2089}.
  {\href{https://arxiv.org/abs/gr-qc/9503013}{{gr-qc/9503013}}}

\bibitem[{Lagarias(2013)}]{EulerConstant}
Lagarias J (2013) {Euler's constant: {Euler}'s work and modern developments}.
  Bull Amer Math Soc 50(4):527--628

\bibitem[{Landau and Lifshitz(1971)}]{LL}
Landau LD, Lifshitz EM (1971) The classical theory of fields, 3rd edn. Pergamon
  Press, Oxford; New York

\bibitem[{Landry(2018)}]{Landry18}
Landry P (2018) Rotational-tidal phasing of the binary neutron star waveform.
  arXiv e-prints {\href{https://arxiv.org/abs/1805.01882}{{1805.01882}}}

\bibitem[{Lang(2014)}]{Lang14}
Lang RN (2014) {Compact binary systems in scalar-tensor gravity. II. Tensor
  gravitational waves to second post-{Newtonian} order}. Phys Rev D
  89(8):084014. \doi{10.1103/PhysRevD.89.084014}.
  {\href{https://arxiv.org/abs/1310.3320}{{arXiv:1310.3320}}} {[gr-qc]}

\bibitem[{Lang(2015)}]{Lang15}
Lang RN (2015) {Compact binary systems in scalar-tensor gravity. III. Scalar
  waves and energy flux}. Phys Rev D 91(8):084027.
  \doi{10.1103/PhysRevD.91.084027}.
  {\href{https://arxiv.org/abs/1411.3073}{{arXiv:1411.3073}}} {[gr-qc]}

\bibitem[{Larrouturou et~al.(2022{\natexlab{a}})Larrouturou, Blanchet, Henry,
  and Faye}]{LBHF22}
Larrouturou F, Blanchet L, Henry Q, Faye G (2022{\natexlab{a}}) {The quadrupole
  moment of compact binaries to the fourth post-{Newtonian} order: II.
  Dimensional regularization and renormalization}. Class Quantum Grav
  39(11):115008. \doi{10.1088/1361-6382/ac5ba0}.
  {\href{https://arxiv.org/abs/2110.02243}{{arXiv:2110.02243}}} {[gr-qc]}

\bibitem[{Larrouturou et~al.(2022{\natexlab{b}})Larrouturou, Henry, Blanchet,
  and Faye}]{LHBF22}
Larrouturou F, Henry Q, Blanchet L, Faye G (2022{\natexlab{b}}) {The quadrupole
  moment of compact binaries to the fourth post-{Newtonian} order: I.
  Non-locality in time and infra-red divergencies}. Class Quantum Grav
  39(11):115007. \doi{10.1088/1361-6382/ac5762}.
  {\href{https://arxiv.org/abs/2110.02240}{{arXiv:2110.02240}}} {[gr-qc]}

\bibitem[{Le~Tiec(2015)}]{LeT15}
Le~Tiec A (2015) {First law of mechanics for compact binaries on eccentric
  orbits}. Phys Rev D 92:084021. \doi{10.1103/PhysRevD.92.084021}.
  {\href{https://arxiv.org/abs/1506.05648}{{arXiv:1506.05648}}} {[gr-qc]}

\bibitem[{Le~Tiec and Blanchet(2010)}]{LB10}
Le~Tiec A, Blanchet L (2010) {The Close-Limit Approximation for Black Hole
  Binaries with Post-{Newtonian} Initial Conditions}. Class Quantum Grav
  27:045008. \doi{10.1088/0264-9381/27/4/045008}.
  {\href{https://arxiv.org/abs/0901.4593}{{arXiv:0901.4593}}} {[gr-qc]}

\bibitem[{Le~Tiec et~al.(2010)Le~Tiec, Blanchet, and Will}]{LBW10}
Le~Tiec A, Blanchet L, Will CM (2010) {Gravitational-Wave Recoil from the
  Ringdown Phase of Coalescing Black Hole Binaries}. Class Quantum Grav
  27:012001. \doi{10.1088/0264-9381/27/1/012001}.
  {\href{https://arxiv.org/abs/0901.4594}{{arXiv:0901.4594}}} {[gr-qc]}

\bibitem[{Le~Tiec et~al.(2011)Le~Tiec, Mrou\'e, Barack, Buonanno, Pfeiffer,
  Sago, and Taracchini}]{Letal11}
Le~Tiec A, Mrou\'e AH, Barack L, Buonanno A, Pfeiffer HP, Sago N, Taracchini A
  (2011) {Periastron Advance in Black-Hole Binaries}. Phys Rev Lett 107:141101.
  {\href{https://arxiv.org/abs/1106.3278}{{arXiv:1106.3278}}} {[gr-qc]}

\bibitem[{Le~Tiec et~al.(2012{\natexlab{a}})Le~Tiec, Barausse, and
  Buonanno}]{LBB12}
Le~Tiec A, Barausse E, Buonanno A (2012{\natexlab{a}}) {Gravitational
  Self-Force Correction to the Binding Energy of Compact Binary Systems}. Phys
  Rev Lett 108:131103. \doi{10.1103/PhysRevLett.108.131103}.
  {\href{https://arxiv.org/abs/1111.5609}{{arXiv:1111.5609}}} {[gr-qc]}

\bibitem[{Le~Tiec et~al.(2012{\natexlab{b}})Le~Tiec, Blanchet, and
  Whiting}]{LBW12}
Le~Tiec A, Blanchet L, Whiting BF (2012{\natexlab{b}}) {First law of binary
  black hole mechanics in general relativity and post-{Newtonian} theory}. Phys
  Rev D 85:064039. \doi{10.1103/PhysRevD.85.064039}.
  {\href{https://arxiv.org/abs/1111.5378}{{arXiv:1111.5378}}} {[gr-qc]}

\bibitem[{Ledvinka et~al.(2008)Ledvinka, Sch{\"a}fer, and
  Bi\v{c}\'{a}k}]{LSB08}
Ledvinka T, Sch{\"a}fer G, Bi\v{c}\'{a}k J (2008) {Relativistic closed-form
  {Hamiltonian} for many-body gravitating systems in the post-{Minkowskian}
  approximation}. Phys Rev Lett 100:251101.
  \doi{10.1103/PhysRevLett.100.251101}.
  {\href{https://arxiv.org/abs/0807.0214}{{arXiv:0807.0214}}} {[gr-qc]}

\bibitem[{Lehner and Pretorius(2014)}]{LP14}
Lehner L, Pretorius F (2014) {Numerical relativity and astrophysics}. Annu Rev
  Astron Astrophys 52:661--694. \doi{10.1146/annurev-astro-081913-040031}

\bibitem[{Leibovich et~al.(2020)Leibovich, Maia, Rothstein, and Yang}]{LMRY20}
Leibovich AK, Maia NT, Rothstein IZ, Yang Z (2020) {Second post-{N}ewtonian
  order radiative dynamics of inspiralling compact binaries in the Effective
  Field Theory approach}. Phys Rev D 101(8):084058.
  \doi{10.1103/PhysRevD.101.084058}.
  {\href{https://arxiv.org/abs/1912.12546}{{arXiv:1912.12546}}} {[gr-qc]}

\bibitem[{{Leibovich} et~al.(2023){Leibovich}, {Pardo}, and {Yang}}]{LPY23}
{Leibovich} AK, {Pardo} BA, {Yang} Z (2023) {Radiation reaction for nonspinning
  bodies at 4.5{PN} in the effective field theory approach}. Phys Rev D
  108(2):024017. \doi{10.1103/PhysRevD.108.024017}.
  {\href{https://arxiv.org/abs/2302.11016}{{arXiv:2302.11016}}} {[gr-qc]}

\bibitem[{Levi(2010{\natexlab{a}})}]{Le10so}
Levi M (2010{\natexlab{a}}) {Next-to-leading order gravitational spin-orbit
  coupling in an effective field theory approach}. Phys Rev D 82:104004.
  \doi{10.1103/PhysRevD.82.104004}.
  {\href{https://arxiv.org/abs/1006.4139}{{arXiv:1006.4139}}} {[gr-qc]}

\bibitem[{Levi(2010{\natexlab{b}})}]{Le10ss}
Levi M (2010{\natexlab{b}}) {Next-to-leading order gravitational spin1-spin2
  coupling with Kaluza-Klein reduction}. Phys Rev D 82:064029.
  \doi{10.1103/PhysRevD.82.064029}.
  {\href{https://arxiv.org/abs/0802.1508}{{arXiv:0802.1508}}} {[gr-qc]}

\bibitem[{Levi(2012)}]{Le12ss}
Levi M (2012) {Binary dynamics from spin1-spin2 coupling at fourth
  post-{Newtonian} order}. Phys Rev D 85:064043.
  \doi{10.1103/PhysRevD.85.064043}.
  {\href{https://arxiv.org/abs/1107.4322}{{arXiv:1107.4322}}}

\bibitem[{Levi(2020)}]{Levirevue}
Levi M (2020) {Effective field theories of post-{Newtonian} gravity: a
  comprehensive review}. Rep Prog Phys 83(7):075901.
  \doi{10.1088/1361-6633/ab12bc}.
  {\href{https://arxiv.org/abs/1807.01699}{{arXiv:1807.01699}}} {[gr-qc]}

\bibitem[{Levi and Steinhoff(2014)}]{LS14}
Levi M, Steinhoff J (2014) {Equivalence of {ADM} {Hamiltonian} and Effective
  Field Theory approaches at next-to-next-to-leading order spin1-spin2 coupling
  of binary inspirals}. J Cosmol Astropart Phys 2014(12):003.
  \doi{10.1088/1475-7516/2014/12/003}.
  {\href{https://arxiv.org/abs/1408.5762}{{arXiv:1408.5762}}} {[gr-qc]}

\bibitem[{Levi and Steinhoff(2015{\natexlab{a}})}]{LS15a}
Levi M, Steinhoff J (2015{\natexlab{a}}) {Leading order finite size effects
  with spins for inspiralling compact binaries}. J High Energy Phys
  2015(06):059. \doi{10.1007/JHEP06(2015)059}.
  {\href{https://arxiv.org/abs/1410.2601}{{arXiv:1410.2601}}} {[gr-qc]}

\bibitem[{Levi and Steinhoff(2015{\natexlab{b}})}]{LS15b}
Levi M, Steinhoff J (2015{\natexlab{b}}) {Spinning gravitating objects in the
  effective field theory in the post-{Newtonian} scheme}. J High Energy Phys
  2015(09):219. \doi{10.1007/JHEP09(2015)219}.
  {\href{https://arxiv.org/abs/1501.04956}{{arXiv:1501.04956}}} {[gr-qc]}

\bibitem[{{Levi} and {Steinhoff}(2016)}]{LS16}
{Levi} M, {Steinhoff} J (2016) {Next-to-next-to-leading order gravitational
  spin-orbit coupling via the effective field theory for spinning objects in
  the post-{Newtonian} scheme}. J Cosmol Astropart Phys 2016(01):011.
  \doi{10.1088/1475-7516/2016/01/011}.
  {\href{https://arxiv.org/abs/1506.05056}{{arXiv:1506.05056}}} {[gr-qc]}

\bibitem[{Lidov(1962)}]{Lidov62}
Lidov ML (1962) {The evolution of orbits of artificial satellites of planets
  under the action of gravitational perturbations of external bodies}. Planet
  Space Sci 9:719. \doi{10.1016/0032-0633(62)90129-0}

\bibitem[{Limousin et~al.(2005)Limousin, Gondek-Rosinska, and
  Gourgoulhon}]{LGG05}
Limousin F, Gondek-Rosinska D, Gourgoulhon E (2005) {Last orbits of binary
  strange quark stars}. Phys Rev D 71:064012. \doi{10.1103/PhysRevD.71.064012}.
  {\href{https://arxiv.org/abs/gr-qc/0411127}{{arXiv:gr-qc/0411127}}} {[gr-qc]}

\bibitem[{Lincoln and Will(1990)}]{LW90}
Lincoln CW, Will CM (1990) {Coalescing binary systems of compact objects to
  (post)$^{5/2}$-{Newtonian} order: Late-time evolution and
  gravitational-radiation emission}. Phys Rev D 42:1123--1143.
  \doi{10.1103/PhysRevD.42.1123}

\bibitem[{Lorentz and Droste(1937)}]{LD17}
Lorentz HA, Droste J (1937) The motion of a system of bodies under the
  influence of their mutual attraction, according to {Einstein}'s theory,
  Nijhoff, The Hague, pp 330--355. \doi{10.1007/978-94-015-3445-1_11},
  translated from Versl. K. Akad. Wet. Amsterdam, 26, 392 and 649 (1917)

\bibitem[{Loutrel and Yunes(2017)}]{LY17}
Loutrel N, Yunes N (2017) {Hereditary effects in eccentric compact binary
  inspirals to third post-{Newtonian} order}. Class Quantum Grav 34(4):044003.
  \doi{10.1088/1361-6382/aa59c3}.
  {\href{https://arxiv.org/abs/1607.05409}{{arXiv:1607.05409}}} {[gr-qc]}

\bibitem[{Love(1911)}]{Love11}
Love AEH (1911) Some problems of geodynamics. Cambridge University Press.
  \urlprefix\url{http://www.archive.org/details/cu31924060184367}

\bibitem[{Madore(1970{\natexlab{a}})}]{MadoreI}
Madore J (1970{\natexlab{a}}) {Gravitational radiation from a bounded source
  {I}}. Ann Inst Henri Poincar{\'e} 12:285.
  \urlprefix\url{http://www.numdam.org/item/AIHPA_1970__12_3_285_0/}

\bibitem[{Madore(1970{\natexlab{b}})}]{MadoreII}
Madore J (1970{\natexlab{b}}) {Gravitational radiation from a bounded source
  {II}}. Ann Inst Henri Poincar{\'e} 12:365.
  \urlprefix\url{http://www.numdam.org/item/AIHPA_1970__12_4_365_0/}

\bibitem[{Maggiore(2008)}]{Maggiore}
Maggiore M (2008) {Gravitational waves: Volume 1: Theory and experiments}.
  Oxford University Press. \doi{10.1093/acprof:oso/9780198570745.001.0001}

\bibitem[{Mandal et~al.(2023{\natexlab{a}})Mandal, Mastrolia, Patil, and
  Steinhoff}]{Mandal2023b}
Mandal MK, Mastrolia P, Patil R, Steinhoff J (2023{\natexlab{a}})
  {Gravitational quadratic-in-spin {Hamiltonian} at {NNNLO} in the
  post-{Newtonian} framework}. J High Energy Phys 2023(7).
  \doi{10.1007/jhep07(2023)128}

\bibitem[{Mandal et~al.(2023{\natexlab{b}})Mandal, Mastrolia, Patil, and
  Steinhoff}]{Mandal2023a}
Mandal MK, Mastrolia P, Patil R, Steinhoff J (2023{\natexlab{b}})
  {Gravitational spin-orbit {Hamiltonian} at {NNNLO} in the post-{Newtonian}
  framework}. J High Energy Phys 2023(3). \doi{10.1007/jhep03(2023)130}

\bibitem[{Mano and Takasugi(1997)}]{MT97}
Mano S, Takasugi E (1997) {Analytic solutions of the {Teukolsky} equation and
  their properties}. Prog Theor Phys 97:213. \doi{10.1143/PTP.97.213}.
  {\href{https://arxiv.org/abs/gr-qc/9611014}{{gr-qc/9611014}}}

\bibitem[{Mano et~al.(1996{\natexlab{a}})Mano, Susuki, and Takasugi}]{MST96b}
Mano S, Susuki H, Takasugi E (1996{\natexlab{a}}) {Analytic solutions of the
  {Regge}-{Wheeler} equation and the post-{Minkowskian} expansion}. Prog Theor
  Phys 96:549. \doi{10.1143/PTP.96.549}.
  {\href{https://arxiv.org/abs/gr-qc/9605057}{{gr-qc/9605057}}}

\bibitem[{Mano et~al.(1996{\natexlab{b}})Mano, Susuki, and Takasugi}]{MST96a}
Mano S, Susuki H, Takasugi E (1996{\natexlab{b}}) {Analytic solutions of the
  {Teukolsky} equation and their low frequency expansions}. Prog Theor Phys
  95:1079. \doi{10.1143/PTP.95.1079}.
  {\href{https://arxiv.org/abs/gr-qc/9603020}{{gr-qc/9603020}}}

\bibitem[{Marchand et~al.(2016)Marchand, Blanchet, and Faye}]{MBF16}
Marchand T, Blanchet L, Faye G (2016) {Gravitational-wave tail effects to
  quartic non-linear order}. Class Quantum Grav 33:244003.
  \doi{10.1088/0264-9381/33/24/244003}.
  {\href{https://arxiv.org/abs/1607.07601}{{arXiv:1607.07601}}} {[gr-qc]}

\bibitem[{Marchand et~al.(2018)Marchand, Bernard, Blanchet, and Faye}]{MBBF17}
Marchand T, Bernard L, Blanchet L, Faye G (2018) {Ambiguity-free completion of
  the equations of motion of compact binary systems at the fourth
  post-{Newtonian} order}. Phys Rev D 97:044023.
  {\href{https://arxiv.org/abs/1707.09289}{{arXiv:1707.09289}}} {[gr-qc]}

\bibitem[{Marchand et~al.(2020)Marchand, Henry, Larrouturou, Marsat, Faye, and
  Blanchet}]{MHLMFB20}
Marchand T, Henry Q, Larrouturou F, Marsat S, Faye G, Blanchet L (2020) {The
  mass quadrupole moment of compact binary systems at the fourth
  post-{N}ewtonian order}. Class Quantum Grav 37(21):215006.
  \doi{10.1088/1361-6382/ab9ce1}.
  {\href{https://arxiv.org/abs/2003.13672}{{arXiv:2003.13672}}} {[gr-qc]}

\bibitem[{Marsat(2015)}]{M15}
Marsat S (2015) {Cubic order spin effects in the dynamics and gravitational
  wave energy flux of compact object binaries}. Class Quantum Grav 32:085008.
  \doi{10.1088/0264-9381/32/8/085008}.
  {\href{https://arxiv.org/abs/1411.4118}{{arXiv:1411.4118}}} {[gr-qc]}

\bibitem[{Marsat et~al.(2013{\natexlab{a}})Marsat, Boh{\'{e}}, Blanchet, and
  Buonanno}]{MBBB13}
Marsat S, Boh{\'{e}} A, Blanchet L, Buonanno A (2013{\natexlab{a}})
  {Next-to-leading tail-induced spin-orbit effects in the gravitational
  radiation of compact binaries}. Class Quantum Grav 31:025023.
  \doi{10.1088/0264-9381/30/5/055007}.
  {\href{https://arxiv.org/abs/1307.6793}{{arXiv:1307.6793}}} {[gr-qc]}

\bibitem[{Marsat et~al.(2013{\natexlab{b}})Marsat, Boh{\'{e}}, Faye, and
  Blanchet}]{MBFB13}
Marsat S, Boh{\'{e}} A, Faye G, Blanchet L (2013{\natexlab{b}})
  {Next-to-next-to-leading order spin-orbit effects in the equations of motion
  of compact binary systems}. Class Quantum Grav 30:055007.
  \doi{10.1088/0264-9381/30/5/055007}.
  {\href{https://arxiv.org/abs/1210.4143}{{arXiv:1210.4143}}}

\bibitem[{Martin and Sanz(1979)}]{MS}
Martin J, Sanz JL (1979) {Slow motion approximation in predictive relativistic
  mechanics. {II}. {N}on-interaction theorem for interactions derived from the
  classical field-theory}. J Math Phys 20:25--34. \doi{10.1063/1.523958}

\bibitem[{Mathews(1962)}]{M62}
Mathews J (1962) {Gravitational multipole radiation}. J Soc Ind Appl Math
  10:768--780. \doi{10.1137/0110059}

\bibitem[{Mathisson(1937)}]{Mathisson37}
Mathisson M (1937) {Neue {M}echanik materieller {S}ysteme}. Acta Phys Polon
  6:163--200

\bibitem[{Mathisson(2010)}]{Mathisson37repub}
Mathisson M (2010) Republication of: New mechanics of material systems. Gen
  Relativ Gravit 42:1011--1048. \doi{10.1007/s10714-010-0939-y}

\bibitem[{McClintock et~al.(2006)McClintock, Shafee, Narayan, Remillard, Davis,
  and Li}]{McClint06}
McClintock JE, Shafee R, Narayan R, Remillard RA, Davis SW, Li LX (2006) {The
  Spin of the Near-Extreme {Kerr} Black Hole GRS 1915+105}. Astrophys J
  652:518--539. \doi{10.1086/508457}.
  {\href{https://arxiv.org/abs/astro-ph/0606076}{{arXiv:astro-ph/0606076}}}

\bibitem[{Memmesheimer et~al.(2004)Memmesheimer, Gopakumar, and
  Sch{\"{a}}fer}]{MGS04}
Memmesheimer R, Gopakumar A, Sch{\"{a}}fer G (2004) {Third post-{Newtonian}
  accurate generalized quasi-Keplerian parametrization for compact binaries in
  eccentric orbits}. Phys Rev D 70:104011. \doi{10.1103/PhysRevD.70.104011}.
  {\href{https://arxiv.org/abs/gr-qc/0407049}{{gr-qc/0407049}}}

\bibitem[{Merritt et~al.(2004)Merritt, Milosavljevi{\'{c}}, Favata, Hughes, and
  Holz}]{Me04}
Merritt D, Milosavljevi{\'{c}} M, Favata M, Hughes SA, Holz DE (2004)
  {Consequences of Gravitational Radiation Recoil}. Astrophys J Lett
  607:L9--L12. \doi{10.1086/421551}.
  {\href{https://arxiv.org/abs/astro-ph/0402057}{{astro-ph/0402057}}}

\bibitem[{Messina and Nagar(2017)}]{MNagar17}
Messina F, Nagar A (2017) {Parametrized-4.5{PN} {TaylorF2} approximant(s) and
  tail effects to quartic nonlinear order from the effective one body
  formalism}. Phys Rev D 96:049907. \doi{10.1103/PhysRevD.95.124001}.
  {\href{https://arxiv.org/abs/1703.08107}{{arXiv:1703.08107}}} {[gr-qc]}

\bibitem[{Mik{\'{o}}czi et~al.(2005)Mik{\'{o}}czi, Vas{\'{u}}th, and
  Gergely}]{MVGer05}
Mik{\'{o}}czi B, Vas{\'{u}}th M, Gergely L (2005) {Self-interaction spin
  effects in inspiralling compact binaries}. Phys Rev D 71:124043.
  \doi{10.1103/PhysRevD.71.124043}.
  {\href{https://arxiv.org/abs/astro-ph/0504538}{{astro-ph/0504538}}}

\bibitem[{Miller and Hamilton(2002)}]{MH02}
Miller MC, Hamilton DP (2002) {Four-Body Effects in Globular Cluster Black Hole
  Coalescence}. Astrophys J 576:894. \doi{10.1086/341788}.
  {\href{https://arxiv.org/abs/astro-ph/0202298}{{astro-ph/0202298}}}

\bibitem[{Mino et~al.(1997{\natexlab{a}})Mino, Sasaki, Shibata, Tagoshi, and
  Tanaka}]{MSSTT97}
Mino Y, Sasaki M, Shibata M, Tagoshi H, Tanaka T (1997{\natexlab{a}}) {Black
  Hole Perturbation}. Prog Theor Phys Suppl 128:1--121.
  \doi{10.1143/PTPS.128.1}.
  {\href{https://arxiv.org/abs/gr-qc/9712057}{{gr-qc/9712057}}}

\bibitem[{Mino et~al.(1997{\natexlab{b}})Mino, Sasaki, and Tanaka}]{MiSaTa}
Mino Y, Sasaki M, Tanaka T (1997{\natexlab{b}}) {Gravitational radiation
  reaction to a particle motion}. Phys Rev D 55:3457--3476.
  \doi{10.1103/PhysRevD.55.3457}.
  {\href{https://arxiv.org/abs/gr-qc/9606018}{{arXiv:gr-qc/9606018}}}

\bibitem[{Mirshekari and Will(2013)}]{MW13}
Mirshekari S, Will CM (2013) {Compact binary systems in scalar-tensor gravity:
  Equations of motion to 2.5 post-{Newtonian} order}. Phys Rev D 87:084070.
  \doi{10.1103/PhysRevD.87.084070}.
  {\href{https://arxiv.org/abs/1301.4680}{{arXiv:1301.4680}}} {[gr-qc]}

\bibitem[{Mishra et~al.(2012)Mishra, Arun, and Iyer}]{MAI12}
Mishra CK, Arun KG, Iyer BR (2012) 2.5pn linear momentum flux from inspiralling
  compact binaries in quasicircular orbits and associated recoil: Nonspinning
  case. Physical Review D 85(4). \doi{10.1103/physrevd.85.044021},
  \urlprefix\url{http://dx.doi.org/10.1103/PhysRevD.85.044021}

\bibitem[{Misner et~al.(1973)Misner, Thorne, and Wheeler}]{MTW}
Misner CW, Thorne KS, Wheeler JA (1973) Gravitation. W. H. Freeman, San
  Francisco

\bibitem[{Moore and Hellings(2002)}]{HM2}
Moore TA, Hellings RW (2002) {Angular resolution of space-based gravitational
  wave detectors}. Phys Rev D 65:062001. \doi{10.1103/PhysRevD.65.062001}

\bibitem[{Mora and Will(2004)}]{MW03}
Mora T, Will CM (2004) {Post-{Newtonian} diagnostic of quasi-equilibrium binary
  configurations of compact objects}. Phys Rev D 69:104021.
  \doi{10.1103/PhysRevD.69.104021}.
  {\href{https://arxiv.org/abs/gr-qc/0312082}{{arXiv:gr-qc/0312082}}}

\bibitem[{Mougiakakos et~al.(2021)Mougiakakos, Riva, and Vernizzi}]{MRV21}
Mougiakakos S, Riva MM, Vernizzi F (2021) {Gravitational Bremsstrahlung in the
  post-{Minkowskian} effective field theory}. Phys Rev D 104(2):024041.
  \doi{10.1103/PhysRevD.104.024041}

\bibitem[{Munna et~al.(2020)Munna, Evans, Hopper, and Forseth}]{MEHF20}
Munna C, Evans CR, Hopper S, Forseth E (2020) {Determination of new
  coefficients in the angular momentum and energy fluxes at infinity to 9PN
  order for eccentric Schwarzschild extreme-mass-ratio inspirals using
  mode-by-mode fitting}. Phys Rev D 102(2):024047.
  \doi{10.1103/PhysRevD.102.024047}.
  {\href{https://arxiv.org/abs/2005.03044}{{arXiv:2005.03044}}}

\bibitem[{Nakamura et~al.(1987)Nakamura, Oohara, and Kojima}]{NakaOo87}
Nakamura T, Oohara K, Kojima Y (1987) {General relativistic collapse to black
  holes and gravitational waves from black holes}. Prog Theor Phys Suppl 90:1.
  \doi{10.1143/PTPS.90.1}

\bibitem[{Newman and Penrose(1962)}]{NP62}
Newman E, Penrose R (1962) {An approach to gravitational radiation by a method
  of spin coefficients}. J Math Phys 3:566--578. \doi{10.1063/1.1724257}

\bibitem[{{Newman} and {Penrose}(1966)}]{NP66}
{Newman} ET, {Penrose} R (1966) {Note on the {Bondi}-{Metzner}-{Sachs} group}.
  J Math Phys 7(5):863--870. \doi{10.1063/1.1931221}

\bibitem[{Newman and Unti(1963)}]{NU63}
Newman ET, Unti T (1963) {A class of null flat-space coordinate systems}. J
  Math Phys 4(12):1467--1469. \doi{10.1063/1.1703927}

\bibitem[{Nichols(2018)}]{N18}
Nichols DA (2018) Center-of-mass angular momentum and memory effect in
  asymptotically flat spacetimes. Phys Rev D 98:064032.
  \doi{10.1103/PhysRevD.98.064032}.
  {\href{https://arxiv.org/abs/1807.08767}{{arXiv:1807.08767}}} {[gr-qc]}

\bibitem[{Nissanke and Blanchet(2005)}]{NB05}
Nissanke S, Blanchet L (2005) {Gravitational radiation reaction in the
  equations of motion of compact binaries to 3.5 post-{Newtonian} order}. Class
  Quantum Grav 22:1007--1031. \doi{10.1088/0264-9381/22/6/008}.
  {\href{https://arxiv.org/abs/gr-qc/0412018}{{gr-qc/0412018}}}

\bibitem[{Nowak et~al.(2012)Nowak, Wilms, Pottschmidt, Schulz, Miller, and
  Maitra}]{Nowak12}
Nowak MA, Wilms J, Pottschmidt K, Schulz N, Miller J, Maitra D (2012) Suzaku
  observations of 4u 1957+11: The most rapidly spinning black hole in the
  galaxy? In: Petre R, Mitsuda K, Angelini L (eds) SUZAKU 2011. Exploring the
  X-ray Universe: Suzaku and Beyond (SUZAKU 2011). AIP Conference Proceedings,
  vol 1427. AIP Publishing, Melville, NY, pp 48--51. \doi{10.1063/1.3696149}

\bibitem[{Ohta et~al.(1973)Ohta, Okamura, Kimura, and Hiida}]{OO73a}
Ohta T, Okamura H, Kimura T, Hiida K (1973) {Physically acceptable solution of
  {Einstein}'s equation for many-body system}. Prog Theor Phys 50:492--514.
  \doi{10.1143/PTP.50.492}

\bibitem[{Ohta et~al.(1974{\natexlab{a}})Ohta, Okamura, Kimura, and
  Hiida}]{OO74b}
Ohta T, Okamura H, Kimura T, Hiida K (1974{\natexlab{a}}) {Coordinate Condition
  and Higher Order Gravitational Potential in Canocical Formalism}. Prog Theor
  Phys 51:1598--1612. \doi{10.1143/PTP.51.1598}

\bibitem[{Ohta et~al.(1974{\natexlab{b}})Ohta, Okamura, Kimura, and
  Hiida}]{OO74a}
Ohta T, Okamura H, Kimura T, Hiida K (1974{\natexlab{b}}) {Higher-order
  gravitational potential for many-body system}. Prog Theor Phys 51:1220--1238.
  \doi{10.1143/PTP.51.1220}

\bibitem[{Okamura et~al.(1973)Okamura, Ohta, Kimura, and Hiida}]{OO73b}
Okamura H, Ohta T, Kimura T, Hiida K (1973) {Perturbation calculation of
  gravitational potentials}. Prog Theor Phys 50:2066--2079.
  \doi{10.1143/PTP.50.2066}

\bibitem[{Oohara and Nakamura(1983)}]{OoNaka83}
Oohara K, Nakamura T (1983) {Energy, momentum and angular momentum of
  gravitational waves induced by a particle plunging into a {Schwarzschild}
  black hole}. Prog Theor Phys 70:757. \doi{10.1143/PTP.70.757}

\bibitem[{Owen et~al.(1998)Owen, Tagoshi, and Ohashi}]{OTO98}
Owen BJ, Tagoshi H, Ohashi A (1998) {Nonprecessional spin-orbit effects on
  gravitational waves from inspiraling compact binaries to second
  post-{Newtonian} order}. Phys Rev D 57:6168--6175.
  \doi{10.1103/PhysRevD.57.6168}.
  {\href{https://arxiv.org/abs/gr-qc/9710134}{{gr-qc/9710134}}}

\bibitem[{Owen et~al.(2023)Owen, Haster, Perkins, Cornish, and Yunes}]{Owen23}
Owen CB, Haster CJ, Perkins S, Cornish NJ, Yunes N (2023) {Waveform accuracy
  and systematic uncertainties in current gravitational wave observations}.
  Phys Rev D 108(4):044018. \doi{10.1103/PhysRevD.108.044018}.
  {\href{https://arxiv.org/abs/2301.11941}{{arXiv:2301.11941}}} {[gr-qc]}

\bibitem[{Pan et~al.(2010)Pan, Buonanno, Buchman, Chu, Kidder, Pfeiffer, and
  Scheel}]{Pan10}
Pan Y, Buonanno A, Buchman LT, Chu T, Kidder LE, Pfeiffer HP, Scheel MA (2010)
  {Effective-one-body waveforms calibrated to numerical relativity simulations:
  Coalescence of non-precessing, spinning, equal-mass black holes}. Phys Rev D
  81:084041. \doi{10.1103/PhysRevD.81.084041}.
  {\href{https://arxiv.org/abs/0912.3466}{{arXiv:0912.3466}}} {[gr-qc]}

\bibitem[{Papapetrou(1951{\natexlab{a}})}]{Papa51}
Papapetrou A (1951{\natexlab{a}}) {Equations of motion in general relativity}.
  Proc Phys Soc London, Sect B 64:57--75

\bibitem[{Papapetrou(1951{\natexlab{b}})}]{Papa51spin}
Papapetrou A (1951{\natexlab{b}}) {Spinning Test-Particles in General
  Relativity. {I}}. Proc R Soc London, Ser A 209:248--258.
  \doi{10.1098/rspa.1951.0200}

\bibitem[{Papapetrou(1962)}]{Papa62}
Papapetrou A (1962) {Relativit\'{e} -- une formule pour le rayonnement
  gravitationnel en premi\`{e}re approximation}. C R Acad Sci Ser II 255:1578

\bibitem[{Papapetrou(1969)}]{Papa69}
Papapetrou A (1969) {Coordonn\'ees radiatives ``cart\'esiennes''}. Ann Inst
  Henri Poincar{\'e} A XI:251.
  \urlprefix\url{http://www.numdam.org/item/AIHPA_1969__11_3_251_0/}

\bibitem[{Papapetrou(1971)}]{Papa71}
Papapetrou A (1971) {\'Etude syst\'ematique du rayonnement gravitationnel
  4-polaire. \'Energie-impulsion et moment cin\'etique du rayonnement}. Ann
  Inst Henri Poincare XIV:79

\bibitem[{Papapetrou and Linet(1981)}]{PapaL81}
Papapetrou A, Linet B (1981) {Equation of motion including the reaction of
  gravitational radiation}. Gen Relativ Gravit 13:335. \doi{10.1007/BF01025468}

\bibitem[{Pati and Will(2000)}]{PW00}
Pati ME, Will CM (2000) {Post-{Newtonian} gravitational radiation and equations
  of motion via direct integration of the relaxed {Einstein} equations:
  Foundations}. Phys Rev D 62:124015. \doi{10.1103/PhysRevD.62.124015}.
  {\href{https://arxiv.org/abs/gr-qc/0007087}{{gr-qc/0007087}}}

\bibitem[{Pati and Will(2002)}]{PW02}
Pati ME, Will CM (2002) {Post-{Newtonian} gravitational radiation and equations
  of motion via direct integration of the relaxed {Einstein} equations. {II}.
  {T}wo-body equations of motion to second post-{Newtonian} order, and
  radiation reaction to 3.5 post-{Newtonian} order}. Phys Rev D 65:104008.
  \doi{10.1103/PhysRevD.65.104008}.
  {\href{https://arxiv.org/abs/gr-qc/0201001}{{gr-qc/0201001}}}

\bibitem[{Paul and Mishra(2023)}]{PM23}
Paul K, Mishra CK (2023) {Spin effects in spherical harmonic modes of
  gravitational waves from eccentric compact binary inspirals}. Phys Rev D
  108(2):024023. {\href{https://arxiv.org/abs/2211.04155}{{arXiv:2211.04155}}}
  {[gr-qc]}

\bibitem[{Penrose(1963)}]{P63}
Penrose R (1963) {Asymptotic Properties of Fields and Space-Times}. Phys Rev
  Lett 10:66--68. \doi{10.1103/PhysRevLett.10.66}

\bibitem[{Penrose(1965)}]{P65}
Penrose R (1965) {Zero Rest-Mass Fields Including Gravitation: Asymptotic
  Behaviour}. Proc R Soc London, Ser A 284:159--203.
  \doi{10.1098/rspa.1965.0058}

\bibitem[{Peres(1962)}]{Peres62}
Peres A (1962) {Classical gravitational recoil}. Phys Rev 128:2471.
  \doi{10.1103/PhysRev.128.2471}

\bibitem[{Peters(1964)}]{Pe64}
Peters PC (1964) {Gravitational Radiation and the Motion of Two Point Masses}.
  Phys Rev 136:B1224--B1232. \doi{10.1103/PhysRev.136.B1224}

\bibitem[{Peters and Mathews(1963)}]{PM63}
Peters PC, Mathews J (1963) {Gravitational Radiation from Point Masses in a
  Keplerian Orbit}. Phys Rev 131:435--440. \doi{10.1103/PhysRev.131.435}

\bibitem[{Petrova(1949)}]{Petrova}
Petrova NM (1949) {Ob Uravnenii Dvizheniya i Tenzore Materii dlya Sistemy
  Konechnykh Mass v Obshchei Teorii Otnositielnosti}. J Exp Theor Phys
  19(11):989--999

\bibitem[{Pfeiffer et~al.(2000)Pfeiffer, Teukolsky, and Cook}]{PfTC00}
Pfeiffer HP, Teukolsky SA, Cook GB (2000) {Quasicircular orbits for spinning
  binary black holes}. Phys Rev D 62:104018. \doi{10.1103/PhysRevD.62.104018}.
  {\href{https://arxiv.org/abs/gr-qc/0006084}{{gr-qc/0006084}}}

\bibitem[{Pirani(1965)}]{Pi64}
Pirani FAE (1965) Introduction to gravitational radiation theory. In: Trautman
  A, Pirani FAE, Bondi H (eds) Lectures on General Relativity, Vol. 1.
  Prentice-Hall, Englewood Cliffs, NJ, pp 249--373

\bibitem[{Pleba{\'{n}}ski and Ba{{z}}a{\'{n}}ski(1959)}]{PB59}
Pleba{\'{n}}ski JF, Ba{{z}}a{\'{n}}ski SL (1959) {The general {Fokker} action
  principle and its application in general relativity theory}. Acta Phys Pol
  18:307--345

\bibitem[{Poisson(1993)}]{P93a}
Poisson E (1993) {Gravitational radiation from a particle in circular orbit
  around a black hole. {I}. {A}nalytic results for the nonrotating case}. Phys
  Rev D 47:1497--1510. \doi{10.1103/PhysRevD.47.1497}

\bibitem[{Poisson(1995)}]{P95}
Poisson E (1995) {Gravitational radiation from a particle in circular orbit
  around a black-hole. {VI}. {A}ccuracy of the post-{Newtonian} expansion}.
  Phys Rev D 52:5719--5723. \doi{10.1103/PhysRevD.52.5719}, {Erratum}: Phys.
  Rev. D, 55, 7980 (1997).
  {\href{https://arxiv.org/abs/gr-qc/9505030}{{gr-qc/9505030}}}

\bibitem[{Poisson(1997)}]{P97quad}
Poisson E (1997) {Gravitational waves from inspiraling compact binaries: The
  quadrupole-moment term}. Phys Rev D 57:5287--5290.
  \doi{10.1103/PhysRevD.57.5287}.
  {\href{https://arxiv.org/abs/gr-qc/9709032}{{gr-qc/9709032}}}

\bibitem[{Poisson and Sasaki(1995)}]{PS95}
Poisson E, Sasaki M (1995) {Gravitational radiation from a particle in circular
  orbit around a black hole. {V}. {B}lack-hole absorption and tail
  corrections}. Phys Rev D 51:5753--5767. \doi{10.1103/PhysRevD.51.5753}.
  {\href{https://arxiv.org/abs/gr-qc/9412027}{{gr-qc/9412027}}}

\bibitem[{Poisson and Will(1995)}]{PW95}
Poisson E, Will CM (1995) {Gravitational waves from inspiraling compact
  binaries: Parameter estimation using second-post-{Newtonian} wave forms}.
  Phys Rev D 52:848--855. \doi{10.1103/PhysRevD.52.848}.
  {\href{https://arxiv.org/abs/gr-qc/9502040}{{arXiv:gr-qc/9502040}}}

\bibitem[{Poisson et~al.(2011)Poisson, Pound, and Vega}]{PoissonLR}
Poisson E, Pound A, Vega I (2011) {The Motion of Point Particles in Curved
  Spacetime}. Living Rev Relativ 14:7. \doi{10.12942/lrr-2011-7}.
  {\href{https://arxiv.org/abs/1102.0529}{{arXiv:1102.0529}}} {[gr-qc]}

\bibitem[{Porto(2006)}]{Po06}
Porto RA (2006) {Post-{Newtonian} corrections to the motion of spinning bodies
  in NRGR}. Phys Rev D 73:104031. \doi{10.1103/PhysRevD.73.104031}.
  {\href{https://arxiv.org/abs/gr-qc/0511061}{{gr-qc/0511061}}}

\bibitem[{Porto(2008)}]{Porto08}
Porto RA (2008) {Absorption effects due to spin in the worldline approach to
  black hole dynamics}. Phys Rev D 77:064026. \doi{10.1103/PhysRevD.77.064026}.
  {\href{https://arxiv.org/abs/0710.5150}{{arXiv:0710.5150}}} {[hep-th]}

\bibitem[{Porto(2010)}]{Po10}
Porto RA (2010) {Next-to-leading-order spin--orbit effects in the motion of
  inspiralling compact binaries}. Class Quantum Grav 27:205001.
  \doi{10.1088/0264-9381/27/20/205001}.
  {\href{https://arxiv.org/abs/1005.5730}{{arXiv:1005.5730}}} {[gr-qc]}

\bibitem[{Porto(2016)}]{Portorevue}
Porto RA (2016) {The effective field theorist’s approach to gravitational
  dynamics}. Phys Rep 633:1--104. \doi{10.1016/j.physrep.2016.04.003}

\bibitem[{Porto and Rothstein(2006)}]{PoR06}
Porto RA, Rothstein IZ (2006) {Calculation of the first nonlinear contribution
  to the general-relativistic spin-spin interaction for binary systems}. Phys
  Rev Lett 97:021101. \doi{10.1103/PhysRevLett.97.021101}.
  {\href{https://arxiv.org/abs/gr-qc/0604099}{{arXiv:gr-qc/0604099}}}

\bibitem[{Porto and Rothstein(2008{\natexlab{a}})}]{PoR08b}
Porto RA, Rothstein IZ (2008{\natexlab{a}}) {Next to leading order
  spin(1)spin(1) effects in the motion of inspiralling compact binaries}. Phys
  Rev D 78:044013. \doi{10.1103/PhysRevD.78.044013}.
  {\href{https://arxiv.org/abs/0804.0260}{{arXiv:0804.0260}}} {[gr-qc]}

\bibitem[{Porto and Rothstein(2008{\natexlab{b}})}]{PoR08a}
Porto RA, Rothstein IZ (2008{\natexlab{b}}) {Spin(1)spin(2) effects in the
  motion of inspiralling compact binaries at third order in the
  post-{Newtonian} expansion}. Phys Rev D 78:044012.
  \doi{10.1103/PhysRevD.78.044012}.
  {\href{https://arxiv.org/abs/0802.0720}{{arXiv:0802.0720}}} {[gr-qc]}

\bibitem[{Porto and Rothstein(2017)}]{PR17}
Porto RA, Rothstein IZ (2017) {Apparent ambiguities in the post-{Newtonian}
  expansion for binary systems}. Phys Rev D 96(2):024062.
  \doi{10.1103/PhysRevD.96.024062}.
  {\href{https://arxiv.org/abs/1703.06433}{{arXiv:1703.06433}}} {[gr-qc]}

\bibitem[{Porto et~al.(2011)Porto, Ross, and Rothstein}]{PRR10}
Porto RA, Ross A, Rothstein IZ (2011) {Spin induced multipole moments for the
  gravitational wave flux from binary inspirals to third Post-{Newtonian}
  order}. J Cosmol Astropart Phys 2011(3):009.
  \doi{10.1088/1475-7516/2011/03/009}.
  {\href{https://arxiv.org/abs/1007.1312}{{arXiv:1007.1312}}} {[gr-qc]}

\bibitem[{Poujade and Blanchet(2002)}]{PB02}
Poujade O, Blanchet L (2002) {Post-{Newtonian} approximation for isolated
  systems calculated by matched asymptotic expansions}. Phys Rev D 65:124020.
  \doi{10.1103/PhysRevD.65.124020}.
  {\href{https://arxiv.org/abs/gr-qc/0112057}{{gr-qc/0112057}}}

\bibitem[{Pound et~al.(2020)Pound, Wardell, Warburton, and
  Miller}]{Poundetal20}
Pound A, Wardell B, Warburton N, Miller J (2020) {Second-order self-force
  calculation of gravitational binding energy in compact binaries}. Phys Rev
  Lett 124(2):021101. \doi{10.1103/PhysRevLett.124.021101}

\bibitem[{Pratten et~al.(2020)Pratten, Husa, Garcia-Quiros, Colleoni,
  Ramos-Buades, Estelles, and Jaume}]{PH20}
Pratten G, Husa S, Garcia-Quiros C, Colleoni M, Ramos-Buades A, Estelles H,
  Jaume R (2020) {Setting the cornerstone for a family of models for
  gravitational waves from compact binaries: The dominant harmonic for
  nonprecessing quasicircular black holes}. Phys Rev D 102(6):064001.
  \doi{10.1103/PhysRevD.102.064001}.
  {\href{https://arxiv.org/abs/2001.11412}{{arXiv:2001.11412}}} {[gr-qc]}

\bibitem[{Press(1977)}]{Press77}
Press WH (1977) {Gravitational Radiation from Sources Which Extend Into Their
  Own Wave Zone}. Phys Rev D 15:965--968. \doi{10.1103/PhysRevD.15.965}

\bibitem[{Pretorius(2005)}]{Pret05}
Pretorius F (2005) {Evolution of Binary Black-Hole Spacetimes}. Phys Rev Lett
  95:121101. \doi{10.1103/PhysRevLett.95.121101}.
  {\href{https://arxiv.org/abs/gr-qc/0507014}{{arXiv:gr-qc/0507014}}}

\bibitem[{Quinn and Wald(1997)}]{QuWa}
Quinn TC, Wald RM (1997) {Axiomatic approach to electromagnetic and
  gravitational radiation reaction of particles in curved spacetime}. Phys Rev
  D 56:3381--3394. \doi{10.1103/PhysRevD.56.3381}.
  {\href{https://arxiv.org/abs/gr-qc/9610053}{{arXiv:gr-qc/9610053}}}

\bibitem[{Racine et~al.(2009)Racine, Buonanno, and Kidder}]{RBK09}
Racine E, Buonanno A, Kidder L (2009) {Recoil velocity at second
  post-{N}ewtonian order for spinning black hole binaries}. Phys Rev D
  80:044010. \doi{10.1103/PhysRevD.80.044010}.
  {\href{https://arxiv.org/abs/0812.4413}{{arXiv:0812.4413}}} {[gr-qc]}

\bibitem[{Rendall(1990)}]{Rend90}
Rendall AD (1990) {Convergent and divergent perturbation series and the
  post-Minkowskian scheme}. Class Quantum Grav 7:803.
  \doi{10.1088/0264-9381/7/5/010}

\bibitem[{Rendall(1992)}]{Rend92}
Rendall AD (1992) {On the definition of post-{Newtonian} approximations}. Proc
  R Soc London, Ser A 438:341--360. \doi{10.1098/rspa.1992.0111}

\bibitem[{Rendall(1994)}]{Rend94}
Rendall AD (1994) {The {Newtonian} limit for asymptotically flat solutions of
  the {Einstein}-{Vlasov} system}. Commun Math Phys 163:89--112.
  \doi{10.1007/BF02101736}.
  {\href{https://arxiv.org/abs/gr-qc/9303027}{{gr-qc/9303027}}}

\bibitem[{Reynolds(2014)}]{Reyn13}
Reynolds CS (2014) {Measuring Black Hole Spin Using X-Ray Reflection
  Spectroscopy}. Space Sci Rev 183:277--294. \doi{10.1007/s11214-013-0006-6}.
  {\href{https://arxiv.org/abs/1302.3260}{{arXiv:1302.3260}}} {[astro-ph.HE]}

\bibitem[{Rieth and Sch{\"{a}}fer(1997)}]{RS97}
Rieth R, Sch{\"{a}}fer G (1997) {Spin and tail effects in the
  gravitational-wave emission of compact binaries}. Class Quantum Grav 14:2357.
  \doi{10.1088/0264-9381/14/8/029}

\bibitem[{Sachs(1961)}]{S61}
Sachs RK (1961) {Gravitational waves in general relativity. {VI}. {T}he
  outgoing radiation condition}. Proc R Soc London, Ser A 264:309--338.
  \doi{10.1098/rspa.1961.0202}

\bibitem[{Sachs(1962)}]{Sachs62}
Sachs RK (1962) {Gravitational Waves in General Relativity. {VIII}. {W}aves in
  Asymptotically Flat Space-Time}. Proc R Soc London, Ser A 270:103--126.
  \doi{10.1098/rspa.1962.0206}

\bibitem[{Sachs and Bergmann(1958)}]{SB58}
Sachs RK, Bergmann PG (1958) {Structure of Particles in Linearized
  Gravitational Theory}. Phys Rev 112:674--680. \doi{10.1103/PhysRev.112.674}

\bibitem[{Sago et~al.(2008)Sago, Barack, and Detweiler}]{SBD08}
Sago N, Barack L, Detweiler S (2008) {Two approaches for the gravitational self
  force in black hole spacetime: Comparison of numerical results}. Phys Rev D
  78:124024. \doi{10.1103/PhysRevD.78.124024}.
  {\href{https://arxiv.org/abs/0810.2530}{{arXiv:0810.2530}}}

\bibitem[{{Saketh} et~al.(2023){Saketh}, {Steinhoff}, {Vines}, and
  {Buonanno}}]{Saketh22}
{Saketh} MVS, {Steinhoff} J, {Vines} J, {Buonanno} A (2023) {Modeling horizon
  absorption in spinning binary black holes using effective worldline theory}.
  Phys Rev D 107(8):084006. \doi{10.1103/PhysRevD.107.084006}.
  {\href{https://arxiv.org/abs/2212.13095}{{arXiv:2212.13095}}} {[gr-qc]}

\bibitem[{Santamar{\'{\i}}a et~al.(2010)Santamar{\'{\i}}a, Ohme, Ajith,
  Br{\"{u}}gmann, Dorband, Hannam, Husa, M{\"{o}}sta, Pollney, Reisswig,
  Robinson, Seiler, and Krishnan}]{Ajith10}
Santamar{\'{\i}}a L, Ohme F, Ajith P, Br{\"{u}}gmann B, Dorband N, Hannam M,
  Husa S, M{\"{o}}sta P, Pollney D, Reisswig C, Robinson EL, Seiler J, Krishnan
  B (2010) {Matching post-{Newtonian} and numerical relativity waveforms:
  Systematic errors and a new phenomenological model for non-precessing black
  hole binaries}. Phys Rev D 82:064016. \doi{10.1103/PhysRevD.82.064016}.
  {\href{https://arxiv.org/abs/1005.3306}{{arXiv:1005.3306}}} {[gr-qc]}

\bibitem[{Sasaki(1994)}]{Sasa94}
Sasaki M (1994) {Post-{Newtonian} Expansion of the Ingoing-Wave
  {Regge}-{Wheeler} Function}. Prog Theor Phys 92:17--36.
  \doi{10.1143/ptp/92.1.17}

\bibitem[{Sasaki and Tagoshi(2003)}]{SasakiLR}
Sasaki M, Tagoshi H (2003) {Analytic Black Hole Perturbation Approach to
  Gravitational Radiation}. Living Rev Relativ 6:6. \doi{10.12942/lrr-2003-6}.
  {\href{https://arxiv.org/abs/gr-qc/0306120}{{arXiv:gr-qc/0306120}}}

\bibitem[{Sch{\"a}fer(1981)}]{Schafer81}
Sch{\"a}fer G (1981) {The gravitational quadrupole formulae and gravitationally
  bound matter systems}. Astrophys J 250:L5--L8

\bibitem[{Sch{\"a}fer(1982)}]{Schafer82}
Sch{\"a}fer G (1982) {The equations of motion for an astrophysical binary with
  accuracy $1/c^5$}. Prog Theor Phys 68(6):2191--2193.
  \doi{10.1143/PTP.68.2191}

\bibitem[{Sch{\"{a}}fer(1984)}]{S84}
Sch{\"{a}}fer G (1984) {Acceleration-dependent Lagrangians in general
  relativity}. Phys Lett A 100:128. \doi{10.1016/0375-9601(84)90947-2}

\bibitem[{Sch{\"{a}}fer(1985)}]{S85}
Sch{\"{a}}fer G (1985) {The Gravitational Quadrupole Radiation-Reaction Force
  and the Canonical Formalism of {ADM}}. Ann Phys (NY) 161:81--100.
  \doi{10.1016/0003-4916(85)90337-9}

\bibitem[{Sch{\"{a}}fer(1986)}]{S86}
Sch{\"{a}}fer G (1986) {The {ADM} {Hamiltonian} at the Postlinear
  Approximation}. Gen Relativ Gravit 18:255--270. \doi{10.1007/BF00765886}

\bibitem[{Sch{\"{a}}fer(1987)}]{S87}
Sch{\"{a}}fer G (1987) {Three-body {Hamiltonian} in general relativity}. Phys
  Lett 123:336--339. \doi{10.1016/0375-9601(87)90389-6}

\bibitem[{Sch{\"{a}}fer(2011)}]{Schaferorleans}
Sch{\"{a}}fer G (2011) Post-{Newtonian} methods: Analytic results on the binary
  problem. In: Blanchet L, Spallicci A, Whiting B (eds) Mass and Motion in
  General Relativity. Fundamental Theories of Physics, vol 162. Springer,
  Dordrecht; New York, pp 167--210. \doi{10.1007/978-90-481-3015-3_6}

\bibitem[{Sch{\"a}fer and Jaranowski(2024)}]{JaraSLRR}
Sch{\"a}fer G, Jaranowski P (2024) {{Hamiltonian} formulation of general
  relativity and post-{Newtonian} dynamics of compact binaries}. Living Rev
  Relativ 27:2. \doi{10.1007/s41114-024-00048-7}.
  {\href{https://arxiv.org/abs/1805.07240}{{arXiv:1805.07240}}} {[gr-qc]}

\bibitem[{Sch{\"{a}}fer and Wex(1993)}]{SW93}
Sch{\"{a}}fer G, Wex N (1993) {Second post-{Newtonian} motion of compact
  binaries}. Phys Lett A 174:196--205. \doi{10.1016/0375-9601(93)90758-R},
  {Erratum}: Phys Lett A, 177, 461 (1993)

\bibitem[{Schott(1915)}]{Schott}
Schott GA (1915) On the motion of the {Lorentz} electron. Phil Mag 29:49.
  \doi{10.1080/14786440108635280}

\bibitem[{Schwartz(1978)}]{Schwartz}
Schwartz L (1978) Th\'eorie des distributions. Hermann, Paris

\bibitem[{Sellier(1994)}]{Sellier}
Sellier A (1994) {Hadamard's finite part concept in dimension $n \ge 2$,
  distributional definition, regularization forms and distributional
  derivatives}. Proc R Soc London, Ser A 445:69--98.
  \doi{10.1098/rspa.1994.0049}

\bibitem[{Sennett et~al.(2016)Sennett, Marsat, and Buonanno}]{SMB16}
Sennett N, Marsat S, Buonanno A (2016) {Gravitational waveforms in
  scalar-tensor gravity at 2{PN} relative order}. Phys Rev D 94(8):084003.
  \doi{10.1103/PhysRevD.94.084003}.
  {\href{https://arxiv.org/abs/1607.01420}{{arXiv:1607.01420}}} {[gr-qc]}

\bibitem[{Shah et~al.(2014)Shah, Friedmann, and Whiting}]{SFW14}
Shah A, Friedmann J, Whiting BF (2014) {Finding high-order analytic
  post-{Newtonian} parameters from a high-precision numerical self-force
  calculation}. Phys Rev D 89:064042. \doi{10.1103/PhysRevD.89.064042}.
  {\href{https://arxiv.org/abs/1312.1952}{{arXiv:1312.1952}}} {[gr-qc]}

\bibitem[{Simon and Beig(1983)}]{SB83}
Simon W, Beig R (1983) {The multipole structure of stationary space-times}. J
  Math Phys 24:1163--1171. \doi{10.1063/1.525846}

\bibitem[{Sopuerta et~al.(2006)Sopuerta, Yunes, and Laguna}]{SYL06}
Sopuerta CF, Yunes N, Laguna P (2006) {Gravitational Recoil from Binary Black
  Hole Mergers: the Close-Limit Approximation}. Phys Rev D 74:124010.
  \doi{10.1103/PhysRevD.74.124010}.
  {\href{https://arxiv.org/abs/astro-ph/0608600}{{astro-ph/0608600}}}

\bibitem[{Steinhoff(2011)}]{St11rev}
Steinhoff J (2011) {Canonical formulation of spin in general relativity}. Ann
  Phys (Berlin) 523:296. \doi{10.1002/andp.201000178}.
  {\href{https://arxiv.org/abs/1106.4203}{{arXiv:1106.4203}}} {[gr-qc]}

\bibitem[{{Steinhoff} and {Puetzfeld}(2010)}]{SP10}
{Steinhoff} J, {Puetzfeld} D (2010) {Multipolar equations of motion for
  extended test bodies in general relativity}. Phys Rev D 81(4):044019.
  \doi{10.1103/PhysRevD.81.044019}.
  {\href{https://arxiv.org/abs/0909.3756}{{arXiv:0909.3756}}} {[gr-qc]}

\bibitem[{Steinhoff et~al.(2008{\natexlab{a}})Steinhoff, Hergt, and
  Sch{\"{a}}fer}]{SHS08a}
Steinhoff J, Hergt S, Sch{\"{a}}fer G (2008{\natexlab{a}}) {Next-to-leading
  order gravitational spin(1)-spin(2) dynamics in {Hamiltonian} form}. Phys Rev
  D 77:081501(R). {\href{https://arxiv.org/abs/0712.1716}{{arXiv:0712.1716}}}
  {[gr-qc]}

\bibitem[{Steinhoff et~al.(2008{\natexlab{b}})Steinhoff, Hergt, and
  Sch{\"{a}}fer}]{SHS08c}
Steinhoff J, Hergt S, Sch{\"{a}}fer G (2008{\natexlab{b}}) {Spin-squared
  {Hamiltonian} of next-to-leading order gravitational interaction}. Phys Rev D
  78:101503(R). {\href{https://arxiv.org/abs/0809.2200}{{arXiv:0809.2200}}}
  {[gr-qc]}

\bibitem[{Steinhoff et~al.(2008{\natexlab{c}})Steinhoff, Sch{\"{a}}fer, and
  Hergt}]{SHS08b}
Steinhoff J, Sch{\"{a}}fer G, Hergt S (2008{\natexlab{c}}) {{ADM} canonical
  formalism for gravitating spinning objects}. Phys Rev D 77:104018.
  \doi{10.1103/PhysRevD.77.104018}.
  {\href{https://arxiv.org/abs/0805.3136}{{arXiv:0805.3136}}} {[gr-qc]}

\bibitem[{Strohmayer(2001)}]{Stroh01}
Strohmayer TE (2001) {Discovery of a 450 Hz quasi-periodic oscillation from the
  microquasar GRO J1655--40 with the Rossi X-ray Timing Explorer}. Astrophys J
  Lett 552:L49--L53. \doi{10.1086/320258}

\bibitem[{Strominger and Zhiboedov(2016)}]{Strom16}
Strominger A, Zhiboedov A (2016) {Gravitational Memory, {BMS} Supertranslations
  and Soft Theorems}. J High Energy Phys 01:086. \doi{10.1007/JHEP01(2016)086}.
  {\href{https://arxiv.org/abs/1411.5745}{{arXiv:1411.5745}}} {[hep-th]}

\bibitem[{Tagoshi and Nakamura(1994)}]{TNaka94}
Tagoshi H, Nakamura T (1994) {Gravitational waves from a point particle in
  circular orbit around a black hole: Logarithmic terms in the post-{Newtonian}
  expansion}. Phys Rev D 49:4016--4022. \doi{10.1103/PhysRevD.49.4016}

\bibitem[{Tagoshi and Sasaki(1994)}]{TSasa94}
Tagoshi H, Sasaki M (1994) {Post-{Newtonian} Expansion of Gravitational Waves
  from a Particle in Circular Orbit around a {Schwarzschild} Black Hole}. Prog
  Theor Phys 92:745--771. \doi{10.1143/ptp.92.745}.
  {\href{https://arxiv.org/abs/gr-qc/9405062}{{gr-qc/9405062}}}

\bibitem[{Tagoshi et~al.(1996)Tagoshi, Shibata, Tanaka, and Sasaki}]{TSTS96}
Tagoshi H, Shibata M, Tanaka T, Sasaki M (1996) {Post-{Newtonian} expansion of
  gravitational waves from a particle in circular orbit around a rotating black
  hole: Up to $O(v^{8})$ beyond the quadrupole formula}. Phys Rev D
  54:1439--1459. \doi{10.1103/PhysRevD.54.1439}

\bibitem[{Tagoshi et~al.(1997)Tagoshi, Mano, and Takasugi}]{TMT97}
Tagoshi H, Mano S, Takasugi E (1997) {Post-{Newtonian} Expansion of
  Gravitational Waves from a Particle in Circular Orbits around a Rotating
  Black Hole}. Prog Theor Phys 98:829. \doi{10.1143/PTP.98.829}.
  {\href{https://arxiv.org/abs/gr-qc/9711072}{{gr-qc/9711072}}}

\bibitem[{Tagoshi et~al.(2001)Tagoshi, Ohashi, and Owen}]{TOO01}
Tagoshi H, Ohashi A, Owen BJ (2001) {Gravitational field and equations of
  motion of spinning compact binaries to 2.5-post-{Newtonian} order}. Phys Rev
  D 63:044006. \doi{10.1103/PhysRevD.63.044006}.
  {\href{https://arxiv.org/abs/gr-qc/0010014}{{gr-qc/0010014}}}

\bibitem[{Tanaka et~al.(1996)Tanaka, Tagoshi, and Sasaki}]{TTS96}
Tanaka T, Tagoshi H, Sasaki M (1996) {Gravitational Waves by a Particle in
  Circular Orbit around a {Schwarzschild} Black Hole: 5.5 Post-{Newtonian}
  Formula}. Prog Theor Phys 96:1087--1101. \doi{10.1143/PTP.96.1087}.
  {\href{https://arxiv.org/abs/gr-qc/9701050}{{gr-qc/9701050}}}

\bibitem[{Taylor(1993)}]{T93}
Taylor JH (1993) {Pulsar timing and relativistic gravity}. Class Quantum Grav
  10:167--174. \doi{10.1088/0264-9381/10/S/017}

\bibitem[{Taylor and Weisberg(1982)}]{TW82}
Taylor JH, Weisberg JM (1982) {A New Test of General Relativity: Gravitational
  Radiation and the Binary Pulsar PSR 1913+16}. Astrophys J 253:908--920.
  \doi{10.1086/159690}

\bibitem[{Taylor et~al.(1979)Taylor, Fowler, and McCulloch}]{TFMc79}
Taylor JH, Fowler LA, McCulloch PM (1979) {Measurements of general relativistic
  effects in the binary pulsar PSR 1913+16}. Nature 277:437--440.
  \doi{10.1038/277437a0}

\bibitem[{Tessmer and Sch{\"{a}}fer(2010)}]{TessS10}
Tessmer M, Sch{\"{a}}fer G (2010) {Full-analytic frequency-domain 1PN-accurate
  gravitational wave forms from eccentric compact binaries}. Phys Rev D
  82:124064. \doi{10.1103/PhysRevD.82.124064}.
  {\href{https://arxiv.org/abs/1006.3714}{{arXiv:1006.3714}}} {[gr-qc]}

\bibitem[{Tessmer and Sch{\"{a}}fer(2011)}]{TessS11}
Tessmer M, Sch{\"{a}}fer G (2011) {Full-analytic frequency-domain gravitational
  wave forms from eccentric compact binaries to 2PN accuracy}. Ann Phys
  (Berlin) 523:813. \doi{10.1002/andp.201100007}.
  {\href{https://arxiv.org/abs/1012.3894}{{arXiv:1012.3894}}} {[gr-qc]}

\bibitem[{Thorne(1980)}]{Th80}
Thorne KS (1980) {Multipole expansions of gravitational radiation}. Rev Mod
  Phys 52:299--339. \doi{10.1103/RevModPhys.52.299}

\bibitem[{Thorne(1983)}]{Thhouches}
Thorne KS (1983) The theory of gravitational radiation: An introductory review.
  In: Deruelle N, Piran T (eds) Gravitational Radiation. North-Holland;
  Elsevier, Amsterdam; New York, pp 1--57

\bibitem[{Thorne(1987)}]{Th300}
Thorne KS (1987) Gravitational radiation. In: Hawking SW, Israel W (eds) Three
  Hundred Years of Gravitation. Cambridge University Press, Cambridge; New
  York, pp 330--458

\bibitem[{Thorne(1992)}]{Th92}
Thorne KS (1992) {Gravitational-wave bursts with memory: The {Christodoulou}
  effect}. Phys Rev D 45:520. \doi{10.1103/PhysRevD.45.520}

\bibitem[{Thorne and Hartle(1985)}]{ThH85}
Thorne KS, Hartle JB (1985) {Laws of motion and precession for black holes and
  other bodies}. Phys Rev D 31:1815--1837. \doi{10.1103/PhysRevD.31.1815}

\bibitem[{Thorne and Kov{\`{a}}cs(1975)}]{ThK75}
Thorne KS, Kov{\`{a}}cs SJ (1975) {Generation of gravitational waves. {I}.
  {W}eak-field sources}. Astrophys J 200:245--262. \doi{10.1086/153783}

\bibitem[{Tichy et~al.(2000)Tichy, Flanagan, and Poisson}]{TFP00}
Tichy W, Flanagan {\'E}{\'E}, Poisson E (2000) {Can the post-{Newtonian}
  gravitational wave form of an inspiraling binary be improved by solving the
  energy balance equation numerically?} Phys Rev D 61:104015.
  \doi{10.1103/PhysRevD.61.104015}.
  {\href{https://arxiv.org/abs/gr-qc/9912075}{{arXiv:gr-qc/9912075}}}

\bibitem[{Tolman(1962)}]{Tolman}
Tolman R (1962) Relativity, Thermodynamics and Cosmology. Clarendon Press,
  Oxford

\bibitem[{Trautman(2002)}]{Traut58}
Trautman A (2002) {Lectures on General Relativity}. Gen Relativ Gravit
  34:721--762. \doi{10.1023/A:1015939926662}, lectures delivered at King's
  College in London in May\,--\,June 1958

\bibitem[{Trestini(2024)}]{Trestini2024}
Trestini D (2024) {Quasi-{Keplerian} parametrization for eccentric compact
  binaries in scalar-tensor theories at second post-{Newtonian} order and
  applications}. arXiv e-prints
  {\href{https://arxiv.org/abs/2401.06844}{{arXiv:2401.06844}}} {[gr-qc]}

\bibitem[{Trestini and Blanchet(2023)}]{TB23}
Trestini D, Blanchet L (2023) {Gravitational-wave tails of memory}. Phys Rev D
  107(10):104048. \doi{10.1103/PhysRevD.107.104048}

\bibitem[{Trestini et~al.(2023)Trestini, Larrouturou, and Blanchet}]{TLB22}
Trestini D, Larrouturou F, Blanchet L (2023) {The quadrupole moment of compact
  binaries to the fourth post-{Newtonian} order: relating the harmonic and
  radiative metrics}. Class Quantum Grav 40(5):055006.
  \doi{10.1088/1361-6382/acb5de}

\bibitem[{Trias and Sintes(2008)}]{TS08}
Trias M, Sintes AM (2008) {LISA observations of supermassive black holes:
  Parameter estimation using full post-{Newtonian} inspiral waveforms}. Phys
  Rev D 77:024030. \doi{10.1103/PhysRevD.77.024030}.
  {\href{https://arxiv.org/abs/0707.4434}{{arXiv:0707.4434}}} {[gr-qc]}

\bibitem[{Tulczyjew(1957)}]{Tulc1}
Tulczyjew W (1957) {On the energy-momentum tensor density for simple pole
  particles}. Bull Acad Polon Sci Cl III 5:279

\bibitem[{Tulczyjew(1959)}]{Tulc2}
Tulczyjew W (1959) {Motion of multipole particles in general relativity
  theory}. Acta Phys Pol 18:37

\bibitem[{Vaidya(2015)}]{Vaidya15}
Vaidya V (2015) {Gravitational spin {Hamiltonians} from the $S$ matrix}. Phys
  Rev D 91:024017. \doi{10.1103/PhysRevD.91.024017}.
  {\href{https://arxiv.org/abs/1410.5348}{{arXiv:1410.5348}}} {[hep-th]}

\bibitem[{{van de Meent}(2017)}]{vdM16}
{van de Meent} M (2017) {Self-force corrections to the periapsis advance around
  a spinning black hole}. Phys Rev Lett 118:011101.
  \doi{10.1103/PhysRevLett.118.011101}.
  {\href{https://arxiv.org/abs/1610.03497}{{arXiv:1610.03497}}} {[gr-qc]}

\bibitem[{Vines et~al.(2011)Vines, Hinderer, and Flanagan}]{VHF11}
Vines J, Hinderer T, Flanagan {\'{E}} (2011) {Post-1-{Newtonian} tidal effects
  in the gravitational waveform from binary inspirals}. Phys Rev D 83:084051.
  \doi{10.1103/PhysRevD.83.084051}.
  {\href{https://arxiv.org/abs/1101.1673}{{arXiv:1101.1673}}} {[gr-qc]}

\bibitem[{Vines and Flanagan(2013)}]{VF13}
Vines JE, Flanagan EE (2013) {First-post-{N}ewtonian quadrupole tidal
  interactions in binary systems}. Phys Rev D 88:024046.
  \doi{10.1103/PhysRevD.88.024046}

\bibitem[{{von Zeipel}(1910)}]{vonZeipel}
{von Zeipel} H (1910) {Sur l'application des s\'eries de {M. Lindstedt} \`a
  l'\'etude du mouvement des com\`etes p\'eriodiques}. Astron Nachr
  183(22-24):345--418. \doi{10.1002/asna.19091832202}

\bibitem[{Wagoner(1975)}]{Wag75}
Wagoner RV (1975) {Test for Existence of Gravitational Radiation}. Astrophys J
  Lett 196:L63--L65. \doi{10.1086/181745}

\bibitem[{Wagoner and Will(1976)}]{WagW76}
Wagoner RV, Will CM (1976) {Post-{Newtonian} gravitational radiation from
  orbiting point masses}. Astrophys J 210:764--775. \doi{10.1086/154886}

\bibitem[{Wald(1973)}]{Wald73}
Wald RM (1973) {On perturbations of a {Kerr} black hole}. J Math Phys
  14:1453--1461. \doi{10.1063/1.1666203}

\bibitem[{Walker and Will(1980)}]{WalkW80}
Walker M, Will CM (1980) {The approximation of radiative effects in
  relativistic gravity: Gravitational radiation reaction and energy loss in
  nearly {Newtonian} systems}. Astrophys J Lett 242:L129--L133.
  \doi{10.1086/183417}

\bibitem[{Warburton et~al.(2021)Warburton, Pound, Wardell, Miller, and
  Durkan}]{WarburtonPound21}
Warburton N, Pound A, Wardell B, Miller J, Durkan L (2021) {Gravitational-wave
  energy flux for compact binaries through second order in the mass ratio}.
  Phys Rev Lett 127(15):151102. \doi{10.1103/PhysRevLett.127.151102}

\bibitem[{Wardell et~al.(2023)Wardell, Pound, Warburton, Miller, Durkan, and
  {Le Tiec}}]{WPWMDL23}
Wardell B, Pound A, Warburton N, Miller J, Durkan L, {Le Tiec} A (2023)
  {Gravitational waveforms for compact binaries from second-order self-force
  theory}. Phys Rev Lett 130(24):241402. \doi{10.1103/PhysRevLett.130.241402}.
  {\href{https://arxiv.org/abs/2112.12265}{{arXiv:2112.12265}}} {[gr-qc]}

\bibitem[{Wen(2003)}]{Wen03}
Wen L (2003) {On the eccentricity distribution of coalescing black hole
  binaries driven by the {Kozai} mechanism in globular clusters}. Astrophys J
  598:419. \doi{10.1086/378794}.
  {\href{https://arxiv.org/abs/astro-ph/0211492}{{astro-ph/0211492}}}

\bibitem[{Westpfahl(1985)}]{Westpf85}
Westpfahl K (1985) {High-speed scattering of charged and uncharged particles in
  general relativity}. Fortschr Physik 33:417. \doi{10.1002/prop.2190330802}

\bibitem[{Westpfahl and Goller(1979)}]{WG79}
Westpfahl K, Goller M (1979) {Gravitational scattering of two relativistic
  particles in post-linear approximation}. Lett Nuovo Cim 26:573.
  \doi{10.1007/BF02817047}

\bibitem[{Westpfahl and Hoyler(1980)}]{WH80}
Westpfahl K, Hoyler H (1980) {Gravitational bremsstrahlung in post-linear
  fast-motion approximation}. Lett Nuovo Cim 27:581. \doi{10.1007/BF02750304}

\bibitem[{Wex(1995)}]{Wex95}
Wex N (1995) {The second post-{Newtonian} motion of compact binary-star systems
  with spin}. Class Quantum Grav 12:983--1005. \doi{10.1088/0264-9381/12/4/009}

\bibitem[{Wiener(1942)}]{Wiener}
Wiener N (1942) {Response of a non-linear device to noise}. Tech. rep., MIT
  Radiation Lab.
  \urlprefix\url{https://apps.dtic.mil/sti/citations/tr/ADA800212}

\bibitem[{Will(1993)}]{W94}
Will CM (1993) Gravitational waves from inspiralling compact binaries: A
  post-{Newtonian} approach. In: Sasaki M (ed) Relativistic Cosmology. NYMSS,
  vol~8. Universal Academy Press, Tokyo, pp 83--98.
  {\href{https://arxiv.org/abs/gr-qc/9403033}{{gr-qc/9403033}}}

\bibitem[{Will(1999)}]{W99}
Will CM (1999) {Generation of post-{Newtonian} gravitational radiation via
  direct integration of the relaxed {Einstein} equations}. Prog Theor Phys
  Suppl 136:158--167. \doi{10.1143/PTPS.136.158}.
  {\href{https://arxiv.org/abs/gr-qc/9910057}{{gr-qc/9910057}}}

\bibitem[{Will(2005)}]{W05}
Will CM (2005) {Post-{Newtonian} gravitational radiation and equations of
  motion via direct integration of the relaxed {Einstein} equations. {III}.
  {R}adiation reaction for binary systems with spinning bodies}. Phys Rev D
  71:084027. \doi{10.1103/PhysRevD.71.084027}.
  {\href{https://arxiv.org/abs/gr-qc/0502039}{{gr-qc/0502039}}}

\bibitem[{Will and Wiseman(1996)}]{WW96}
Will CM, Wiseman AG (1996) {Gravitational radiation from compact binary
  systems: Gravitational waveforms and energy loss to second post-{Newtonian}
  order}. Phys Rev D 54:4813--4848. \doi{10.1103/PhysRevD.54.4813}.
  {\href{https://arxiv.org/abs/gr-qc/9608012}{{gr-qc/9608012}}}

\bibitem[{Wilson and Matthews(1989)}]{WM80}
Wilson J, Matthews G (1989) Relativistic hydrodynamics. In: Evans C, Finn L,
  Hobill D (eds) Frontiers in Numerical Relativity. Cambridge University Press,
  p 306

\bibitem[{Wiseman(1992)}]{Wi92}
Wiseman AG (1992) {Coalescing binary systems of compact objects to
  (post)$^{5/2}$-{Newtonian} order. {II}. {H}igher-order wave forms and
  radiation recoil}. Phys Rev D 46:1517--1539. \doi{10.1103/PhysRevD.46.1517}

\bibitem[{Wiseman(1993)}]{Wi93}
Wiseman AG (1993) {Coalescing binary systems of compact objects to
  (post)$^{5/2}$-{Newtonian} order. {IV}. {T}he gravitational wave tail}. Phys
  Rev D 48:4757--4770. \doi{10.1103/PhysRevD.48.4757}

\bibitem[{Wiseman and Will(1991)}]{WW91}
Wiseman AG, Will CM (1991) {Christodoulou's nonlinear gravitational-wave
  memory: Evaluation in the quadrupole approximation}. Phys Rev D
  44:R2945--R2949. \doi{10.1103/PhysRevD.44.R2945}

\bibitem[{{Zel'dovich} and Polnarev(1974)}]{Zeldovich74}
{Zel'dovich} YB, Polnarev A (1974) {Radiation of gravitational waves by a
  cluster of superdense stars}. Sov Astron 18:17

\bibitem[{Zeng and Will(2007)}]{ZW07}
Zeng J, Will CM (2007) {Application of energy and angular momentum balance to
  gravitational radiation reaction for binary systems with spin-orbit
  coupling}. Gen Relativ Gravit 39:1661. \doi{10.1007/s10714-007-0475-6}.
  {\href{https://arxiv.org/abs/0704.2720}{{arXiv:0704.2720}}}

\end{thebibliography}

\end{document}